\documentclass[12pt,a4]{report}
\large
\usepackage{epsfig}
\usepackage{colortbl}
\usepackage{textcomp}
\usepackage{mathrsfs}
\usepackage{abstract}
\usepackage[small, bf]{caption} 

\newcommand{\ETAL}{{\it et al.}}
\newcommand{\PT}{$p_{T}$~}

\newcommand{\PM}{$\pm$~}
\newcommand{\DNDY}{$d\sigma/dy$~}
\newcommand{\GEVC}{GeV/$c$~}

\title{Production of charm and bottom quarks in $p+p$ collisions at 200GeV}
\author{Yuhei Morino\\
\\
Center for Nuclear Study\\
Graduate School of Sience\\
The University of Tokyo\\
\\
A Dissertation Submitted in Partial Fulllment of\\
the Requirements for the Degree of Doctor of Science}
\date{}

\begin{document}
\pagenumbering{roman}
  \maketitle	
 \begin{abstract}

The measurement of single electrons from semi-leptonic decay of heavy flavor 
in $p+p$ collisions at $\sqrt{s}$=200~GeV has been carried out with the 
PHENIX detector in the RHIC Year-2005 and Year-2006 run.
Measured single electrons include contribution from both charm and bottom.
This measurement provides a good test of the perturbative QCD.
The separate measurement of $p_{\rm{T}}$ distribution about single electrons from bottom
is important for a precise test of not only perturbative QCD but also fragmentation process.

The measurement of heavy flavor in $p+p$ collisions also provides an important base line to interpret
the measurement of heavy flavor in Au+Au collisions at RHIC, since heavy flavor is only produced in
the initial collisions. 
A strong suppression of the single electrons from heavy flavor at high $p_{\mathrm{T}}$ 
has been observed in central Au+Au collisions compared with $p+p$ collisions.
The suppression indicates charm quarks lose a large fraction of their energy in the hot and dense medium,
since the magnitude of energy loss of bottom in the medium is expected to be much smaller
than that of charm due to the large difference of their masses.
The contribution from bottom in the single electrons from heavy flavor must be determined for the discussion of
the energy loss of bottom.

The first separate measurement of charm and bottom via a new method, partial reconstruction 
of $D/\bar{D} \rightarrow e^{\pm} K^{\mp} X$ decay, has been also carried out in $p+p$ collisions 
with the PHENIX detector  in the RHIC Year-2005 and Year-2006 run. 
The measured contribution from bottom in the single electrons provides
more precise test of perturbative QCD. 
It also provides the important base line to discuss at the behavior
of bottom quarks in the medium created in Au+Au collisions.

It is found that there is the considerable contribution from bottom in the single
electrons from heavy flavor above 3~GeV/$c$.
The first separate spectra of the single electrons from charm and bottom are measured 
based on the fraction of bottom.
The $p_{\rm{T}}$ distribution of the single electrons predicted in pQCD agrees with the measured spectra 
within its uncertainty and 
the ratio, (data/pQCD) is $\sim$ 2 for charm production and  $\sim$ 1 for bottom production.
The total cross section of bottom is also determined 
to  $\sigma_{b\bar{b}}= 3.16 ^{+1.19}_{-1.07}({\rm stat}) {}^{+1.37}_{-1.27}({\rm sys}) \mu {\rm b}$.
The existence of energy loss of bottom quarks in the medium created in Au+Au collisions is found
based on the measured $(b\rightarrow e)/(c\rightarrow e+ b\rightarrow e)$ in $p+p$ collisions.
\end{abstract}


  \tableofcontents
  \listoffigures
  \addcontentsline{toc}{chapter}{List of Figures}
  \listoftables
  \addcontentsline{toc}{chapter}{List of Tables}
  \chapter{Introduction}
\pagenumbering{arabic}
\section{Quantum Chromodynamics}
Quantum ChromoDynamics (QCD) is the theory of strong interactions between the quarks and gluons.
QCD was developed as an extension of Quantum Electrodynamics~(QED) via the imposition of a 
local SU(3) symmetry in 'color' space.
The most important difference between QCD and QED is that QCD is the non-abelian gauge theory
and as a consequence has gluon self-interaction.
This nature of QCD leads 'asymptotic freedom' which is the most important feature of QCD.
The strong coupling constant, $\alpha_s$, can be expressed as a function of the momentum transfer, 
$Q^2$ as follows~\cite{bib:qcd1}.
\begin{equation}
\alpha_s (Q^2) \sim \frac{12\pi}{(33-2N_f) \ln(Q^2/\lambda_{QCD}^2)},
\end{equation}
\noindent where  $N_f$ is the number of quark flavors and $\lambda_{QCD} \sim 0.2$~GeV  is the typical 
QCD scale.
When the momentum transfer $Q^2$ is much larger than $\lambda_{QCD}$, 
$\alpha_s$ becomes small enough to allow us to use the perturbative method for QCD calculation~(pQCD) 
as is the case in QED. 
On the other hand, when the momentum transfer $Q^2$ is not large, QCD is in non-perturbative 
regime and many approaches have been proposed to compute the non-perturbative effect.

Another important feature of QCD is 'confinement hypothese' that all observable states are 
color-singlets, implying directly the non-existence of colored free quarks and states.
This hypothese is based on that particles with color content have never been observed.

\section{Quark Gluon Plasma}
The environment of extremely high temperature and/or density can also reduce $\alpha_s$.
The color confinement may be broken with increase of the temperature 
and/or density of a many body system consisted of hadrons.
This results in a phase transition 
from the confined nuclear matter~(ordered phase) to the deconfined state~(disordered phase). 
The deconfined state is called 'Quark Gluon Plasma'~(QGP)~\cite{bib:qgp1}.

The existence and the thermodynamical properties of the QGP have been studied for long time using
phenomenological models~(Bag model, Hagedorn gas and so on)~\cite{bib:qgp2,bib:qgp3}.
Recently, the lattice QCD calculation, which is numerical approach based on the first principle, 
show that a phase transition is realized from confined nuclear matter to the QGP at extreme high temperature
$T_c \sim 170$~MeV and high energy density 
$\epsilon_c \sim 1$~GeV/fm$^3$~\cite{bib:lat1,bib:lat2,bib:lat3}.
\begin{figure}[htb]
  \begin{center}
    \includegraphics[width=14cm]{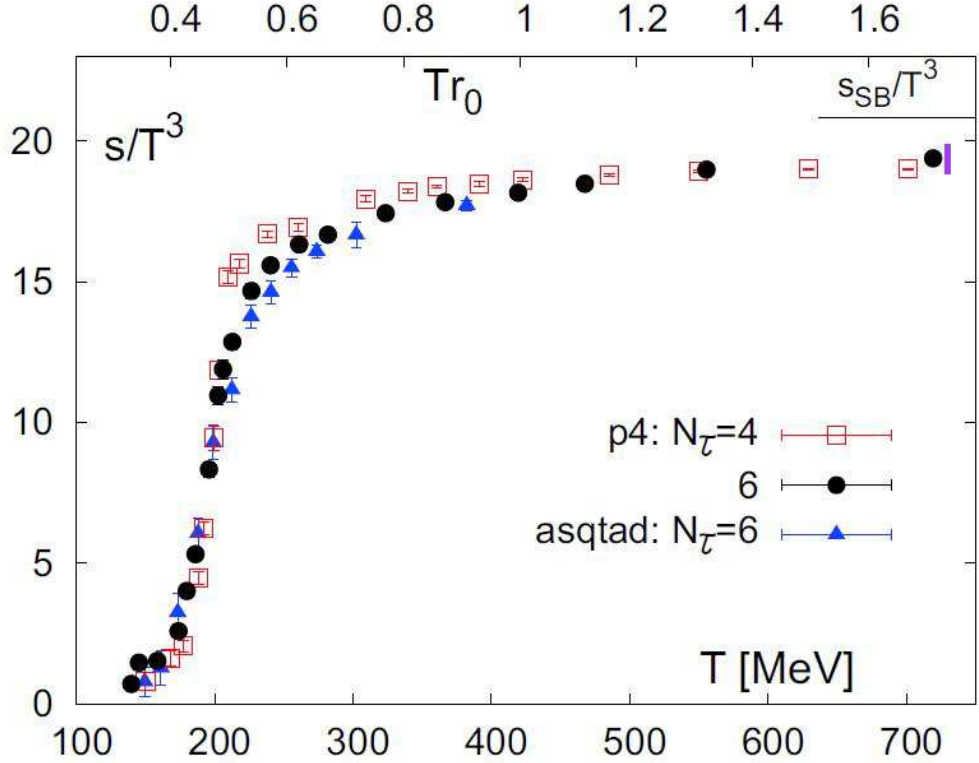}
    \caption{The entropy density~($s=\epsilon+p$) in units of $T^3$ as a function of
      $T$ calculated by lattice QCD~\cite{bib:lat3}.
    }
    \label{fig:chap1_lat}
  \end{center}
\end{figure}
Figure~\ref{fig:chap1_lat} shows the entropy density~($s=\epsilon+p$) in units of $T^3$ as a function of
$T$ calculated by lattice QCD~\cite{bib:lat3}.
This calculation indicates that the entropy density increases rapidly around the critical temperature
$T_c \sim 170$~MeV due to the increase of the degree of freedom, which is associated by the deconfinement 
of the matter.

\begin{figure}[htb]
  \begin{center}
    \includegraphics[width=14cm]{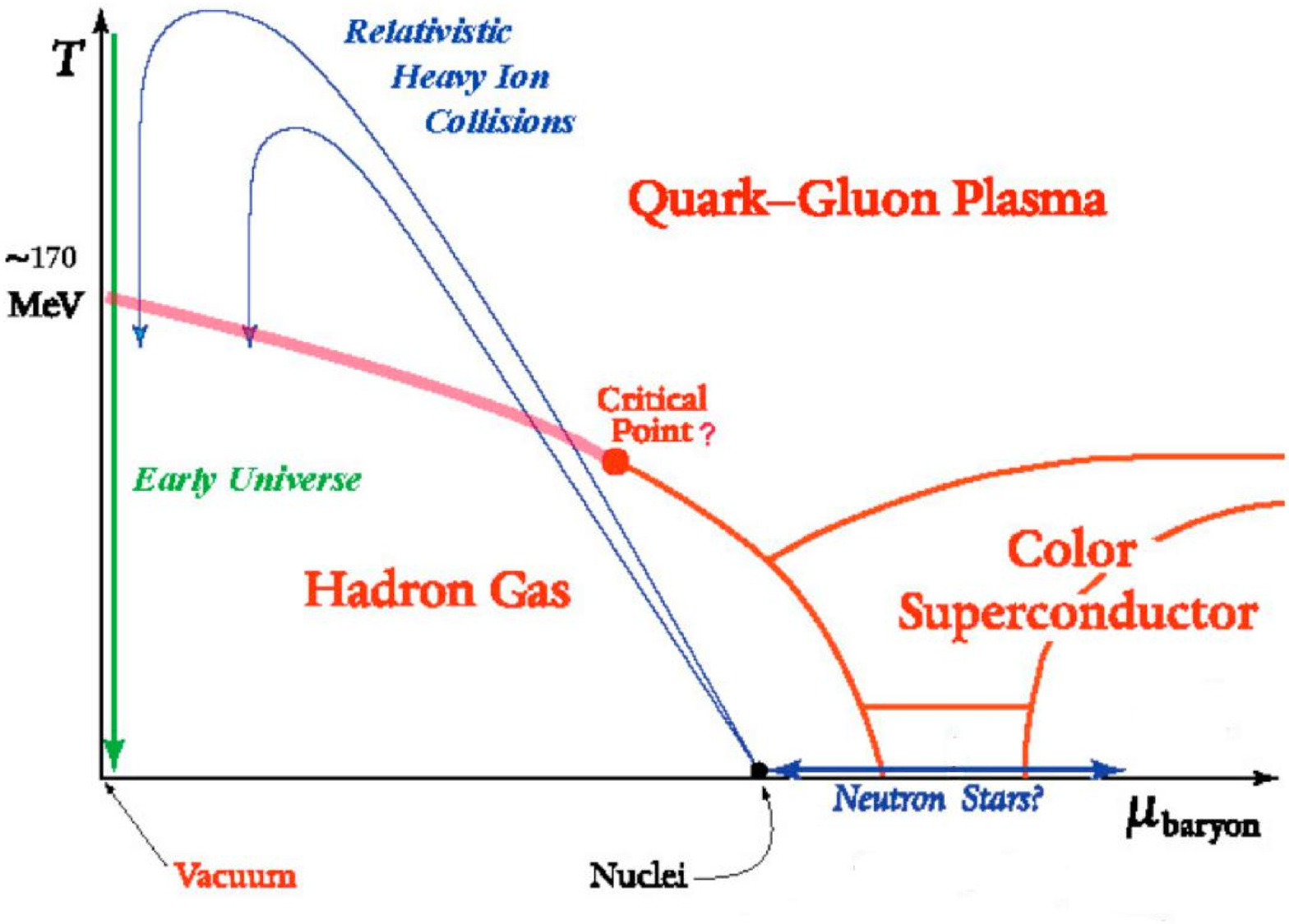}
    \caption{A predicted schematic phase diagram  of hadronic matter including 
      QGP~\cite{bib:phase1,bib:phase2}.
    }
    \label{fig:chap1_qcd}
  \end{center}
\end{figure}

Figure~\ref{fig:chap1_qcd} shows a schematic  picture of the expected phase diagram of hadronic matter 
including QGP~\cite{bib:phase1,bib:phase2}.
The horizontal axis is the baryon chemical potential, $\mu_{baryon}$ and the vertical axis is the temperature. 

\section{High Energy Heavy Ion Collisions}
High energy heavy ion collision is a powerful and unique tool to realize high energy temperature enough to
create QGP~\cite{bib:lee1,bib:lee2}.
Fixed target experiments with high-energy heavy-ion collisions began at Bevalac at
Lawrence Berkeley with $\sim$ 2A GeV beams in the middle of 1970's. 
The Relativistic Heavy Ion Collider (RHIC) at the Brookhaven National Laboratory~(BNL) is the first 
colliding-type accelerator which can collide heavy nuclei up to gold (197Au) with the center of mass energy 
per nucleon pair of $\sqrt{s_{NN}}$ = 200 GeV and started its operation in 2000. The energy density 
achieved by the collisions at RHIC is expected to be well above the critical temperature.

The most important feature of high energy heavy ion collisions is that 
the matter created in heavy ion collisions undergoes space-time evolution.
Particles are produced at each stage in the space-time evolution and interact with the matter.
Since all experimentally observables are particles after the space-time evolution, the integrated information 
of the interaction with the matter is only measured.
This fact leads that it is important for the study of the QGP via high energy heavy ion collisions to know 
when the measured particles are produced.
Remarkable features of high energy heavy ion collisions are summarized in Sec.~\ref{sec:rhic} and 
important results measured in RHIC are  summarized in Sec.~\ref{sec:rrhic}.

\section{Heavy Flavor Produciton}
Heavy quarks~(charm, bottom) are primarily produced in hard partonic 
scattering in nucleon-nucleon collisions, since the initial content of 
heavy flavor in nuclei is negligibly small.
The energy scale for the production of heavy quarks~(charm and bottom)
is significantly higher than the typical QCD scale, $\lambda_{QCD}\sim 0.2$~GeV. 
This gives us a coupling constant of the order of $\alpha_s \sim 0.3$, which is 
small enough to apply pQCD calculation for the production of heavy quarks. 
Measurement of charm and bottom production in $p+p$ collisions provides 
a good test of the perturbative QCD calculations.
The $p_{\rm T}$ distribution of bottomed hadrons in perturbative QCD calculations becomes compatible with 
the results at Tevatron because of the improvement of the fragmentation process recently.
Therefore, measurement of $p_{\rm T}$ distribution of bottomed hadrons at RHIC also provides 
the important cross check for pQCD due to the large mass and a test of theoretical treatments of fragmentation process. 

Heavy quarks are believed to be a special probe of the medium created in 
heavy ion collisions due to their large mass.
Since heavy quarks are only produced in the initial stage in the heavy ion collision,
heavy quark spectra in $p+p$ collisions provide a well defined initial state, even for
low-momentum heavy quarks. 
Then, generated heavy quarks propagate through the hot and dense medium  created in heavy ion collisions and 
their propagation probes characteristics of the medium.

Experiments at the RHIC has measured single electrons from semi-leptonic decay of heavy flavor 
at mid-rapidity~($\mid\eta\mid<0.35$) in $p+p$ and Au+Au collisions at $\sqrt{s_{NN}} = 200$~GeV.
A strong suppression at high $p_{\mathrm{T}}$ and azimuthal anisotropy of the single electrons have 
been observed in central Au+Au collisions~\cite{bib:hq1}.
Measured single electrons include contribution from both charm and bottom.
The magnitude of energy loss of bottom in the hot and dense medium is expected to be much smaller
than that of charm due to the large difference of their masses~\cite{bib:bdmp}.
In addition, since the thermalization time of bottom should be larger than that of charm, 
the magnitude of flow of bottom is expected to be much smaller than that of charm.
This fact indicates charm quarks lose a large fraction of their energy and flow in the matter created in Au+Au collisions.
On the other hands, the existence of bottom modification~(energy loss and flow) in the medium 
has been an open question without the determination of $(b\rightarrow e)/(c\rightarrow e+ b\rightarrow e)$.
Details of production of heavy flavor are summarized in Sec.~\ref{sec:heavy}.

\section{Organization of This Thesis}
Measurement of the fraction of bottom in single electrons from heavy flavor  
is crucial in $p+p$ collisions at RHIC for the pQCD test and the interpretation of results of single electrons 
in Au+Au collisions.
For this purpose, production of charm and bottom quarks in $p+p$ collisions at $\sqrt{s}=200$GeV at RHIC Year-2005 RUN
and Year-2006 RUN has been studied via electrons from their semi-leptonic decay using PHENIX detectors at RHIC. 
At first, the $p_{\mathrm{T}}$ distribution of single electrons from heavy flavor in $p+p$ collisions is measured.
In addition, a new analysis method using electron-hadron correlation is established to determine the fraction 
of bottom in the single electrons in this thesis.
The new method provides the first result of the $p_{\rm T}$ distribution of single electrons from bottom quarks at RHIC.

The organization of this thesis is as follows. \\
Chapter 2 introduces the theoretical
and experimental background for the heavy ion collisions and heavy flavor production at RHIC. 
The motivation to separate the contribution of charm and bottom quarks in electron measurement 
in $p+p$ collisions is also described here.\\
In chapter 3, the RHIC accelerator complex and the PHENIX detectors are described. \\
In chapter 4, the conditions of beam and trigger in the p + p runs in Year-2005 and Year-2006 are summarized. \\
In chapter 5, the analysis of measurement of the electrons from semi-leptonic decay of charm and bottom 
is explained. The analysis to separate the contribution of charm and bottom quarks in electron is also described here.\\
In chapter 6, the result of measurement of charm and bottom is shown. \\
Interpretations of the results are discussed in chapter 7. \\
Chapter 8 is the conclusion of  this thesis.

  \newlength{\minitwocolumn}
\setlength{\minitwocolumn}{0.5\textwidth}
\addtolength{\minitwocolumn}{0.1\columnsep}
\chapter{Theoretical and Experimental Background}
In this chapter, theoretical and experimental approaches of relativistic heavy ion collision  are summarized.
The important feature and description of relativistic heavy ion collision are introduced in Sec.~\ref{sec:rhic}.
Current experimental results about 'jet quenching' and 'azimuthal anisotropy' measured in RHIC, 
which are relevant to the motivation of this this thesis, are summarized in Sec.~\ref{sec:rrhic}.
Theoretical and experimental approaches of production of heavy quarks in nucleus-nucleus collisions
are described in Sec.~\ref{sec:heavy}.

\section{Relativistic Heavy Ion Collision} \label{sec:rhic}
Relativistic heavy ion collisions provide a unique method to realize a high-temperature 
and high-density state which is required for QGP formation.
Since there are many nucleons in a heavy nucleus, many nucleon-nucleon collisions 
are involved in heavy ion collisions.
Longitudinal energies of the colliding nuclei are dissipated by the collisions
and a huge amount of energy are released into a tiny colliding region. Then, the matter which has
energy density large enough to form QGP will be created.

\subsection{Collision Geometry}
The number of hard scattering and nuclei participating in the collisions is largely determined by the collision 
geometry.
The geometrical aspects of high-energy heavy-ion collisions play an important role in 
collision dynamics. 
\subsubsection{The Participant-Spectator Model}
\begin{figure}[htb]
  \begin{center}
    \epsfig{figure=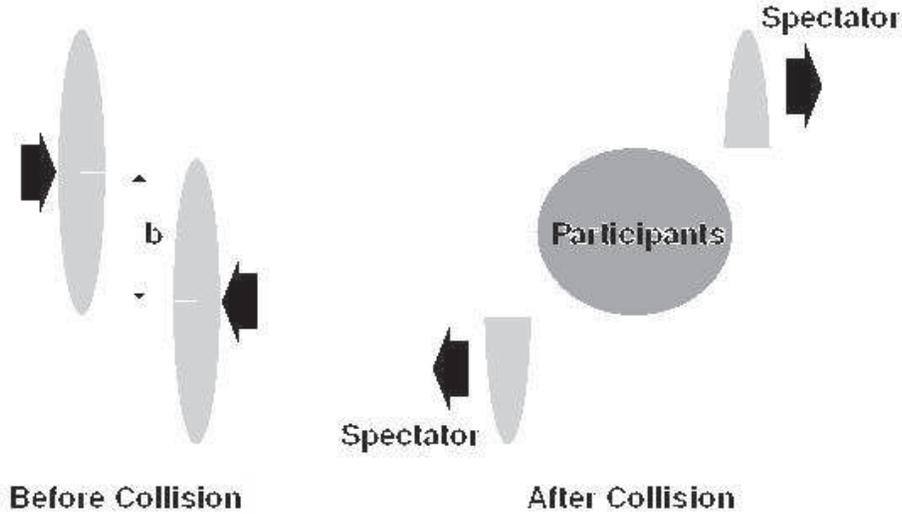,width=12cm}
    \caption { The sketch  of the  colliding nuclei before and after collision.
      They approach each other with impact parameter $b$ before collision.
      After the  collision, the system consists of two components:
      participants and spectators.}
    \label{fig:impact_parameter}
  \end{center}
\end{figure}

\begin{figure}[htb]
  \begin{center}
    \includegraphics[angle=0,width=12cm]{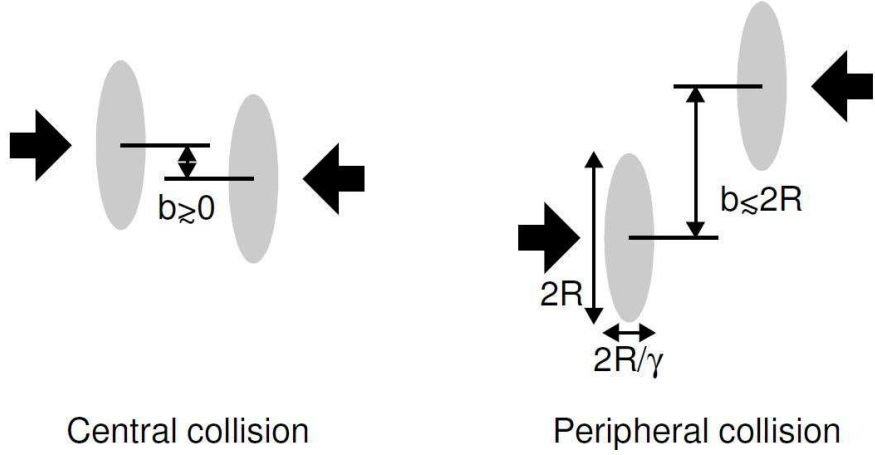}
    \caption { A cartoon of central (left) and peripheral (right) collisions.}
    \label{fig:chap1_cent}
  \end{center}
\end{figure}

The participant-spectator model is a simple geometrical picture which describes the collision characteristics using the
transverse distance between the colliding nuclei, impact parameter $b$. As schematically sketched in 
Figure~\ref{fig:impact_parameter}, the colliding  nuclei looks like thin pancakes in the center of mass frame because 
of the Lorentz contraction. 
Only the overlapping region of nuclei participate in  the collision.
The nucleons in this region are called participants. The nucleons in other region of nuclei, which are called  spectators or
projectile  fragments, do not participate in the collision and pass through  the collision region with the  same velocity as beam.
The spectators are unstable and evaporate the nucleons which also have the almost  same velocity as beam.  
Figure~\ref{fig:chap1_cent} illustrates a central collision and a peripheral collision of nuclei with radius of the nucleus R.
Information about the impact parameter $b$ is obtained by measurement of the observables related to the sizes of 
the spectators and/or the participants.

\subsection{The Glauber Model}    \label{sec:Glauber}
The Glauber Model describes the heavy ion collisions based on the Participant-Spectator Model, the nuclear density distribution, 
the geometry of the colliding the nucleus and the interaction between constituent nucleons~\cite{bib:Glauber,bib:glau}.
This model provides the estimation of  total inelastic cross section of collisions of nucleus A and nucleus B, starting from
nucleon-nucleon inelastic cross section $\sigma_{NN}$.
From the Glauber Model, the number of nucleon-nucleon collisions occurred inside the participant 
region~($N_{coll}$) and the number of participants ($N_{part}$) 
in a collision with impact parameter $b$ are also obtained.
This picture of collisions is useful to study the scaling properties of particle production in heavy ion collisions. 
The coherent interaction of each nucleus scales $N_{part}$, where the momentum
transfer, $Q^2$, is small. They are called soft process.
On the other hand, $N_{coll}$ scale is applicable when $Q^2$ is large and interaction  can be considered as the incoherent sum of 
nucleon-nucleons  binary collision.
They are called hard process.

The nucleons in each colliding nucleus are distributed according to the Woods-Saxon distribution.
\begin{equation}
  \rho(r) = \rho_{0} \cdot \frac{1}{1+{\rm exp}(\frac{r-R}{a})},
  \label{eq:wood-saxon}
\end{equation}
\noindent where $\rho_0$ stands for the normal nuclear density, $R$ is the radius, and $a$ is diffuseness parameter. 
In case of Au nucleus,  $R\simeq$ 6.64~fm and a $\simeq$ 0.53~fm. The probability for occurrence of nucleon-nucleon collision 
between the nucleus  A and B along $z$-axis at an impact parameter $b$ is expressed in the integral form.
\begin{equation}
  T(b)\sigma_{NN} = \int \rho^{z}_{A}(b_A)db_A
  \rho^{z}_{B}(b_B)db_B
  t(b-b_A+b_B)\sigma_{NN},
\end{equation}
\noindent where $ \rho^{z}_{A}(b_A)$ and  $\rho^{z}_{B}(b_B)$ are the $z$-integrated densities of nucleus  A and B, 
$t(b)db$ is the probability for having a nucleon-nucleon collisions within the transverse element $db$ when A and B collide with an 
impact parameter $b$.
Up to A$\times$B collisions can be occurred. The probability having $n$ nucleon-nucleon collisions can be written
using the binomial relation.
\begin{equation}
  P(n,b) = \left(
  \begin{array}{c}
    AB\\
    n
  \end{array}
  \right)
  (1-s)^n(s)^{AB-n},
\end{equation}
\noindent where $s= 1- T(b)\sigma_{NN}$. The probability for having at least one nucleon-nucleon
collision in the collision of nucleus A and B at impact parameter b is

\begin{equation}
  \frac{d\sigma_{AB}}{db} = \sum^{AB}_{n=1}P(n,b) = 
  \sum^{AB}_{n=0}P(n,b) - P(0,b) =  1- s^{AB}.
\end{equation}

\noindent
The total inelastic  cross section $\sigma_{AB}$ can be written as

\begin{equation}
  \sigma_{AB} = 2 \pi \int bdb(1-s^{AB}).
\end{equation}

The average number of inelastic nucleon-nucleon collisions $N_{coll}$ at the impact parameter
b is expressed as follows.
\begin{equation}
N_{coll}(b) = <n(b)> = \sum^{AB}_{n=1} np(n,b) = ABT(b)\sigma_{NN}.
\end{equation}
The results of Glauber calculation are summarized at Appendix~\ref{sec:gtable}.

\subsection{Space Time Evolution of the Matter}
\begin{figure}[htb]
  \begin{center}
    \includegraphics[angle=0,width=15cm]{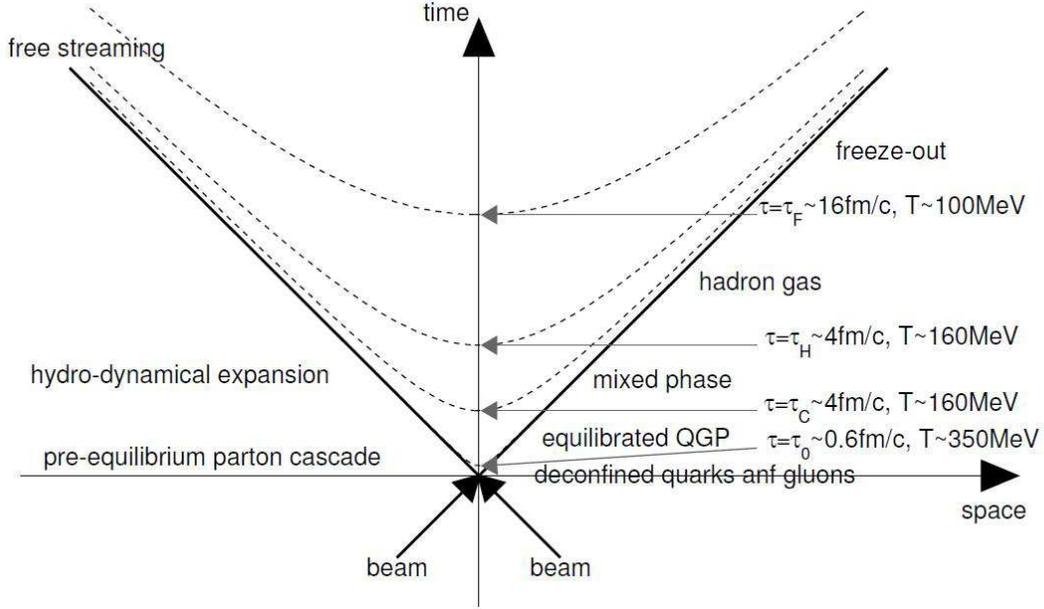}
    \caption { A space-time picture of a nucleus-nucleus collision. The times and temperatures
    Mixed phase would exist only if the transition is first order.}
    \label{fig:chap2_space}
  \end{center}
\end{figure}
The matter created in heavy ion collisions undergoes space-time evolution.
The evolution of the matter can be described based on the Bjorken's space-time 
scenario~\cite{bib:Bjorken}. In the cylindrically symmetric collision at center 
of mass frame, the longitudinal proper time, $\tau$,  is expressed as:
\begin{equation}
  \tau =\sqrt{t^{2}-z^{2}}.
\end{equation}
\noindent 
The evolution of the system is characterized with proper time $\tau$.
Figure~\ref{fig:chap2_space} shows the diagram of the space-time evolution, 
which is according to the picture established by Bjorken~\cite{bib:Bjorken}.
It is assumed that the space-time evolution depends on only the proper time in the 
high-energy limit.
The space-time evolution can be separated into 4 region as shown in Fig.~\ref{fig:chap2_space}
\begin{itemize}
\item{ pre-equilibrium~($\sim \tau_0$)\\
         A huge amount of energy is released in a tiny colliding volume. Free partons, 
	 mainly gluons, are produced by a collision between the two nuclei. The
	 system is initially not in thermal equilibrium.  
	 The subsequent multiple parton scattering brings the
	 matter to local equilibrium.}
\item{Deconfined state~(QGP) in thermal equilibrium~($\tau_0 \sim \tau_C$)\\
  If the deposited energy is large enough and exceeds the critical energy density, the
  QGP will be formed at $\tau_0$.
  The matter created at RHIC is expected to expand and cool down according to hydrodynamics.
}

\item{Mixed phase~($\tau_C \sim \tau_H$) \\
  When the matter reaches the critical temperature $T_c$ between QGP and ordinary hadronic matter, 
  QGP begins to hadronize and the matter becomes the mixed phase consisting of the quarks, gluons 
  and hadrons.
  This state would exist only if the transitions is first order.
}
\item{Hadron gas~($\tau_H \sim \tau_F$) \\
  The system finishes hadronization and produced hadrons keep interacting with each other until 
  the temperature drops to the kinematical freeze-out temperature~($\tau_F$).
  Hadrons cease to interact and move away after $T_F$.
}
\end{itemize}
What we really want to know is the information of QGP, while what we can observe is the integrated 
information from 0 to $\tau_F$.
Therefore, the understanding of each stage via various measurements is important in heavy ion 
collisions.

\subsection{Initial Energy Density}
Bjorken has provided a way to estimate the energy density in a collision system based on the experimental observables: the 
multiplicity of particles and the transverse energy~\cite{bib:Bjorken}. 
We take $\mathscr{A}$ as a transverse overlapping area in the collision of the two nuclei and $\Delta z$ as a  longitudinal 
length of overlapping region. Then the  colliding volume is expressed as  $\mathscr{A}\Delta z$.
Taking $\Delta N$ as a number of particles in this volume, following
relation is derived.

\begin{equation}
  \frac{\Delta N}{\mathscr{A}\Delta z} = \frac{1}{\mathscr{A}}\frac{dN}{dy}\frac{dy}{dz}
  = \frac{1}{\mathscr{A}}\frac{dN}{dy}\frac{1}{\tau_0 {\rm cosh}y},
  \label{eq:energy1}
\end{equation}
The energy of a particle with a rapidity $y$ is $m_T \cosh y$, with  $m_{T} = \sqrt{m_{0}^{2} + p_{T}^{2}}$. 
Therefore, the initial energy density is expressed as,
\begin{equation}
  \epsilon_0 = m_T \cosh y \frac{\Delta N}{\mathscr{A}\Delta z}.
  \label{eq:energy2}
\end{equation}
\noindent Making use of Eq.~\ref{eq:energy1} and Eq.~\ref{eq:energy2}, the energy density $\epsilon_{0}$ at mid-rapidity 
region is expressed as,  
  \begin{equation}	
     \left. \epsilon_{0} = \frac{m_{T}}{\tau_{0}\mathscr{A} } \frac{dN}{dy} \right|_{y=0},
     \end{equation}	
  At RHIC, the energy density reaches to $\sim$5.5~GeV/fm$^{3}$ in Au+Au collisions~\cite{bib:phet}.

\clearpage

\section{Experimental Results at RHIC}  \label{sec:rrhic}
Experiments at the Relativistic Heavy-Ion Collider~(RHIC) have indeed provided
convincing evidence that a thermalized medium is produced in heavy ion collisions at $\sqrt{s_{NN}}$ = 200 GeV.
In this section, a brief summary of the important observations and pertinent interpretations which is relevant to the motivation
of this thesis is presented~\cite{bib:wphenix,bib:wstar,bib:wphobos,bib:wbrah}.
At RHIC, many hadron $p_{\rm{T}}$ spectra and its azimuthal angle dependence with respect to reaction plane have 
been measured. 
Two major findings at RHIC may be classified by their $p_{\rm{T}}$ regime.
\begin{itemize}
\item{Azimuthal Anisotropy~($p_{\rm{T}}<5$GeV/$c$)}
\item{Jet Quenching~($p_{\rm{T}}>5$GeV/$c$)}
\end{itemize}

\subsection{Initial Nuclear Effect}\label{sec:ini}
There are known normal nuclear effects~(initial state effect) which  modify the yield and $p_{{\rm T}}$ distribution 
of produced particles.
When one wants to extract the information of the matter created with heavy ion collisions, these nuclear effects should be
taken into account.
\subsubsection{Cronin Effect}
Incident partons suffer multiple scatterings while passing through the nucleus~(A) before the hard collision. 
Partons from the projectile nucleus collide with various target nucleons exchanging a transverse momentum in each collision. 
As a result, the $p_{{\rm T}}$ distribution of partons becomes wider compared to that in p + p collisions and is known 
as the Cronin effect~\cite{bib:cronin}.
The $p_{{\rm T}}$ distribution in p+A collisions is parameterized as
\begin{equation}
E\frac{d^3\sigma}{dp^3}(p_{{\rm T}},A) = E\frac{d^3\sigma}{dp^3}(p_{{\rm T}},p)\times A^{\alpha(p_{{\rm T}})}.
\end{equation}
$\alpha(p_{{\rm T}})$ becomes greater than 1 for the $p_{{\rm T}}$ region of $p_{{\rm T}} > \sim 1$~GeV/$c$ in
$\sqrt{s}=200$~GeV collisions.
\subsubsection{Nuclear (Anti-)Shadowing}
\begin{figure}[htb]
  \begin{center}
    \includegraphics[angle=0,width=10cm]{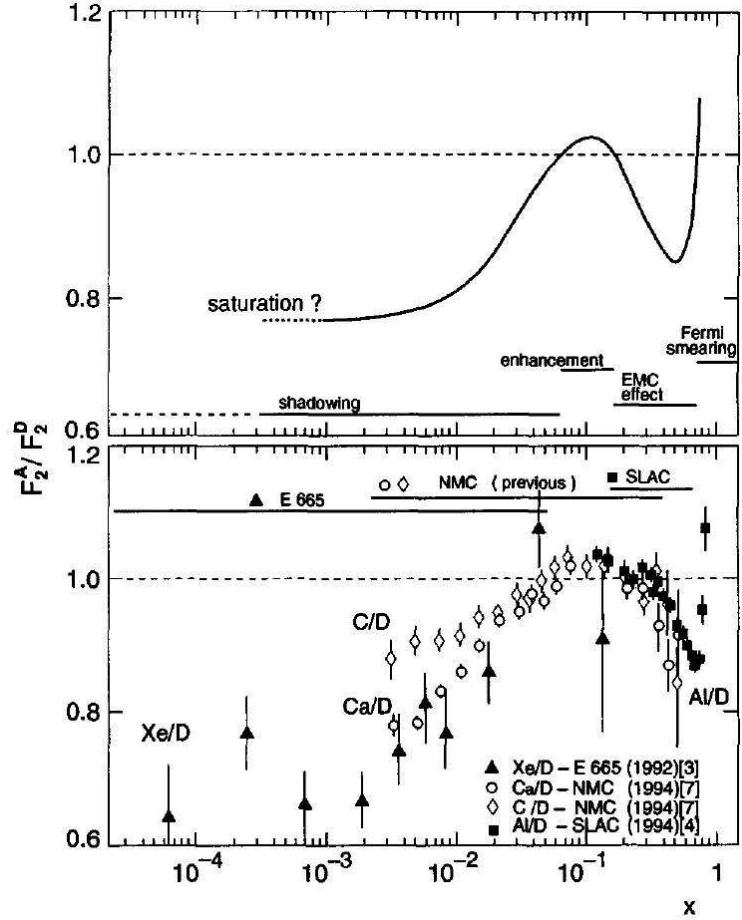}
    \caption {Top:A phenomenological curve of the ratio of the parton structure. Bottom:the ratio of the parton structure 
      functions~($F_2^A(x,Q^2)/F_2^D(x,Q^2)$) for different nuclei~\cite{bib:nshadow2}.}
    \label{fig:chap2_shadow}
  \end{center}
\end{figure}

It was found by the EMC group in $\mu$+Fe scattering that parton distribution
in free protons is modified when partons are bound in the nucleus~\cite{bib:nshadow1}.
Modification of parton distribution affects the yield of the particles.
For momentum fractions $x < 0.1$ and $0.3 < x < 0.7$~(called as EMC region), a depletion is observed in the nuclear parton 
distributions. Momentum fractions at mid-rapidity can be expressed as 
the low x, or shadowing region and the larger x, or EMC region, is bridged by an enhancement
known as anti-shadowing for $0.1 < x < 0.3$. Figure~\ref{fig:chap2_shadow} shows the ratio of the parton structure 
functions~($F_2^A(x,Q^2)/F_2^D(x,Q^2)$) for different nuclei~\cite{bib:nshadow2}.

\subsection{Jet Quenching}
\begin{figure}[htb]
  \begin{center}
    \includegraphics[angle=0,width=14cm]{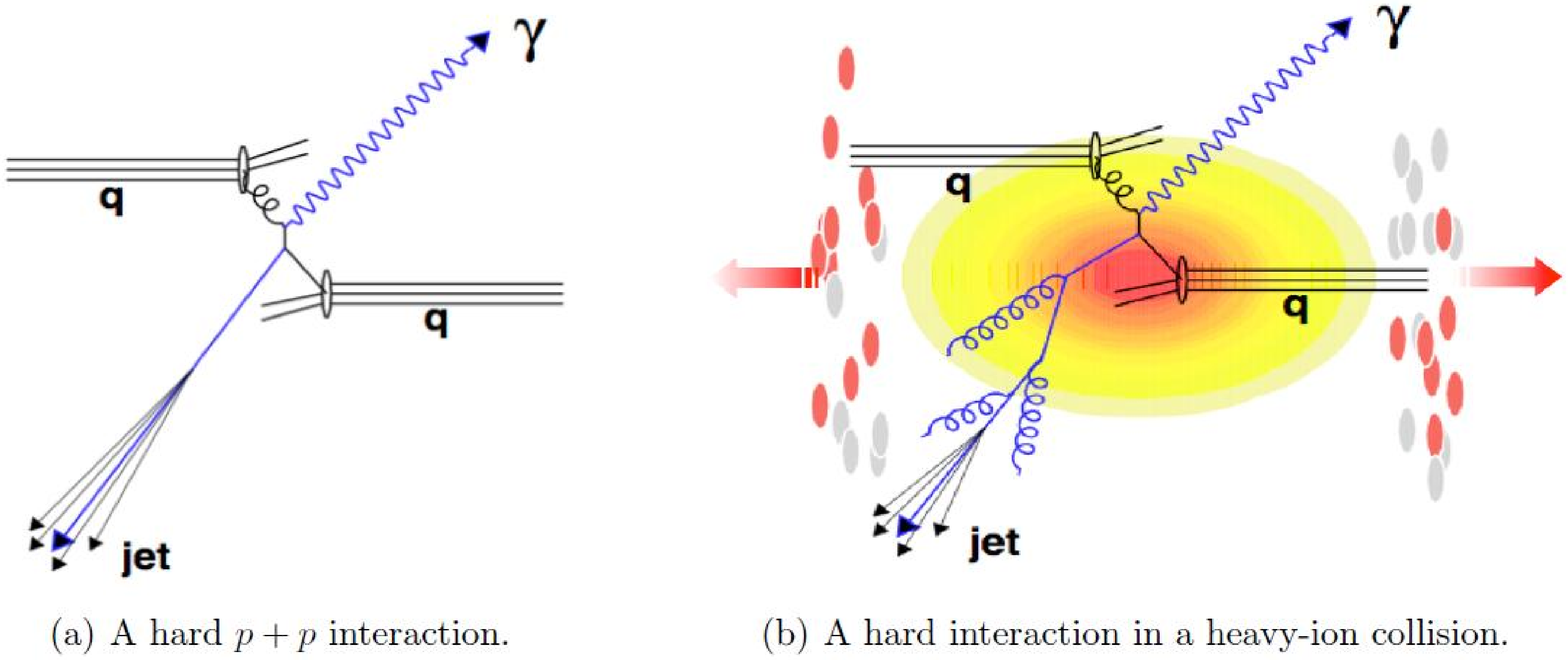}
    \caption {Illustration of interactions in $p+p$ and heavy ion collisions. Particles produced in
      heavy ion collisions will pass through the medium before being detected.}
    \label{fig:chap2_viewjet}
  \end{center}
\end{figure}
Particles with high $p_{\rm{T}}$~($p_{\rm{T}}>5$GeV/$c$) are producted from hard processes in the initial collisions.
The produced high-energy partons in the initial collision subsequently fragment into a spray of hadrons, called jet. 
Figure~\ref{fig:chap2_viewjet} illustrates  interactions in $p+p$ and heavy ion collisions. 
The effect of the medium on these hard collision probes can be studied by comparing the yield of hard collision probes 
in heavy ion collisions to that in $p+p$ collisions. 

Back-to-back jets are observed in high energy collisions of elementary particles, but are difficult to identify in 
the high-multiplicity environment in a heavy ion collision. However, a jet typically contains a leading particle which carries 
most of the momentum of the parent parton.
Therefore, the modification of high $p_{\rm{T}}$ spectra in heavy-ion collisions essentially provides the information 
of the matter which high energy parton propagates through.
It has been predicted that the yields of high $p_{\rm{T}}$ particles are suppressed compared with the binary scaled yield 
in p+ p collisions at RHIC due to the energy loss of partons, which called jet quenching~\cite{bib:jet1,bib:jet2}.
\subsubsection{Experimental Result of Jet Quenching}
\begin{figure}[htb]
  \begin{center}
    \includegraphics[angle=0,width=10.5cm]{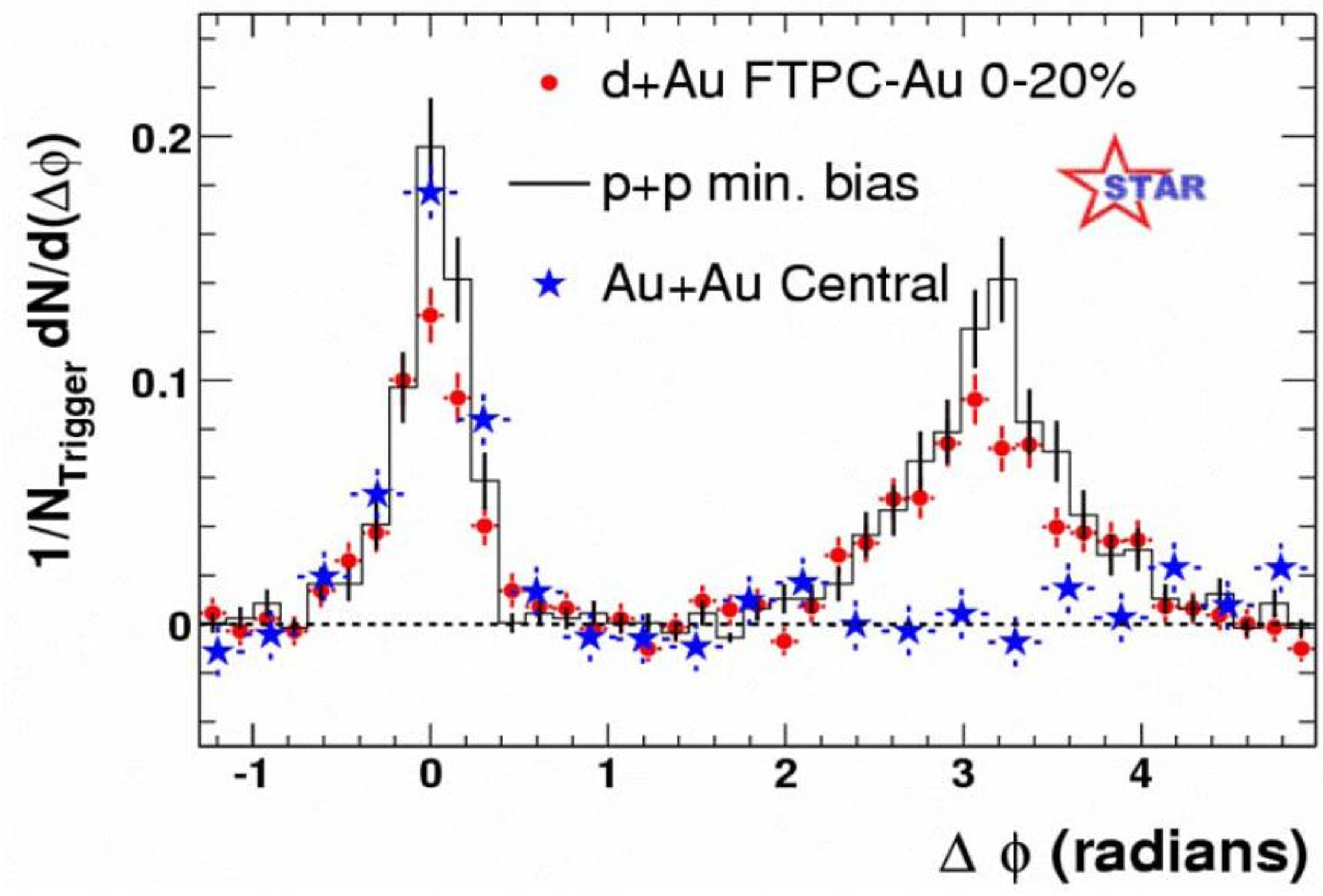}
    \caption {Comparison of two-particle azimuthal angular correlations of charged particles
      for central Au + Au, central d + Au and p+ p collisions, where $N_{Trigger}$ is the number of
      high $p_{\rm{T}}$ particles.}
    \label{fig:chap2_star}
  \end{center}
\end{figure}
The observed 'jet quenching' at RHIC is nicely demonstrated by the two particle azimuthal angular correlations. 
Since the high $p_{\rm{T}}$ particles are produced as back-to-back jets, peaks around $\Delta \phi \sim 0$ 
(near side) and $\Delta \phi \sim \pi$(away side) are expected, where $\Delta \phi $ is the azimuthal angle between 
the leading particle  and the associated particle. 
Figure~\ref{fig:chap2_star} shows the two particles 
azimuthal angular correlations~($\Delta \phi$) for inclusive charged hadrons in p+ p, d+Au and Au+Au 200 GeV 
collisions~\cite{bib:starjet}. 
One of the two particles is the trigger particle~($4<p_{\rm{T}}<6$~GeV/$c$) and the other is the associated 
particle~($2<p_{\rm{T}}<p_{\rm{T}}^{trig}$~GeV/$c$).
Clear peaks around at $\Delta \phi \sim 0$  could be seen in p+ p, d+Au and Au+Au which is expected from the high $p_{\rm{T}}$ 
particles production. On the other hand, the away side peak at $\Delta \phi \sim \pi$ is vanished in Au+Au collisions, 
while $\Delta \phi \sim \pi$ peaks exist in p+ p and d+Au collisions.
Absence to the peak at $\Delta \phi \sim \pi$ in Au+Au collisions indicates the suppression of the away
side yields  due to the final state interactions with the medium created in Au+Au collisions.

Nuclear modification factor~($R_{AA}$) is a good observable to quantify the magnitude of the yield suppression.
$R_{AA}$ is the ratio of the yield in heavy ion collisions over the binary scaled~($N_{coll}$) yield in $p+p$ collisions and 
is defined as bellows.
\begin{equation}
R_{AA}(p_{\rm{T}}) \equiv \frac{dN_{AA}/dp_{\rm{T}}}{N_{coll}dN_{pp}/dp_{\rm{T}}} \label{eq:raa}.
\end{equation}
Thus, $R_{AA}=1$ indicates there is no modification of the leading hadron spectrum in the heavy-ion collision.

\begin{figure}[htb]
  \begin{tabular}{c c}
    \begin{minipage}{\minitwocolumn}
      \begin{center}
	\includegraphics[width=7cm]{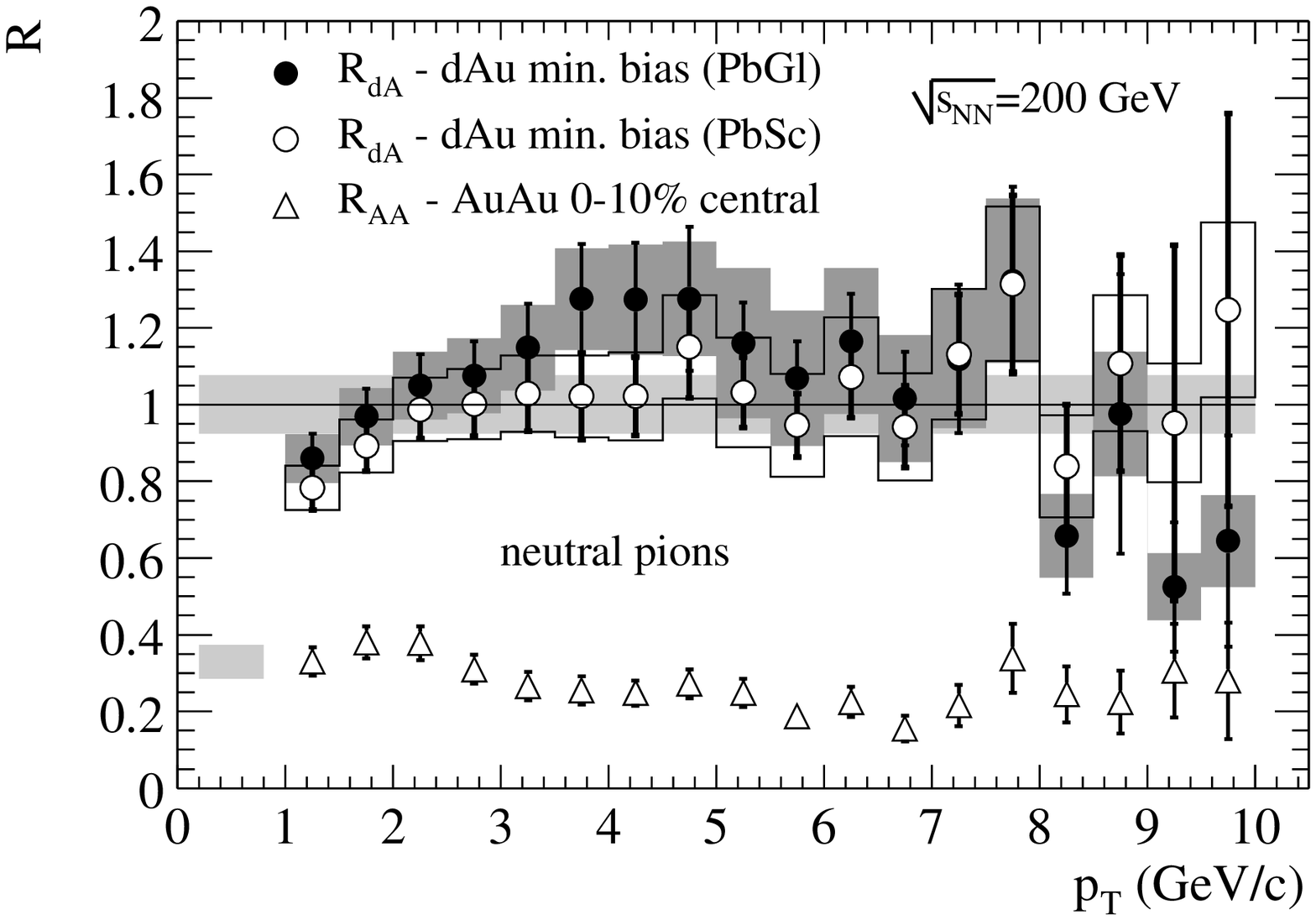}
        \caption{$R_{dAu}$ in minimum bias d+Au collisions for $\pi^0$ and $R_{AA}$ for $\pi^0$ in most central
	  Au+Au collisions.
	}
        \label{chap2:raa_dau}
      \end{center}
    \end{minipage}
    &
    \begin{minipage}{\minitwocolumn}
      \begin{center}
	\includegraphics[width=7cm]{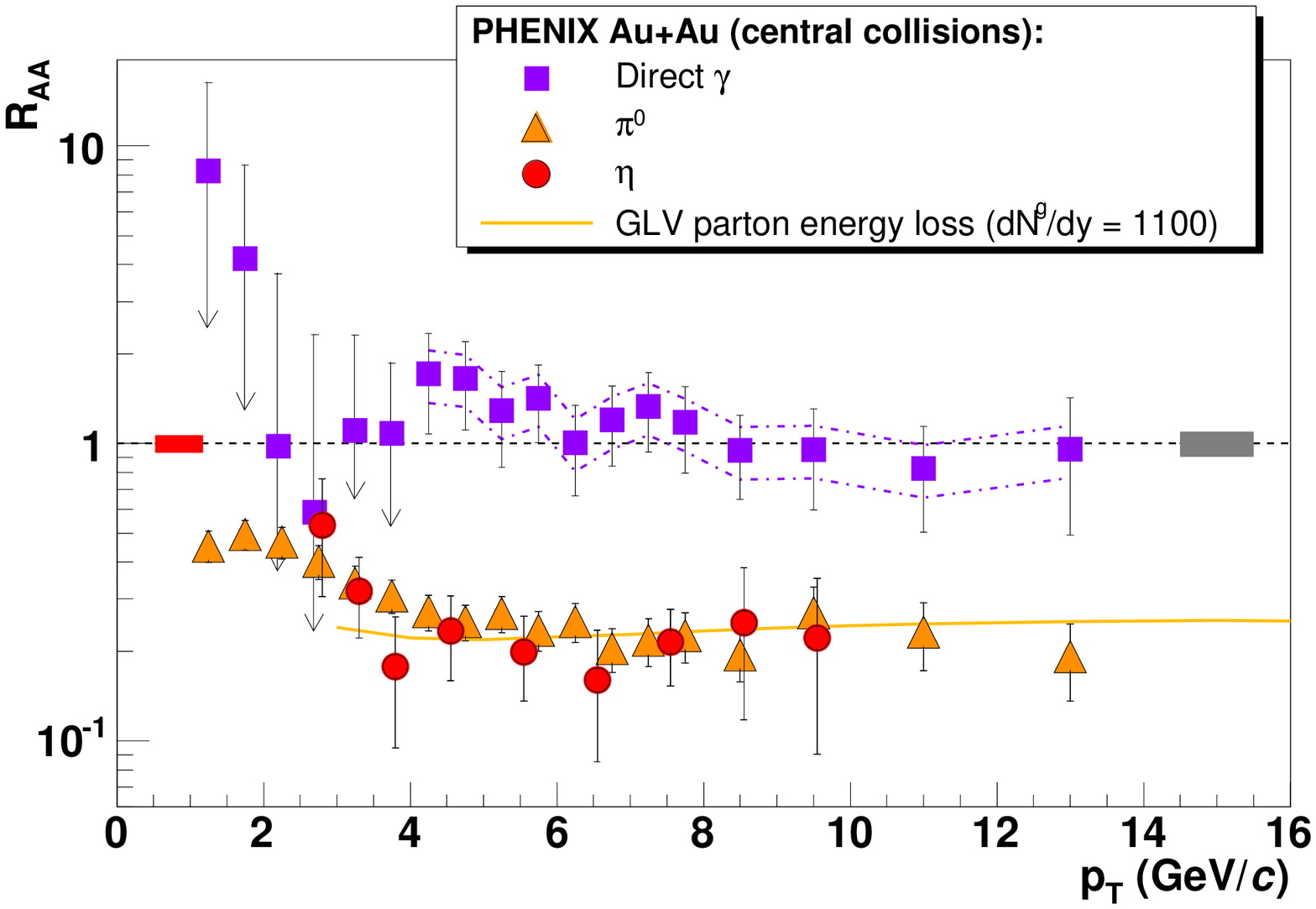}
        \caption{$R_{AA}$ for $\pi^0$, $\eta$ and direct $\gamma$ in most central Au+Au collisions.}
        \label{chap2:raa_val}
      \end{center}
    \end{minipage}
  \end{tabular}
\end{figure}

Figure~\ref{chap2:raa_dau} shows $R_{dAu}$ in minimum bias d+Au collisions, which mean any cut for the impact parameter 
is not applied,
 for $\pi^0$ and $R_{AA}$ for $\pi^0$ in most central Au+Au collisions~\cite{bib:pidau}. 
The data clearly indicates that the $\pi^0$ yield at high $p_{\rm{T}}$ is strongly suppressed 
in Au+Au collisions, while there is no suppression of high $p_{\rm{T}}$ particles in d+Au collisions.
The data suggests that the suppression of high $p_{\rm{T}}$  hadrons in Au+Au is from not an initial nuclear effect but the effect 
of the produced dense medium.
Figure~\ref{chap2:raa_val} shows $R_{AA}$ for $\pi^0$, $\eta$ and direct $\gamma$ in most central Au+Au 
collisions~\cite{bib:ppg051}.
The data indicates that $\pi^0$ and $\eta$ have a similar suppression pattern, while direct $\gamma$ is not suppressed.
This fact can be understood that light quarks lose a large fraction of their energy and direct $\gamma$ does not lose their energy
in the medium, since $\gamma$ does not have color charge.
Absence to the modification of $\gamma$ yield also indicates the initial state effect is not large.

\subsubsection{Theoretical Interpretation for Jet Quenching} \label{sec:jet}
For fast partons going through  QGP, the most important microscopic process for their energy loss is  radiative
gluon bremsstrahlung radiation induced by the static gluon fields with a particular (e.g. plasma) screening configuration,
similar to QED.
The two effects complicate the theoretical treatment of the energy loss.
The one is the Landau Pomerancuk-Migdal (LPM) effect, which destructively interferes to the bremsstrahlung process.
The similar process is also found at QED.
The LPM effect occurs when the characteristic formation length (1/$\Delta p$) of the emitted gluon becomes large
compensable to its mean free path in the medium~\cite{bib:lpmqcd}.
The other is space time evolution of the matter. Since the energy loss obviously depends on the energy density of the medium,
the space time dependence of the energy density should be taken into account for realistic calculations.

Two theoretical approaches are discussed here.
One is PQM model which is quantified with the average squared transverse momentum transferred from the medium
to the parton per mean free path~($\hat{q}$). 
The PQM model is based on BDMPS model~\cite{bib:pqm,bib:bdmp}. BDMPS is a perturbative
calculation explicitly including only coherent radiative energy loss for the parton via gluon bremsstrahlung. 
In BDMPS models, the mean of energy loss for the parton~($<\Delta E>$) is expressed in terms of $\hat{q}$ in the limit 
of large parton initial energy~($E$) as following.
\begin{equation}
<\Delta E> \propto \alpha_{s} \hat{q} L^2,
\end{equation}
where $L$ is the in-medium path length of the parton.
The PQM model incorporates a realistic transverse collision geometry, though with a static medium. 
The PQM model does not include initial 
state multiple scattering or modified nuclear parton distribution functions.

The other is a more realistic approach, GLV model, which is quantified with the gluon density~($dN^g/dy$)~\cite{bib:glv0,bib:glv}.
The GLV model employs an operator product formalism in which probability amplitude for gluon emissions is calculated. 
An analytic expression is derived for the single gluon emission spectrum to all orders in opacity~(the ratio of the length traversed 
to mean free path), assuming an infrared cutoff is given by the plasma frequency.
A realistic transverse collision geometry and Bjorken expansion of the medium are taken into account in the GLV model.
In GLV models, the fraction of energy loss for the parton~($<\Delta E>/E$) is expressed in terms of $dN^g/dy$ in the limit 
of large parton energy~($E$) as following.
\begin{equation}
\frac{\Delta E}{E} \propto \alpha_{s}^3 \frac{dN^g}{dy} \frac{L}{E} \ln\frac{2E}{\mu^2 L},
\end{equation}
where $\mu$ the Debye screening scale in the plasma.
The calculation also incorporates the Cronin effect and the nuclear shadowing.

\begin{figure}[htb]
  \begin{center}
    \includegraphics[angle=0,width=14cm]{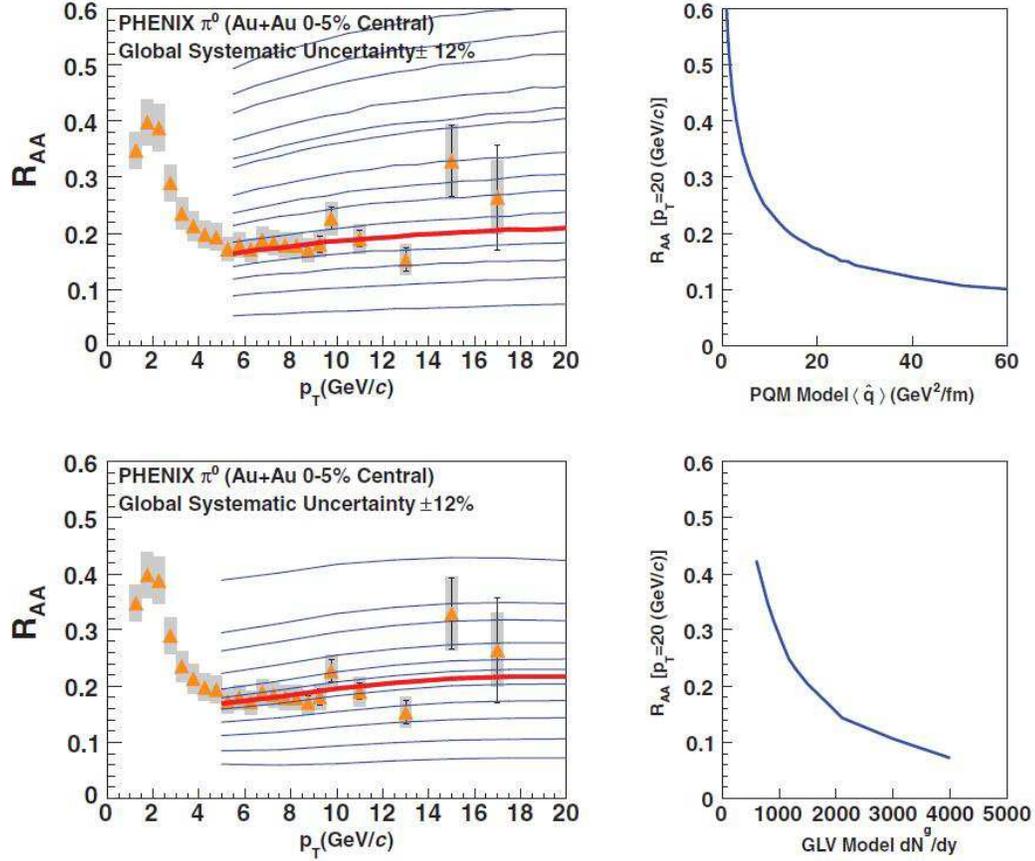}
    \caption {Left panels:
      $R_{AA}$ for $\pi^0$ in 0-5\% central Au+Au collisions
      and predictions from PQM~\cite{bib:bdmp} and GLV  models~\cite{bib:glv} with various free parameters.
      Right panels: $R_{AA}$ at $p_{\rm{T}}= 20$~GeV/$c$ predicted from PQM and GLV  models.
    }
    \label{fig:chap2_pifit}
  \end{center}
\end{figure}

Figure~\ref{fig:chap2_pifit} shows $R_{AA}$ for $\pi^0$ in 0-5\% central Au+Au collisions
and predictions from PQM~\cite{bib:bdmp} and GLV  models~\cite{bib:glv} with various values of free parameter~(left panels), and 
$R_{AA}$ at $p_{\rm{T}}= 20$~GeV/$c$ predicted from PQM and GLV  models~(right panels).
The property of the created medium can be determined via the comparison with the predicted $R_{AA}$ between measured $R_{AA}$.
The results are as follows~\cite{bib:phfit}.
\begin{eqnarray}
PQM:\quad \hat{q} &=& 13.2^{+2.1}_{-3.2} GeV^2/fm\\
GLV:\quad \frac{dN^g}{dy} &=& 1400^{+270}_{-150}.
\end{eqnarray}
These values indicate a large medium density is achieved.

\subsection{Azimuthal Anisotropy}
\begin{figure}[htb]
  \begin{center}
    \includegraphics[angle=0,width=12.5cm]{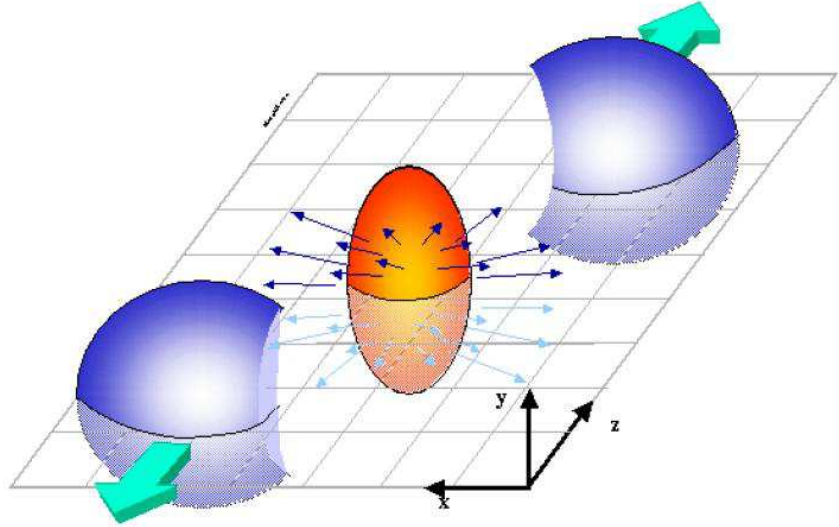}
    \caption {An illustration of non-central Au-Au collisions. The plane defined as the direction of the impact parameter (x) 
    and direction of the beam axis (z) is called 'reaction plane'.
    }
    \label{fig:chap2_arm}
  \end{center}
\end{figure}

The distribution of the particle yield in the azimuthal space provides information about collective motion of the partons in the 
medium created in heavy ion collisions. 
In the non-central collisions, the spatial shape of initial medium created by the collisions is an almond like shape as shown in
Figure~\ref{fig:chap2_arm}. In Fig~\ref{fig:chap2_arm}, x-axis is the direction
of the impact parameter and z-axis is the  direction of the beam axis.
In the hydrodynamical framework, since the pressure gradient in x-z plane, which is the driving force of the collective flow, 
is larger than y direction in the initial stage of the collisions, the particles produced by collisions 
are expected to be emitted more to the x-z plane than to y direction.
A large anisotropy can be generated only if the thermalization of the medium is rapid enough.
In this way, the magnitude of collective flow (and its $p_{\rm{T}}$ dependence) is, in principle, a quantitative index 
of the thermalization time,~$\tau_0$.

\subsubsection{Experimental Result of Azimuthal Anisotropy}
\begin{figure}[htb]
  \begin{center}
    \includegraphics[angle=0,width=16cm]{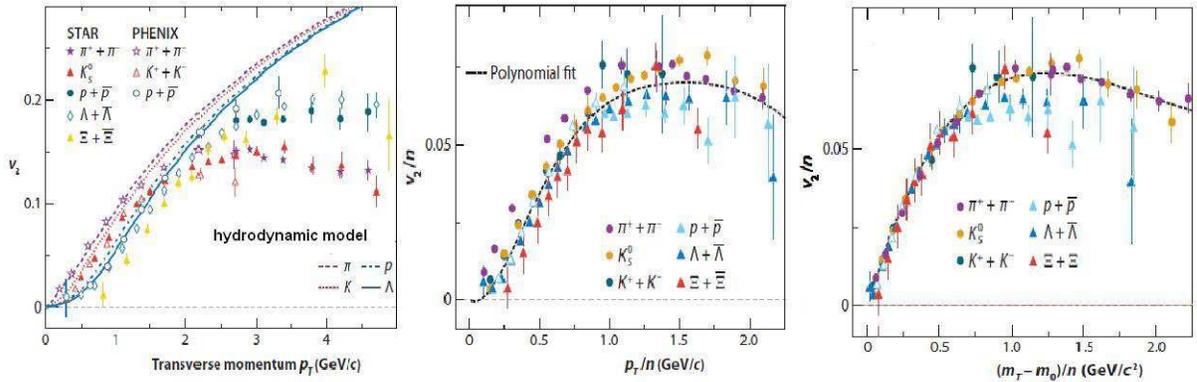}
    \caption {Left:  $v_2(p_{\rm{T}})$ of identified hadrons in minimum bias Au+Au  collisions hydrodynamic model 
    calculations~\cite{bib:starv2,bib:phv2,bib:hyde}. 
    Middle: $v^q_2(p_{\rm{T}}/n)$ of identified hadrons.
    Right: $v^q_2(KE_{\rm{T}}/n)$ of identified hadrons.
    }
    \label{fig:chap2_hadv2}
  \end{center}
\end{figure}

Experimentally, the magnitude of azimuthal anisotropy has been quantified using Fourier expansion of the azimuthal distribution of 
emitted particles. 
The particle distribution is expanded according to Fourier expansion~(at mid rapidity, the system is mirror symmetric 
in the x-y plane and odd Fourier components vanish.)
\begin{equation}
\frac{d^2N}{d{\bf p}^{2}_{\rm{T}}} \propto \frac{d^2N}{dp^2_{\rm{T}}}[1+2v_2(p_{\rm{T}})\cos(2\phi)+....],
\end{equation}
\noindent where $\phi$ is the azimuthal angle of particles with respect to the reaction plane.
Especially the second harmonic coefficient of the Fourier expansion of the azimuthal distribution is called as elliptic
flow. $v_2(p_{\rm{T}})$ is defined as the magnitude of the elliptic flow.
Left panel of Figure~\ref{fig:chap2_hadv2} shows the $v_2(p_{\rm{T}})$ of identified hadrons in minimum bias Au+Au 
collisions with hydrodynamical model calculations~\cite{bib:starv2,bib:phv2,bib:hyde}.
Large $v_2(p_{\rm{T}})$ has been observed in Au+Au collisions at RHIC.
Applications of relativistic hydrodynamics have shown that the experimentally measured $v_2(p_{\rm{T}})$ for
various hadrons is well described when a thermalization time of $\tau_0$=0.5-1~fm and a small viscosity of 
the matter~($\eta/s \sim 0$)  are assumed.
Therefore, rapid thermalization and a small viscosity of the matter are suggested from the 
measurement of $v_2(p_{\rm{T}})$.
The disagreement of  hydrodynamic model with the data above 2~GeV/$c$ indicates hydrodynamic model is only valid for
low $p_{\rm{T}}$ particles as is expected, 
since another process, for example jet fragmentation, becomes dominant at high $p_{\rm{T}}$ region.

The other remarkable feature of the elliptic flow is a constituent-quark number scaling of $v^h_2(p_{\rm{T}})$, as determined by
the number~(n) of constituent quarks in each hadron~(h).
$v^q_2(p_{\rm{T}})$ which can be interpreted as $v_2$ of quarks is defined as
\begin{equation}
v^q_2(p_{\rm{T}}/n) \equiv  \frac{1}{n} v^h_2(p_{\rm{T}}).
\end{equation}
$v^q_2(p_{\rm{T}}/n)$ of identified hadrons is shown at middle panel in Fig.~\ref{fig:chap2_hadv2}.
It is found that $v^q_2(p_{\rm{T}}/n)$ of identified hadrons has the almost same shape.
Empirically, the better scaling is found when use transverse kinetic energy~$(KE_{\rm{T}})$, which is defined as
$\sqrt{p^2_{\rm{T}} +(M^h)^2} - (M^h)$, is used instead of $p_{\rm{T}}$.
$v^q_2(KE_{\rm{T}}/n)$ of identified hadrons is shown at right panel in Figure~\ref{fig:chap2_hadv2}.

Quark coalescence model is motivated by this observation and the measurement of  baryon to meson ratios~\cite{bib:bmratio}.
Successful description by the quark coalescence model implies a large thermalized source of quarks and anti-quarks.
Then it may be a strong evidence for a QGP formation at RHIC, since the quark coalescence model is assumed that hadrons are 
produced from coalescence thermalized quark source in this model.
\subsubsection{Quark Coalescence Model}\label{sec:coal}
Quark coalescence~(recombination) model is one of the models for hadron production in heavy ion collisions~\cite{bib:recom1}. 
In this model, hadrons are produced by valence quarks in the thermal medium when they are close together in 
the phase space. 
The basic equation in the coalescence model for the formation of meson and baryon from quarks can be written as
\begin{eqnarray}
  \frac{d^3N_{M}}{d^3{\bf p}_{M}} &=& g_{M} \int \prod_{i=1}^{2}\left[ d^3{\bf x}_i  d^3{\bf p}_i f_i({\bf x}_i,{\bf p}_i) \right]
  \times f_{M}({\bf x}_1,{\bf x}_2:{\bf p}_1,{\bf p}_2) \delta^3({\bf p}_M-{\bf p}_1-{\bf p}_2)\\
  \frac{d^3N_{B}}{d^3{\bf p}_{B}} &=& g_{B} \int \prod_{i=1}^{3}\left[ d^3{\bf x}_i  d^3{\bf p}_i f_i({\bf x}_i,{\bf p}_i) \right]
  \times f_{B}({\bf x}_1,{\bf x}_2,{\bf x}_3:{\bf p}_1,{\bf p}_2,{\bf p}_3) \delta^3({\bf p}_B-{\bf p}_1-{\bf p}_2-{\bf p}_3).
  \nonumber \\
  & &
\end{eqnarray}
\noindent  The functions $f_i({\bf x}_i,{\bf p}_i)$ is distribution functions of quarks and antiquarks in the phase space, and 
they are normalized to their numbers, $\int d^3{\bf x}_i  d^3{\bf p}_if_i({\bf x}_i,{\bf p}_i) = N_i$.
The factor $g_{M(B)}$ takes into account the probability of forming a colorless meson~(baryon) from spin 1/2 colored quarks.
$f_{M(B)}$ is Wigner function for forming meson~(baryon) and depends on the overlap of the spatial and momentum distribution of 
its constituent quarks.

Let us consider elliptic flow in quark coalescence model at mid-rapidity~($\mid{\bf p}\mid \sim \mid{\bf p}_{\rm T}\mid$).
The momentum space distribution of quark~(a) can be written in terms of the azimuthal angle $\phi$.
\begin{equation}
f_a({\bf p}_{\rm T}) \sim \tilde{f}_a(p_{\rm T})(1+2v^a_2(p_{\rm T})\cos 2\phi).
\end{equation}
The elliptic flow of mesons~($v^M_2$) becomes as bellow when the elliptic flow of mesons is small compared with unity~($v2\ll1$)
\begin{eqnarray}
v^M_2(p_{\rm T}) &=& \frac{\int d\phi \cos2\phi dN_M/d^2{\bf p}_{T}}{\int d\phi dN_M/d^2{\bf p}_{T}} \nonumber \\
                 &\sim& \int dp_{\rm Ta} dp_{\rm Tb}  f_{M} [v^a_2(p_{\rm Ta})+ v^b_2(p_{\rm Tb})] 
\delta(p_{\rm T}-p_{\rm Ta}-p_{\rm Tb}) \label{eq:v2coal}
\end{eqnarray}
When we take $f_{M}$ to have a uniform distribution in momentum space as in Ref.~\cite{bib:dcoal4}, i.e. ,
\begin{equation}
f_M \propto \Theta(\Delta_p-\mid {\bf p}_a -  {\bf p}_b \mid)
\end{equation}
and the momentum window is very narrow~($\Delta_p\rightarrow0$), Eq.~\ref{eq:v2coal} leads to the following expression.
\begin{equation}
v^M_2(p_{\rm T}) = v^a_2(p_{\rm T}/2) + v^b_2(p_{\rm T}/2)
\end{equation}
Therefore, when quarks have the same elliptic flow before hadronization,
we arrive at a simple scaling law as follows.
\begin{equation}
v^M_2(p_{\rm T}) = 2v^q_2(p_{\rm T}/2) \label{eq:v2meson}
\end{equation}
We also obtain the following expression about the elliptic flow of baryons by the similar way.
\begin{equation}
v^M_3(p_{\rm T}) = 3v^q_2(p_{\rm T}/3) \label{eq:v2bar}
\end{equation}
Eq.~\ref{eq:v2meson} and Eq.~\ref{eq:v2bar} represent the constituent-quark number scaling which is found at RHIC.

\subsubsection{AdS/CFT Correspondence}
The observation of large azimuthal anisotropy at RHIC and the successful description of it by hydrodynamics with 
very small viscosity~($\eta/s \sim 0$) also have a great interest for super-string~(brane) theorists, while it looks that
there is little relation between the experiments at RHIC and super-string theory.
The interest is motivated by so called, 'Anti-de Sitter space/Conformal Field Theory~(AdS/CFT) Correspondence' which represents
the equivalence between $N=4$ supersymmetric Yang-Mills~(SYM) gauge theory and the string theory on 5-dimensional 
anti-deSitter space~\cite{bib:ads0}. 
Especially, the gauge theory at finite temperature in the strong coupling limit corresponds to the classical gravity theory
for the black-hole on 5-dimensional anti-deSitter space.
Since the calculation of the gauge theory in the strong coupling is difficult due to its non-perturbative nature,
this correspondence has a possibility to be a useful method of the calculation of the  the strong coupling gauge theory.
\begin{figure}[htb]
  \begin{center}
    \includegraphics[width=13cm]{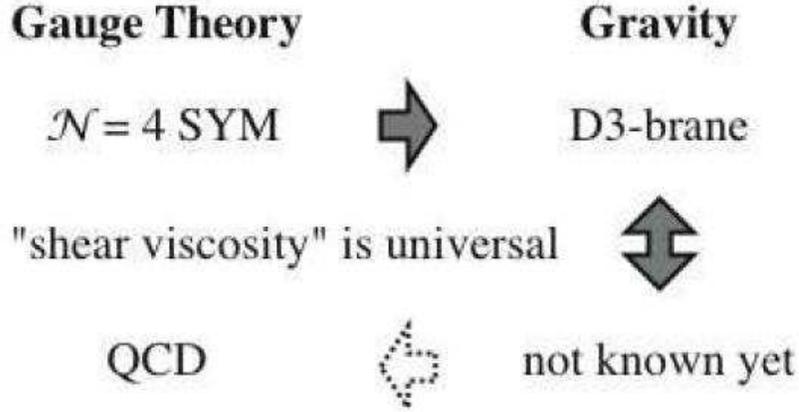}
    \caption{The black hole quantity corresponding to the 'viscosity' 
    is universal, so probably the results for supersymmetric gauge 
    theories are directly applicable to the real QCD.}
    \label{fig:chap2_cft}
  \end{center}
\end{figure}

The most famous benefit of AdS/CFT correspondence is the result of the ratio of shear viscosity over the entropy 
density~($\eta/s$).
$\eta/s$ of the classical black-hole on 5-dimensional anti-deSitter space is determined as bellow~\cite{bib:adsdrag3}.
\begin{equation} \label{eq:eta_black}
\frac{\eta}{s} = \frac{1}{4 \pi}.
\end{equation}
This $\eta/s$ is significantly small compared with various matters and the conjectured quantum lower bound.
Based on the AdS/CFT correspondence, $\eta/s$ of the gauge theory matter in the strong coupling limit becomes very small, $1/4 \pi$.
This claim can not be compared directly with the observation of small viscosity at RHIC, since QCD is not 
$N=4$ supersymmetric Yang-Mills gauge theory. Black-hole which corresponds to QCD has been not found.
However, it is known Eq.~\ref{eq:eta_black} is valid for various type of black-holes and rather general.
Therefore, it is expected that Eq.~\ref{eq:eta_black} is 'universal value' of black-hole.
When Eq.~\ref{eq:eta_black} is 'universal', $\eta/s$ of the QCD matter in the strong coupling limit also becomes $1/4 \pi$ which
can be compared with the observation of small viscosity at RHIC.
Figure~\ref{fig:chap2_cft} shows a conceptual view of above discussion.

In above way, AdS/CFT correspondence has a possibility to provide a useful method for the calculation of 
non-perturbative~(strong coupling) QCD matter which is created at RHIC.
In addition, the experiments at RHIC have a possibility to provide a first test of  the super-string theory.
It is worth to note that the QCD calculation using AdS/CFT correspondence still has several ambiguous points, 
for example, 'universal' assumption, AdS/CFT correspondence itself and the correction of finite coupling constant in QCD.
Therefore, further study is necessary for AdS/CFT correspondence to be a reliable method for  
non-perturbative QCD.

\clearpage
\section{Heavy Flavor} \label{sec:heavy}
Heavy quarks~(charm, bottom) are primarily produced in hard partonic scattering in nucleon-nucleon collisions,
since the initial content of heavy flavor in nuclei is negligibly small.
In addition, the energy scale for the production of heavy quarks $Q^2 \sim M^2_{c(b)}$
is significantly higher than $\lambda_{QCD}$. This gives us a coupling constant of the order of $\alpha_s \sim 0.3$,
which is small enough to apply perturbative QCD calculation for the production of heavy quarks. 
Therefore, measurement of heavy quarks in $p+p$ collisions provides a good test of the perturbative QCD calculation.

Heavy quarks are also expected to be a special probe of the medium created in heavy ion collisions,
since their mass is significantly larger than the typical temperature of the created medium~($\sim$ 200~MeV in Au+Au collisions 
at RHIC) and $\lambda_{QCD}$. The expectations for the heavy quark production in heavy ion collisions and the motion 
inside the medium are summarized as bellow.
\begin{itemize}
\item{Heavy quarks are only produced in the initial stage in the heavy ion collision. 
  Thus heavy quark spectra in $p+p$ collisions provide a well defined initial state, even for
  low-momentum heavy quarks. Then, generated heavy quarks propagate through the hot and dense medium 
  created in heavy ion collisions. This feature makes the measurement of heavy quark a
  prime tool to extract properties of the medium.
}
\item{The magnitude of the suppression of heavy quarks yield at high $p_{\rm{T}}$ region is expected much smaller compared with 
  that of light quarks due to their large mass.
  This expectation is based on the 'dead cone' effect~\cite{bib:dead1}.
  The gluon radiation from heavy quarks, which is a dominant source of energy loss in the medium in the case of light quark, 
  is hindered by the angular screening. That is, the soft gluon emission in the forward direction of a heavy quark is
  suppressed within the angle $\Theta=M_{c(b)}/E_{c(b)}$ due to the causality. \\
  Thermalization of heavy quarks is expected to be 'delayed' relative to light quarks by a factor of $M_{c(b)}/T$.
  The magnitude of the elliptic flow of heavy quarks is also expected much smaller to be compared with that of light quarks.\\
  In addition, the magnitude of the energy loss and the elliptic flow of bottom is expected to be smaller to 
  be compared with that of charm due to the large difference of their masses.
}
\end{itemize}

\subsubsection{Method of Measurement of Heavy Flavor at RHIC}
\begin{figure}[htb]
  \begin{center}
    \includegraphics[angle=0,width=14cm]{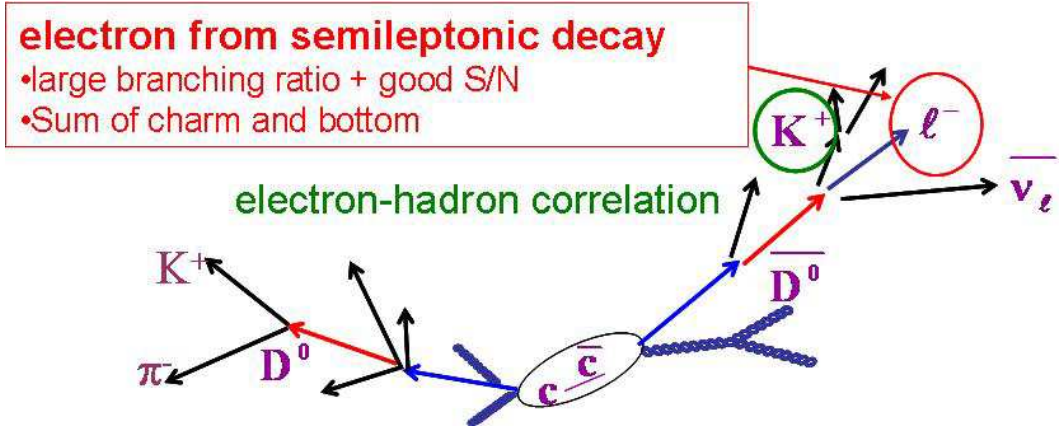}
    \caption{A conceptual view of measurement of heavy flavor at RHIC.}
    \label{fig:chap2_concep}
  \end{center}
\end{figure}

Measurement of heavy flavor is carried out via electrons from semi-leptonic decay of heavy flavored hadron, so called 'single non-photonic
electron' 
in this thesis. 
Figure~\ref{fig:chap2_concep} shows a conceptual view of measurement of heavy flavor at RHIC.
The entire process from the production to measurement can be schematically represented as
\begin{equation}
p+p\quad(A+A) \stackrel{pQCD}{\longrightarrow} c(b) \stackrel{N.P.frag}{\longrightarrow} D(B)  \stackrel{decay}{\longrightarrow} lepton
\label{eq:hob}
\end{equation}
where N.P. frag stands for non-perturbative fragmentation process and lepton represents the final-state observable.

The advantage of the measurement of single electrons from heavy flavor at PHENIX is a good signal 
to noise ratio because of well controlled material budget, while we can not detect a displaced vertexes of single non-photonic 
electrons.
However, the observables are the mixture of single electrons from charm and bottom.
The determination of the fraction of the contribution from bottom in the single non-photonic electrons is important to 
interpret the result of heavy flavor, since 
the behavior of bottom in the medium is expected to be quite different from that of charm due to the large difference of 
their mass.

A new analysis method is introduced in this thesis to measure the fraction of bottom in single non-photonic electrons.
The correlation in unlike charge-sign electron-hadron pairs from weak decay of 
charmed hadrons is utilized, which is based on partial reconstruction of the $D/\bar{D} \rightarrow e^{\pm} K^{\mp} X$ decay as 
shown in Fig.~\ref{fig:chap2_concep}.
The new analysis method are described in Sec.~\ref{sec:method}.

In this section, the each step in Eq.~\ref{eq:hob} is briefly reviewed and the results of measurements of heavy quarks at RHIC are also
described.
\subsection{Heavy Flavor Production}
\begin{figure}[htb]
  \begin{center}
    \includegraphics[angle=0,width=14cm]{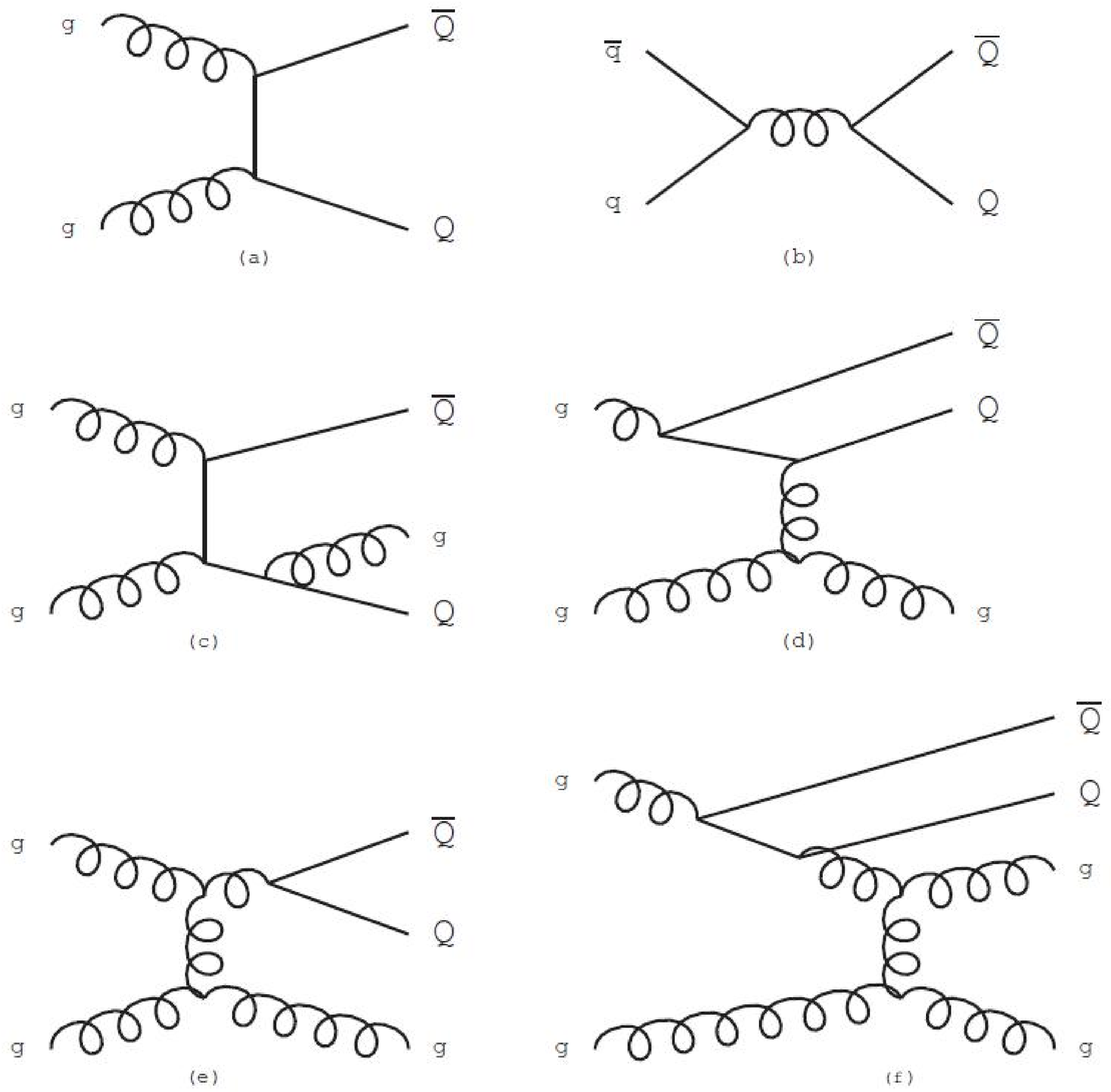}
    \caption {LO and most important NLO heavy quark production diagrams. LO - a) 'gluon fusion' b) 'quark-antiquark annihilation' 
    NLO - c) Pair creation with gluon emission in output channel d) 'flavor excitation' e) 'gluon splitting' 
    f) 'gluon splitting' but of 'flavor excitation' character.}
    \label{fig:chap2_hqfey}
  \end{center}
\end{figure}
The general perturbative calculation for the total cross section of quark pair production in the partonic level can be expressed 
by the following equation.
\begin{equation}
\sigma_{ij}(\tilde{s},M^2_{Q},\mu^2_R) = \frac{\alpha^2_s(\mu^2_R)}{M^2_{Q}} \sum_{k=0}^{\inf} (4\pi\alpha_s(\mu^2_R))^k
\sum_{l=0}^{k} f_{ij}^{(k,l)}(\eta) \ln^l \left( \frac{\mu^2_R}{M^2_{Q}} \right),
\end{equation}
\noindent where $\mu^2_R$ is called renormalization scale usually assumed to be $0.5-2\times M^2_{Q}$ and 
$\tilde{s}$ is the partonic energy in center of mass frame. The dimensionless parameter $\eta=\tilde{s}/4M^2_{Q}-1$
reflects the phase space of the heavy quark pair production~($\sqrt{\tilde{s}}$ should be at least 2$M_{Q}$ to create a 
quark-antiquark pair).
$i$ and $j$ are the partonic indexes. $f_{ij}^{(k,l)}$ is a dimensionless scaling function representing the amplitude of a given 
partonic scattering diagram. $k$ shows the order of the process diagram, $k=0$ is called as Leading-Order~(LO) and $k=1$ is 
called as Next-Leading-Order~(NLO).
Figure~\ref{fig:chap2_hqfey} shows the Feynman diagram of LO and important NLO process.
Using the parton distribution function in proton~(PDF), we can write the total cross section for heavy flavor production 
in term of $\sigma_{ij}(\tilde{s},M^2_{Q},\mu^2_R)$ in $p+p$ collisions as follows.
\begin{equation}
\sigma_{pp}(\tilde{s},M^2_{Q},\mu^2_R,\mu^2_F)=\sum_{i,j=q,\bar{q},g}\int_{\frac{4M^2_Q}{s}}^{1} d\tau \int_{\tau}{1} \frac{dx_1}{x_1}
f^p_i(x_1,\mu^2_F)f^p_j(\tau/x_1,\mu^2_F) \sigma_{ij}(\tau s,M^2_{Q},\mu^2_R) \label{eq:cross}
\end{equation}
\noindent $\mu^2_F$ a momentum transfer scale (factorization scale) of the PDF factorization and usually assumed to be 
$0.5-2\times M^2_{Q}$. $f^p_i(x_1,\mu^2_F)$ is parton distribution function in term of a momentum fraction~($x$) 
and factorization scale.
Eq.~\ref{eq:cross} have three free parameters, $M^2_{Q},\mu^2_R$ and $\mu^2_F$. The uncertainty of the perturbative 
calculation is usually determined by varying these parameters.

\subsection{Fragmentation} \label{sec:ufrag}
Colored heavy quarks pick up light quarks in order to create color singlet hadrons, which is called as fragmentation process. 
The differential cross section of heavy flavor hadrons~($\frac{d\sigma^H}{dp^H_{\rm T}}$) can be written as follows using 
the factorization theorem.
\begin{equation}
\frac{d\sigma^H}{dp^H_{\rm T}} = \int d\tilde{p}_{\rm T} dz \frac{d\sigma_Q}{d\tilde{p}_{\rm T}} D^{H}_{Q}(z)
\delta(p^H_{\rm T}-z\tilde{p}_{\rm T}),
\end{equation}
\noindent where $\tilde{p}_{\rm T}$ is the transverse momentum of heavy quarks and $p^H_{\rm T}$ is the transverse momentum of 
heavy flavored hadrons. $\frac{d\sigma_Q}{d\tilde{p}_{\rm T}}$ is the differential cross section of heavy quarks, and 
$z$ is the momentum fraction of the quark carried by the hadron. $D^{H}_{Q}(z)$ is called as fragmentation function 
and determines the the probability of producing hadron with given momentum fraction~($z$).

The fragmentation function of heavy quarks should be much harder than that of a light hadron. In the limit of a very heavy
quark, one expects that the fragmentation function for a heavy quark to go into any heavy hadron to be peaked near 1.

The fragmentation function can be split into a perturbative part and non-perturbative part.
Non-perturbative effect in the calculation of the heavy quark fragmentation function is done in practice by convolving 
the perturbative result with a phenomenological non-perturbative form.
There are various parameterizations for the non-perturbative part which have free parameters.

The free parameters in the non-perturbative parameterizations  are determined by the experimental results in $e^+ e^-$ collisions 
based on 'universality of the fragmentation process' which is the assumption that the fragmentation function is independent 
of the hard-scattering process. In general, the parameters entering the non-perturbative forms do not have 
any absolute meaning, since these depend on the order of the perturbative calculation in the fragmentation function.

\begin{figure}[htb]
  \begin{center}
    \includegraphics[angle=0,width=9cm]{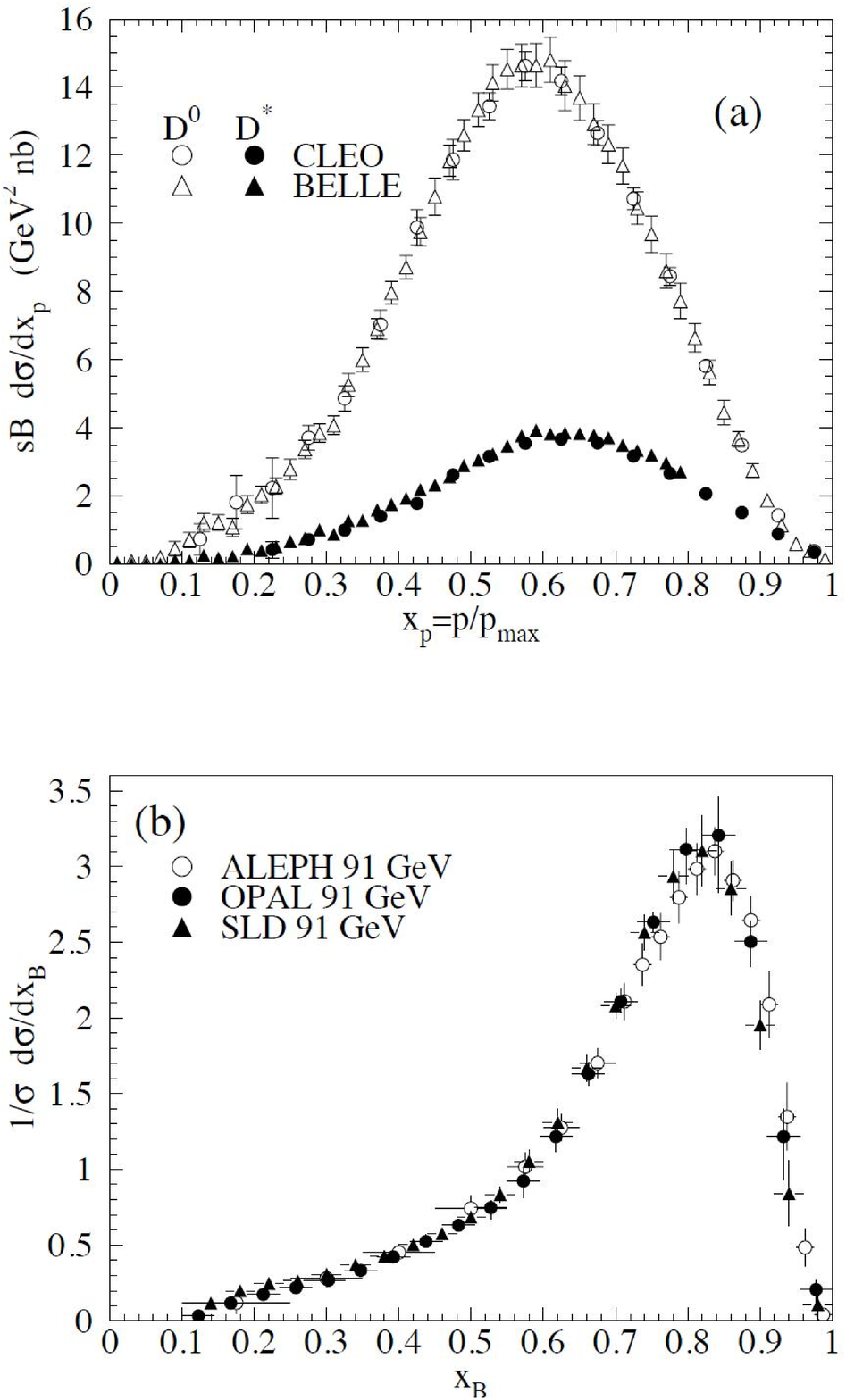}
    \caption { (a) Inclusive cross-section measurements of $D^0,D^{\ast+}$ in CLEO and BELLE as a function of $x_p$ which 
    approximates the momentum fraction $z$~\cite{bib:cleo6,bib:belle2}.
    (b) Fragmentation function for b quarks studied at LEP and SLD~\cite{bib:lep1}.
    }
    \label{fig:chap2_hqfrag}
  \end{center}
\end{figure}

Figure~\ref{fig:chap2_hqfrag}(a) shows inclusive cross-section measurements of $D^0,D^{\ast+}$ in CLEO and BELLE as a function of
$x_p$ which approximates the momentum fraction $z$~\cite{bib:cleo6,bib:belle2}.
Figure~\ref{fig:chap2_hqfrag}(b) shows fragmentation a function for b quarks studied at LEP and SLD~\cite{bib:lep1}.
The most accurate approach to derive the fragmentation function is to use the Mellin transforms of the fragmentation function 
and obtain the momenta of this transform from the experimental data.

The treatment of the fragmentation process discussed above is expected to be valid in $p+p$ collisions.
In the case of heavy ion collisions, the coalescence process becomes important in the fragmentation 
of heavy flavor.

\subsection{Semi-Leptonic Decay}
\begin{figure}[htb]
  \begin{center}
    \includegraphics[angle=0,width=8cm]{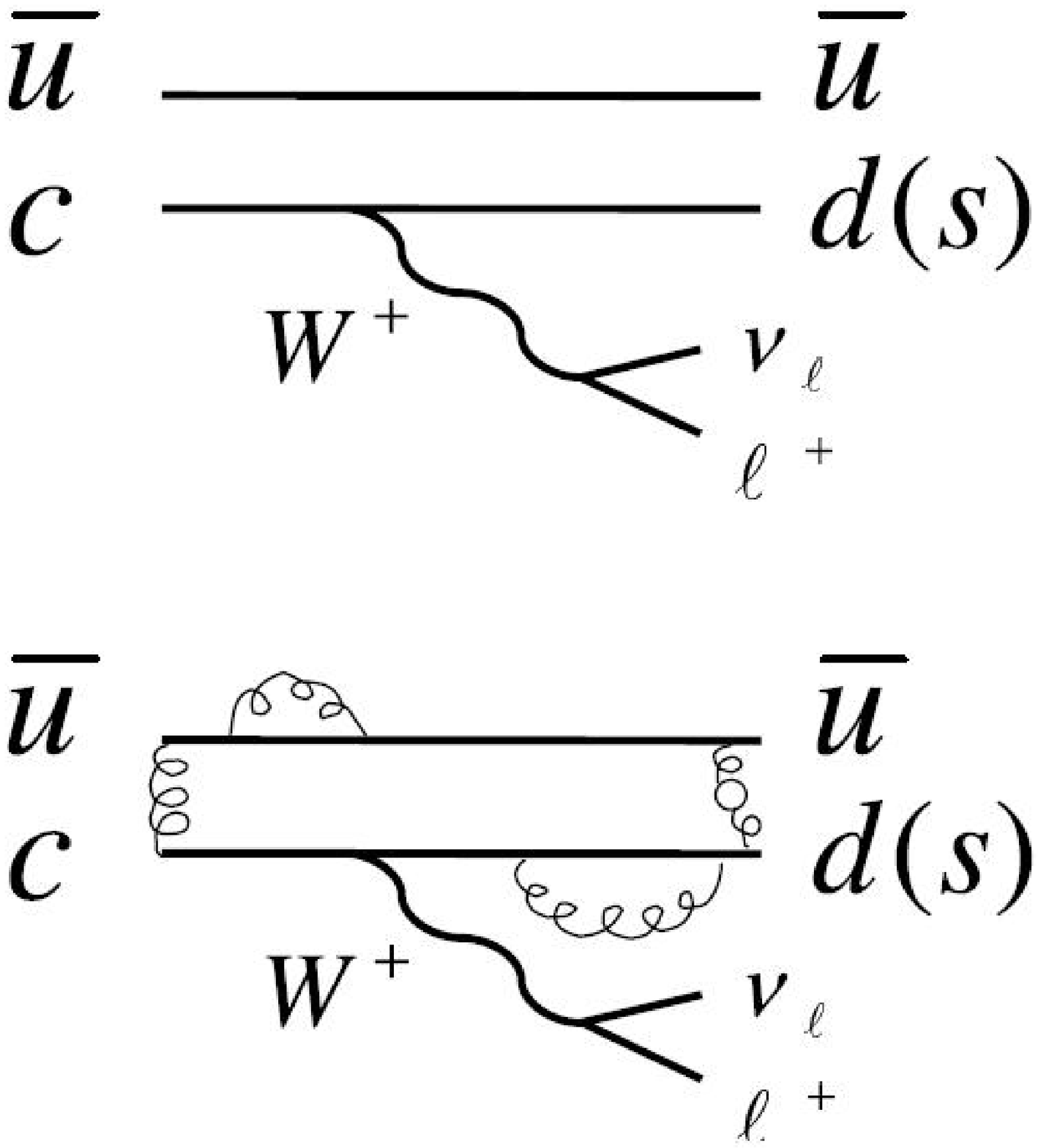}
    \caption {Top: Simplified feynman diagrams for $D^0\rightarrow \pi^-(K^-) l^+ \nu$. Bottom: A slightly more realistic diagram.}
    \label{fig:chap2_semi}
  \end{center}
\end{figure}

In decays of heavy flavored hadrons, semi-leptonic modes are generally accessible experimentally, because semi-leptonic branching 
ratios are large.
Semi-leptonic decay is also more accessible theoretically than hadronic decays because of their relative simplicity which is 
a consequence of the fact that the effects of the strong interactions can be isolated.
Therefore, they are the primary tool for the study of the CKM matrix and are well studied at CLEO, BELLE, BABAR and so on.
Figure~\ref{fig:chap2_semi} shows an example of a Feynman diagram for the semi-leptonic decay. Top panel is simplified Feynman 
diagram and bottom panel is a slightly more realistic diagram which includes the contributions from complex interactions of gluons.

For processes where the momentum transfer is much less than the W boson mass, to a very good approximation, the amplitude for 
the semi-leptonic decay of a quark of type $Q$ to one of type $q$~($Y_{Qq'}\rightarrow X_{qq'}l^+\nu$) can be given by
\begin{equation}
M(Y_{Qq'}\rightarrow X_{qq'}l^+\nu) = -i\frac{G_F}{\sqrt{2}} V_{qQ} L^{\mu} H_{\mu} \label{eq:semi}.
\end{equation}
\noindent Here, $G_F$ is the Fermi constant of weak interaction and $V_{qQ}$ is an element of the CKM matrix.
$L^{\mu}$ is the leptonic current and $H_{\mu}$ is the hadronic current.
In Eq.~\ref{eq:semi}, only $H_{\mu}$  is difficult to calculate from first principles since it includes non-perturbative 
QCD effect as shown in Fig.~\ref{fig:chap2_semi}.
The hadronic current is usually parameterized with Lorentz invariant functions called 'form factors'.
When one knows the form factors, the decay dynamics of semi-leptonic decays is determined according to Eq.~\ref{eq:semi}.
The form factors are functions of momentum transfer~($q^2$). In addition, the number of form factors 
and the parametrization form depend on the spin type of parent~($Y_{Qq'}$) and daughter~($X_{qq'}$) hadrons.
As the simplest example, the hadronic current of pseudoscalar to pseudoscalar meson decays can be written as
\begin{equation}
H_{\mu} = F(q^2)(p+p')_{\mu}.
\end{equation}
\noindent Here, $p$ and $p'$ are four momenta of the initial and final hadrons and $F(q^2)$ is form factor.

The form factors have been calculated by many theoretical models~\cite{bib:semirev}.
In this thesis, ISGW2 model is often used for the semi-leptonic decay of heavy flavored hadrons, which
is based on quark model with the application of heavy quark symmetry~\cite{bib:isgw1,bib:isgw2}.

\subsection{Fixed-Order plus Next-to-Leading-Log Calculation}
Fixed-Order plus Next-to-Leading-Log~(FONLL) calculation is the theory based on perturbative QCD calculation about heavy flavor
production~\cite{bib:fonll1,bib:fonll2,bib:fonll3,bib:fonll4}. 
FONLL can be compared directly with the experimental results, specially $p_{\rm{T}}$ distribution.
The each process in Eq.~\ref{eq:hob} is implemented in FONLL as follows.
\begin{equation}
E\frac{d^3\sigma^l}{dp^3} = E^Q \frac{d^3\sigma^Q}{dp^3} \otimes D(Q\rightarrow H_Q) \otimes f(H_Q \rightarrow l),
\end{equation}
where the symbol $\otimes$ denotes a generic convolution, the leptonic decay spectrum 
term $f(H_Q \rightarrow l)$ also implicitly accounts for the proper branching ratio and  $D(Q\rightarrow H_Q)$ 
denotes fragmentation process.

The distribution of heavy quarks, $Ed^3\sigma^Q/dp^3$ is evaluated at the
Fixed-Order plus Next-to-Leading-Log~(FONLL) level pQCD calculation, that is, FONLL includes the full fixed-order NLO result~(FO) 
and  re-summation perturbative terms proportional to $\alpha_s^n(\log^k(p_{\rm{T}}/m))$ to all orders with next-to-leading
logarithmic~(NLL) accuracy~(i.e. $k = n, n-1$), where $m$ is mass of heavy quark.
NLL terms take an important role to converge of the perturbative series for high $p_{\rm{T}}$~($p_{\rm{T}}>m$) region.

Heavy quark fragmentation is implemented within the FONLL formalism that merges the
FO + NLL calculations. The NLL formalism is used to extract the non-perturbative fragmentation effects 
from the experimental data in $e^+e^-$ collisions using Mellin transforms~\cite{bib:fonll6}.
The decay of the D and B mesons into leptons is controlled by the experimentally measured decay spectra
and branching ratios~\cite{bib:babar,bib:cleo7}. 

\begin{figure}[htb]
  \begin{center}
    \includegraphics[angle=0,width=12cm]{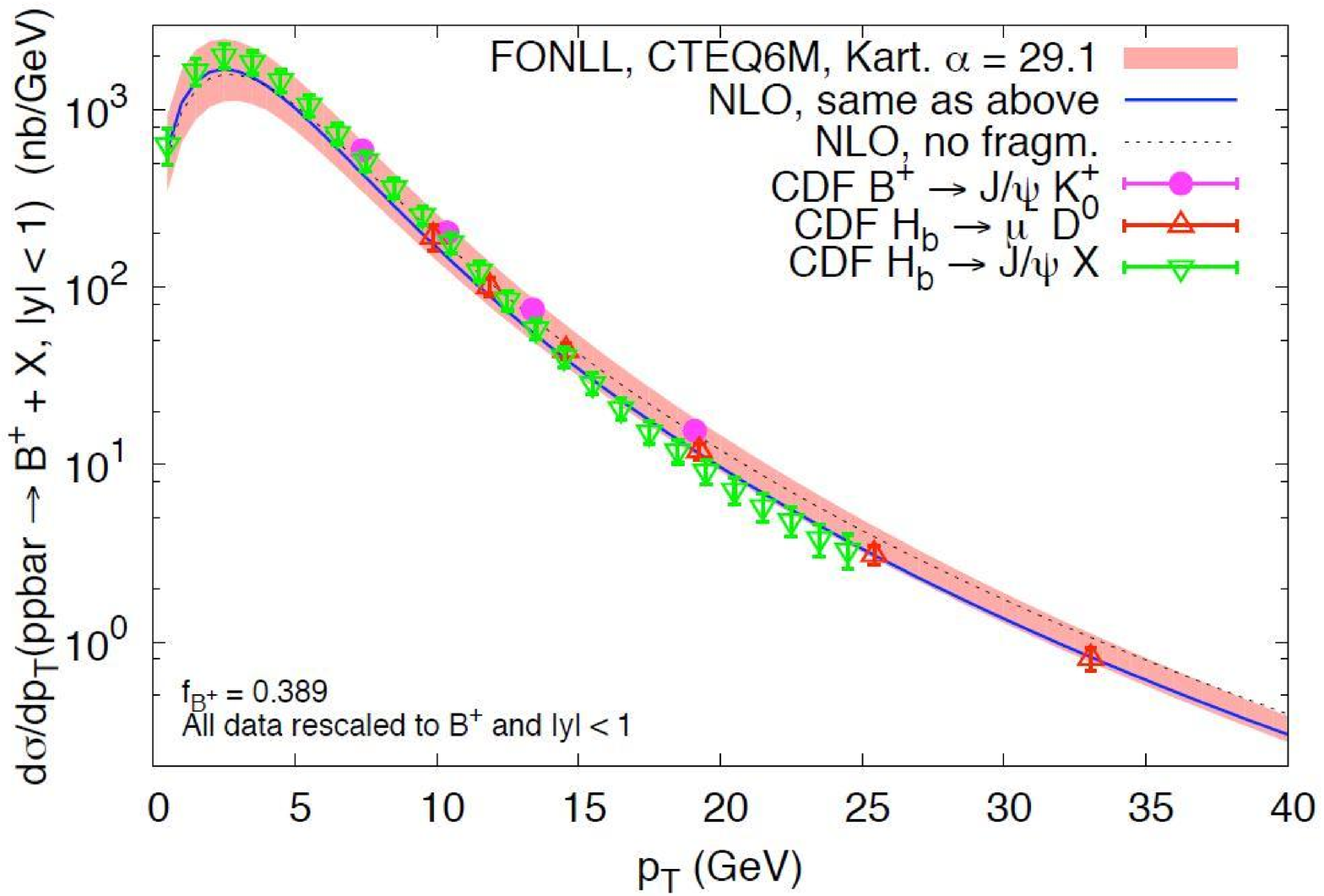}
    \caption {The $p_{\rm{T}}$ distributions B hadron measured in CDF with FONLL 
    predictions in p+$\bar{{\rm p}}$ collisions at $\sqrt{s}=1960GeV$~\cite{bib:cdf3,bib:cdf4,bib:cdf5,bib:fonll3}}
    \label{fig:fonll1}
  \end{center}
\end{figure}

\begin{figure}[htb]
  \begin{center}
    \includegraphics[angle=0,width=11cm]{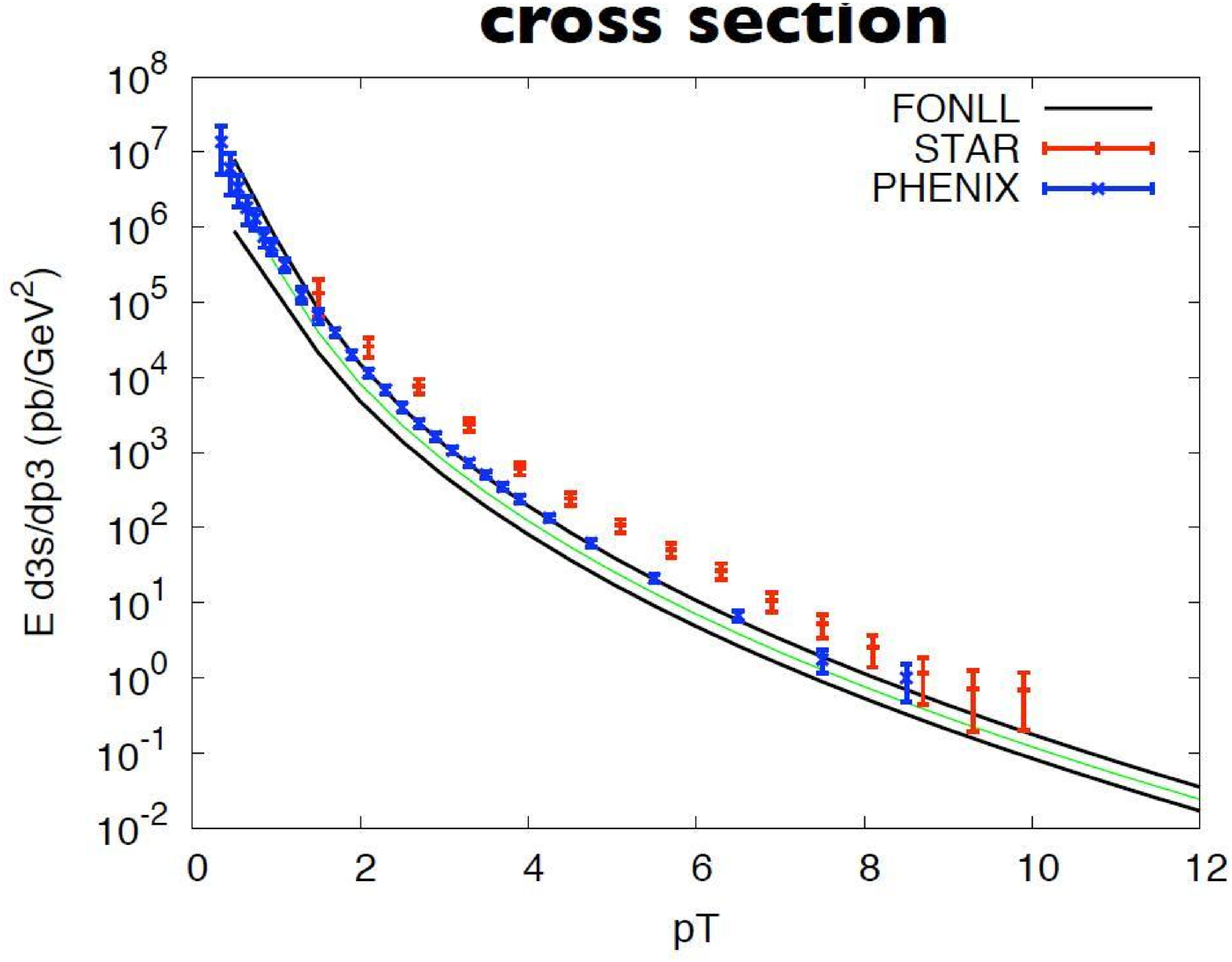}
    \caption {The differential cross sections of non-photonic electrons from heavy flavor measured in RHIC 
    with FONLL predictions in $p+p$ collisions at $\sqrt{s}=200GeV$~\cite{bib:hq3,bib:hq6,bib:fonll1}}
    \label{fig:fonll2}
  \end{center}
\end{figure}

Figure~\ref{fig:fonll1} shows the $p_{\rm{T}}$ distributions of B hadron measured in CDF with FONLL 
predictions in p+$\bar{{\rm p}}$ collisions at $\sqrt{s}=1960$~GeV~\cite{bib:cdf3,bib:cdf4,bib:cdf5,bib:fonll3}.
Figure~\ref{fig:fonll2} shows the differential cross sections of non-photonic electrons from heavy flavor measured in RHIC
with FONLL predictions in $p+p$ collisions at $\sqrt{s}=200$~GeV~\cite{bib:fonll1,bib:hq3,bib:hq6}.
FONLL calculation provides a successful description for the experimental $p_{\rm{T}}$ distributions  of heavy flavor.
However, there is large theoretical uncertainty for the absolute value of cross section of heavy flavor at even FONLL.
For example, FONLL predicts total cross section of charm, $\sigma_{c\bar{c}}$ to be $256^{+400}_{-146} \mu b$
and total cross section of bottom $\sigma_{b\bar{b}}$ to be $1.87^{+0.99}_{-0.67} \mu b$ in $p+p$ collisions at $\sqrt{s}=200GeV$.
%

\subsection{Initial Nuclear Effect for Heavy Flavor}
\begin{figure}[htb]
  \begin{center}
    \includegraphics[angle=0,width=11cm]{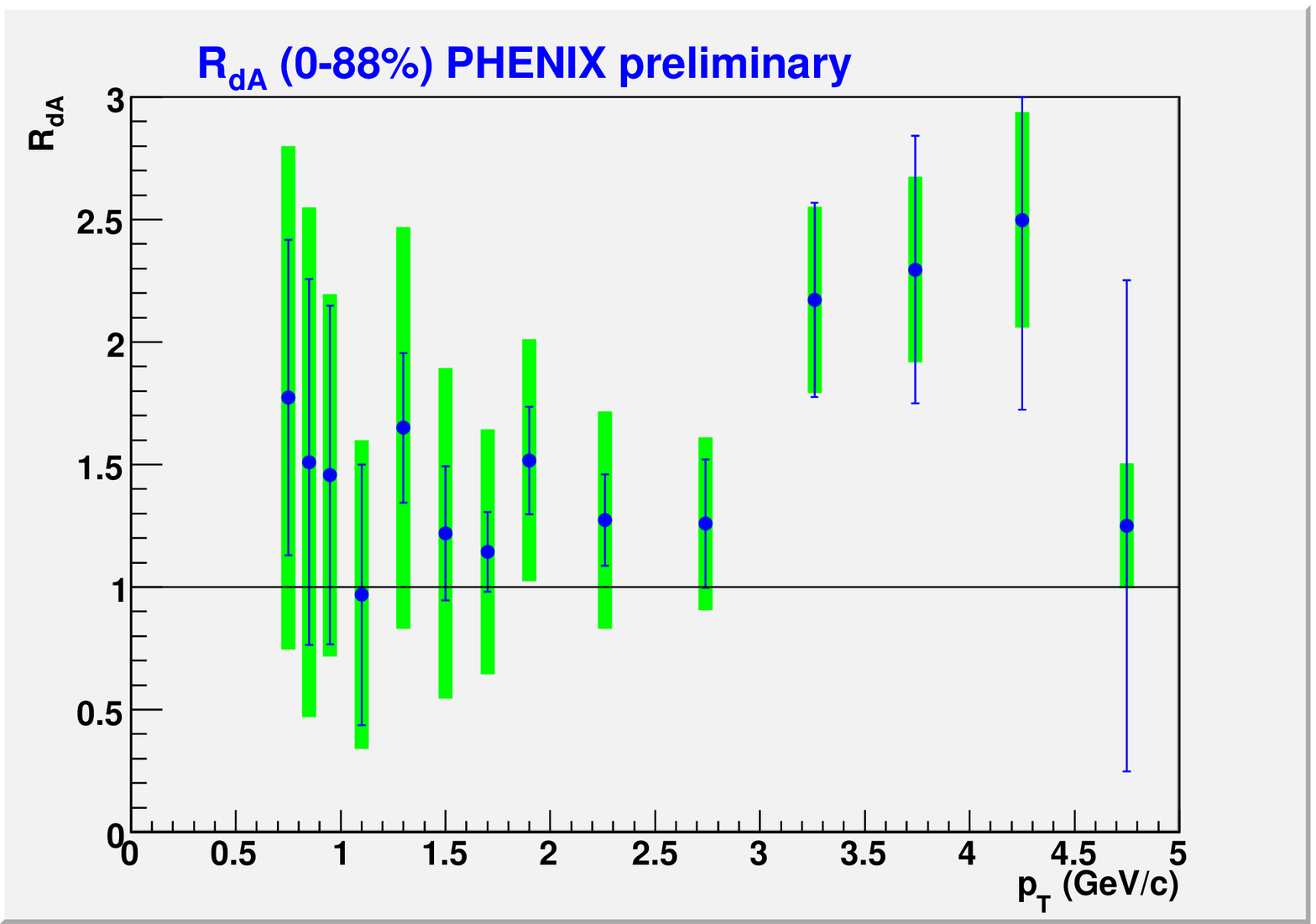}
    \caption {The nuclear modification factor of the electron from heavy flavor in d+Au collisions~($R_{dAu}$)}
    \label{fig:chap2_dau}
  \end{center}
\end{figure}

Initial nuclear modification of heavy flavor production is studied by the measurement of the electrons from heavy flavored 
hadrons in d+Au collisions at $\sqrt{s_{NN}}$ = 200~GeV at PHENIX~\cite{bib:phdau_e}.
Figure~\ref{fig:chap2_dau} shows the nuclear modification factor of the electrons from heavy flavor in d+Au collisions~($R_{dAu}$) 
defined in Eq.~\ref{eq:raa}.
The measured $R_{dAu}$ indicates the yield of the electrons from heavy flavor is slightly enhanced in d+Au collisions for the 
measured $p_{\rm{T}}$ range, while it is almost consistent with unity due to large uncertainty.

\begin{figure}[htb]
  \begin{center}
    \includegraphics[width=13cm]{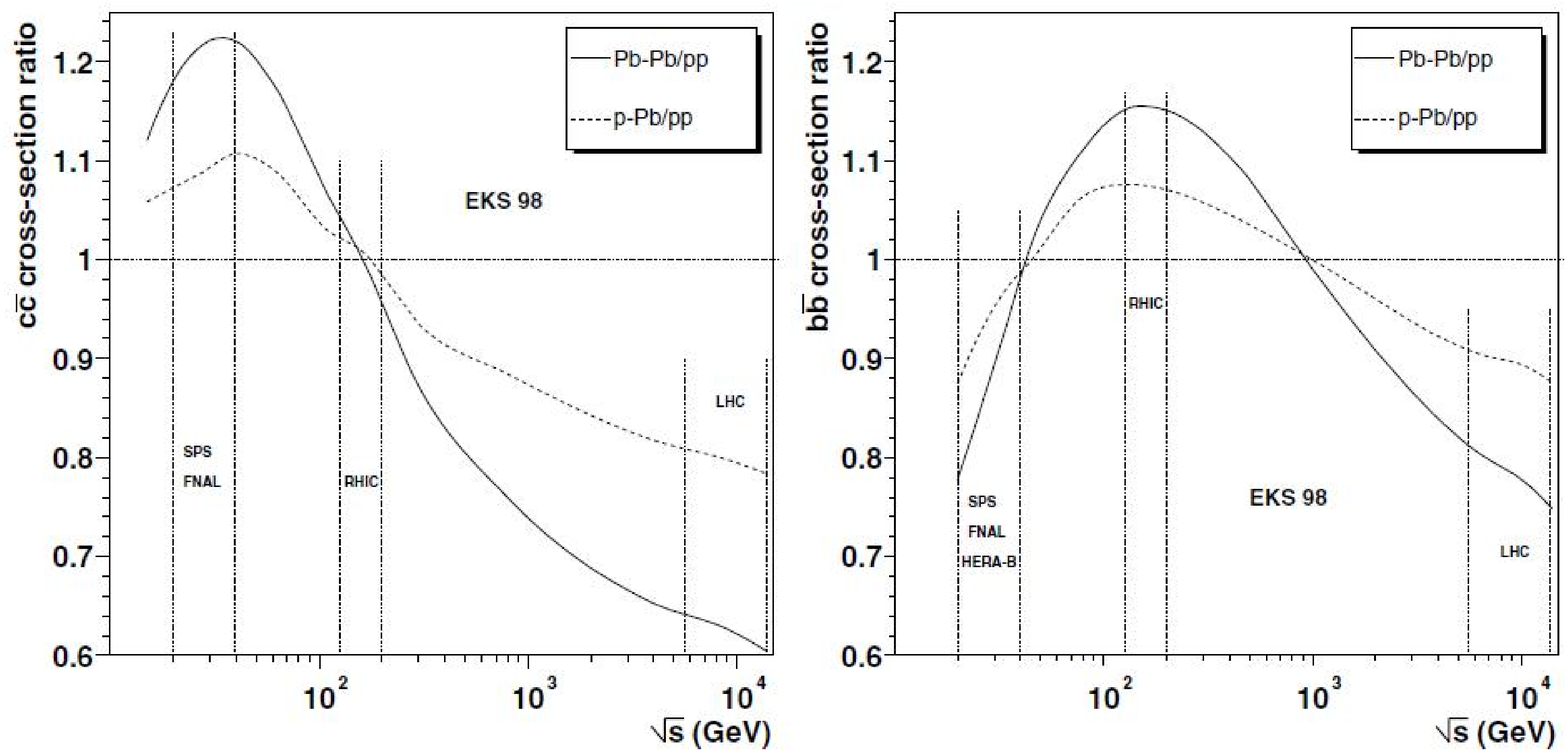}
    \caption{changes induced on charm~(left) and bottom~(right) cross-sections by the nuclear effects of
      the PDFs, calculated using the EKS 98 nuclear weight functions~\cite{bib:heavy_nlo}.}
    \label{fig:chap2_pdfh}
  \end{center}
\end{figure}

Figure~\ref{fig:chap2_pdfh} shows changes induced on charm~(left) and bottom~(right) cross-sections at mid-rapidity 
by the nuclear effects of the PDFs, calculated using the EKS 98 nuclear weight functions~\cite{bib:heavy_nlo}.
As shown in Fig.~\ref{fig:chap2_pdfh}, charm production is not modified and bottom production is slightly enhanced~(anti-shadowing)
by the PDF modification.
Therefore, the slight enhancement of the measured $R_{dAu}$ could be interpreted as the cronin effect.
In near future, the uncertainty of $R_{dAu}$ will be significantly reduced and initial nuclear effect for heavy flavor production will be
revealed precisely by the data of d+Au collisions at RHIC in Year 2008 RUN.

\subsection{Medium Modification of Heavy Flavor} \label{sec:heavy_mod}
\begin{figure}[htb]
  \begin{center}
    \includegraphics[width=11cm]{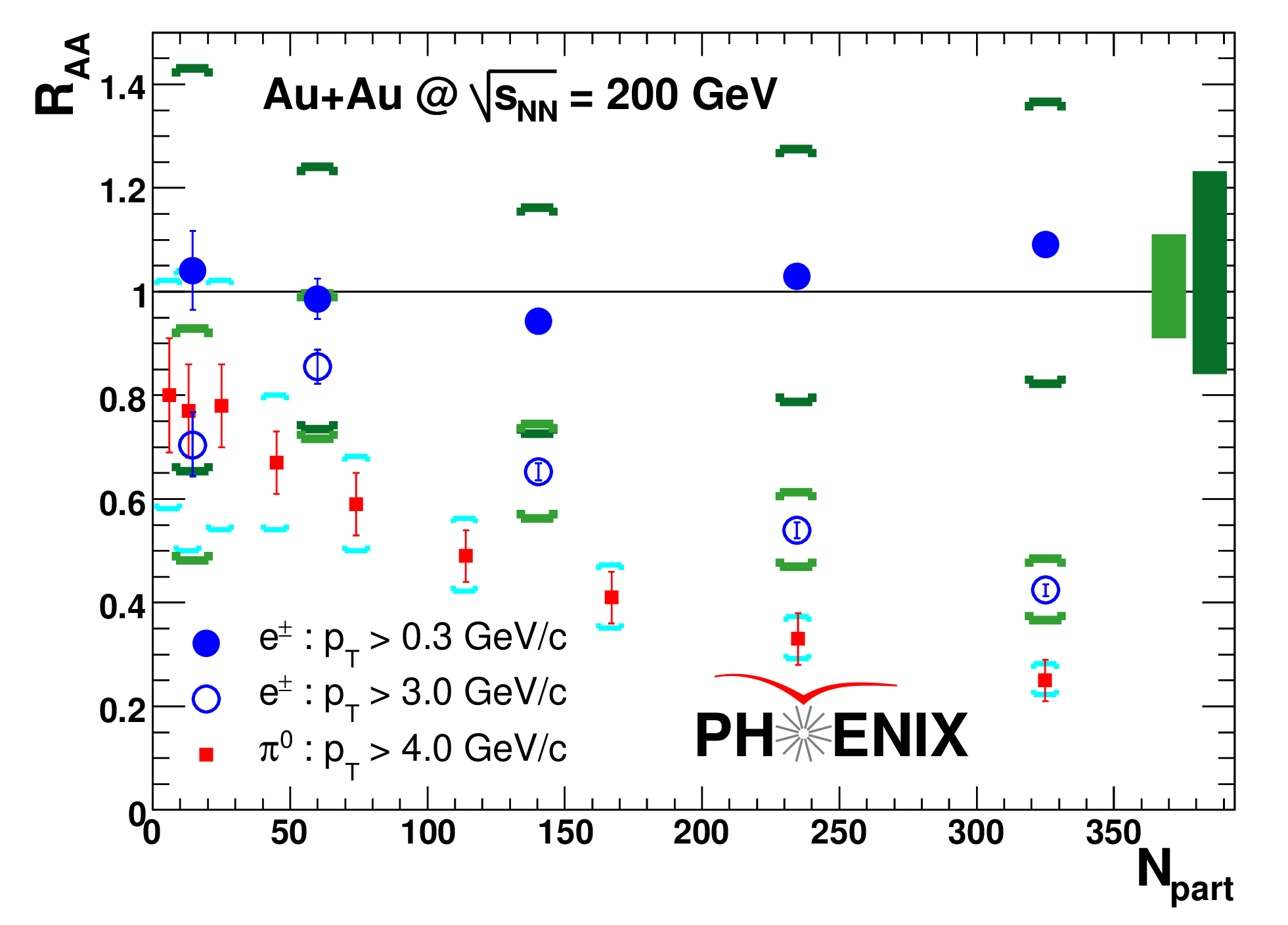}
    \caption{$R_{AA}$ of heavy flavor electrons with $p_{\rm{T}}$ above 0.3 and 3~GeV/$c$ and of $\pi^0$ with $p_{\rm{T}}>4$~GeV/$c$ 
      as a function of  $N_{part}$.The right (left) box at $R_{AA}=1$ shows the relative uncertainty from the
      $p+p$ reference common to all points for $p_{\rm{T}}>0.3(3)$~GeV/$c$}
    \label{fig:chap2_inte}
  \end{center}
\end{figure}

Medium modification of heavy quarks is studied by the measurement of the electrons from heavy flavored 
hadrons in Au+Au collisions at $\sqrt{s_{NN}}$ = 200 GeV at PHENIX~\cite{bib:hq1}.
Figure~\ref{fig:chap2_inte} shows $R_{AA}$ of the electrons from heavy flavor decays for two different $p_{\rm{T}}$ ranges as a function 
of the number of participant nucleons $N_{part}$.  For the integration interval $p_{\rm{T}}>0.3$~GeV/$c$ containing more than half 
of the heavy flavor decay electrons, $R_{AA}$ is consistent with unity for all $N_{part}$ in accordance with the binary scaling of
the total heavy-flavor yield. This fact supports the expectation that heavy flavor is only produced in the initial hard scattering. 
For the integration with $p_{\rm{T}}>3.0$~GeV/$c$, the heavy flavor electron $R_{AA}$ decreases systematically with $N_{part}$, 
while that is larger than $R_{AA}$ of $\pi^0$ with $p_{\rm{T}}>4.0$~GeV/$c$~\cite{bib:phpi1}.

\begin{figure}[htb]
  \begin{center}
    \includegraphics[width=14cm]{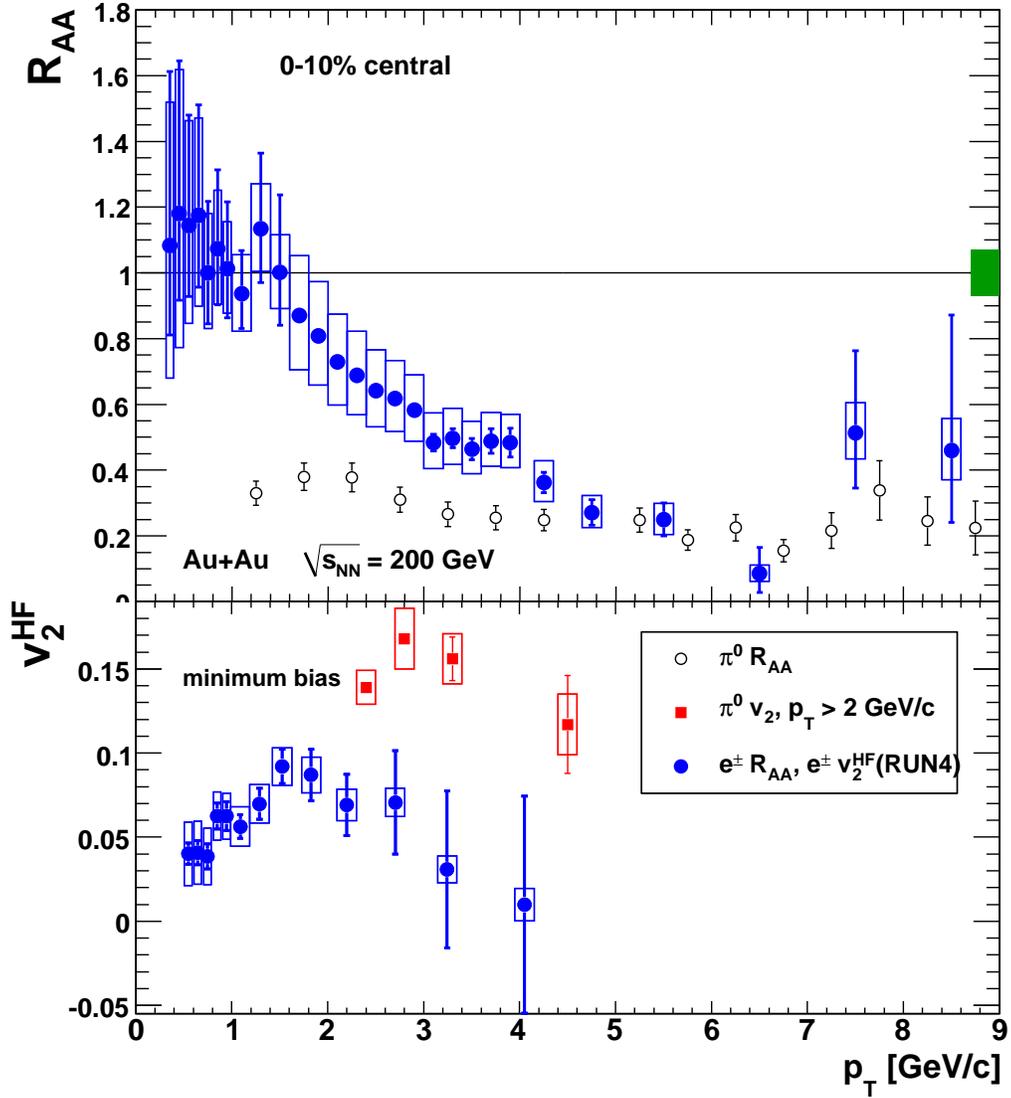}
    \caption{Top:$R_{AA}$ of heavy flavor electrons in 0-10\% central collisions compared with $\pi^0$ data and
      The box at  $R_{AA}=1$  shows the uncertainty in the number of binary collisions. Bottom: $v^{HF}_2$ of heavy flavor
      electrons in minimum bias collisions compared with $\pi^0$.
    }
    \label{fig:chap2_raae}
  \end{center}
\end{figure}

Figure~\ref{fig:chap2_raae} shows the measured $R_{AA}$ and $v^{HF}_2$ of the electrons from heavy flavor in 0-10\% 
central~(most central) and minimum bias collisions, and corresponding $\pi^0$ data at PHENIX~\cite{bib:phpi1,bib:phpi2}. 
While at low $p_{\rm{T}}$ the suppression is smaller than that of $\pi^0$, $R_{AA}$ of heavy flavor decay electrons approaches the same
magnitude of $\pi^0$ for $p_{\rm{T}}>4.0$~GeV/$c$.  The observed large $v^{HF}_2$ indicates that the charm relaxation time is 
comparable to the short time scale of QGP lifetime. 
Such behavior of heavy flavor is far from the early expectation.  Therefore, the understanding of the behavior of heavy flavor is
experimentally and theoretically challenging.

\clearpage
\subsubsection{Radiative Energy Loss}
\begin{figure}[htb]
  \begin{center}
    \includegraphics[width=10cm]{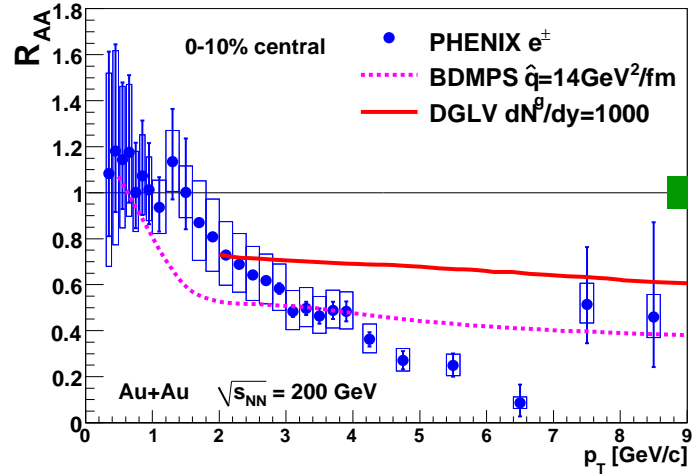}
    \caption{The comparison of the measured $R_{AA}$ with DGLV and BDMPS models~\cite{bib:dglv,bib:bdmp_e}.
    }
    \label{fig:chap2_rad}
  \end{center}
\end{figure}

Figure~\ref{fig:chap2_rad} shows the comparison of the measured $R_{AA}$ with DGLV and BDMPS models~\cite{bib:dglv,bib:bdmp_e}.
DGLV model is based on GLV model. Radiative gluon emission and 'dead cone effect' is implemented as the source of energy 
loss in the medium in these models. 

As already described, the single non-photonic electrons originate from charm and bottom whose behaviors in the medium 
should be different due to their large mass difference. 
The fraction of the contribution from bottom in all single non-photonic electrons from heavy flavor is calculated 
using perturbative QCD in Fig~\ref{fig:chap2_rad}. 
The spectra of single non-photonic electrons from charm and bottom are merged according to the fraction, after 
the suppression pattern of the single electrons from charm and bottom is calculated separately.
In Fig~\ref{fig:chap2_rad}, the parameter in BDMPS $\hat{q}$ is 14~GeV$^2$/fm and that in DGLV $dN^g/dy$ is 1000.
These models with the chosen parameters provide a successful description of the measured $R_{AA}$ of $\pi^0$ as shown 
in Sec.~\ref{sec:jet}.
The predicted $R_{AA}$ of heavy flavor electrons are larger than that of $\pi^0$ due to 'dead cone' effect and 
larger than the measured $R_{AA}$ of heavy flavor electrons.
This fact indicates radiative gluon emission is not enough to describe the energy loss mechanism of heavy flavor in the medium.

\subsubsection{Collisional Energy Loss}
\begin{figure}[htb]
  \begin{center}
    \includegraphics[width=12cm]{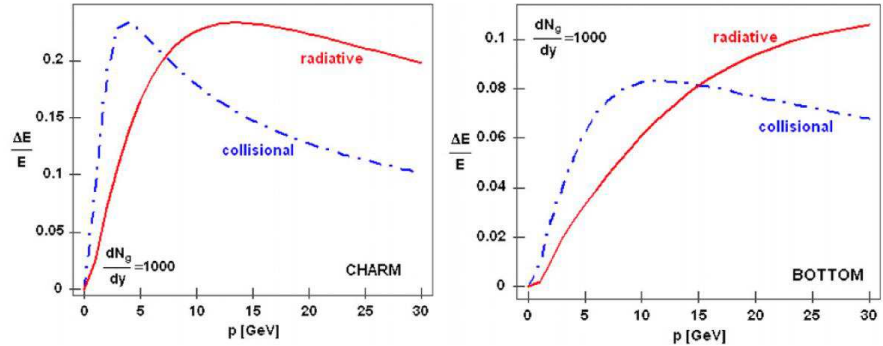}
    \caption{Comparison between collisional and medium-induced radiative fractional energy loss
      is shown as a function of momentum for charm and bottom quark jets~\cite{bib:hqcol2}.
    }
    \label{fig:chap2_colrad}
  \end{center}
\end{figure}

\begin{figure}[htb]
  \begin{center}
    \includegraphics[width=12cm]{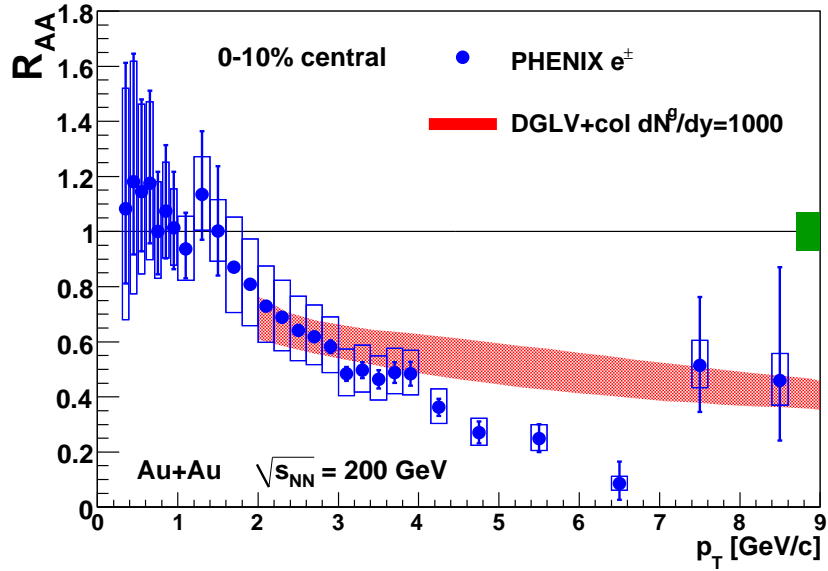}
    \caption{The comparison of the measured $R_{AA}$ with the extended DGLV including elastic scattering~\cite{bib:exdglv}.
    }
    \label{fig:chap2_radcol}
  \end{center}
\end{figure}

It has been pointed out that elastic scattering is an important source of energy loss of heavy flavor in the medium, while
the effect of elastic scattering is negligible for the energy loss of light quark~\cite{bib:hqcol1,bib:hqcol2}.
It seems to be natural because most of produced heavy flavor are not ultra-relativistic. It may be worth noting here that
for an electron traversing a hydrogen target, bremsstrahlung losses first exceed ionization losses when $\gamma \beta \sim 700$.
Figure~\ref{fig:chap2_colrad} shows the comparison of the energy loss from radiative emission and elastic scattering of charm and 
bottom~\cite{bib:hqcol2}.
The energy loss from elastic scattering is comparable to that from radiative emission.
Figure~\ref{fig:chap2_radcol} shows the comparison of the measured $R_{AA}$ with the extended DGLV including elastic 
scattering~\cite{bib:exdglv}.
The parameter in the extended DGLV, $dN^g/dy$, is 1000.
The extended DGLV still underpredicts the magnitude of the suppression of electrons from heavy flavor.

\subsubsection{Charm Alone Model}
\begin{figure}[htb]
  \begin{center}
    \includegraphics[width=12cm]{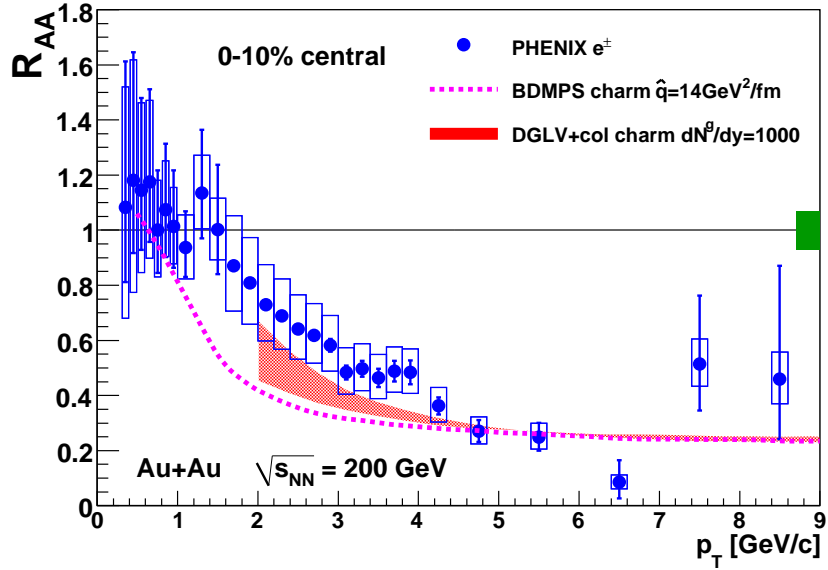}
    \caption{The comparison of the measured $R_{AA}$ with the extended DGLV and BDMPS models when the contribution of 
bottom quark is neglected~\cite{bib:bdmp_e,bib:exdglv}.
    }
    \label{fig:chap2_charm}
  \end{center}
\end{figure}

There is a simple and robust solution for the puzzle of the energy loss of heavy flavor.
When we assume all single electrons are from charm quarks, the predicted $R_{AA}$ agrees with the experimental result because
the suppression magnitude of bottom quark is expected to be much smaller than that of charm quarks.
Figure~\ref{fig:chap2_charm}  the comparison of the measured $R_{AA}$ with the extended DGLV and BDMPS models when the contribution of 
bottom quark is neglected~\cite{bib:bdmp_e,bib:exdglv}.
In Fig~\ref{fig:chap2_charm}, the parameter in BDMPS $\hat{q}$ is 14~GeV$^2$/fm and that in extended DGLV $dN^g/dy$ is 1000.

Of course, this assumption is too extreme  and may be incorrect.
However, this model suggests that determination of the fraction of bottom in single non-photonic electrons is necessary to 
interpret the result of energy loss about heavy flavor and to extract the property of the medium from the result.
The determination of the fraction of bottom in single non-photonic electrons is the most important subject in order to
understand the behavior of heavy quarks in the hot and dense matter produced in Au+Au collisions at RHIC.
It is the motivation of this thesis.

  \chapter{The Experimental Setup}\label{chap3}
	
The data analyzed in this thesis are $p+p$ collisions at $\sqrt{s} = 200$~GeV at the BNL Relativistic Heavy Ion Collider
and are collected with the PHENIX detector using its two central arm spectrometers.
Each spectrometer covers $\mid \eta \mid <$~0.35 in pseudorapidity 
and $\Delta \phi$~=~$\pi/2$ in azimuth in a nearly back-to-back configuration.
The capability of charged particle tracking and electron identification is necessary to measure single electrons and 
a correlation in electron-hadron pairs.
The arms include drift chambers~(DC) and pad chambers~(PC1,2,3) for charged particle tracking, 
a ring imaging $\check{C}$erenkov detector~(RICH) for electron identification, 
and an electromagnetic calorimeter~(EMCal) for electron identification and 
triggering~(ERTLL1).
Beam-beam counters~(BBCs), positioned at pseudorapidity
3.1~$<\mid \eta \mid <$ ~3.9, measure the position of the collision
vertex along the beam ($z_{vtx}$) and provide the interaction
trigger~(BBCLL1). 
In this chapter, the accelerator complex at BNL and the details of PHENIX detectors  are presented.
 \section{Accelerator Complex at BNL}
 
\begin{figure}[p]
  \begin{center}
    \includegraphics[width=13cm]{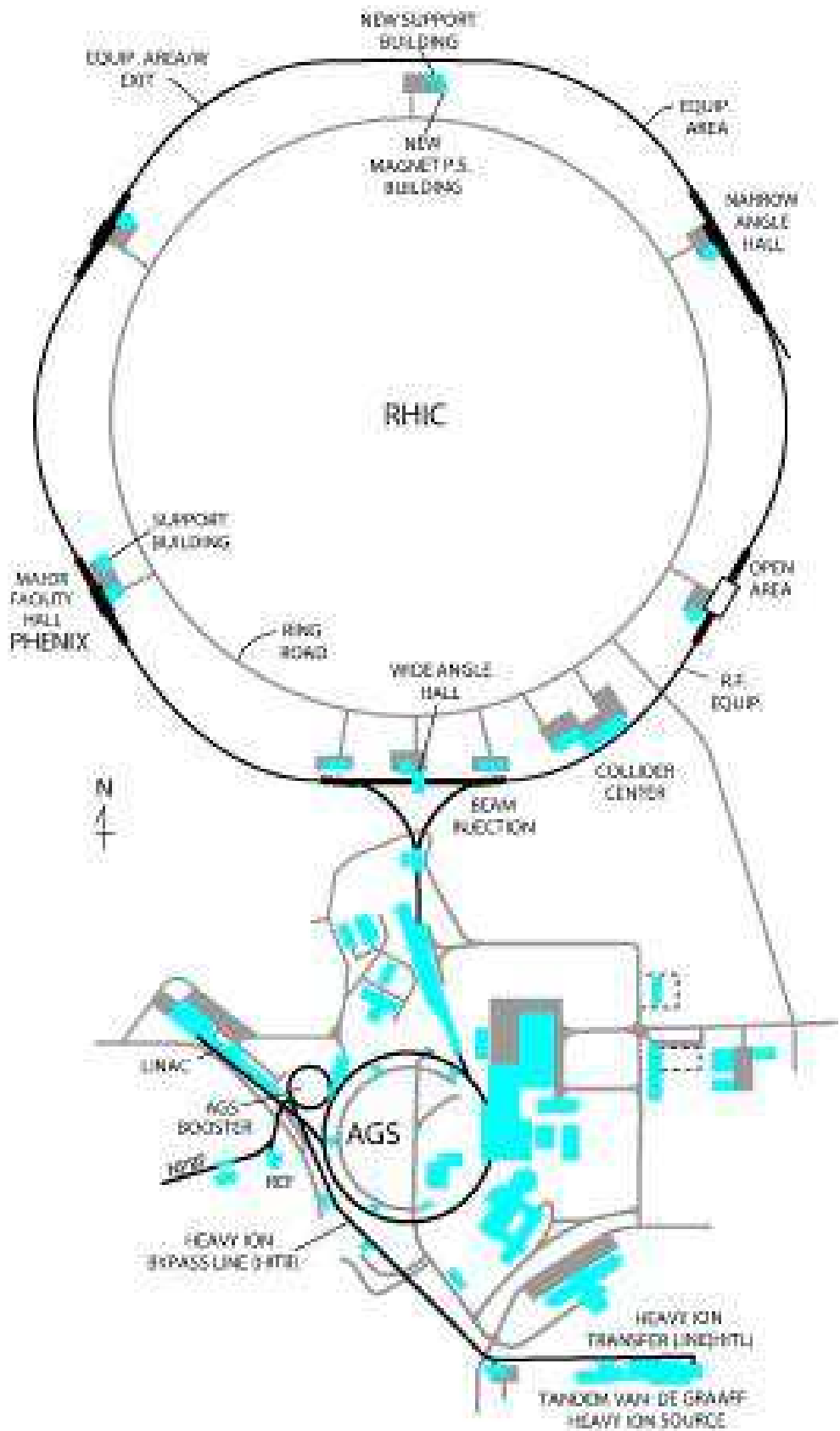}
   \caption{Accelerator Complex at Brookhaven National Laboratory.}
     \label{fig:RHIC}
  \end{center}
 \end{figure}

The Relativistic Heavy Ion Collider~(RHIC) is a colliding-type accelerator 
at BNL  to study the extreme hot and dense matter.
The RHIC started its operation in 2000~\cite{bib_rhic}. 
Figure~\ref{fig:RHIC} shows the layout of the RHIC accelerator complex.
The accelerator complex consists of  Tandem Van de Graaff facility, 
Linear Accelerator (LINAC) facility, Booster Accelerator, Alternating
Gradient Synchrotron (AGS), and Relativistic Heavy Ion Collider (RHIC).
 RHIC can accelerate form protons~(p) to gold~(Au)ions at the maximum center 
of mass energy of 500~GeV in $p+p$ collisions and 200~GeV per nucleon pair 
in Au+Au collisions. 
 
  \subsubsection{Tandem Van de Graaff Facility}
  
  The heavy ion beam is supplied by the Tandem Van de Graaff facility.
  The facility consists of two 15 MeV  electro-static
  accelerators (MP-7 and MP-8), each of which  is about 24 meters long and 
  aligned end-to-end.    

   A pulsed sputter ion source is used as an injector of 
   Tandem Van de Graaff and provides a 500~$\mu$s long pulse of Au$^{-}$
   with the peak intensity of 290~$\mu$A.
   Then the gold beam  are accelerated from ground to +14 MV potential. They pass through
   a stripping foil in the high voltage terminal in the middle of Tandem. The partially
   stripped ions are accelerated back to ground potential and are selected charge state
   of Au$^{12+}$. When Au ions are accelerated, another carbon foil at the exit of
   Tandem is used to strip electrons and to make higher charged ion, Au$^{32+}$.
    
   \subsubsection{Linear Accelerator (LINAC) Facility}
  
  The polarized (70--80~\%) or un-polarized proton beam is supplied by Linear 
  Accelerator (LINAC) facility,
  which consists of Optically Pumped Polarized H$^-$ Ion Sources (OPPIS),
  a radio-frequency quadrupole (RFQ)
  pre-injector,  and nine radio-frequency cavities.
  The LINAC is capable to produce up to  35~mA proton beam at
  the energy of 200~MeV.
  The beam intensity is 15$\times 10^{11}$ proton/pulse at the ion source
  and 6$\times 10^{11}$ at  the end of LINAC.
  The beam is injected into the Booster Accelerator for further
  acceleration.

  \subsubsection{Booster Accelerator}
   
   The Booster Accelerator facility accepts the beam  from Tandem Van de
   Graaff facility,  or the beam from LINAC.
   It is used as a  pre-accelerator.
   The machine  circumference is 200~m.
   In  case of gold beam, the booster has the capability to accelerate the beam to the energy of 
   72~MeV/nucleon and strip the beam to +77 charge state.
      
  \subsubsection{Alternating Gradient Synchrotron (AGS)}
   
   The Alternating Gradient Synchrotron (AGS) accepts the beam from Booster
   and is served  as an injector for the RHIC.
   The AGS has the   circumference of  800~m.
   
   The AGS employed the concept of alternating gradient focusing, 
   in which the field gradients of the accelerator's 240 magnets are
   successively alternated inward and outward, permitting particles 
   to be propelled and focused in both the horizontal and vertical 
   plane at the same time.
   
   In the AGS, the gold beam is stripped to +79 charge state and is
   accelerated to  an energy of 10~GeV/nucleon.
   The maximum energy of polarized proton beam is 24.3~GeV.
   The beam is delivered to RHIC via AGS to RHIC Line (ATRL). 
   
  \subsection{Relativistic Heavy Ion Collider (RHIC)}
   
   There are two rings in RHIC, each of which  has circumference of 3.83~km.
   They are  called Blue Ring  (circulating clockwise) and Yellow Ring 
   (circulating counter-clockwise).
   Rings are kept to be the vacuum of 5$\times$10$^{-11}$ Torr.
   Each ring has 192 superconducting dipole magnets with the magnet
   field of 3.46~T.
   Also, 12 common dipole magnets, 492 quadrupole magnets and 852 trim
   or corrector magnets are used.
   For $p+p$ collisions, 2 superconducting Helical Siberian Snakes were
   installed in each   ring to avoid the depolarization during the acceleration of the
   polarized protons~\cite{bib:siberian}. 

   The RHIC accepts the heavy ion or proton beams from AGS and store them. 
   The designed maximum  energy is $\sqrt{s_{NN}}$ = 200~GeV for Au+Au and 
   $\sqrt{s}$ =500~GeV  for $p+p$, respectively.
   The designed maximum luminosity is
   2$\times10^{26}$cm$^{-2}$sec$^{-1}$ for gold beam and 
   2$\times10^{32}$cm$^{-2}$sec$^{-1}$ for proton beam.
   Each bunch has the length of  5~ns.
   In $p+p$ run, the polarization of $\sim$60~\% was achieved at Year~2006/2007.
   Table.~\ref{tab:luminosiy} summarize the major parameters
   achieved in Year-2005 and Year-2006 Run in $p+p$ collisions.
   
   The RHIC collides two beams head-on using DX dipole magnet 
   at six interaction regions.
   Four of the interaction regions are equipped with the experiments: 
   PHENIX, STAR, BRAHMS, and PHOBOS.
   
   \begin{table}[hbt]
     \begin{center}
       \caption{The parameters of RHIC accelerator in Year-2005 and Year-2006 $p+p$ RUN.}
       \label{tab:luminosiy}
       \begin{tabular}{c|cc}
	 \hline
	 parameter & Year-2005 & Year-2006 \\
	 \hline\hline
	 beam energy~(GeV) & 100 & 100 \\
	 revolution frequency~(kHz) & 78&  \\ 
	 number of bunches &106 & 111\\
	 number of particles/bunch~($10^{11}$) & 0.9 & 1.35\\
	 emittance~($\beta^*=1$)(mm~mrad) & 28 & 18\\
	 peak luminosity~($10^{30}$~cm$^{-2}$s$^{-1}$) & 10 & 35\\
	 average luminosity~($10^{30}$~cm$^{-2}$s$^{-1}$) & 6 & 20\\
	 \hline
       \end{tabular}
     \end{center}
   \end{table}

   \clearpage
 \section{PHENIX Experiment}
 The PHENIX experiment is one of the large-scale  experiments 
 at RHIC~\cite{bib:cdr, bib:phenix98, bib:phenix_nim}.
 The PHENIX is designed to measure a wide variety of physics observables 
 of QGP formation as  possible.
 \subsection{PHENIX Global Coordinate System}\label{sec:coordinate}
 
 The PHENIX global coordinate system defines the geometrical center of 
 the interaction point as the origin (0,0,0).
 Figure~\ref{fig_axis} shows the definition of global coordinate 
 system used in the PHENIX experiment.
 Taking the  beam-line as a $z$-axis (North is positive $z$ direction),
 the direction to west arm is defined as $x$-axis, and upward is defined
 as $y$-axis.
 The azimuthal angle $\phi$ is measured counter-clockwise 
 relative to the  positive $x$ direction, and the 
 negative $x$ direction is $\phi$ = 180 degrees.
 The polar angle $\theta$ is defined as the angle relative to $z$-axis.
 Using the polar angle $\theta$, the  pseudo-rapidity variable is
 expressed  as:
 \begin{equation}
  \eta = - \ln\tan(\frac{\theta}{2}).
 \end{equation}

 \subsection{Detector Overview}\label{sec:detec}
\begin{figure}[p]
   \begin{center}
     \includegraphics[width=13cm]{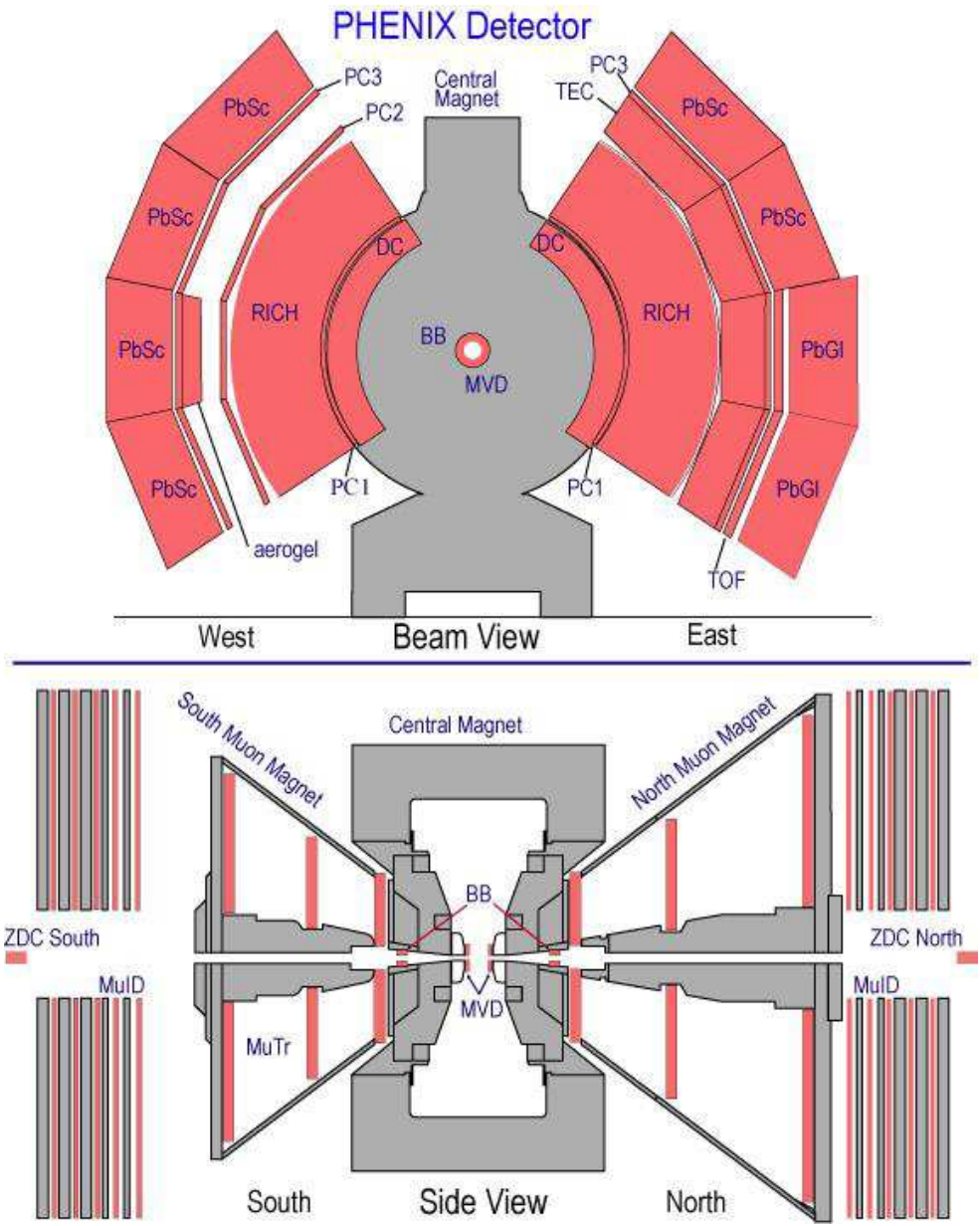}
    \caption{Experimental Layout of PHENIX Detector in Year-2005 and 2006 Run.
      TOP) Beam view : Inner detectors, two Central Arms, and Central Magnet
    are shown. BOTTOM)Side view : Inner detectors, two Muon Arms, Central
    Magnet, and Muon Magnets are shown.  \label{fig:phenix}}
   \end{center}
  \end{figure}

\begin{figure}[h]
   \begin{center}
    \epsfig{figure=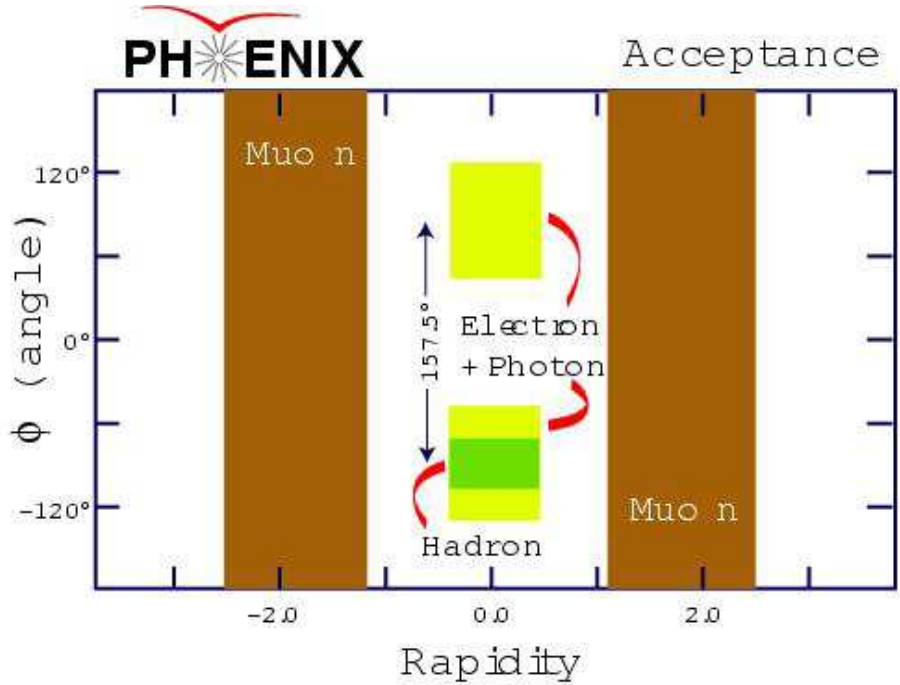,width=12cm}
    \caption{The PHENIX acceptance in terms of rapidity and $\phi$ angle.
    \label{fig:acceptance}}
   \end{center}
  \end{figure}

\begin{figure}[h]
   \begin{center}
     \includegraphics[width=13cm]{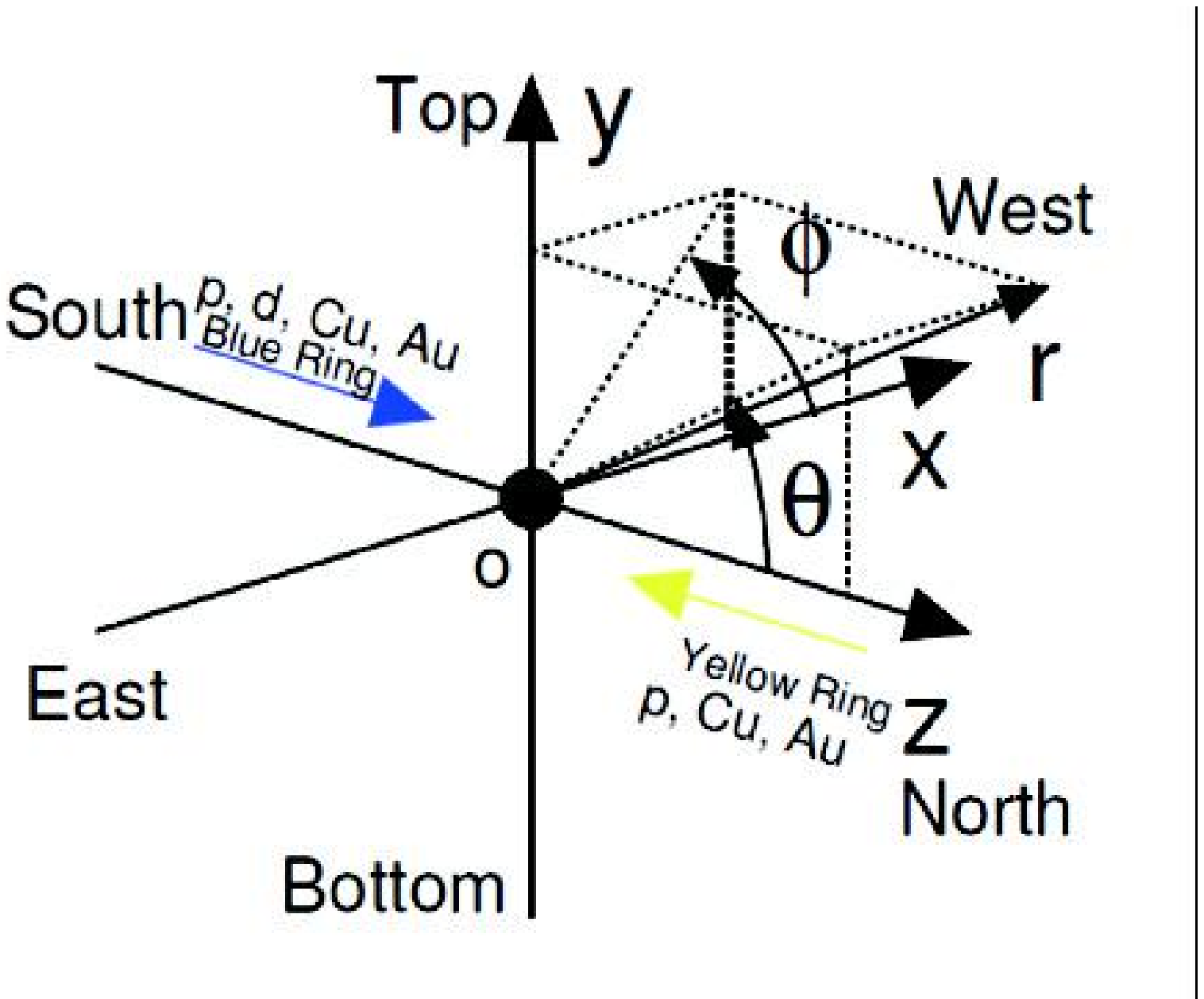}
    \caption{Definition of global coordinate system used in the PHENIX}
    \label{fig_axis}
   \end{center}
  \end{figure}

 Figure~\ref{fig:phenix} illustrates the experimental layout of PHENIX
 detectors in RUN Year-2005 and Year-2006.
 Both beam view (top) and side view (bottom) are shown.
 The PHENIX consists of trigger counters, a central magnet, two muon
 magnets, two central arms, and two muon arms.
 The acceptance coverage, and the function of the detector 
 subsystems are briefly summarized in Table~\ref{tab:detector}.
 The details about detectors are presented in the following  sections.
 The pseudo-rapidity coverage of PHENIX Central Arm and Muon Arm are 
 $\mid \eta \mid  < 0.35$ and $1.2<\mid \eta \mid<2.4$, respectively. 
 Around the interaction point, the beryllium beam pipe is used and its
 diameter and thickness are 76mm and 1mm, respectively.
 \noindent 
 Figure~\ref{fig:acceptance} shows the acceptance of both central arm 
 and muon arm in the pseudo rapidity - $\phi$ angle plane. 
 The measurement of electron is performed by using 
 the central arm.

  \begin{table}[p]\small
   \begin{center}
   \caption{Summary of the PHENIX Detector Subsystems.}
   \label{tab:detector}
    \begin{tabular}{|c|c|c|c|}
     \hline
     \hline
     Element & $\Delta\eta$ & $\Delta\phi$ & Purpose and Special Feature\\
     \hline
     \hline
     Central Magnet (CM) & $\pm$0.5 & ${\rm 360^o}$ & Up to 1.0 T$\cdot$m.\\
     Muon Magnet (MM)    & $\pm$1.1-2.5 & ${\rm 360^o}$ & 0.72 T$\cdot$m for $\eta$=$
     \pm$2, 0.36 T$\cdot$m for $\eta$=$\pm$1.3.             \\
     \hline
     BBC & $\pm$(3.0-3.9) & ${\rm 360^o}$ & Start timing, fast vertex.\\
     Drift chambers (DC) & $\pm$0.35 & ${\rm 90^o \times 2}$ & Good momentum and mass
     resolution,\\
     &&& $\Delta p/p = 1.0\% \; at\; p=1GeV/c$.\\
     Pad Chamber (PC) & $\pm$0.35 & ${\rm 90^o \times 2}$ & Pattern recognition,\\
     &&& tracking for non-bend direction.\\
     TEC & $\pm$0.35 & ${\rm 90^o \times 2}$ & Pattern recognition, $dE/dx$.\\
     \hline
     RICH & $\pm$0.35 & ${\rm 90^o \times 2}$ & Electron identification.\\
     \hline
     TOF & $\pm$0.35 & ${\rm 30^o}$ & Good hadron identification, $\sigma <100 ps$.\\
     \hline
     PbSc & $\pm$0.35 & ${\rm 90^o \times 1.5}$ & Photon detection.\\
     PbGl & $\pm$0.35 & ${\rm 45^o}$ & Photon detection.\\
     \hline
     Muon chambers ($\mu$T) & $\pm$1.2-2.4 & ${\rm 360^o}$ & Tracking for muons.\\
     Muon identifier ($\mu$ID)& $\pm$1.2-2.4 & ${\rm 360^o}$ & Concrete absorbers and
     chambers for\\
     &&& $\mu$/hadron separation.\\
     \hline
    \end{tabular}
   \end{center}
  \end{table}


\clearpage
 
  \section{The Trigger Counters}\label{sec:trigger}
  
  The  PHENIX trigger system consists of Beam-Beam Counters (BBC)
  and Zero-Degree Calorimeters (ZDC).
  The ZDC is the common trigger device among four RHIC experiments.
  The details about BBC and ZDC are presented  in this section.
  
  \subsection{Beam-Beam Counters (BBC)}\label{sec:bbc}
  The Beam-Beam Counters (BBC) provide the measurement of 
  collision point and start signal for time of flight 
  measurement~\cite{bib:bbc}.
  
\begin{figure}[p]	
   \begin{center}
    \epsfig{figure=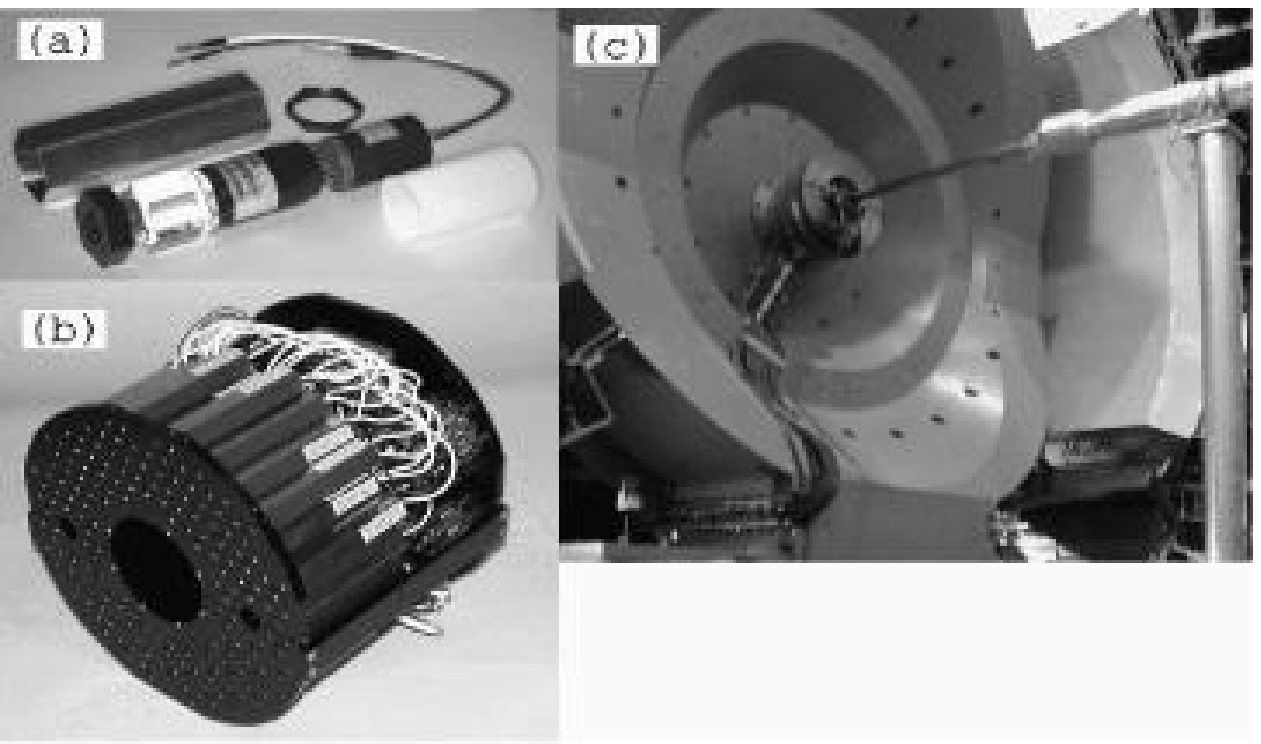,width=14cm}
    \caption{(a) Single BBC. (b) A BBC array comprising 64 BBC elements.
    (c) The BBC is shown mounted around beam pipe and 
    just behind the Central Magnet.
    \label{fig:bbc}}
    \vspace{1.5cm}
    \epsfig{figure=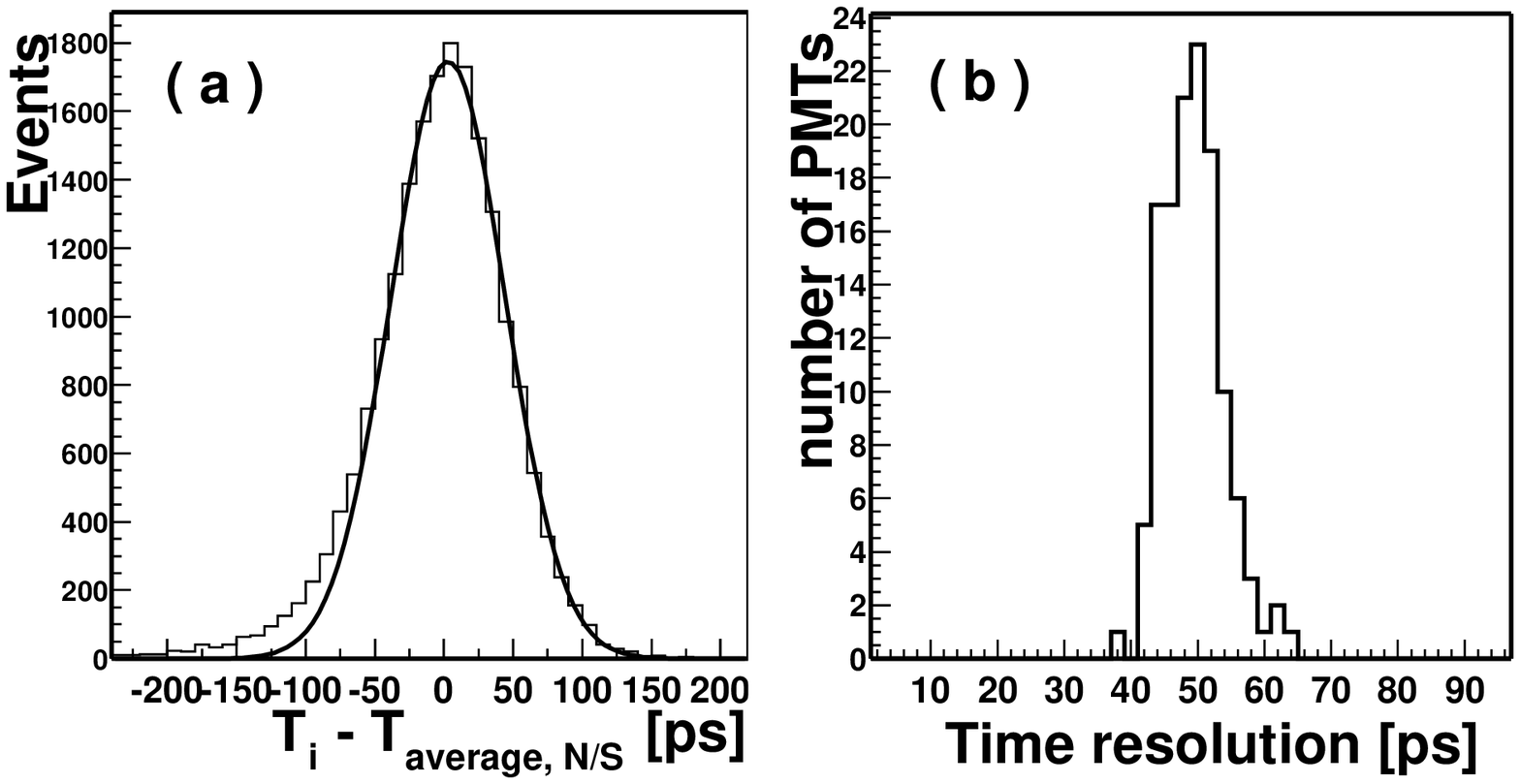,width=17cm}
    \caption{(a) The distribution of timing deviation for a typical
    BBC element from averaged hit timing of all BBC element.
    (b) The profile of the timing resolution for each BBC element. 
    \label{fig:bbc2}}
   \end{center}
  \end{figure}

  There are two arrays of BBC in PHENIX along the beam line.
  They are  quartz Cherenkov detectors which locate at 144~cm from the 
  interaction point and surround the beam axis with the 10~cm of inner 
  diameter and 30~cm of outer diameter.
  The pseudo-rapidity coverage
  is  $\eta = \pm (3.0 - 3.9)$.
  Each counter consists of 64 one-inch diameter mesh-dynode (15 step)
  Photo Multiplier Tubes (PMT:Hamamatsu R6178)  equipped with 3~cm quartz
  on the head of PMT as a Cherenkov radiator.
  Figure~\ref{fig:bbc} shows the pictures of the (a) single BBC, (b) a
  BBC array,   and (c) BBC location.

  The BBC detects the particles such as $\pi^{+/-}$ from collisions. 
  The start signal for timing measurement ($T_{0}$) and 
  the beam-beam collision point along beam axis ($z$-vertex) are provided
  by using the average arriving time of  particles between North
  ($T_{1}$) and South BBC array ($T_{2}$) and  their difference.
  They are calculated as:
  
  \begin{eqnarray}
   T_{0} &=& \frac{(T_{1} + T_{2})}{2} - \frac{|z_{bbc}|}{c} + t_{offset},\\
   z  {\rm -vertex}& =& \frac{c(T_{1} - T_{2})}{2},
  \end{eqnarray}
  
  \noindent where $z_{bbc}$ is the BBC position of 144~cm,
  $c$ is the velocity of light, and $t_{offset}$ is the time offset
  intrinsically introduced by devices.
  
  Figure~\ref{fig:bbc2}~(a) shows the distribution of timing deviation   
  of a typical BBC element from the BBC average time. 
  Figure~\ref{fig:bbc2}~(b) shows the distribution of time resolution 
  over all BBC elements.  
  The time resolution of a single BBC element was 52$\pm$4 ps (rms) and
  the $z$-vertex resolution was $\sim$0.5~cm under the experimental condition.

  \subsection{Zero-Degree Calorimeters (ZDC)}
  The Zero-Degree Calorimeters~(ZDC) is a hadron calorimeter designed to
  detect  the forward neutrons and measure their total 
  energy~\cite{bib:zdc, bib:zdc2}.
  The ZDC's have the angular acceptance of $|\theta| < 2$~mrad.
  There are two ZDC's in PHENIX. 
  
\begin{figure}[h]	
   \begin{center}
    \epsfig{figure=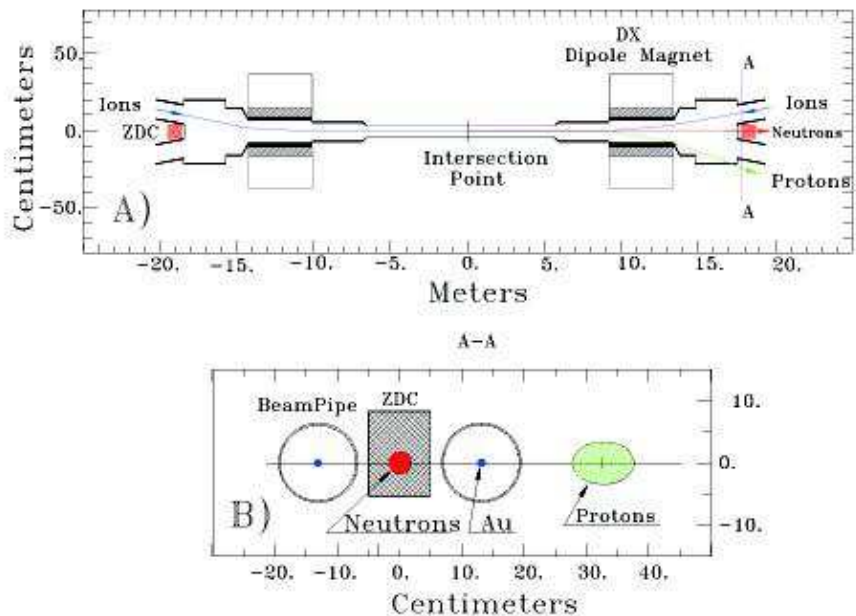,width=14cm}
    \caption{The ZDC location. A)Top view  and B)beam view of ZDC location.
    Charged fragments are bended by Dipole Magnets towards the
    outside of the acceptance of ZDC.	
    \label{fig:zdc}}
   \end{center}
  \end{figure}

  They are sampling type hadron calorimeters which
  positioned at  18~m from the interaction point and sit just behind 
  the DX dipole magnet as shown in Fig.~\ref{fig:zdc}.
  The DX dipole magnets serve to bend the incoming beams to the
  colliding region   and outgoing beams to the collider beam
  line~\cite{bib:DX}.
  Because of the magnetic bending by  DX dipole magnet, only the
  neutrons can  reach to ZDC.
  
  Each ZDC consists of 27 layers of tungsten absorbers and 
  fibers which are connected to a  PMT (Hamamatsu R329-2).
  Taking the correlation between North and South provides the background 
  rejection due to single-beam interaction with the residual gas in 
  the beam pipe. 
  The energy resolution of ZDC was obtained to be  $\delta E/ E \simeq
  218 /\sqrt{E({\rm GeV})}$~\% from the test beam results.

  \section{The Central Magnet}\label{sec:magnet}
\begin{figure}[htb]
   \begin{center}
     \includegraphics[width=13cm]{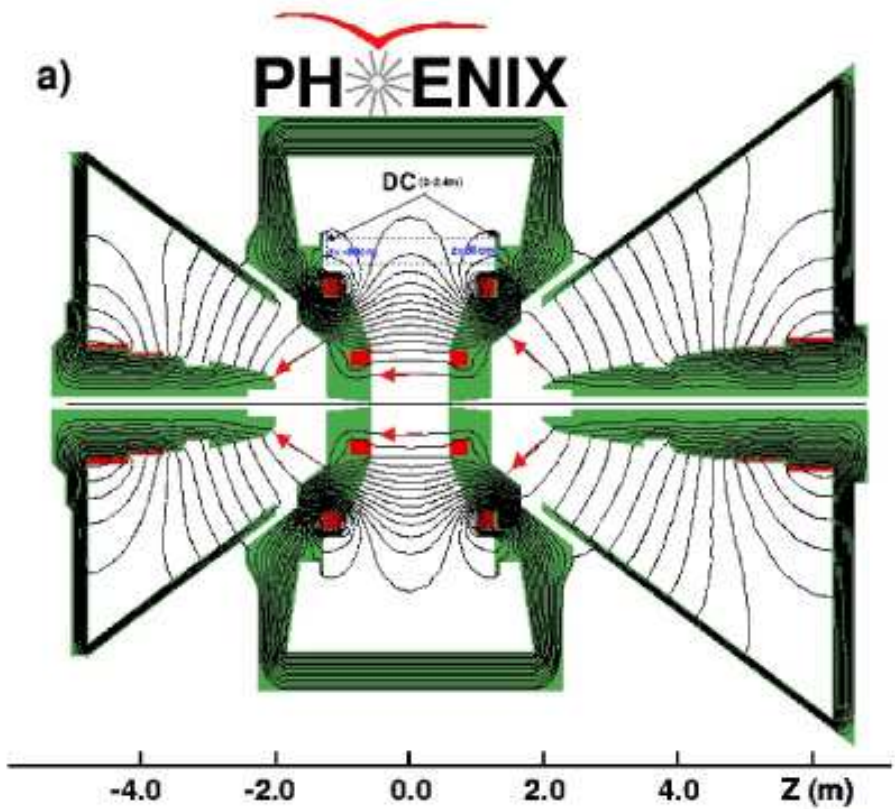}
     \caption{The field lines of the central magnet and muon magnets shown on a vertical
       cutaway drawing of the PHENIX magnets. The beams travel along the $r=0$-axis in
       this figure and collide at $r=z=0$. Arrows indicate the field direction}
     \label{fig:mag1}
   \end{center}
 \end{figure}

 The PHENIX Central Magnet (CM) provides the 
 axial field of $\int B\cdot dl$ = 0.78 T$\cdot$m at $\theta =$ 90 degrees.
 It is used to determine the momentum of charged particles using
 magnetic bending.
 The CM is energized by two, inner and outer, pairs of concentric coils,
 which can be operated separately, with the same polarities or opposite polarities.  
 In RUN Year-2005 and Year-2006, the same polarity operation
 was chosen.
 The operation is called as CM++ or CM-~- according to the polarity
 of the magnetic field.
 The magnetic field produced by the magnets is shown in Figure~\ref{fig:mag1}.
 The pole faces of the magnet are positioned at $\pm$45~cm in $z$ direction
 covering the rapidity range of  $\pm$0.35.
 The CM pole tips also serve as the hadron absorbers
 for the muon spectrometers.
 \clearpage
 \section{The Central Arms}\label{sec:central_arm}
 The PHENIX central arms are designed to detect charged particles and photons.
 The central arm consists of tracking devices, particle identification
 devices, and calorimetry devices.
 In this section, the details about central arm devices are reviewed.
 
 \subsection{Drift Chambers (DC)}
  
  The Drift Chambers (DC) are used to measure the charged particle 
  trajectories in $r-\phi$ plane and to provide the high resolution
  momentum determination.
  The requirements for DC are
  (1) single wire resolution better than 150~$\mu$m in r-$\phi$,
  (2) single wire two track separation better than 1.5~mm,
  (3) single track efficiency greater than 99~\%, and 
  (4) spatial resolution in $z$-direction better than 2~mm.
  
  The DC system consists of two independent gas volumes located in the west 
  and east arms, respectively.
  DC's are filled with the gas mixture of 50~\% Argon and 50~\% Ethane.
  The DC's are placed between 2.02 and 2.46~m  in radial
  distance from the interaction point for both West and East arm.
  They occupy 180~cm in $z$-direction and 90 degrees per arm in azimuth.
  
\begin{figure}[h]	
   \begin{center}
    \epsfig{figure=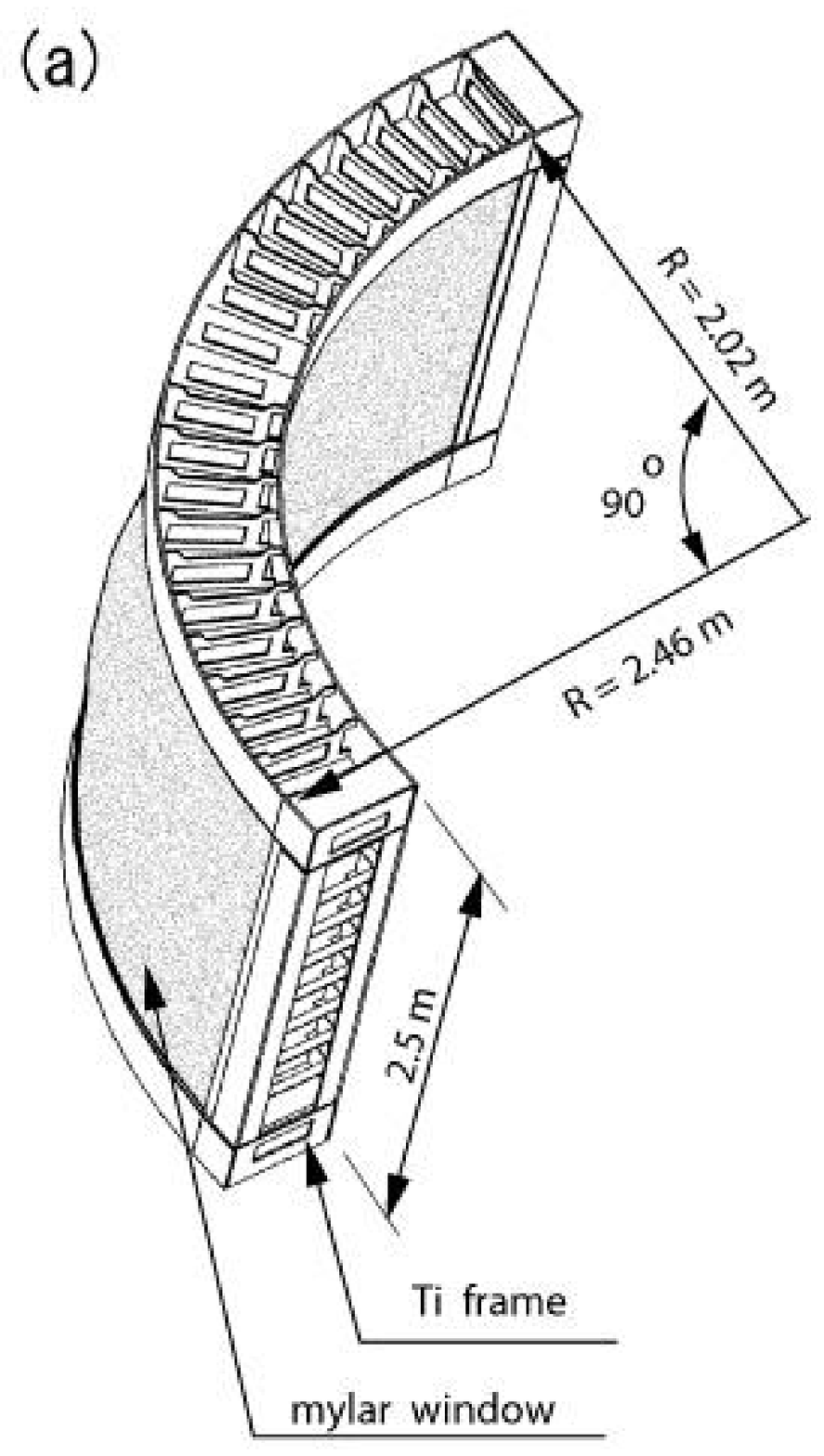,width=5.5cm}
    \epsfig{figure=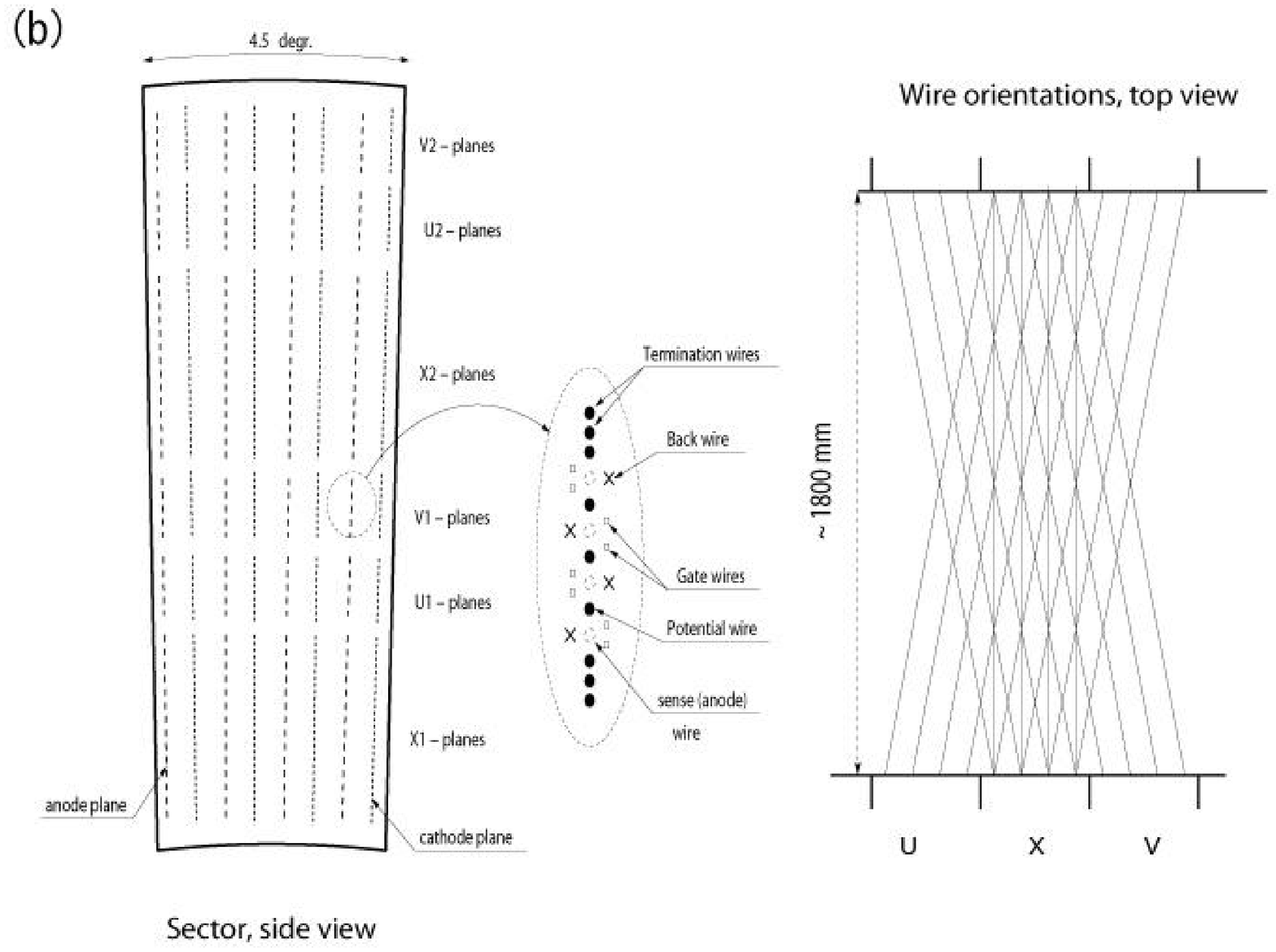,width=11cm}
    \caption{(a) The construction of a DC frame.
    (b)The layout of wire position within one sector of DC.	
    \label{fig:dc2}}
   \end{center}
  \end{figure}

  Figure~\ref{fig:dc2} (a) shows the construction of a cylindrical
  titanium frame of a DC.
  Each DC consists of 20 sectors, each of which  covers 
  4.5 degrees in azimuth.
  In each sector, there are six types of wire modules stacked radially.
  They are called X1, U1, V1, X2, U2, V2.
  The sketch of a sector and the layout of wire position are shown in
  Figure~\ref{fig:dc2} (b).

  The X1 and X2 wires run in parallel to the beam axis in order 
  to perform the track measurements in $r-\phi$ plane.
  They are followed by two sets of small angle U, V wire planes.
  U1, V1, U2, and V2 wires have stereo angle of  about 6 degrees relative to 
  the X wires in order to measure  the $z$-coordinate of the track.   
  The X wire modules contain 12 sense (anode) planes and 4 cathode
  planes. 
  Both U and V wire modules contain 4 sense (anode) planes and 4 cathode
  planes. 
  They form the cells with a 2 $\sim$ 2.5~cm drift space in $\phi$
  direction.
  In this scheme, 40 drift cells are located at different radii in the
  DC frame.

  The DC system contains roughly 6,500 anode wires.
  Each wires are separated into two halves by the Kapton support at $z =
  0$, and the signals are  independently extracted.
  Thus, the number of total readout channel is 13,000.    
  The anode wires are separated by Potential (P) wires and surrounded by 
  Gate (G) and Back (B) wires.
  The P wires form a strong electric field and separate sensitive regions of
  individual anode wires.
  The G wires limit the track sample length to roughly 3~mm and terminate the 
  unwanted drift line.
  The B wire has a rather low potential and terminates most of the 
  drift lines from side.

  With a 50-50 mixture of argon-ethane gas, the stable drift velocity
  plateau   at 53~mm/$\mu$s is achieved for the field gradation from 800
  V~cm up  to 1.4~kV~cm.
  Therefore, the maximum drift time in a cell is approximately 470~ns.

  \subsection{Pad Chambers (PC)}
  The Pad Chambers (PC) is used to determine the space points which are used
  for momentum determination in the z direction($p_{\mathrm{z}}$).
  The PC's are the multi-wire proportional chambers with cathode readout 
  that form three separate layers (PC1, PC2, PC3) in the central arm.
  PC1 layer is innermost chamber located between DC and RICH, occupying
  2.47~m through 2.52~m in radial distance from the interaction point,
  PC2 layer is placed behind RICH occupying 4.15~m through 4.21~m, and
  PC3 is located in front of EMCal occupying 4.91~m through 4.98~m.
  PC1 is essential for determination of the three dimensional momentum by
  by providing the z coordinate at the exit of the DC.
  The combination of DC and PC1 information provides  the direction 
  vector through RICH.
  PC2 and PC3 are needed to resolve ambiguities in outer detectors
  where about 30~\% of the particle striking the EMCal are produced by 
  either secondary interactions and the particle decays outside the aperture
  of DC and PC1. 
  
\begin{figure}[p]	
   \begin{center} 
    \epsfig{figure=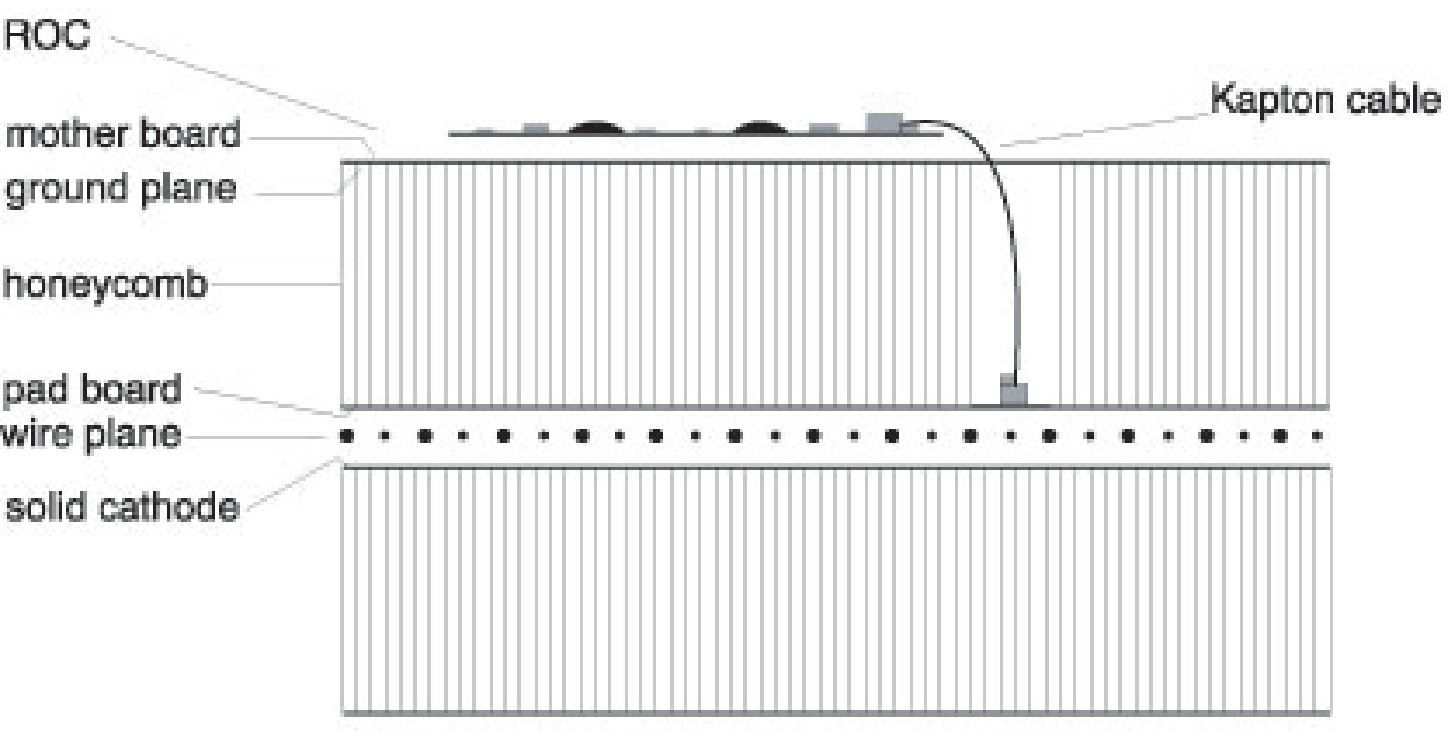,width=16cm}
    \caption{The vertical cut through a PC.\label{fig:pc2}}
    \vspace{2cm}
    \epsfig{figure=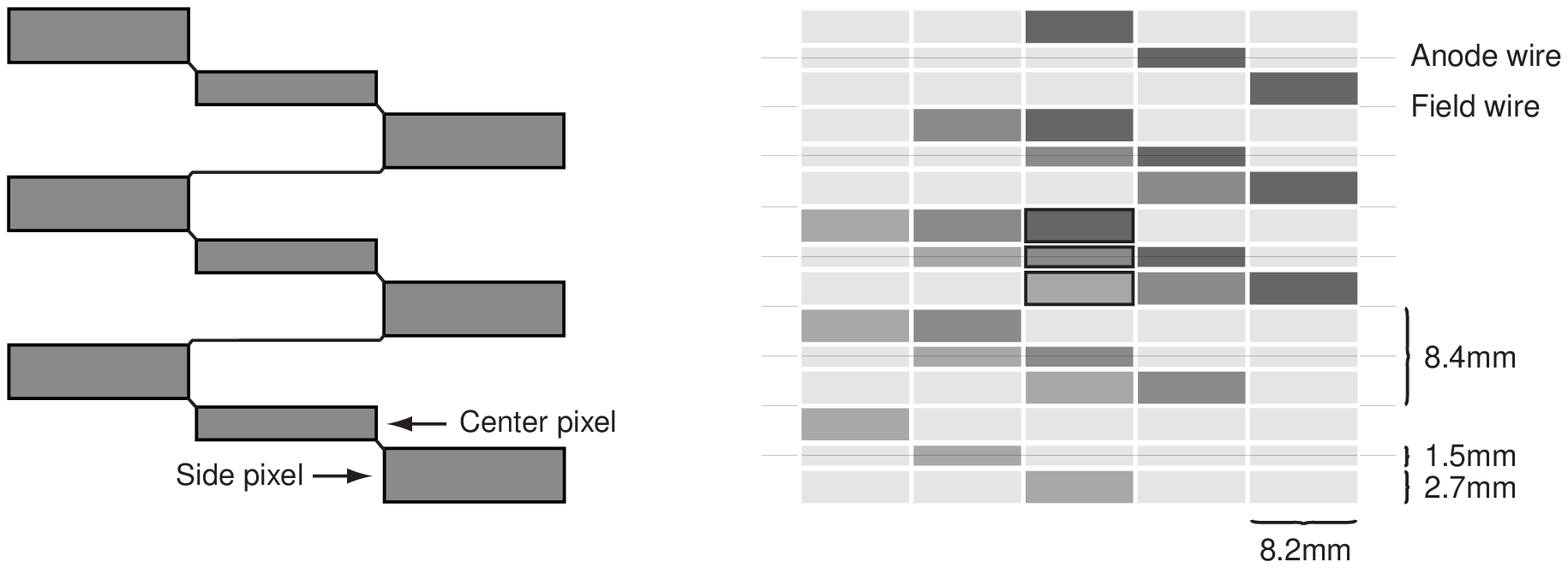,width=16cm}
    \caption{The pad and pixel geometry(left).
    A cell defined by three pixels is at the center of the 
    right picture\label{fig:pc1}}
   \end{center}
  \end{figure}

  Each detector consists of a cathode panel, single plane of anode 
  and field wires.
  Figure~\ref{fig:pc2} shows the vertical cut through of a PC.
  Each panel is fabricated as an FR4-honeycomb-FR4 sandwich.
  The wires are lying in a gas volume between two cathode planes.
  The gas was chosen to be the mixture of 50~\% Argon and 50~\% Ethane at 
  atmospheric pressure.
  One cathode is finely segmented into an array of pixels as 
  shown in Fig.~\ref{fig:pc1}.
  The position resolution was measured to be $\pm1.7$~mm for PC1 along the 
  wire ($z$-direction). 

  \subsection{Ring Imaging Cherenkov detectors (RICH)}
  
\begin{figure}
    \begin{center}
     \epsfig{figure=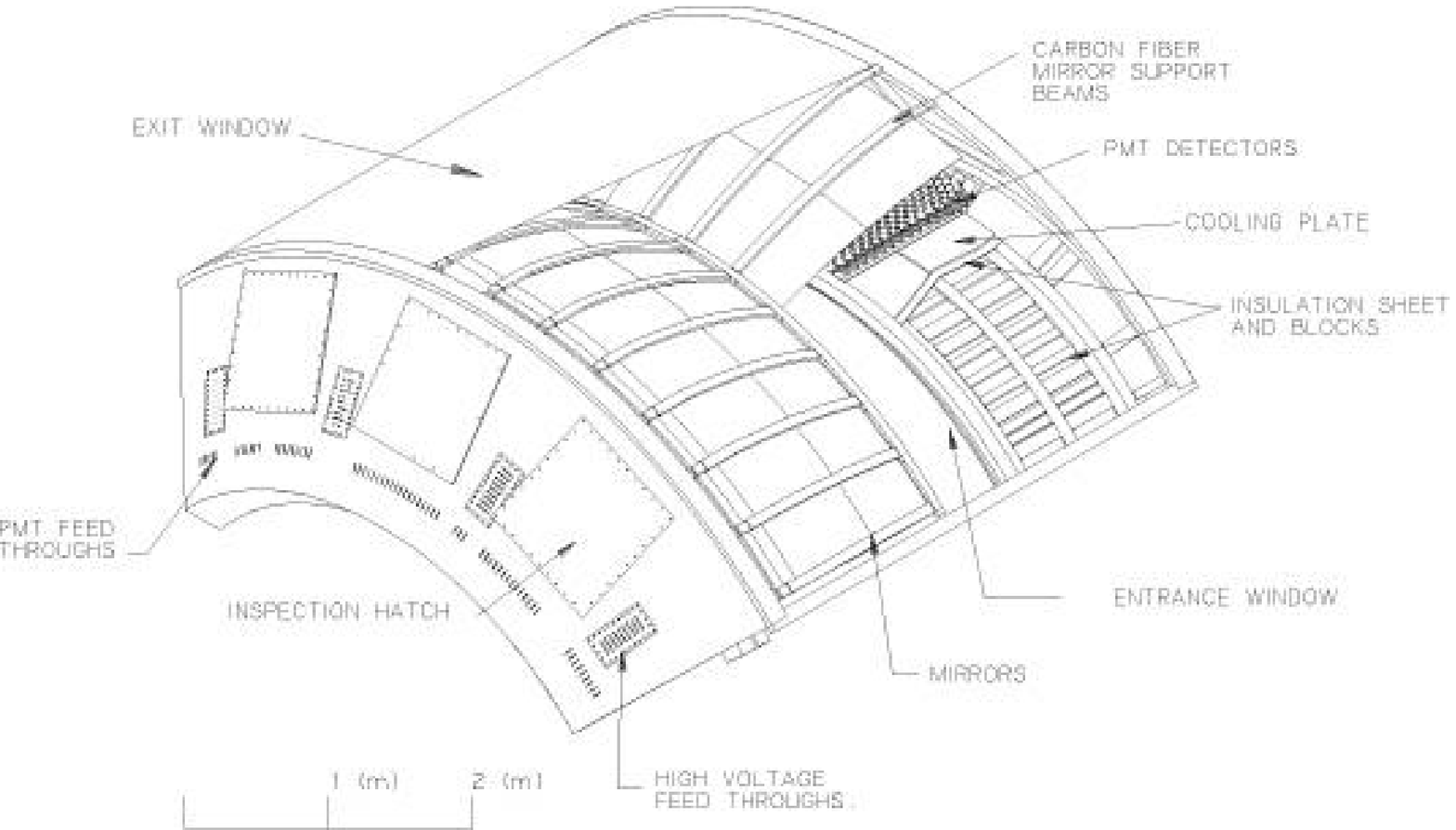,width=14cm}
     \caption{The cut-away view of RICH detector.
     \label{fig:RICH}}
     \vspace{2cm}
     \epsfig{figure=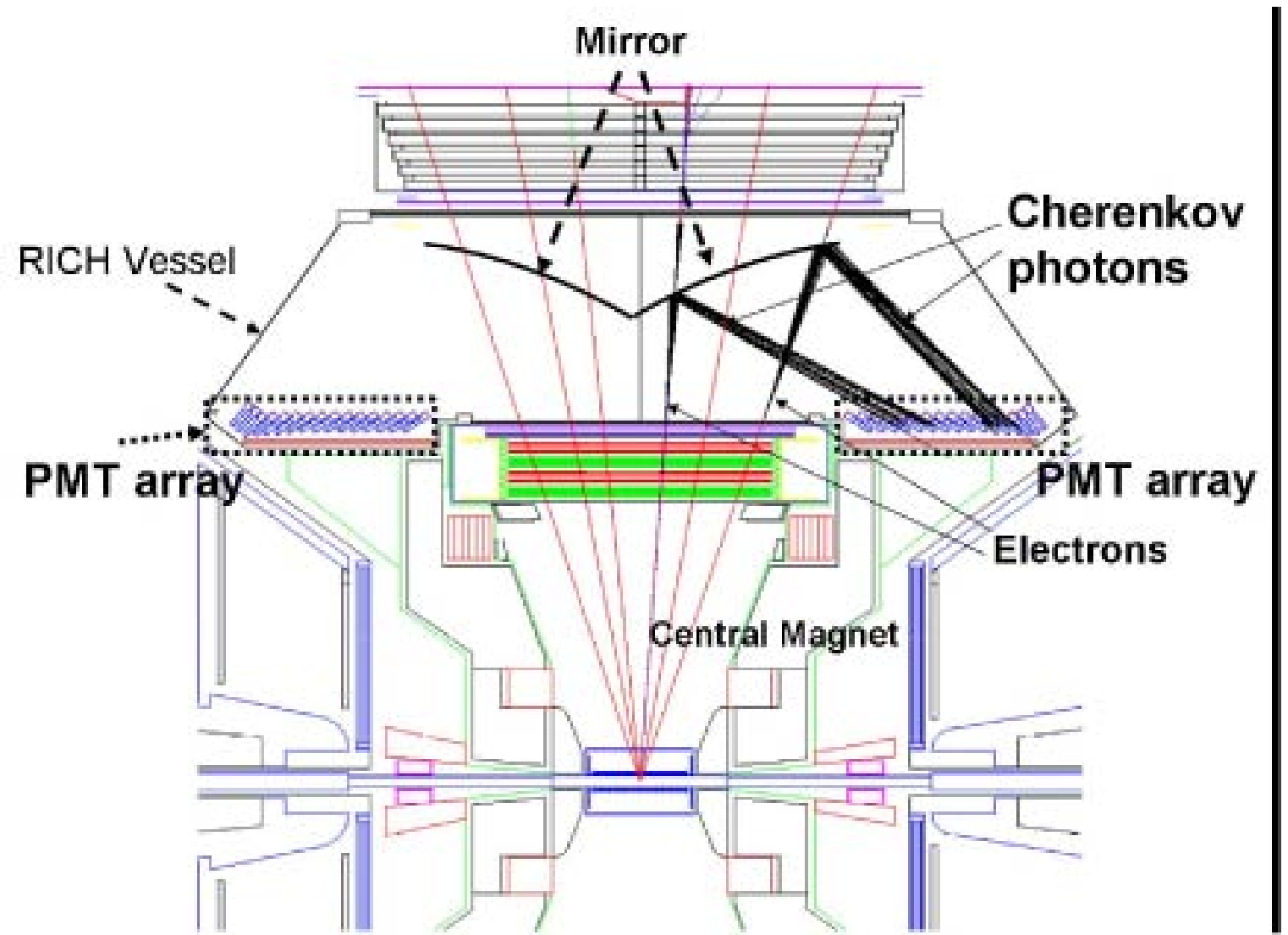,width=14cm}
     \caption{A schematic view of the RICH detector\label{fig:phot}}
    \end{center}
   \end{figure}

  The Ring Imaging Cherenkov detectors (RICH) are the primary devices  
  for the electron identification in PHENIX~\cite{bib:Akiba}.
  They are threshold gas type Cherenkov detector and provides $e/\pi$
  separation below the pion Cherenkov threshold, which is set at 4.85~\GEVC.   
  Figure~\ref{fig:RICH} shows the cut-away view of a RICH detector.
  The RICH is  placed between 2.5~m and 4.1~m in radial distance from
  the interaction point for both West and East arm. 
  They cover the 70 to 110 degrees in polar angle, and 90 degrees per arm
  in azimuth.
  \noindent
  Figure~\ref{fig:phot} shows a schematic view of the RICH.
  In the RICH gas vessel, charged particles moving faster than the speed of light
  in the gas emit Cherenkov photons. The emitted photons are reflected and focused
  by the spherical mirror on the plane of phototube array. Radiator gas filled in 
  the RICH is CO$_{2}$ gas at 1~atm.
  RICH have the pion rejection power at the order of 10$^4$ for single track.
  In high multiplicityenvironment, there are the source of miss identification such as ring
    sharing.
  
  Each  gas vessel is  fabricated from aluminum, and has a volume of about 
  40~m$^{3}$.
  The entrance and exit windows are made of aluminized Kapton with
  125~$\mu$m thick.
  
  In each gas vessel, two arrays of 24 aluminum-evaporated mirror panels
  are mounted on the graphite-epoxy mirror support.
  The arrays are located symmetrically about $z$ = 0.
  The shape of each mirror is a section of a sphere of 
  radius 4.0~m, whose center is located at $|z| = 2.0$~m.
  
  The Cherenkov photons emitted by the charged particles are reflected by
  the mirrors, and detected by two arrays of 1280 Hamamatsu H3171S UV
  photomultiplier tubes which are placed behind the Central Magnet in
  order to avoid the direct hit of the particles~\cite{bib:tamagawa}. 
   
  The PMT has a bi-alkaline photocahode and a linear-focussed 10 stage
  dynode.
  It is equipped with 2 inch $\phi$ Winston cones and magnetic
  shields that allow the operation under the magnetic field of 100~G.
  In total, RICH have 5120 PMT's (2 (arm) $\times$ 2 (side) $\times$
  16 ($\theta$) $\times$ 80 ($\phi$)).
  The angular segmentation is approximately
  1\textdegree ~$\times$ 1\textdegree~ in $\theta$ and $\phi$.
  
  The quantum efficiency is about 20~\% (5~\%) at the wavelength of
  300~nm (200~nm).
  The typical dark current is 10~nA.
  The typical operation voltage of PMT is 1.4~kV $\sim$ 1.8~kV
  and the gain is $\sim$10$^7$.

   The RICH Frond-End Electronics~(FEE) processes the signal from  5120 PMT's in total
   at each bunch crossing~(9.4MHz) and transmits digitized data to the PHENIX
   data acquisition system on the trigger signal supplied by the PHENIX global
   trigger system~($\sim$25 kHz)\cite{bib:sakag}.
   The acceptable charge range is from 0 to 10 photoelectrons, which corresponds
   to the input charge from 0 to 160 pC preceded by the preamplifier.
   Charge resolution is $\sim$1/10 photoelectron and timing resolution is  $\sim$ 240 ps.
   Both of the charge and timing outputs are stored in Analog Memory Unit~(AMU) clocked 
   at the RHIC bunch crossing frequency.
   The analog data stored in the AUM are digitized only after the receipt of an 
   acceptance from the PHENIX global trigger system.

  \subsection{Electro-Magnetic Calorimeters (EMCal)}
  The primary role of the  Electro-Magnetic Calorimeters (EMCal) is to
  provide a measurements of energy and hit position of both photons and
  electrons~\cite{bib:emcal, bib:emcal_2}.
  
  Two kinds of EMCal are installed in PHENIX as shown in Figure~\ref{fig:phenix}.
  One is lead-scintillator calorimeter (PbSc) and the other is
  lead-glass calorimeter (PbGl).
  Each arm ($\Delta \phi = 90$\textdegree) is divided into four sectors in
  azimuth. 
  The PbGl's occupy the lower two sectors of East Arm and the PbSc occupy
  other six  sectors. 
  The EMCal surface is placed at 510~cm (PbSc) and 550~cm (PbGl) in
  radial distance from the interaction  point.  
  \subsubsection{Lead-scintillator calorimeter~(PbSc)}
  The PbSc is a shashlik type sampling calorimeter made of alternating tiles of lead and
  scintillator.
  It consists of 15552 individual towers
  (5.2~cm $\times$ 5.2~cm $\times$ 37.5~cm).
  Total radiation length of PbSc is 18.2~$X_0$ and Moliere radius is
  $\sim6$~cm.
  Each tower contains 66 sampling cells: 1.5~mm of lead and 4~mm of
  injection molded scintillator, ganged together by penetrating optical
  fibers doped with wave length shifter for light collection.
  Lights are read out by 30~mm$\phi$ PMT's (FEU115, MELS, Russia) which are 
  implemented at the back of the towers.
  Figure~\ref{fig:pbsc_tower} shows a PbSc calorimeter module which is
  assembled from four towers.
  The energy resolutions were evaluated as
  follows from the tests using electron beams:
  
  \begin{eqnarray}
    \delta E/E &=& 8.1~\%/\sqrt{E ({\rm GeV})} + 2.1~\%~~~({\rm PbSc}),\\
  \end{eqnarray}

  The measured position resolution of PbSc depends on both  energy $E$ (in the
  unit of GeV) and impact angle  $\theta$ ($\theta=0$ means orthogonal
  impact).
  It is expressed as:

  \begin{eqnarray}
   \sigma_{x}(E,\theta) &=& \sigma_{0}(E)+\Delta \times {\rm sin}\theta ,\\
   \sigma_{0}(E) &=& 1.55~{\rm mm} + \frac{5.7~{\rm mm}}{\sqrt{E({\rm GeV})}},
  \end{eqnarray}
  
  \noindent where, $\sigma_{0}$ is the position resolution for
  orthogonal incidence  and $\Delta$ is given by the  radiation length
  of $\sim20$~mm.

   \subsubsection{Lead-glass calorimeter~(PbGl)}
\begin{figure}[p]
    \begin{center}
     \epsfig{figure=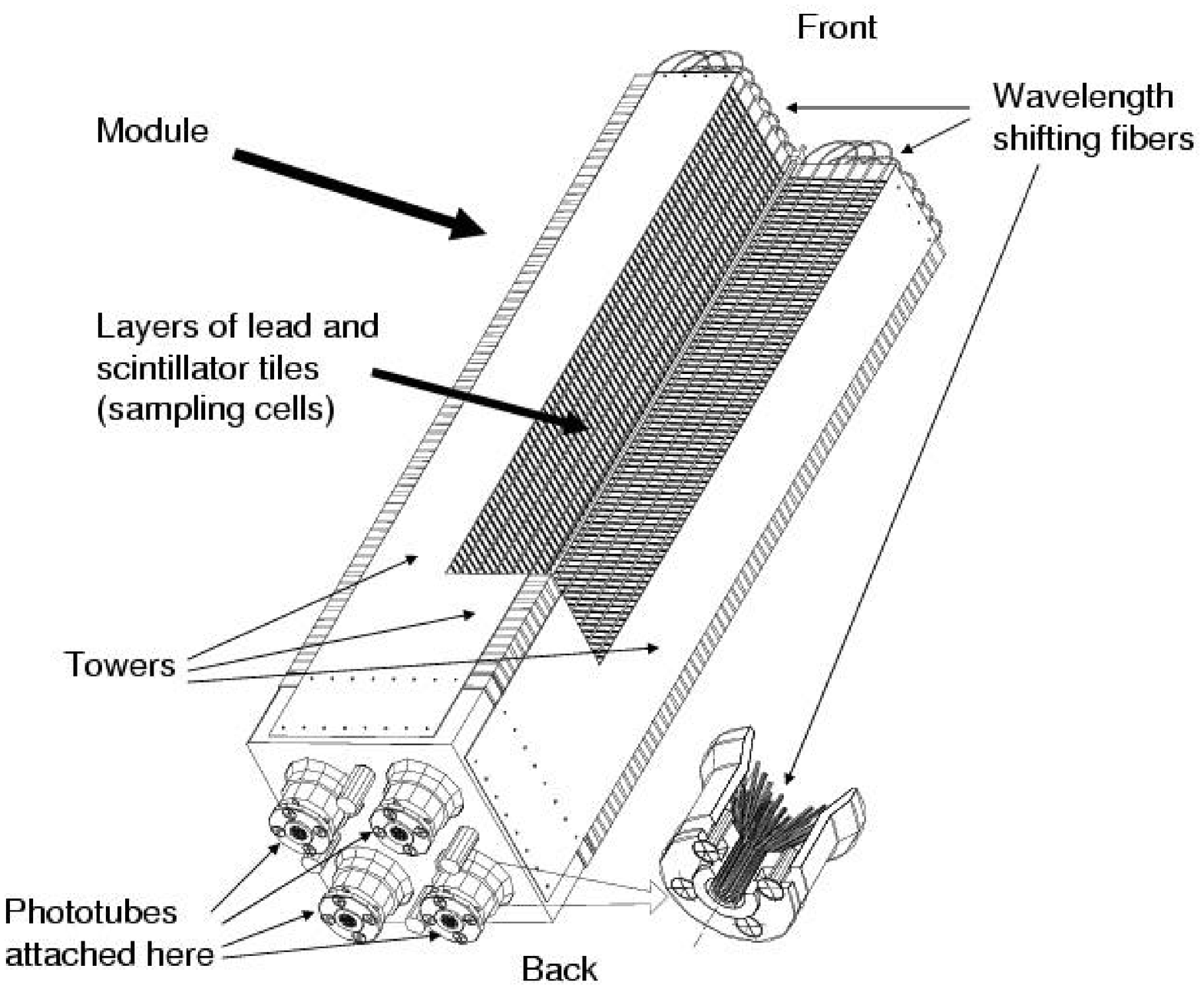 ,angle=-90,width=10cm}
     \caption{Cut-away  view of a PbSc module showing a stack of
     scintillator and lead plates, wavelength shifting
     fiber.\label{fig:pbsc_tower}}     
     \vspace{1cm}
     \epsfig{figure=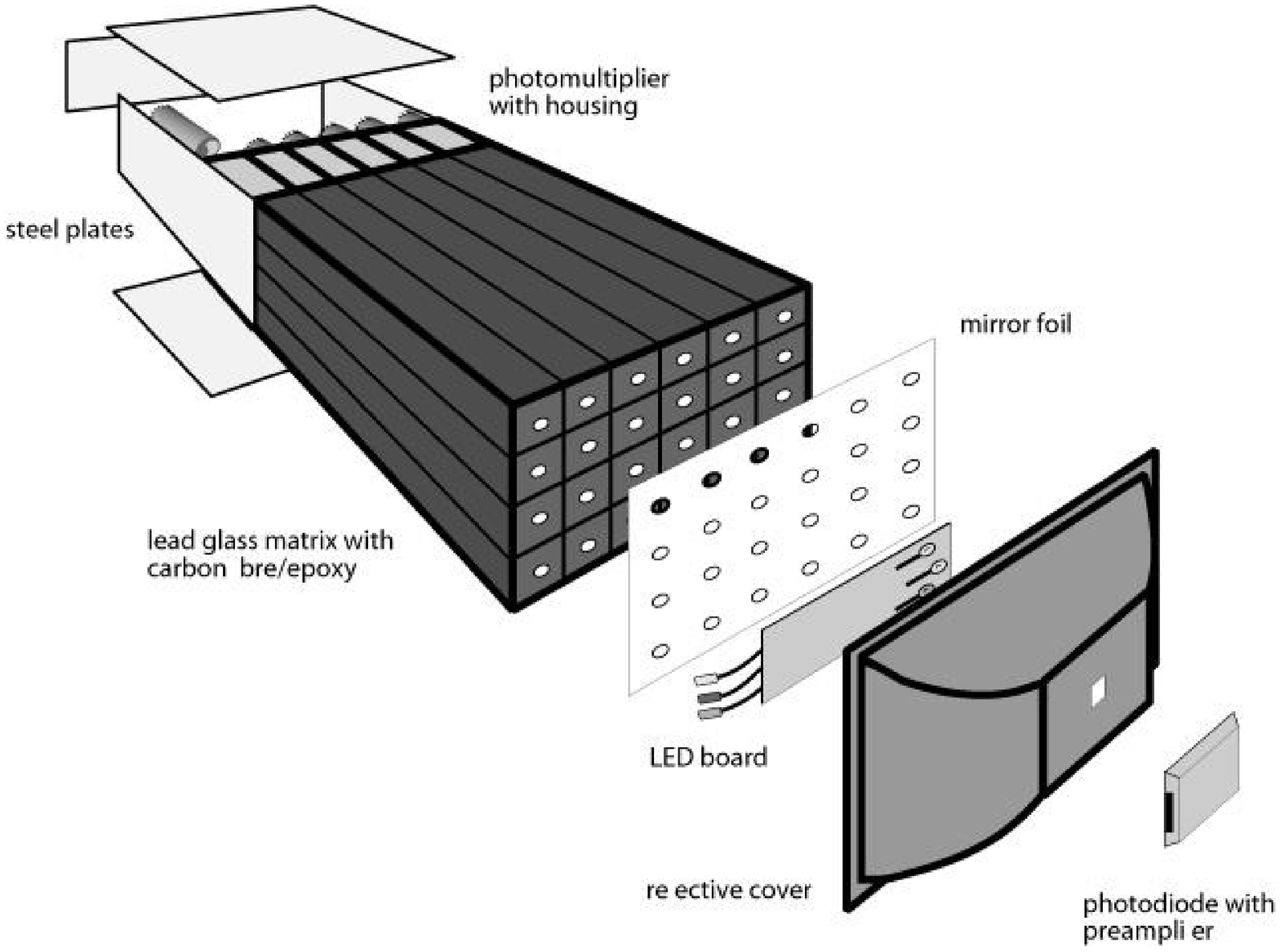, width=10cm}
     \caption{A PbGl super-module\label{fig:pbgl}}
    \end{center}
   \end{figure}

   The PbGl is a Cherenkov calorimeter with 1.648 of index of refraction.  
   It consists of  9216 individual towers
   (4~cm $\times$ 4~cm $\times$ 40~cm), 
   which were previously used in WA98 experiment at CERN.
   Total radiation length of PbGl is 14.4~$X_0$ and Moliere radius is
   $\sim4$~cm.
   Each PbGl sector comprises 192 super-modules (SM) in an array of
   16 (wide) by 12 (high).
   Figure~\ref{fig:pbgl} shows a PbGl super-module which consists of 24 
   lead-glass towers in an array of 6 (wide) by 4 (high). 
   At the back of the towers,  PMT's (FEU84) are implemented for readout.

    The energy resolutions of PbGl were also evaluated as
    follows from the tests using electron beams:
  \begin{eqnarray}
    \delta E/E &=& 6.0~\%/\sqrt{E ({\rm GeV})} + 0.8~\%~~~({\rm PbGl}).
  \end{eqnarray}
  
  The measured position resolution of PbGl was:
  \begin{equation}
    \sigma_{x}(E) = 0.2~{\rm mm} + \frac{8.4~{\rm mm}}{\sqrt{E({\rm GeV})}}.
  \end{equation}

  \subsection{Time of Flight Counter (TOF)}
  The Time of Flight Counter (TOF) provides the measurements of 
  the flight time of the particles.
  It is a primary particle identification device for charged hadrons and 
  achieves the $\pi$/K separation up to 2.4~\GEVC and 
  K/p separation up to 4~\GEVC.
  
\begin{figure}[h]
   \begin{center}
    \epsfig{figure=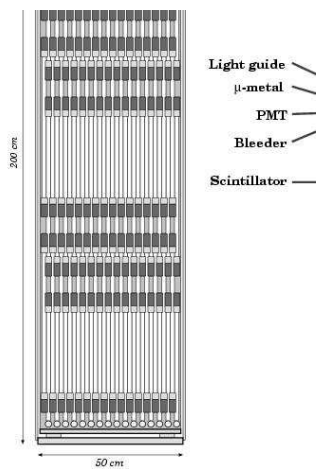,width=5.5cm}
    \caption{The components of  single TOF panel.
    \label{fig:tof2}}
   \end{center}
  \end{figure}

  The TOF consists of 10 TOF panels.
  They are mounted on the east central arm at 510~cm in radial distance
  from the interaction point,  covering  the whole  $\eta$ range of the
  central arm  30 degrees in   azimuth.
  Figure~\ref{fig:tof2} shows the sketch  of a single TOF panel.
  A TOF panel consists of 96 segments, each of which is equipped with
  a plastic scintillator slat and two 19~mm$\phi$ PMT's (Hamamatsu R3478S) at
  both ends of the slat.
  The PMT has a bi-alkaline photocathode and a 8 stage linear-focused dynode.
  The magnetic shielding is provided by $\mu$-metal with an inner
  diameter of 23~mm.
  The slat is oriented along the $r$-$\phi$ direction and provides the
  information of time and longitudinal position. 



  \subsection{Aerogel Cherenkov Counters (ACC)}
  The Aerogel Cherenkov Counters (ACC) is used to identify charged K at high
  momentum region.
  A cell of the aerogel Cherenkov counters consists of 
  $22(z) \times 11(\phi) \times 12(r)$ cm$^3$ aerogel with  refractive index of
  $\sim$ 1.01, an integration cube and two 3-inch phototubes.
  The aerogel Cherenkov counter covers the region of 22.5\textdegree
  $(\phi) \times 0.7(\eta)$.

  \subsection{Time Expansion Chambers (TEC)}
  The Time Expansion Chambers~(TEC) is a transition radiation detector
  and gives information of charged particle tracking and electron identification
  by $dE/dx$ and transition radiation.
  One TEC sector has an active area of $3.1-3.5m(z) \times 1.7-1.9m(\phi)$ and
  consists of 6 individual chambers. Each chamber is buled in two layers:
  one is window support and radiator foils and the other is the active elements
  of the wire chamber. The chamber is filled with a Xe/CO$_2$ gas mixture.
  The TEC covers the region of 90\textdegree$(\phi) \times 0.7(\eta)$.

\clearpage
 \section{The Muon Arm}
 
  \subsection{Muon Magnet}

\begin{figure}
   \begin{center}
    \epsfig{figure=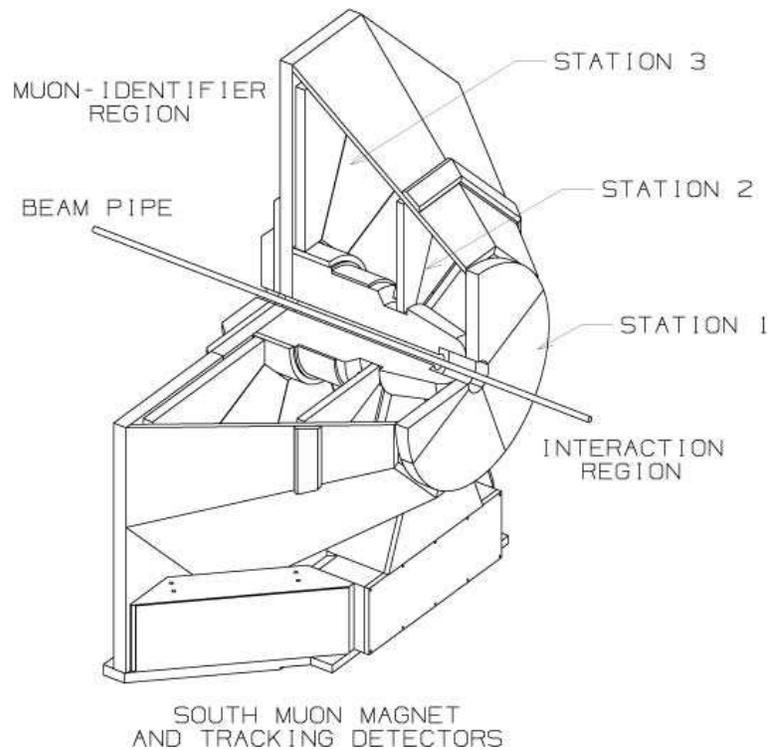,width=10cm}
    \caption{The cut-away view  of the Muon magnet and the Muon Tracker.
    \label{fig:mu_tracker}}
   \end{center}
  \end{figure}

  The Muon Magnet provides the radial magnetic field.
  In Figure~\ref{fig:mu_tracker}, the cut-away view of the Muon Magnet is
  shown as well as the Muon Tracker which is described in the next section.
  The central iron ``piston'' 
  defines the minimum polar angle of spectrometer.
  The rest of iron yoke consists of an eight-sided ``lamp shade'', which
  defines the maximum polar angle.
  The resulting radial magnetic field has an integral that is roughly
  proportional to the polar angle.

  \subsection{Muon Tracker (MuTr)}
  
  Muon Tracker (MuTr) provides the measurement of the muon track.
  Figure~\ref{fig:mu_tracker} shows the schematic diagram of the 
  MuTr. 
  The MuTr consists of three stations of cathode strip chambers, each of
  which is the shape of octant built with a 3.175~mm half gap, 5~mm
  cathode strip and with alternate strips readout.
  The honeycomb construction is used for station 1 and 3, and thin foil
  construction is used for station 2 in order to produce a cathode
  pattern.

  \subsection{Muon Identifier (MuID)}
  
  Muon Identifier (MuID) is the primary device for  muon identification.
  MuID consists of  five layers of chambers interleaved with steel
  absorbers.

  In oder to set the punch-through probability for pions of up to
  4~\GEVC to be 3~\% or less, a total steel depth of 90~cm,  
  corresponding to 5.4 hadronic interaction length, is required.
  Subtracting the thickness of the muon magnet backplate, a total depth
  of 60~cm of steel is required in MuID itself.
  A muon at the vertex must have a energy of at least 1.9~GeV to reach
  MuID, and 2.7~GeV to penetrate through MuID .
 \section{Photon Converter}
\begin{figure}
   \begin{center}
    \epsfig{figure=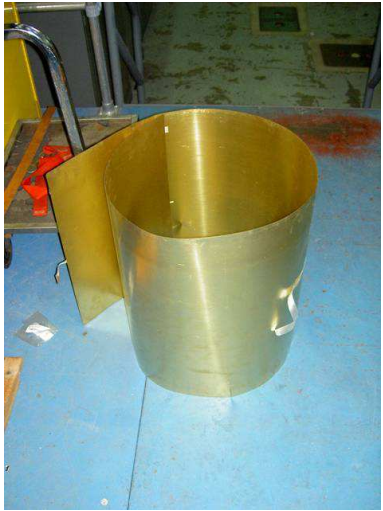,width=5cm}
    \caption{ A picture of the converters.
    \label{fig:piconv}}
   \end{center}
  \end{figure}

 The converter was installed around beam pipe at a part of physics run.
 The aim of the installation of the converter is to determine the amount of 
 the electrons from $\gamma$ conversion.
 Figure~\ref{fig:piconv} shows a picture of the converters.

  The converter is a thick brass sheet which material budget is well known. 
  The composition of the brass converter is Cu (70 \%), Zu (29.88 \%), Fe (0.05 \%) 
  and Pb (0.07 \%). The ideal mass density is 8.5 [g/cm$^3$].
  The tube shape with 3.85 (3.94) [cm] inner (outer) radius, 60.96 [cm] height 
  and 0.0254 [cm] thickness is placed at PHENIX center along with beam axis.
  The density of the converter was checked by measurement of the area and the
  weight of a piece of the converter that was installed. The determined area 
  density is 0.215313[g/cm$^2$]. Thus, the density of the converter is 8.4769 
  [g/cm$^3$] because the thickness is 0,0254[cm].

  \section{Trigger }
  The Level-1 trigger~(LVL1) have been designed to ensure a significant 
  fraction of the physics events of interest are recorded, due to the limited 
  speed of PHENIX data acquisition system. The LVL1 trigger is a pipelined 
  and dead timeless system.
  \noindent 
  The triggers consist of two type: One is the Local Level-1~(LL1) system
  and the other is the Global Level-1~(GL1) system. LL1 system communicates
  directly with the detectors such as BBC, EMCal and RICH. The input data 
  from these detectors is processed by the following LL1 algorithms to 
  produce a set of reduced input data for each RHIC beam crossing.
  In this section, BBC LL1 trigger~(BBCLL1) and EMCal-RICH LL1 trigger~(ERTLL1)
  are presented in this section.
  These triggers are designed to select  the physics events of interest 
  in this thesis.
   \subsection{BBC LL1 trigger~(BBCLL1)}
   The main trigger for events in PHENIX relies on a coincidence between
   the two BBC modules.
   The timing information of BBC is used to determine the position of 
   the collision point.
   BBC LL1 trigger~(BBCLL1) in an event trigger, which requires the event
   to occur within the nominal interaction region~($\mid z \mid<37.5$ cm).
   The BBCLL1 is defined as our minimum bias trigger condition for 
   the data in $p+p$ collisions.
    \subsection{EMCal-RICH LL1 trigger~(ERTLL1)}
    EMCal-RICH LL1 trigger~(ERTLL1) is used as the electron and high
    energy photon events. For the photon triggers, on the EMCal
    information is used. Acceptance coverage of each
    of the EMCal and RICH is divided into 16 trigger segments.
    Each segment consists of 9(PbSc)/16(PbGl, RICH) trigger tiles. 
    Each trigger tile consists of 144 EMCal towers
    (20 RICH phototubes). EMCal has two different methods, 2X2 tower sum 
    and 4$\times$4 tower sum, to sum the energy of towers. 
    The energy threshold value of EMCal for the hit dentition can be changed. 
    If there is a hit 
    tile defined by 4$\times$4 sum (2$\times$2 sum) in the EMCal part, 
    ERTLL1 4$\times$4 (ERTLL1 2x2) is issued.
    There are 3 versions of this trigger, which differ 
    only in the threshold energy. 4$\times$4a, 4$\times$4b, and 4$\times$4c, 
    as they are named, require
    respectively a 2.1, 2.8 and 1.4 GeV energy deposit in a 4$\times$4 
    tower block made up of 4 neighboring basic tiles.
    These triggers are photon triggers which provide enough rejection power to record all triggered events.

    If there are an EMCal hit tile defined by 2$\times$2 sum and 
    an associated RICH
    hit tile, an electron trigger, ERTLL1 E is issued. Association of EMCal 
    and RICH tiles
    is performed using the look-up table in the ERTLL1 module.
    The GL1 receives and combines the LL1 data to make a trigger decision. 
    The GL1 also manages busy signals.

 \section{The Data Acquisition System}    \label{sec:DAQ}
\begin{figure}[t]
  \begin{center}
   \epsfig{figure=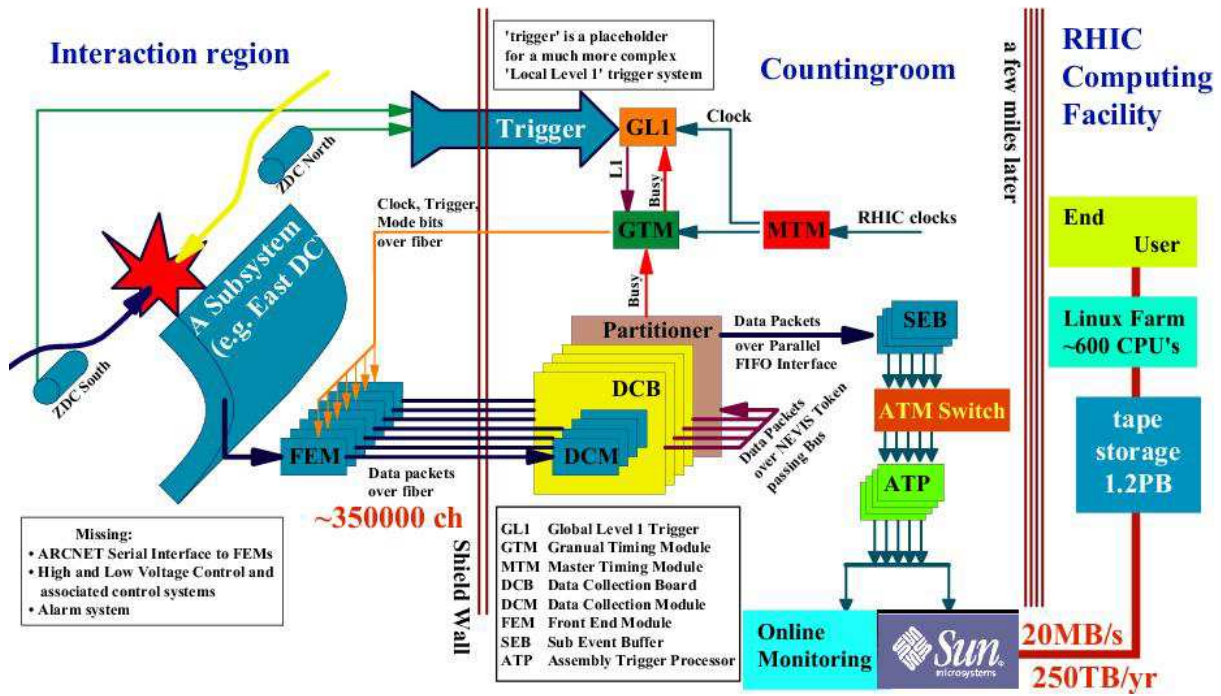,width=16cm}
   \caption{Schematic diagram of the data acquisition flow.
   \label{fig:daq}}
  \end{center}
 \end{figure}

 The PHENIX data acquisition system processes the signals  from each
 detector, produces  the trigger decision, and  stores the  triggered
 data.
 The typical data logging  rate of PHENIX was $\sim$1~kHz for Au+Au
 collisions and  $\sim$5~kHz for $p+p$ collisions.
 The zero-suppressed event sizes are 160~kbytes for Au+Au and 40~kbytes
 for $p+p$, respectively.
 The block diagram of the data acquisition flow is shown in
 Fig.~\ref{fig:daq}.

 The data acquisition system employs the concept of granule and
 partition.
 A granule is smallest unit, which consists of individual timing
 control and data collection for  each detector.
 The partition is the combination of granules, that share  busy
 signals and accept signals.
 This configuration makes it possible to run the data acquisition in
 desired combination of detectors.

 Overall control of the data acquisition is provided  by the
 Master Timing Module (MTM), the Granule Timing Module (GTM),  and the
 Global Level-1 Trigger System (GL1).
 The MTM receives 9.4~MHz RHIC clock and deliver it to GTM and GL1.
 The MTM also receives LVL1 accept signal.
 The GTM deliver the clock, the control commands (Mode Bits), event
 accept signal to each detector's FEM.
 The GTM is capable of a fine delay tuning of the clock in $\sim$50~ps
 step, in order to  compensate the timing difference among FEM's.
 The GL1 produces the first Level-1 (LVL1) trigger decision,  combining
 LVL1 signal from detector components.

 The FEM of each detector is designed to convert the analog response of
 the  detectors into the digitized signal.
 The LVL1 trigger signals  are simultaneously generated.
 The  generation of  global decision,  whether event should be taken or not, 
 takes $\sim$30~B.C.
 While the GL1 system is making decision, the event data is stored in
 AMU.
 After receiving the accept signal, each FEM starts to digitize the data.

 The data collection from each FEM is performed  by Data Collection
 Modules (DCM) via  G-LINK.
 The DCM's  have the capability to receive 100 Gbytes of uncompressed 
 event data per second at the highest trigger rate.
 The DCM's  provide  data buffering,  zero suppression, error checking,
 and data formatting.
 The DCM send the compressed  data to PHENIX Event Builder (EvB). 

 The EvB is the system which consists of 32 Sub Event Buffers (SEB),
 Asynchronous Transfer Mode (ATM) switch, and 28 Assembly Trigger
 Processors (ATP).
 The SEB's are the front end of EvB and communicate with each granule.
 The SEB's  transfer the data from  granules to ATP via ATM,
 where event assemble is performed. 
 The combined data is once stored to the disk and used for  
 online monitoring and for generation of trigger decision by the
 second level (LVL2) software  trigger.

 The data storage is finally provided by HPSS-based tape storage robot
 system with maximum transfer rate of  20~Mbytes/s.
 Combining the buffering to local disk, the maximum data logging rate
 become $\sim$60~Mbytes/s.

  \chapter{Run Conditions}\label{chap:run_condition}
  \section{Overview}
  The recorded data by the PHENIX is summarized at Table~\ref{chap4_table1}.
  The analysis in this thesis has been performed  using  the $p+p$ data sample obtained 
  during the RHIC in Year-2005 run~(RUN5) and  Year-2006 run~(RUN6) period.
  The beam and trigger conditions are described briefly in this chapter.

\begin{table}[hbt]
    \begin{center}
     \caption{The recorded data summary at the PHENIX}
     \label{chap4_table1}
     \begin{tabular}{c|cccc}
        & Year & Species & $\sqrt{s}(GEV)$ & recorded luminosity\\
       \hline \hline
       RUN1 & 2000 & Au-Au &130 & 1$\mu$b$^{-1}$ \\
       \hline
       RUN2 & 2001-02 & Au-Au &200 & 24~$\mu$b$^{-1}$ \\
            &         & p-p   &200 & 0.15~pb$^{-1}$ \\
       \hline
       RUN3 & 2002-03 & d-Au &200  & 2.74~nb$^{-1}$ \\
            &         & p-p  &200  & 0.35~pb$^{-1}$ \\
       \hline
       RUN4 & 2003-04 & Au-Au &200 & 241~$\mu$b$^{-1}$ \\
            &         & Au-Au &62.4& 9~$\mu$b$^{-1}$ \\
       \hline
       RUN5 & 2005    & Cu-Cu &200 & 3~nb$^{-1}$ \\
            &         & Cu-Cu &62.4& 0.19~nb$^{-1}$ \\
            &         & Cu-Cu &22.4& 2.7~$\mu$b$^{-1}$ \\
            &         & p-p   &200 & 3.8~pb$^{-1}$ \\
       \hline
       RUN6 & 2006    & p-p   &200 & 10.7~pb$^{-1}$ \\
            &         & p-p   &62.4 & 0.1~pb$^{-1}$ \\
       \hline
       RUN7 & 2007    & Au-Au &200 & 813~$\mu$b$^{-1}$ \\
     \end{tabular}
    \end{center}
\end{table}

  \section{Collisions in $p+p$ at $\sqrt{s}$=200GeV in 2005 and 2006}
\begin{figure}[p]
  \begin{center}
    \includegraphics[width=14cm]{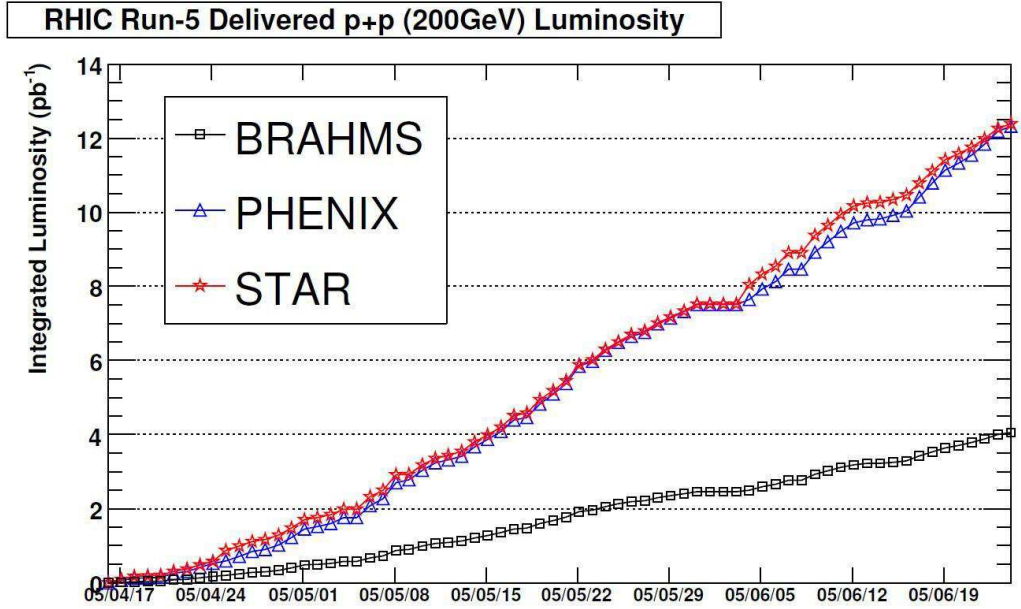}
    \caption{The delivered integrated luminosities of p + p collisions 
      in RUN5 as a function of date \label{fig:run5lumi}}
  \end{center}
\end{figure}

\begin{figure}[p]
  \begin{center}
    \includegraphics[width=14cm]{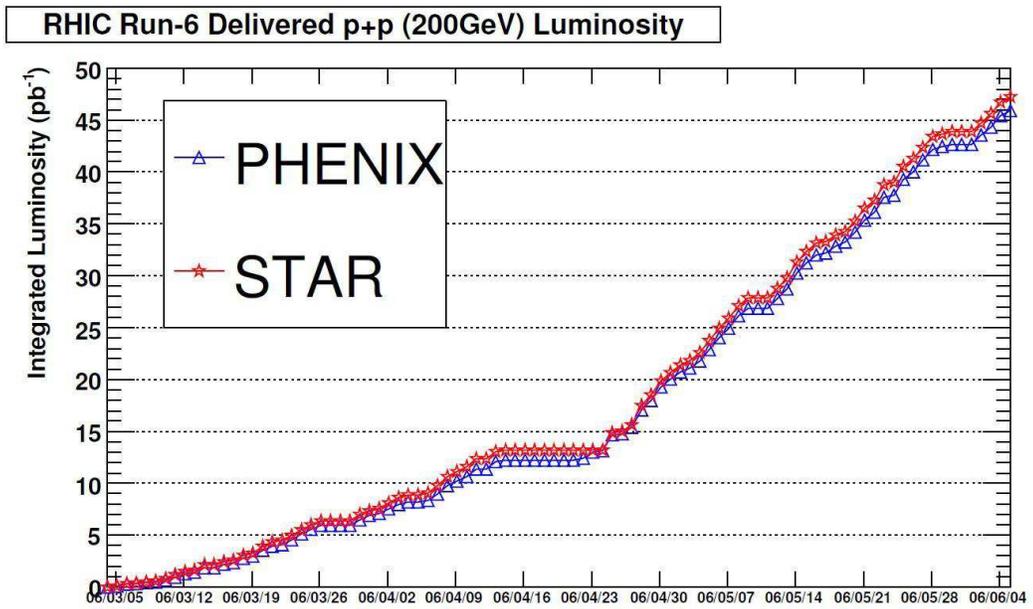}
    \caption{The delivered integrated luminosities of p + p collisions 
      in RUN6 as a function of date \label{fig:run6lumi}}
  \end{center}
\end{figure}

  During the polarized proton run period in the RHIC RUN5 (April 16, 2005-
  June 24,2005) and the RHIC RUN6 (March 4, 2006-June 5, 2006), 
  p + p collisions at $\sqrt{s}$ =200 GeV were collected with the PHENIX 
  detector. The mode of 111 bunch was used and there were 1.3$\times 10^{11}$
  protons in each bunch. The peak luminosity was $3.5 \times 10^{31}$ cm$^{-2}$
  s$^{-1}$. The delivered integrated luminosities of p + p collisions 
  in RUN5 and RUN6 are shown as a function of date in Figure~\ref{fig:run5lumi}
  and Figure~\ref{fig:run6lumi}, respectively. 
  The recorded integrated luminosities are 3.8 pb$^{-1}$  (RUN5) and 
  10.7 pb$^{-1}$ (RUN6).
  
The data for the analysis was taken with the 'Minimum Bias (MB)' trigger and the 'ERT'
trigger. 
The 'ERT' trigger consists of ERT electron and ERT photon trigger.
The trigger logic of the MB, ERT electron and ERT photon 
triggers for the $p+p$ collisions
are defined as:

\begin{eqnarray}
  {\rm Minimum\quad Bias} &\equiv&  {\rm BBCLL1}(> 0{\rm tubes}), \\
  {\rm ERT\quad electron} &\equiv&  {\rm ERTLL1\_E}\cap{\rm BBCLL1}(> 0{\rm tubes}), \\
  {\rm ERT\quad photon} &\equiv&  {\rm ERTLL1\_{4x4i}}\cap{\rm BBCLL1}(> 0{\rm tubes}),
  ({\rm i} = {\rm a,b,c})
\end{eqnarray}
where BBCLL1($>$ 0 tubes) means that at least one hit is required in each BBC and the
vertex position obtained by BBCLL1 online in z direction, zBBCLL1, is
required to be less than 37.5 cm. 
ERTLL1\_E denotes the coincidence of a EMCal hit with energy deposit
above the ERTLL1 2x2 threshold and a RICH hit of $\geq$ 3 photoelectrons.
ERTLL1\_{4x4i}  denotes the EMCal trigger with 4x4 tower sum.
4x4a, 4x4b and 4x4c mean different energy thresholds. The threshold values
of these ERTLL1 4x4 triggers are 2.1 GeV (4x4a), 2.8 GeV (4x4b) and 1.4 (4x4c)
during RUN5 and RUN6 p + p 200 GeV periods. The threshold
value of ERTLL1 2x2 trigger is shown in Table~\ref{chap4_table2}.

\begin{table}[hbt]
    \begin{center}
     \caption{RUN summary of the RUN5 and RUN6 $p+p$ 200GeV periods}
     \label{chap4_table2}
     \begin{tabular}{c|ccccc}
       RUN & polarization & RUN number & Magnet   & ERT 2x2   & Converter\\
           &              &            & polarity & threshold & \\
       \hline
       RUN5 & longitudinal & 166030-171594 & CM-~- & 0.4GeV & Without\\
       RUN5 & longitudinal & 171595-172080 & CM-~- & 0.4GeV & With\\
       RUN5 & longitudinal & 172081-179846 & CM-~- & 0.4GeV & Without\\
       \hline
       RUN6 & transverse   & 188216-197795 & CM++ & 0.4GeV & Without\\
       RUN6 & longitudinal & 198061-199767 & CM++ & 0.4GeV & Without\\
       RUN6 & longitudinal & 200240-204639 & CM++ & 0.6GeV & Without\\
     \end{tabular}
    \end{center}
\end{table}

  \chapter{Data Analysis }
 This chapter describes the data analysis to extract the electron samples from
 semi-leptonic decay of charm and bottom, so called as 'single non-photonic electron', 
 in $p+p$ collisions at $\sqrt{s}$ = 200~GeV.
 The data taken with the 'Minimum Bias' trigger~(MB data) and the data taken with 
 the 'ERT' trigger~(ERT data) is used to the analysis.
 Specially, the data taken with 4x4c ERT photon trigger~(PH data) 
 in ERT trigger is used to determine the total yield of the electrons from semi-leptonic 
 decay of heavy flavor.

 The analysis procedure is following.
 \begin{enumerate}
 \item{ Inclusive electron spectrum is measured in RUN5 and RUN6 data. (Sec.~5.1 - Sec.~5.5)}
 \item{ The spectrum of single electrons from heavy flavor is measured by subtraction of background electrons.
   The contribution of background electrons is evaluated by 'converter method' and 'cocktail method'.
   (Sec.~5.6)}
 \item{ The fraction of bottom contribution in single electrons is determined by using the correlation in 
 electron-hadron pairs. (Sec.~5.7 - Sec.~5.10)}
 \end{enumerate} 

 The formalization of invariant yield is described in Sec.~5.1.
 Sec.~5.2 describes reconstruction and momentum determination of charged particles by using DC and PC.
 Sec.~5.3 describes the method of electron identification by using EMCal and RICH.
 Selection of runs used for this analysis is described in Sec.~5.4.
 In Sec.~5.5, reconstruction efficiency of electrons and PH trigger efficiency are determined. Then 
 inclusive electron spectrum is measured in RUN5 and RUN6 data.

 The spectrum of single electrons from heavy flavor is determined by 'converter method' and 'cocktail method'
 in RUN5 and RUN6 data in Sec.~5.6.

 Sec.~5.7 describes a new analysis method to determine the fraction of bottom contribution in single electrons.
 The method uses the correlation in electron-hadron pairs in real data and simulation.
 The correlation in electron-hadron pairs in real data is studied in Sec.~5.8 and 
 that in simulation is studied in Sec.~5.9.
 In Sec.~5.10, the correlations in real data and simulation are compared to obtain the fraction of bottom.
 %
 %
 \section{Invariant Yield}
The invariant yield in Lorentz invariant form can be written as,

 \begin{equation}
  E\frac{d^3\sigma}{dp^3} = \frac{d^3\sigma}{p_{\mathrm{T}} 
    dy dp_{\mathrm{T}} d\phi} =
  \frac{d^2\sigma}{2\pi p_{\mathrm{T}} dp_{\mathrm{T}} dy},
 \end{equation}

 \noindent where $p$ is the momentum of the particle, $y$ is the
 rapidity, $p_{\mathrm{T}} = \sqrt{p_x^2 + p_y^2}$ is the transverse momentum,
 and $\phi$ is azimuthal angle.
 
 The goal of this analysis is to obtain the invariant yield \DNDY, 
 differential cross section $d^2\sigma/dydp_{\mathrm{T}}$ of the electrons 
 from semi-leptonic decay of charm and bottom at  mid-rapidity.
 Using the number of measured electrons, $d^2\sigma/dydp_{\mathrm{T}}$ of the electrons 
 can be extracted  experimentally  as follows.
 
 \begin{equation} \label{eq:inv}
   \frac{1}{2\pi p_{\mathrm{T}}}\frac{d^2\sigma}{dydp_{T}} 
     =   \frac{1}{2\pi p_{\mathrm{T}}} \frac{N_e(p_{\mathrm{T}})}
       {\int L dt \epsilon(p_{\mathrm{T}}) \epsilon_{bias} 
	 \Delta p_{\mathrm{T}}\Delta y },
 \end{equation}
 \noindent where
 \begin{itemize}
 \item{$N_e(p_{\mathrm{T}})$ is the number of reconstructed electrons in 
   a $p_{\mathrm{T}}$ bin}
 \item{$\epsilon_{bias}$  is BBC trigger bias}
 \item{$\epsilon(p_{\mathrm{T}})$  is the overall efficiency including acceptance,
 reconstruction efficiency and trigger efficiency}
 \item{$\Delta y$ is the rapidity bin width and is set to $\Delta y$ = 1}
 \item{$\Delta p_{\mathrm{T}}$ is the $p_{\mathrm{T}}$ bin width}
 \item{$\int L dt$ is the integrated luminosity.}
 \end{itemize}
 \subsection{Integrated Luminosity }\label{sec:lumi}
 The integrated luminosity can be expressed using the number of minimum bias~(MB)
 triggered events~($N_{MB}$).
 \begin{equation}
 \int L dt = \frac{N_{MB}} {\sigma_{p+p}\epsilon_{BBC}^{p+p} },
 \end{equation}
 where
 $\sigma_{p+p}$ is the cross section of inelastic $p+p$ collisions at 
 $\sqrt{s}$ = 200GeV, and $\epsilon_{BBC}^{p+p} $ the BBC trigger~(MB trigger)
 efficiency.
 The MB trigger cross section in $p+p$ collisions, which is defined as 
$\sigma_{p+p}\epsilon_{BBC}^{p+p}$,
 has been determined to be 21.8$\pm$ 2.1~mb, by using a van der Merr scan measurement 
 in RUN2~\cite{bib:ph_bbrun2}.
 The MB trigger cross section in RUN5 and RUN6 is determined to be the 23.0$\pm$ 2.2~mb
 by making correction of
 BBC efficiency to take into account the change of BBC mask~\cite{bib:ph_bbc,bib:ph_bbc2}.
 
 The equivalent number of sampled minimum bias events in the data taken with 
 4x4c ERT photon trigger~(PH data), $N_{MB}^{sample}$
 instead of $N_{MB}$ is used to obtain the integrated luminosity of PH data set.
 The $scale\_down\_factor$ of 4x4c photon~(PH) trigger, which represents the 
fraction of recorded
 MB events in triggered PH events, is determined and recorded at each run.
 $N_{MB}^{sample}$ is determined as $scale\_down\_factor \times N_{MB}$.
 \subsection{BBC Trigger Bias}\label{sec:biasbbc}
 BBC trigger bias, $\epsilon_{bias}$ is PHENIX-specific term referring 
 to the probability that the BBC counter makes MB trigger 
 for an event containing specific particle of interest due to the acceptance of the BBCs.
 It is obvious that events with a hard
 parton scattering have higher probability of making BBC MB trigger 
 because the track multiplicity
 in the BBC is higher for these events.
 This means that of all events that contain a hard scattering process, 
 The apparent cross section of events which contain hard scattering
 will be higher than the BBC trigger cross section, 
 $\sigma_{p+p}\epsilon_{BBC}^{p+p}$. 
 The fact that the trigger cross section depends upon the physics process 
 is what we term 'bias'.
 
 $\epsilon_{bias}$ is determined to be 0.79$\pm$0.02 as the $p_{\mathrm{T}}$ 
 independent fraction for hard scattering process, from the yield ratio of high 
 $p_{\mathrm{T}}$ $\pi^0$ with and without the BBC trigger~\cite{bib:ph_bbrun2}.
 This measured value of the constant BBC trigger bias is in good agreement
 with PYTHIA calculations of the BBC efficiency for hard pQCD partonic
 scattering processes. 
 \clearpage
 \section{Reconstruction of Track and Momentum }\label{sec:trk_reco} 
 
 \subsection{Variables for  Particle Trajectory Measurements }

 \begin{figure}[htb]
     \begin{center}
      \epsfig{figure=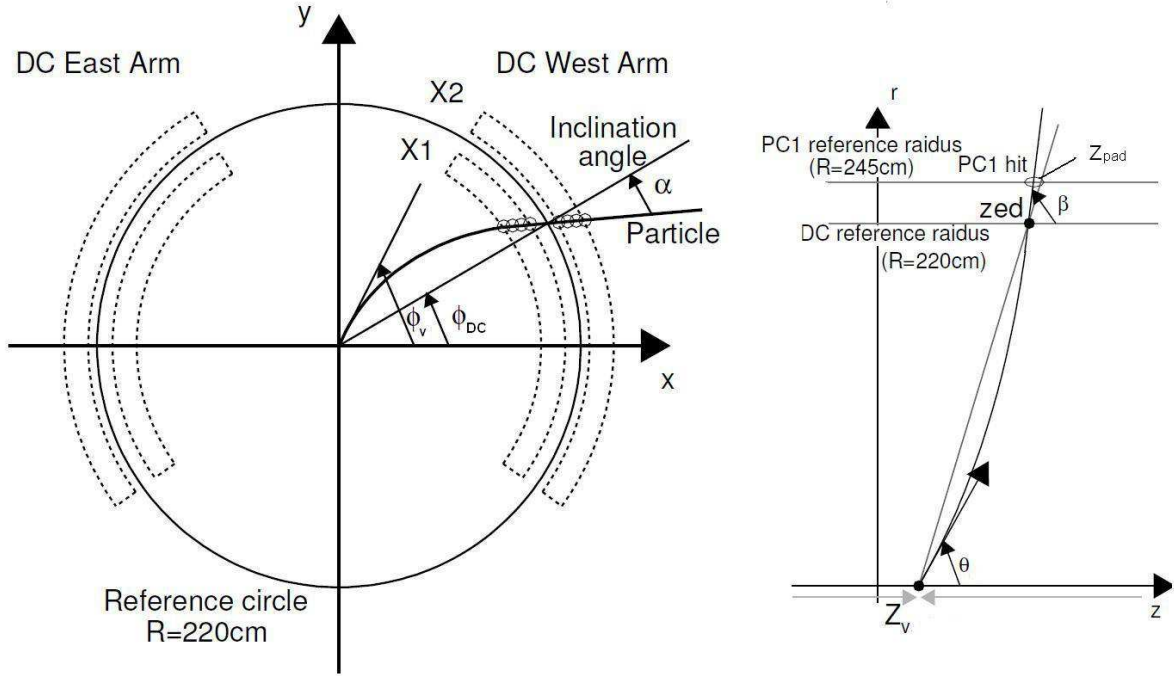,width=16cm}
      \caption{A particle trajectory and the parameters used in the track
    reconstruction are shown in
    Left: beam view and  RIGHT:side view  of PHENIX central}
    \label{fig:parameters}
     \end{center}
    \end{figure}

 The detailed explanation about the track reconstruction technique in
 PHENIX  can be found in published papers~\cite{bib:reco,
 bib:reco_2}.
 The essential parts are  explained in this section.

 Figure~\ref{fig:parameters} shows a particle trajectory in PHENIX up to
 PC1 and the definition of parameters, which are introduced  
 in order to describe the charged  particle trajectory traveling through
 the axial magnetic field.
 The variables are summarized as follows:

   \begin{itemize}	
    \setlength{\itemsep}{0.8mm}

    \item Measured variables

    \begin{itemize}
     \item{{\bf $\alpha$} : The angle between the projection of 
	  trajectory in the $x-y$ plane and the radial
	  direction,  at the intersection point of the
	  trajectory with the circle of reference radius
	  $R_{DC}=$2.2~m.}.
     \item{{\bf $\beta$} : Considering  the plane which includes the
	  $z$-axis and  $z_{pad}$,
	  $\beta$ is defined as the angle between the projection of
	  trajectory onto that plane and the $z$-axis. }
     \item{{\bf $z_{pad}$} : The intersection point of the trajectory
	  with  PC1 surface radius  $R_{PC1}=2.45~m$,  .}
     \item{{\bf $\phi_{DC}$} : The $\phi$-angle of intersection point
	  of particle trajectory with the circle of radius
	  with $R_{DC}$.}
    \end{itemize}	

    \item Variables to be reconstructed
    \begin{itemize}
     \item{{\bf $\theta_v$} : The angle between the initial direction of the 
	  particle trajectory and $z$-axis.}
     \item{{\bf $\phi_{v}$} : The initial azimuthal angle  of the 
	  particle trajectory.}	
     \item{{\bf $p_{\mathrm{T}}$} : The transverse momentum.}
    \end{itemize}

   \end{itemize}
   
   \subsection{Track and Momentum Reconstruction Technique}

    \begin{figure}[htb]
     \begin{center}
      \epsfig{figure=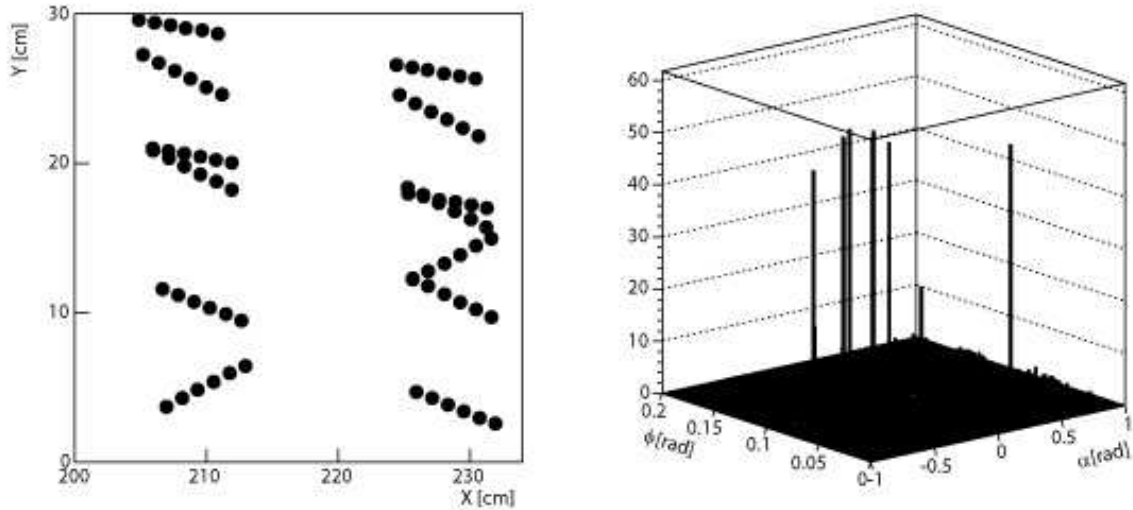,width=16cm}
      \caption{The Hough transformation of the DC hits in the x-y plane to the feature
	space of $\alpha$ and $\phi$. The left panel shows simulated hits 
	for a small physical region of the drift chamber. The right panel shows 
	the Hough transform feature space for this region. 
	Tracks appear as peaks in this plot.
      \label{fig:hough_example}}
     \end{center}
    \end{figure}
   
    The intersection of particle trajectory with various detector planes
    is uniquely determined by three variables: $\theta_v$, $\phi_{v}$,
    and $p_{\mathrm{T}}$.
    They are  reconstructed from  the measured
    variables: $\alpha$, $\beta$, $z_{pad}$, $\phi_{DC}$ and vertex information.
    The collision vertex is assumed to be (0,0) in x-y plane in PHENIX and 
    z position of the vertex is determined by the timing information of BBC as described
    at Sec.~\ref{sec:bbc}.
    Hough Transform technique is used for the track reconstruction.
    The Hough Transform  is a general algorithm for finding straight lines.
    It is popular in image processing. 
    Figure~\ref{fig:hough_example} shows a simple example of Hough
    Transform in a part of the DC hits.
    In case of PHENIX,  track reconstruction started from finding the
    hits in X1 and X2 wires.
    
    \begin{enumerate}
     \item  Project the drift chamber hits onto $x-y$ ($r-\phi$) plane at $z=0$ .
     \item  Perform the  Hough Transform using all  possible X1-X2 hit combinations
	    taking $\alpha$  and $\phi_{DC}$ as the parameters in  Hough Space.
	    For example if there are 6 hits in X1-wires and 6 hits in X2-wires,
	    total 36 combination are taken.
     \item  Associate X-wire hits with the reconstructed track.
     \item  Perform a Hough Transform of the UV-wires and 
	    associate the resulting UV hits with the reconstructed track
	    in order to obtain the  $z$ information.
     \item  Associate PC1 hit.
    \end{enumerate}

    The $\alpha$ measured in the drift chamber is closely related to the 
    field integral along the track trajectory.
    The transverse momentum, $p_{\mathrm{T}}$(\GEVC) and  the $\alpha$-angle 
    (mrad) have the following  approximate relation:

   \begin{equation} \label{eq:mom}
     p_{\mathrm{T}} \sim \frac{K}{\alpha},
   \end{equation}

   \noindent where $K \sim 0.10$ [rad \GEVC] is the effective  field
   integral, expressed as :

   \begin{equation}
    K = \frac{e}{R}\int lBdl.
    \label{eq:field_int}
   \end{equation}
   \noindent
   Here, $e$ is the elementary charge in the hybrid unit
   ($e$=0.2998~\GEVC T$^{-1}$m$^{-1}$) and $R$ is the DC reference
   radius.

   However, due to the small non-uniformity of the focusing magnetic field 
   along the flight path of charged particles, an accurate analytical expression 
   for the momentum of the particles is not possible.
   Therefore, the non-linear grid interpolation 
   technique is used~\cite{bib:runge-kutta}.
   A four-dimensional field integral grid is constructed within the entire 
   radial extent of the central arm for momentum determination based on 
   drift chamber hits.
   The parameters of the field integral are momentum $p$, polar angle
   at the vertex, $\theta_v$, z-vertex, and radial distance $r$ at which
   the filed integral $f(p,r,\theta_v, z)$ is calculated.
   The field integral grid is generated by explicitly swimming particles 
   through the magnetic field map from survey measurement and numerically 
   integrating to obtain for $f(p,r,\theta_v, z)$  each grid point. 
   An iterative procedure is used to determine the momentum for 
   reconstructed tracks, using Eq.~\ref{eq:mom} as an initial guess.
   The initial estimate of $\theta_v$ is given by the $z$-vertex and 
   DC hit position.
   
   Each reconstructed track is associated with hit information of outer 
   detectors (PC2,PC3, EMCal and RICH). In the association with the outer 
   detectors, the residual magnetic field is not taken into account and the 
   track is assumed to be a straight line.
   
   The momentum resolution depends on (1) the intrinsic  angular
   resolution of the DC and (2) the multiple  scattering of a charged
   particle as it travels up to DC  due to the intervening material.
   As a result, the momentum resolution is about 1\% for tracks with 
   $p_{\mathrm{T}}$=1 GeV/$c$ and the reconstruction efficiency is above 99\% for a single track.
   
    \subsection{Track Quality}
    The quality of a reconstructed charged track is defined using hit 
    information of X and UV wires in the DC and the associated PC1 hit. 
    This information is implemented in the data of reconstructed tracks 
    as the 6-bit variable, 
    {\sf quality}, for each track. Table~\ref{tab:quality} is the definition of quality.
    The best case is {\sf quality} = 63 and the second best case is {\sf quality} = 31, 
    where the PC1 hit is ambiguous, but the UV hit is unique.
    \begin{table}[htb]\small
      \begin{center}
      \caption{The bit definition of {\sf quality} variable.\label{tab:quality}}
	\begin{tabular}{c|c|c|c }
	  \hline
	  \hline
	  \multicolumn{2}{c|}{bit} & decimal & description\\
	  \hline
	  LSB & 0 &1   & X1 hits used\\
	  &1  &2   & X2 hits  used\\
	  &2  &4   & UV hits found\\
	  &3  &8   & Unique UV. (No hit sharing) \\
	  &4  &16  & PC1 hits found\\
	  MSB &5 &32  & Unique PC1 (No hit sharing). \\
	  \hline
	  \hline
	\end{tabular}
      \end{center}
    \end{table}

    \subsection{Analysis Variables}
    The information of the reconstructed track and the associated hits of 
    the detectors are recorded in so called nano Summary Data Tape~(nDST) in PHENIX. 
    Parameters which characterize events, such as the collision vertex and 
    centrality, are also recorded in the nDST. 
    In this thesis, the variables in the nDST are written in {\sf Sans serif font}. 
    For example, there are the total momentum {\sf mom}, 
    initial polar angle {\sf the0} and initial azimuthal angle {\sf phi0} etc.
    The $z$ position of the collision point measured at BBC counter is 
    defined as {\sf bbcz}.
    The variables about track reconstruction which are used in this thesis
    are summarized at Table~\ref{tab:track}

  \begin{table}[htb]
   \begin{center}
 \caption{The track variables used in the
 analysis.\label{tab:track}}
    \begin{tabular}{c|c}
      \hline
      Variable  & Description    \\
      \hline
      {\sf bbcz} &  The $z$ position of the collision point measured at BBC counter.\\
      {\sf mom}  &   Total momentum of the reconstructed track.\\
      {\sf phi0} &   This is the initial $\phi$ direction of the track at the vertex. \\
      {\sf the0} &   This is the initial $\theta$ direction of the track at the vertex. \\
      {\sf alpha}&   The angle between the track in the DC $x-y$ plane
      and the radial direction.\\
      {\sf beta} &  $\theta$ angle of the track vector as it passes through the 
      DC reference radius.\\
      {\sf quality} & The quality of the Drift Chamber Tracks.\\
      {\sf phi} & $\phi$ coordinate at which the track crosses the DC reference radius.\\
      {\sf zed} & $z$ coordinate at which the track crosses the DC reference radius.\\
      {\sf pc1phi} & $\phi$ coordinate of the measured hit in PC1.\\
      {\sf pc1z} & $z$ coordinate of the measured hit in PC1.\\
      {\sf pc3phi} & $\phi$ coordinate of the measured hit in PC3.\\
      {\sf pc3z} & $z$ coordinate of the measured hit in PC3.\\
      {\sf emcphi} & $\phi$ coordinate of the measured hit in EMCal.\\
      {\sf emcz} & $z$ coordinate of the measured hit in EMCal.\\
      {\sf pemcphi} & $\phi$ coordinate of the reconstructed track projection at EMCal surface.\\
      {\sf pemcz} & $z$ coordinate of the reconstructed track projection at EMCal surface.\\
      \hline
    \end{tabular}
   \end{center}
\end{table}

  \section{Electron Identification}
  Electron identification is performed for the reconstructed particles 
  by RICH and EMCal and is described in this section.
  
  \subsection{Summary of Variables}

  The electron identification (eID) is performed by  combining  the
  information from tracking, RICH, and EMCal.
  The variables used in the analysis are summarized in
  Table~\ref{tab:eid_variables}.
  The details are explained in succeeding sections.
  

  \begin{table}[htb]
   \begin{center}
 \caption{The  eID variables used in the
 analysis.\label{tab:eid_variables}}
    \begin{tabular}{c|c}
      \hline
      Variable  & Description    \\
      \hline
      {\bf RICH} &  \\
      {\sf n0} & The number of fired phototubes in the nominal ring area \\
      &                   (3.8[cm]$<$r$<$ 8.0[cm])                               \\ 
      {\sf n1} & The number of fired phototubes in the larger ring   \\
      &                   (r$<$ 11.0[cm])                                        \\ 
      {\sf npe0}& The number of photo electrons detected in nominal ring radius\\
      {\sf disp}& Displacement between the projection point on the RICH phototube\\
      & plane and the centroid of the associated fired phototubes  \\
      {\sf chi2}& Ring shape calculated from hit PMT's in the nominal ring radiusZ\\
      \hline
      {\bf EMCAL} & \\
      {\sf ecore} & The shower energy detected at EMCal(summed up 3$\times$3 towers)(GeV)\\
      {\sf emcsdphi\_e}& The difference between the track projection and the EMCal cluster\\
      &  position  in the $\phi$ direction at EMCal surface normalized by $\sigma$ \\
      {\sf emcsdz\_e}& The difference between the track projection and the EMCal cluster\\
      &  position  in the $z$ direction at EMCal surface normalized by $\sigma$ \\
      {\sf prob} &The probability for a shower being a EM shower from the shower shape in EMCal\\
	\hline
	\end{tabular}
   \end{center}
\end{table}

  \subsection{Electron Identification with RICH}
  \subsubsection{RICH Calibration}
\begin{figure}[htb]
   \begin{center}
    \epsfig{figure=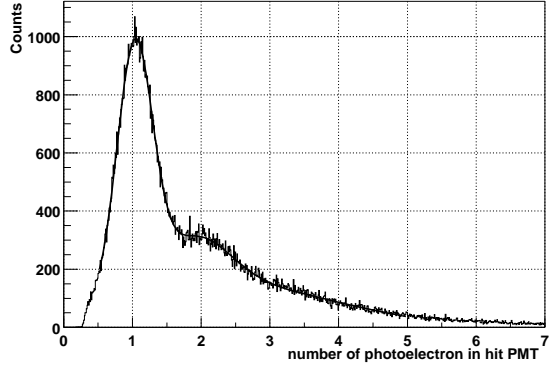,width=8cm}
    \caption{
    The distribution of the number of photo-electrons in each hit phototube.
    \label{fig:npe_PMT}}
   \end{center}
  \end{figure}

  Gain calibration is performed for each phototube by fitting the 
  raw ADC spectrum.
  Gaussian functions are used to fit the pedestal peak and 
  the one photo-electron peak, and those peak positions ($ADC_{pedestal}$ and 
  $ADC_{1p.e.}$) are obtained.
  Using these values, the number of photo-electrons ($N_{p.e.}$) of the 
  phototube is calculated from its ADC value (ADC) as follows:
  \begin{equation}
  N_{p.e.} = \frac{ADC - ADC_{pedestal}}{ADC_{1p.e.}-ADC_{pedestal}}.
  \end{equation}
  Figure~\ref{fig:npe_PMT} shows the distribution of the number of photo-electrons 
  in each phototube.
  The simultaneous Gaussian fit to the peaks gives  the one
  photo-electron peak position of  1.034 \PM 0.003 and the width of
  0.259 \PM 0.003, respectively.
  The hit phototube in RICH  is defined to have greater than  0.2 photo-electron.

  \subsubsection{Track - RICH Ring Association : {\sf n0, npe0, disp}}\label{sec:ring_ass}
  
  After the track reconstruction with DC and PC1 (DC-PC1 track), 
  the tracks are associated with the PC2, PC3 and EMCal. 
  The hit positions of the outer detectors or projection points of the DC-PC1 track 
  if any associated hit is not found are used for the association 
  with RICH. The tracks are projected with respect to the RICH mirror and 
  the reflected tracks are projected onto the RICH phototube plane. 
  Then, fired phototubes around the projection points of the reflected 
  tracks are associated with the tracks.
  
\begin{figure}[p]
   \begin{center}
    \vspace{1cm}
    \epsfig{figure=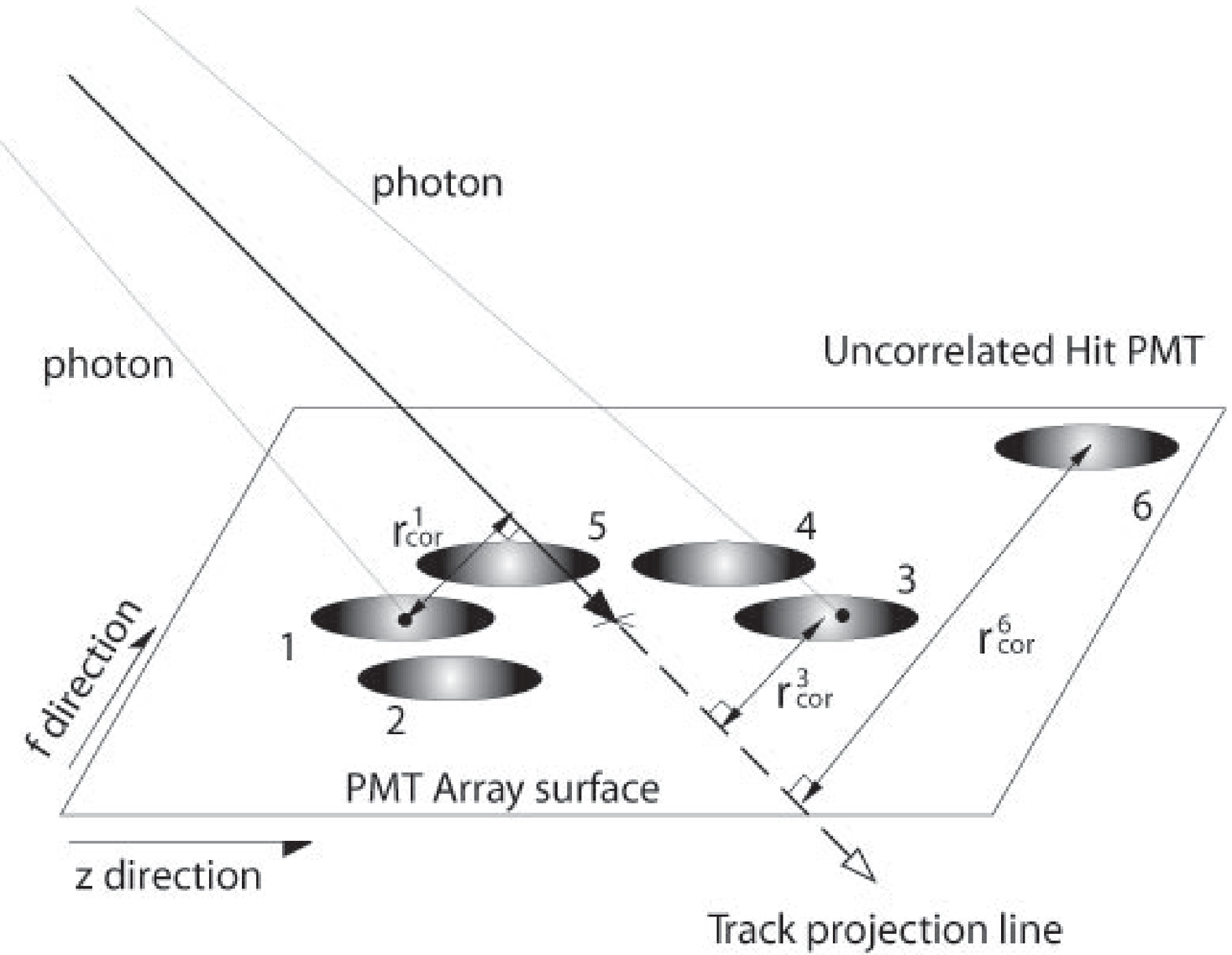,width=11cm}
    \caption{The schematic description of the 
    definition of variable which characterize the RICH ring.
    The track projection vector and five hit phototube are shown as an example.
    The distance between the center of  hit phototube 1, 3 and 
    the track projection vector are represented as 
    $r^{1}_{cor}$ and $r^{3}_{cor}$, respectively.
    \label{fig:r_def}}
\vspace{1cm}
    \epsfig{figure=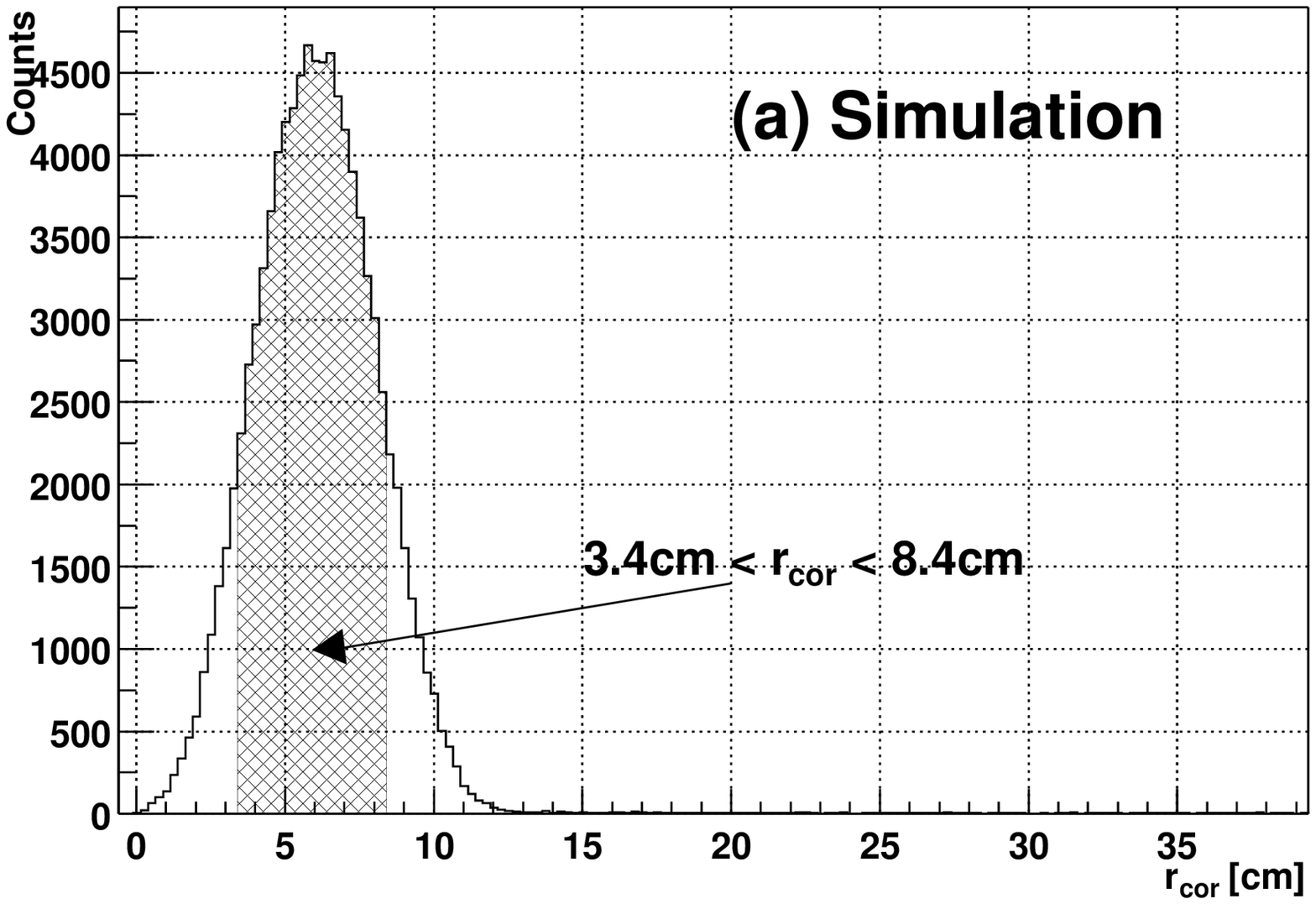,width=8cm}
    \epsfig{figure=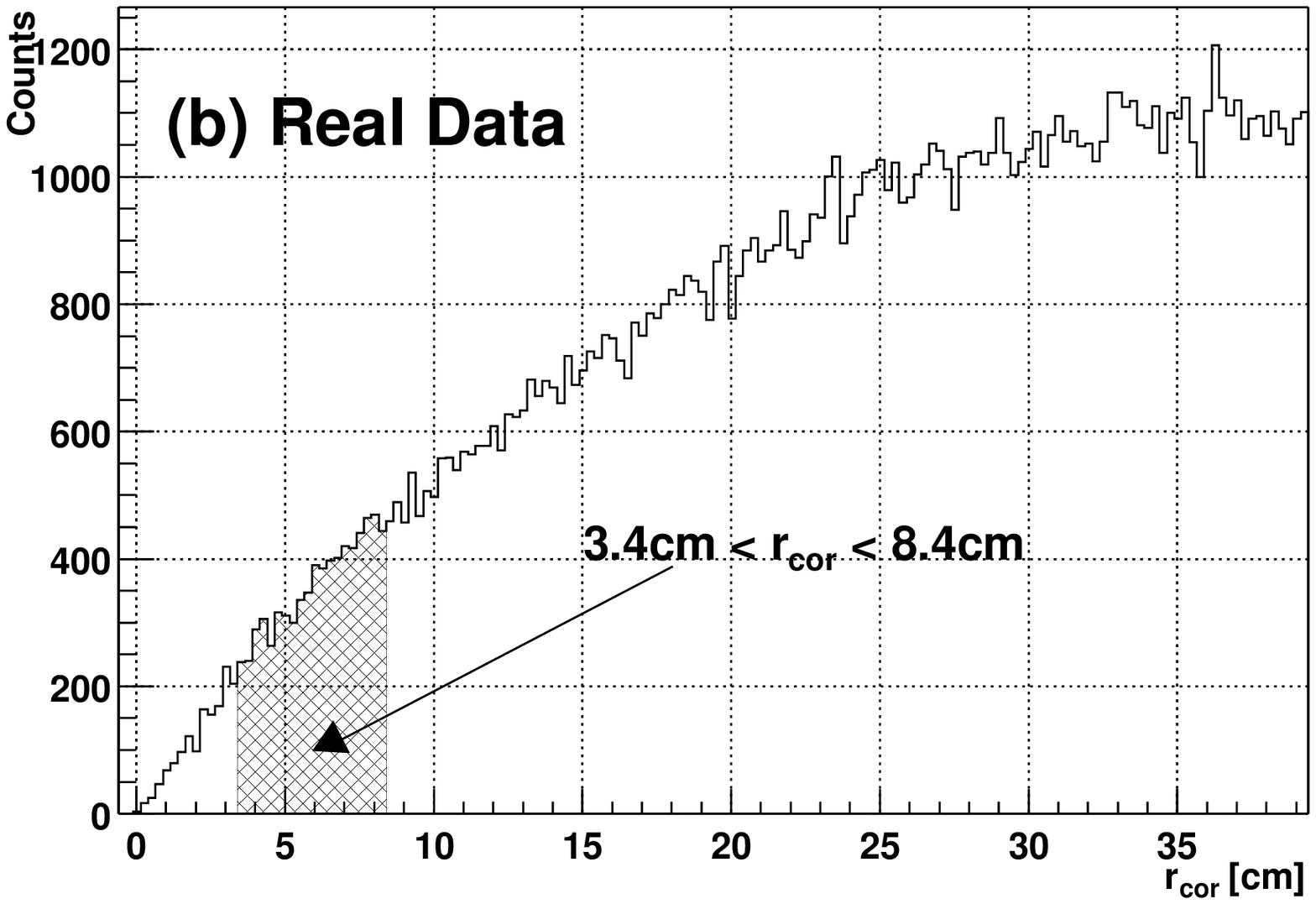,width=8cm}
    \caption{The $r_{cor}$ distribution for 
    (a) single electron simulation and (b) real data.
    The shaded area shows the $r_{cor}$ region from 3.4~cm to 8.4~cm.
    Since $r_{cor}$ is calculated for all hit phototube,
    there is large random association in real data.
    \label{fig:r_corr_hit}}

   \end{center}
  \end{figure}

  Figure~\ref{fig:r_def} shows a part of the phototube array surface
  with the  definition of the variables which characterizes the 
  association between a track and hit phototubes in RICH.
  A reconstructed track projection vector which is reflected by the mirror,
  track projection position at the phototube array surface ($R_{0}$),
  hit phototube's 1--5, and hit phototube position vector ($R_{i}$) are shown.

  The  distances between the center of hit phototube $R_{i}$ and the 
  track projection vector is calculated as $r_{cor}^{i}$ ($r_{cor}^1$
  and $r_{cor}^{3}$ in Fig.~\ref{fig:r_def}, for example).
  Figure~\ref{fig:r_corr_hit} shows the $r_{cor}$ distribution for the
  charged tracks for (a) simulation and (b) real data.
  The  $\left<r_{cor}\right>$ of 5.9~cm is ideal ring radius.
  The shaded area shows the $r_{cor}$ range of 5.9 \PM 2.5~cm corresponding $\pm 1\sigma$ region.

  The number of the fired phototubes with the association of the track, 
  {\sf n0, n1} is defined as follows.
  \begin{eqnarray}
    {\sf n0}&\equiv&{\rm number\quad of\quad fired \quad phototube\quad in} \quad
    3.4<r_{cor}<8.4 {\rm cm}, \\
    {\sf n1}&\equiv&{\rm number\quad of\quad fired\quad phototube\quad in}\quad
    r_{cor}<11.0 {\rm cm}.
  \end{eqnarray}
  The variable which represents the association between track and
  the number of photo-electrons in an associated RICH ring, {\sf npe0} is defined as 
  \begin{equation}
     {\sf npe0} \equiv \sum_{3.4~{\rm cm} < r_{i} < 8.4~{\rm cm}}N_{p.e}(i), 
  \end{equation}
  \noindent where $N_{p.e}(i)$ is the number of photo-electron in the  
   hit phototube denoted as $i$.
   
   The position of ring center, $R_{center}$, is derived as 
   the weighted average of hit phototube position $R_{i}$, where the weights
   are  taken to be  $N_{p.e}(i)$ 
   \begin{equation}
    R_{center} \equiv \frac {\sum_{i} N_{p.e}(i) \times R_{i}}
    {{\sf npe0}}. 
   \end{equation}
   The distance between the  RICH ring center,  $R_{center}$, and
   the reconstructed  track projection line is defined as {\sf disp}.
  \subsubsection{RICH Ring Shape : {\sf chi2/npe0} }\label{sec:sring}

  The variable which represents the ring shape is  {\sf chi2/npe0}.
  It is the weighted average of the deviation of the hit phototube position
  from the ideal ring radius, $r_0$.
  The weight is taken to be the number of photo-electrons in each phototube.
  The definition of   {\sf chi2/npe0}  is
  \begin{equation}
   {\sf chi2/npe0} \equiv  \frac{\sum_{3.4~cm<r_{i}<8.4~cm}(r_{i}-r_{0})^{2}
    \times N_{p.e}(i)}{ {\sf npe0}},
  \end{equation}	
  \noindent where $r_{i}$ represents the distance between phototube hit position and
  track projection point in the plane perpendicular to the track 
  projection line, and $r_{0}$ = 5.9~cm represents the mean ring radius
  $\left<r_{cor}\right>$  as shown in Fig.~\ref{fig:r_corr_hit}.
  
  \subsection{Electron Identification with EMCal}
  EMCal measures the energy and hit position of electrons and photons. 
  EMCal consists of  eight sectors, W0-W3 (in the west arm, from bottom to top) 
  and E0-E3 (in the east arm, from bottom to top). The sectors E0 and E1 
  are PbGl and the rest are PbSc. Energy calibration of each EMCal tower 
  is performed using the $\pi^0$ peak mass reconstructed from two photons.

  \subsubsection{Track - EMCal Hit Position Association : {\sf emcsdphi\_e, 
  emcsdz\_e}}\label{sec:demc}
\begin{figure}[htb]
   \begin{center}
    \epsfig{figure=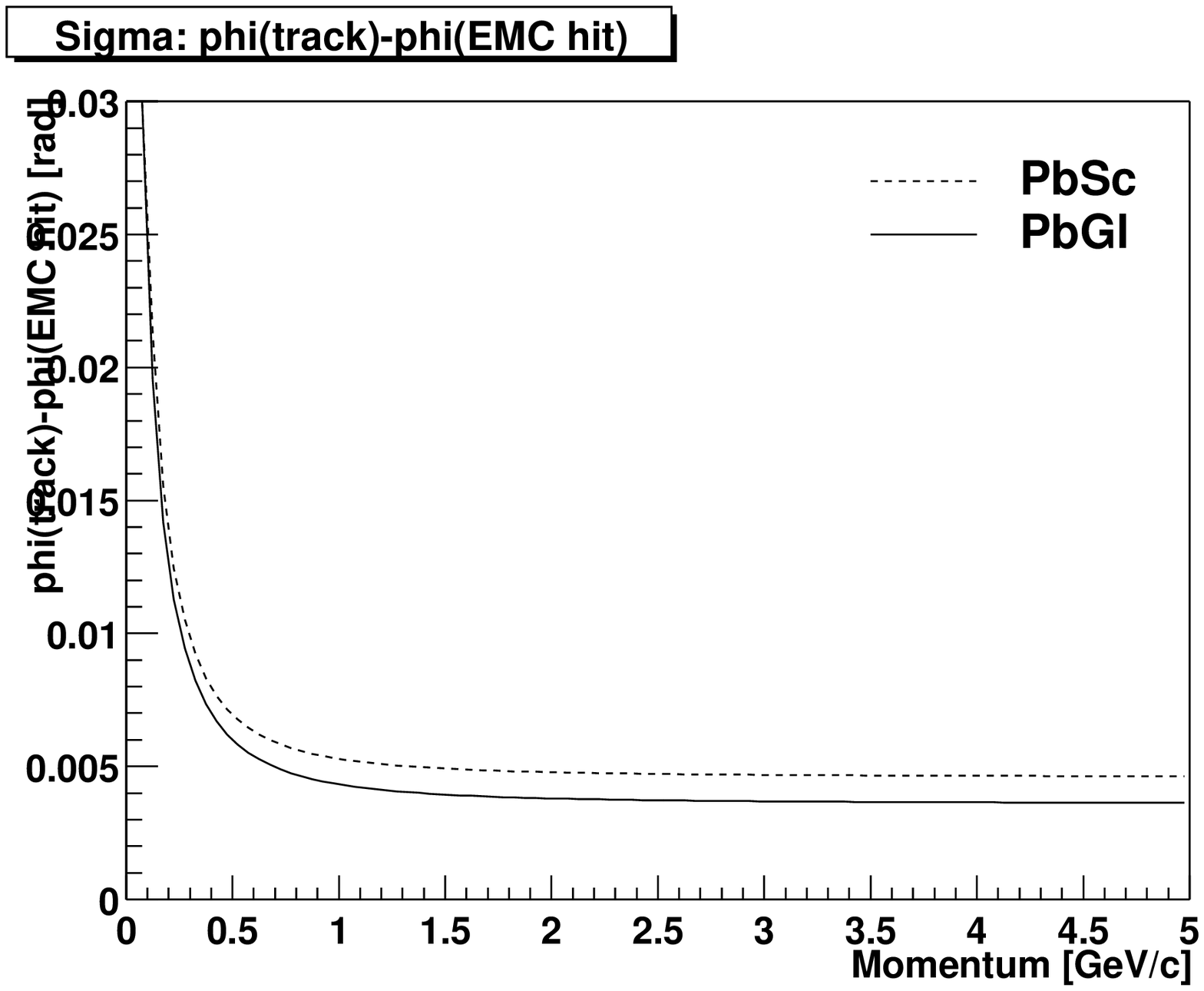,width=8.2cm}
    \epsfig{figure=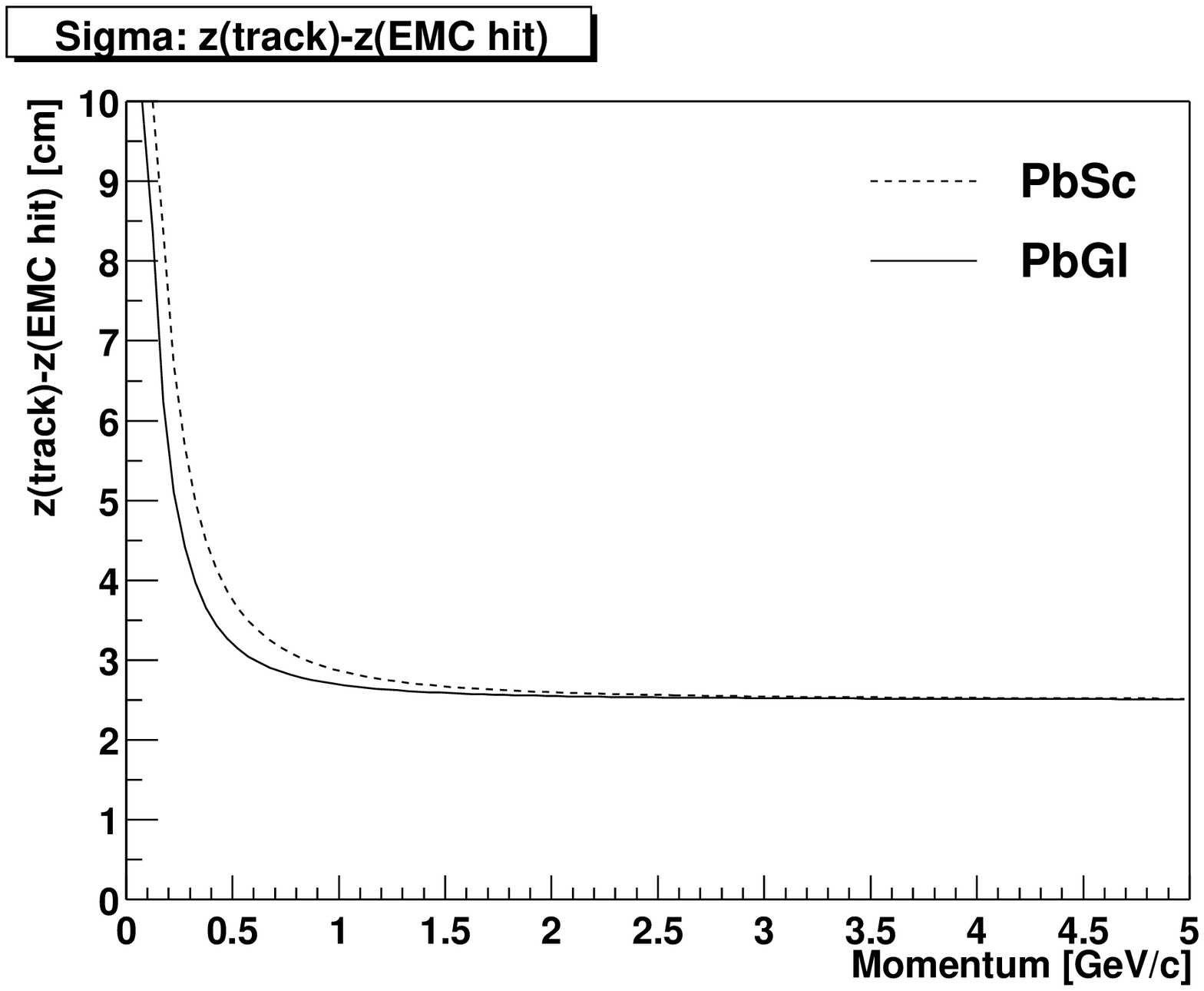,width=8.2cm}
    \caption{ Left : The momentum dependence of $\sigma_{\phi}$. 
     Right : The momentum dependence of $\sigma_{z}$.     
    The dashed line and solid line correspond to the 1 $\sigma$ value of 
    PbSc and PbGl, respectively.
    \label{fig:phi_z_sigma}}
   \end{center}
  \end{figure}

  The distance between the reconstructed track projection point 
  at the surface of the EMCal and the hit position (shower center)
  of EMCal is  parametrized by {\sf emcdphi} variable in $\phi$ direction, 
  {\sf emcdz} variable in $z$-direction.
  \begin{eqnarray}  
   {\sf emcdz} &=& {\sf emcz} - {\sf pemcz} ,\\
   {\sf emcdphi} &=& {\sf emcphi} - {\sf pemcphi}, 
  \end{eqnarray}	
  \noindent where {\sf pemcz} and {\sf pemcphi} are the projected $z$ and $\phi$
  position  of the reconstructed track at EMC surface,  {\sf emcz} and {\sf emcphi} 
  is the $z$ and $\phi$ position of EMC hit.
   {\sf emcdz} and {\sf emcdphi} are normalized by 
  $\sigma_{\sf emcdz}(p)$ and $\sigma_{\sf emcdphi}(p)$, which are the typical 
  momentum dependent width of {\sf emcdphi} and  {\sf emcdz} values.
  Figure.~\ref{fig:phi_z_sigma} shows $\sigma_{\sf emcphi}(p)$~(Left)
  and $\sigma_{\sf emcz}(p)$~(Right) 
  as a function of momentum.
  The normalized variables, {\sf emcsdphi\_e, emcsdz\_e} are determined
  as bellow.
  \begin{eqnarray}  
    {\sf emcsdz\_e} &=& \frac{{\sf emcdz} -<{\sf emcdz}>}{\sigma_{\sf emcsdz}(p)}, \\
    {\sf emcsdphi\_e} &=& \frac{{\sf emcdphi} -<{\sf emcdphi}>}{\sigma_{\sf emcsdphi}(p)},
  \end{eqnarray}	
  {\sf emcsdz\_e} and {\sf emcsdphi\_e} depend on the total momentum,
  the momentum direction and the electric charge of electrons/positrons 
  and the sector and position of EMCal due to the residual field.
  These variables are calibrated to the standard normal distribution 
  with a mean of 0 and a $\sigma$ of 1 for convenience of the analysis.

  \subsubsection{Energy - Momentum  Matching :  {\sf ecore/mom}}\label{sec:ep}
  Since the electron deposits all of its energy in EMCal and the 
  mass is small compared to its momentum, the ratio of the momentum, {\sf mom}
  to the shower energy  detected at EMCal, {\sf ecore}
  is  around  one  in case of the electron track.
  In the analysis, the ratio {\sf ecore/mom} is used  to select electron
  candidate.
  \subsubsection{Probability for Shower Shape Matching:  {\sf prob}}\label{sec:prob}
  To reduce charged hadron background in the electron candidates, the shower 
  shape information at the EMCal is used.
  This hadron rejection comes from the $\chi^2$ variable,
  \begin{equation}
  \chi^2 = \sum_i \frac{E_i-E'_i}{\sigma_i^2},
  \end{equation}
  \noindent where $E_i$ is the energy measured at tower number~(i) of EMCal,
  $E'_i$ is the predicted energy for an electromagnetic particle of total
  energy~$\sum E_i$, and $\sigma_i$ is the standard deviation for $E'_i$.
  Both $E'_i$ and  $\sigma_i$ are obtained from the electron test beam data.
  {\sf prob} is defined as the probability for a shower being an EM shower, 
  which is obtained by mapping  $\chi^2$ to the probability.

  \subsection{Used Cut for the Electron Reconstruction} \label{sec:ecut}
  The following cuts are used to select the electron candidate.
   \begin{itemize}
   \item{{\bf Event Cut}: $-25<{\sf bbcz}<25$ (cm)}
   \item{{\bf Track Quality}: {\sf quality}$>$15~(PC1 hit is required)}
   \item{{\bf RICH Hit}: {\sf n0}$>=2$ }
   \item{{\bf RICH Associate}: {\sf disp}$<6$ }
   \item{{\bf RICH Shape}: {\sf npe0/chi2}$<25$ }
   \item{{\bf EMCal E/p}: \\
     1.4$>${\sf ecore/mom}$>$ 0.65 ($p_{\mathrm{T}}<0.7$  GeV/$c$) \\
     1.4$>${\sf ecore/mom}$>$ 0.7 ($0.7<p_{\mathrm{T}}<1$  GeV/$c$)\\
     1.4$>${\sf ecore/mom}$>$ 0.75 ($1<p_{\mathrm{T}}<2$  GeV/$c$)\\
     1.4$>${\sf ecore/mom}$>$ 0.8 ($2<p_{\mathrm{T}}$  GeV/$c$)}
   \item{{\bf EMCal Associate}: $\mid${\sf emcsdphi\_e}$\mid$$<$4\&
     $\mid${\sf emcsdz\_e}$\mid$$<$4}
   \item{{\bf EMCal Shower Shape}: {\sf prob}$>0.01$ }\\ 
   \end{itemize}
   These cuts are called as 'standard eID cut' in this thesis.
   Hadron contamination in the selected electron samples with standard eID cut is bellow 1\% level
   for $p_{\mathrm{T}}<5$~GeV/$c$, as described at Sec.~\ref{sec:hadc}.
   
   Since pion emits Cerenkov light for $p_{\mathrm{T}}>$4.85 GeV/$c$, the pion rejection power of 
   RICH is reduced significantly. Therefore, 'tight electron cut' is applied to identify 
   high $p_{\mathrm{T}}$~($>$5.0 GeV/$c$) electrons.
   'Tight electron cut' is defined as follows.
    \begin{itemize}
   \item{{\bf Event Cut}: $-25<{\sf bbcz}<25$ (cm)}
   \item{{\bf Track Quality}: {\sf quality}$>$15~(pc1 hit was requite)}
   \item{{\bf RICH Hit}: {\sf n0}$>=2$\&\&  {\sf n1}$>=5$}
   \item{{\bf RICH Associate}: {\sf disp}$<6$ }
   \item{{\bf RICH Shape}: {\sf npe0/chi2}$<25$ }
   \item{{\bf EMCal E/p}: 1.4$>${\sf ecore/mom}$>$ 0.8 }
   \item{{\bf EMCal Associate}: $\mid${\sf emcsdphi\_e}$\mid$$<$4\&$\mid${\sf emcsdz\_e}$\mid$$<$4}
   \item{{\bf EMCal Shower Shape}: {\sf prob}$>0.1$} \\ 
   \end{itemize}
  \subsection{Detector Response}\label{sec:pisa}
  The variables used for the track reconstruction and the electron identification  
  are studied in the real data and the simulation.
  \subsubsection{PISA Simulation}
  Detector simulation is performed PISA, using the GEANT3 simulator of the PHENIX detector~\cite{bib:geant}.
  The PISA simulation is important in this analysis, since some of the correction factors
  are obtained from PISA simulation.
  To study the detector response with the PISA simulation, a single particle simulation for 
  a sample of electrons is performed with the  PISA simulation.
  CM-~- magnetic field is used as the RUN5 configuration and 
  CM++ magnetic field is used as the RUN6 configuration.
  Kinematic conditions of the generated single electron, which are determined to have wider region
  than analyzed region, are as follows.
  \begin{itemize}
  \item{Transverse momentum: $0.<p_{\mathrm{T}}<12.0$ GeV/$c$ (flat)}
  \item{Rapidity: $\mid$y$\mid$$<$ 0.5 (flat) }
  \item{Azimuthal angle: $0 < \phi < 2\pi$ (flat) }
  \item{vertex $z$: $\mid$vertex $z$$\mid$ $<$ 40 cm } 
  \end{itemize}
  Distributions of each variable in simulation which characterizes detector response are weighted
  according to the input $p_{\mathrm{T}}$, so that the distribution of the input $p_{\mathrm{T}}$ 
  have a realistic $p_{\mathrm{T}}$ distribution of electrons.
  \subsubsection{Comparison Between Real Data and Simulation}
   \begin{table}[htb]
     \begin{center}
     \caption{The eID cut used for the comparison of the variables.\label{tab:ecut_comp}}
       \begin{tabular}{c|cccccc}
	 Used cut&  Compared variable & & & & &   \\
	 \hline \hline
	 & {\sf n0} & {\sf disp} & {\sf npe0/chi2} & {\sf emcsdphi(z)\_e} &
         {\sf ecore/mom} & {\sf prob}\\
	 {\sf n0}$>=2$ & $\times(${\sf n1}$>=1$) &$\bigcirc$&$\bigcirc$&$\bigcirc$
	 &$\bigcirc$&$\bigcirc$\\
	 {\sf disp}$<6$ &$\bigcirc$  &$\times$&$\bigcirc$&$\bigcirc$
	 &$\bigcirc$&$\bigcirc$\\
	 {\sf npe0/chi2}$<25$ &$\bigcirc$ &$\bigcirc$& $\times$&$\bigcirc$
	 &$\bigcirc$&$\bigcirc$\\
	 $\mid${\sf emcsdphi(z)\_e}$\mid$$<$4&$\bigcirc$ &$\bigcirc$&$\bigcirc$& $\times$
	 &$\bigcirc$&$\bigcirc$\\
	 {\sf ecore/mom} cut $\bigcirc$ &$\bigcirc$&$\bigcirc$& $\bigcirc$
	 &$\times$&$\bigcirc$\\
	 {\sf prob}$>0.01$&$\bigcirc$ &$\bigcirc$&$\bigcirc$& $\bigcirc$
	 &$\bigcirc$&$\times$\\
       \end{tabular}
     \end{center}
   \end{table}
   The distributions of the variables for the electron identification in the 
   PISA simulation are compared to those of the real data.
   The applied cuts for the comparison of each variable are summarized at Table~\ref{tab:ecut_comp}.
   Electron sample with 0.5$<p_{\mathrm{T}}<$5 GeV/$c$ is selected for this comparison.
   Electron sample for the comparison is selected by the the 'standard eID' without the cut 
   for the compared variable.
   {\sf n1}$>=1$ and 0.8$<${\sf ecore/mom}$<$ 1.4 cut is used instead of 
   ${\sf n0}$ cut for the ${\sf n0}$ comparison.
   
\begin{figure}[htb]
     \begin{center}
       \epsfig{figure=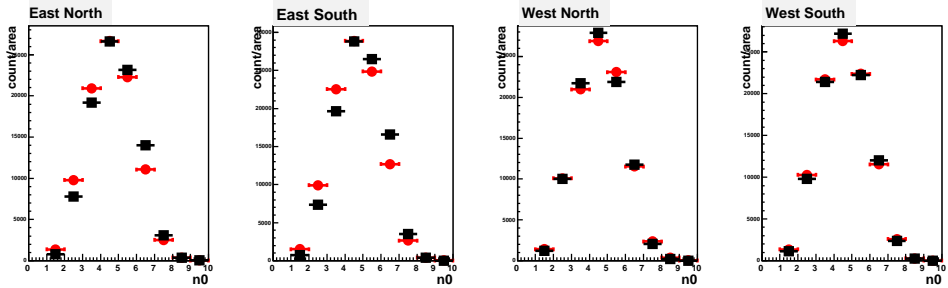,width=13cm}
       \caption{
	 The distribution of {\sf n0} with the standard eID cut without {\sf n0} cut 
	 and the 0.5$<p_{\mathrm{T}}<$5 GeV/$c$ cut in the real data~(square) and 
	 simulation~(circle).
	 \label{fig:n0_run5}}
     \end{center}
   \end{figure}

\begin{figure}[htb]
     \begin{center}
       \epsfig{figure=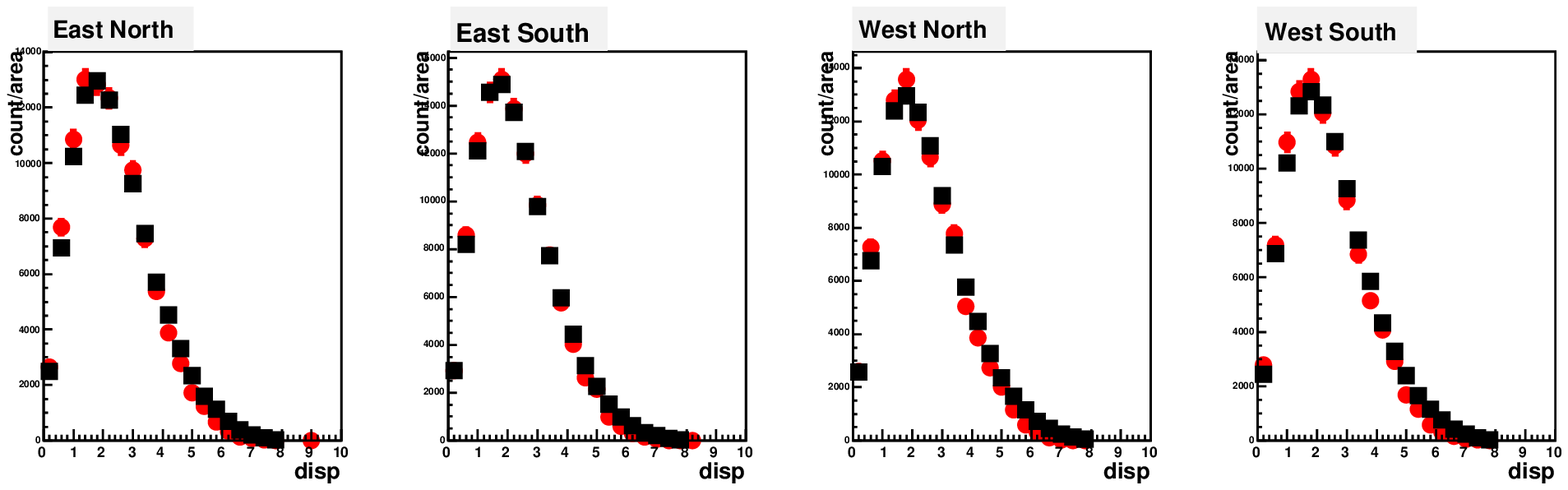,width=13cm}
       \caption{
	 The distribution of {\sf disp} with the standard eID cut without {\sf disp} cut 
	 and the 0.5$<p_{\mathrm{T}}<$5 GeV/$c$ cut in the real data~(square) and 
	 simulation~(circle).
	 \label{fig:disp_run5}}
     \end{center}
   \end{figure}

\begin{figure}[htb]
     \begin{center}
       \epsfig{figure=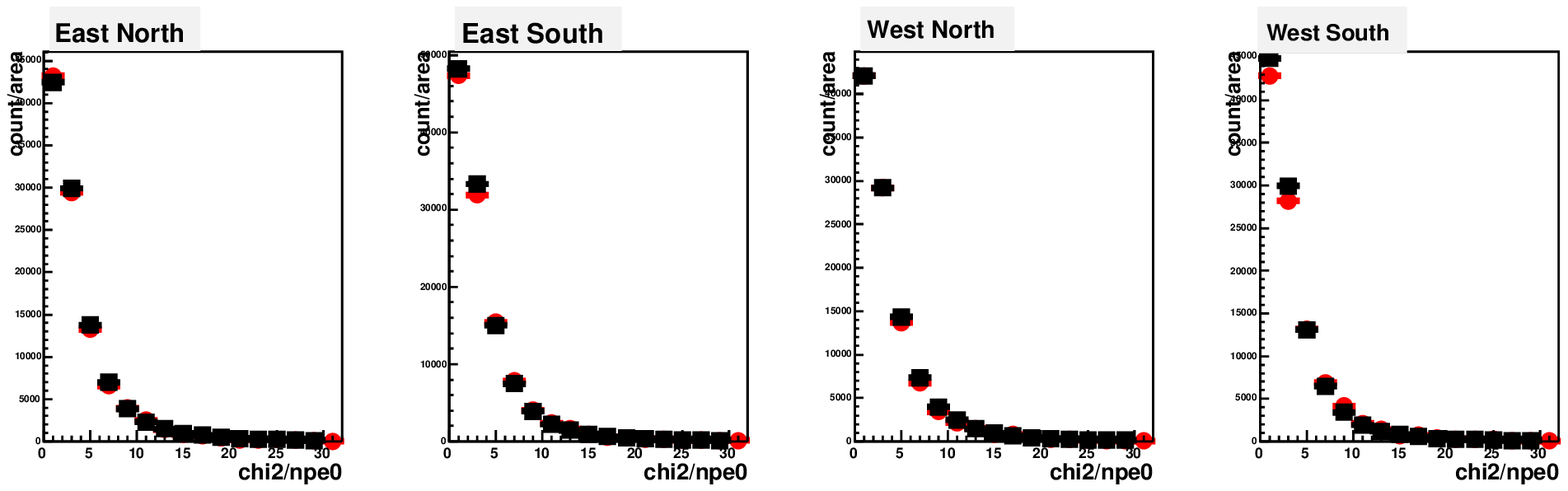,width=13cm}
       \caption{
	 The distribution of {\sf chi2/npe0} with the standard eID cut without {\sf chi2/npe0} 
	 cut and the 0.5$<p_{\mathrm{T}}<$5 GeV/$c$ cut in the real data~(square) and 
	 simulation~(circle).
	 \label{fig:chi_run5}}
     \end{center}
   \end{figure}

\begin{figure}[htb]
     \begin{center}
       \epsfig{figure=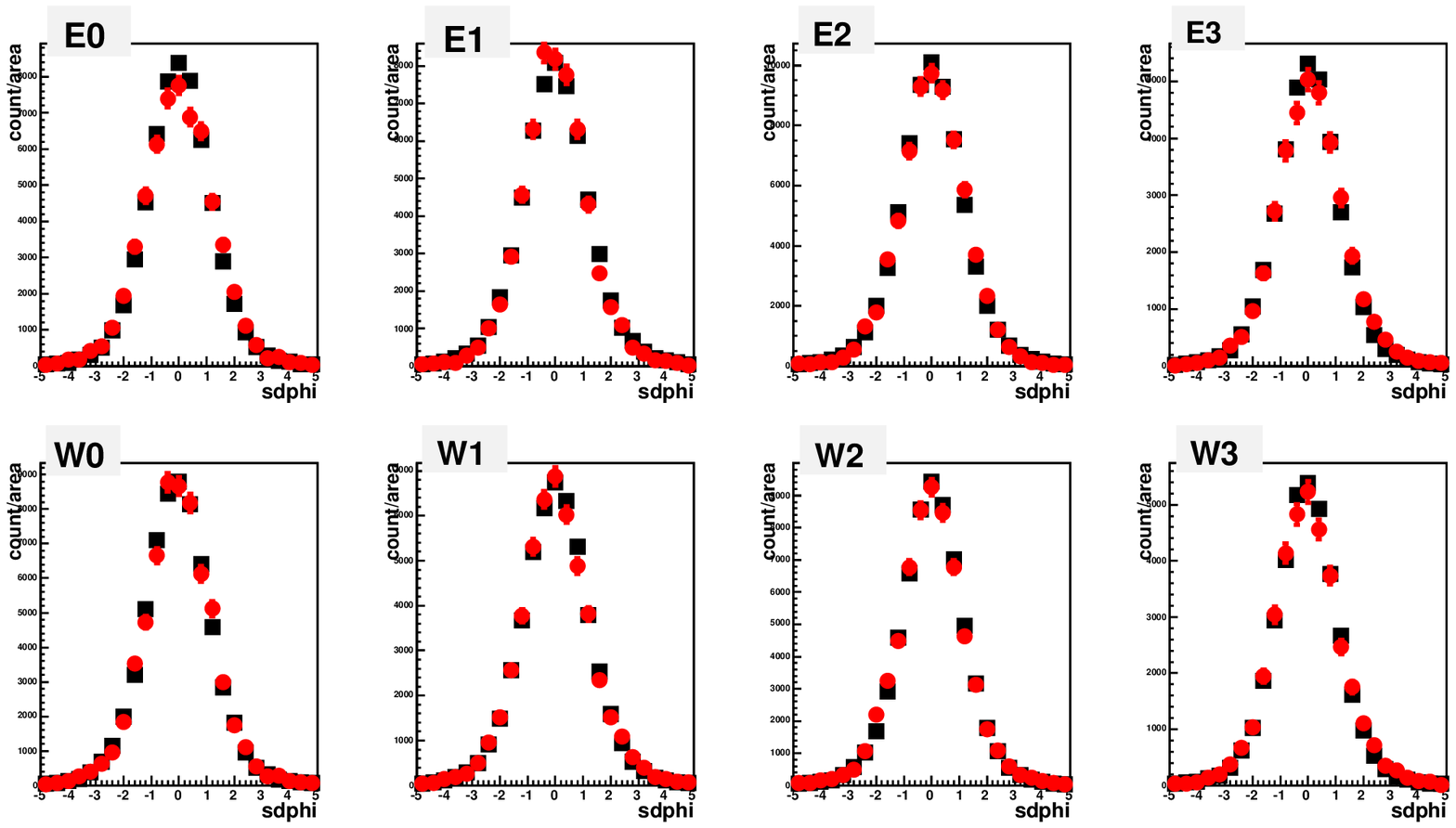,width=13cm}
       \caption{
	 The distribution of {\sf emcsdphi\_e} with the standard eID cut without 
	 {\sf  emcsdphi(z)\_e} cut and the 0.5$<p_{\mathrm{T}}<$5 GeV/$c$ 
	 cut in the real data~(square) and simulation~(circle).
	 \label{fig:sdphi_run5}}
     \end{center}
   \end{figure}

\begin{figure}[htb]
     \begin{center}
       \epsfig{figure=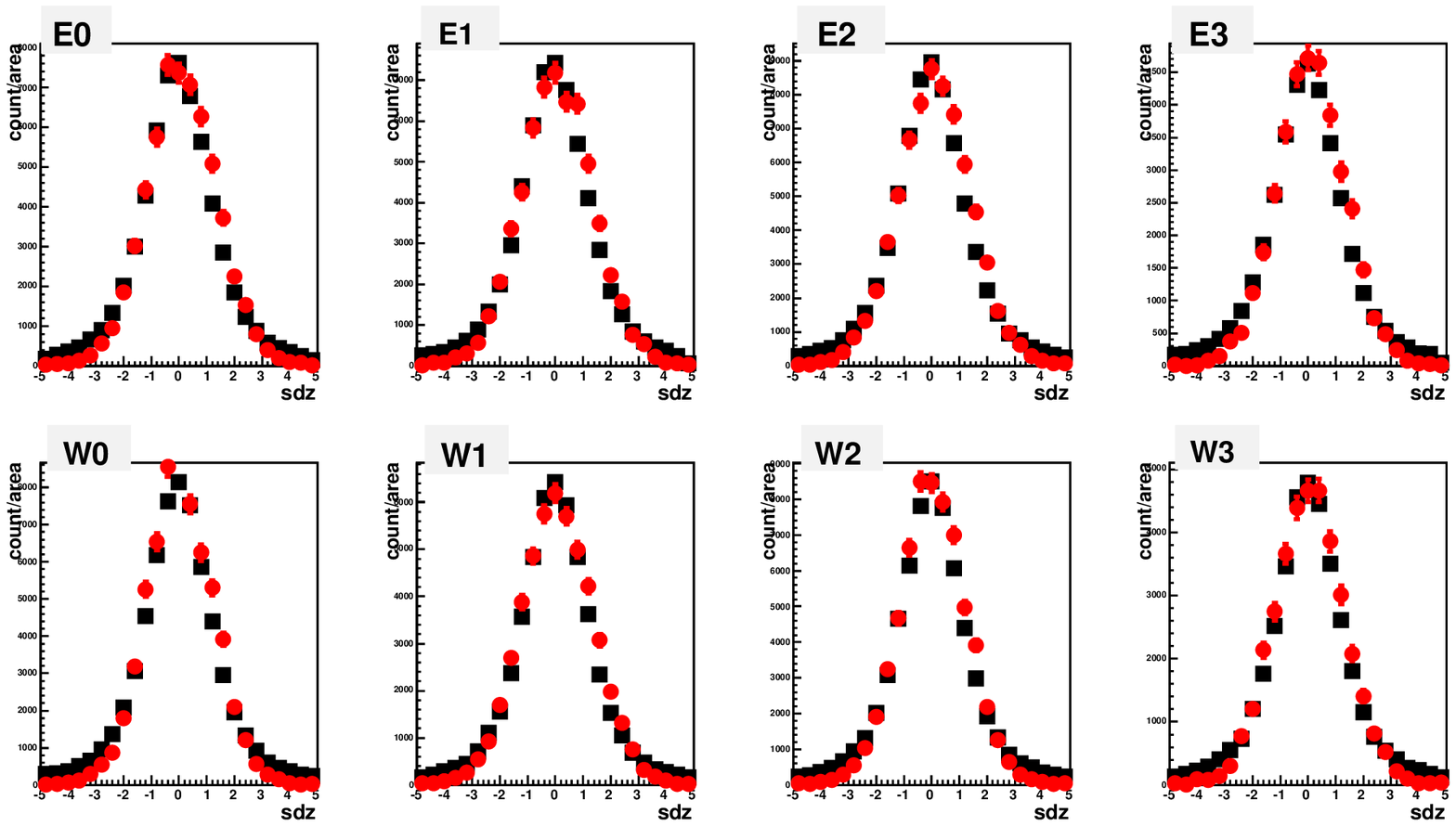,width=13cm}
       \caption{
	 The distribution of {\sf emcsdz\_e} with the standard eID cut without 
	 {\sf  emcsdphi(z)\_e}  cut 
	 and the 0.5$<p_{\mathrm{T}}<$5 GeV/$c$ cut in the real data~(square) and 
	 simulation~(circle).
	 \label{fig:sdz_run5}}
     \end{center}
   \end{figure}

\begin{figure}[htb]
     \begin{center}
       \epsfig{figure=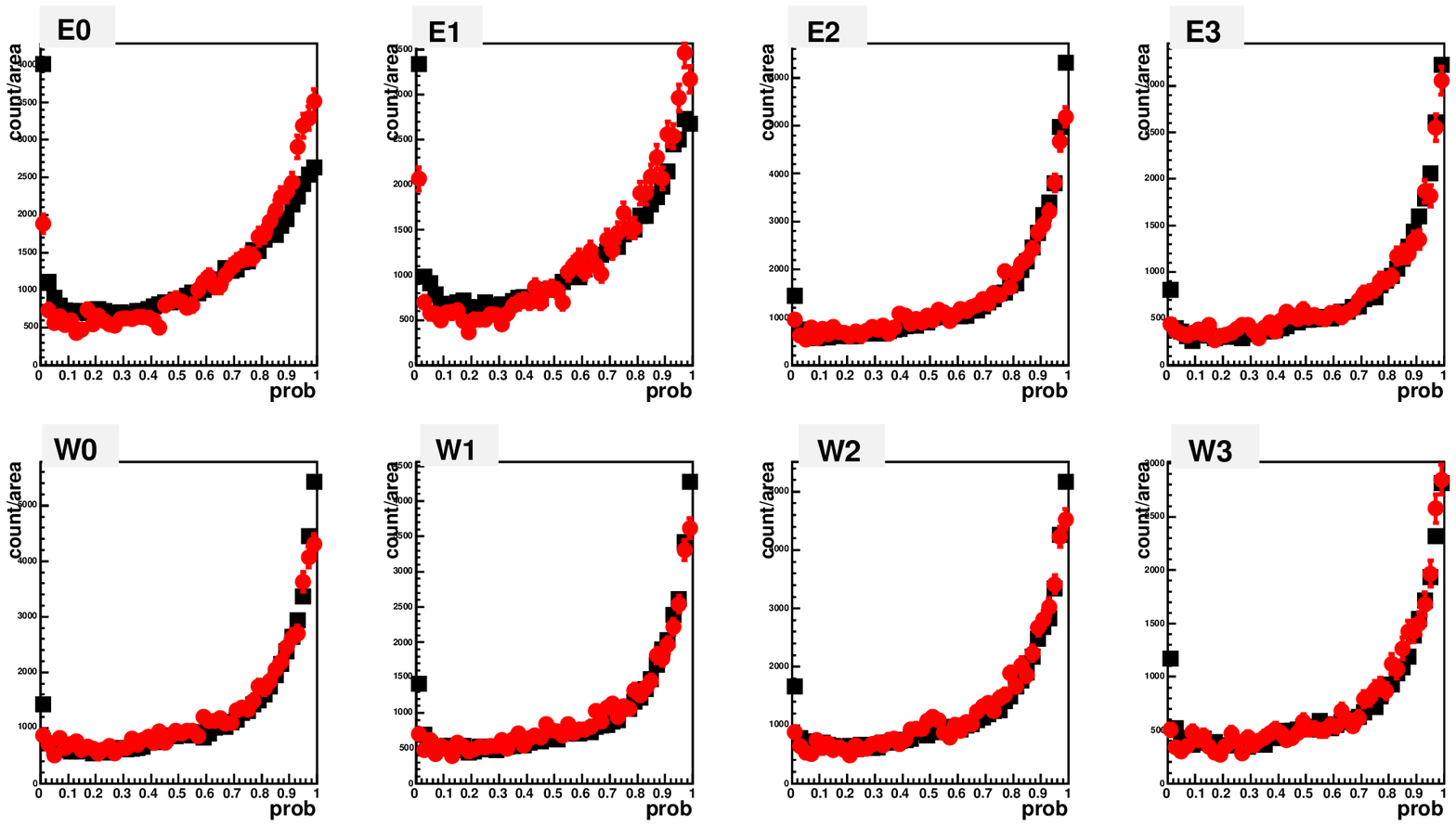,width=13cm}
       \caption{
	 The distribution of {\sf prob} with the standard eID cut without {\sf prob} cut 
	 and the 0.5$<p_{\mathrm{T}}<$5 GeV/$c$ cut in the real data~(square) and 
	 simulation~(circle).
	 \label{fig:prob_run5}}
     \end{center}
   \end{figure}

\begin{figure}[htb]
       \begin{center}
	 \epsfig{figure=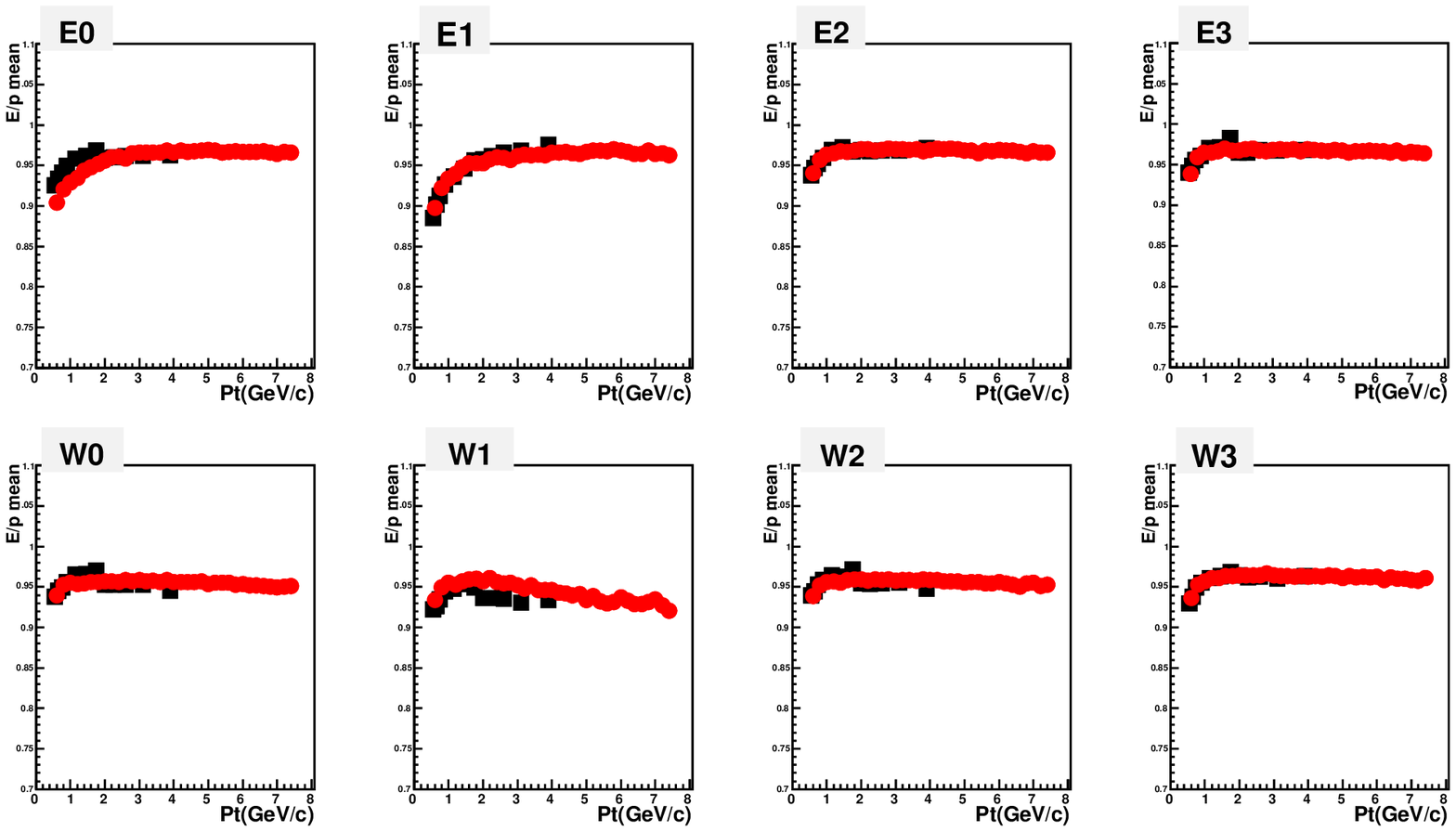,width=13cm}
	 \caption{
	   The mean value of {\sf ecore/mom} distribution with the standard eID cut 
	   as a fiction of electron $p_{\mathrm{T}}$ in the real data~(square) and 
	   simulation~(circle).
	   \label{fig:epmean_run5}}
       \end{center}
     \end{figure}

\begin{figure}[htb]
       \begin{center}
       \epsfig{figure=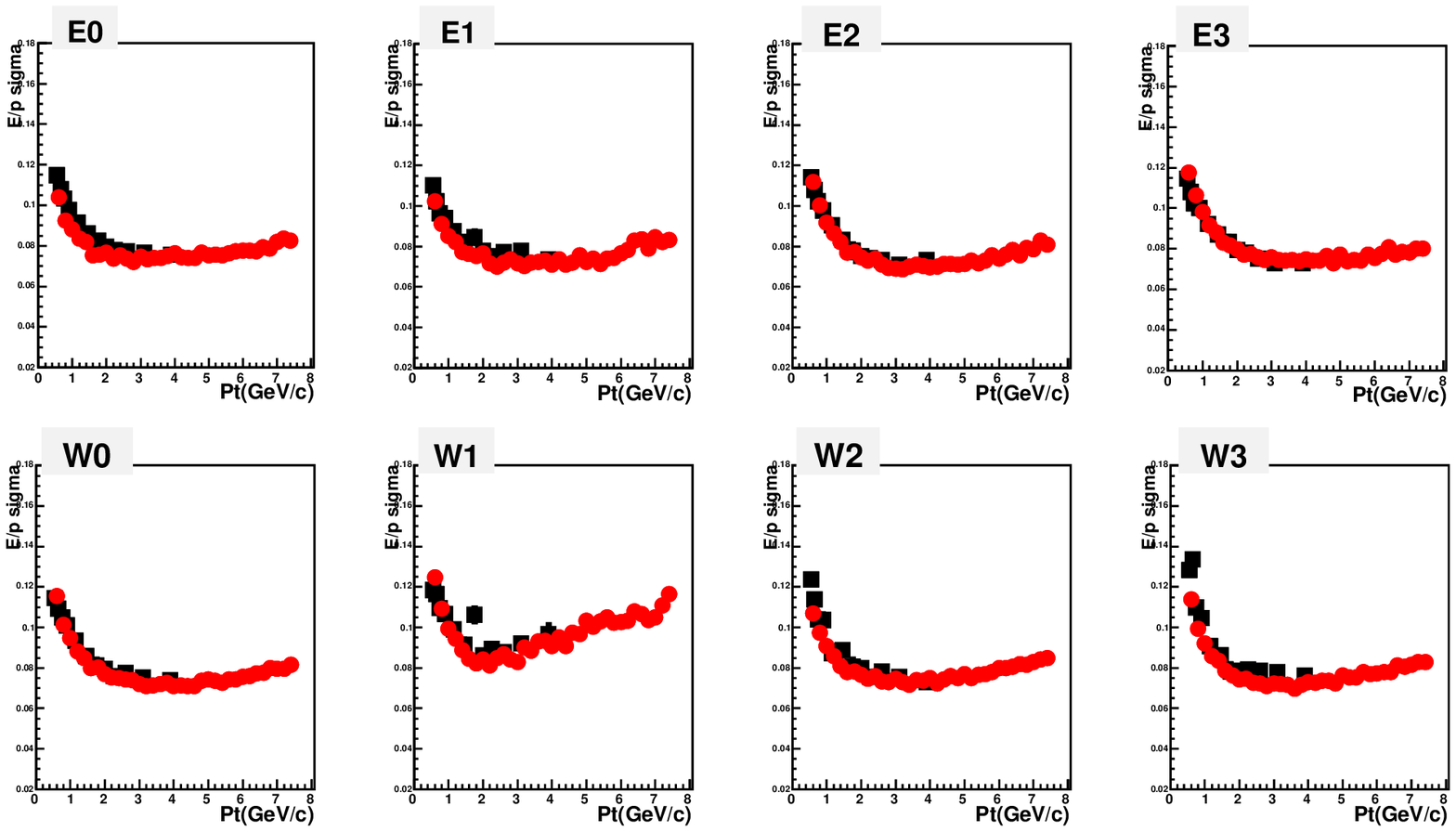,width=13cm}
       \caption{
	 The sigma value of {\sf ecore/mom} distribution with the standard eID cut 
	 as a fiction of electron $p_{\mathrm{T}}$ in the real data~(square) and 
	 simulation~(circle).
	 \label{fig:epsigma_run5}}
     \end{center}
   \end{figure}

   Figure~\ref{fig:n0_run5}, \ref{fig:disp_run5} and \ref{fig:chi_run5} show 
   the distributions of RICH variables, {\sf n0}, {\sf disp} and {\sf chi2/npe0} 
   at each RICH sector, respectively.
   In addition, Figure~\ref{fig:sdphi_run5}, ~\ref{fig:sdz_run5} and 
   ~\ref{fig:prob_run5} show the distributions of EMCal variables at each sector, 
   {\sf emcsdphi\_e}, {\sf emcsdz\_e} and {\sf prob}, respectively.
   Figure~\ref{fig:epmean_run5} and \ref{fig:epsigma_run5} show mean and sigma values
   of {\sf ecore/mom} distributions as a function of electron $p_{\mathrm{T}}$.
   In Fig.~\ref{fig:n0_run5}-\ref{fig:epsigma_run5}, black squares show the results from the real data
   in RUN5 and red circles show these from the PISA simulation with RUN5 tuning parameters and 
   CM-~- field.
   The distribution in simulation is normalized by the number of entries at 
   each sector.
   The distributions of the simulation and these of the real data match well.
   The difference of the efficiency of the cut for the each variable between the real data
   and the simulation is less than 1\%, as describled in Sec.\ref{sec:sysinv}.
   The comparison between real data and simulation in RUN6 is described in Sec.\ref{sec:eidrun6}

   \subsubsection{Acceptance Evaluation for eID}
\begin{figure}[htb]
       \begin{center}
	 \epsfig{figure=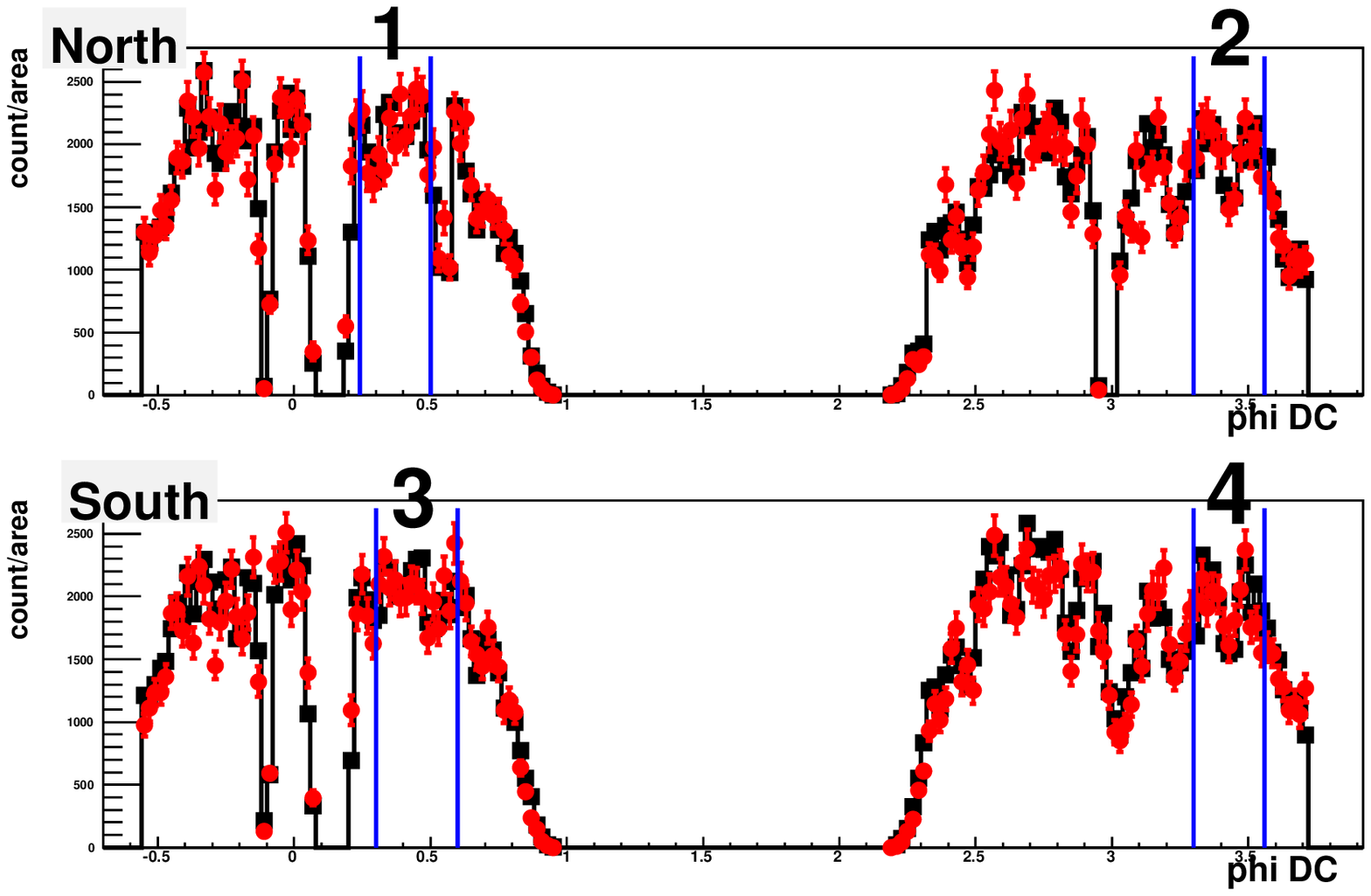,width=13cm}
	 \caption{
	   The distribution of {\sf phi} with the standard eID cut and 
	   the 0.5$<p_{\mathrm{T}}<$5 GeV/$c$ cut in the real data in RUN5~(square) and 
	   simulation~(circle).
	   \label{fig:dcphi_run5}}
       \end{center}
     \end{figure}

\begin{figure}[htb]
       \begin{center}
       \epsfig{figure=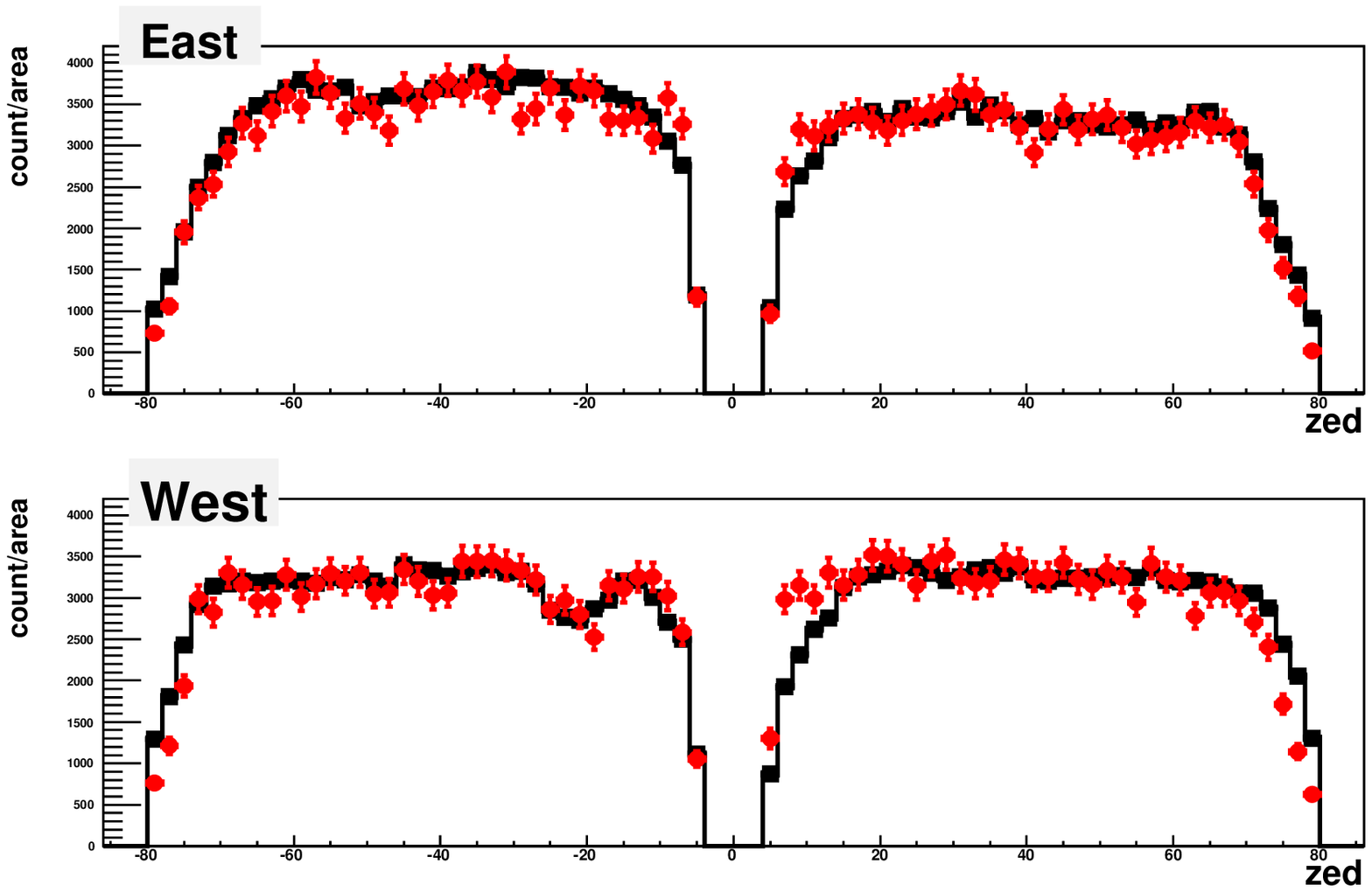,width=13cm}
       \caption{
	 The distribution of {\sf zed} with the standard eID cut and 
	 the 0.5$<p_{\mathrm{T}}<$5 GeV/$c$ cut in the real data in RUN5~(square) and 
	 simulation~(circle).
	 \label{fig:dczed_run5}}
     \end{center}
   \end{figure}

   Detector area with low efficiency, dead or noisy is removed from
   the data analysis by fiducial cut. 
   The same fiducial cut is also applied for the simulation to make the geometrical
   acceptance of the simulation identical as that of the real data, so that
   the reconstruction efficiency is evaluated from the simulation.
   The distributions of {\sf phi}, {\sf zed} of the simulation are compared 
   with these of the real data for the electron samples selected by the standard eID 
   and a transverse momentum  with  0.5$<p_{\mathrm{T}}<$5 GeV/$c$.
   Figure~\ref{fig:dcphi_run5} shows the distributions of  {\sf phi} at North~(top panel)
   and South~(bottom panel)sector, and 
   Figure~\ref{fig:dczed_run5} shows the distributions of  {\sf zed} at East~(top panel)
   and West~(bottom panel) sector.
   In Fig.~\ref{fig:dcphi_run5} and \ref{fig:dczed_run5}, black squares show 
   the real data in RUN5 and red circles show the PISA simulation with 
   RUN5 tuning parameters and CM-~- field.
   The distributions of simulation are normalized by number of entries in 
   the reference regions, where are little low efficiency, dead or noisy area.
   In Fig~\ref{fig:dcphi_run5} and \ref{fig:dczed_run5}, the used reference region  
   to normalize is region 1.
   The ratio of the number of entries in the simulation over that in the real data
   except for the reference region used for the normalization is calculated
   for each reference region.
   The same procedure is done for the real data in RUN6 and the simulation with
   RUN6 tuning parameters and CM++ field.
   The distribution of {\sf phi}, {\sf zed} in RUN6 is described at Sec.~\ref{sec:georun6}
   The results of the ratios in RUN5 and RUN6 are summarized at Table~\ref{tab:geoacc}.
   The geometrical acceptance of the PISA simulation agrees with the real data within 3\%.
   \begin{table}[hbt]
     \begin{center}
       \caption{The ratio of the number of entries at the simulation over that at real data}
       \label{tab:geoacc}
       \begin{tabular}{c|cc}
	 \hline
	 reference region & the simulation/real~(RUN5)&  the simulation/real~(RUN6)\\
	 \hline\hline
	 region~1 & 0.98 & 1.01 \\
	 region~2  & 0.97 &0.97 \\
	 region~3 & 0.98 &1.01 \\
	 region~4 &0.99 &0.98 \\
	 \hline
       \end{tabular}
     \end{center}
   \end{table}

   \clearpage
    \section{Run Selection}
     \begin{table}[hbt]
     \begin{center}
       \caption{Summary of run group}
       \label{tab:rung}
       \begin{tabular}{c|c}
	 \hline
	 RUN group & RUN Number   \\
	 \hline\hline
	 RUN5& \\
	 G5A &   166030-171594 \& 172081-176573 \\
	 C   & 171595-172080\\
	 G5B & 176574-178936\\
	 G5L & 178936-179846\\
	 
	 \hline\hline
	 RUN6& \\
	 G6   & 188216-202500\\
	 G6HBD  & 202500-204639\\
	 \hline
       \end{tabular}
     \end{center}
   \end{table}
     Electron yield and hadron yield per event are checked for each run taken during RUN5 and RUN6, 
     which tells the stability of the track reconstruction and the electron identification performance 
     as a function of time period.
     Furthermore, some runs, which have low electron yield due to inefficient detector response, 
     are removed in the analysis.  
    
    Physics runs in RUN5 and RUN6 are divided into  several run groups
    according to detector and trigger configuration as listed at Table~\ref{tab:rung}.
    The PH trigger mask (4x4c) is changed after RUN176574 in RUN5. 
    As a result, the 4x4c trigger efficiency in G5A and G5B is slightly different.
    Electron yield in the G5L group is reduced, since two of RICH data packets
    are disabled during this period.
    The G5L group is not used for this analysis.
    The photon converter is installed in the C group.

    A new detector, Hadron Blind Detector~(HBD) is installed at the front of
    East DC in the G6HBD group for the performance test of the HBD.
    The G6HBD group is not used for this analysis, since the amount of 
    the electrons from $\gamma$ conversion increases due to the material budget 
    of the HBD.
    The fiducial cut and offline mask for the 4x4c trigger are applied
    so that the detector and trigger configuration become identical during
    the G6 group.
    
    The good runs in the G5A, G5B, G6 and C groups are selected as follows.
    \begin{itemize}
    \item{Select the good runs according to the hadron yield~(0.4$<p_{\mathrm{T}}<$5
      GeV/$c$) in each run in MB trigger data}
    \item{Select the good runs according to the electron yield~(0.5$<p_{\mathrm{T}}<$5
      GeV/$c$) in each run in MB trigger data}
    \item{Select the good runs according to the electron yield~
      (1.6$<p_{\mathrm{T}}<$5~GeV/$c$) in each run in PH trigger data} 
    \end{itemize}

    The procedure to select good runs according to the hadron yield is 
    as follows.
    The hadron is selected by the {\sf quality}$>15$, {\sf n0}$<0$ cut and 
    with 0.4$<p_{\mathrm{T}}<$5 GeV/$c$.
    \begin{enumerate}
    \item  Mean of $N_{ch}(run)$ in each good run~($<N_{ch}>$) and RMS of $N_{ch}(run)$ in
      good runs~($\sigma$) are obtained.
    \item calculate the following ratio,
      \begin{equation}
      \frac{\mid N_{ch}(run) - <N_{ch}>\mid}{\sigma}.
      \end{equation}
    \item Remove runs which the ratio is above 2.5 from good runs.
    \end{enumerate}  
    The above procedure is continued, until no run is removed by the procedure.
    
    The procedure to select good runs according to the electron yield is 
    as follows.
    The electron is selected by the standard eID cut and 
    with 0.5~(1.6)$<p_{\mathrm{T}}<$5~GeV/$c$ for MB~(PH) data.
    \begin{enumerate}
    \item  Mean of $N_{ele}(run)$ in each good run~($<N_{ele}>$) and statistical errors
      of $N_{ele}(run)$ in good runs~($\sigma(run)$) are obtained.
    \item calculate the following ratio,
      \begin{equation}
      \frac{\mid N_{ele}(run) - <N_{ele}>\mid}{\sigma(run)}.
      \end{equation}
    \item Remove runs with the ratio is larger than 3 from good runs.
    \end{enumerate}  
    The above procedure is also continued, until no run is removed by the procedure.
    
\begin{figure}[htb]
    \begin{center}
      \includegraphics[width=15cm]{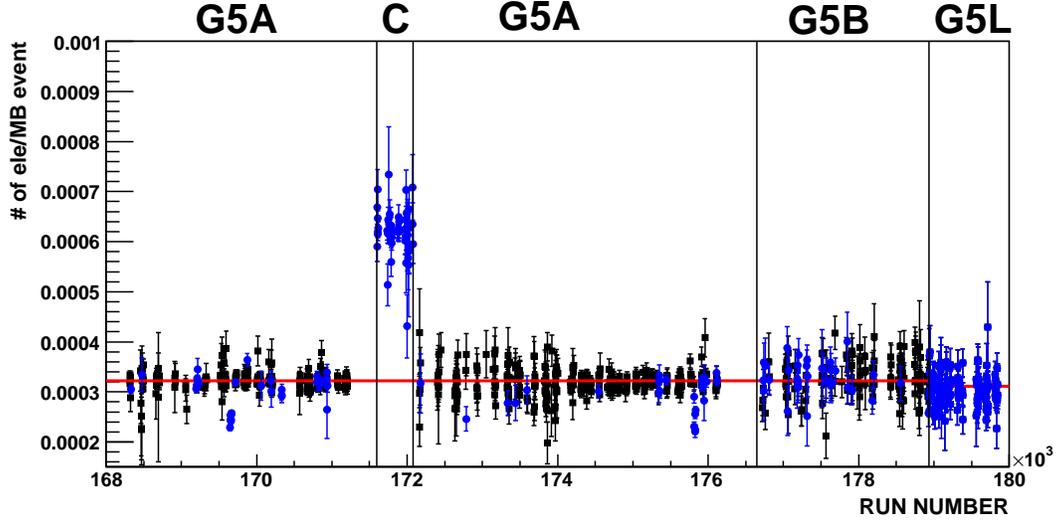}
      \caption{Electron yield in 0.5-5.0 GeV/$c$ per MB event in RUN5 as 
	a function of run number.
	Black circles show good runs and blue squares show bad runs.}
      \label{fig:run5good}
    \end{center}
    \end{figure}

\begin{figure}[htb]
      \begin{center}
	\includegraphics[width=15cm]{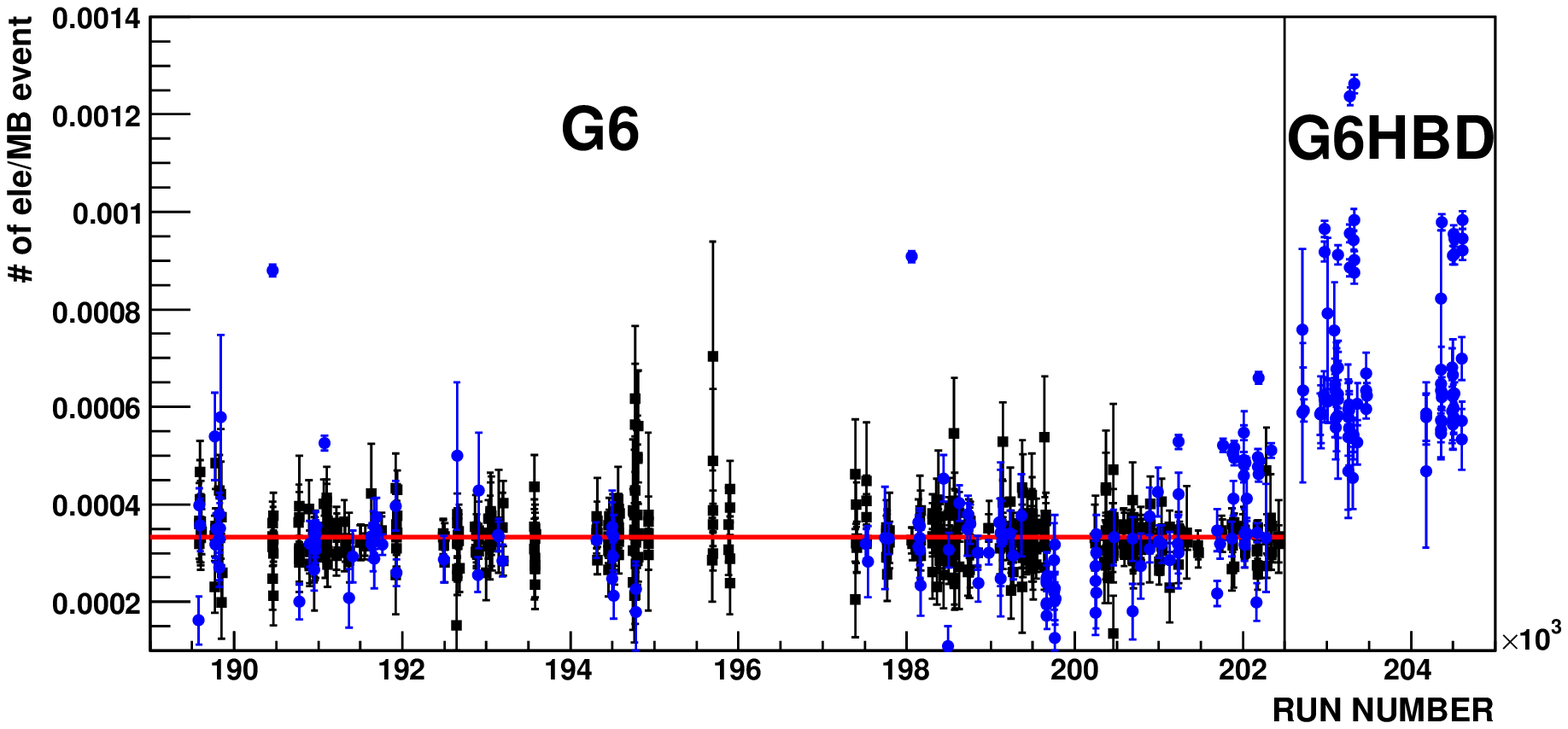}
	\caption{Electron yield in 0.5-5.0 GeV/$c$ per MB event in RUN6 as 
	  a function of run number.
	  Black circles show good runs and blue squares show bad runs.}
	\label{fig:run6good}
    \end{center}
    \end{figure}

    As a result, 596 runs out of 722 runs in the G5A and G5B group are selected
    in RUN5 and 501 runs out of 643 runs in the G6 group are selected in RUN6.
    51 runs out of  58 runs in the C group in RUN5 are selected as good runs
    for the converter analysis.
    As an example, Figure~\ref{fig:run5good} and Figure~\ref{fig:run6good}
    show the electron yield in 0.5-5.0 GeV/$c$ per MB event as 
    a function of run number in RUN5 and RUN6, respectively.
    Black circles show the selected good runs and blue squares show runs rejected
    for this analysis.

    \section{Inclusive Electron Spectrum}
    Invariant yield of inclusive electron is obtained using Eq.~\ref{eq:inv} and this procedure 
    is described in this section. 
    In Eq~\ref{eq:inv}, $\epsilon(p_{\mathrm{T}})$, which  is the overall efficiency 
    including acceptance, reconstruction efficiency and trigger efficiency, 
    can be written as follows in MB and PH data.
    \begin{eqnarray}  
      \epsilon(p_{\mathrm{T}}) &=& A\times \epsilon_{eff}(p_{\mathrm{T}})\quad\quad
      ({\rm MB data}),\\
      \epsilon(p_{\mathrm{T}}) &=& A\times \epsilon_{eff}(p_{\mathrm{T}})\times 
      \epsilon_{trig}(p_{\mathrm{T}})\quad\quad
      ({\rm PH data}),
    \end{eqnarray}	
    \noindent
    where $A\times \epsilon_{eff}(p_{\mathrm{T}})$ is the acceptance times 
    the reconstruction efficiency for electrons and 
    $\epsilon_{trig}(p_{\mathrm{T}})$ is the 4x4c~(PH) trigger efficiency.
    \subsection{Reconstruction Efficiency for Electron}\label{sec_eid}
    \subsubsection{Reconstruction Efficiency with Standard eID Cut}
\begin{figure}[htb]
	\begin{center}
	  \includegraphics[width=13cm]{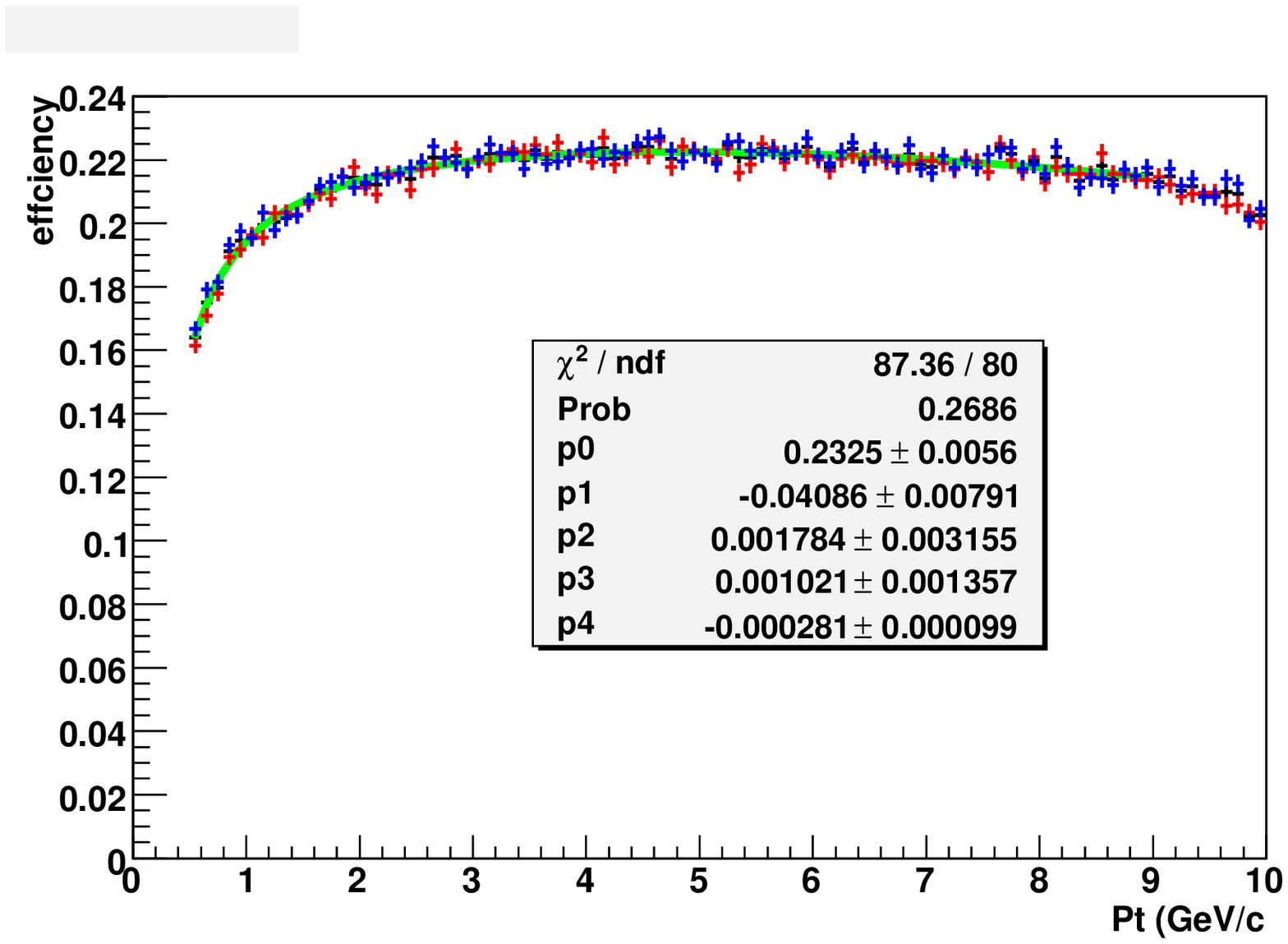}
	  \caption{Electron efficiency as a function of electron$p_{\mathrm{T}}$.
	    Red points are electron efficiency and blue points are
	    positron in RUN5.
	    Black points are efficiency of electron and positron.
	    Green line is a fit function of efficiency of electron and positron.
	  }
	  \label{fig:effrun5}
	\end{center}
      \end{figure}

\begin{figure}[htb]
	\begin{center}
	  \includegraphics[width=13cm]{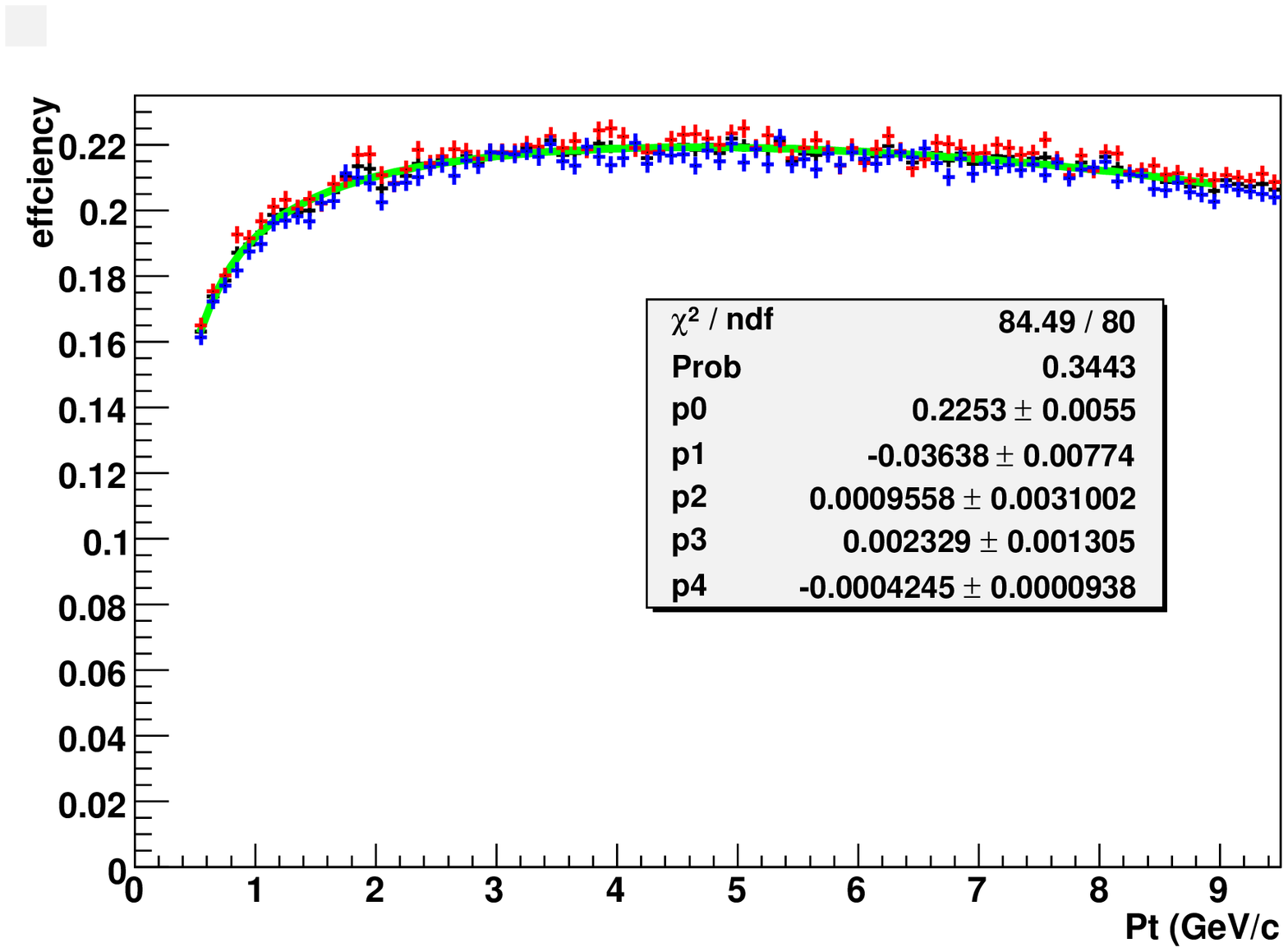}
	  \caption{Electron efficiency as a function of electron$p_{\mathrm{T}}$.
	    Red points are electron efficiency and blue points are
	    positron in RUN6.
	    Black points are efficiency of electron and positron.
	    Green line is a fit function of efficiency of electron and positron.
	  }
	  \label{fig:effrun6}
	\end{center}
      \end{figure}

    $A\times \epsilon_{eff}(p_{\mathrm{T}})$ with the standard eID cut is determined by the 
    PISA simulation.
    The simulation sample described in Sec.~\ref{sec:pisa} is used.
    $A\times \epsilon_{eff}(p_{\mathrm{T}})$ is determined as bellow.
    \begin{equation}
    A\times \epsilon_{eff}(p_{\mathrm{T}}) = 
    \frac{{\rm output\quad with\quad standard\quad eID\quad cut}(p_{\mathrm{T}}) }
	 {{\rm input}(p_{\mathrm{T}})\times w(p_{\mathrm{T}})},
    \end{equation}
    where $w(p_{\mathrm{T}})$ is the weighting factor which is used so that
    the input distribution of $p_{\mathrm{T}}$ in the simulation have the realistic 
    form for inclusive electrons.
    Figure~\ref{fig:effrun5} shows the result of  the geometrical acceptance times electron 
    reconstruction efficiency as a function of electron $p_{\mathrm{T}}$ in RUN5.
    Figure~\ref{fig:effrun6} also shows the result of the geometrical acceptance times electron 
    reconstruction efficiency in RUN6.
    Red points show electron efficiency and blue points show
    positron.
    Black points show efficiency of electron and positron.
    Green line is a fit function of efficiency of electron and positron.
    Fit function is 
    \begin{equation}
    p_0+\frac{p_1}{p_{\mathrm{T}}}+\frac{p_2}{p_{\mathrm{T}}^2}
    +p_3\times p_{\mathrm{T}} + p_4\times p_{\mathrm{T}}^2.
    \end{equation}
    The fit function is used as the efficiency curve of electron and 
    positron.
    
      \subsubsection{Reconstruction Efficiency with Tight eID}

      In $p_{\mathrm{T}}$ above 4.85~GeV/$c$, pions start emitting Cerenkov light
      in CO2 gas in the RICH detector. Since rejection power of RICH is reduced,
      the tight eID cut,  as is defined at Sec~\ref{sec:ecut} 
      is applied above 5~GeV/$c$.
      
\begin{figure}[htb]
	\begin{tabular}{c c}
	  \begin{minipage}{\minitwocolumn}
	    \begin{center}
	      \includegraphics[width=7cm]{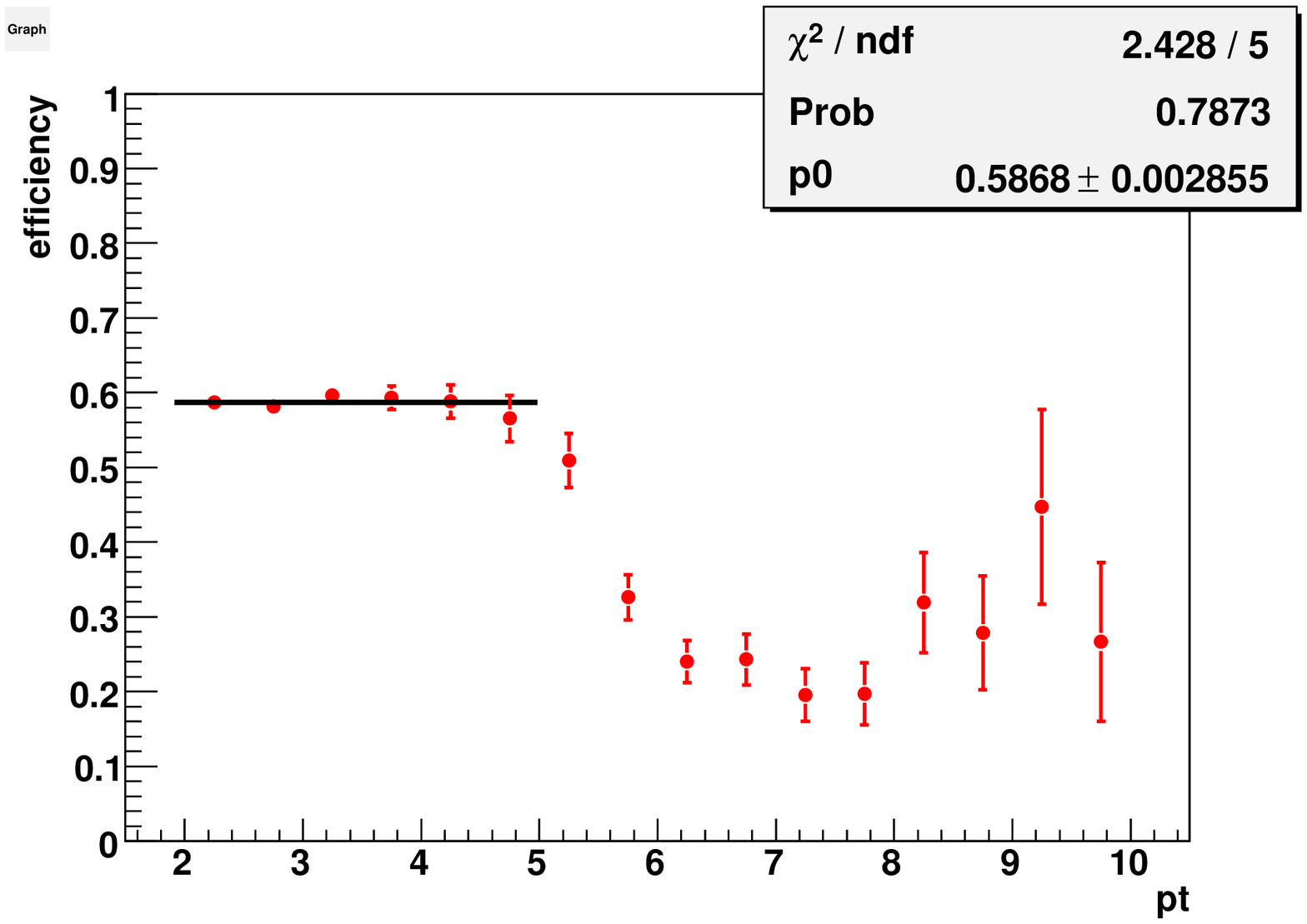}
              \caption{ The ratios the number of electron 
		in 0.8$<$ecore/mom$<$1.4 with tight eID cut over 
		that with standard eID cut
		as a function of electron $p_{\mathrm{T}}$ in RUN5.
	      }
              \label{fig:tcutrun5}
	    \end{center}
	  \end{minipage}
	  &
	  \begin{minipage}{\minitwocolumn}
	    \begin{center}
	      \includegraphics[width=7cm]{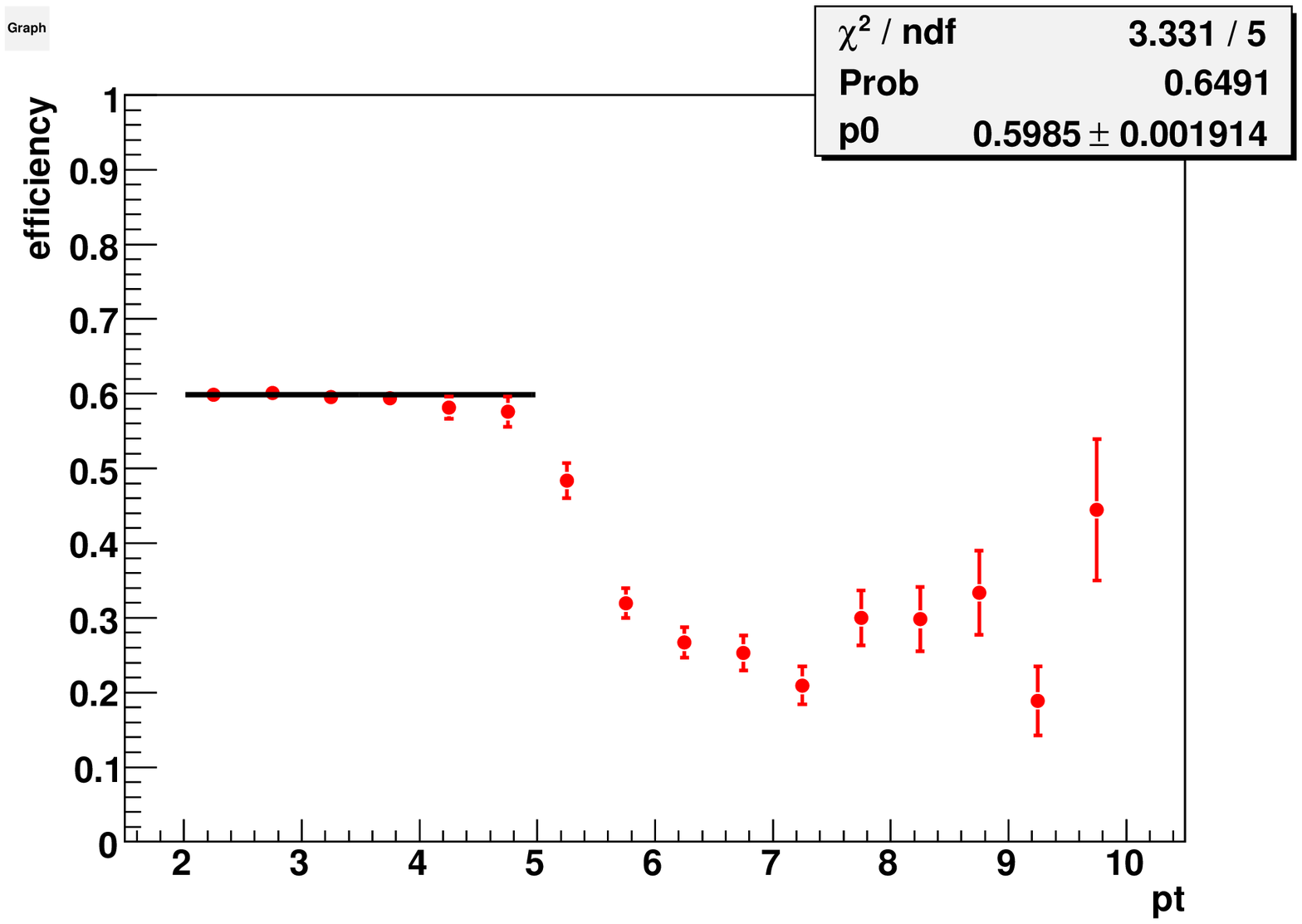}
              \caption{The ratios the number of electron 
		in 0.8$<$ecore/mom$<$1.4 with tight eID cut over 
		that with standard eID cut
		as a function of electron $p_{\mathrm{T}}$ in RUN6.
	      }
              \label{fig:tcutrun6}
	    \end{center}
	  \end{minipage}
	\end{tabular}
      \end{figure}

      The tight eID cut requires {\sf n1}$>$4 and {\sf prob}$>$0.1 in addition to the standard eID cut.
      Efficiency of tight eID cut is calculated as bellow.\\
      \begin{equation}
      \epsilon_{tight}(0.8<{\sf ecore/mom}<1.4) = 
      \epsilon_{standard}(0.8<{\sf ecore/mom}<1.4) \times R_{tight},
      \end{equation}
      where $R_{tight}$ is the efficiency corresponding to the  additional cuts in the
      tight eID cut~({\sf prob}$>$0.1 and {\sf n1}$>$4).
      $R_{tight}$ is determined from the ratio of the number of electron 
      in 0.8$<${\sf ecore/mom}$<$1.4 with the tight eID cut over that with the standard eID cut 
      in the real data.
      Figure~\ref{fig:tcutrun5} and Figure~\ref{fig:tcutrun6} show the ratios as 
      a function of electron $p_{\mathrm{T}}$ in RUN5 and RUN6, respectively.
      These figures indicate the ratio is independent of the $p_{\mathrm{T}}$
      in 2.0$<p_{\mathrm{T}}<$ 5.0~GeV/$c$ and then it drops.
      The constant behavior below 5.0~GeV/$c$ is due to 
      independence of {\sf n1} and {\sf prob} cut on electron $p_{\mathrm{T}}$.
      The drop is due to large hadron contamination in electron with 
      standard eID cut. Therefore, $R_{tight}$ itself is expected to be independent 
      of $p_{\mathrm{T}}$ even above 5.0~GeV/$c$.
      In Fig.\ref{fig:tcutrun5} and Fig.\ref{fig:tcutrun6}, black line is a constant 
      value fit to the ratios in 2.0$<p_{\mathrm{T}}<$ 5~GeV/$c$.
      The fitted values are used as relative efficiency.
      The values are 0.587$\pm$0.003 and 0.599$\pm$0.002 for RUN5 and RUN6, respectively.\\

      \subsection{Trigger Efficiency}\label{sec_tri}
\begin{figure}[htb]
	\begin{tabular}{c c}
	  \begin{minipage}{\minitwocolumn}
	    \begin{center}
	      \includegraphics[width=7cm]{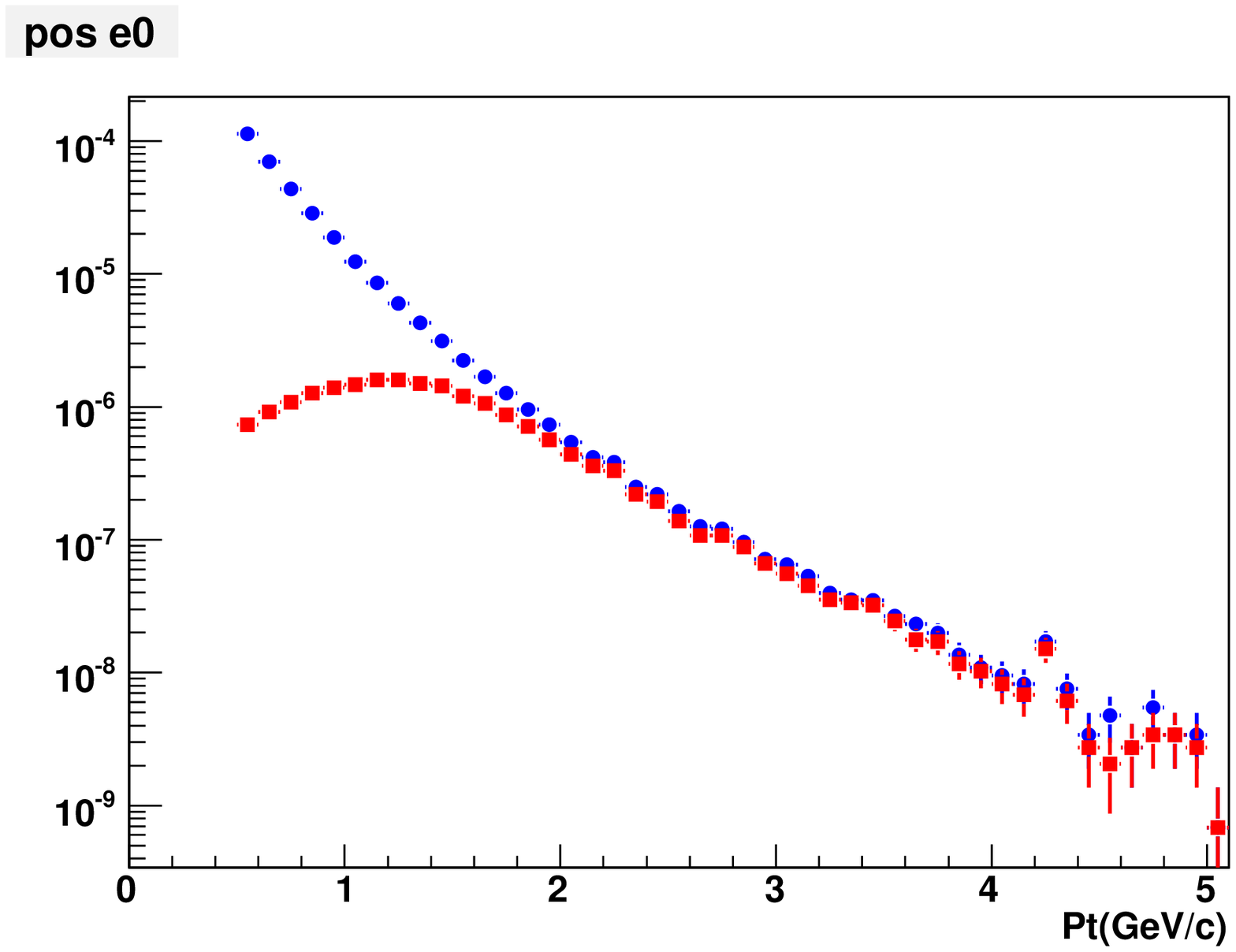}
              \caption{Raw spectra of electrons in RUN5 MB data.
		Blue circles show all electrons in MB data and red squares
		show 4X4c fired electrons.
	      }
              \label{fig:erawrun5}
	    \end{center}
	  \end{minipage}
	  &
	  \begin{minipage}{\minitwocolumn}
	    \begin{center}
	      \includegraphics[width=7cm]{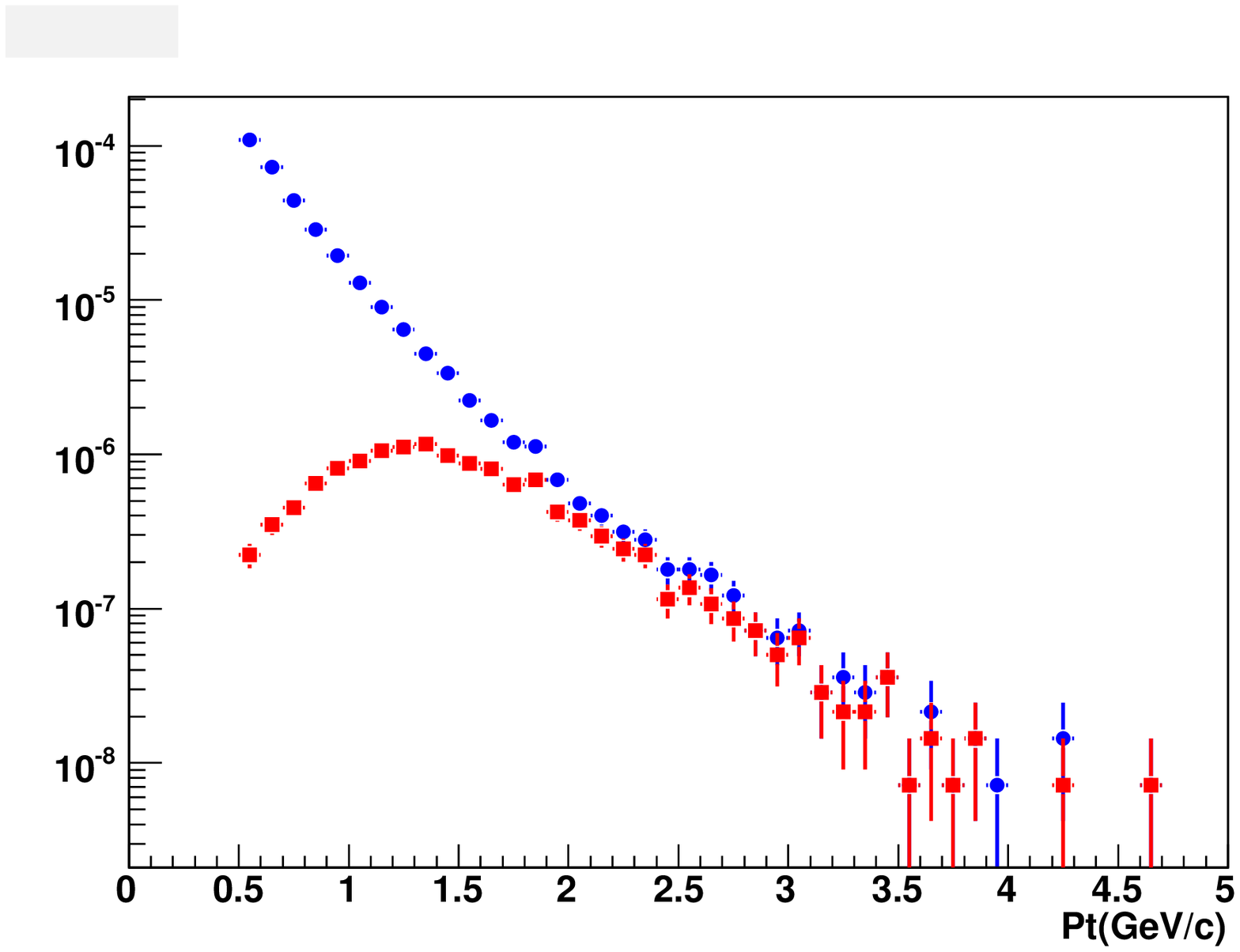}
              \caption{Raw spectra of electrons in RUN6 MB data.
		Blue points show all electrons in RUN6 MB data and red squares
		show 4X4c fired electrons.
	      }
              \label{fig:erawrun6}
	    \end{center}
	  \end{minipage}
	\end{tabular}
      \end{figure}

\begin{figure}[htb]
	\begin{tabular}{c c}
	  \begin{minipage}{\minitwocolumn}
	    \begin{center}
	      \includegraphics[width=7cm]{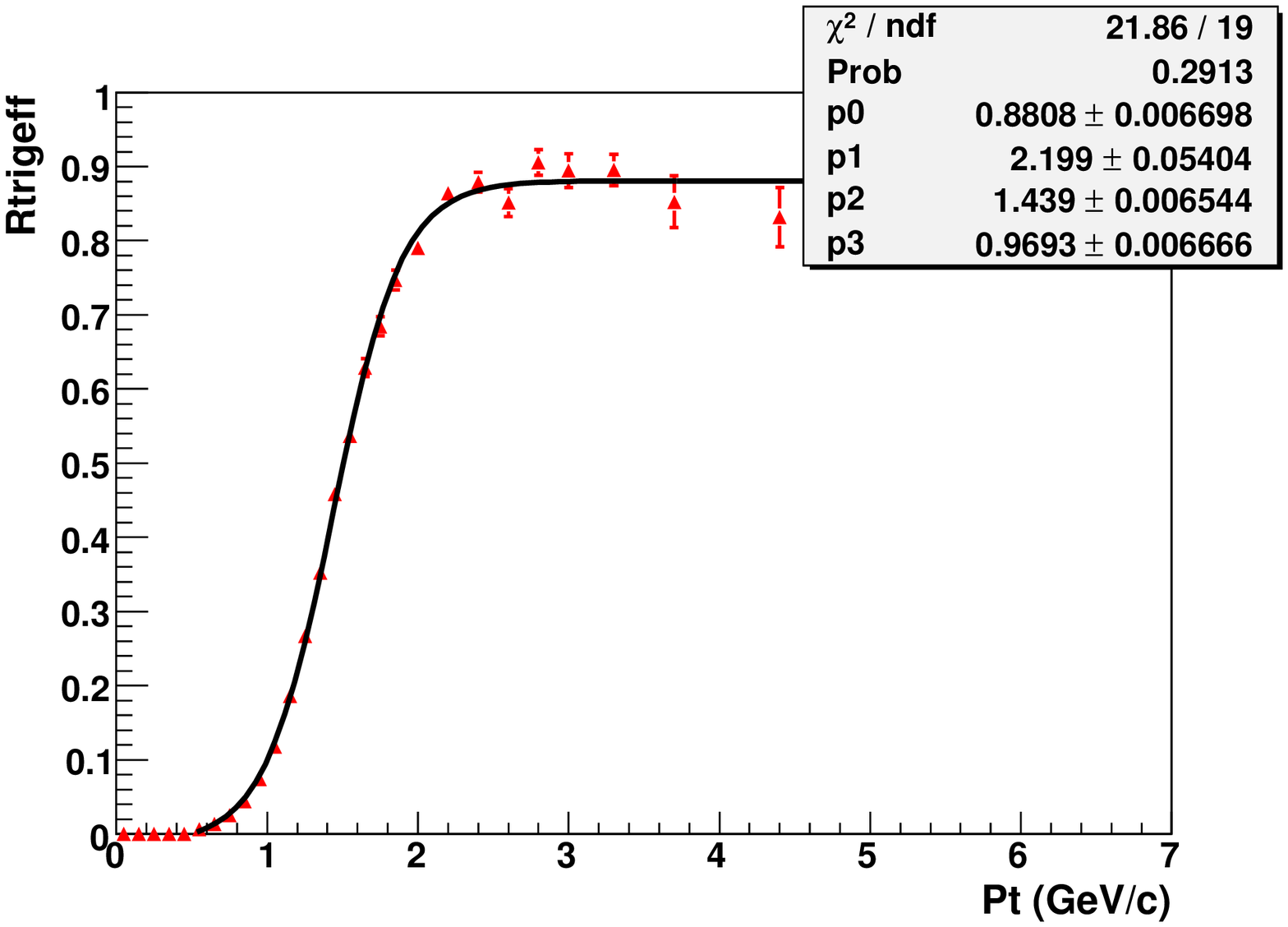}
              \caption{Trigger efficiency of 4x4c trigger in RUN5.
	      }
              \label{fig:pheff5}
	    \end{center}
	  \end{minipage}
	  &
	  \begin{minipage}{\minitwocolumn}
	    \begin{center}
	      \includegraphics[width=7cm]{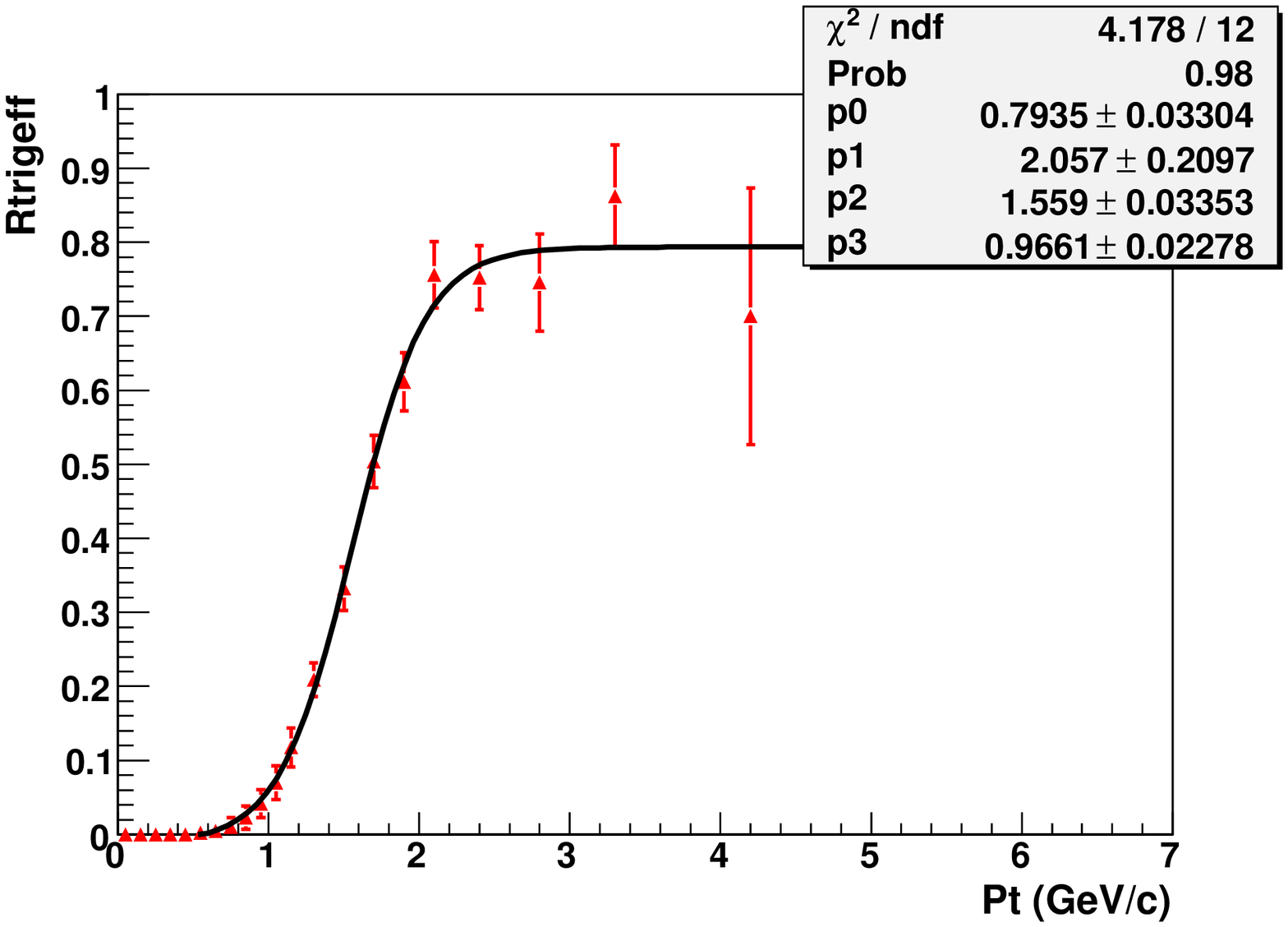}
              \caption{Trigger efficiency of 4x4c trigger in RUN6.
	      }
              \label{fig:pheff6}
	    \end{center}
	  \end{minipage}
	\end{tabular}
      \end{figure}

      Figure~\ref{fig:erawrun5} and \ref{fig:erawrun6} show raw spectra 
      of electrons with the standard eID cut in RUN5 MB and RUN6 MB data,
      respectively.
      In Fig.~\ref{fig:erawrun5} and \ref{fig:erawrun6}, blue circles show 
      all electron in MB data and red squares show 4x4c fired electrons.
      4x4c trigger efficiency is determined as a ratio of 
      PH fired electrons over measured electrons in MB data.
      Figure~\ref{fig:pheff5} and \ref{fig:pheff6} shows the determined 
      efficiency of PH trigger in RUN5 and RUN6.\\
      \noindent
       The solid curves in Fig.~\ref{fig:pheff5} and Fig.~\ref{fig:pheff6} are 
       the fitted functions with the following parameterization:
       \begin{equation}
       \frac{p_0}{1+p_3} \times (p_3+\tanh(p_1\times (p_{\mathrm{T}} -p_2))).
       \end{equation}
     
      These fitted functions are used for the efficiency of PH trigger in RUN5 and RUN6.
      
      \subsection{Hadron Contamination} \label{sec:hadc}
      Hadron contamination in the electrons selected with the standard eID cut 
      using estimated via {\sf ecore/mom} distribution.
      The distribution of {\sf ecore/mom} has a peak around one in the case of
      the electron peak, while the hadron track has a small {\sf ecore/mom} value.

      \subsubsection{Hadron Contamination below 5GeV/$c$}
      The idea is to use {\sf prob} cut is to enrich the hadron contamination. 
      The {\sf prob}$>0.01$ cut has about 50\% hadron efficiency for 
      $p_{\mathrm{T}}>$ 1~GeV/$c$ and 99\% efficiency to electrons. 
      Therefore, the hadron contamination is increased by a factor of 100 
      if we reverse the {\sf prob} cut~({\sf prob}$<0.01$).
      Then we can look at the {\sf ecore/mom} distribution to see much
      enhanced hadron contamination.
      
    The procedure we used is the following:\\
    Two {\sf ecore/mom} distributions of inclusive charged particles are prepared as the 
    distributions of hadrons. 
    $(H_a)$ is the {\sf ecore/mom} distribution of hadrons with {\sf prob}$<0.01$~
    (rejected sample) and $(H_b)$ is the {\sf ecore/mom} distribution of hadrons 
    with {\sf prob}$>0.01$~(accepted sample).\\
    We also make  {\sf ecore/mom} distribution of electron candidate with the reverse 
    cut ({\sf prob}$<0.01$) and with normal cut ({\sf prob}$>0.01$~). 
    The former $(E_a)$ contains large hadron contamination,  and the latter $(E_b)$
    is the normal electron candidate sample. For both samples, the standard eID cuts
    except the {\sf prob} cut is applied.
    The distribution of rejected hadron sample~$(H_a)$ is scaled by a factor of $f_h$ 
    and the distribution of the accepted electron candidate~$(E_b$) by a factor of 
    $f_e$ corresponding to the efficiency of the {\sf prob} cut,  
    so that sum of these two distribution reproduces the  {\sf ecore/mom} 
    distribution of the rejected electron sample~$(E_a)$.
    \begin{equation}
    E_a \sim f_e \times E_b + f_h \times H_a \label{eq:contami},
    \end{equation}
    \noindent
    When $f_e \times E_b + f_h \times H_a$ is roughly consistent with 
    $E_a$, the $f_e \times E_b$ term corresponds to the real electron 
    component in $E_a$, and the $f_h \times H_a$ term is the hadron component in $E_a$.
    Since $f_h \times H_a$ presents the hadron contamination in the rejected 
    electron sample,  the hadron contamination in accepted electron sample 
    should be presented  as the distribution $f_h \times H_b$ by using the fixed 
    normalization factor $f_h$ . 
    In this way, the hadron contamination in the accepted
    electron sample can be determined as $(f_h \times H_b)/E_b$

\begin{figure}[htb]
      \begin{center}
	\includegraphics[width=13cm]{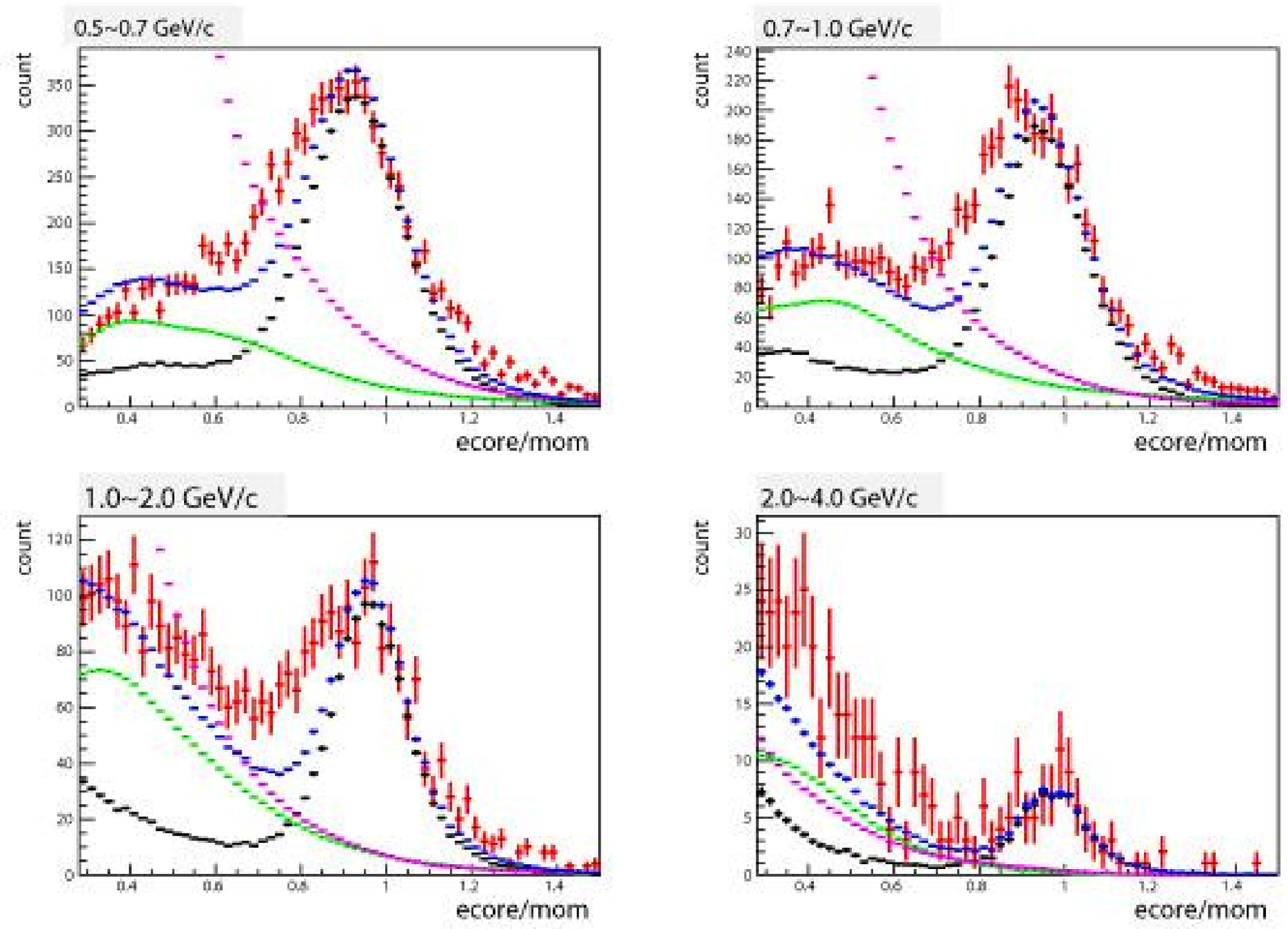}
	\caption{The distributions of {\sf ecore/mom} in RUN5 MB data were 
	  used to study 
	  hadron contamination in $0.5<p_{\mathrm{T}}<4.0$ GeV/$c$
	}
	\label{fig:ephad}
      \end{center}
    \end{figure}

    Figure~\ref{fig:ephad} shows the comparison of the {\sf ecore/mom} distribution 
    in RUN5 MB data produced by this procedure.
    The four panels in the figure correspond to four different  $p_{\mathrm{T}}$ bins.
    In each panel, the green histogram is the  {\sf ecore/mom} distribution of the
    rejected hadron $(H_a)$, the black histogram is the distribution of the accepted 
    electron sample $(E_b)$ scaled by a factor of $f_e$ = 0.02. 
    The blue histogram is the sum of the two. The red histogram
    is the distribution of rejected electrons $(E_a)$ and the magenta histogram
    is the distribution of accepted hadrons $(H_b)$, which represents
    hadron contamination in the selected electrons with standard eID cut.
    The same rescaling factor $f_h$ is used for all panels. 
    The sum of the two distribution roughly reproduces the rejected electron 
    distribution as described in Eq.~\ref{eq:contami}.
    
    The same comparison is done for ERT data to study the hadron contamination
    in the electrons with $2.0<p_{\mathrm{T}}<5.0$ GeV/$c$.
    The results of the estimation of hadron contamination in MB and ERT data 
    are summarized at Table~\ref{tab:hadcont}.
    
    Hadron contamination is less than 1\% for 0.7~GeV/$c$ $<$electron 
    $p_{\mathrm{T}}$ $<$ 4.5~GeV/$c$. In 4.5-5.0~GeV/$c$ range, hadron contamination 
    becomes about 2\%, since pions start emitting Cerenkov light in CO2 gas in
    RICH detector above 4.85~GeV/$c$.
    \begin{table}[hbt]
      \begin{center}
	\caption{Estimated hadron contamination}
	\label{tab:hadcont}
	\begin{tabular}{c|cc}
	  \hline
	  electron $p_{\mathrm{T}}$ range & hadron contamination(RUN5) & hadron contamination(RUN6)\\
	  \hline\hline
	  0.5-0.7~GeV/$c$ & 0.0123 &0.0147\\
	  0.7-1.0~GeV/$c$ & 0.0074 &0.0089 \\
	  1.0-2.0~GeV/$c$ & 0.0042 &0.0052\\
	  2.0-2.5~GeV/$c$ & 0.0045 &0.0044\\
	  2.5-3.0~GeV/$c$ & 0.0036 &0.0039\\
	  3.0-3.5~GeV/$c$ & 0.0033 & 0.0039\\
	  3.5-4.0~GeV/$c$ & 0.0035 &0.0039\\
	  4.0-4.5~GeV/$c$ & 0.0057 &0.0061\\
	  4.5-5.0~GeV/$c$ & 0.0155 &0.0201\\
	  \hline
	\end{tabular}
      \end{center}
    \end{table}

    \subsubsection{Hadron Contamination above 5GeV/$c$}

    Hadron background is estimated by similar 'reverse prob method'.
    Hadron background is not negligible above $p_{\mathrm{T}}>$5.0~GeV/$c$,
    even when the tight eID is applied.
    
    $R_{tight}$ shown at Fig.~\ref{fig:tcutrun5} drops into about half
    for electron $p_{\mathrm{T}}>$5.0~GeV/$c$. This represents hadron contamination
    with the standard eID cut becomes $\sim$50\% above  $p_{\mathrm{T}}>$5.0~GeV/$c$.
    Therefore, when we apply the reverse {\sf prob} cut~({\sf prob}$<0.01$),
    the selected particles are hadrons with  $\sim$99\% purity since
    the reverse {\sf prob} cut increases the hadron contamination 
    by a factor of 100.
    We use the  {\sf ecore/mom} distributions of the particles which
    is selected by the standard eID cut and the reverse {\sf prob} cut
    to estimate hadron background in electron samples.
    
    \begin{figure}[htb]
      \begin{center}
	\includegraphics[width=14cm]{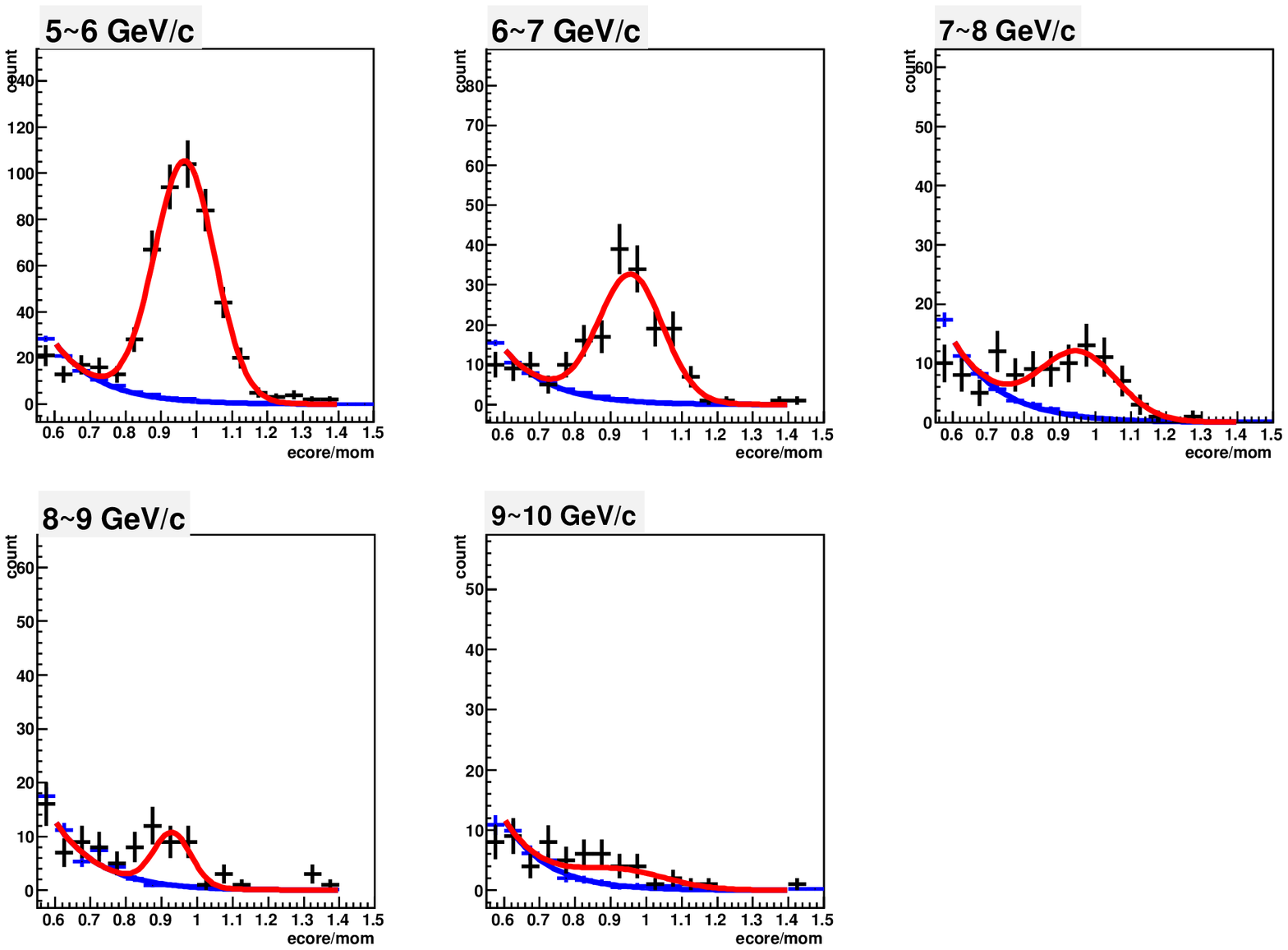}
	\caption{
	  The ecore/mom distribution of electrons
	  with tight eID cut and estimated that of hadron as described above in RUN5.
	  Black points show the distribution of electron and 
	  blue points show estimated that of hadron.
	}
	\label{fig:highep}
      \end{center}
    \end{figure}
    Figure~\ref{fig:highep} shows the {\sf ecore/mom} distribution of electrons
    with tight eID cut and estimated that of hadron as described above from RUN5 data.
    Black points show the distribution of electron and 
    blue points show estimated that of hadron. The estimated distribution of hadron is
    normalized by number of entries in 0.6$<${\sf ecore/mom}$<$0.75.\\
    Blue lines are exponential fit to the {\sf ecore/mom} distribution of hadron.
    Red lines are gauss + exponential fit to the distribution of 
    electrons in the condition that exponential parts are fixed at 
    blue lines.
    
    Signals are counted as number of entries in 0.8$<${\sf ecore/mom}$<$1.4.
    Hadron background is estimated from fit functions.
    The fitting error is counted into the statistical error of the signals.
    The results for RUN5 and RUN6 are summarized in Table~\ref{tab:highhad}.
    
    \begin{table}[hbt]
      \begin{center}
	\caption{Estimated hadron background}
	\label{tab:highhad}
	\begin{tabular}{c|cc}
	  \hline
	  electron $p_{\mathrm{T}}$ range & hadron background(RUN5)& hadron background(RUN6)\\
	  \hline\hline
	  5.0-6.0~GeV/$c$& 0.033 & 0.038 \\
	  6.0-7.0~GeV/$c$ & 0.051 &0.066 \\
	  7.0-8.0~GeV/$c$ & 0.137 &0.146 \\
	  8.0-9.0~GeV/$c$ &0.259 &0.156 \\
	  9.0-10.0~GeV/$c$ &0.257 &0.250 \\
	  \hline
	\end{tabular}
      \end{center}
    \end{table}
    
    \subsection{Invariant Cross Section of Inclusive Electron}
    The overall efficiency, $\epsilon(p_{\mathrm{T}})$ can be determined from 
    the obtained electron reconstruction efficiency and trigger efficiency. 
    Therefore, we are ready to determine invariant cross section of inclusive electron 
    according to Eq~\ref{eq:inv}

\begin{figure}[htb]
      \begin{center}
	\includegraphics[width=12cm]{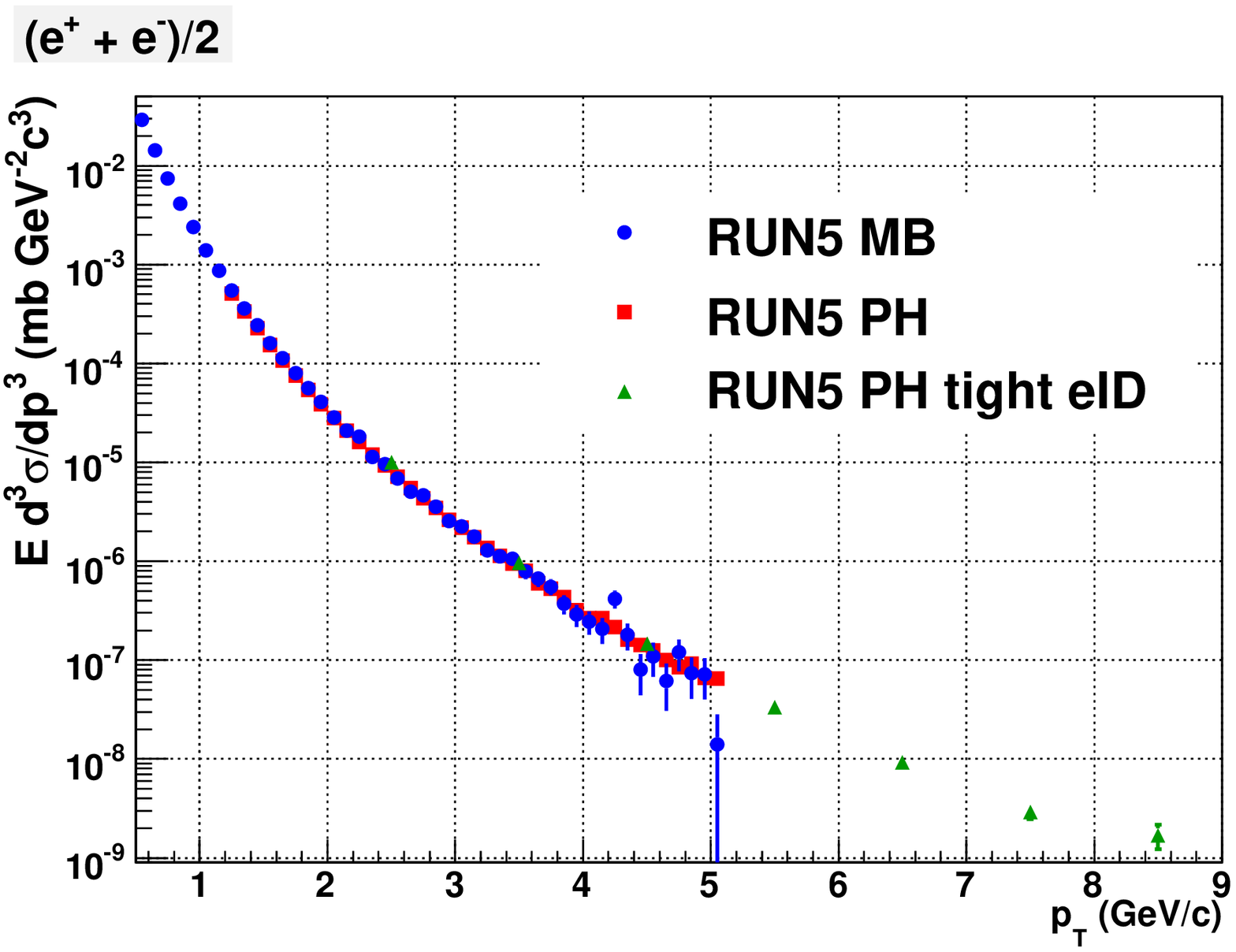}
	\caption{Invariant cross section of inclusive electrons in RUN5 MB and 
	  PH triggered events. 
	  Blue circles show electrons in  MB events and red squares show
	  electrons in PH triggered events with standard eID cut.
	  Green triangles show electrons in PH triggered events 
	  with tight eID cut.
	}
	\label{fig:inclrun5}
      \end{center}
    \end{figure}

\begin{figure}[htb]
      \begin{center}
	\includegraphics[width=12cm]{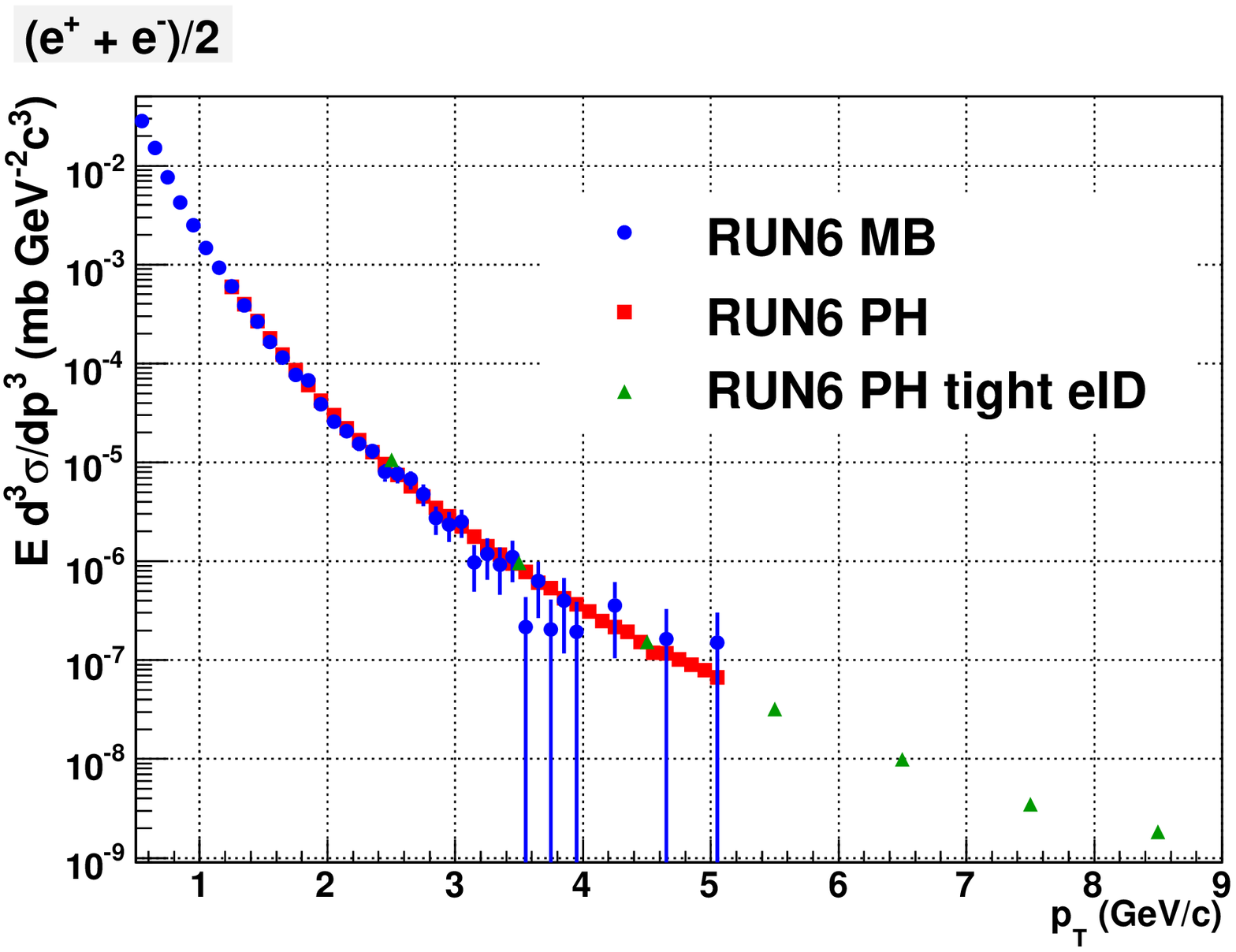}
	\caption{Invariant cross section of inclusive electrons in RUN6 MB and 
	  PH triggered events. 
	  Blue circles show electrons in  MB events and red squares show
	  electrons in PH triggered events with standard eID cut.
	  Green triangles show electrons in PH triggered events 
	  with tight eID cut.
	}
	\label{fig:inclrun6}
      \end{center}
    \end{figure}

    Figure~\ref{fig:inclrun5} and \ref{fig:inclrun6} show
    invariant cross sections of inclusive electrons for MB and PH 
    triggered events in RUN5 and RUN6, respectively.
    Blue circles show the spectrum of electrons in  MB data and red squares show
    that of electrons in PH data with the standard eID cut.
    Green triangles show electrons for PH data with tight eID cut.
    These cross sections of inclusive electron are
    consistent with each other among three cases.
    
     \subsection{Systematic Errors} \label{sec:sysinv}
     \subsubsection{Geometrical Acceptance}
      Since the simulation reproduces the real data about
      the {\sf phi} distribution within 3\% as shown 
      at Sec.~\ref{sec:pisa}, 
      3\% systematic error is assinged for geometical acceptance for RUN5 and RUN6.

      \subsubsection{eID Parameters}
        \begin{figure}[htb]
	\begin{tabular}{c c}
	  \begin{minipage}{\minitwocolumn}
	    \begin{center}
	      \includegraphics[width=7cm]{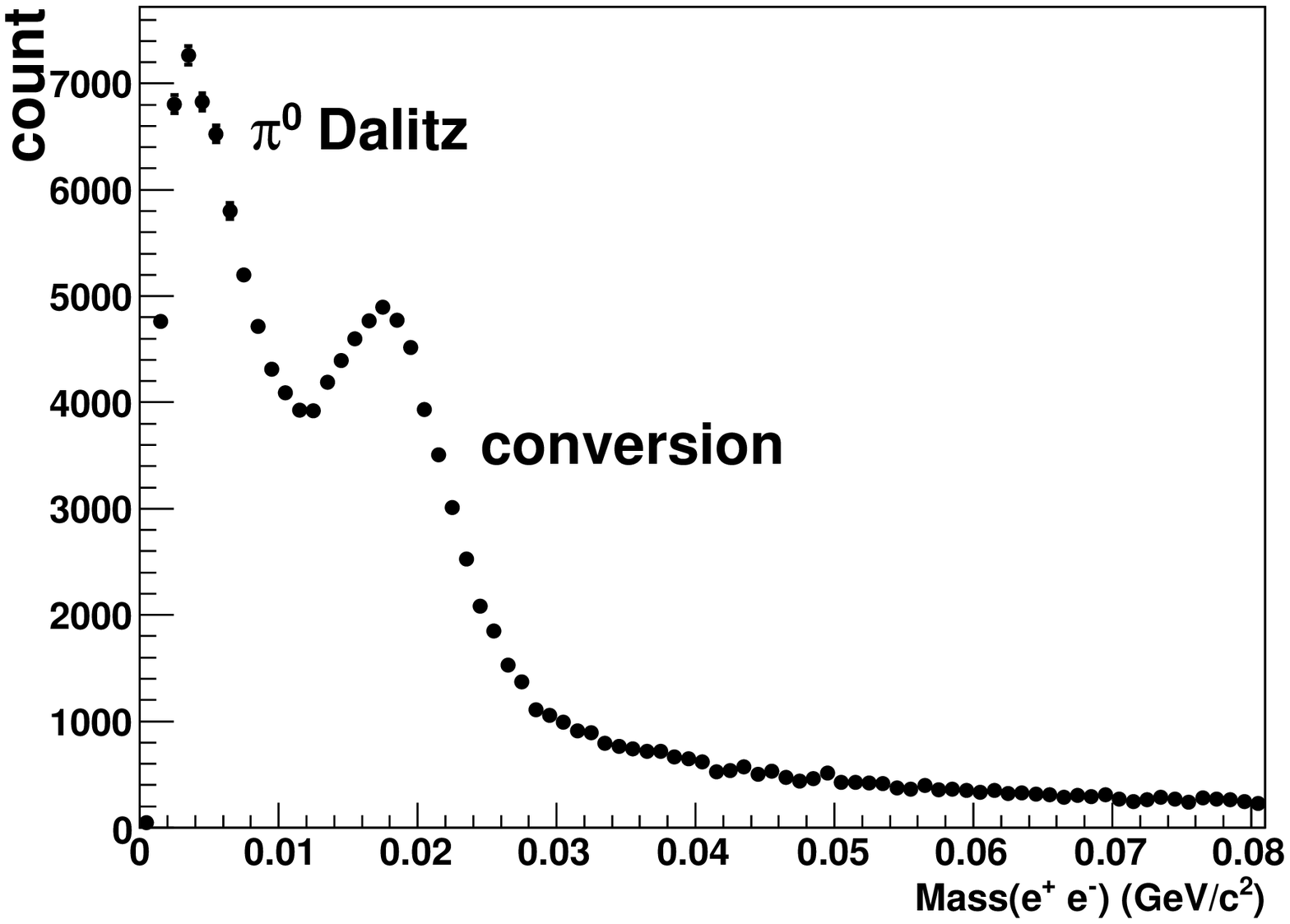}
              \caption{The invariant $e^+$ $e^-$mass peak in data sample.
	      }
              \label{fig:lowpeak}
	    \end{center}
	  \end{minipage}
	  &
	  \begin{minipage}{\minitwocolumn}
	    \begin{center}
	      \includegraphics[width=7cm]{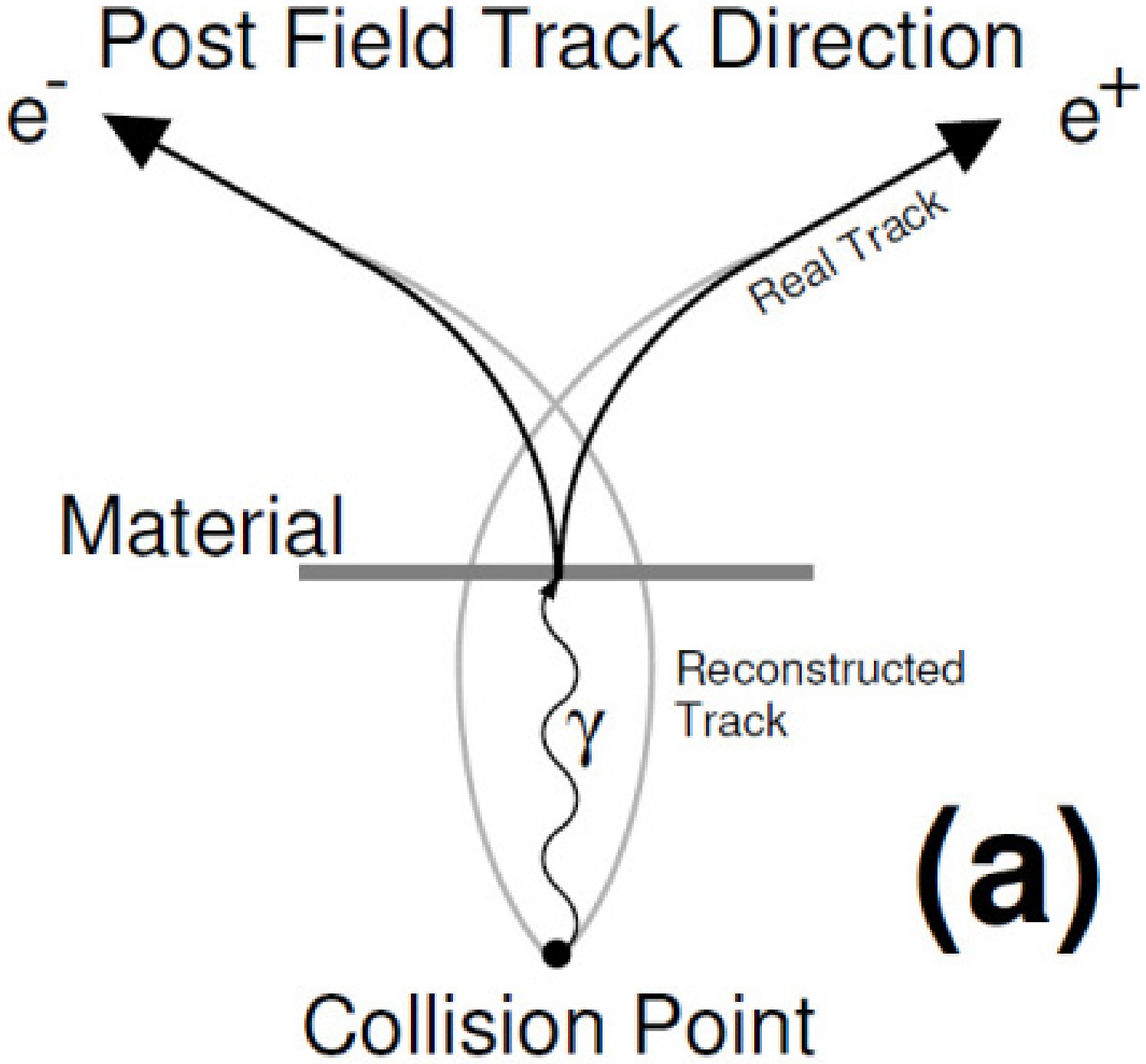}
              \caption{The production of  conversion  electrons.
	      }
              \label{fig:conv_mass}
	    \end{center}
	  \end{minipage}
	\end{tabular}
      \end{figure}

      Systematic error for eID parameters is determined by the comparison 
      of the efficiency of each eID parameter between the real data and the 
      PISA simulation.

      Efficiency of each eID parameter in the simulation is determined from the 
      distribution of the parameter with other cuts being applied.
      For example, efficiency of {\sf n0} cut in the simulation is determined as follows.
      \begin{equation}
      \frac{\int_2 N(n0) dn0}{\int_0 N(n0) dn0},
      \end{equation}
      where $N$ is the distribution of {\sf n0} with the standard cut except {\sf n0}  being applied
      
      The efficiency of each cut for eID parameter in the real data is determined 
      by tagging the electrons from conversion and dalitz decay.
      The reconstructed invariant mass distribution of $e^+$ $e^-$~pair  has the 
      peaks at the low mass region as shown in Figure~\ref{fig:lowpeak}.
      It is a useful tool to tag pure electrons.
      The sources of the peaks are $\pi^0$ Dalitz decay~($\pi^0 \rightarrow e^+ e^- \gamma$)
      and $\gamma$ conversion at the beam pipe.
      Since the track reconstruction algorithm  assumes that all tracks come
      from the collision vertex, the electron pairs produced at the point with
      $R>0$ are reconstructed to have incorrect momentum.
      This is schematically shown in Fig.~\ref{fig:conv_mass}.
      As a result, each of the conversion pairs acquires the fake \PT and 
      the invariant mass is approximately proportional to $\int Bdl$.
      Therefore, the reconstructed mass of conversion electron pairs
      is determined by the location of the conversion sources.
      The peak position of the pairs from $\gamma$ conversion at the beam pipe 
      is around 20~MeV/$c^2$.
      
\begin{figure}[htb]
	\begin{center}
	  \includegraphics[width=7cm]{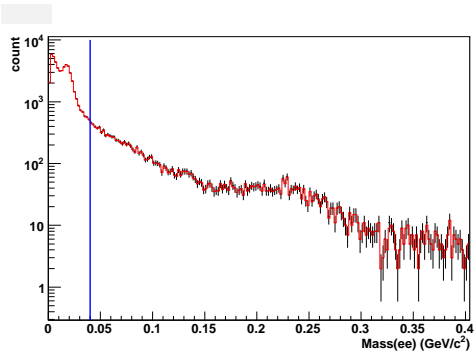}
	  \caption{$e^+ e^-$ mass distribution in RUN5 PH fired 
	    events. Black points show the mass distribution when 
	    both electrons are applied all cuts and red points show the one when
	    one electron is applied all cuts and the other is applied all cuts except
	    {\sf n0} cut.
	  }
	  \label{fig:n0mass}
	\end{center}
      \end{figure}

      Using these clearly tagged pairs of electrons, the  efficiency of each eID cut in the
      real data could be evaluated. One electron is selected by the standard eID cut and 
      the other electron is selected by the standard eID cut except the cut for
      the parameter whose efficiency will be evaluated.
      Figure~\ref{fig:n0mass} shows the invariant mass distribution of $e^+ e^-$ 
      in RUN5 PH fired events.
      Black points show the mass distribution when both electrons are selected by the standard 
      eID cut and red points show that when one electron is selected by the standard eID cut 
      and the other is selected the cut without {\sf n0}~(RICH fire~({\sf n1}$>$1) is required)\\
      Efficiency of n0 cut in real data is determined as follows.
      \begin{equation}
      \frac{\int_{0}^{0.04} N(mass) dmass~(n0>1)}{\int_{0}^{0.04} N(mass) dmass},
      \end{equation}
      where, $N$ is the distribution of the invariant mass.  \\
      The efficiencies of other parameters are also determined in the same way.
      The results are summarized in Table~\ref{tab:eideff}.
      The efficiencies in the simulation agrees well with these in real data.
      
      Systematic error of 1\% is assigned for RICH paramenters from Table~\ref{tab:eideff},
      since the efficiencies of RICH parameter are expected not to depend on 
      electron $p_{\mathrm{T}}$.
      Systematic error of 2\% is assigned for EMC parameters to be conservative,
      since the efficiencies of EMCal parameters may have small $p_{\mathrm{T}}$ dependence.
      
      \begin{table}[hbt]
	\begin{center}
	  \caption{The efficiency of eID parameter at real data and the simulation}
	  \label{tab:eideff}
	  \begin{tabular}{c|cc|cc}
	    \hline
	    eID parameter & real~(RUN5)&  simulation~(RUN5)& real~(RUN6)&  simulation~(RUN6)\\
	    \hline\hline
	    RICH & & & & \\
	    n0 & 99.4\% &98.5\% & 99.3\%& 98.5 \% \\
	    disp & 99.0\% &99.3\% & 99.2\%& 99.3 \% \\
	    chi$^2$ & 99.7\% &99.4\% & 99.6\%& 99.4 \% \\
	    \hline
	    EMC & & & & \\
	    e/p & 97.7\% &97.6\% & 96.1\%& 97.1 \% \\
	    prob & 98.8\% &98.7\% & 98.4\%& 98.7 \% \\
	    $\delta \phi_{EMC}$ & 99.3\% &99.5\% & 99.1\%& 99.4 \% \\
	    $\delta$z$_{EMC}$& 99.3\% &99.5\% & 99.4\%& 99.5 \% \\
	    \hline
	  \end{tabular}
	\end{center}
      \end{table}
      \subsubsection{Trigger Efficiency}
      The systematic error of the PH trigger efficiency is evaluated based on 
      the error of the fit in Fig~\ref{fig:pheff5} and \ref{fig:pheff6}.
      We assign the systematic error of the PH trigger efficiency in RUN5 as
      3\% $\bigoplus$ 5\% $\times \frac{1}{\epsilon_{trig}-1}$
      and 
      4\% $\bigoplus$ 10\% $\times \frac{1}{\epsilon_{trig}-1}$.
      
      \subsubsection{Tight eID Efficiency}
      The relative efficiency~($R_{tight}$) is independent of $p_{\mathrm{T}}$  
      in 2$<p_{\mathrm{T}}<$5~GeV/$c$ as shown in Fig.~\ref{fig:tcutrun5} and 
      \ref{fig:tcutrun6}. The efficiency of the tighter RICH cut({\sf n1}$>$4) 
      is $\sim$ 70\% and this part should be  $p_{\mathrm{T}}$ independent. 
      The efficiency of the {\sf prob} cut is approximately 90\%. The 10\% loss due to 
      the {\sf prob} may have some small $p_{\mathrm{T}}$ dependence.
      We assign 20\% of the 10\% loss as possible $p_{\mathrm{T}}$ dependence of the
      {\sf prob} cut, which is enough conservative compared with the result in PISA simulation.
      Therefore, the systematic error for the relative efficiency~($R_{tight}$) 
      for high $p_{\mathrm{T}}$ extension is 10\%$\times$ 20\% $=$ 2\%.
      \subsection{Absolute Normalization}
      Systematic error for absolute normalization is described in Sec.~\ref{sec:lumi}
      and \ref{sec:biasbbc}.
      \begin{itemize}
      \item We use $\sigma_{BBC} = 23.0 \pm 2.2$ mb. Thus, systematic error is 9.6\%
      \item We use $\epsilon_{bias} = 0.79 \pm 0.02$. Thus, systematic error is 2.5\%
      \end{itemize} 
      The systematic error for the absolute normalization is assigned to 9.9\% from
      the quadratic sum of the two components.
      
      \clearpage
      \section{Single Electron From Heavy Flavor}
      The inclusive electron spectrum consists of three components:
      \begin{enumerate}
	\item{ 'non-photonic' electrons from semi-leptonic decay of heavy-flavor~(single non-photonic electron). 
	This component is what we want to measure.}
	\item{ 'photonic' background from Dalitz decays of light neutral mesons
	  and external photon conversions (mainly in the beam pipe).}
	\item{ 'non-photonic' background from $K \rightarrow e\pi\nu$ ($K_{e3}$),
	  dielectron decays of vector mesons and quarkonium~(J/$\psi$ and $\Upsilon$) and 
	  Drell-Yan process.}
      \end{enumerate} 
      The photonic background~(2) is much larger than the non-photonic background except 
      at highest $p_{\mathrm{T}}$~($>$5GeV/$c$).
      The signal of electrons from heavy-flavor decays is small compared to the
      background at low $p_{\mathrm{T}}$~($S/N < 0.2$ for $p_T < 0.5$~GeV/$c$) but rises
      with increasing $p_{\mathrm{T}}$~($S/N > 1$ for $p_T > 2$~GeV/$c$).

      In order to extract the heavy-flavor signal, the background has to be 
      subtracted from the inclusive electron spectrum.
      'cocktail method' and 'converter method' are used in this 
      analysis to subtract the electron background~\cite{bib:hq1,bib:hq3,bib:hq4,bib:hq2}.
      \subsection{Cocktail Method}\label{sec:cock}
      One technique to accomplish this task is the so-called 'cocktail subtraction'
      method.
      A cocktail of electron spectra from all background sources
      is calculated using a Monte Carlo event generator of hadron decays and 
      then subtracted from the inclusive electron spectra~\cite{bib:exodus}.
      This technique relies  on the fact that the $p_{\mathrm{T}}$ distributions
      of the relevant background sources are known well enough.
      It turns out that the PHENIX measurements of the relevant electron sources
      are precise enough to allow for cocktail calculations that constrain
      the background within a systematic uncertainty better than 15~\% for all
      $p_{\mathrm{T}}$~\cite{bib:hq2}.
      This uncertainty is in the same order with the signal to background ratio at the
      lowest $p_{\mathrm{T}}$ and, therefore, it is not sufficiently small to extract the 
      heavy-flavor signal via the cocktail subtraction over the full $p_{\mathrm{T}}$ range.
      The cocktail method is useful at  
      high $p_{\mathrm{T}}$, {\it e.g.} for $p_{\mathrm{T}} > 2$~GeV/$c$, where  signal to background
      ratio is large and the cocktail input is known with small systematic
      uncertainties as discussed in the following.
      
      \subsubsection{Neutral Pions}
      The most important background source is the $\pi^0$. 
      $\pi^0$ decays contribute to the photonic background in two ways.
      First, the Dalitz decay of $\pi^0$~($\pi^0 \rightarrow e^+e^-\gamma$)
      is a primary source of electrons from the collision vertex and, second, the
      conversion of photons from the decay $\pi^0 \rightarrow \gamma\gamma$ in 
      material in the PHENIX central arm aperture~(mainly the beam pipe) gives
      a source of electrons originating not from the original 
      collision vertex.
      The contribution from photon conversions is small compared to the contribution 
      from Dalitz decays, since material budget in the PHENIX central arms 
      is well controlled.
      
      The $p_{\mathrm{T}}$ distribution of $\pi^0$ is obtained via simultaneous fit
      to $\pi^{\pm}$~(low $p_{\mathrm{T}}$) and $\pi^0$~(high $p_{\mathrm{T}}$) spectra 
      at PHENIX~\cite{bib:ph_pip,bib:ph_pizero}.  This approach is only valid under 
      the assumption that the invariant $\pi^0$
      spectrum and the averaged charged pion spectrum $(\pi^+ + \pi^-)/2$ are the same.
      This assumption is justified with a few \% presicion at PHENIX, while 
      at low $p_{\mathrm{T}}$, {\it i.e.} for $p_T < 1$~GeV/$c$, the decay of $\eta$ mesons 
      into three $\pi^0$ creates a tiny charge asymmetry.

      \begin{figure}[htb]
	\begin{center}
	  \includegraphics[width=7.3cm]{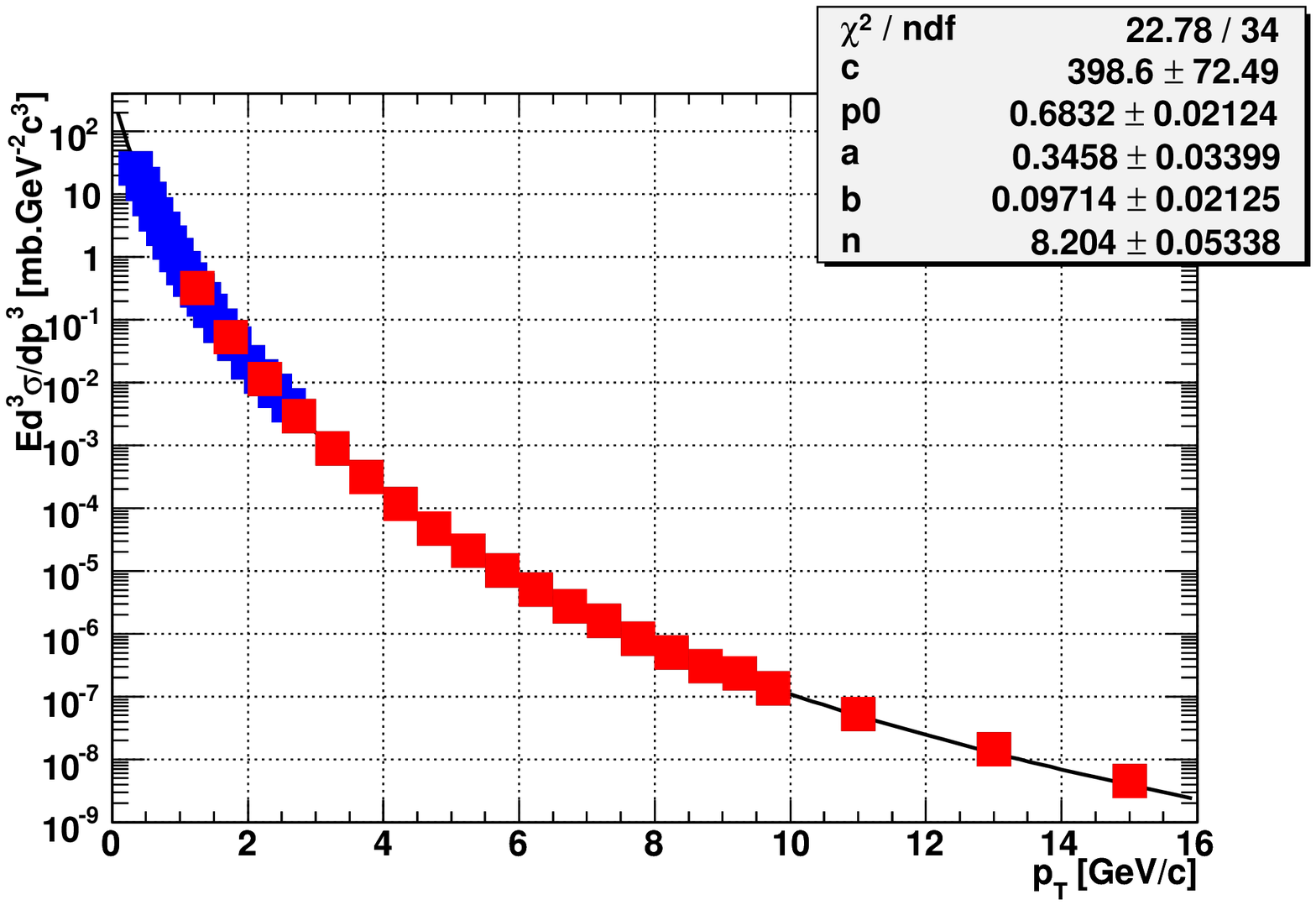}
	  \includegraphics[width=7.3cm]{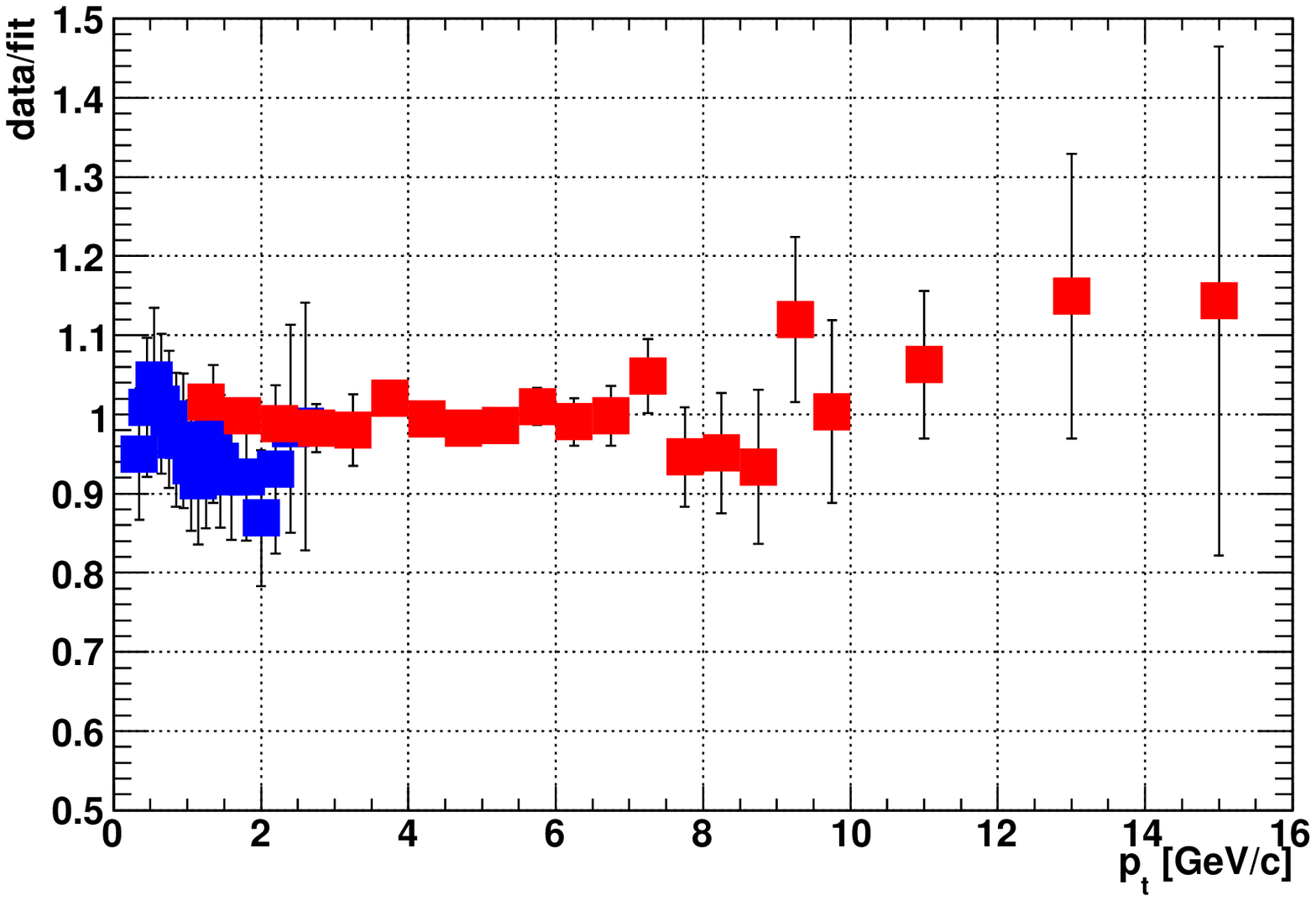}
	  \caption{Invariant differential cross section of 
	    (blue symbols at low $p_T$) and $\pi^0$s (red symbols) 
	    together with a fit according to $\pi^{\pm}$
	    Eq.~\ref{fitfunc} (left panel). Ratio of the data to the fit (right panel).}
	  \label{fig:pion_input_pp}
	\end{center}
      \end{figure}
      
      Figure.~\ref{fig:pion_input_pp} shows the comparison of the neutral and charged 
      averaged invariant differential cross sections of pions in $p+p$ collisions 
      at $\sqrt{s} = 200$~GeV in comparison with a simultaneous fit to the data 
      with a modified Hagedorn parameterization:
      
      \begin{equation}
	E \frac{d^3\sigma}{d^3p} = 
	\frac{c}{(exp(-a p_T - b p_T^2) + p_T / p_0)^n}.
	\label{fitfunc}
      \end{equation}
      Both an absolute comparison as well as the ratio of the data to the fit
      are shown to demonstrate the good quality of the parameterization.
      

      \subsubsection{Other Light Mesons}
      Other light mesons contributing to the electron cocktail are the $\eta$, 
      $\rho$, $\omega$, $\eta'$, and $\phi$ mesons.
      The $\eta$ meson has the largest 
      contribution among these mesons.
   
      For the cocktail calculation, the shape of the invariant $p_{\mathrm{T}}$ distributions,
      and the relative yield to the $\pi^0$ yield are required as input parameter.
      The $p_{\mathrm{T}}$ spectra are derived from the pion spectrum assuming the $m_T$ scaling, 
      {\it i.e.} the same modified Hagedorn parameterizations are used 
      (Eq.~\ref{fitfunc}), only $p_{\mathrm{T}}$ is replaced by 
      $\sqrt{p_{\mathrm{T}}^2 + m_{meson}^2 - m_{\pi^0}^2}$. 
            
      Since this approach of $m_T$ scaling ensures that at high $p_{\mathrm{T}}$ the 
      spectral shapes of all meson distributions are the same, the normalization of 
      the meson spectra relative to the pion spectrum can be given by the ratios 
      meson-to-pion at high $p_{\mathrm{T}}$ (5~GeV/$c$ is used). 
      The following values are used.
      
      \begin{itemize}
      \item $\eta   / \pi^0 = 0.48 \pm 0.03$  ~\cite{bib:ppg051}
      \item $\rho   / \pi^0 = 1.00 \pm 0.30$  ~\cite{bib:an089}
      \item $\omega / \pi^0 = 0.90 \pm 0.06$  ~\cite{bib:ryabov}
      \item $\eta'  / \pi^0 = 0.25 \pm 0.075$ ~\cite{bib:an089}
      \item $\phi   / \pi^0 = 0.40 \pm 0.12$  ~\cite{bib:an089}
      \end{itemize}
      The resulting $\eta /\pi^0$ ratio agrees within 
      experimental uncertainties for $p_{\mathrm{T}} > 2$~GeV/$c$ with the 
      corresponding PHENIX data for $p+p$ collisions~\cite{bib:ppg051}.

      \subsubsection{$K_{e3}$ Decay}
      The contribution from the $K_{e3}$ decay of kaons in flight is 
      evaluated via the PISA simulation to take into account the effect of the exact
      analysis cuts~(specially {\sf ecore/mom} cut). 
      The measured yield of electrons originating not from the 
      collisions vertex depends on the analysis cut.
      The input kaon spectrum is parameterized  based on the 
      measured charged kaon spectrum in $p+p$ collisions at PHENIX~\cite{bib:chk}.
      The contribution from kaon decays is only relevant ({\it i.e.} 
      larger than 5~\%) for electrons with $p_{\mathrm{T}} < 1$~GeV/$c$.
      The contribution becomes negligible for electrons with $p_{\mathrm{T}} > 2$~GeV/$c$.

      \subsubsection{Photon Conversions}
      The contribution from $\gamma$ conversions depends almost entirely on 
      the material present in the detector aperture.
      Apart from the beam pipe, which is made of Beryllium and contributes 
      less than 0.3~\% of a radiation length to the material budget, Helium bags
      constitute the only material between the beam pipe and the tracking and 
      electron identification detectors in RUN5 setup.
      As is verified by the PISA simulation
      of $\pi^0$ decays, the ratio of
      electrons from the conversion of photons from $\pi^0 \rightarrow \gamma\gamma$
      decays to electrons from $\pi^0$ Dalitz decays is 0.403 with a systematic
      uncertainty of about 10~\%, independent of $p_{\mathrm{T}}$ in the relevant range.
      For heavier mesons, this ratio is rescaled in the cocktail to properly account 
      for the fact that the branching ratio of the Dalitz decay relative to the
      $\gamma\gamma$ decay increases slightly with increasing parent meson mass.
      
      The material budget  between the beam pipe and the tracking detector
      in RUN6 setup increase slightly, since there are not Helium bags at 
      RUN6 to install HBD.
      The effect is also estimated by the PISA simulation.
      It is found that the ratio of the electrons from air conversion 
      due to the absence of He bag to the electrons from $\pi^0$ Dalitz decays is
      7$\pm$1\%.
      
      It is crucial to note that the contribution from photon conversion
      to the background electron spectra is less than half of the contribution
      from direct Dalitz decays.
      For a reliable measurement of single non-photonic electrons, this is essential.
      
      \subsubsection{Direct Photon}
      Contributions to the background electrons from direct radiation have two process.
      First, real photons produced in initial hard scattering processes, {\it i.e.}
      direct photons convert to electron pairs in material in the PHENIX detector as photons from 
      light neutral meson decays.
      Second, every source of real photons also presents a source of virtual photons.
      In the case of the $\pi^0$ these two sources are the the $\gamma\gamma$
      decay of $\pi^0 $and the corresponding Dalitz decays, which is also called
      an internal conversion.
      Similarly, direct real photon production is accompanied by direct virtual
      photon production, {\it i.e.} the emission of $e^+e^-$ pairs.
      \begin{figure}
	\begin{center}
	  \includegraphics[width=13cm]{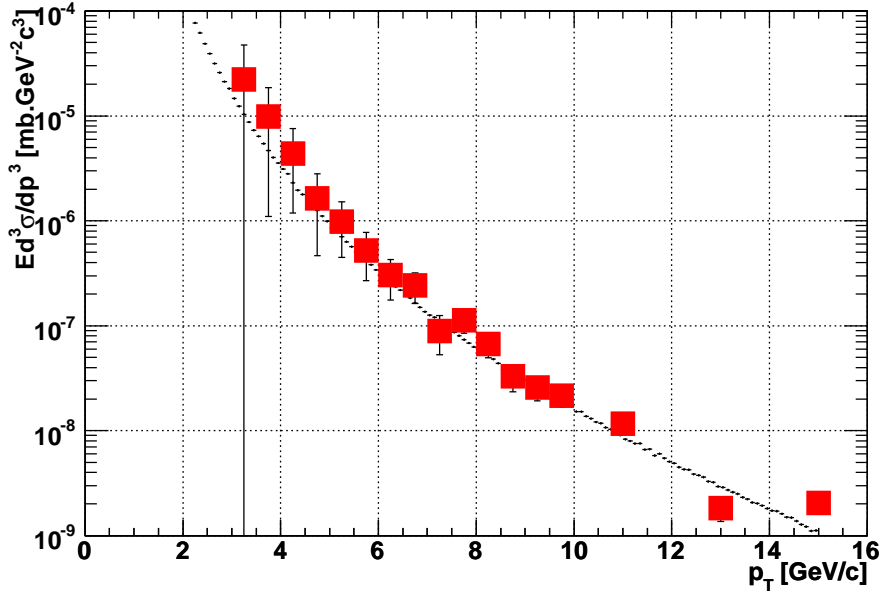}
	  \caption{Measured direct photon spectrum (large symbols shown in red) compared 
	    with the cocktail parameterization (histogram indicated by small 'datapoints')
	    for $p+p$ collisions.}
	  \label{fig:direct_photon_pp}
	\end{center}
      \end{figure}
      
      The measured real direct photon spectrum is parameterized. The corresponding conversion
      electron spectrum of these is added to the electron cocktail.
      Figure~\ref{fig:direct_photon_pp} shows the measured direct photon spectrum 
      with the cocktail parameterization~\cite{bib:direct}.

      The ratio of virtual direct photons to real direct photons depends on 
      $p_{\mathrm{T}}$ because the phase space for dielectron emission increases 
      with increasing $p_{\mathrm{T}}$~\cite{bib:wada}.
      The very same effect is seen in the Dalitz decays of light neutral mesons, 
      {\it i.e.} the Dalitz decay branching ratio relative to the two photon decay 
      branching ratio is larger for the $\eta$ meson than for the $\pi^0$. 
      Consequently, the ratio of virtual and real direct photon emission increases 
      with $p_{\mathrm{T}}$ or, to be more precise, a logarithmic dependence. 
      Such dependence is implemented for internal conversion of virtual photons
      based on the theory~\cite{bib:wada}.
      
      \subsubsection{Quarkonium and Drell-Yan}
      The contribution from di-electron decay of J/$\psi$ and  $\Upsilon$ becomes significant 
      above $\sim$2~GeV/$c$ due to their large mass, while the contribution is negligible  at
      low $p_{\mathrm{T}}$.
      The $p_{\mathrm{T}}$ spectrum of J/$\psi$ is measured up to 9~GeV/$c$ via di-electron decay 
      at mid-rapidity in $p+p$ collisions at PHENIX~\cite{bib:phjpsi}.
      The $p_{\mathrm{T}}$ spectrum of J/$\psi$ is fitted with a power-law function and $m_{\rm T}$ scaling.
      The average shape of these two function is used for the cocktail calculations.
      The deviation for each function is taken into account as systematic error.
      
      Unlike the case of J/$\psi$, there is not a measured $p_{\mathrm{T}}$ spectrum of $\Upsilon$
      at mid-rapidity in $p+p$ collisions at $\sqrt{s}=200$~GeV.
      Therefore,  $p_{\mathrm{T}}$ spectrum of $\Upsilon$ is taken from NLO pQCD 
      calculation~\cite{bib:upsi1}. The total cross section at 
      mid-rapidity~($d\sigma/dy\mid_{y=0}$) in NLO pQCD is 6.89$\times 10^{-6}$~mb.
      This value is compatible with the measured cross section at PHENIX and STAR and 
      it is found the contribution of $\Upsilon$ is not significant~\cite{bib:upsi2,bib:upsi3}.
      
      LO pQCD calculation is used for the estimation of the contribution of Drell-Yan process.
      The result from LO pQCD calculation is scaled by a factor of 1.5 to take into account
      the higher order effect.
      The contribution of Drell-Yan process becomes important as electron $p_{\mathrm{T}}$
      increases. However, the contribution from Drell-Yan process is found not to 
      be significant for the measured $p_{\mathrm{T}}$ range~(up to 9GeV/$c$) 
      compared to other background sources.

      %
      %
      \subsubsection{Implemented Cocktail in RUN5 and RUN6}
\begin{figure}[htb]
	\begin{center}
	  \includegraphics[width=16cm]{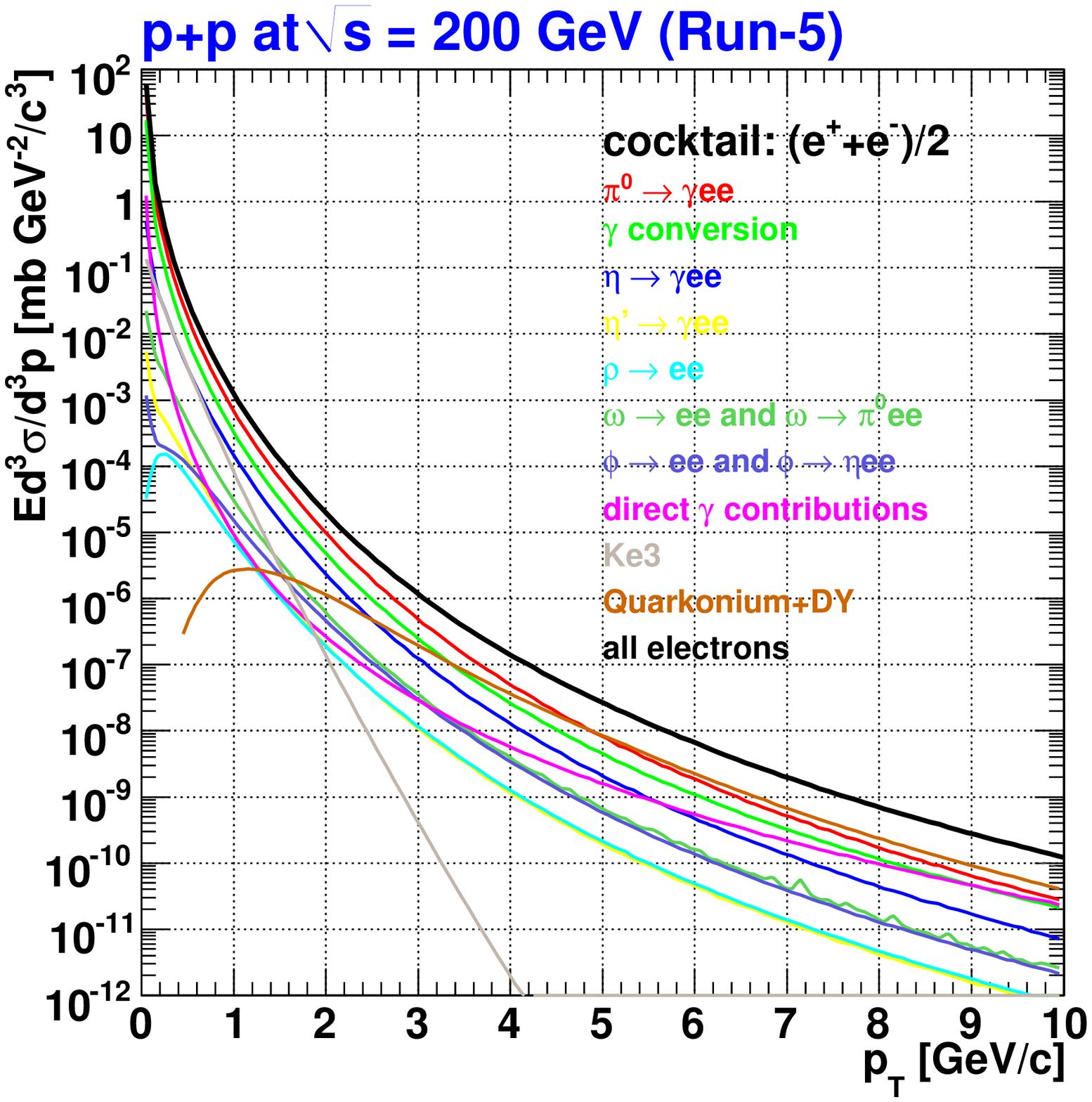}
	  \caption{Invariant cross section of electrons from all sources 
	    considered in the RUN5 $p+p$ cocktail
	  }
	  \label{fig:run5cock}
	\end{center}
      \end{figure}

\begin{figure}[htb]
	\begin{center}
	  \includegraphics[width=16cm]{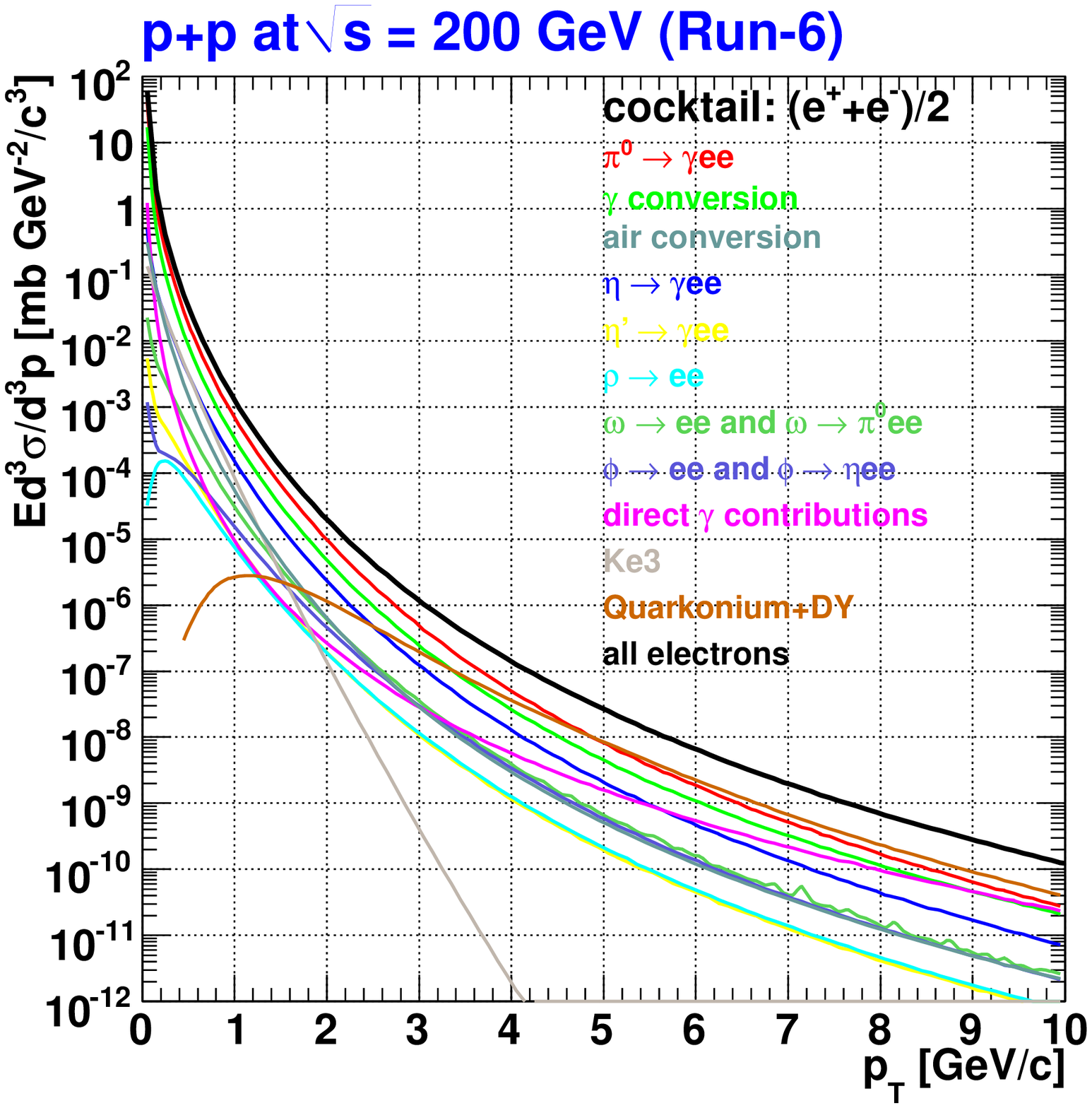}
	  \caption{Invariant cross section of electrons from all sources 
	    considered in the RUN6 $p+p$ cocktail
	  }
	  \label{fig:run6cock}
	\end{center}
      \end{figure}

      Figure~\ref{fig:run5cock} and ~\ref{fig:run6cock} show the invariant cross section for 
      background electrons  calculated by cocktail method
      in the $p+p$ collisions in RUN5 and RUN6, respectively.
      
      \subsubsection{Systematic Errors}
      Systematic errors are estimated for all cocktail ingredients, propagated to
      the corresponding electron spectra, and then added in quadrature to determine
      the total cocktail systematic error. 
      
      The following systematic errors are evaluated and listed up 
      as follows.
      
      \begin{itemize}
      \item {pion spectra: To evaluate this uncertainty the full cocktail calculation is repeated
	in $\pm 1 \sigma$ uncertainty bands for the pion input, propagating the
	uncertainty in the pion spectra to the electron cocktail. 
	With systematic uncertainty of ~10~\% almost independent on $p_{\mathrm{T}}$, the
	pion input represents the largest contributor to the electron cocktail
	uncertainty up to $\sim$ 5~GeV/$c$.}
      \item {light mesons: Since the contributions from all other mesons are much smaller than the 
	contribution from $\eta$ decay only $\eta$ is of practical
	relevance.  The systematic  uncertainties are calculated from particle ratios listed above.
	This contribution is small compared to the uncertainty in
	the pion spectra and it depends on $p_{\mathrm{T}}$ only slightly.}
      \item {conversion material: The contribution from photon
	conversions obviously depends on the material present in the aperture.
	An analysis of fully reconstructed dielectrons from photon
	conversions suggests that this uncertainty is not larger than 10~\%.
	Therefore, 10\% systermatic error is assigned. }
      \item {$K_{e3}$ decay: This contribution is estimated via the PISA 
	simulation. Given the limited statistics of this calculation a 50~\% 
	systematic error is assigned, which is only relevant at low $p_{\mathrm{T}}$, 
	{\it i.e.} below 1~GeV/$c$.
      }
      \item {direct radiation: This contribution is directly propagated from the 
	systematic error quoted for the direct photon measurement. It is
	relevant only at high $p_{\mathrm{T}}$.
      }
      \item {quarkonium and Drell-Yan: The contribution from J/$\psi$ di-electron decay 
	among dominant in these contributions and becomes significant above 2~GeV/$c$.
	The $p_{\mathrm{T}}$ distribution of J/$\psi$ is well measured at PHENIX~\cite{bib:phjpsi} and 
	10\% systematic error is assinged for the abusolute value for J/$\psi$ contribution.
	In addition, the differece from two fit function~(power-law and $m_{\rm T}$ scaling) is 
	taken into accout as systematic error.
	40\% systematic error is assinged for the contribution from $\Upsilon$ based on
	the comparison of the total cross section between NLO pQCD and the result from PHENIX
	and STAR.
	The uncertainty of the contribution from Drell-Yan process is unclear. Therefore,
	100\% systematic error is assined for the contribution from Drell-Yan process to be
	conservative.
      }
      \end{itemize}

\begin{figure}[htb]
	\begin{center}
	  \includegraphics[width=14cm]{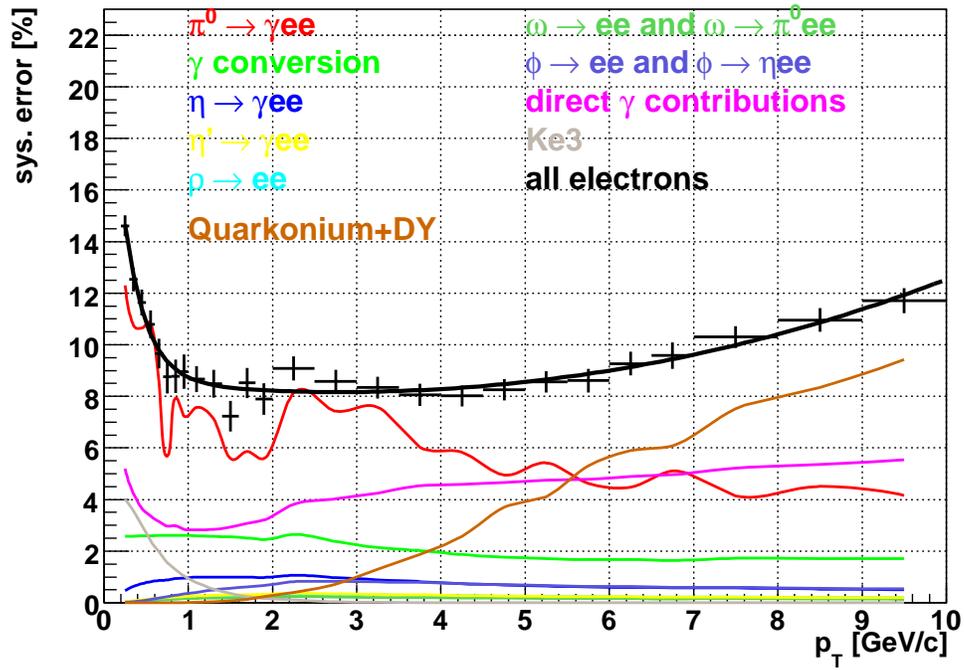}
	  \caption{
	    Individual contributions to the cocktail systematic error. The total
	    error is depicted by the data points which are shown together with a fit.
	  }
	  \label{fig:cocksys}
	\end{center}
      \end{figure}

      Figure~\ref{fig:cocksys} shows individual contributions to the cocktail systematic error and 
      the resulting total systematic error.
      A fit of the total systematic error is shown 
      in Fig.~\ref{fig:cocksys} , where the fitting function is parameterized as follows:
      \begin{equation}
      SE[\%] = p_0\times \exp(p_1\times p_{\mathrm{T}}) + p_2 + p_3 \times p_{\mathrm{T}}
      +p_4 \times p_{\mathrm{T}}^2 + p_5 \times p_{\mathrm{T}}^3.
      \end{equation}

      \clearpage
      \subsection{Converter Method}
      The 'converter subtraction' method  is used, which directly measures the photonic 
      background and, thus, allows to extend the heavy-flavor measurement to the 
      low $p_{\mathrm{T}}$ with good precision. 
      Photonic and non-photonic electrons  are obtained   by measuring the difference 
      of inclusive electron yields with and without a photon converter with precise and well 
      known thickness: a brass sheet of $1.680~\%$ radiation length ($X_0$). 
      
\begin{figure}[htbp]
	\begin{center}
	\includegraphics[width=\linewidth]{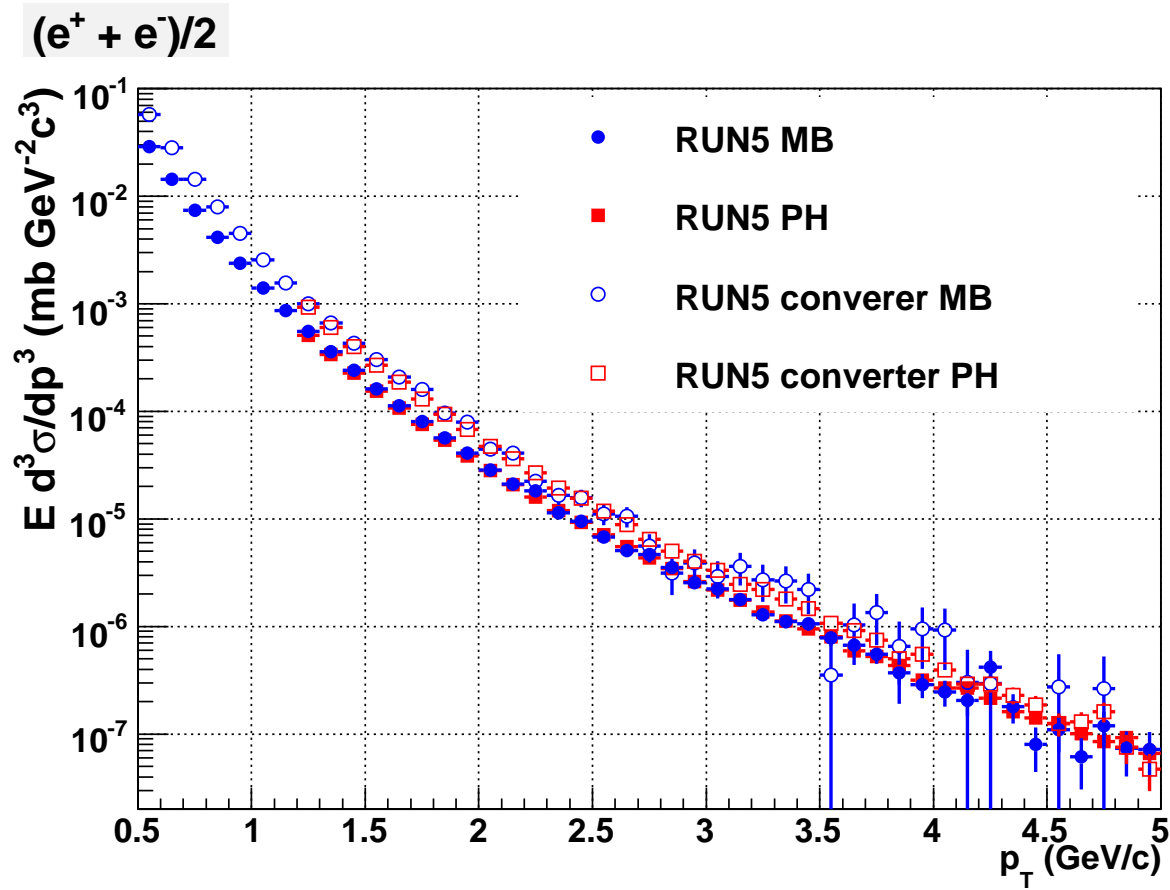}
	\caption{Invariant yields of inclusive electrons in coveter and non-converter runs.
	   Open symbols show the spectra in the converter run
	   and closed symbols show the spectra in the non-converter run.
	   Red squares show the results in PH data set and blue circles show the results 
	   in MB data set.\label{fig:inc_wwo_conv_mb}}
	\end{center}
      \end{figure}

      The C run group in RUN5 is the physics run with the converter.
      The G5A run group in RUN5 is used to compare with the electron yield with that in 
      C run group.
      Figure~\ref{fig:inc_wwo_conv_mb} shows the corresponding inclusive electron spectra.
      In Fig.~\ref{fig:inc_wwo_conv_mb}, open symbols show the spectra in the converter run
      and closed symbols show the spectra in the non-converter run.
      Red squares show the results in the PH data set and blue circles show the results 
      in the MB data set.
      \subsubsection{Method to Subtract Photonic Electrons}
      Raw yields in coveter and non-converter runs can be expressed 
      as the following relations:
      \begin{eqnarray}
	N^\mathrm{Conv\--out}_e &=& N^{\gamma}_e + N^\mathrm{Non\--\gamma}_e, \\
	N^\mathrm{Conv\--in}_e  &=& R_{\gamma} N^{\gamma}_e 
	+ (1-\epsilon) N^\mathrm{Non\--\gamma}_e. 
      \end{eqnarray}
      Here, $N^\mathrm{Conv\--in}_e$ ($N^\mathrm{Conv\--out}_e$) is the measured 
      electron raw yield with (without) the converter. $N^{\gamma}_e$ 
      ($N^\mathrm{Non\--\gamma}_e$) is the photonic (non-photonic) electron yields. 
      $\epsilon$ represents  the blocking factor of the converter which is a small 
      loss of $N^\mathrm{Non\--\gamma}_e$ due to the converter. 
      $R_{\gamma}$ is the multiplication factor of the photonic electron 
      due to the existence of  the converter.
      Then, $N^{\gamma}_e$ and $N^\mathrm{Non\--\gamma}_e$ are determined as follows.
      \begin{eqnarray}
	N^{\gamma}_e &=& \frac{N^\mathrm{Conv\--in}_e-N^\mathrm{Conv\--out}_e}
	{R_{\gamma}-1+\epsilon},\label{eq:photrun5}, \\
	N^{Non\--\gamma}_e &=& \frac{R_{\gamma} N^\mathrm{Conv\--out}_e
	  -N^\mathrm{Conv\--in}_e} {R_{\gamma}-1+\epsilon} \label{eq:nphrun5}.
      \end{eqnarray}
      Non-photonic electrons still include a small background which needs to be subtracted to 
      obtain the electrons from semi-leptonic decay of heavy flavor.
      These are K$_{e3}$ electrons~($E\frac{d^3\sigma^{K_{e3}}}{dp^3}$), 
      $\rho \rightarrow e^+ e^-$~($E\frac{d^3\sigma^{\rho \rightarrow e^+ e^-}}{dp^3}$) 
      ,$\omega \rightarrow e^+ e^-$~($E\frac{d^3\sigma^{ \omega \rightarrow 
	  e^+ e^-}}{dp^3}$), J/$\psi$,$\Upsilon \rightarrow  e^+ e^-($~$E\frac{d^3\sigma^{ 
	   {\rm J}/\psi~\Upsilon \rightarrow e^+ e^-}}{dp^3}$) and Drell-Yan process.
      The spectrum of the electrons from semi-leptonic decay of heavy flavor~(single non-photonic electrons) is determined as follows.
      \begin{eqnarray}
      E\frac{d^3\sigma^{HQ}}{dp^3} &=& E\frac{d^3\sigma^{non-\gamma}}{dp^3} -
      E\frac{d^3\sigma^{K_{e3}}}{dp^3} - E\frac{d^3\sigma^{\rho \rightarrow e^+ e^-}}
      {dp^3}
      -E\frac{d^3\sigma^{ \omega \rightarrow e^+ e^-}}{dp^3} \nonumber \\
      &-& E\frac{d^3\sigma^{{\rm J}/\psi,\Upsilon \rightarrow e^+ e^-}}{dp^3}
       - E\frac{d^3\sigma^{DY}}{dp^3}.
      \end{eqnarray}
      The yield of K$_{e3}$ electrons, $\rho \rightarrow e^+ e^-$, 
      $\omega \rightarrow e^+ e^-$, J/$\psi$~$\Upsilon \rightarrow  e^+ e^-$ and Drell-Yan process 
      are determined at the cocktail calculation.
      Obtained $E\frac{d^3\sigma^{HQ}}{dp^3}$ still have little background, di-electron decay
      of light mesons.
      Such background is negligible.
      \subsubsection{$R_{\gamma}$ and the Blocking Factor}
      The blocking factor is determined to be 2.1\% $\pm$ 1\% from the comparision of
      the conversion peak at the beam pipe between the simulation and real 
      data~\cite{bib:hq1,bib:hq2}.
      
      $R_{\gamma}$ is the crucial parameter in the  converter subtraction method.
      The source of photonic electron is a mixture of 
      mesons~($\pi^0,~\eta,~\eta^{\prime},~\omega$, and $\phi$) decaying into 
      real or virtual photons with their different $p_{\mathrm{T}}$ slopes.     
      However, the photonic electron contributions from $\pi^0$ decays occupies 
      almost of all photonic electrons and determine $R_{\gamma}$.  
      \begin{table}[htbp]
	\begin{center}
	  \caption{Radiation length ($L$) of each material near the interaction point.  
            Conversion probability ($P^\mathrm{Conv}$) is calculated for the 
	    case of electrons emitted from photon with $p_{\mathrm{T}}
	    =1.0$ GeV/$c$~\cite{bib:tai}.~\label{tab:conv_prob01}}
	  \begin{tabular}{c|c|c}
	    \hline
	    Material          & $L$ ($X_0$ [g/cm$^2$]) & $P^\mathrm{Conv}$ \\ 
	    \hline
	    \hline
	    Beam pipe ($Be$)  & 0.288 \%    	       & 0.201 \% \\ 
	    Air ($r<30~cm$)   & 0.099 \%               & 0.069 \% \\ 
	    \hline
	    Total	      & 0.387 \%               & 0.270 \% \\ 
	    \hline
	    Converter (brass) & 1.680 \%    	       & 1.226 \% \\ 
	    \hline
	  \end{tabular}
	\end{center}
      \end{table}

      To calculate $R_{\gamma}$, it is necessary to know exactly the amount of material amounts 
      near the interaction point. Table \ref{tab:conv_prob01} shows the list of 
      each material thickness.       
      The converter sheet is rolled just around beam pipe in converter runs.     
      Conversion probability ($P^\mathrm{Conv}$) in Tab.~\ref{tab:conv_prob01} is 
      calculated for the case of electrons emitted from photon 
      with $p_{\mathrm{T}} =1.0$ GeV/$c$~\cite{bib:tai}.
      The equivalent conversion probability  of a virtual photon in $\pi^0$ Dalitz 
      decay ($P^\mathrm{Dalitz}$) is 0.598\%~\cite{bib:tai}.    
      $R_{\gamma}$ can be estimated with these values for the photon with 
      $p_{\mathrm{T}} = 1.0$ GeV/$c$.
      \begin{eqnarray}
	R_{\gamma}= \frac{P^\mathrm{Conv} + P^\mathrm{Dalitz}~(\mathrm{with~converter})}
            {P^\mathrm{Conv} + P^\mathrm{Dalitz}~(\mathrm{without~converter})} 
	    \sim 2.41.
      \end{eqnarray}
      To obtain more realistic $R_{\gamma}$ for considering geometrical effects and 
      $p_{\mathrm{T}}$ dependence of the conversion provability, the PISA simulations 
      for photon conversions from $\pi^0$ are performed with (without) the converter. 
      We use the spectra of the light mesons which are used cocktail calculation. 
      The $R_{\gamma}$ for $\pi^0$~($R_{\gamma}^{\pi^0}$)  is determined from 
      the the simulation as bellow.
      \begin{equation}
      R_{\gamma}^{\pi^0} = 2.37+0.07\tanh(0.6p_{\mathrm{T}}).
      \end{equation}
      The $\eta$ meson is the second dominant source of the photonic electrons.
      Since $\eta$ mass is larger than $\pi^{0}$ mass, the phase space of $\eta$ 
      Dalitz decay is slightly than $\pi^{0}$. The relative branching 
      ratio (Dalitz decay)/(two $\gamma$ decay)  is 1.2\% for $\pi^{0}$ and 1.5 \% 
      for $\eta$ ~\cite{bib:PDG}.  This difference makes $R^{\eta}_{\gamma}$ 
      smaller than $R^{\pi^{0}}_{\gamma}$.   
      $R_{\gamma}$ for $\eta$~($R_{\gamma}^{\eta}$) is determined as bellow.
      \begin{eqnarray}
	R_{\gamma}^{\eta}&= &\frac{P_{\rm bp} +P_{\rm air} + P_{\rm Dalitz}^{\eta}+
	P_{\rm conv}}{P_{\rm bp} +P_{\rm air} + P_{\rm Dalitz}^{\eta}} \nonumber \\
	&\sim& 1+ (R_{\gamma}^{\pi^0}-1) \times \frac{0.87\%}{1.1\%}.
      \end{eqnarray}
      Contributions from other mesons which undergo Dalitz decay ($\eta^{\prime},
      ~\rho,~\omega,$~and~$\phi$) are small (6 \% at $p_{\mathrm{T}}=3$ GeV/$c$, and smaller 
      at lower $p_{\mathrm{T}}$). The particle ratios used in the cocktail 
      calculation  are used to calculate total $R_{\gamma}$. The uncertainties in the 
      particle ratios are included in the systematic uncertainties of $R_{\gamma}$. 
      
      In this method, it is essential that the  amount of material is accurately 
      modeled in the simulation.  We compare the yield of identified photon 
      conversion pairs in the data and in the simulation, and conclude  that the 
      simulation reproduces  $R_{\gamma}$ within $\pm 2.7 \%$. 
      Figure.~\ref{fig:rcn_rgmb} and \ref{fig:rcn_rgph} show
      $R_{\gamma}$ as a solid curve, 
      which is compared  with  the ratio of inclusive electron yield with/without 
      photon converter~($R_{CN}$)

      \subsubsection{$R_{CN}$}
\begin{figure}[htb]
      \begin{tabular}{c c}
	\begin{minipage}{\minitwocolumn}
	  \begin{center}
	    \includegraphics[width=7.5cm]{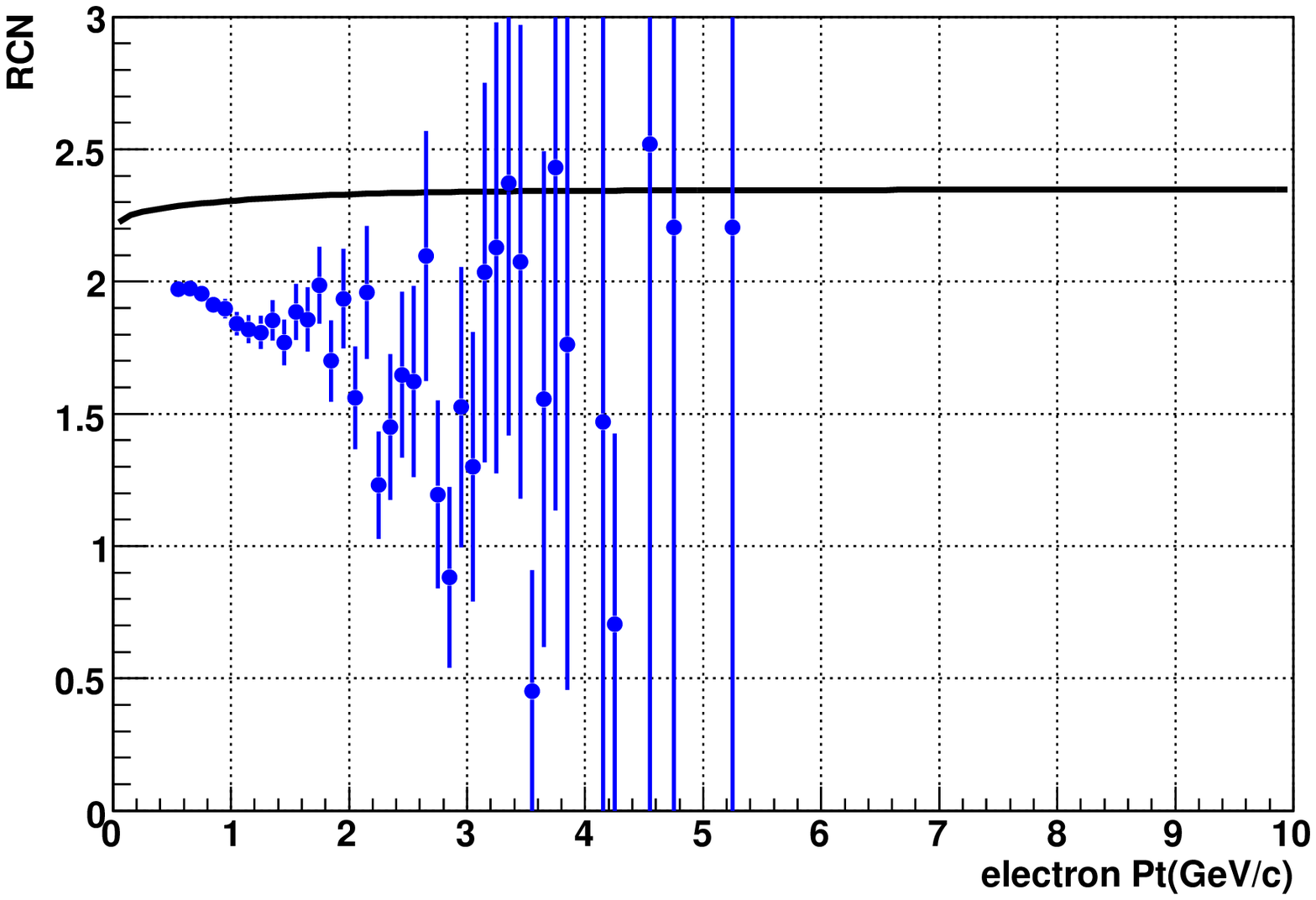}
            \caption{The ratios of the electron 
	      yield in the converter run over the non-converter run~($R_{CN}$)
	      as a function of 
	      electron $p_{\mathrm{T}}$ in RUN5 MB data.
	      The black line is  $R_{\gamma}(p_{\mathrm{T}})$
	    }
            \label{fig:rcn_rgmb}
	  \end{center}
	\end{minipage}
	&
	\begin{minipage}{\minitwocolumn}
	  \begin{center}
	    \includegraphics[width=7.5cm]{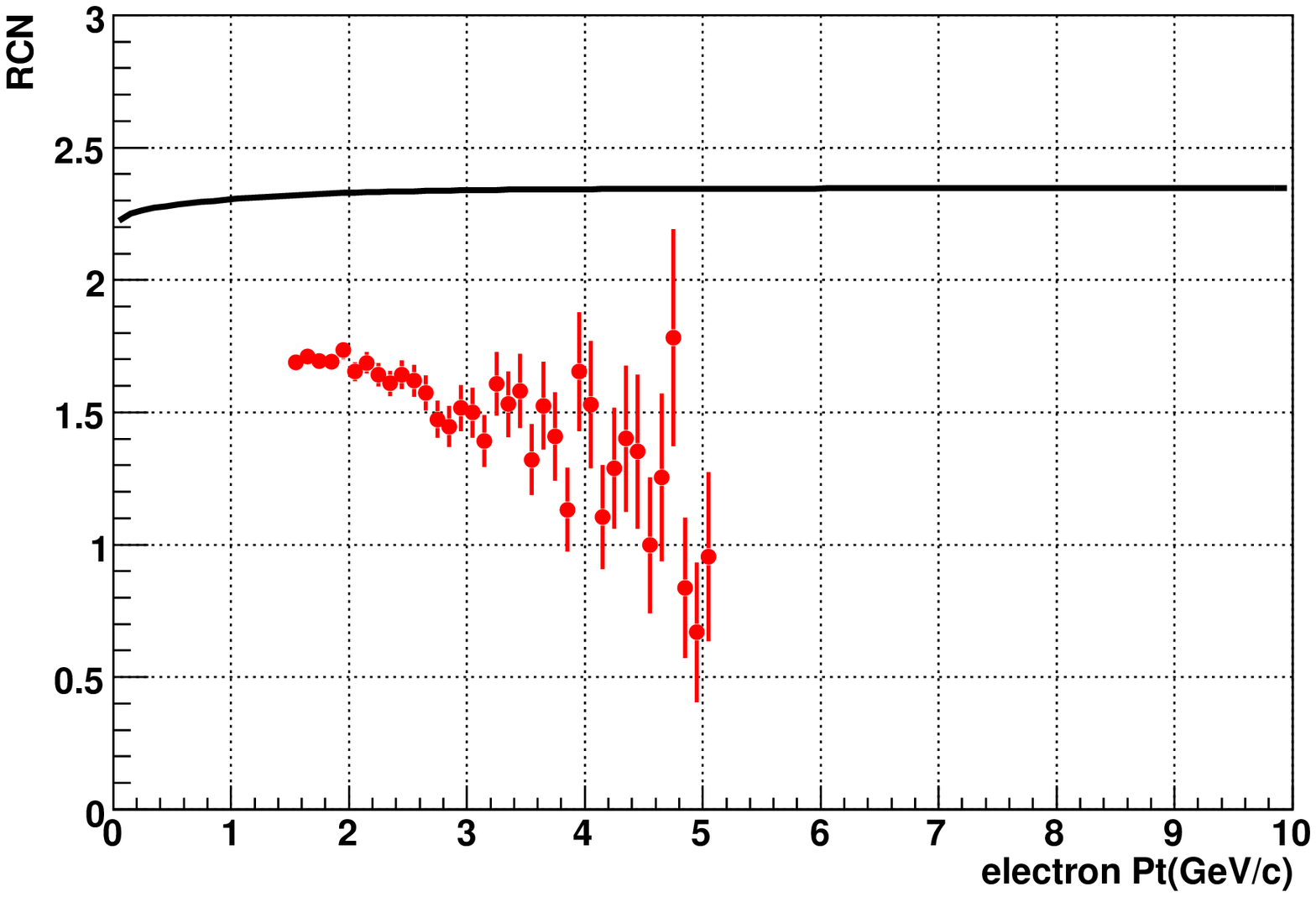}
            \caption{The ratios of the electron 
	      yield in the converter run over the non-converter run~($R_{CN}$) as a function of 
	      electron $p_{\mathrm{T}}$ in RUN5 PH data.
	      The black line is  $R_{\gamma}(p_{\mathrm{T}})$
	    }
            \label{fig:rcn_rgph}
	  \end{center}
	\end{minipage}
      \end{tabular}
      \end{figure}

      $R_{CN}$ is defined as the ratio of inclusive electron yield with/without 
      photon converter.
      Figure.~\ref{fig:rcn_rgmb} and \ref{fig:rcn_rgph} show $R_{CN}$
      measured in RUN5 MB and PH data, respectively.

      If there are no non-photonic contribution , then $R_{CN}=R_{\gamma}$. 
      Figure~\ref{fig:rcn_rgmb} and \ref{fig:rcn_rgph}  show that $R_{CN}$ gradually 
      decreases  with 
      increasing $p_{\mathrm{T}e}$, while $R_{\gamma}$ slightly increases with $p_{\mathrm{T}}$. 
      The difference between $R_{CN}$ and $R_{\gamma}$ proves  the existence of 
      non-photonic electrons.   
      The systematic error of $R_{CN}$ is originated from the instability of the
      efficiency of electron reconstruction during the C run group and 
      the G5A run group.
      We assign 1\% systematic error for $R_{CN}$.
      \subsubsection{Converter Method for RUN6 }
      Since there are no converter run during RUN6, we use $R_{CN}$ measured in RUN5 for 
      RUN6 converter analysis.
      Statistics is improved in RUN6 and this is the great advance in RUN6 data analysis. 
      However, the statistical error of non-photonic electron yield in the converter method 
      is not improved in RUN6, since the statistics in the converter method is determined by
      the RUN5 converter run.
      Thus, we obtain only photonic electron spectrum to compare with cocktail.
      The comparison with photonic electrons in the cocktail between the measured 
      photonic electrons is used to determine the normalization factor of cocktail.
      The difference of the photonic electron between RUN5 and RUN6 due to 
      the absence of the Helium bags is taken into account as follows.
      \begin{eqnarray}
	E\frac{d^3\sigma^{\gamma}}{dp^3} = & \left( E\frac{d^3\sigma^{incl}}{dp^3}  - 
	E\frac{d^3\sigma^{air}}{dp^3} \right)\times \frac{R_{CN}(RUN5)-1+\epsilon}
	{R_{\gamma}(p_{\mathrm{T}})-1+\epsilon} \label{eq:photrun6},  \\
	E\frac{d^3\sigma^{non-\gamma}}{dp^3} = & \left( E\frac{d^3\sigma^{incl}}{dp^3}  - 
	E\frac{d^3\sigma^{air}}{dp^3} \right)\times \frac{R_{\gamma}(p_{\mathrm{T}})-R_{CN}(RUN5)}
	{R_{\gamma}(p_{\mathrm{T}})-1+\epsilon} \label{eq:nphrun6} .
      \end{eqnarray}
      Here,
      \begin{itemize}
      \item  $E\frac{d^3\sigma^{incl}}{dp^3}$ is the spectrum of inclusive electrons 
	in RUN6.
      \item  $E\frac{d^3\sigma^{air}}{dp^3}$ is the spectrum of electrons 
	from air conversion, which is determined by the cocktail calculation without He bag.
      \item  $R_{CN}(RUN5)$ is the $R_{CN}$ which is measured in RUN5.
      \end{itemize}
      \subsubsection{Systematic Errors}
      The systematic error of converter analysis is determined as follows.
      The details of each systematic error are already described.
      \begin{itemize}
      \item   $R_{\gamma}(p_{\mathrm{T}})$: The systematic error of  
	$R_{\gamma}(p_{\mathrm{T}})$ is assigned 0.062. 
      \item  $R_{CN}$: 1\% systematic error is assinged to $R_{CN}$.
      \item  $\epsilon$: 0.01 is assigned as systematic error.
      \end{itemize}
      The systematic error is defined as the quadratic sum of the deviation from
      the above change of each parameters.
      \subsection{Comparison of the Results from Two Methods}
      The spectra of photonic and non-photonic electrons are obtained from the 
      two methods, cocktail method and converter method.
      The results from these methods should be consistent with each other.
      This comparison can be used to reduce the uncertainty of the cocktail.
      The spectrum shape of the cocktail is determined by the spectrum shape of 
      the parent mesons, dominated by $\pi^0$. 
      At high $p_{\mathrm{T}}$, the acceptance curve for parent mesons becomes almost 
      constant in $p_{\mathrm{T}}$. The shape and slope of the spectrum is well 
      determined, while it is more difficult to determine the absolute normalization of the data.
      Therefore in the cocktail calculation, the shape of the spectrum 
      can well be determined.
      It is useful to tune the absolute normalization of the cocktail from the comparison between 
      the measure photonic electrons and the cocktail at high $p_{\mathrm{T}}$.
      \subsubsection{Photonic Electrons}
\begin{figure}[htb]
	\begin{tabular}{c c}
	  \begin{minipage}{\minitwocolumn}
	    \begin{center}
	      \includegraphics[width=7.5cm]{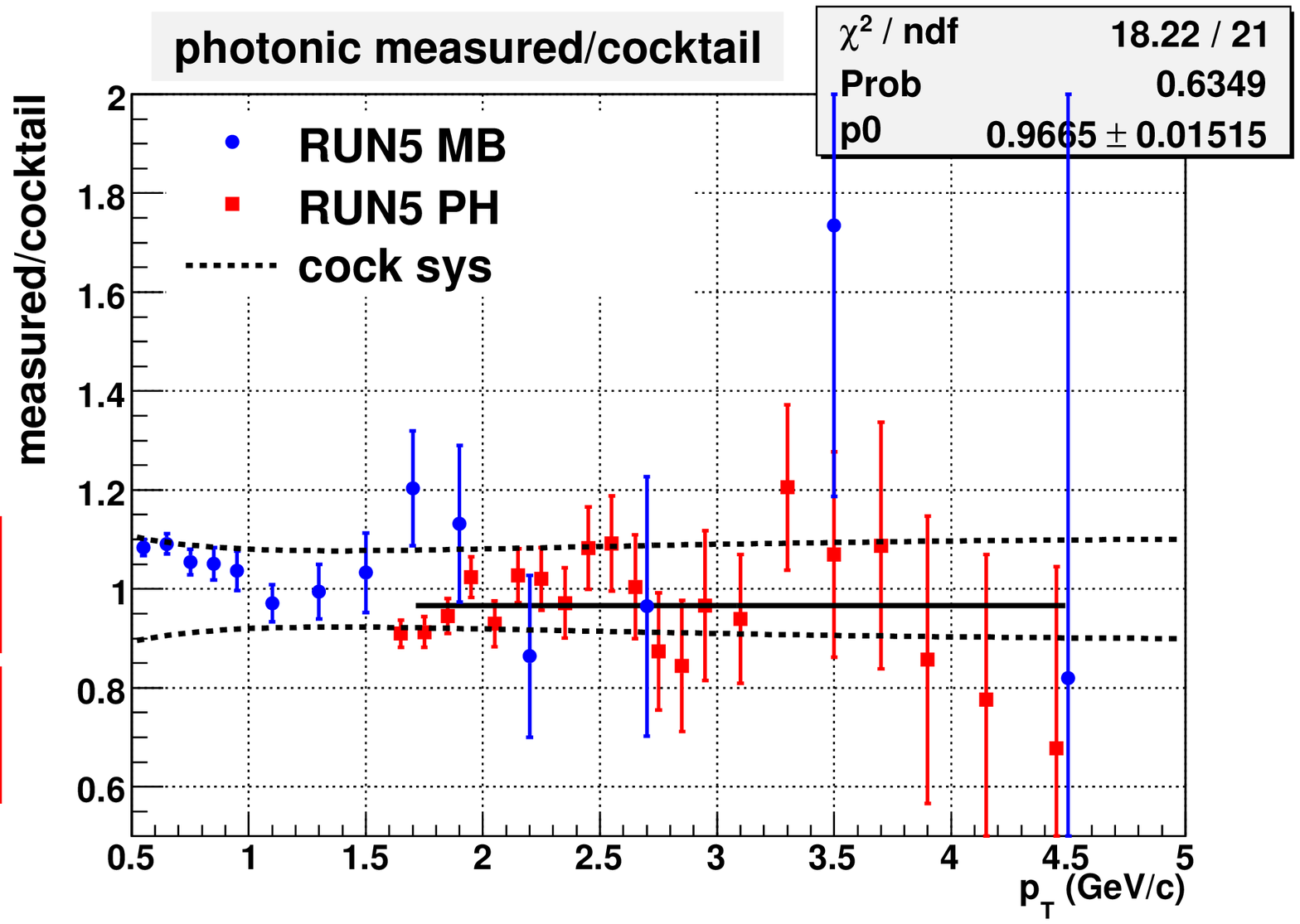}
              \caption{The ratio of measured/cocktail photonic electron spectra in RUN5.
		Blue circles show the ratios at MB data and red squares show the ratios in
		PH data.
		Dotted line show systematic error of the cocktail.
	      }
              \label{fig:phratiorun5}
	    \end{center}
	  \end{minipage}
	  &
	  \begin{minipage}{\minitwocolumn}
	    \begin{center}
	      \includegraphics[width=7.5cm]{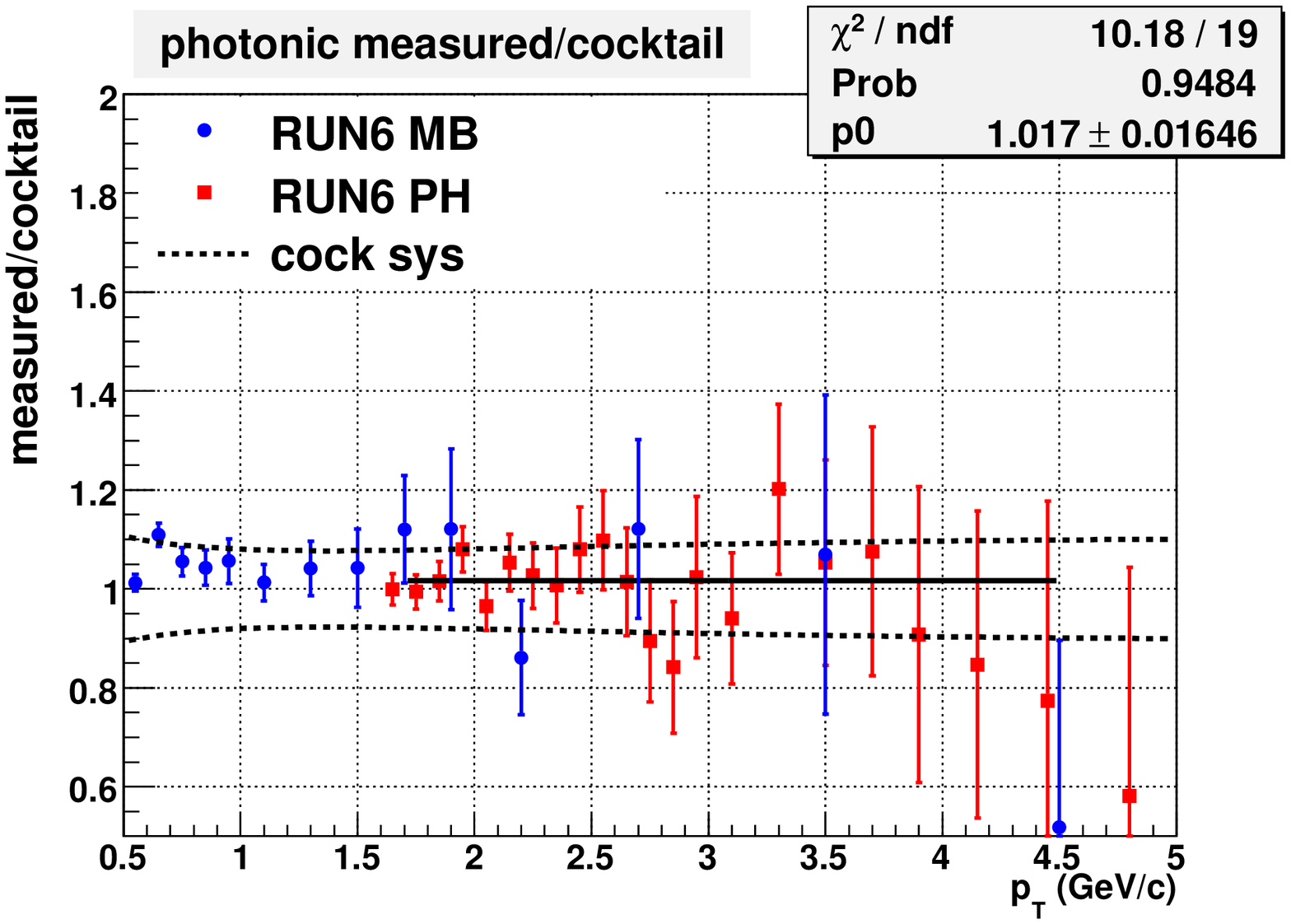}
              \caption{The ratio of measured/cocktail photonic electron spectra in RUN6.
		Blue circles show the ratios at MB data and red squares show the ratios in
		PH data.
		Dotted line show systematic error of the cocktail.
	      }
              \label{fig:phratiorun6}
	    \end{center}
	  \end{minipage}
	\end{tabular}
      \end{figure}

      The photonic electrons are obtained according to Eq.~\ref{eq:photrun5} and 
      Eq.~\ref{eq:photrun6}.
      The spectra of the measured photonic electrons are compared with the photonic
      component in the cocktail.
      Figure~\ref{fig:phratiorun5} and \ref{fig:phratiorun6} show
      the ratio of measured/cocktail photonic electron spectra in RUN5 and 
      RUN6, respectively.
      In Fig.~\ref{fig:phratiorun5} and \ref{fig:phratiorun6}, 
      blue circles show the ratios in MB data and red squares show the ratios in
      PH data.
      Systematic error of the cocktail is also shown as the dotted line in 
      these figures.
      The spectra of the cocktail are consistent with 
      those of the measure photonic electrons within the systematic error of the cocktail.
      
      \subsubsection{Normalization of Cocktail}
      We tune the absolute normalization of the photonic component in the cocktail from 
      the comparison with the measured photonic electrons and the photonic electrons in the cocktail.
      The shape and slope of the meson spectra, which is used as the input of cocktail, 
      is determined with the best precision for high $p_{\mathrm{T}}$ at PHENIX.
      
      The ratios of measured/cocktail photonic electron spectra above 1.7~GeV/$c$ 
      are fitted with a constant, which is expected behavior.
      The fitted lines are shown as striate lines in Fig.~\ref{fig:phratiorun5} 
      and Fig.~\ref{fig:phratiorun6}.
      The fitted values are 0.97$\pm$0.02 and 1.017 $\pm$ 0.02 in RUN5 and RUN6, 
      respectively.
      We calculate the re-normalization factor of the cocktail as 0.992$\pm$ 0.025 for RUN5 and RUN6, 
      since the normalization factor should be common within RUN5 and RUN6.
     
      \begin{figure}[htb]
	\begin{center}
	  \includegraphics[width=12cm]{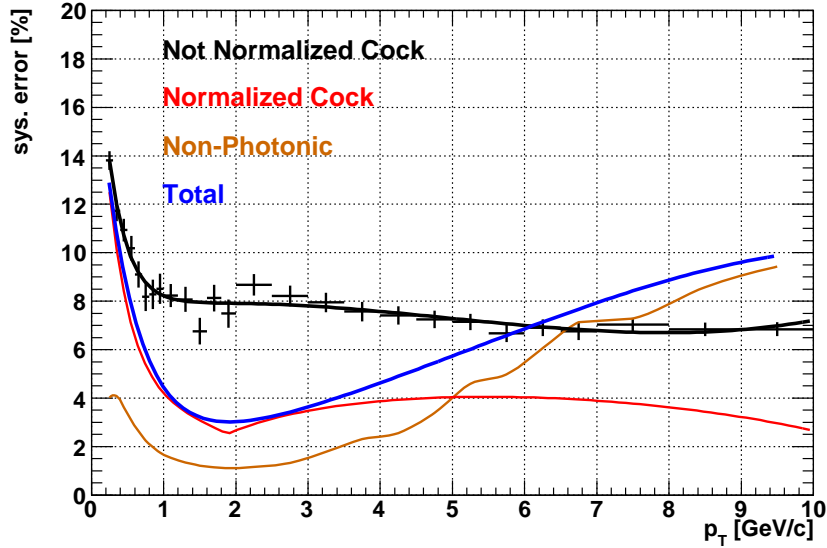}
	  \caption{The systematic error of the cocktail.
	    Black line show the total systematic error of the cocktail 
	    before the normalization
	    of the cocktail.Red line shows the systematic error after the normalization.
	  }
	  \label{chap4_fig20}
	\end{center}
      \end{figure}

	After the rescaling, systematic error of the photonic component in the cocktail 
	is determined as follows.
	\begin{equation}
	SE^{photo}_R(p_{\mathrm{T}}) = \sqrt{(SE^{photo}(p_{\mathrm{T}})-SE^{photo}(1.9GeV/c))^2
	+(\frac{0.025}{0.992})^2 },
	\end{equation}
	where $SE^{photo}(p_{\mathrm{T}})$ is the systematic error of the photonic component before the 
	normalization and $SE^{photo}_R(p_{\mathrm{T}})$ is the systematic error of the photonic 
	component after the normalization.
	Since the normalization point is 1.9GeV/$c$, the deviation from 1.9GeV/$c$ 
	is taken into account as the systematic error.
	Total systematic error is defined as the quadratic sum of $SE^{photo}_R(p_{\mathrm{T}})$
	and the systematic error of non-photonic background that is dominated by J/$\psi$.
	Figure~\ref{chap4_fig20} show the systematic error of the cocktail.
	Black line shows the systematic error of the photonic component in the cocktail 
	before the normalization.
	Red line shows the systematic error of the photonic component 
	after the normalization.
	Orange line shows the systematic error of non-photonic background and 
	blue line shows the calculated total systematic error.
	\subsection{Results}
	The spectra of the single non-photonic electron are determined
	via two independent method, cocktail method and converter method in
	RUN5 and RUN6.
	Converter method  could determine the spectrum of the single non-photonic electrons at low $p_{\mathrm{T}}$ with good precision 
	as already described.
	On the other hand, cocktail method provides better precision than
	converter method towards high $p_{\mathrm{T}}$, {\it e.g.} for $p_{\mathrm{T}} \sim
	1.5$~ GeV/$c$, since the converter method starts to suffer from a lack of 
	statistical precision and the cocktail input is known with small systematic
	uncertainties at high $p_{\mathrm{T}}$.
	Therefore, we use cocktail method at high $p_{\mathrm{T}}$ and 
	converter method at low $p_{\mathrm{T}}$.

	\subsubsection{RUN5 and RUN6 Results}
\begin{figure}[htb]
	  \begin{center}
	    \includegraphics[width=14cm]{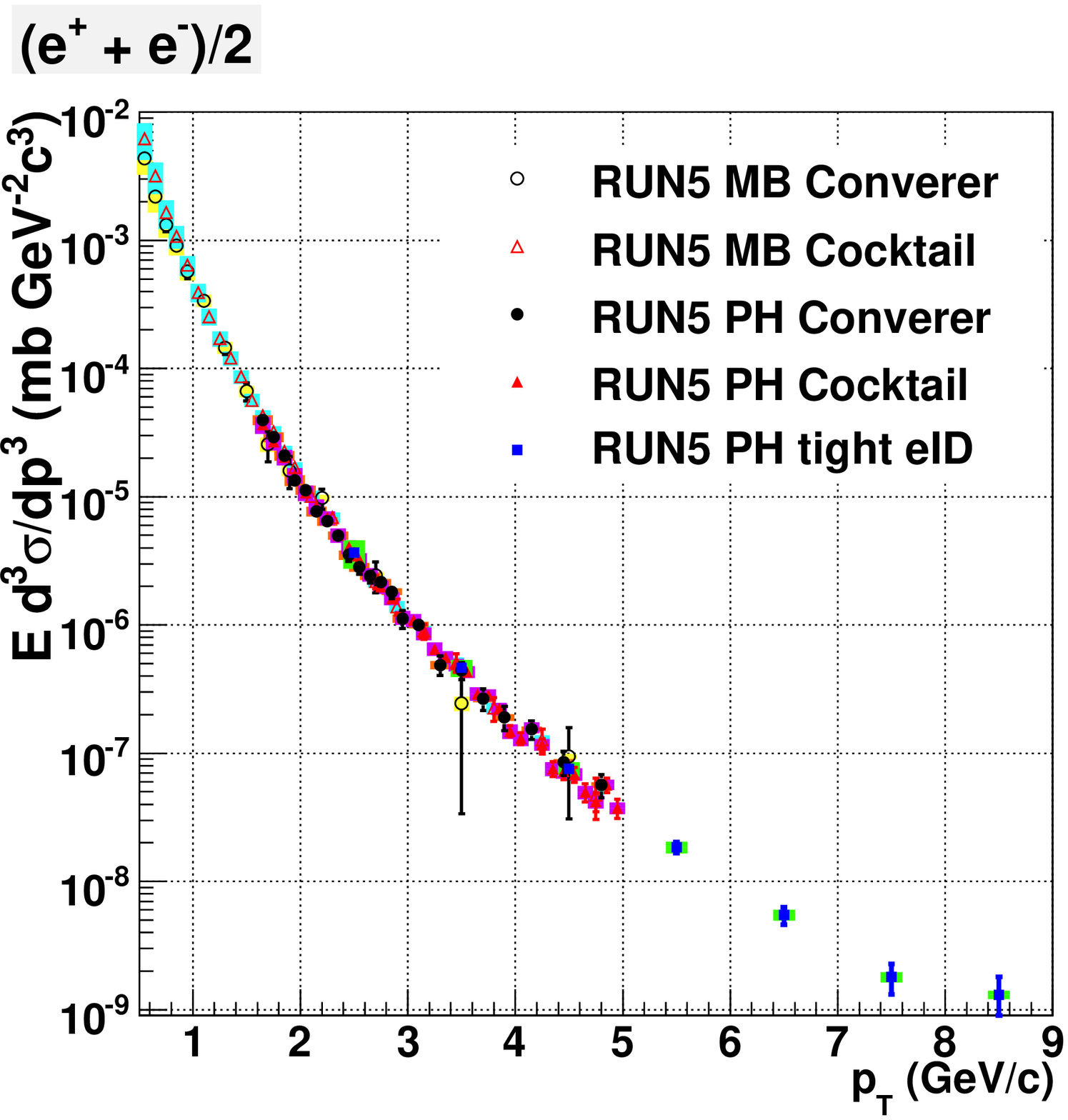}
	    \caption{
	      The invariant cross section of electrons from heavy flavor decay
	      in RUN5 MB and PH data.
	      Circle points show the result from converter method and
	      triangle points show the result from cocktail method.
	      Open symbols show the result at MB data and 
	      closed symbols show the result at PH data.
	      Closed squares show the result at PH data with tight eID cut. 
	    }
	    \label{chap4_fig21}
	  \end{center}
	\end{figure}

\begin{figure}[htb]
	  \begin{center}
	    \includegraphics[width=14cm]{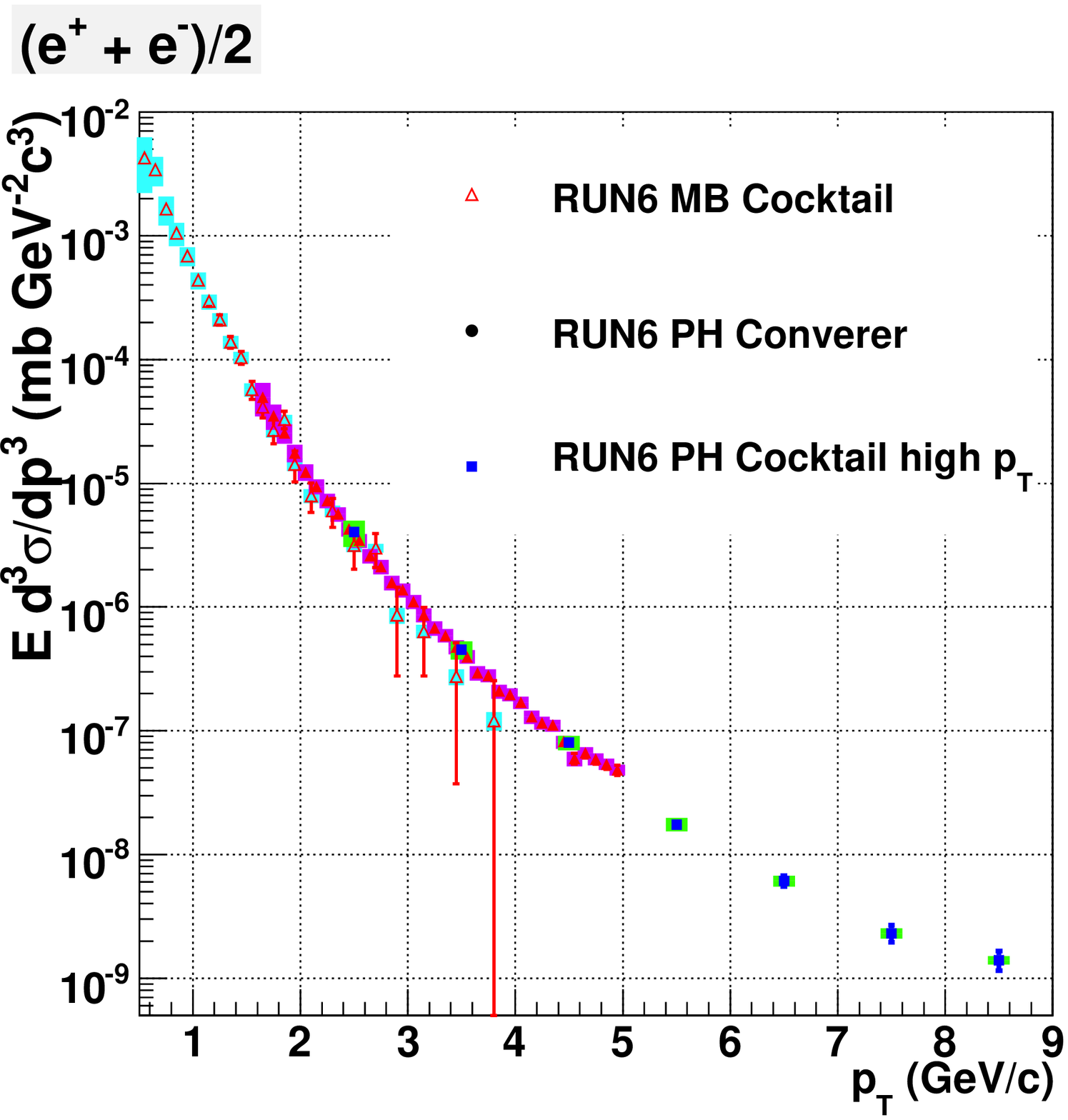}
	    \caption{
	      The invariant cross section of electrons from heavy flavor decay
	      in RUN6 MB and PH data.
	      Circle points show the result from converter method and
	      triangle points show the result from cocktail method.
	      Open symbols show the result at MB data and 
	      closed symbols show the result at PH data.
	      Closed squares show the result at PH data with tight eID cut. 
	    }
	    \label{chap4_fig22}
	  \end{center}
	\end{figure}

	Figure~\ref{chap4_fig21} and \ref{chap4_fig22} show the obtained invariant cross
	section of the single non-photonic electrons with systematic errors in RUN 5 and 
	RUN6, respectively.
	9.9\% systematic error for the abusolute normalization  is {\bf NOT} included
	in Fig.\ref{chap4_fig21} and \ref{chap4_fig22}.
	Circle points show the result from converter method and
	      triangle points show the result from cocktail method.
	      Open symbols show the result from MB data and 
	      closed symbols show the result from PH data.
	      Closed squares show the result from PH data with tight eID cut. 
	
	\subsubsection{Combined Result} \label{hq_spe}
	The result from converter method in RUN5 is used for low $p_{\mathrm{T}}$ and 
	the combined result in RUN5 and RUN6 from cocktail method is used for high $p_{\mathrm{T}}$.
	Since the precision of the converter analysis is determined by the statistics 
	at the converter runs in RUN5, we use only the result from converter method at 
	RUN5 MB data for low $p_{\mathrm{T}}$.
	When the results from RUN5 and RUN6 with cocktail method are combined, 
	the results from PH data are used to improve the statistics for high $p_{\mathrm{T}}$.
	BLUE~(Best Linear Unbiased Estimate) method is applied to combine the results 
	of RUN5 and RUN6~\cite{bib:blue1,bib:blue2}, since a part of systematic errors of 
	RUN5 and RUN6 are correlated.
	Error sources are summarized at Table~\ref{chap6_table3}.
	Error sources are divided into three types in this thesis as follows
	according to the nature of the error.
	 \begin{itemize}
	 \item   {\bf TYPE A} Point-to-point errors.
	 \item   {\bf TYPE B} momentum-correlated errors.
	 \item   {\bf TYPE C} Absolute normalization errors.
	 \end{itemize}
	The averages and errors were determined according to BLUE as bellow. \\
	\begin{eqnarray}
	  <r> &   =& \frac{r_{run5}(\sigma_{run6}^2-\rho\sigma_{run5}\sigma_{run6})+
	    r_{run6}(\sigma_{run5}^2-\rho\sigma_{run5}\sigma_{run6})}
	  {\sigma_{run5}^{2}+\sigma_{run6}^{2}-2 \rho \sigma_{run5} \sigma_{run6}},\\
	  \sigma &= &\sqrt{\frac{\sigma_{run5}^2\sigma_{run6}^2(1-\rho^2) }
	    {\sigma_{run5}^2+\sigma_{run6}^2-2\rho\sigma_{run5}\sigma_{run6}} } .
	\end{eqnarray}
	Here,$r_{runi}$ and $\sigma_{runi}$ are respectively the average of 
	the yield of the single non-photonic electrons and total error in RUNi~(i=5 or 6).
	$\rho$ is the correlation coefficient between RUN5 and RUN6.
	$\rho$ is defined as\\
	\begin{equation}
	\rho = \frac{\sum_{\alpha} \rho^{\alpha} \sigma_{run5}^{\alpha}
	  \sigma_{run6}^{\alpha}} {\sigma_{run5}\sigma_{run6}} ,
	\end{equation}
	where $\alpha$ is the type of error. $\alpha=$ A, B or C.
	Total errors were determined as below.\\
	\\
	\begin{equation}
	\sigma_{runi} = \sqrt{(\sigma_{runi}^{stat})^2 + (\sigma_{runi}^{sys})^2}.
	\end{equation}
	
	\begin{table}[hbt]
	  \begin{center}
	    \caption{Summary of error source}
	    \label{chap6_table3}
	    \begin{tabular}{|c|c|}
	      \hline
	      error source & correlation run5/6 ~(Type) \\
	      \hline \hline
	      statistics & 0 \% (A)\\
	      \hline
	      PISA geometries & 0\%(B) \\
	      \hline
	      eID cut & 0\%(B)\\
	      \hline
	      cocktail calculation & 100\%(B)\\
	      \hline
	      trigger efficiency & 0\% (B)\\
	      \hline
	    \end{tabular}
	  \end{center}
	\end{table}
	
\begin{figure}[htb]
	  \begin{tabular}{c c}
	    \begin{minipage}{\minitwocolumn}
	    \begin{center}
	      \includegraphics[width=7.5cm]{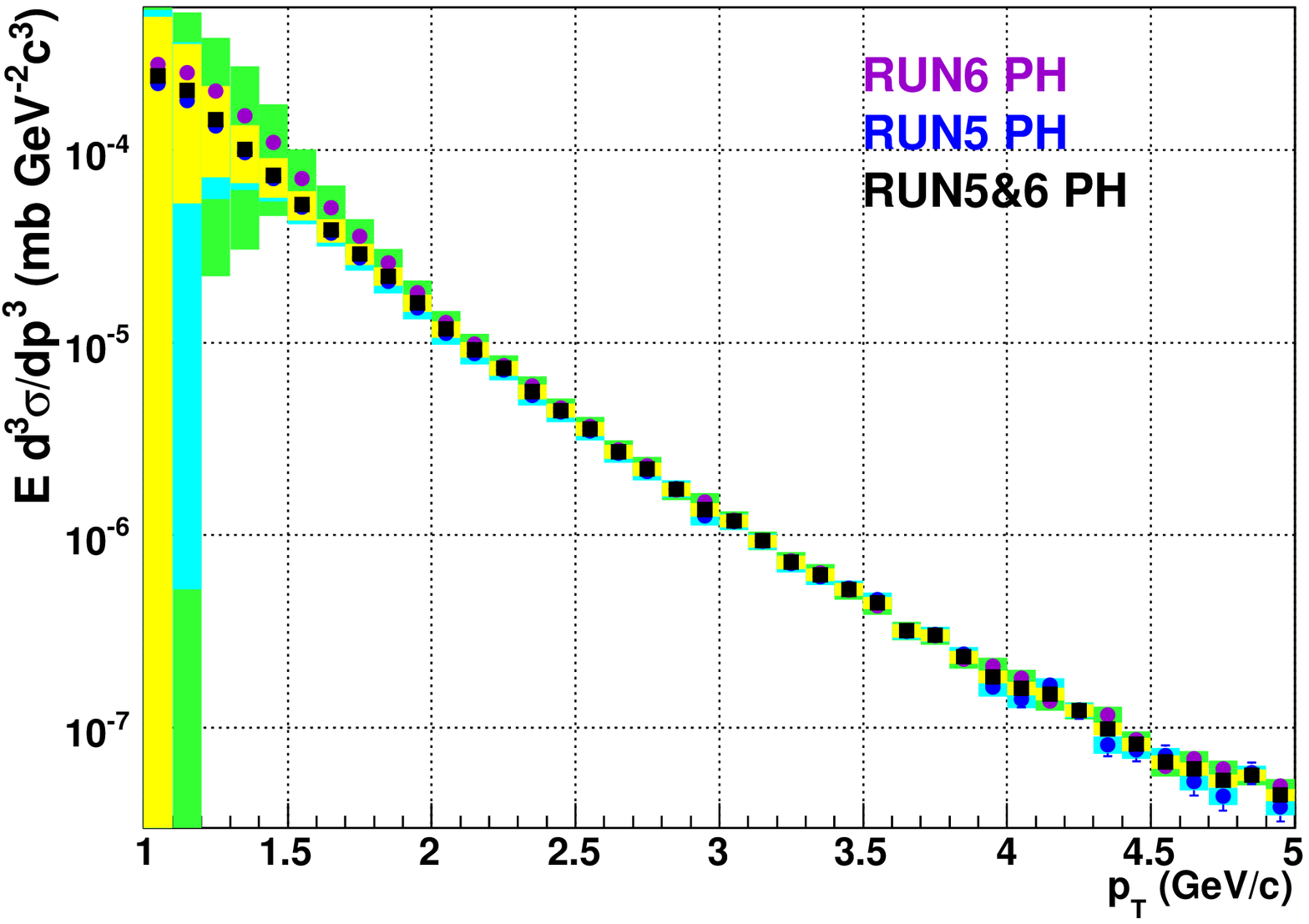}
	      \caption{
		The invariant cross section of electrons from heavy flavor decay
		in RUN6 PH data.
		Red points show the results from cocktail method and black points show
		the result at high $p_{\mathrm{T}}$ extension.
	      }
	      \label{chap4_fig26}
	    \end{center}
	    \end{minipage}
	    &
	    \begin{minipage}{\minitwocolumn}
	      \begin{center}
		\includegraphics[width=7.5cm]{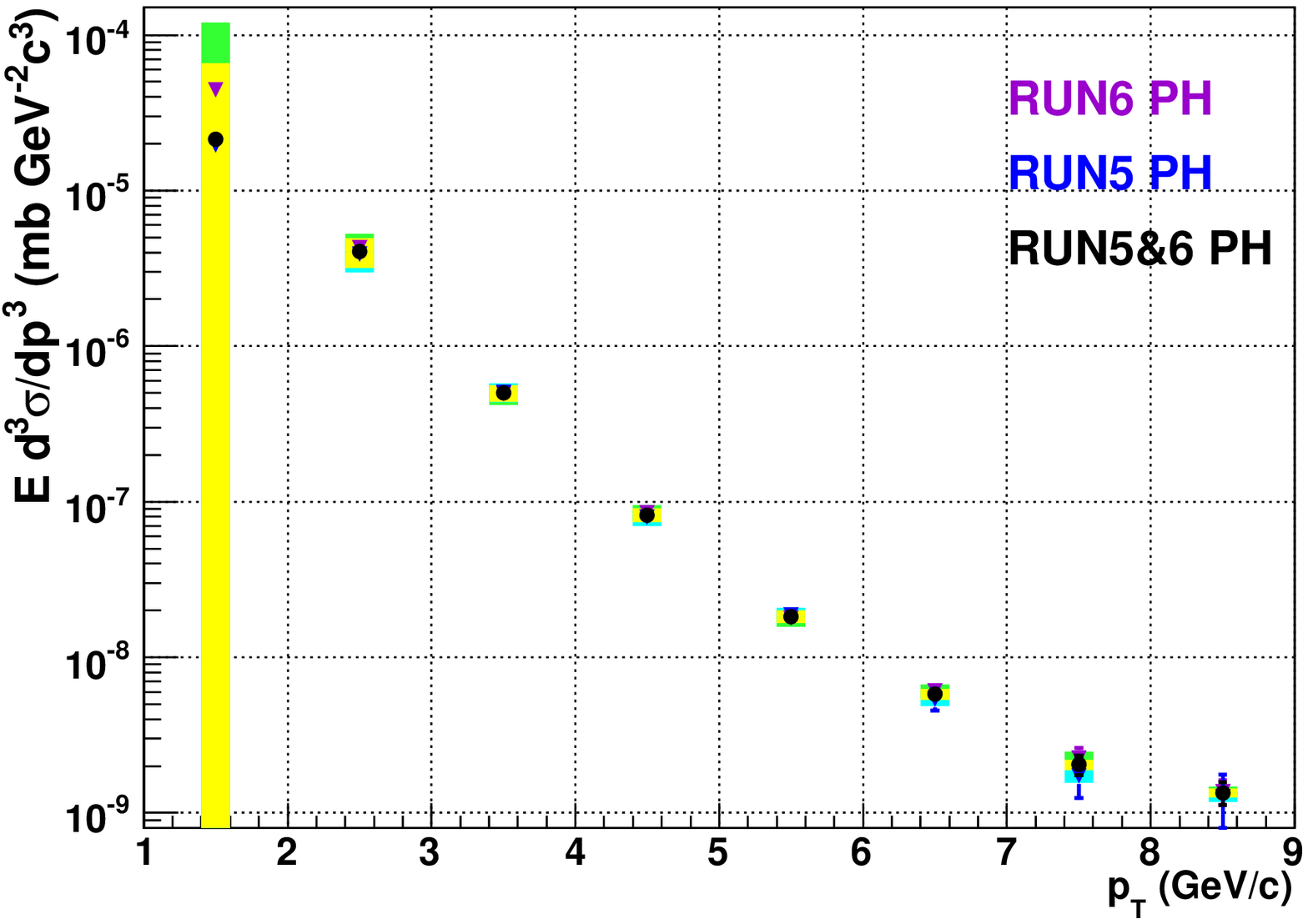}
		\caption{
		  The invariant cross section of electrons from heavy flavor decay
		  in RUN6 PH data.
		  Red points show the results from cocktail method and black points show
		  the result at high $p_{\mathrm{T}}$ extension.
		}
		\label{chap4_fig27}
	      \end{center}
	    \end{minipage}
	  \end{tabular}
	\end{figure}

	The combined results from cocktail method are shown in Figure~\ref{chap4_fig26}
	and \ref{chap4_fig27}.
	$\chi^2$/ndf is 17.2/33 with the standard eID cuts from 1.7GeV/$c$ to 5GeV/$c$ 
	and is  3.2/7 with the tight eID cut from 1.7GeV/$c$ to 9GeV/$c$.
	The values indicate the results in RUN5 and RUN6 are consistent.

\begin{figure}[htb]
	  \begin{center}
	    \includegraphics[width=14cm]{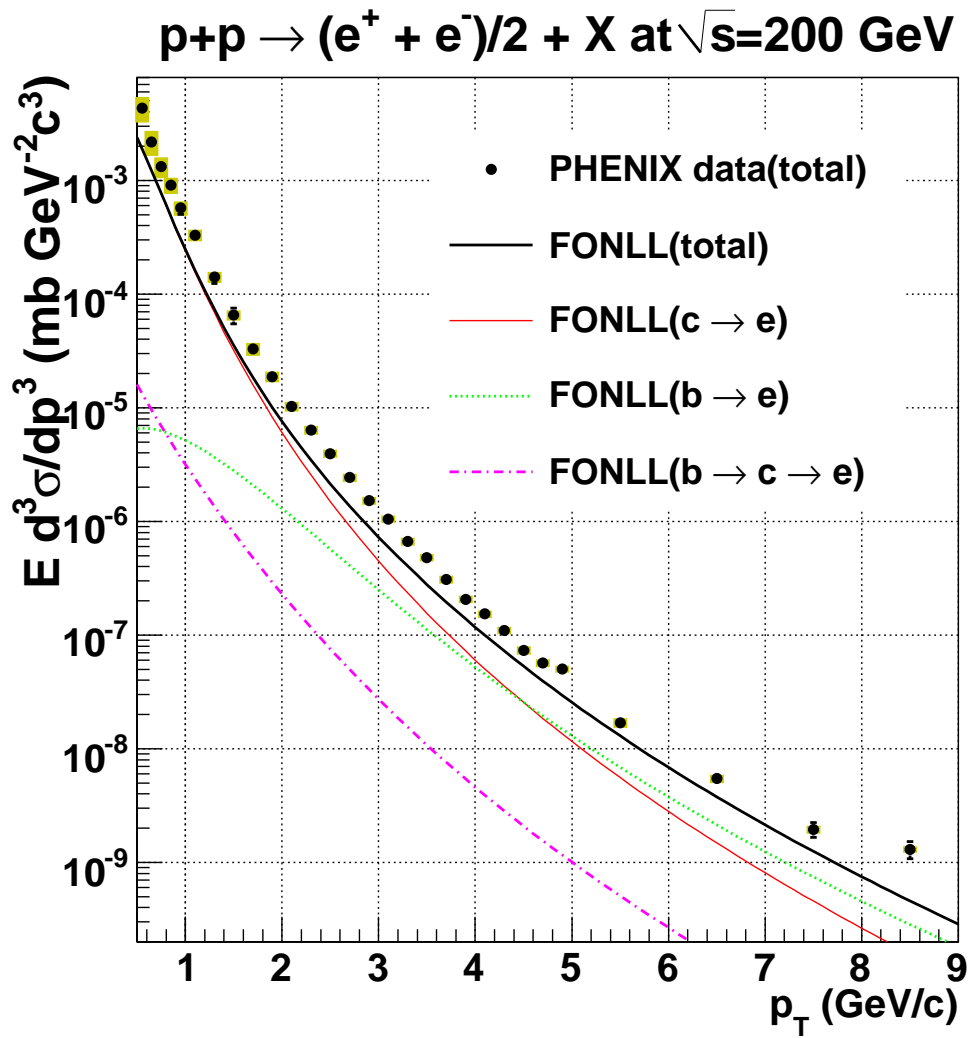}
	    \caption{The spectrum of the single non-photonic electrons
	    in RUN5 and RUN6 with FONLL calculation~\cite{bib:fonll1}.}
	    \label{chap4_fig28}
	  \end{center}
	\end{figure}

	The spectrum of the single non-photonic electron is 
	shown in Figure~\ref{chap4_fig28}.
	FONLL calculation, which is Fixed-Order plus Next-to-Leading-Log 
	perturbative QCD calculation~\cite{bib:fonll1}, is also shown in 
	Fig~\ref{chap4_fig28}.
	
	\clearpage

	\section{Overview of Extraction of $(b\rightarrow e)/(c\rightarrow e+ b\rightarrow e)$}\label{sec:method}
	The spectrum of the electron from semi-leptonic decay of charm and bottom
	is obtained in the previous analysis.
	The fraction of the single electrons from bottom in single non-photonic 
	electrons~($(b\rightarrow e)/(c\rightarrow e+ b\rightarrow e)$) is crucial 
	parameter to understand the behavior of the heavy quarks in the hot and dense matter.
	In this section, the method to extract the fraction utilizing the correlation of 
	the single non-photonic electrons and the associated hadrons is described.
	ERT triggered data in RUN5 and RUN6 is used in the correlation analysis.

	\subsection{Extraction Method} \label{sec:overview}
	The extraction of $(b\rightarrow e)/(c\rightarrow e+ b\rightarrow e)$ by utilizing the correlation of 
	the single non-photonic electrons and the associated hadrons is based on partial reconstruction 
	of $D^0$$\rightarrow e^+  K^-  \nu_e$ decay.
	Unlike charge sign pairs of trigger electrons for $2.0 < p_{\mathrm{T}}<7.0$~GeV/$c$ 
	and associated hadrons for $0.4 < p_{\mathrm{T}}<5.0$~GeV/$c$ are
	reconstructed as partial reconstruction of $D^0$$\rightarrow e^+  K^-  \nu_e$  decay.
	Since most of charged kaons do not reach the hadron identification detector~(TOF and EMCal) due to 
	their short life time, the reconstruction efficiency of identified charged kaon is rather small.
	Therefore, kaon identification is not performed in the analysis and inclusive hadrons 
	are assigned to be kaons.
	As a result, this analysis is NO Particle IDentified~(NO PID) partial 
	reconstruction of $D^0$.
	
	Determination of the background is crucial for this analysis, since
	the signal to background ratio is not good~($\sim$1/10).
	There are two main sources in the background.
	The one is the combinatorial background from electrons and hadrons, where
	the selected trigger electron is not from semi-leptonic decay of heavy flavor.
	The other is the combinatorial background, where the trigger electron is the single non-photonic electron
	and the associated hadron is not from heavy flavor decay.
	The best way to subtract these backgrounds is to use like sign charge pairs of electrons 
	and hadrons. 
	This subtraction method is essential in this analysis.
	Since electron hadron pairs with opposite charge signs are produced only by weak decay, 
	the background subtraction using like sign pairs cancel out the combinatorial background completely for
	the contribution of the trigger electron from $e^+ e^-$ pair creation.
	The electrons from $e^+ e^-$ pair creation are most in all background of the trigger
	electrons.
	Moreover, most of the associated hadrons not from heavy flavor decay are
	from jet fragmentation. The background subtraction using like sign pair cancel out 
	most of contribution from the combination of the single non-photonic electrons 
	and the hadrons from jet fragmentation, since jet is basically charge independent.
	
	\begin{figure}[htb]
	  \begin{center}
	    \includegraphics[width=10cm]{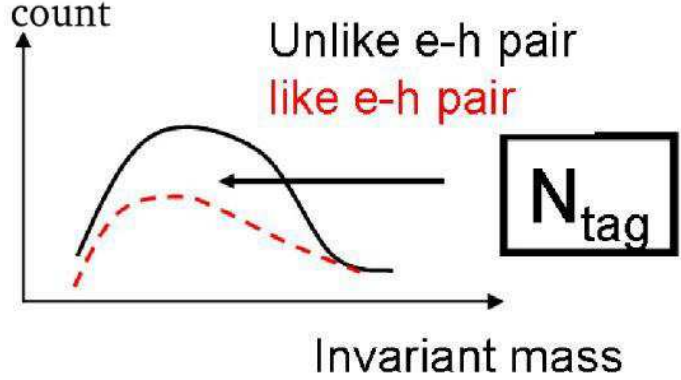}
	    \caption{A conceptual view of invariant mass distributions of
	      unlike sign pairs and like sign pairs.}
	    \label{chap4_fig1}
	  \end{center}
	\end{figure}
	Figure~\ref{chap4_fig1} shows a conceptual view of invariant mass
	distributions of unlike sign pairs and like sign pairs.
	N$_{tag}$ is defined as the number of unlike sign electron-hadron pair entries~ 
	(N$_{unlike}$) minus number of like sign electron-hadron pair entries~(N$_{like}$).
	As already described, extracted signals~N$_{tag}$ are interpreted as the electron-hadron pairs 
	mostly from heavy flavor decays, which are reconstructed partially such as  
	$D/\bar{D} \rightarrow e^{\pm} K^{\mp} X$ decay. 
	N$_{tag}$ contains inclusive signals from other heavy flavored hadons 
	($D^+$,$B^+$,$B^0$ etc) and the remaining contribution from
	the associated hadron which is not from heavy flavor decay.
	These effect are evaluated by using the Monte-Carlo event generators.
	
	The analysis procedure is as follows.\\
	Tagging efficiency~($\epsilon_{data}$), which is a similar variable as a conditional 
	probability of the detection of an associated hadron in PHENIX detector when the electron from 
	semi-leptonic decay of heavy flavored hadron is detected, is defined as below.
	\begin{equation}
	\label{eq:tagdata}
	\epsilon_{data} \equiv \frac{N_{tag}}{N_{e(HF)}} =
	\frac{N_{c\rightarrow tag}+N_{b\rightarrow tag}}
	     {N_{c\rightarrow e}+N_{b\rightarrow e}},
	\end{equation}
	where $N_{e(HF)}$ is the number of electrons from semi-leptonic decay of heavy flavor.
	N$_{c(b)\rightarrow e}$ is the number of electrons from semi-leptonic decay of
	charmed (bottomed) hadrons. 
	N$_{c(b)\rightarrow tag}$ is the number of reconstructed signals~(N$_{tag}$)  for 
	charm (bottom) production.
	Since N$_{tag}$ include the contribution only from the single electrons from heavy flavor, 
	$\epsilon_{data}$ could be written by only charm and 
	bottom terms.
	$\epsilon_{data}$ is determined from real data analysis.
	The analysis detail to obtain $\epsilon_{data}$ is written at Sec.~\ref{sec:cor_real}.

	As a next step, tagging efficiency in the case of  charm production  $\epsilon_{c}$ 
	and tagging efficiency in the case of  bottom production  $\epsilon_{b}$ 
	are defined as bellow.
	\begin{equation}
	\epsilon_{c} \equiv \frac{N_{c\rightarrow tag}}{N_{c\rightarrow e}} ,\quad \quad
	\epsilon_{b} \equiv \frac{N_{b\rightarrow tag}}{N_{b\rightarrow e}}.
	\end{equation}
	$\epsilon_{c(b)}$ is determined from the Monte-Carlo event generators.
	Since the extracted signal~N$_{tag}$ is dominated by decay products of 
	heavy flavored hadrons, tagging efficiency is determined by decay kinematics in
	the first order.
	Therefore, we can determine $\epsilon_{c(b)}$ with good precision using the simulation.
	The analysis detail to obtain $\epsilon_{c(b)}$ is written at Sec.~\ref{sec:cor_sim}.

	Then, the fraction of bottom contribution to the electrons from heavy flavor 
	is determined as,
	\begin{equation}
	\frac{N_{b\rightarrow e}}{N_{c\rightarrow e}+N_{b\rightarrow e}} =
	\frac{\epsilon_c -\epsilon_{data}}{\epsilon_c-\epsilon_b},
	\end{equation}
	
	\subsection{Electrons from $e^+ e^-$ Creation}
\begin{figure}[htb]
	  \begin{center}
	    \includegraphics[width=13cm]{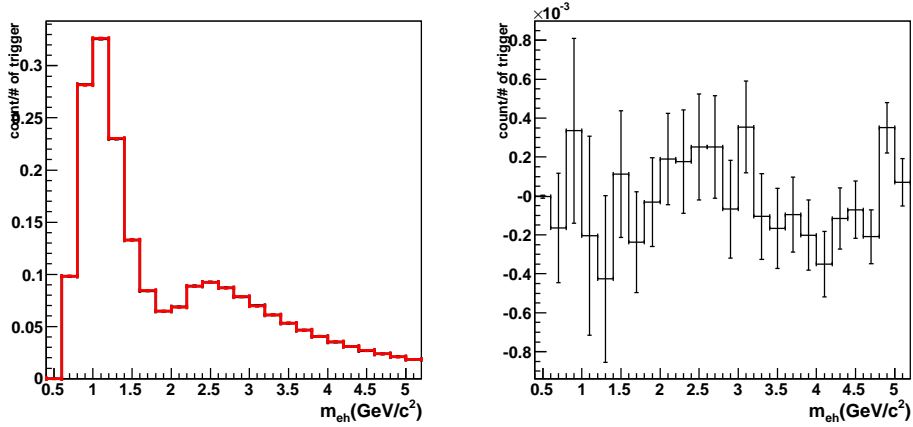}
	    \caption{The invariant mass distribution of electron-hadron pairs in 
	      $\mid y\mid<0.4$, when the trigger electron is photonic electron.
	      In left panels, black lines are unlike 
	      charge sign pairs and red lines are like charge sign pairs.
	      Subtracted invariant mass distribution of electron-hadron pairs was shown 
	      in the right panels.}
	    \label{chap5_fig-1}
	  \end{center}
	\end{figure}

	The contribution of the trigger electrons from $e^+ e^-$ creation must be canceled out 
	in the subtraction of like sign electron-hadron pairs.
	This is the most important issue in this analysis.
	This fact is confirmed by PYTHIA event generator~\cite{bib:pythia1,bib:pythia2}. 
	Figure~\ref{chap5_fig-1} shows invariant mass distributions of unlike 
	sign electron-hadron pairs~(black) and like sign electron-hadron pairs~(red)
	in $\mid y\mid<$0.4, where the trigger electron is from $e^+ e^-$ creation in PYTHIA events.
	Subtracted invariant mass distribution of electron-hadron pairs is  shown 
	in the right panels.
	Tagging efficiency for the electrons from $e^+ e^-$ creation, $\epsilon_{photo}$ 
	is -0.00051$\pm$ 0.00097 in $\mid y\mid <$0.4.
	This result confirms the issue  that the contribution of the electrons from 
	$e^+ e^-$ creation is canceled out completely.

	\section{Correlation Analysis at Real Data} \label{sec:cor_real}
	In this section, tagging efficiency in the real data analysis, $\epsilon_{data}$, is 
	obtained.
	ERT triggered data in RUN5 and RUN6 is used in this correlation analysis.
	\subsection{Used Cut for the Correlation Analysis} \label{sec:ecutcorr}
	The following cuts are used to select the trigger electrons and the associated hadrons 
	in real data and the simulation.
	
	\begin{figure}[htb]
	  \begin{center}
	    \includegraphics[width=10cm]{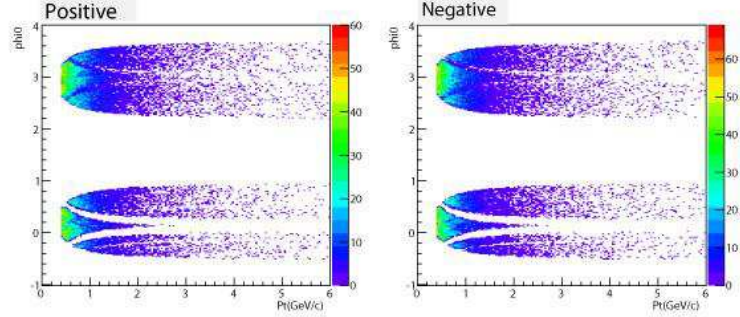}
	    \caption{Phase spaces of positive charged hadron 
              with the geometrical cut in RUN5.}
	    \label{fig:phase}
	  \end{center}
	\end{figure}

\begin{figure}[htb]
	  \begin{center}
	    \includegraphics[width=7cm]{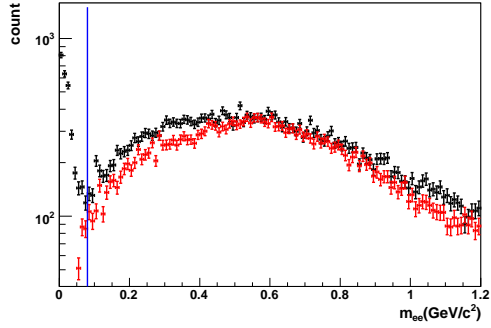}
	    \caption{M$_{ee}$ distribution of unlike sign and like sign
	      pairs of the selected electron and hadron in RUN6.
	      Black points show unlike charge sign pairs
	      and red points show like charge sign pairs.}
	    \label{fig:massee}
	  \end{center}
	\end{figure}

	\begin{itemize}
	\item{{\bf Event Cut}:$-25<{\sf bbcz}<25$ (cm)}
	  
	\item{{\bf Electron Cut}: The standard electron cut is applied for the tracks with
	  $2<p_{\mathrm{T}}<5$~GeV/$c$
	  and the tight electron cut is applied above 5~GeV/$c$. The details of this cut are described
	in Sec.~\ref{sec:ecut}}
	\item{{\bf Hadron Cut}:  {\sf quality}$>15$ and {\sf n0}$<$0~(RICH veto) cut is applied
	  to select hadron for the charged particles with $0.4<p_{\mathrm{T}}<5.0$ GeV/$c$.
	  The selected hadron tracks are analyzed with the kaon hypothesis, that is,
	  the selected particles have kaon mass.}

	\item{{\bf Acceptance Filter}: Since the acceptance~(phase space) for positive charged particles 
	  and negative charged particles is different due to the detector geometry of PHENIX, 
	  the effect of the difference in the phase space needs to be corrected for the subtraction
	  of the like charge sign pairs.
	  The fiducial cut is applied to make the phase space of negative and positive 
	charged tracks identical as the correction for the phase space effect.
	Figure~\ref{fig:phase} shows the phase spaces of associated negative charged particle 
	and positive charged particle with the geometrical cut.
	}
	\item{{\bf  Electron Pair Cut}:  RICH veto cut~({\sf n0}$<$0)  used in hadron cut
	  does not reject electron contamination in the selected hadrons completely
	  due to dead area and limited acceptance of RICH.
	  Since about a half of the measured electrons above 2~GeV/$c$ is produced 
	  via the $e^+$ $e^-$ pair creation, there are strong charge correlation of electron pairs
	  in the events where the trigger electron is found.
	  %
	  %
	  It is found that the effect of such electron contamination in the hadron tracks is not negligible.
	  The electron contamination is rejected using M$_{ee}$, which is the invariant mass between 
	  identified trigger electrons and the associated tracks where their mass is assigned 
	  to be electron mass~(0.511MeV).
	  Most of the electron pairs are produced from $\pi^0$ Dalitz decay and $\gamma$ conversion 
	  at the beam pipe. They could be identified via the reconstructed invariant mass 
	  distribution of $e^+$ $e^-$~pair as the peaks at the low mass region.
	  Figure~\ref{fig:massee} shows the M$_{ee}$ distribution of unlike and like 
	  pairs of the selected electron and hadron in RUN6.
	  In Fig~\ref{fig:massee}, black points show unlike charge sign pairs
	  and red points show like charge sign pairs.
	  The clear peak is shown at the low mass region and 
	  M$_{ee}$ $>$0.08GeV is required for the rejection of these electron pairs.
	}
	  
	\end{itemize}
	
	\subsection{Calculation of $\epsilon_{data}$}
	Tagging efficiency in the real data, $\epsilon_{data}$ is calculated  
	with  the trigger electron for $2.0<p_{\mathrm{T}}<7.0$~GeV/$c$.
	\subsubsection{Count of N$_{tag}$}
\begin{figure}[htb]
	  \begin{center}
	    \includegraphics[width=9cm]{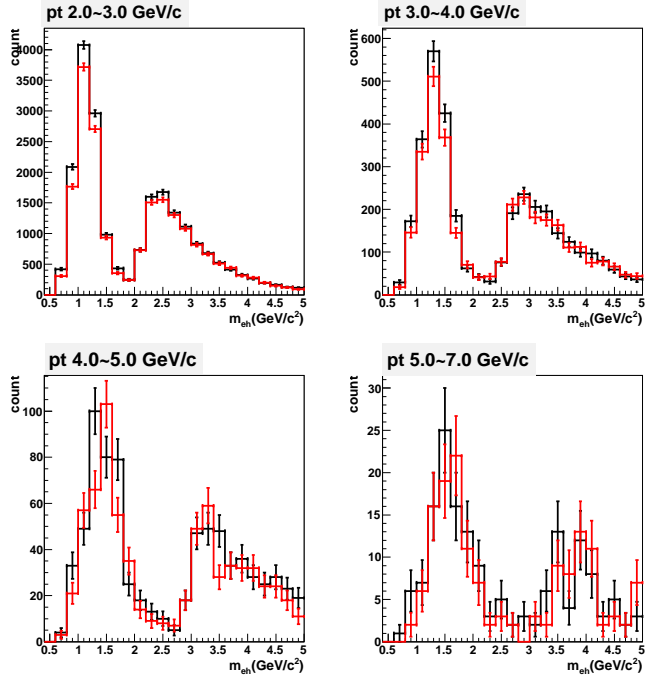}
	    \caption{Invariant mass distribution from 
	      trigger electrons and associated hadrons in RUN5.
	      Black lines are unlike charge sign pairs
	      and red lines are like charge sign pairs.}
	    \label{chap5_fig4}
	  \end{center}
	\end{figure}

\begin{figure}[htb]
	  \begin{center}
	    \includegraphics[width=9cm]{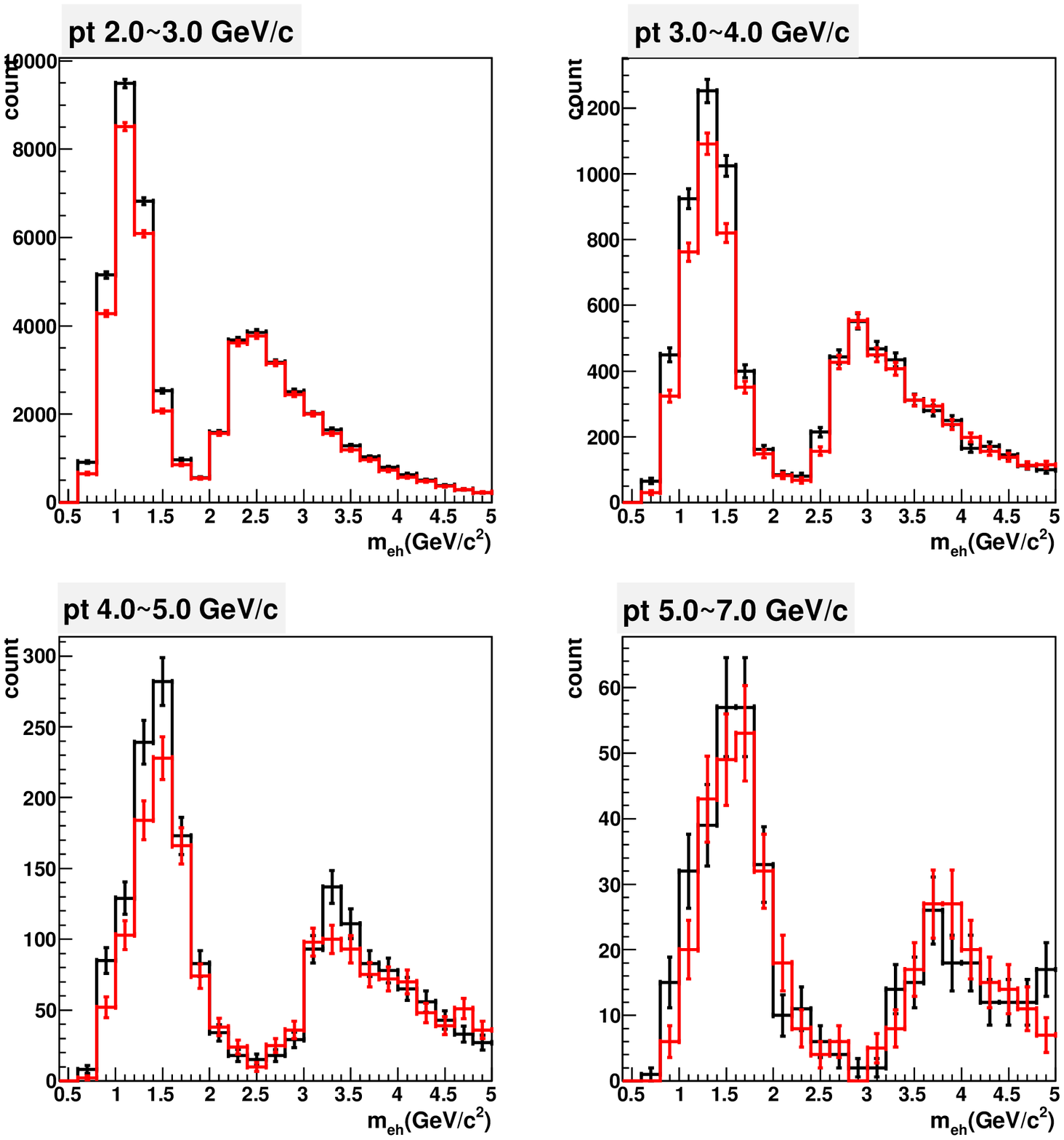}
	    \caption{Invariant mass distribution from 
	      trigger electrons and associated hadrons in RUN6.
	      Black lines are unlike charge sign pairs
	      and red lines are like charge sign pairs.}
	    \label{chap5_fig5}
	  \end{center}
	\end{figure}

	M$_{eh}$ is defined as the invariant mass of particle pairs when the trigger particle is assumed to be 
	electron and the associated particle is assumed to be kaon.
	Figure~\ref{chap5_fig4}  and Figure~\ref{chap5_fig5} show invariant mass 
	distributions~(M$_{eh}$) of unlike and like sign pairs of the trigger electrons and 
	the associated hadrons at each electron $p_{\mathrm{T}}$ range in RUN5 and RUN6, respectively.
	In Fig~.\ref{chap5_fig4} and \ref{chap5_fig5}, black lines are 
	unlike charge sign pairs and red lines are like charge sign pairs.
	Title in each panel shows the trigger electron $p_{\mathrm{T}}$ range.
	Clear excess of unlike sign pairs can be seen. The excess indicates the existence of 
	the $D^0$$\rightarrow$e$^+$ K$^-$ $\nu_e$ signals.
	The distributions of like sign pairs are subtracted from the distributions of
	unlike sign pairs to utilize the effect of semi-leptonic decay 
	of D and B hadrons.
	
\begin{figure}[htb]
	  \begin{center}
	    \includegraphics[width=9cm]{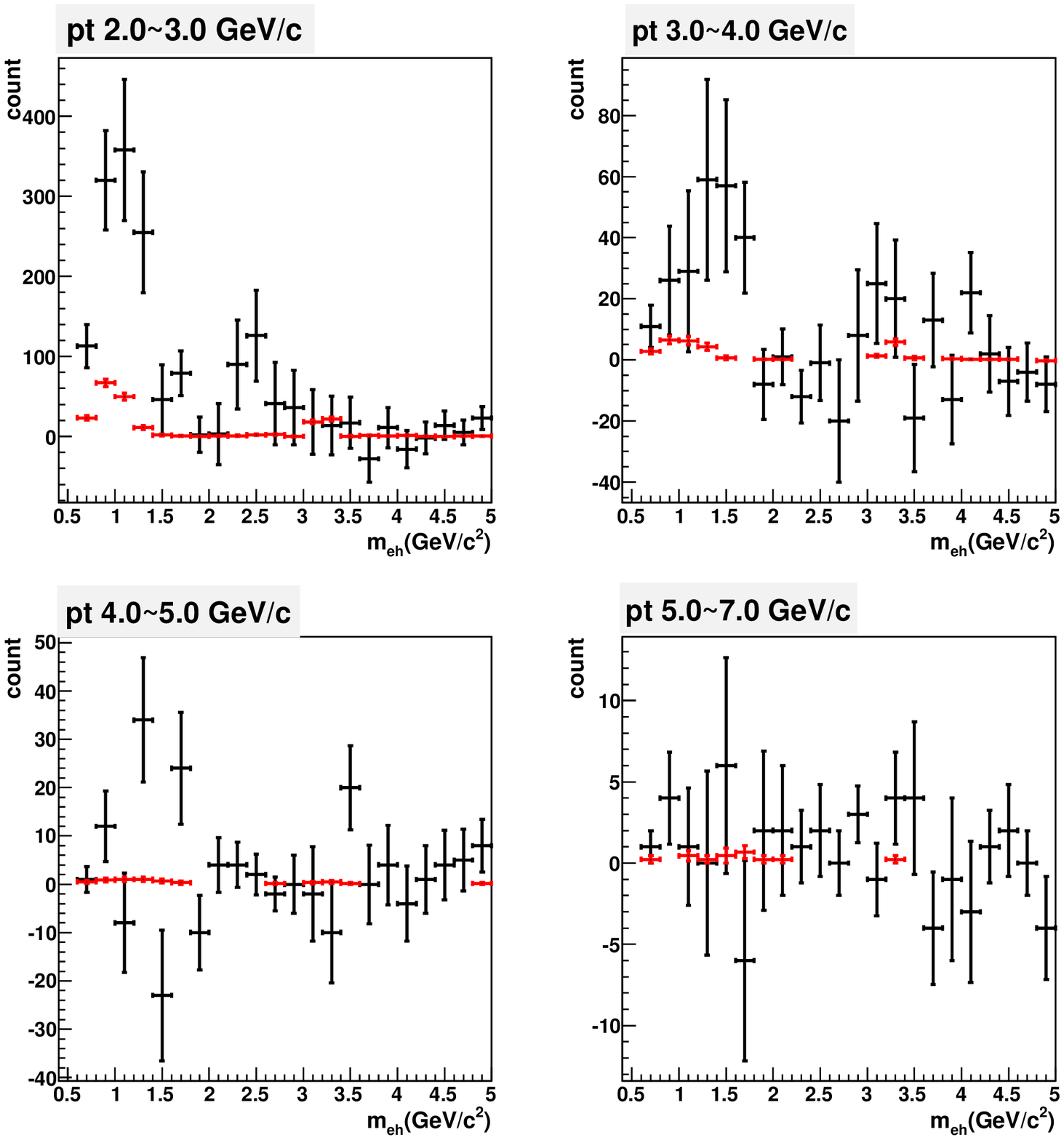}
	    \caption{Subtracted M$_{eh}$ distribution of electron-hadron pairs~(black points)
	      and estimated  M$_{eh}$ distributions of the remaining electron 
	      pairs~(red points) in RUN5.}
	    \label{chap5_fig6}
	  \end{center}
	\end{figure}

\begin{figure}[htb]
	  \begin{center}
	    \includegraphics[width=9cm]{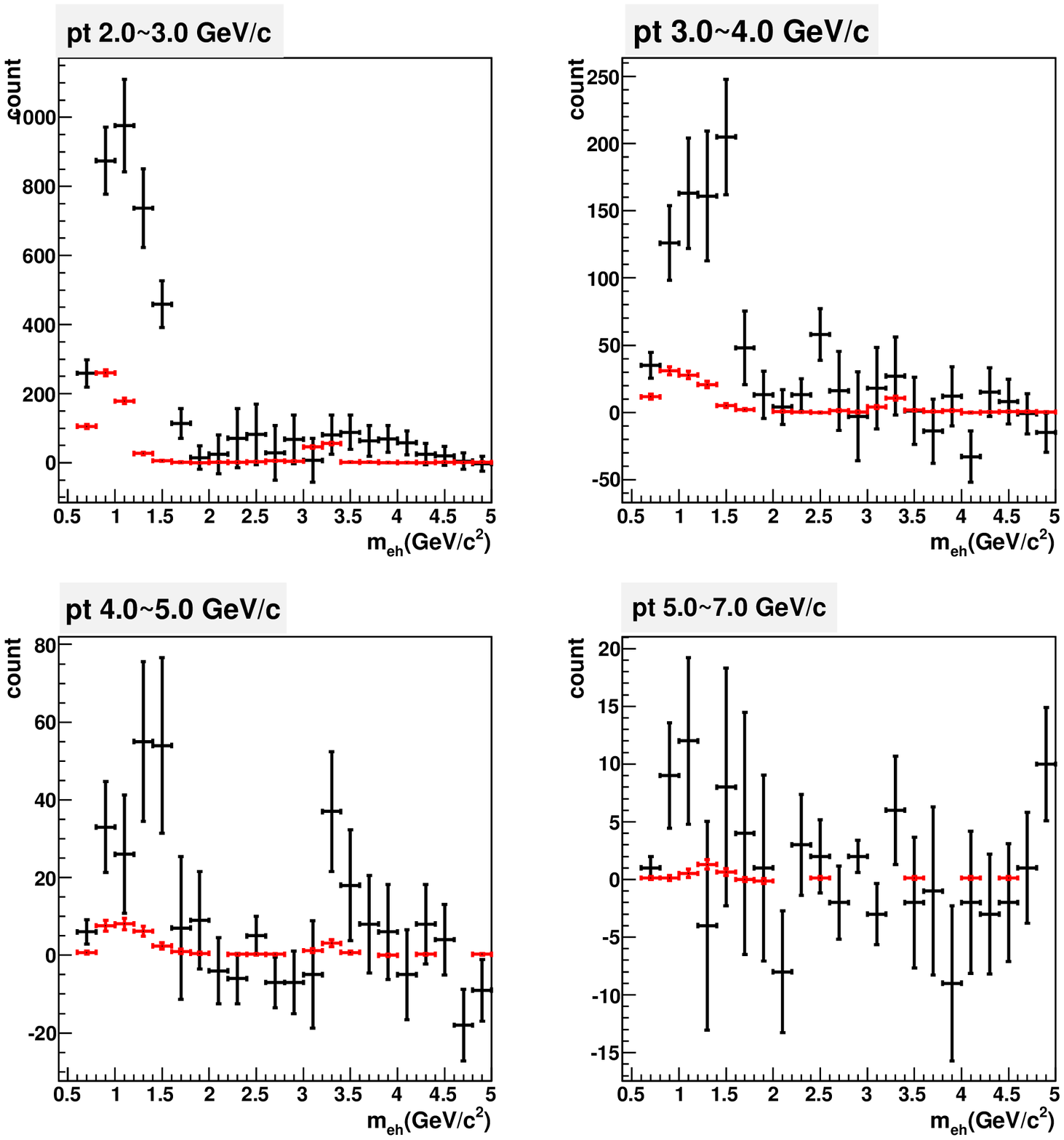}
	    \caption{Subtracted M$_{eh}$ distribution of electron-hadron pairs~(black points)
	      and estimated  M$_{eh}$ distributions of the remaining electron 
	      pairs~(red points) in RUN6.}
	    \label{chap5_fig7}
	  \end{center}
	\end{figure}

	Subtracted invariant mass distributions still include the contribution of the remaining 
	electron pairs which have M$_{ee}$ $>$ 0.08~GeV.
	These remaining electron pairs must be estimated and subtracted to count signals.
	Identified electron pairs are used to estimate the amount of the remaining electron pairs.
	One of the electron pair is the trigger electron  and the other is associated electron 
	with 0.4$< p_{\mathrm{T}} <$ 5.0~GeV/$c$.
	The contribution of the remaining electron pairs is estimated by the normalized M$_{eh}$ 
	distribution of the identified electron pairs in M$_{ee}$ $>$0.08~GeV.
	Normalization of the M$_{eh}$ distribution of the identified electron pairs, where the associated
	electron is assigned as kaon mass, is determined by
	the number of entries in M$_{ee}$ $<$0.08~GeV of the electron and 
	hadron pairs~(the number of entries in the peaks from $\pi^0$ Dalitz and beam pipe conversion).
	Figure~\ref{chap5_fig6} and \ref{chap5_fig7} show the subtracted M$_{eh}$ 
	distributions of electron hadron pairs and the estimated M$_{eh}$ distributions 
	of the remaining electron pairs at each electron $p_{\mathrm{T}}$ range in RUN5 and 
	RUN6, respectively.
	In Fig~.\ref{chap5_fig6} and \ref{chap5_fig7}, black points show
	the subtracted M$_{eh}$ distributions and red points show the estimated M$_{eh}$ 
	distributions of the remaining electron pairs.
	
\begin{figure}[htb]
	  \begin{center}
	    \includegraphics[width=9cm]{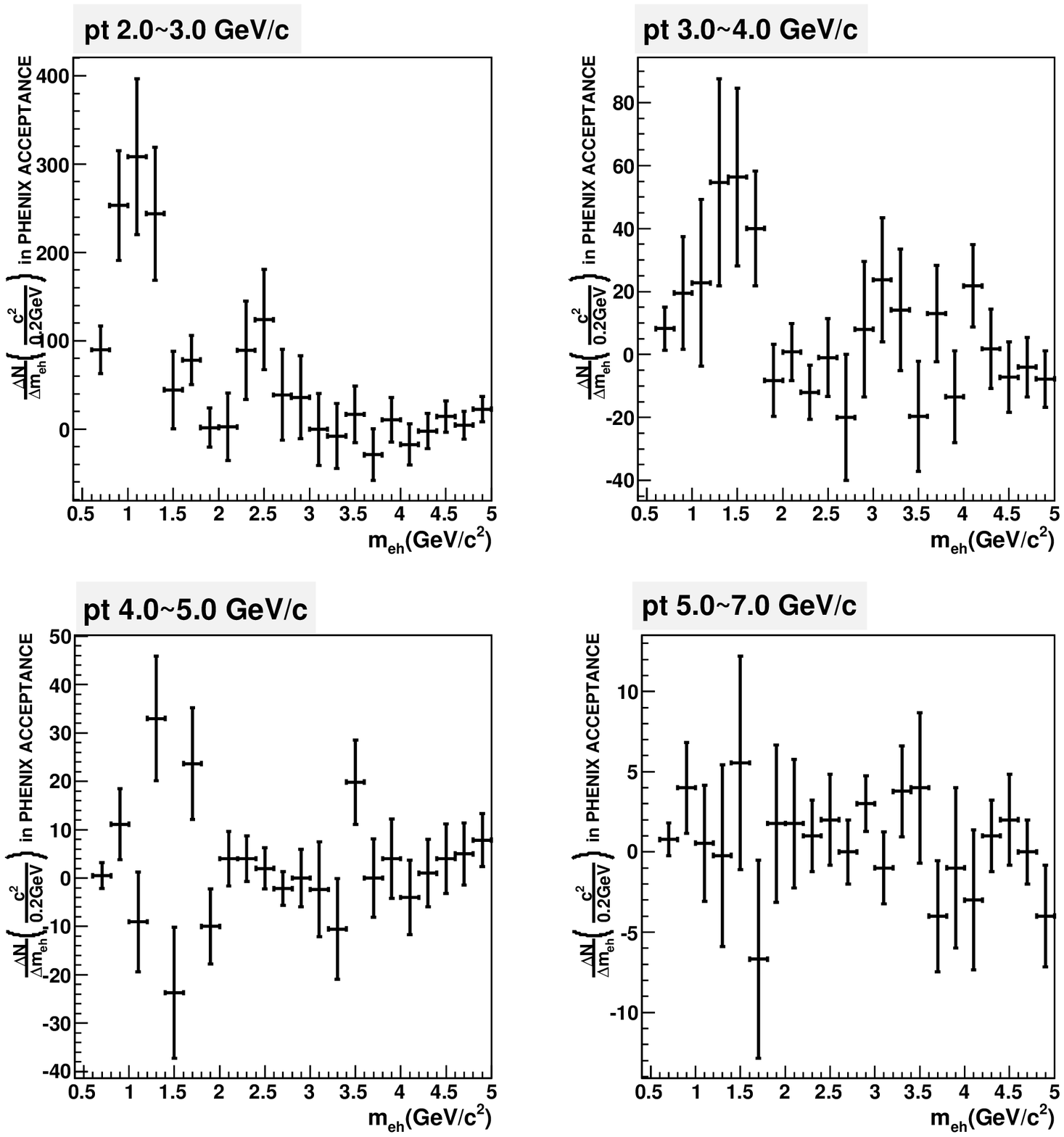}
	    \caption{Subtracted invariant mass distribution of electron-hadron pairs after 
	      subtraction of estimated remaining electron pairs in RUN5.}
	    \label{chap5_fig8}
	  \end{center}
	\end{figure}

\begin{figure}[htb]
	  \begin{center}
	    \includegraphics[width=9cm]{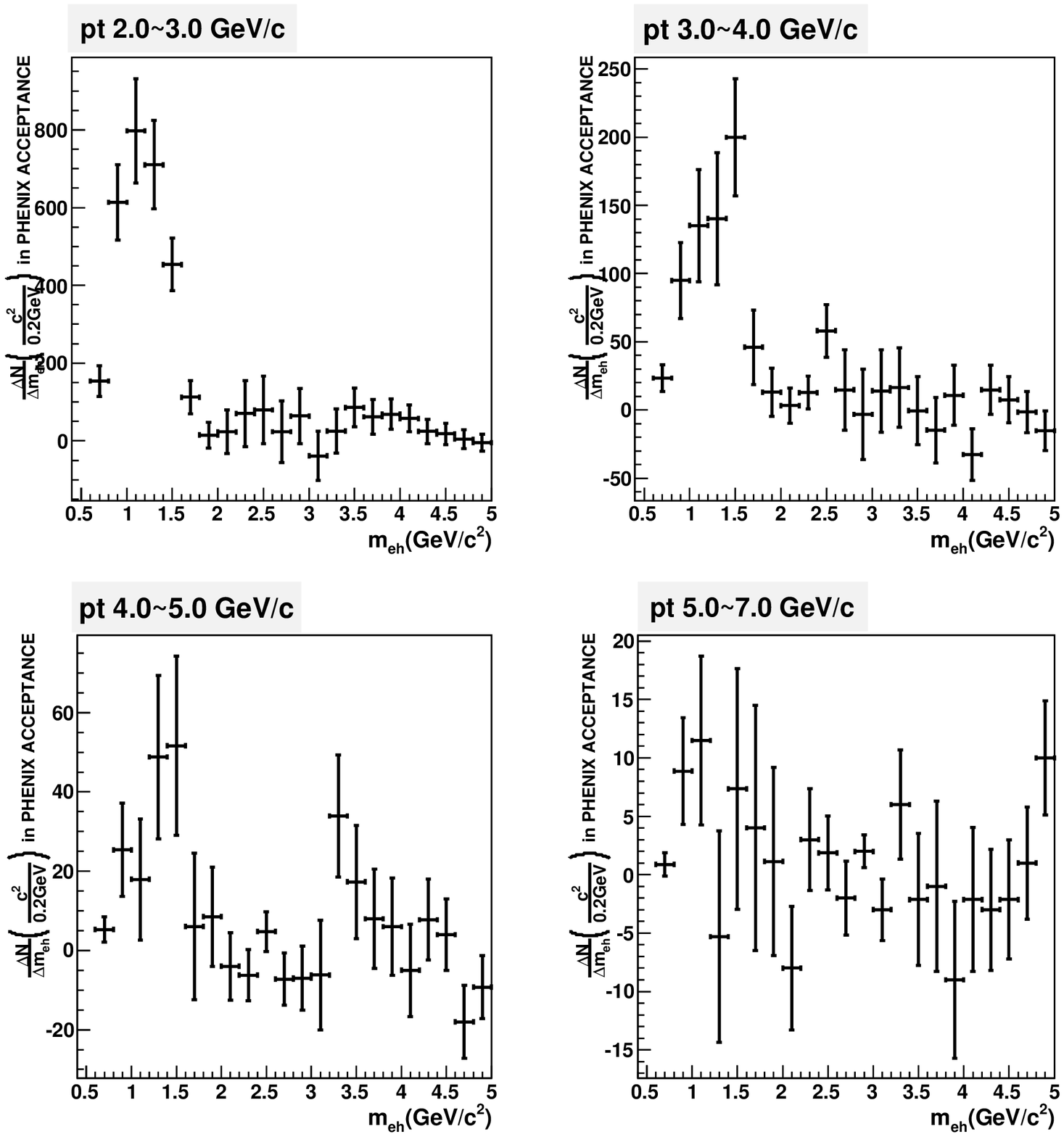}
	    \caption{Subtracted invariant mass distribution of electron-hadron pairs after 
	      subtraction of estimated remaining electron pairs in RUN6.}
	    \label{chap5_fig9}
	  \end{center}
	\end{figure}

	The estimated M$_{eh}$ distributions of the remaining electron pairs are subtracted 
	from the M$_{eh}$ distributions of electron hadron pairs.
	After this subtraction, the M$_{eh}$ distributions are regarded as 
	the extracted signals.
	Figure~\ref{chap5_fig8} and \ref{chap5_fig9} show the extracted reconstruction
	signals in RUN5 and RUN6 respectively. 
	In Fig~.\ref{chap5_fig8} and \ref{chap5_fig9}, numbers of entries in 
	0.4$<$ M$_{eh}$ $<$ 1.9 GeV are counted as N$_{tag}$, since this analysis
	is partial reconstruction of $D$ and it is not necessary to require tight mass cut around
	$D$ region.
	The results of N$_{tag}$ are summarized in Table~\ref{chap5_table1} and Table~\ref{chap5_table2}.

	\subsubsection{Number of Electrons from Heavy Flavor}
\begin{figure}[htb]
	  \begin{center}
	    \includegraphics[width=10cm]{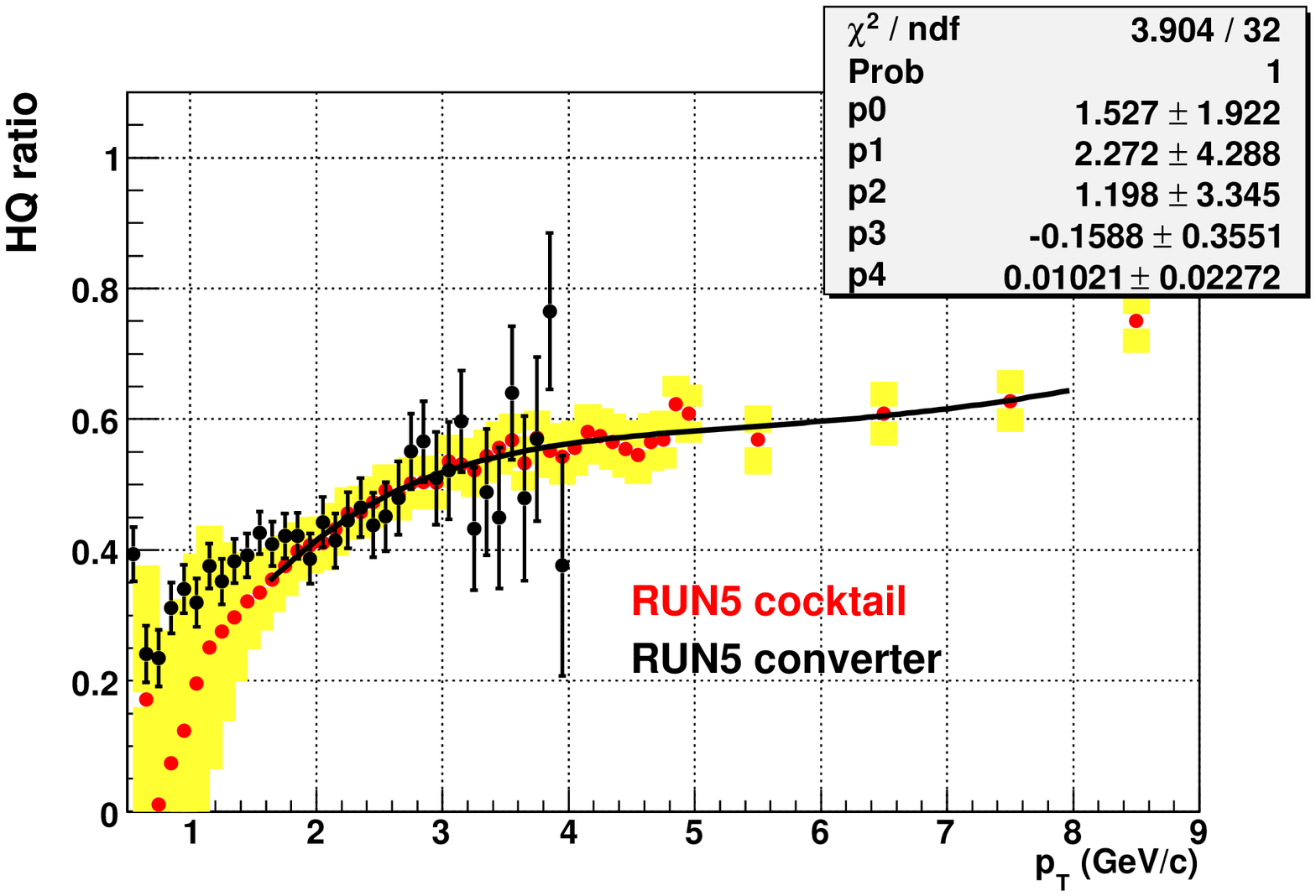}
	    \caption{The fraction of electron from heavy flavor decay
	      in inclusive electrons in RUN5 as a function of electron $p_{\mathrm{T}}$.}
	    \label{chap5_fig10}
	  \end{center}
	\end{figure}

\begin{figure}[htb]
	  \begin{center}
	    \includegraphics[width=10cm]{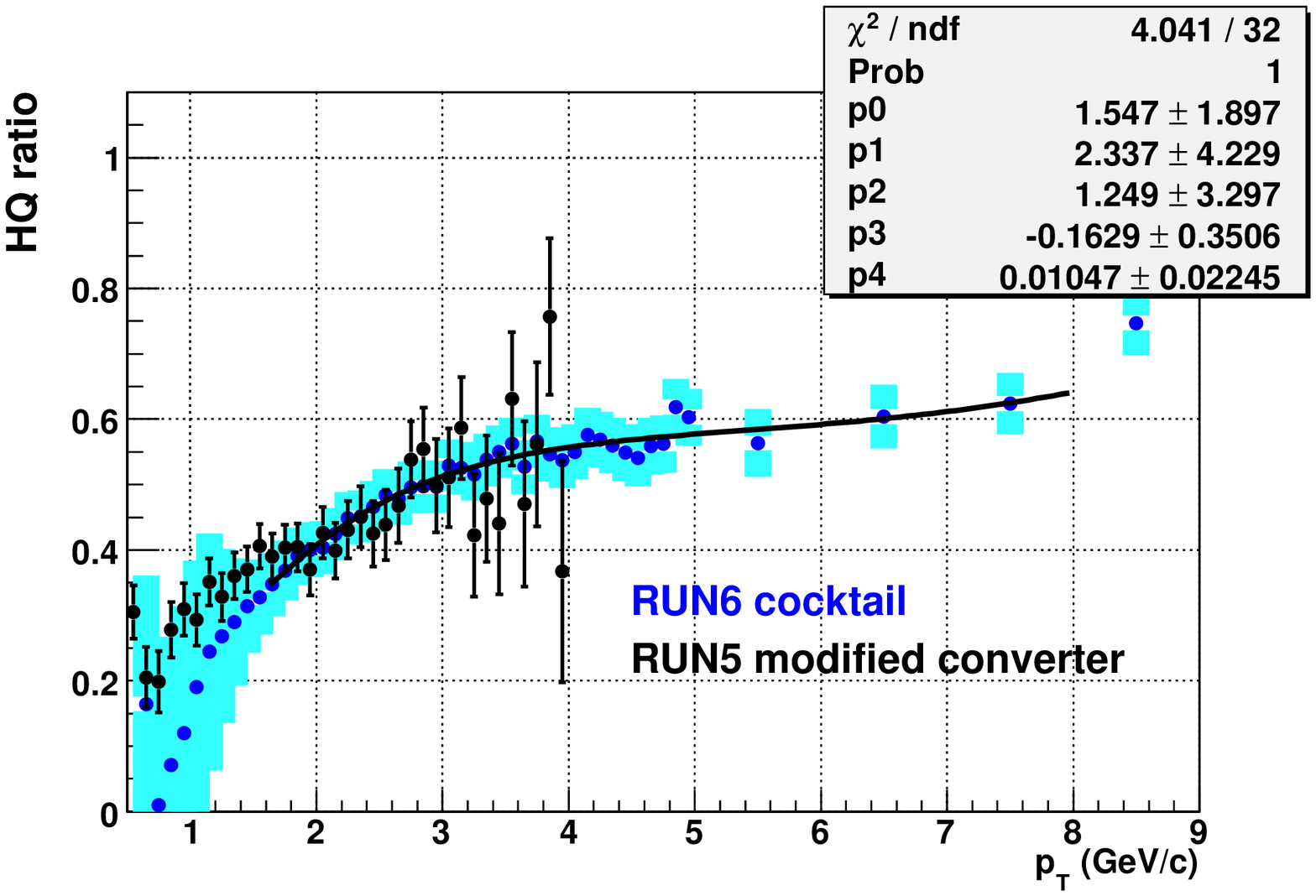}
	    \caption{The fraction of electron from heavy flavor decay
	      in inclusive electrons in RUN6 as a function of electron $p_{\mathrm{T}}$.}
	    \label{chap5_fig11}
	  \end{center}
	\end{figure}

	The number of single non-photonic electrons, $N_{e(HF)}$ in Eq.~\ref{eq:tagdata} 
	is counted according to the following equation.
	\begin{equation}
	N_{e(HF)} = \int dp_{\mathrm{T}} N_e(p_{\mathrm{T}}) \times R_{HF}(p_{\mathrm{T}}),
	\end{equation}
	where,  $N_e(p_{\mathrm{T}})$ is the number of measured electrons and 
	$R_{HF}(p_{\mathrm{T}})$ is the fraction of single non-photonic electrons
	in measured inclusive electrons as a function of electron $p_{\mathrm{T}}$.
	$R_{HF}(p_{\mathrm{T}})$ is determined as the ratio of the spectrum of single non-photonic electrons
	obtained at Section\ref{hq_spe} over the sum of spectrum of single non-photonic electrons 
	and the background electrons in the cocktail.
	Figure~\ref{chap5_fig10} and Figure~\ref{chap5_fig11} show the 
	obtained $R_{HF}(p_{\mathrm{T}})$ in RUN5 and RUN6, respectively.
	The obtained $R_{HF}(p_{\mathrm{T}})$ is fitted, which is shown black line
	in Fig.\ref{chap5_fig10} and \ref{chap5_fig11}.
	The number of single non-photonic electrons is calculated from the fitted line.
	
	\subsection{Systematic Error of $\epsilon_{data}$}
	The following factors are considered.
	\begin{itemize}
	  \item{The subtraction of like sign pairs}
	  \item{The subtraction of the remaining electron pairs}
	  \item{The counting of single non-photonic electrons}
	  \item{Other contributions to N$_{tag}$ background}
	\end{itemize}
	\subsubsection{Subtraction of Like Sign Entries}
	Systematic error associated with the subtraction of like sign entries is determined based on the effect of
	the difference in the phase space with the acceptance filter described 
	in Sec.~\ref{sec:ecutcorr}.
	The effect of the difference in the phase space on the extracted signals~(N$_{tag}$)
	is evaluated by using un-correlated electron hadron pairs.
	Event mixing method is used to create the pairs of un-correlated electrons and hadrons.
	
\begin{figure}[htb]
	  \begin{center}
	    \includegraphics[width=9cm]{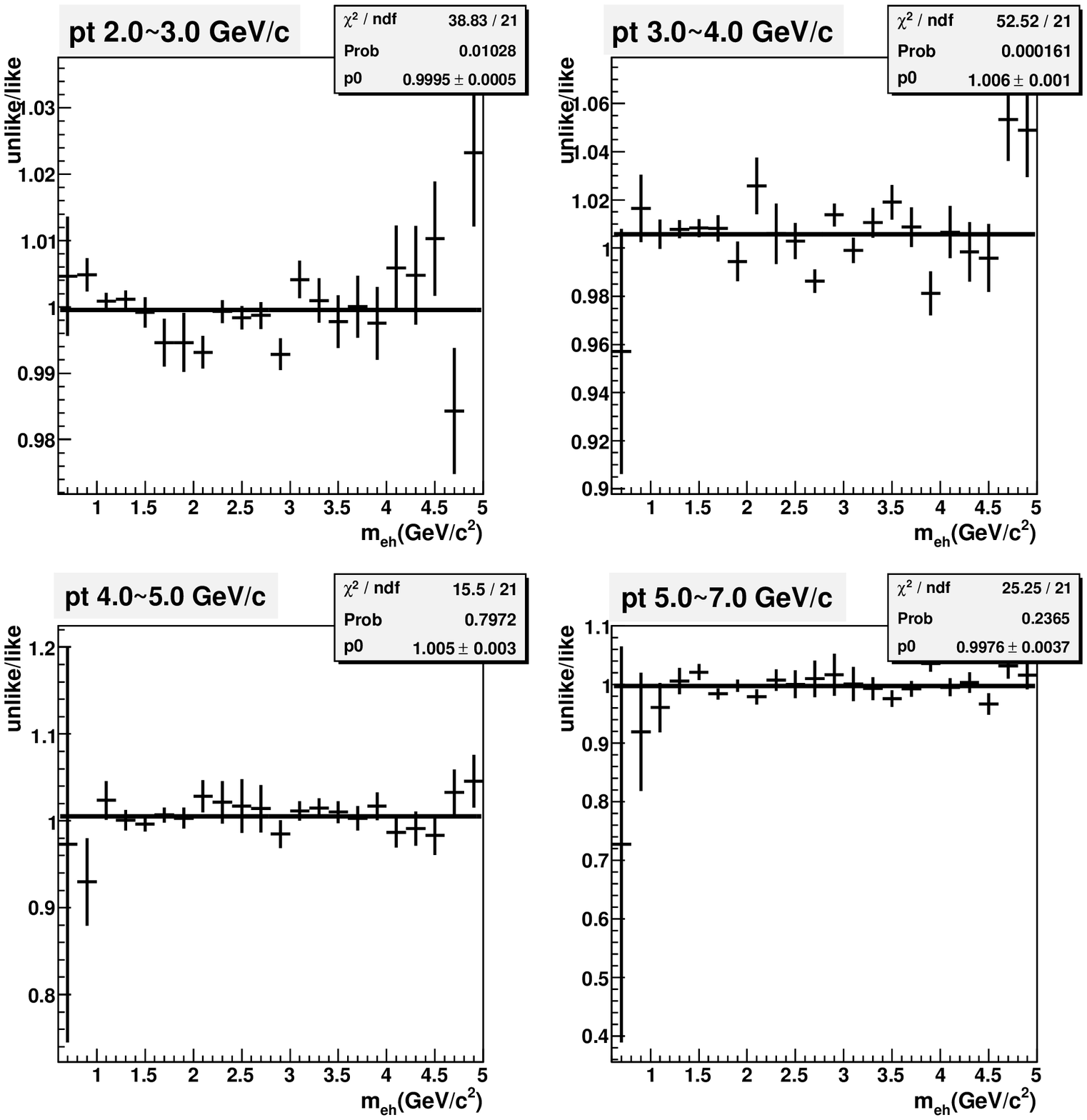}
	    \caption{The ratio of (unlike sign)/(like sign)
	      M$_{eh}$ distribution in mixing events in RUN5.
	    }
	    \label{chap5_fig12}
	  \end{center}
	\end{figure}

	The M$_{eh}$ distributions of the un-correlated like sign pairs are expected to be identical
	as these of the un-correlated unlike sign pairs, if 
	the phase space of negative and positive charged tracks is identical.
	Therefore, the discrepancy between unity and the ratio of (unlike sign)/(like sign)
	M$_{eh}$ distribution in mixing events is used for the estimation of 
	the systematic error.
	Figure~\ref{chap5_fig12} shows the ratio of (unlike sign)/(like sign)
	M$_{eh}$ distribution in mixing events in RUN5.
	The ratios of (unlike sign)/(like sign) M$_{eh}$ distributions 
	are fitted by a constant as shown at Fig.~\ref{chap5_fig12}.
	Result of fit is summarized at Table~\ref{chap5_table12}.
	
	When the fit result is consistent with unity within the fitting error,
	(error of the fit)$\times$(M$_{eh}$ distribution of like sign pairs) is 
	assigned as the systematic error for the subtraction of like sign entries.
	When the fit result is not consistent with unity within fitting error,
	(the deviation of the fitted value from unity) $\times$(M$_{eh}$ distribution)
	is assigned as the systematic error for the subtraction.
	
	\begin{table}[hbt]
	  \begin{center}
	    \caption{Result of fit for (mixing unlike sign)/(mixing like sign) by constant}
	    \label{chap5_table12}
	    \begin{tabular}{c|cc|cc}
	      electron $p_{\mathrm{T}}$ range & mean(RUN5) & error(RUN5) & mean(RUN6)& error(RUN6) \\
	      \hline\hline
	      2.0-3.0~GeV/$c$ & 0.9995 & 0.0005&1.0009&0.0002 \\
	      \hline
	      3.0-4.0~GeV/$c$ & 1.006 & 0.001 &1.002& 0.0004\\
	      \hline
	      4.0-5.0~GeV/$c$ & 1.005 & 0.003 &1.002& 0.001\\
	      \hline
	      5.0-7.0~GeV/$c$ & 0.998 & 0.004& 1.002&0.001\\
	    \end{tabular}
	  \end{center}
	\end{table}
	\subsubsection{Subtraction of Remaining Electron Pairs}
	The systematic error for the subtraction of the remaining electron pairs
	is evaluated by the error of the normalization factors for the
	M$_{ee}$ distribution of identified electron pairs.
	The error of the normalization factors is determined by the statistical
	uncertainty of the numbers of entries  of identified electron pairs and 
	electron hadron pairs in M$_{ee}$ $<$0.08~GeV.
	The uncertainty of normalization is assigned as 
	the systematic error for subtraction of the remaining electron pairs.
	\subsubsection{Count of Electrons from Heavy Flavor }
	This uncertainty is the largest source of the systematic error of N$_{tag}$.
	Systematic error for the number of electrons from heavy flavor is calculated 
	based on the systematic error of spectra of electrons from heavy flavor decay.
	The systematic error of spectra are shown in Fig.\ref{chap5_fig10} and \ref{chap5_fig11}.
	\subsubsection{Other Contributions to N$_{tag}$ Background}
	Following sources are possible to make correlation of electrons and hadrons.
	\begin{itemize}
	\item{{\bf K$_{e3}$ decay}: \\
	  Since K$_L$ $\rightarrow e^{\pm} \pi^{\mp}$ is weak decay, the subtraction of 
	  like sign entries can not cancel out this contribution.
	  Therefore, K$_L$ $\rightarrow e^{\pm} \pi^{\mp}$ is possible to be background source 
	  of charge correlation of electrons and hadrons. 
	  This contribution is estimated by the PISA simulation which is used at the
	  cocktail calculation.
	  It is found the contribution of K$_{e3}$ decay to N$_{tag}$ is 0.5\% level.
	  Therefore, this  contribution can be neglected.
	}
	\item{{\bf Hadron-hadron correlation}: \\
	  The charge correlation of hadron hadron pairs becomes the background source of
	  charge correlation of electron and hadron pairs, since there is small hadron 
	  contamination in trigger electrons .
	  
	  The amount of hadron contamination in the trigger electrons is less than 0.5\% 
	  at $2.0<p_{\mathrm{T}} <5.0$ GeV/$c$, which is estimated in Sec.~\ref{sec:hadc}.
	  The tagging efficiency of hadron hadron pairs correlation~($\epsilon_{had}$) is
	  determined from real data analysis. 
	  As a result, the contribution of hadron hadron correlation to N$_{tag}$ 
	  is 0.5\% level at $2.0<p_{\mathrm{T}} <5.0$ GeV/$c$.
	  This  contribution  can be also neglected at $2.0<p_{\mathrm{T}} <5.0$ GeV/$c$.
	  
	  Hadron background is not negligible at high $p_{\mathrm{T}}$~($>$5.0~GeV/$c$) as estimated in 
	  Sec.~\ref{sec:hadc}, while tight eID cut is applied.
	  The number of hadron contamination is calculated according to 
	  Table~\ref{tab:hadcont}.
	  The tagging efficiency of hadron hadron pairs~($\epsilon_{had}$) 
	  at  high $p_{\mathrm{T}}$ is also determined from real data analysis. 
	  The amount of hadron contamination is calculated from the estimated number of 
	  hadron contamination in the trigger electrons and the tagging efficiency of hadron hadron pairs.
	  This contribution is subtracted from extracted signals~(N$_{tag}$) at 
	  high $p_{\mathrm{T}}$~($>$5.0~GeV/$c$) region.
	  Tagging efficiency of hadrons depends on the distribution of 
	  hadron $p_{\mathrm{T}}$ .
	  The $p_{\mathrm{T}}$ distribution of hadron with above cuts may differ
	  from that of hadron background in electrons with tight eID cut.
	  50\% systematic error is assiged for this subtraction.\\
	}
	\end{itemize}

	\subsection{Results of $\epsilon_{data}$}
	$\epsilon_{data}$ is calculated from N$_{tag}$ and the number of electrons 
	from heavy flavor decay.
	$\epsilon_{data}$ and used values are summarized in Table~\ref{chap5_table1} and 
	\ref{chap5_table2}.
	\begin{table}[hbt]
	  \begin{center}
	    \caption{$\epsilon_{data}$ and used values at each electron $p_{\mathrm{T}}$ range(RUN5) }
	    \label{chap5_table1}
	    \begin{tabular}{|c|c|}
	      \hline
	      electron $p_{\mathrm{T}}$ 2.0-3.0~GeV/$c$ & \\
	      number of unlike sign entries & 11050. \\
	      number of like sign entries   & 9872. \\
	      number of (unlike -like)   & 1178.0 $\pm$ 144.6 $\pm$ 5.0\\
	      remaining e-e pair &  153.3 $\pm$ 7.4 $\pm$ 6.1 \\
	      N$_{tag}$ &1024.7 $\pm$ 144.8 $\pm$ 7.9 \\
	      number of heavy flavor electron & 31402.2 $\pm$ 262.5 $\pm$ 2783. \\
	      $\epsilon_{data}$ & 0.0326 $\pm$ 0.0046 $\pm$ 0.0029 \\
	      \hline \hline
	      electron $p_{\mathrm{T}}$ 3.0-4.0~GeV/$c$ & \\
	      number of unlike sign entries & 1770. \\
	      number of like sign entries   & 1548. \\
	      number of (unlike -like) & 222.0 $\pm$ 57.6 $\pm$ 9.3\\
	      remaining e-e pair & 20.7 $\pm$ 2.5 $\pm$ 2.5 \\
	      N$_{tag}$ &  201.3 $\pm$ 57.7 $\pm$ 9.6 \\
	      number of heavy flavor electron & 5310.1 $\pm$ 99.4$\pm$ 402.5 \\
	      $\epsilon_{data}$ & 0.0379 $\pm$ 0.0109 $\pm$ 0.0034 \\
	      \hline \hline
	      electron $p_{\mathrm{T}}$ 4.0-5.0~GeV/$c$ & \\
	      number of unlike sign entries & 353. \\
	      number of like sign entries   &323. \\
	      number of (unlike -like) &  30.0 $\pm$ 26.00 $\pm$1.6 \\
	      remaining e-e pair & 4.5 $\pm$ 0.9 $\pm$ 1.2 \\
	      N$_{tag}$ & 25.5 $\pm$ 26.0 $\pm$ 2.0 \\
	      number of heavy flavor electron &  1181.9 $\pm$ 45.5 $\pm$  89.2 \\
	      $\epsilon_{data}$ & 0.0216 $\pm$ 0.0220  $\pm$ 0.0023 \\
	      \hline \hline
	      electron $p_{\mathrm{T}}$ 5.0-7.0~GeV/$c$ & \\
	      number of unlike sign entries & 78. \\
	      number of like sign entries   & 71. \\
	      number of (unlike -like) &  7. $\pm$ 12.2  $\pm$0.3 \\
	      remaining e-e pair &  2.0$\pm$ 0.7 $\pm$ 1.1 \\
	      number of background hadron &  17.8 $\pm$ 3.7(sys)  \\
	      signal from hadron &  1.5 $\pm$ 0.8(sys)  \\
	      N$_{tag}$ & 3.5 $\pm$ 12.2 $\pm$ 1.4 \\
	      number of heavy flavor electron &  269.9 $\pm$ 21.8 $\pm$  23.5\\
	      $\epsilon_{data}$ & 0.0131 $\pm$ 0.0457  $\pm$ 0.0052 \\
	      \hline
	    \end{tabular}
	  \end{center}
	\end{table}

	\begin{table}[hbt]
	  \begin{center}
	    \caption{$\epsilon_{data}$ and used values at each electron $p_{\mathrm{T}}$ range(RUN6) }
	    \label{chap5_table2}
	    \begin{tabular}{|c|c|}
	      \hline
	      electron $p_{\mathrm{T}}$ 2.0-3.0~GeV/$c$ & \\
	      number of unlike sign entries & 26066. \\
	      number of like sign entries   & 22630. \\
	      number of (unlike -like)   & 3436. $\pm$220.7  $\pm$ 5.0\\
	      remaining e-e pair &  578.4 $\pm$ 13.6 $\pm$ 15.5 \\
	      N$_{tag}$ &2857.6 $\pm$ 221.1 $\pm$ 16.1 \\
	      number of heavy flavor electron & 76408. $\pm$ 412.5 $\pm$  6763.9\\
	      $\epsilon_{data}$ & 0.0374 $\pm$ 0.0029 $\pm$ 0.0033 \\
	      \hline \hline
	      electron $p_{\mathrm{T}}$ 3.0-4.0~GeV/$c$ & \\
	      number of unlike sign entries &  4191.\\
	      number of like sign entries   & 3447. \\
	      number of (unlike -like) & 744.0 $\pm$ 87.4 $\pm$ 7.0\\
	      remaining e-e pair &  98.3 $\pm$ 5.5 $\pm$ 7.2 \\
	      N$_{tag}$ &  645.7 $\pm$ 87.6 $\pm$ 7.2 \\
	      number of heavy flavor electron & 12897.0 $\pm$ 155.7 $\pm$ 977.4 \\
	      $\epsilon_{data}$ & 0.0501 $\pm$ 0.0068 $\pm$ 0.0039 \\
	      \hline \hline
	      electron $p_{\mathrm{T}}$ 4.0-5.0~GeV/$c$ & \\
	      number of unlike sign entries & 951. \\
	      number of like sign entries   & 774. \\
	      number of (unlike -like) &  177.0 $\pm$ 41.5 $\pm$0.5 \\
	      remaining e-e pair & 26.2 $\pm$ 2.6 $\pm$ 4.5 \\
	      N$_{tag}$ & 150.8 $\pm$ 41.9 $\pm$ 4.7 \\
	      number of heavy flavor electron &  2933.0 $\pm$ 72.0 $\pm$  222. \\
	      $\epsilon_{data}$ & 0.0514 $\pm$ 0.0142  $\pm$ 0.0042 \\
	      \hline \hline
	      electron $p_{\mathrm{T}}$ 5.0-7.0~GeV/$c$ & \\
	      number of unlike sign entries & 216.00 \\
	      number of like sign entries   & 183.0 \\
	      number of (unlike -like) &   33.0 $\pm$ 20.0  $\pm$0.4 \\
	      remaining e-e pair &  2.7$\pm$ 0.7 $\pm$ 1.1 \\
	      number of background hadron &  51.5 $\pm$ 10(sys)  \\
	      signal from hadron &  4.5 $\pm$ 2.3(sys)  \\
	      N$_{tag}$ & 25.8 $\pm$ 20.0 $\pm$ 2.6 \\
	      number of heavy flavor electron &  638.4 $\pm$ 33.6 $\pm$ 54.6 \\
	      $\epsilon_{data}$ & 0.0404 $\pm$ 0.0314  $\pm$ 0.0057 \\
	      \hline
	    \end{tabular}
	  \end{center}
	\end{table}
	
	\clearpage

	\section{Simulation Study for Correlation Analysis} \label{sec:cor_sim}
	This  section describes the evaluation of $\epsilon_{c}$ and $\epsilon_{b}$.
	$\epsilon_{c}$ and $\epsilon_{b}$ are determined by using Monte-Carlo event generator
	as outlined in Sec.~\ref{sec:overview}.
	\subsection{Simulation Overview}
\begin{figure}[htb]
	  \begin{center}
	    \includegraphics[width=16cm]{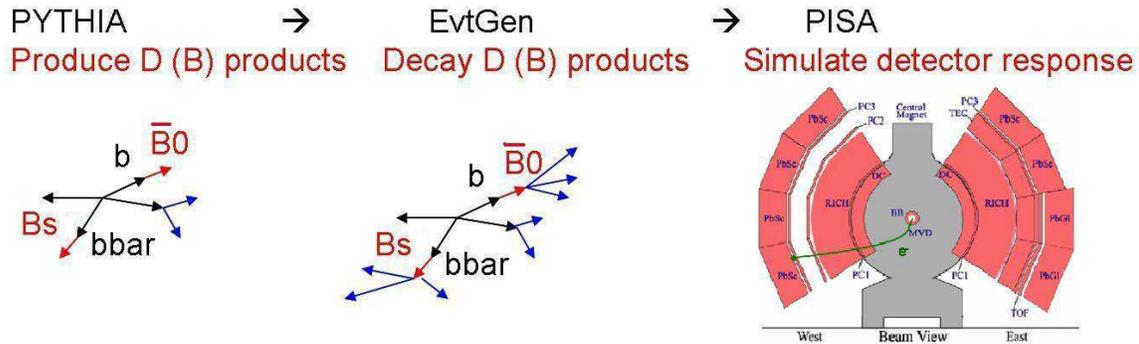}
	    \caption{A conceptual view of the simulation study}
	      \label{fig:simcon}
	  \end{center}
	\end{figure}

	Figure~\ref{fig:simcon} shows a conceptual view of the simulation study.
	The simulation is performed in three steps.
	First, $p+p$ collision at 200~GeV in the center of mass system is generated by PYTHIA
	event generator~\cite{bib:pythia1,bib:pythia2}.
	As a next step, the decay of D and B hadrons in the generated event is 
	simulated by using EvtGen event generator~\cite{bib:evt,bib:evt2,bib:evt3}.
	Therefore, the event which contains D and B hadrons  is generated by the 
	combination of PYTHIA and EvtGen.
	Finally, all stable particles in the generated event are put into the PISA simulation
	to evaluate the detector response.
	\subsubsection{PYTHIA Simulation}
	PYTHIA simulation~(version 6.403) is used to generate $p+p$ collision 
	at 200~GeV in the center of mass system.
	PYTHIA parameters are tuned to reproduce previous results of heavy flavor 
	production measured by PHENIX~\cite{bib:hqtune,bib:hq3,bib:hq5,bib:hq2} and jet production measured by 
	CDF~\cite{bib:cdf1,bib:cdf2}.
	Since $\epsilon_{c}$ and $\epsilon_{b}$ contain inclusive signals from 
	various heavy flavored hadons, the production ratios of D or B mesons 
	and baryons~($D^+/D^0$,$B^+/B^0$ etc) are most important parameters to 
	determine $\epsilon_{c}$ and $\epsilon_{b}$.
	Therefore, the production ratios are also tuned according to 
	the experimental results~\cite{bib:cratio1,bib:cratio2,bib:cratio3}.
	Tuning parameters of PYTHIA are summarized at Table~\ref{tab:parpythia}.
	Tuning status of PYTHIA is described in Appendix.~\ref{sec:c_hq}
	\begin{table}[hbt]
	  \begin{center}
	    \caption{PYTHIA tuning parameters}
	    \label{tab:parpythia}
	    \begin{tabular}{|c|c|}
	      \hline
	      parameter name  & value\\
	      \hline\hline
	      charm mass & 1.25~GeV\\
	      \hline
	      bottom mass & 4.3~GeV\\
	      \hline
	      $k_{\mathrm{T}}$ & 1.5~GeV/$c$\\
	      \hline
	      PDF    & CTEQ5L \\
	      \hline
	      PARJ(13) & 0.55\\
	      (charm production) & \\
	      \hline
	      PARJ(2) & 0.36\\
	      (charm production) & \\
	      \hline
	      PARJ(2) & 0.44\\
	      (bottom production) & \\
	      \hline
	      MSTP(82) &  4\\
	      \hline
	      PARP(81) &  1.9\\
	      \hline
	      PARP(82) &  2.0\\
	      \hline
	      PARP(83) &  0.5\\
	      \hline
	      PARP(84) &  0.4\\
	      \hline
	      PARP(85) &  0.9\\
	      \hline       
	      PARP(86) &  0.95\\
	      \hline
	      PARP(89) &  1800\\
	      \hline
	      PARP(90) &  0.25\\
	      \hline
	      PARP(67) &  4.0\\
	      \hline
	    \end{tabular}
	  \end{center}
	\end{table} 

	\subsubsection{EvtGen Simulation}
\begin{figure}[htb]
	\begin{center}
	  \includegraphics[width=11cm]{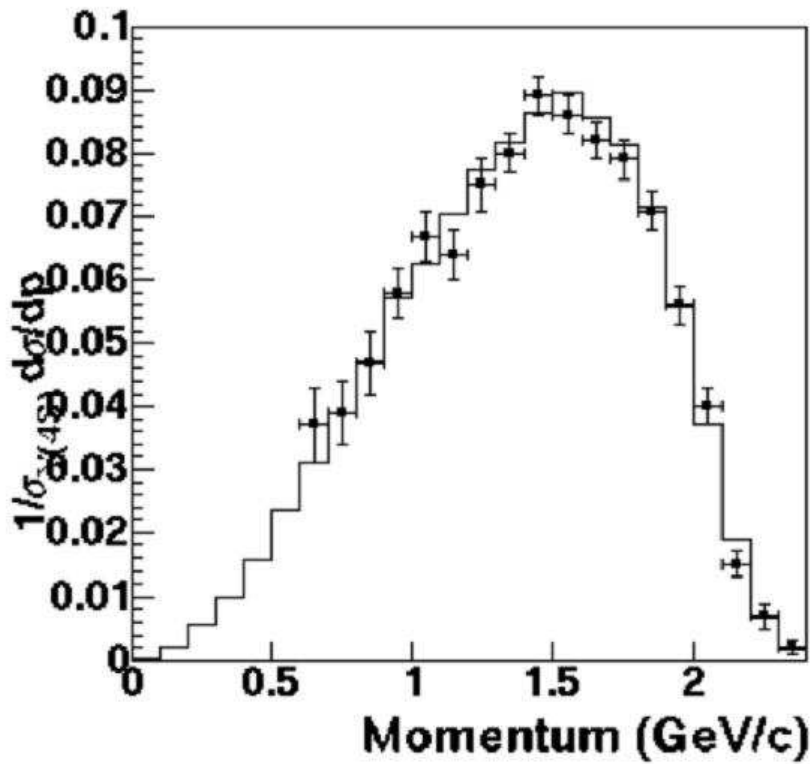}
	  \caption{Electron energy spectrum at $B \rightarrow e \nu X$
	    of EvtGen and CLEO~\cite{bib:cleo1}.
	  }
	  \label{fig:exaevt}
	\end{center}
	\end{figure}

	EvtGen~(version alpha-00-14-05) is used to simulate the decay of D and B 
	hadrons~\cite{bib:evt2,bib:evt3}.
	EvtGen simulation provides a framework for implementation of the decay process
	of D and B hadrons.
	EvtGen simulation is tuned to reproduce the results of heavy flavor decay
	at CLEO, BaBar and  
	Belle~\cite{bib:cleo1,bib:cleo2,bib:cleo3,bib:cleo4,bib:cleo5,bib:belle1}.
	Semi-leptonic decay of D and B hadrons is main interest in this analysis.
	Most of semi-leptonic decay is simulated based on the ISGW2 model 
	in EvtGen~\cite{bib:isgw1,bib:isgw2}.
	For example, Figure~\ref{fig:exaevt} shows the electron energy spectrum 
	of inclusive semi-leptonic decay of B meson~($B \rightarrow e \nu X$) 
	in EvtGen simulation and that in CLEO data~\cite{bib:cleo1}.

	\subsubsection{PISA Simulation}
	All stable particles in the generated event are put into the PISA simulation
	to evaluate the detector response.
	The PISA simulation is tuned for the RUN5 and RUN6 detector response
	as described in Sec.~\ref{sec:ecut}.
	
	\subsection{Calculation of $\epsilon_{c}$ and $\epsilon_{b}$}
	$\epsilon_{c}$ and $\epsilon_{b}$ are determined via the simulation 
	as outlined in the previous subsection.
	PYTHIA with MSEL of 4 and 5 are used to produce charm and bottom and to 
	determine $\epsilon_{c}$ and $\epsilon_{b}$,
	respectively.
	For the calculation of $\epsilon_{c}$ and $\epsilon_{b}$, electron hadron pairs in the simulation 
	are processed  in a similar way to evaluate $\epsilon_{data}$ in the real data analysis.
	In the simulation, the rejection of the estimated remaining electron pairs
	and the calculation of the number of single non-photonic electrons are not performed, 
	since we can reject background for the trigger electron by looking the parent particle 
	in the simulation.
	Only the subtraction of the M$_{eh}$ distribution of like sign pairs is
	performed to extract signals from heavy flavor.
	As a next step, the M$_{eh}$ distribution is normalized by the number of 
	trigger electrons.
	
\begin{figure}[htb]
	  \begin{tabular}{c c}
	    \begin{minipage}{\minitwocolumn}
	      \begin{center}
		\includegraphics[width=7.5cm]{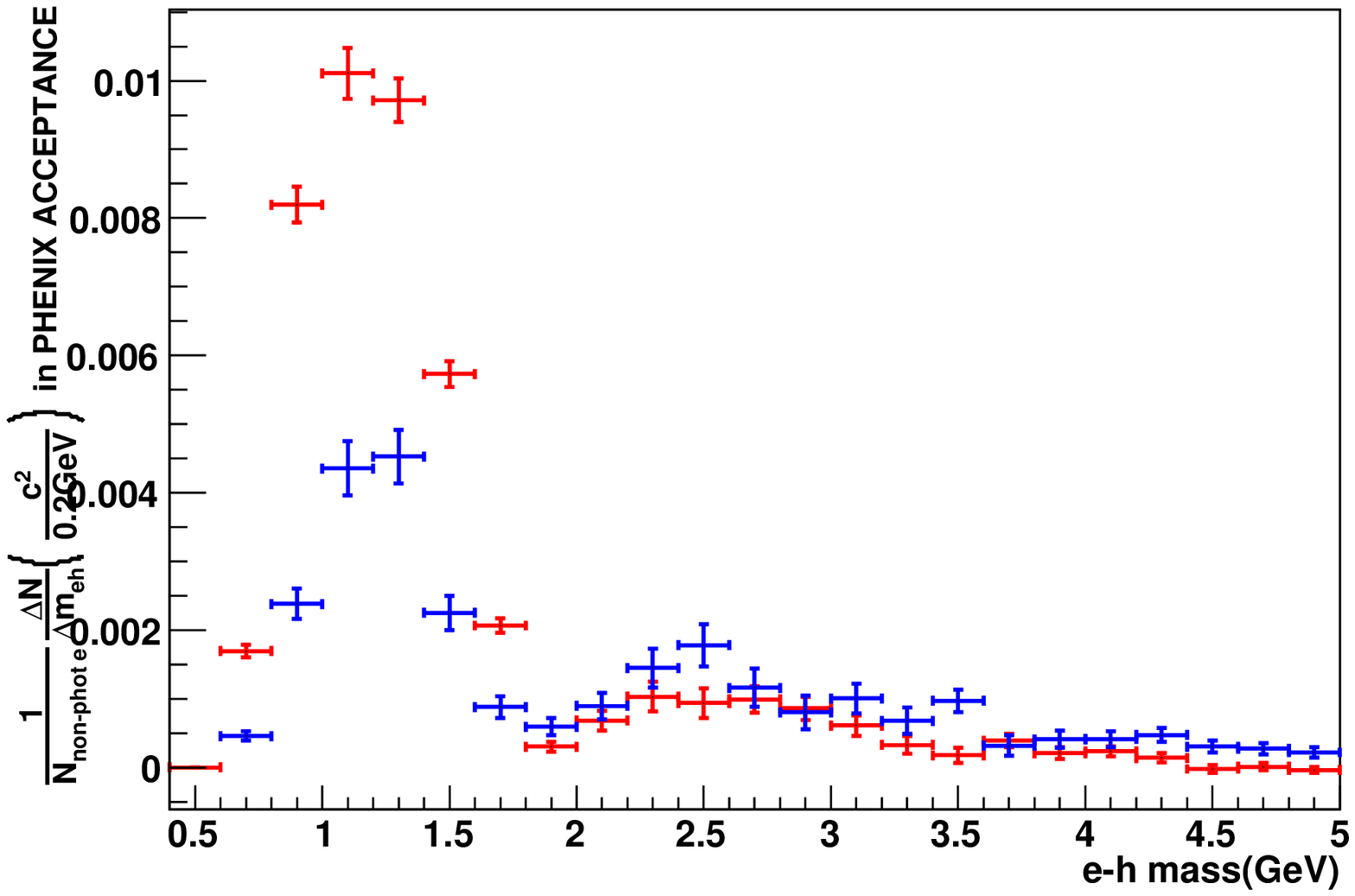}
		\caption{Subtracted and normalized invariant 
		  mass distributions of electron-hadron pairs in charm and bottom production.
		  Red points show charm case and blue points show bottom case.
		  $p_{\mathrm{T}}$ range of trigger electrons is 2.0-3.0~GeV/$c$.
		}
		\label{chap6_fig7}
	      \end{center}
	    \end{minipage}
	    &
	    \begin{minipage}{\minitwocolumn}
	      \begin{center}
		\includegraphics[width=7.5cm]{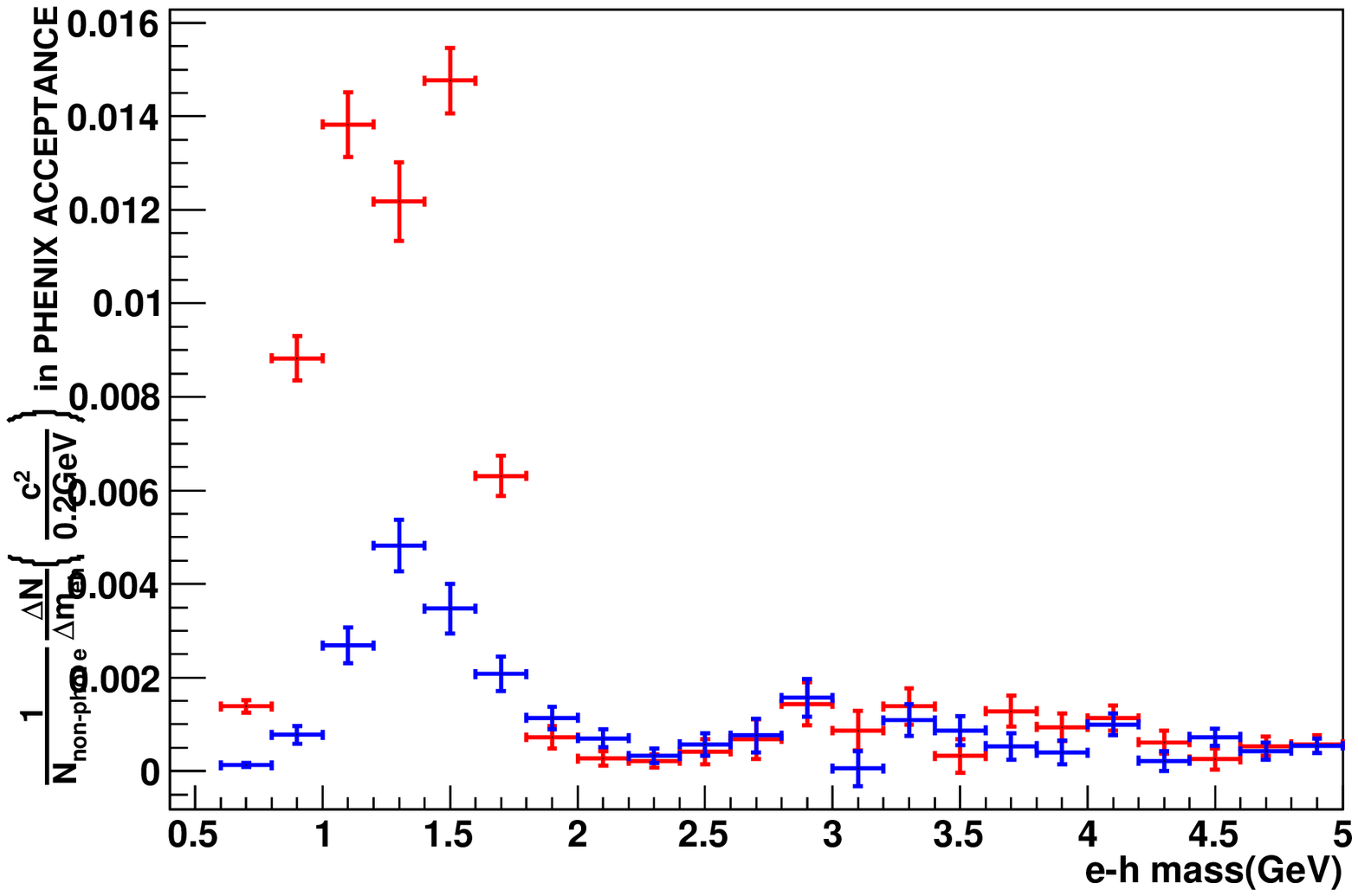}
		\caption{Subtracted and normalized invariant 
		  mass distributions of electron-hadron pairs in charm and bottom production.
		  $p_{\mathrm{T}}$ range of trigger electrons is 3.0-4.0~GeV/$c$.
		}
		\label{chap6_fig8}
	      \end{center}
	    \end{minipage}
	  \end{tabular}
	  \begin{tabular}{c c}
	    \begin{minipage}{\minitwocolumn}
	      \begin{center}
		\includegraphics[width=7.5cm]{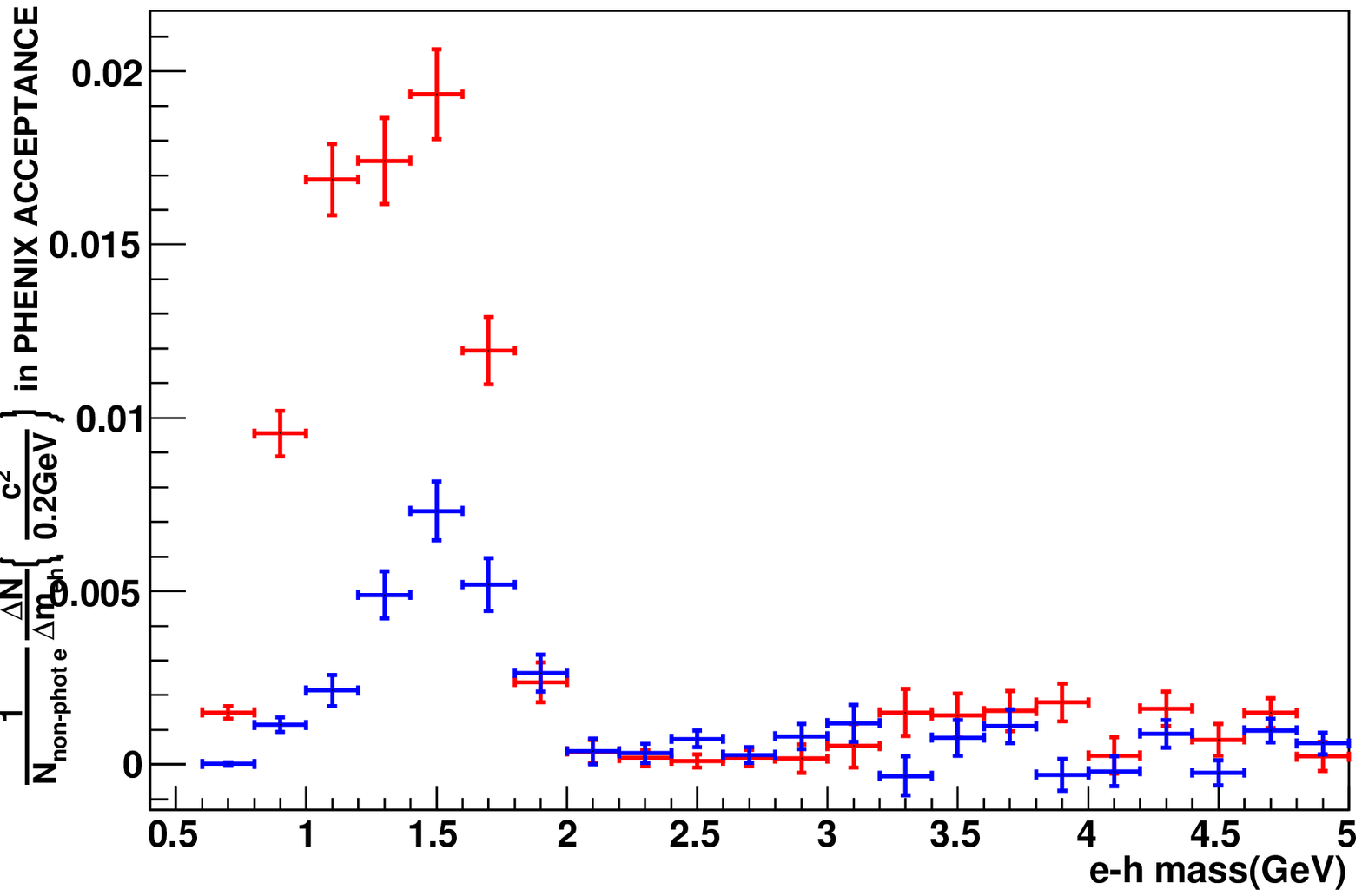}
		\caption{Subtracted and normalized invariant 
		  mass distributions of electron-hadron pairs in charm and bottom production.
		  $p_{\mathrm{T}}$ range of trigger electrons is 4.0-5.0~GeV/$c$.
		}
		\label{chap6_fig9}
	      \end{center}
	    \end{minipage}
	    &
	    \begin{minipage}{\minitwocolumn}
	      \begin{center}
		\includegraphics[width=7.5cm]{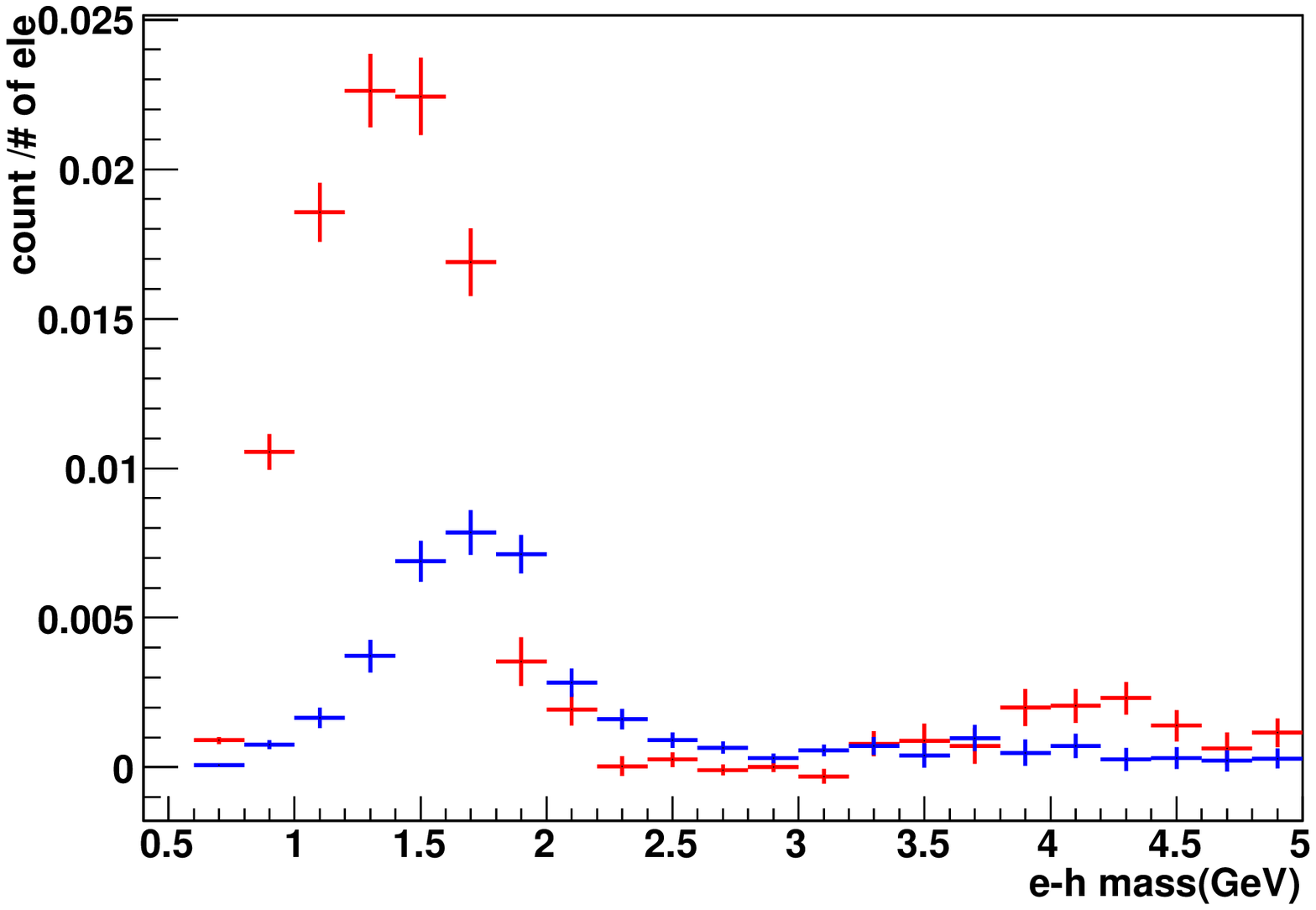}
		\caption{Subtracted and normalized invariant 
		  mass distributions of electron-hadron pairs in charm and bottom production.
		  $p_{\mathrm{T}}$ range of trigger electrons is 5.0-7.0~GeV/$c$.
		}
		\label{chap6_fig10}
	      \end{center}
	    \end{minipage}
	  \end{tabular}
	\end{figure}

	Figure~\ref{chap6_fig7} to \ref{chap6_fig10} show the normalized 
	reconstruction signals in charm and bottom production at each electron 
	$p_{\mathrm{T}} $ range in RUN5 configuration.
	Red points show reconstruction signals in charm production and blue points 
	show these in bottom production.

	$\epsilon_{c}$ and $\epsilon_{b}$ at each electron $ p_{\mathrm{T}} $ range 
	are determined as the number of entries in 0.4$<$ M$_{eh}$ $<$1.9~GeV 
	in Fig.\ref{chap6_fig7} to \ref{chap6_fig10}.
	Results of $\epsilon_{c}$ and $\epsilon_{b}$ are summarized in 
	Table~\ref{chap6_table2}.
	$\epsilon_{c}$ increases as electron $p_{\mathrm{T}}$ increases by kinematic reason. 
	\begin{table}[hbt]
	  \begin{center}
	    \caption{Result of $\epsilon_{c}$}
	    \label{chap6_table2}
	    \begin{tabular}{c|cc|cc}
	      \hline
	      electron $p_{\mathrm{T}}$ range & $\epsilon_{c}$ &$\epsilon_{b}$& $\epsilon_{c}$ &$\epsilon_{b}$\\
              &RUN5&RUN5&RUN6&RUN6\\
	      \hline\hline
	      2.0-3.0~GeV/$c$ & 0.0378&0.0162 & 0.0371&0.156\\
	      
	      3.0-4.0~GeV/$c$ & 0.0566& 0.0160&0.0563&0.0168\\
	      
	      4.0-5.0~GeV/$c$ & 0.0810& 0.0210 &0.0781&0.0198\\
	      
	      5.0-7.0~GeV/$c$ & 0.0921& 0.0275 &0.0913&0.0270\\
	      \hline
	    \end{tabular}
	  \end{center}
	\end{table}
	\clearpage
	\subsection {Systematic Errors for $\epsilon_{c}$ and $\epsilon_{b}$}
	Systematic error for $\epsilon_{c}$ and $\epsilon_{b}$ can be categorized into two
	components.
	One is systematic error of the difference in reconstruction efficiency including geometrical 
	acceptance between real data and PISA simulation.
	This component is common factor for $\epsilon_{c}$ and $\epsilon_{b}$.
	The other is from the uncertainty of the event generator~(PYTHIA and EvtGen).
	The uncertainty of $\epsilon_{c}$ and $\epsilon_{b}$ originated from the uncertainty of PYTHIA and EvtGen
	needs to  be assigned as systematic error.
	These errors are estimated for $\epsilon_{c}$ and $\epsilon_{b}$ separately.
	The following factors are considered.
	\begin{itemize}
	  \item{Production ratios of charmed and bottomed hadrons}
	  \item{Branching ratios of charmed and bottomed hadrons}
	  \item{Momentum distribution of charmed and bottomed hadrons}
	  \item{PYTHIA parameters}
	\end{itemize}
	\subsubsection{Geometrical Acceptance}
	3\% systematic error is assigned for geometical acceptnace, 
	as described in Sec.~\ref{sec:pisa} for RUN5 and RUN6 configuration.
	\subsubsection{Production Ratios of Charmed and Bottomed Hadrons}
	Production ratios of D and B hadrons~($D^+/D^0$, $D_s/D^0$, $B^+/B^0$, $B_s/B^0$...)
	are one of the most important parameters to determine  $\epsilon_c$ and $\epsilon_b$.
	Although the production ratios in the generated events are tuned based on the 
	experimental results as already described, the ratios have considerable
	uncertainty~\cite{bib:cratio1,bib:cratio2,bib:cratio3}.
	Therefore, the uncertainty of the production ratios should be considered as the 
	systematic error source.

	$D^+/D^0$, $D_s/D^0$ and $\Lambda_c/D^0$ ratios in PYTHIA are summarized in Table~\ref{chap6_table13}.
	The assigned uncertainties of $D^+/D^0$,$D_s/D^0$,$\Lambda_c/D^0$ based on experimental 
	results are also listed in Table~\ref{chap6_table13}.
	\begin{table}[hbt]
	  \begin{center}
	    \caption{$D^+/D^0$,$D_s/D^0$,$\Lambda_c/D^0$ ratios from other experiments
	       ~\cite{bib:cratio1,bib:cratio2,bib:cratio3} and PYTHIA~(default and tuned)}
	    \label{chap6_table13}
	    \begin{tabular}{|c|cccc|}
	      \hline
              &PYTHIA  & CDF & P.D.G & PYTHIA \\
              &(default)&($p+p$)&($e^+e^-$@$\sqrt{s}=91$GeV) & (tuned) \\
	      \hline
	      $D^+/D^0$ & 0.3 & 0.45 & & 0.45 $\pm$ 0.1  \\
	      $D_s/D^0$ & 0.2 & 0.23 &0.29 & 0.25$\pm$ 0.1  \\
	      $\Lambda_c/D^0$ & 0.1 &  &0.17 & 0.1 $\pm$ 0.05 \\
	      \hline
	    \end{tabular}
	  \end{center}
	\end{table} 
	$B^+/B^0$, $B_s/B^0$ and B baryons/$B^0$ ratios and uncertainty are summarized 
	in Table~\ref{chap6_table14}.
	$B^+/B^0$ is fixed to 1, since there are no reason to break isospin symmetry.
	$B_s/B^0$ and B baryons/$B^0$ ratios and their uncertainty are summarized 
	in Table~\ref{chap6_table14}.
	Since there are little experimental results of $B_s/B^0$ and B baryons/$B^0$ ratios, 
	50\% uncertainty is assigned for $B_s/B^0$ ratio and  75\% uncertainty is
	assigned for B baryons/$B^0$ ratio.
	\begin{table}[hbt]
	  \begin{center}
	    \caption{$B^+/B^0$,$B_s/B^0$,B baryons/$B^0$ ratios from other 
	      experiment~\cite{bib:PDG}
	      and PYTHIA~(tuned)}
	    \label{chap6_table14}
	    \begin{tabular}{|c|cc|}
	      \hline
              & P.D.G & PYTHIA \\
              &($e^+e^-$@$\sqrt{s}=91$GeV) & (tuned) \\
	      \hline
	      $B^+/B^0$ &  1 & 1      \\
	      \hline
	      $B_s/B^0$ & 0.35 & 0.4$\pm$ 0.2  \\
	      \hline
	      B baryons/$B^0$ & 0.2 & 0.2 $\pm$ 0.15 \\
	      \hline
	    \end{tabular}
	  \end{center}
	\end{table}

	The effect of the assigned uncertainty on the tagging efficiency is
	regarded as the systematic error of the tagging efficiency.
	For the study of this effect,
	details of $\epsilon_{c}$ and  $\epsilon_{b}$ are evaluated for each decay channel 
	at each trigger electron $p_{\mathrm{T}}$ range.
	For example, the results at the trigger electron with $2<p_{\mathrm{T}}<3$~GeV/$c$ are summarized in
	Table~\ref{chap6_table5}.
	For the results at other electron $p_{\mathrm{T}}$ are shown in Appendix.~\ref{sec:detail}.
	\begin{table}[hbt]
	  \begin{center}
	    \caption{Detail of charm and bottom decay for electron $p_{\mathrm{T}}$ 2-3 GeV/$c$}
	    \label{chap6_table5}
	    \begin{tabular}{c|c c c}
	      \hline
	      channel & $N_{tag}$ (part)/(all) & $N_{ele}$ (part)/(all) & $\epsilon$ \\
	      \hline
	      $D^0 $ & & & \\
	      $D^0\rightarrow e^+ K^- \nu_e$&38.96\%&29.64\%&4.68 $\pm$ 0.09\% \\
	      $D^0\rightarrow e^+ K^{\ast -} \nu_e$&15.24\%&3.73\%&14.57 $\pm$ 0.34\% \\
	      $D^0\rightarrow e^+ \pi^- \nu_e$&4.34\%&5.24\%&2.95 $\pm$ 0.19\% \\
	      $D^0\rightarrow e^+ \rho^- \nu_e$&2.04\%&0.52\%&13.88 $\pm$ 0.81\% \\
	      $D^0\rightarrow e^+ other$&1.23\%&0.51\%&8.67 $\pm$ 0.83\% \\
	      \hline
	      $D^+ $ & & & \\
	      $D^+\rightarrow e^+ \bar{K^0} \nu_e$&25.23\%&38.55\%&2.33 $\pm$ 0.07\% \\
	      $D^+\rightarrow e^+ \bar{K^{\ast 0}} \nu_e$&6.00\%&4.70\%&4.55 $\pm$ 0.31\% \\
	      $D^+\rightarrow e^+ \pi^0 \nu_e$&2.00\%&3.32\%&2.15 $\pm$ 0.21\% \\
	      $D^+\rightarrow e^+ \rho^0 \nu_e$&0.25\%&0.36\%&2.52 $\pm$ 1.14\% \\
	      $D^+\rightarrow e^+ other$&1.91\%&2.02\%&3.37 $\pm$ 0.37\% \\
	      \hline
	      $D_s $ & & & \\
	      $D_s\rightarrow e^+ \phi \nu_e$&0.45\%&0.89\%&1.80 $\pm$ 0.73\% \\
	      $D_s\rightarrow e^+ \eta \nu_e$&3.70\%&7.66\%&1.72 $\pm$ 0.14\% \\
	      $D_s\rightarrow e^+ \eta' \nu_e$&0.30\%&0.67\%&1.60 $\pm$ 0.69\% \\
	      $D_s\rightarrow e^+ other$&0.35\%&0.82\%&1.52 $\pm$ 0.47\% \\
	      \hline
	      $\Lambda_c $ & & & \\
	      $\Lambda_c\rightarrow e^+ \Lambda \nu_e$&-0.40\%&0.31\%&-4.63 $\pm$ 2.67\% \\
	      $\Lambda_c\rightarrow e^+ other$&-1.60\%&1.05\%&-5.44 $\pm$ 1.25\% \\
	      \hline
	      $B^0 $ & & & \\
	      $B^0\rightarrow e^+ D^- \nu_e$&3.27\%&6.20\%&0.89 $\pm$ 0.15\% \\
	      $B^0\rightarrow e^+ D^{\ast -} \nu_e$&0.50\%&21.21\%&0.04 $\pm$ 0.09\% \\
	      $B^0\rightarrow e^+ other$&4.39\%&5.88\%&1.26 $\pm$ 0.17\% \\
	      \hline
	      $B^+ $ & & & \\
	      $B^+\rightarrow e^+ D^0 \nu_e$&7.16\%&6.72\%&1.80 $\pm$ 0.15\% \\
	      $B^+\rightarrow e^+ D^{\ast 0} \nu_e$&21.30\%&22.93\%&1.57 $\pm$ 0.08\% \\
	      $B^+\rightarrow e^+ other$&2.95\%&6.47\%&0.77 $\pm$ 0.17\% \\
	      \hline
	      $B_s $ & & & \\
	      $B_s \rightarrow e^+ $total  &15.87 \%&10.81\%&2.49 $\pm$ 0.12\% \\
	      \hline
	      $B had\rightarrow e^+  $&10.51\%&10.10\%&1.76 $\pm$ 0.12\% \\
	      \hline
	      $B \rightarrow c \rightarrow e $&34.05\%&9.69\%&5.95 $\pm$ 0.18\% \\
	      \hline
	    \end{tabular}
	  \end{center}
	\end{table}

	The effect on the $\epsilon_c$ and $\epsilon_c$ are calculated by changing the production ratios
	of D and B hadrons according to the assigned uncertainties.
	The results are summarized in Table\ref{chap6_table15} and \ref{chap6_table16}.
	\begin{table}[hbt]
	  \begin{center}
	    \caption{The effect of $D^+/D^0$,$D_s/D^0$,$\Lambda_c/D^0$ changes 
	      on $\epsilon_{c}$ }
	    \label{chap6_table15}
	    \begin{tabular}{|c|cccc|}
	      \hline
              & $\Delta(\epsilon_{c})/\epsilon_{c}$ & $\Delta(\epsilon_{c})/\epsilon_{c}$ 
	      & $\Delta(\epsilon_{c})/\epsilon_{c}$ & $\Delta(\epsilon_{c})/\epsilon_{c}$ \\
	      electron $p_{\mathrm{T}}$& $2<p_{\mathrm{T}}<3$~GeV/$c$& $3<p_{\mathrm{T}}<4$~GeV/$c$
	      &  $4<p_{\mathrm{T}}<5$~GeV/$c$& $5<p_{\mathrm{T}}<7$~GeV/$c$\\
	      \hline \hline
	      $D^{\pm}/D^0  $& 2.7\%&  3.6\%& 3.3\%&4\%  \\
	      $D_s/D^0$ & 2.0\% & 1.9\% & 2.0\%& 2.0\%  \\
	      $\Lambda_c/D^0$& 1.7\% &1.5\% & 1.0\%& 0.4\% \\
	      \hline 
	    \end{tabular}
	  \end{center}
	\end{table}

	\begin{table}[hbt]
	  \begin{center}
	    \caption{The effect of $B_s/B^0$ and B baryons/$B^0$ changes on $\epsilon_{b}$ }
	    \label{chap6_table16}
	    \begin{tabular}{|c|cccc|}
	      \hline
	      & $\Delta(\epsilon_{c})/\epsilon_{c}$ & $\Delta(\epsilon_{c})/\epsilon_{c}$ 
	      & $\Delta(\epsilon_{c})/\epsilon_{c}$ & $\Delta(\epsilon_{c})/\epsilon_{c}$ \\
	      electron $p_{\mathrm{T}}$& $2<p_{\mathrm{T}}<3$~GeV/$c$& $3<p_{\mathrm{T}}<4$~GeV/$c$
	      & $4<p_{\mathrm{T}}<5$~GeV/$c$& $5<p_{\mathrm{T}}<7$~GeV/$c$\\
	      \hline \hline
	      $B_s/B^0$ & 2.5\% & 2.5\% &  3.7 \% &  4.5 \% \\
	      B baryons/$B^0$& 0.5\%& 1\% & 2.5 \%& 3 \%\\
	      \hline
	      \hline 
	    \end{tabular}
	  \end{center}
	\end{table} 

   \subsubsection{Branching Ratio}
   Branching ratios in EvtGen simulation are implemented  according to
   P.D.G and the results from CLEO, BarBar etc~\cite{bib:evt2,bib:evt3}.
   However, branching ratios listed in P.D.G have uncertainty and 
   there is small discrepancy in the branching ratios between P.D.G values and implemented values 
   in EvtGen for some decay channels.
   For these decay channels, these 
   discrepancy are taken as the uncertainty of the branching ratios for corresponding channel.
   The implemented and P.D.G values of branching ratios are summarized in Table~\ref{chap6_table17}.
   The assigned uncertainty of branching ratios are also summarized in Table~\ref{chap6_table17}.
   
   We calculate the effect of the uncertainty of the branching ratios 
   on the $\epsilon_c$ and $\epsilon_c$ by the similar way used to estimate  the
   systematic errors of the production ratio.
   That is, the effect on the $\epsilon_c$ and $\epsilon_c$ are calculated when the branching ratios
   of D and B hadrons are changed according to the assigned uncertainty.
   The results are summarized in Table~\ref{chap6_table17}
   and assigned as 
   a systematic error for the branching ratio.
    \begin{table}[hbt]
     \begin{center}
       \caption{Branching ratio of D and B hadrons in P.D.G and EvtGen.
	 The assigned uncertainties for the branching ratios and these effect 
	 on the $\epsilon_c$ and $\epsilon_b$  electron $p_{\mathrm{T}}$ 2-3 GeV/$c$ 
       }
       \label{chap6_table17}
       \begin{tabular}{c|c c|cc }
       \hline
       charmed hadrons & Branching Ratio& Branching ratio & $\delta(Br)/Br $ & $ \delta(\epsilon_c)/\epsilon_c $  \\
       channel   & EvtGen & P.D.G & & \\
       \hline
       $D^0 $ & &  & & \\
       $D^0\rightarrow e^+ K^- \nu_e$ & 3.50$\pm$0.11\% & 3.51\%$\pm$0.11\% & 3.10\% & 0.29\%\\
       $D^0\rightarrow e^+ K^{\ast -} \nu_e$ & 2.25$\pm$0.16\% & 2.17\%$\pm$0.16\%  & 7.10\%& 0.81\%\\
       $D^0\rightarrow e^+ \pi^- \nu_e$ &0.34$\pm$0.06\% & 0.28\%$\pm$0.02\%& 17.60\%  & 0.16\% \\
       $D^0\rightarrow e^+ \rho^- \nu_e$ & 0.22$\pm$0.03\% & 0.19\%$\pm$0.04\% & 13.60\% & 0.21\%  \\
       $D^0\rightarrow e^+ other$ &  0.45$\pm$0.35\% & 0.56\%$\pm$0.35\% & 77.80\% & 0.56\%  \\
       $D^0\rightarrow e^+ $ total&  &  &  & 1.06\% \\
       \hline
       $D^+ $ & &  & & \\
       $D^+\rightarrow e^+ \bar{K^0} \nu_e$ & 9.00$\pm$0.50\% & 8.60\%$\pm$0.50\% & 5.60\% & 0.73\% \\
       $D^+\rightarrow e^+ \bar{K^{\ast 0}} \nu_e$ & 5.50$\pm$0.50\% & 8.60\%$\pm$0.50\% & 5.40\% & 0.07\% \\
       $D^+\rightarrow e^+ \pi^0 \nu_e$ & 0.44$\pm$0.07\% & 0.44\%$\pm$0.07\% & 16.30\% & 0.21\% \\
       $D^+\rightarrow e^+ \rho^0 \nu_e$ & 0.28$\pm$0.06\% & 0.22\%$\pm$0.04\%& 21.40\% & 0.02\% \\
       $D^+\rightarrow e^+ other$ & 1.46$\pm$0.7\% & 1.12\%$\pm$0.7\% & 48.70\% & 0.05\% \\
       $D^+\rightarrow e^+ $ total& & &  & 0.77\% \\
       \hline
       $D_s $  & & & & \\
       $D_s\rightarrow e^+ \phi \nu_e$ & 2.42$\pm$0.50\% & 2.50\%$\pm$0.30\%($l^+$) & 20.70\% & 0.09\%  \\
       $D_s\rightarrow e^+ \eta \nu_e$ &  3.07$\pm$0.8\% & 3.10\%$\pm$0.60\% & 26.10\%  & 1.01\% \\
       $D_s\rightarrow e^+ \eta' \nu_e$  &  1.06$\pm$0.4\% & 1.08\%$\pm$0.35\% & 47.20\% & 0.17\%   \\
       $D_s\rightarrow e^+ other$ & 0.37$\pm$0.37\% & 100.00\% & 0.46\% \\
       $D_s\rightarrow e^+ $ total&  & &   & 1.13\% \\
       \hline
       $\Lambda_c $  & &  & & \\
       $\Lambda_c\rightarrow e^+ \Lambda \nu_e$ & 1.8$\pm$0.6\% & 2.1\%$\pm$0.6\% & 33.30\% & 0.24\% \\
       $\Lambda_c\rightarrow e^+ other$ & 2.7$\pm$1.8\% & 2.4\%$\pm$1.8\%& 66.70\% & 1.76\% \\
       $\Lambda_c\rightarrow e^+ $ total & & &   & 1.77\% \\
       \hline
       $total  $   &  &  & & 2.48\% \\
       \hline \hline
       bottomed hadrons & Branching Ratio& Branching ratio& $\delta(Br)/Br $ & $ \delta(\epsilon_b)/\epsilon_b $\\
       channel   & EvtGen & P.D.G  & & \\
       \hline
       $B^0 $  & &  & & \\
       $B^0\rightarrow e^+ D^- \nu_e$ & 2.07$\pm$0.2\% & 2.12\%$\pm$0.2\%  & 9.70\% & 0.28\% \\
       $B^0\rightarrow e^+ D^{\ast -} \nu_e$ & 5.70$\pm$0.35\% & 5.35\%$\pm$0.2\%  & 6.10\% & 1.25\% \\
       $B^0\rightarrow e^+ other$ & 2.6$\pm$0.5\% & 2.93\%$\pm$0.5\%& 19.20\% & 0.28\% \\
       $B^0 \rightarrow e^+ $ total &   & & & 1.31\% \\
       \hline
       $B^+ $  & &  & &\\
       $B^+\rightarrow e^+ D^0 \nu_e$ & 2.24$\pm$0.22\% & 2.15\%$\pm$0.22\% & 10.20\% & 0.04\% \\
       $B^+\rightarrow e^+ D^{\ast 0} \nu_e$ & 6.17$\pm$0.5\% & 6.50\%$\pm$0.5\% & 7.70\% & 0.12\%\\
       $B^+\rightarrow e^+ other$ & 2.49$\pm$0.7\% & 2.25\%$\pm$0.67\% & 26.90\% & 0.93\% \\
       $B^+ \rightarrow e^+ $ total & & &   & 0.94\% \\
       \hline
       $B_s $  & & & &\\
       $B_s \rightarrow e^+ $total   &7.9$\pm$3\% & 7.9\%$\pm$2.4\% & 37.00\% & 1.80\% \\
       \hline
       $B had\rightarrow e^+ others $  & 7.5$\pm$4.0\% & 8.5\%$\pm$4.0\%& 40.0\% & 0.2 \%  \\
       \hline
       $total  $ &  & & & 2.42\% \\
       \hline
     \end{tabular}
    \end{center}
\end{table}

   \subsubsection{$b \rightarrow c \rightarrow e$ Process}
   There is discrepancy in the production ratios of D hadrons which originates from inclusive 
   $b \rightarrow c$ process between P.D.G and EvtGen, which is the known problem of EvtGen.
   The production ratios of D hadrons in inclusive 
   $b \rightarrow c$ process in P.D.G and EvtGen are summarized at Table~\ref{chap6_table21}.
   As a result, $D^+/D^0$ in B decay are 0.30 and 0.41 in P.D.G value and EvtGen, respectively.
   The effect of this discrepancy in $b \rightarrow c \rightarrow e$
   needs to be considered as systematic error 
   of tagging efficiency of $\epsilon_b$.
   The difference in $D^+/D^0$ in B decay makes $\sim$ 4\% effect on the tagging efficiency 
   in $c \rightarrow e$ process, which is estimated in  the same way to estimate 
   the uncertainty from the production ratio.
   The 4\% effect of the tagging efficiecy in $c \rightarrow e$ process changes the 
   tagging efficiency of bottom production via $b \rightarrow c \rightarrow e$ process, which is evaluated 
   from Table~\ref{chap6_table5}.
   It is found that such effect is less than 2\%.

\begin{table}[hbt]
  \begin{center}
    \caption{Inclusive resonance D production in B decays at PDG and EvtGen}
    \label{chap6_table21}
    \begin{tabular}{|c|c|c|}
      \hline
              & P.D.G 06 (\%) & EvtGen (\%)\\
      \hline \hline
      $B \rightarrow e \nu X$ & $10.24 \pm 0.15$ & $10.6$ \\
      \hline 
      $B \rightarrow D^{\pm} X$ & $22.8 \pm 1.4 $ & $32.4$ \\
      \hline 
      $B \rightarrow D^0 X$ & $63.7 \pm 1.4 $ & $68.2$ \\
      \hline 
      $B \rightarrow D^{*\pm} X$ & $22.5 \pm 1.5 $ & $26.2$ \\
      \hline 
      $B \rightarrow D^{*0} X$ & $26.0 \pm 2.7 $ & $25.7$ \\
      \hline 
      $B \rightarrow D^{(*)} D^{(*)}bar  X$ & $7.1 + 2.7 -1.7 $ & $7.7$ \\
      \hline 
    \end{tabular}
  \end{center}
\end{table}


\subsubsection{Momentum Distribution of Charmed and Bottomed Hadrons}
\begin{figure}[htb]
  \begin{center}
    \includegraphics[width=12cm]{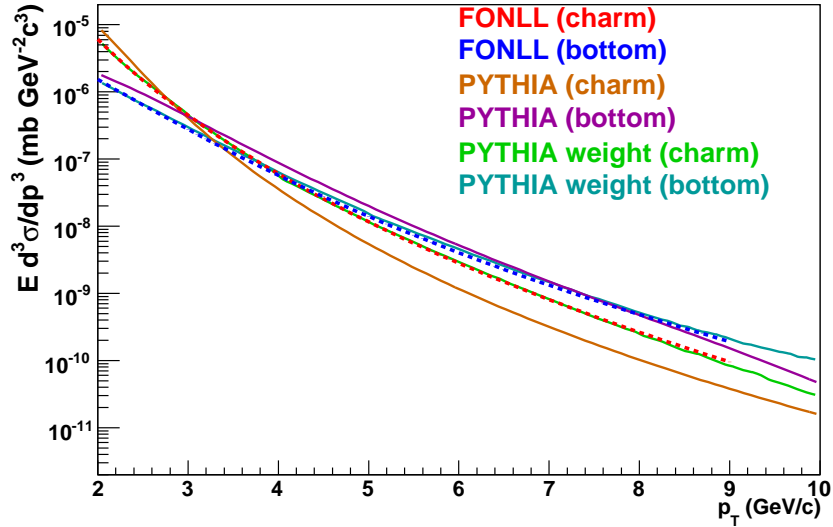}
    \caption{The spectra of the single electrons from charm and bottom at PYTHIA and 
      FONLL~\cite{bib:fonll1}.
      Dark orange line and magenta line show the spectra from charm and bottom at PYTHIA.
      Green line and cyan line show the spectra of the electrons from
      charm and bottom at PYTHIA with weighting factor.
      Red line and blue lines show the spectra from charm and bottom at FONLL.
      }
    \label{chap6_fig102}
  \end{center}
\end{figure}
D and B hadrons in the PYTHIA simulation have the uncertainty in their momentum distribution.
Tagging efficiency as a function of trigger electron $p_{\mathrm{T}}$ depends on the momentum distributions 
of parent D and B hadrons.
Therefore, the systematic error for the momentum distribution of the parent particles should 
be estimated.

The shape of the $p_{\mathrm{T}}$ spectra of the electrons from charm and bottom reflects
the momentum distribution of the parent particles.
The difference of the momentum distribution of the parent particles between PYTHIA and the experimental results
can be  estimated by comparing the shape of electron $p_{\mathrm{T}}$ spectra
obtained by PYTHIA and real data.
Figure~\ref{chap6_fig102} shows the $p_{\mathrm{T}}$ spectra of the electrons 
from charm~(Dark Orange) and bottom~(Magenta) at PYTHIA.
The spectra of the electrons from PYTHIA are compared with the spectra from FONLL~\cite{bib:fonll4}, where
the shape of the spectra from FONLL almost agrees with the experimental results including PHENIX
~\cite{bib:fonll1,bib:fonll2,bib:fonll3}.
Weighting factor, $w(p_{\mathrm{T}})$, is defined as follows.
\begin{equation}
  w(p_{\mathrm{T}}) = FONLL(p_{\mathrm{T}})/PYTHIA(p_{\mathrm{T}}).
\end{equation}
This weighting factor is used to correct the difference of the momentum distribution.
In Fig~\ref{chap6_fig102}, green line and cyan line show the spectra of the electrons from
charm and bottom produced at PYTHIA with weighting factor.
The difference of $\epsilon_{c}$ and  $\epsilon_{b}$ between with and 
without the weighting reflects the correction of the momentum distribution.
As a result, about 1.5\% systematic error is assinged for $\epsilon_{c}$ and about 2\% systematic 
error is assigned for  $\epsilon_{b}$


   \subsubsection{PYTHIA Uncertainty}
    N$_{tag}$ includes the remaining contribution from the associated hadron which is not 
    from heavy flavor decay~(jet fragmentation).
    Such effect has the dependence of the PYTHIA parameters and this dependence should be 
    included into the systematic error.
    Study is done to estimate the amount of the contribution from the associated 
    hadron which is from jet fragmentation.

\begin{figure}[htb]
      \begin{center}
	\includegraphics[width=10.5cm]{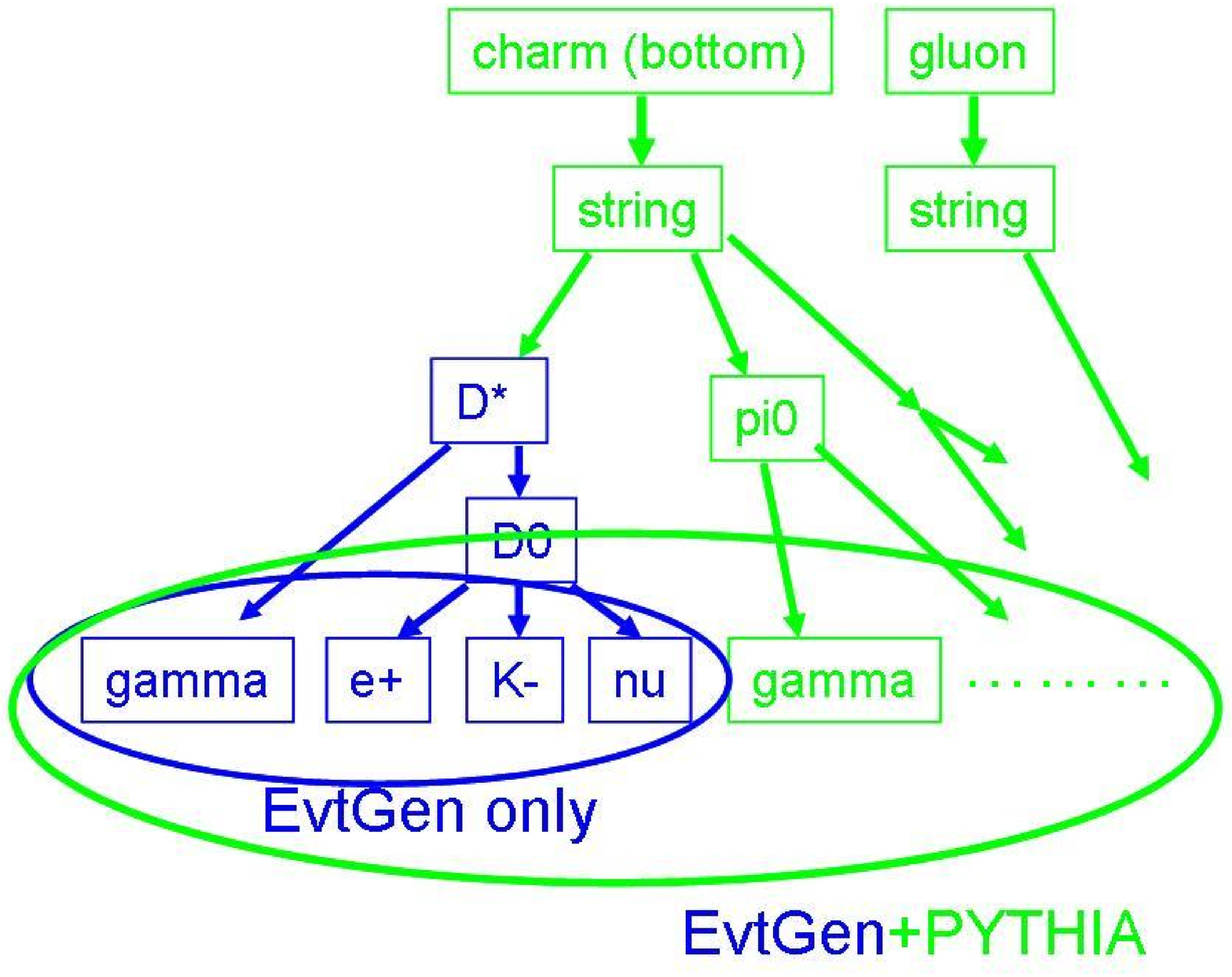}
	\caption{
      Conceptual view of the procedure to estimate contribution of jet fragmentation 
      simulated by PYTHIA.
      Blue particles are D mesons ans baryons simulated by EvtGen and
      green particles are jet fragmentation simulated by PYTHIA.
    }
    \label{chap6_fig22}
  \end{center}
\end{figure}

    $\epsilon_c$ and $\epsilon_b$ are calculated in the two cases for all stable particles
    and only decay daughters of D and B hadrons  in generated events.
    Figure~\ref{chap6_fig22} shows a conceptual view of this procedure.
    In Fig.\ref{chap6_fig22}, blue particles are  decay daughters of D and B hadrons and 
    green particles are from jet fragmentation simulated with PYTHIA.
    The contribution of jet fragmentation generated by PYTHIA can be estimated by 
    comparison between the above two cases~(with and without hadron from jet fragmentation). 
    It is found that the contribution of jet fragmentation to $\epsilon_c$ 
    and $\epsilon_b$ is less than 15\%.
    Therefore, the effect of uncertainty of jet fragmentation on the 
    tagging efficiency is expected to be small.
    For a example, if the uncertainty of the contribution of jet fragmentation in PYTHIA is 20\%, 
    the uncertainty of  $\epsilon_c$ and $\epsilon_b$ becomes $15\% \times 20 \% = 3\%$.

\begin{figure}[htb]
  \begin{center}
    \includegraphics[width=14cm]{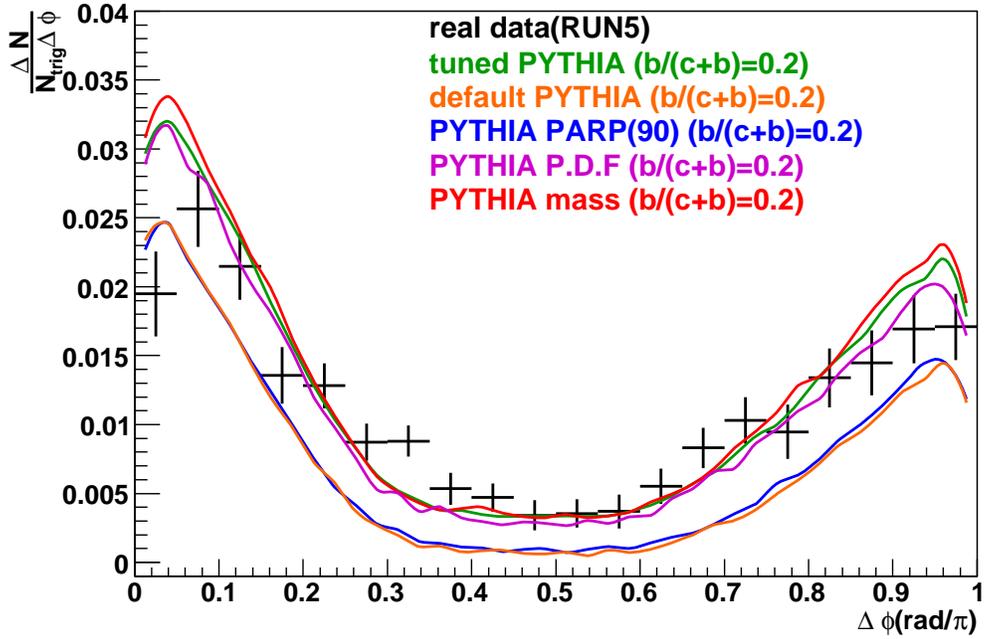}
    \caption{
      The correlation function of electrons and hadrons, when
      the trigger electrons were from heavy flavor.
      Black points show the result in RUN5 data obtained at Section\ref{sec:c_hq} and 
      various lines show the result at PYTHIA with the parameter sets.
    }
    \label{chap6_fig23}
  \end{center}
\end{figure}

    More precisely, the uncertainty from the PYTHIA dependence on the contribution of jet fragmentation
    is estimated by looking at the measured yield of associated hadrons 
    as a function of azimuthal angle between the trigger non-photonic electrons and the associated 
    hadron~(correlation function) in charm and bottom production.
    Since the contribution of jet fragmentation is not canceled out for the inclusive 
    multiplicity, the inclusive multiplicity is a good observable to study the PYTHIA
    dependence of the jet fragmentation.
    The  multiplicity in RUN5 is obtained in Appendix.~\ref{sec:c_hq}.
    Some parameter sets, which are expected to affect the contribution of jet fragmentation, are prepared to estimate the effect 
    of PYTHIA uncertainty on the $\epsilon_c$ and $\epsilon_b$ as follows.
    \begin{itemize}
    \item default PYTHIA~(1)
    \item PARP(90) 0.25$\rightarrow$0.16~(2)
    \item P.D.F CTEQ5L$\rightarrow$GRV94L~(3)
    \item charm mass 1.2$\rightarrow$1.4GeV/$c^2$, bottom mass 4.3$\rightarrow$4.5GeV/$c^2$~(4)
    \end{itemize}
    Figure~\ref{chap6_fig23} shows the inclusive multiplicity as a function of azimuthal angle 
    between the trigger single non-photonic electrons and the associated hadron.
    Black points show the result in RUN5 data obtained at Appendix.~\ref{sec:c_hq} and 
    various lines show the results from PYTHIA with the parameter sets.
    Since (1) and (2) parameter sets are NOT consistent with real data, the deviation of 
    $\epsilon_c$ and $\epsilon_b$ with the (1) and (2) parameter sets from tuned PYTHIA
    gives enough conservative systematic error.
    We assign 6\% systematic error as PYTHIA uncertainty, since the deviations are
    5\% for $\epsilon_c$ and $\epsilon_b$.
    
\

\subsubsection{Summary of Systematic Error for PYTHIA and EvtGen}
Systematic error of $\epsilon_c$ and $\epsilon_b$ are summarized in Table~\ref{chap5_table14}.

\begin{table}[htb]
  \begin{center}
 \caption{Summary of $\epsilon_c$ and $\epsilon_b$}
     \label{chap5_table14}
     \begin{tabular}{|c|c|c|c|c|}
       \hline
       $\epsilon_c$ & & & & \\
       electron $p_{\mathrm{T}}$ & simulation statistics & EvtGen+PYTHIA & geometrical acceptance & total\\
       \hline \hline
        2.0-3.0 GeV/$c$ & 1.7\%  &  7.6\% & 3\% & 8.3\% \\
        \hline
        3.0-4.0 GeV/$c$ & 2.2\%  &  7.9\% & 3\% & 8.7\% \\
        \hline
        4.0-5.0 GeV/$c$ & 1.6\%  &  7.5\% & 3\% & 8.3\% \\
	\hline
        5.0-7.0 GeV/$c$ & 2.1\%  &  8.0\% & 3\% &  8.8\%\\
        & & & & \\
	\hline \hline
	$\epsilon_b$ & & & & \\
	 electron $p_{\mathrm{T}}$ & simulation statistics & EvtGen+PYTHIA & geometrical acceptance & total\\
       \hline \hline
        2.0-3.0 GeV/$c$ & 3.5\%  &  7.5\% & 3\% & 8.8\% \\
        \hline
        3.0-4.0 GeV/$c$ & 5.0\%  &  7.3\% & 3\% & 9.4\%\\
        \hline
        4.0-5.0 GeV/$c$ & 4.8\%  &  9.1\% & 3\% & 10.7\% \\
	\hline
        5.0-7.0 GeV/$c$ & 3.8\%  &  9.6\% & 3\% &  10.7\% \\
	\hline
     \end{tabular}
    \end{center}
\end{table}

\section{$\epsilon_{data}$, $\epsilon_{c}$ and $\epsilon_{b}$} \label{sec:eps}
$\epsilon_{data}$, $\epsilon_{c}$ and $\epsilon_{b}$ are obtained in Sec.~\ref{sec:cor_real} 
and \ref{sec:cor_sim}.
The location of effective bin center of electron $p_{\mathrm{T}}$ at each $p_{\mathrm{T}}$ range 
is determined as weighted mean of electron $p_{\mathrm{T}}$ at each $p_{\mathrm{T}}$ range.
Since the bin center of $\epsilon_{c(b)}(p_{\mathrm{T}})$  is different from the bin center of 
$\epsilon_{data}(p_{\mathrm{T}})$, it is necessary to correct to $\epsilon_{c(b)}(p_{\mathrm{T}})$ 
the value at the same $p_{\mathrm{T}}$ as used in $\epsilon_{data}$. 
This is done as follows. 
\begin{equation}
\epsilon_{c(b)}(p_{\mathrm{T}}^{real}) = \frac{f_{c(b)}(p_{\mathrm{T}}^{real}) }
	{f_{c(b)}(p_{\mathrm{T}}^{c(b)})} \times \epsilon_{c(b)}(p_{\mathrm{T}}^{c(b)}),
\end{equation}
where 
\begin{itemize}
\item $p_{\mathrm{T}}^{real}$ and  $p_{\mathrm{T}}^{c(b)}$ are the effective bin center of 
electron $p_{\mathrm{T}}$ in real data and simulation for charm~(bottom) production.
\item $f_{c(b)}(p_{\mathrm{T}})$ is the fit function for the obtained $\epsilon_{c(b)}$.
\end{itemize}
\begin{figure}[htb]
  \begin{center}
    \includegraphics[width=11.5cm]{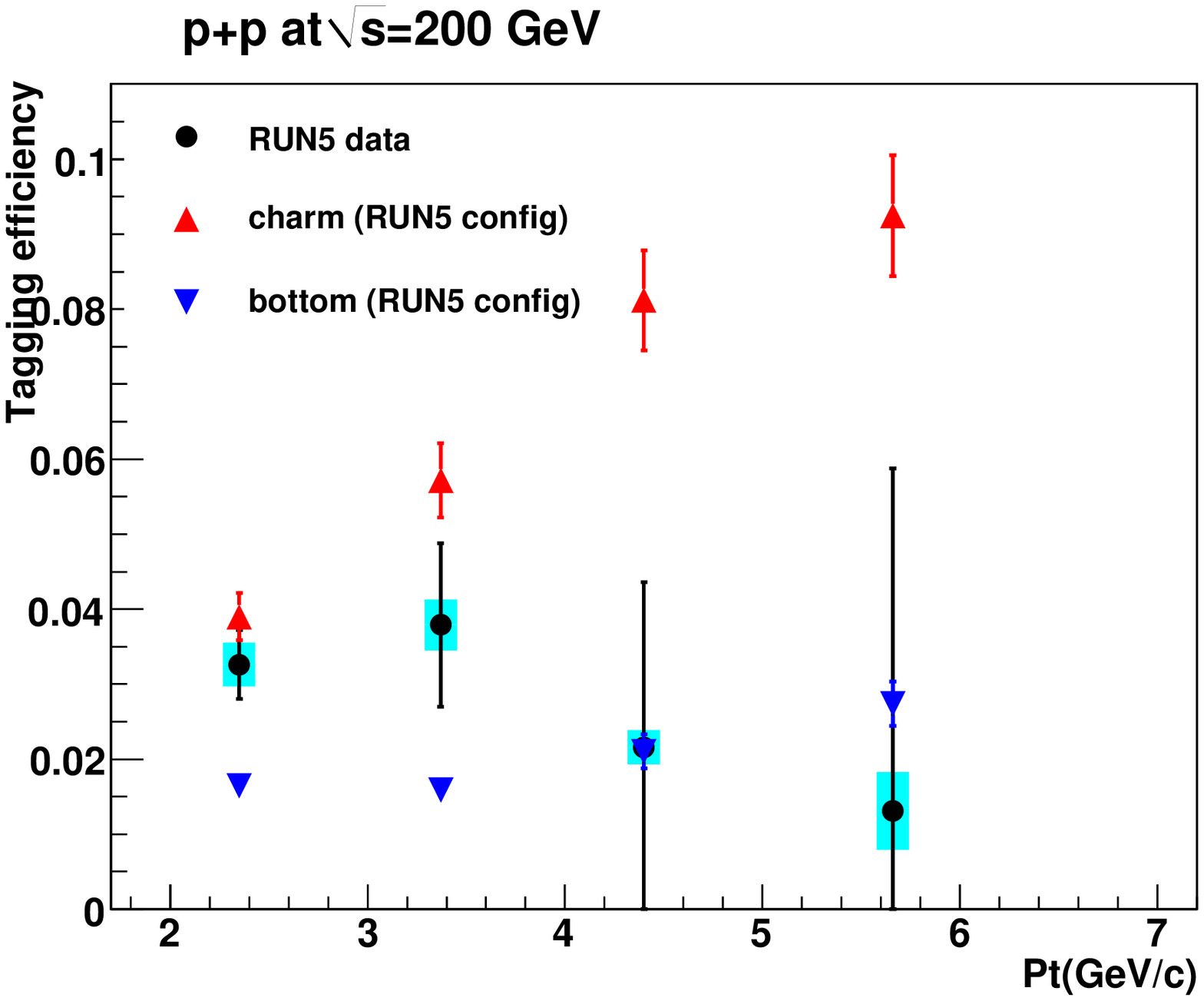}
    \caption{$\epsilon_{c}$, $\epsilon_{b}$ and $\epsilon_{data}$ as
      a function of electron $p_{\mathrm{T}}$ in RUN5.}
   \label{chap7_fig3}
  \end{center}
\end{figure}

\begin{figure}[htb]
  \begin{center}
    \includegraphics[width=11.5cm]{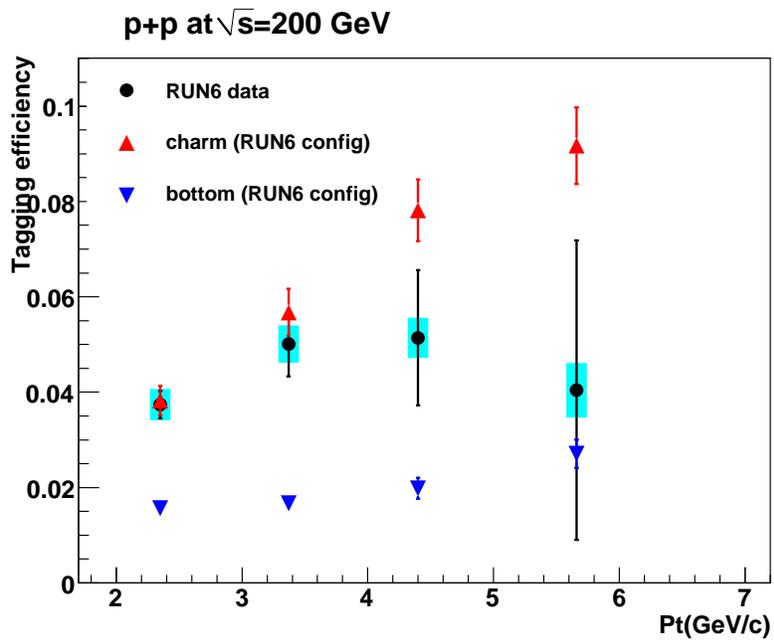}
    \caption{$\epsilon_{c}$, $\epsilon_{b}$ and $\epsilon_{data}$ as
      a function of electron $p_{\mathrm{T}}$ in RUN6.}
   \label{chap7_fig4}
  \end{center}
\end{figure}

Figure~\ref{chap7_fig3} and \ref{chap7_fig4} show $\epsilon_{data}$, 
$\epsilon_{c}$ and $\epsilon_{b}$ as  a function of electron $p_{\mathrm{T}}$ 
in RUN5 and RUN6, respectively.
Here, black points correspond to $\epsilon_{data}$, 
red points correspond to $\epsilon_{c}$ and blue points correspond to $\epsilon_{b}$.
In Fig.~\ref{chap7_fig3} and \ref{chap7_fig4}, data points move near bottom values
as electron $p_{\mathrm{T}}$ increases. This fact indicates the fraction of the electrons 
from bottom increases with electron $p_{\mathrm{T}}$.
Results are summarized in Table~\ref{chap7_table1} 
\begin{table}[hbt]
    \begin{center}
     \caption{$\epsilon_{data}$, $\epsilon_{c}$ and $\epsilon_{b}$}
     \label{chap7_table1}
     \begin{tabular}{|c|ccc|}
       \hline
       electron $p_{\mathrm{T}}$ & $\epsilon_{data}$ &$\epsilon_{c}$ & $\epsilon_{b}$\\
       \hline \hline
       RUN5 & & & \\
       2.35GeV/$c$ & $0.0326\pm 0.0046\pm 0.0029$&$0.0390 \pm 0.0031$&$0.0164\pm0.0014$ \\
       3.37GeV/$c$ & $0.0379\pm 0.0109\pm 0.0034$&$0.0571 \pm 0.0050$&$0.0159\pm0.0015$ \\
       4.40GeV/$c$ & $0.0216\pm 0.0220\pm 0.0023$&$0.0812 \pm 0.0067$&$0.0210\pm0.0023$ \\
       5.66GeV/$c$ & $0.0131\pm 0.0457\pm 0.0052$ &$0.0924 \pm 0.0081$&$0.0274\pm0.0030$ \\
       \hline
       RUN6 & & & \\
       2.35GeV/$c$ & $0.0374\pm 0.0029\pm 0.0033$&$0.0382 \pm 0.0031$&$0.0156\pm0.0014$ \\
       3.37GeV/$c$ & $0.0501\pm 0.0068\pm 0.0039$&$0.0567 \pm 0.0049$&$0.0167\pm0.0015$ \\
       4.40GeV/$c$ & $0.0514\pm 0.0142\pm 0.0042$&$0.0781 \pm 0.0065$&$0.0198\pm0.0022$ \\
       5.66GeV/$c$ & $0.0404\pm 0.0314\pm 0.0057$ &$0.0917 \pm 0.0080$&$0.0271\pm0.0029$ \\
       \hline
     \end{tabular}
    \end{center}
\end{table}

  \chapter{Result}
The spectrum of electrons from semi-leptonic decay of heavy flavor and tagging efficiency,
$\epsilon_{data}, \epsilon_{c}$ and $\epsilon_{b}$ are obtained as described in the previous chapter.
$(b\rightarrow e)/(c\rightarrow e+ b\rightarrow e)$ is determined using 
$\epsilon_{data}, \epsilon_{c}$ and $\epsilon_{b}$ in Sec.~6.1.
The reconstructed signals in real data are compared with these in simulation in Sec.~6.2.
The electron spectra for charm and bottom are obtained from the ratio,
$(b\rightarrow e)/(c\rightarrow e+b\rightarrow e)$, and the spectrum of single 
electrons in Sec.~6.3.
Total cross section of bottom is also measured by integrating this spectrum.

\section{$(b\rightarrow e)/(c\rightarrow e+ b\rightarrow e)$ Results}
The fraction of the contribution of bottom quark in the single non-photonic electrons
~($(b\rightarrow e)/(c\rightarrow e+ b\rightarrow e)$)
is obtained from $\epsilon_{data}$, $\epsilon_{c}$ and $\epsilon_{b}$ using
following equation.
\begin{equation}
\frac{b\rightarrow e}{c\rightarrow e+ b\rightarrow e} = \frac{\epsilon_c - \epsilon_{data}} {\epsilon_c - \epsilon_b}.
\end{equation}

The obtained values of $(b\rightarrow e)/(c\rightarrow e+ b\rightarrow e)$ in 
RUN5 and RUN6 are combined.
When the results in RUN5 and RUN6 are combined, following two issues should be taken carefully.
First, a part of systematic errors of RUN5 and RUN6 are correlated.
Second, a physical boundary exits in the value of  
$(b\rightarrow e)/(c\rightarrow e+ b\rightarrow e)$, that is 
$0<(b\rightarrow e)/(c\rightarrow e+ b\rightarrow e)<1$.
BLUE~(Best Linear Unbiased Estimate) method and Bayes' principle are applied
to take into account such conditions~\cite{bib:blue1,bib:blue2}. The combined mean values 
and standard deviations are determined using the BLUE method under the condition that 
there is no physical boundary. Then, Bayes' principle is applied to take into account
the physical constraint.
\subsection{BLUE Method in Correlation Analysis} \label{sec:comb}
The BLUE  method is applied to combine the results in RUN5 and RUN6.
Error sources are summarized in Table~\ref{chap7_table3}.
The definition of the types of errors is described in Sec.\ref{hq_spe}.
\begin{table}[hbt]
    \begin{center}
     \caption{Summary of error source}
     \label{chap7_table3}
     \begin{tabular}{|c|c|}
       \hline
       error source & run5/6 correlation~(Type) \\
       \hline \hline
       statistics & 0 \% (A)\\
       \hline
       signal count & 0\% (A)\\
       \hline
       cocktail calculation & 100\%(B)\\
       \hline
       PISA geometry & 0\% (B)\\
       \hline
       simulation statistics & 0\% (A)\\
       \hline
       Event generator & 100\%(B)\\
       \hline
     \end{tabular}
    \end{center}
\end{table}
In Table~\ref{chap7_table3}, the correlated systematic errors is tagged as B and 
they are assumed to have 100\% correlation.
With this assumption, the relation between correlated and uncorrelated errors in BLUE to become simple
as follows.
\begin{eqnarray}
  (\sigma^{corr})^2 & =& \rho\sigma_{run5}\sigma_{run6}, \\
  \sigma^{corr}  &=& \sqrt{\sigma_{run5}^{sys-B}\sigma_{run6}^{sys-B}}, \\
  \sigma_{runi}^{uncorr}  &=& \sqrt{(\sigma_{runi})^2 - (\sigma^{corr})^2}. 
\end{eqnarray}
From the above equations, the weighted average and combined error are obtained 
as follows.\\
\begin{eqnarray} 
  <r> &   =& \frac{r_{run5}(\sigma_{run6}^{uncorr})^2+
    r_{run6}(\sigma_{run5}^{uncorr})^2}
  {(\sigma_{run5}^{uncorr})^2+ (\sigma_{run6}^{uncorr})^2 },\\
\sigma &=& \sqrt{\frac{\sigma_{run5}^2\sigma_{run6}^2-(\sigma^{corr})^4}
  {(\sigma_{run5}^{uncorr})^2+ (\sigma_{run6}^{uncorr})^2 }},\\
\sigma^{stat} &=& \frac{\sqrt{(\sigma_{run6}^{stat})^2
    (\sigma_{run5}^{uncorr})^4+(\sigma_{run5}^{stat})^2
    (\sigma_{run6}^{uncorr})^4 }}
      {(\sigma_{run5}^{uncorr})^2+ (\sigma_{run6}^{uncorr})^2 },\\
      \sigma^{sys}&=& \sqrt{\sigma^2-(\sigma^{stat})^2}.
\end{eqnarray}
($\chi^2/ndf$) of the combination is 3.0/4, which indicates that the results in
RUN5 and RUN6 are consistent.

\subsection{Physical Constraint}
The ratio, $(b\rightarrow e)/(c\rightarrow e+ b\rightarrow e)$ which we want to determine in this analysis 
has physical boundary, $0 \le$ $(b\rightarrow e)/(c\rightarrow e+ b\rightarrow e)$ $\le 1$.
Bayes' principle is applied to take into account the effect of  
the physical constraint~\cite{bib:PDG,bib:beys}.
Bayes' principle is \\
\begin{equation}
\label{eq:boverc}
f(r \mid \epsilon) = \frac{L(\epsilon \mid r) g(r)} 
{\int_{r'} L(\epsilon \mid r') g(r') dr'}.
\end{equation}
Here, $\epsilon$ is the outcome of experiment~(tagging efficiency in this analysis) and 
$r$ is an unknown parameter that we want to 
determine~($(b\rightarrow e)/(c\rightarrow e+ b\rightarrow e)$ in this analysis).
$f(r \mid \epsilon)$ is the posterior probability density function when 
experimental value $ \epsilon$ is given.
Since $f(r \mid \epsilon)$ includes all knowledge about $r$,
we can determine the error of $r$ when we get $f(r \mid \epsilon)$. 
$L(\epsilon \mid r)$ is the likelihood function, that is the joint 
probability density function for the data given a certain value of $r$.
$g(r)$ is the prior probability density function. 
Since the statistics does not give us any information about $g(r)$, we must assume the
distribution of $g(r)$ reasonably.
We assume that $(b\rightarrow e)/(c\rightarrow e+ b\rightarrow e)$ has uniform 
distribution from 0 to 1 in this analysis.
Since the obtained error for $\epsilon$ obey Gaussian, $f(r \mid \epsilon)$ becomes
\begin{eqnarray}
  \label{bay}
f(r \mid \epsilon)& = A \times \exp \{-(r-r_0)/2\sigma_{r}^2 \}   & (0\le r \le1) ,\\
f(r \mid \epsilon)& =0  &   (0>r \mid \mid r>1) \nonumber.
\end{eqnarray}
Here, $r_0$ is the obtained $r$ from the combination of RUN5 and RUN6 analysis 
and $\sigma_{r}$ is the obtained deviation from $r$ the combination.
$A$ is a normalization factor so that the integral value of $f(r \mid \epsilon)$ becomes 1.

The variables, $x_1$ and $x_2$ are defined as following equations.

\begin{eqnarray}
\label{region}
\int_{0}^{x_1} f(r \mid \epsilon) dr& =&\int_{x_2}^{1} f(r \mid \epsilon) dr  = (1-\alpha)/2  ,\\
\int_{x_1}^{x_2} f(r \mid \epsilon) dr& =&\alpha   \nonumber .\\
\alpha& =& 0.6827  \nonumber 
\end{eqnarray}
Deviation from $x_1$ and $x_2$ to mean value are considered as the standard deviation.

\subsection{Lower and Upper Limit}
90\% C.L is determined for the highest and the lowest elecctron $p_{\mathrm{T}}$ 
range~($2.0<p_{\mathrm{T}}<3.0$~GeV/$c$ and $5.0<p_{\mathrm{T}}<7.0$~GeV/$c$), 
since the mean value obtained by the BLUE analysis is close to the boundary.

Probability density function is defined as bellow.\\
\begin{eqnarray}
\label{low}
 f(r) &   = A\times \exp(-(r-r_{0})^2/2\sigma^2) & (0<r<1),\\
 r_0 &    = \frac{\epsilon_c - \epsilon_{data}} {\epsilon_c - \epsilon_b} &, \\
 \sigma & = \sqrt{\sigma_{stat}^2 + \sigma_{sys}^2} &.
\end{eqnarray}
Here, A is a normalization factor to have integrated value becomes 1.
$\sigma_{stat}$ is statistical error of $\epsilon_{data}$ without
the consideration of the boundary.
$\sigma_{sys}$ is also systematic errors of $\epsilon_{data}$, $\epsilon_{c}$
and  $\epsilon_{b}$ without the consideration of the boundary.

90\% C.L is determined from this probability density fucntion. 
The values at which integrated probability density function becomes 50\% from the boundary 
are also determined as the mean values.
\subsection{The Combined Result} \label{sec:bc_r}
\begin{figure}[htb]
  \begin{center}
    \includegraphics[width=17.cm]{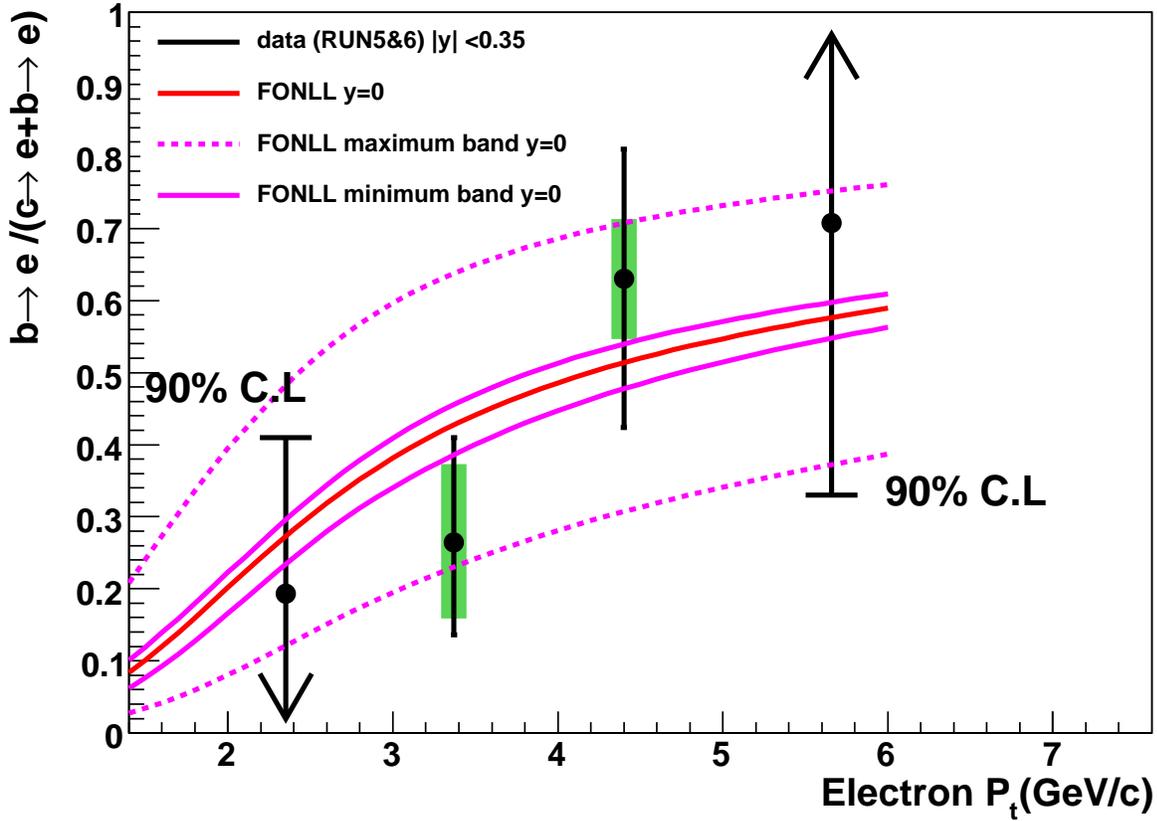}
    \caption{$(b\rightarrow e)/(c\rightarrow e+ b\rightarrow e)$ in the electrons from heavy flavor 
      as a fiction of electron $p_{\mathrm{T}}$
      in RUN6 and RUN5 with FONLL calculation. 
      Black points show the result in RUN6 and RUN5.
      Red lines are FONLL prediction and pink solid and dotted lines are
      uncertainty of FONLL prediction.
    }
    \label{chap7_fig5}
  \end{center}
\end{figure}

Figure~\ref{chap7_fig5} shows the combined result about the bottom fraction,
$(b\rightarrow e)/(c\rightarrow e+ b\rightarrow e)$
as a function of electron $p_{\mathrm{T}}$ with FONLL prediction~\cite{bib:fonll1,bib:fonll5}.
In this figure, black points show the the obtained $(b\rightarrow e)/(c\rightarrow e+ b\rightarrow e)$.
Red line show the central value in FONLL prediction
and pink solid and dotted lines show the uncertainty of FONLL calculation.
Pink solid lines in Fig.~\ref{chap7_fig5} show $(b\rightarrow e)/(c\rightarrow e+ b\rightarrow e)$
of the FONLL prediction when the correlation of the uncertainty about cross sections of charm and bottom 
are maximum.
Pink dotted lines show $(b\rightarrow e)/(c\rightarrow e+ b\rightarrow e)$
of the FONLL prediction when the correlation of the uncertainty about cross sections of charm and bottom 
are anti-maximum.
The results are also summarized in Table~\ref{chap7_table20}.
FONLL is almost consistent with the obtained result within the theoretical uncertainty.

It is worth to note that the point at lowest $p_{\mathrm{T}}$  has a small value.
This suggests majority of interested yield of 'single non-photonic electron', that is the electron after the subtraction of
all possible background and what we have been measured, can be explained as semi-leptonic $c\rightarrow e$
decay.
It provides the proof that the indirect measurement of heavy flavor via electrons performed at PHENIX
is really measurement of heavy flavor.

\begin{table}[hbt]
    \begin{center}
     \caption{Result of $(b\rightarrow e)/(c\rightarrow e+ b\rightarrow e)$ in RUN5 and RUN6 }
     \label{chap7_table20}
     \begin{tabular}{|c|c|}
       \hline
        2.35~GeV/$c$ & $<0.41$~(90\% C.L) 0.19~(50\% point) \\
       \hline
       3.37~GeV/$c$ & $0.26^{+0.14}_{-0.13}({\rm stat}) ^{+0.11}_{-0.11} ({\rm sys})  $     \\
       \hline
       4.40~GeV/$c$ & $0.63^{+0.18}_{-0.21} ({\rm stat}) \pm 0.08 ({\rm sys})  $     \\
       \hline
       5.66~GeV/$c$ & $>0.33$~(90\% C.L) 0.71~(50\% point)  \\
       \hline
     \end{tabular}
    \end{center}
\end{table}

\section{Comparison of the Data with Simulation}
The invariant mass~(M$_{eh}$) distributions of extracted signals in the data shown at Fig~\ref{chap5_fig8} and 
\ref{chap5_fig9} are compared with those generated by PYTHIA and EvtGen simulation.
The distributions are normalized by number of the single non-photonic electrons.
An agreement of the simulation results with the data provide  the confidence 
for this analysis method and the result.

\begin{figure}[htb]
  \begin{center}
    \includegraphics[width=10.5cm]{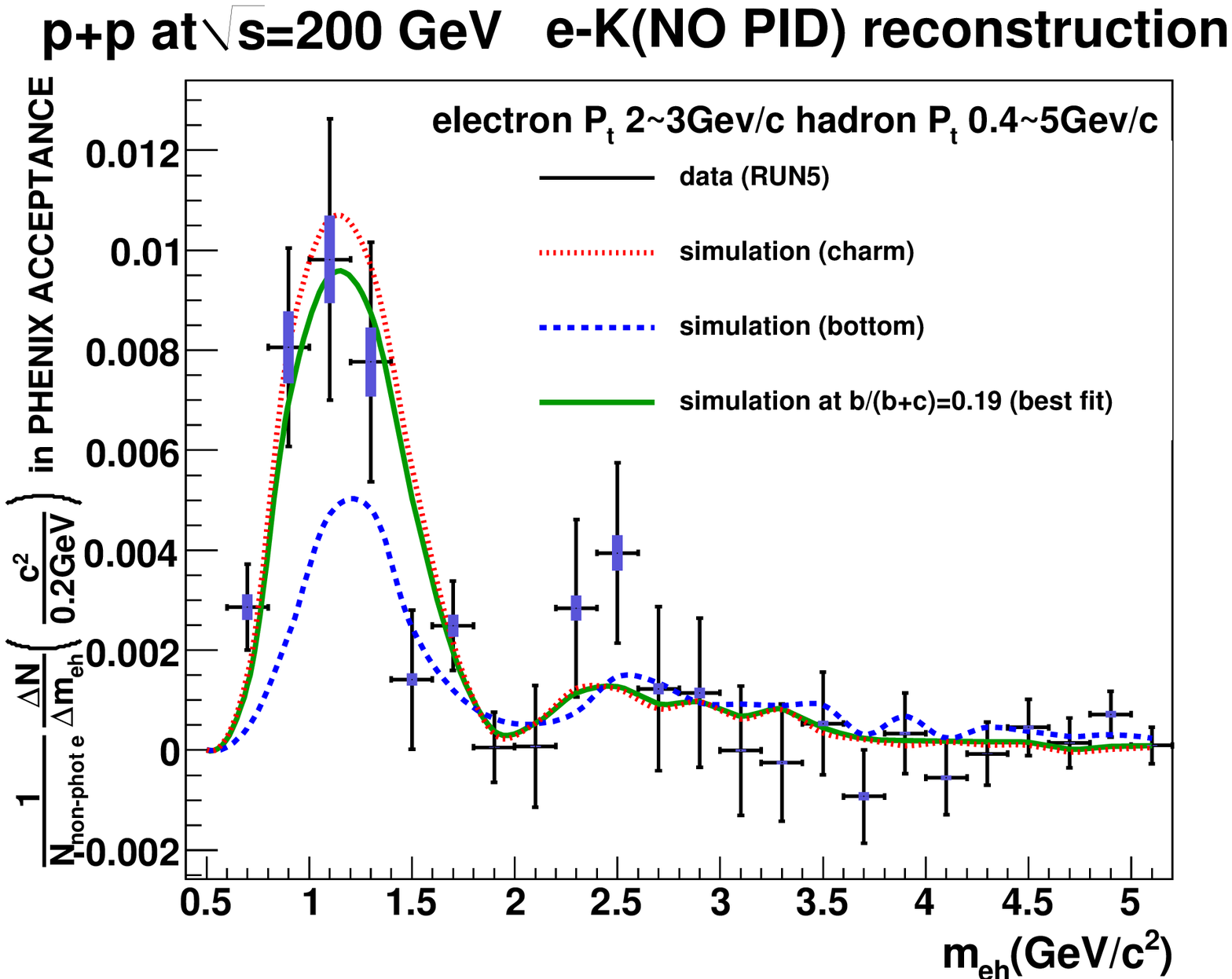}
    \caption{Comparison of the data with PYTHIA and EvtGen simulation 
      about subtracted invariant mass distributions in RUN5.
      Electron $p_{\mathrm{T}}$ range is 2.0-3.0~GeV/$c$.
    }
    \label{chap7_fig6}
  \end{center}
\end{figure}

\begin{figure}[htb]
  \begin{center}
    \includegraphics[width=10.5cm]{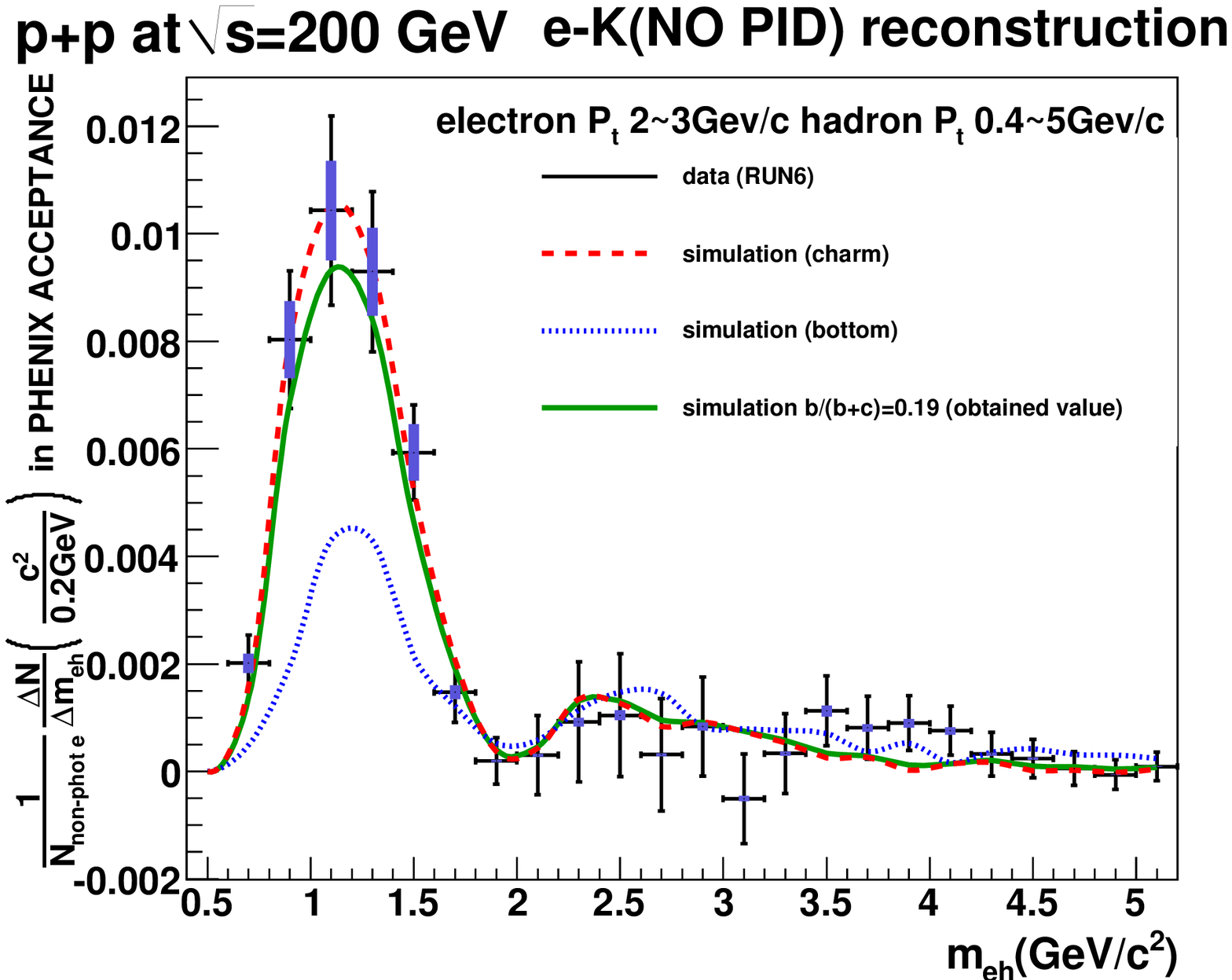}
    \caption{Comparison of the data with PYTHIA and EvtGen simulation 
      about subtracted invariant mass distributions in RUN6.
      Electron $p_{\mathrm{T}}$ range is 2.0-3.0~GeV/$c$.
    }
    \label{chap7_fig7}
  \end{center}
\end{figure}

\begin{figure}[htb]
  \begin{center}
    \includegraphics[width=10.5cm]{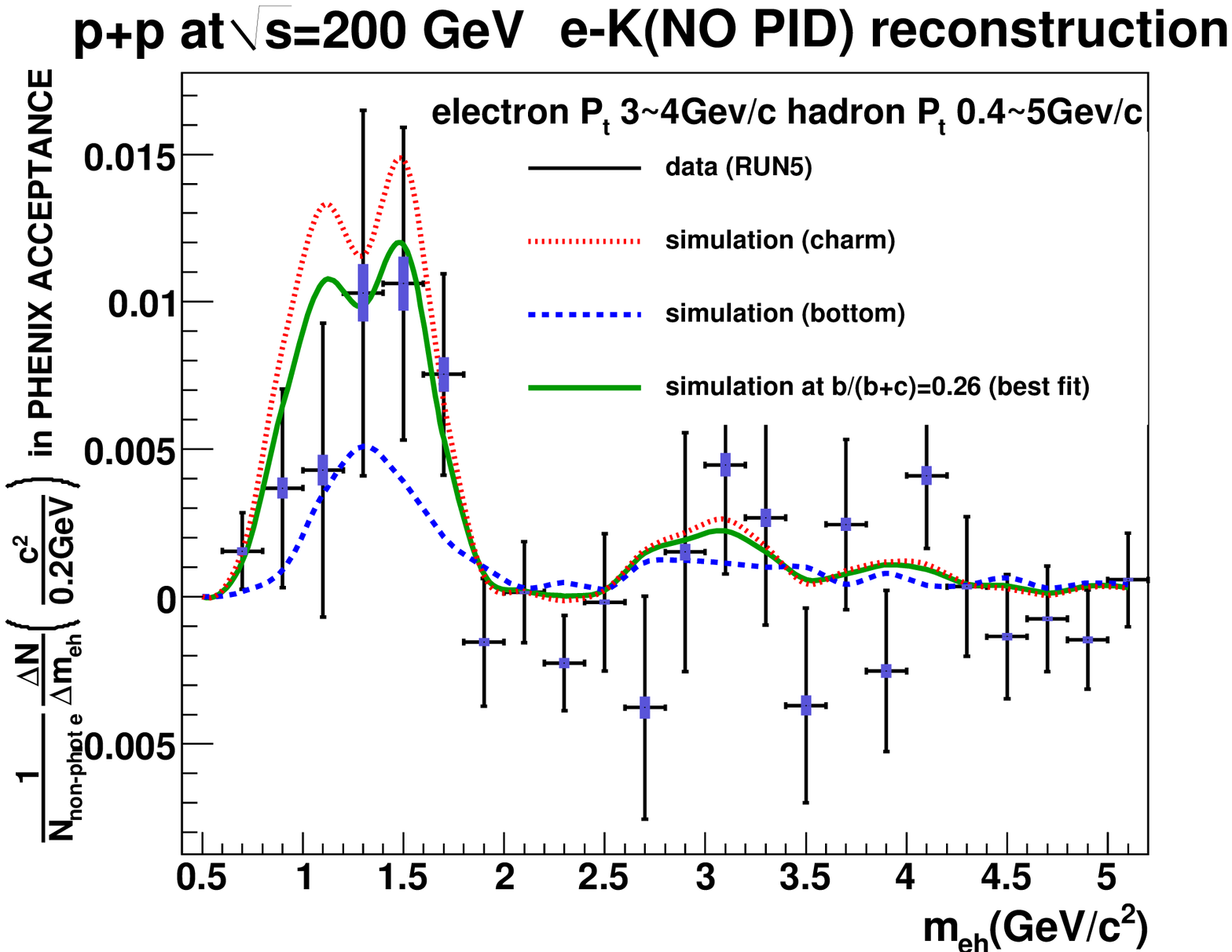}
    \caption{Comparison of the data with PYTHIA and EvtGen simulation 
      about subtracted invariant mass distributions in RUN5.
      Electron $p_{\mathrm{T}}$ range is 3.0-4.0~GeV/$c$.
    }
    \label{chap7_fig8}
  \end{center}
\end{figure}

\begin{figure}[htb]
  \begin{center}
    \includegraphics[width=10.5cm]{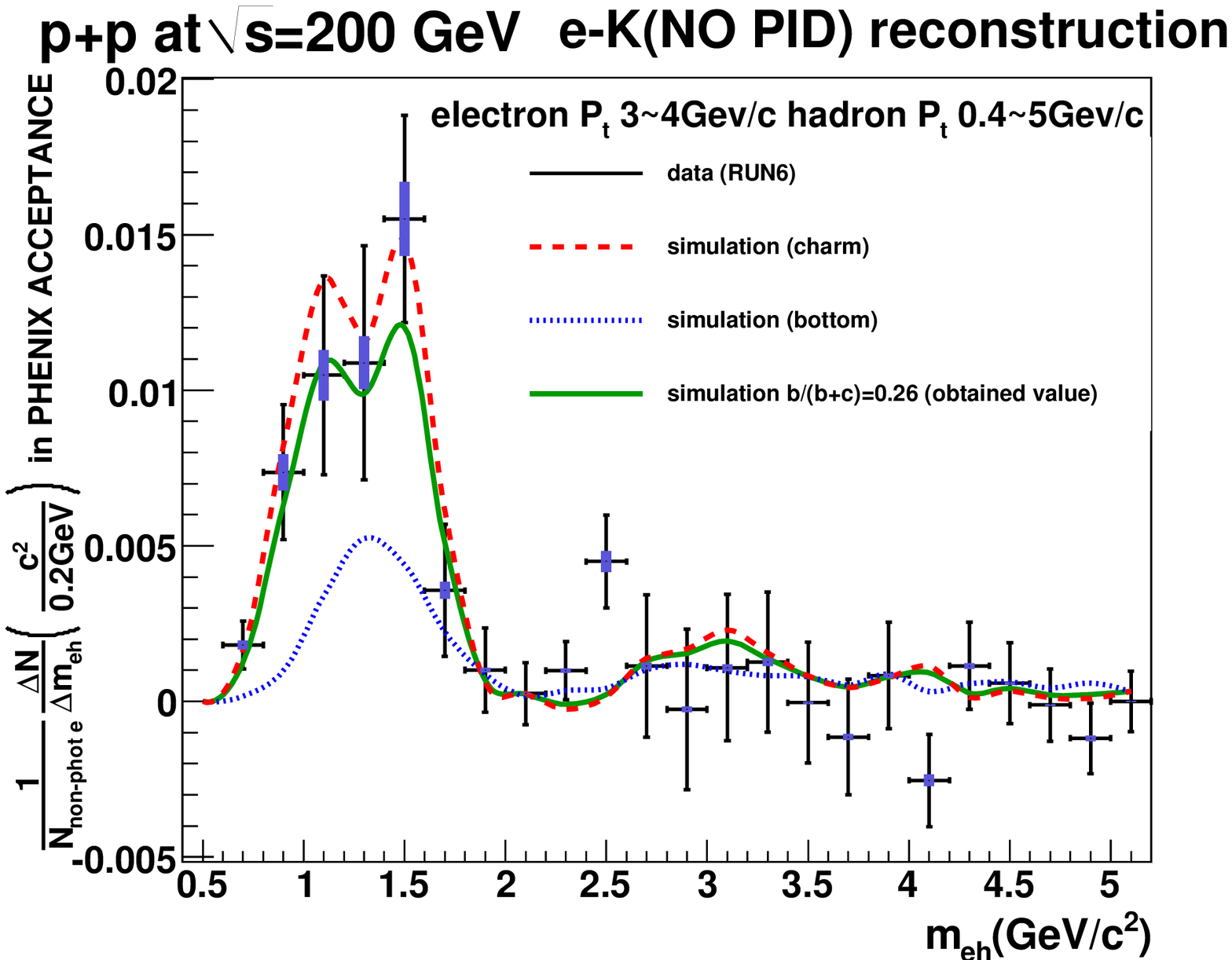}
    \caption{Comparison of the data with PYTHIA and EvtGen simulation 
      about subtracted invariant mass distributions in RUN6.
      Electron $p_{\mathrm{T}}$ range is 3.0-4.0~GeV/$c$.
    }
    \label{chap7_fig9}
  \end{center}
\end{figure}

\begin{figure}[htb]
  \begin{center}
    \includegraphics[width=10.5cm]{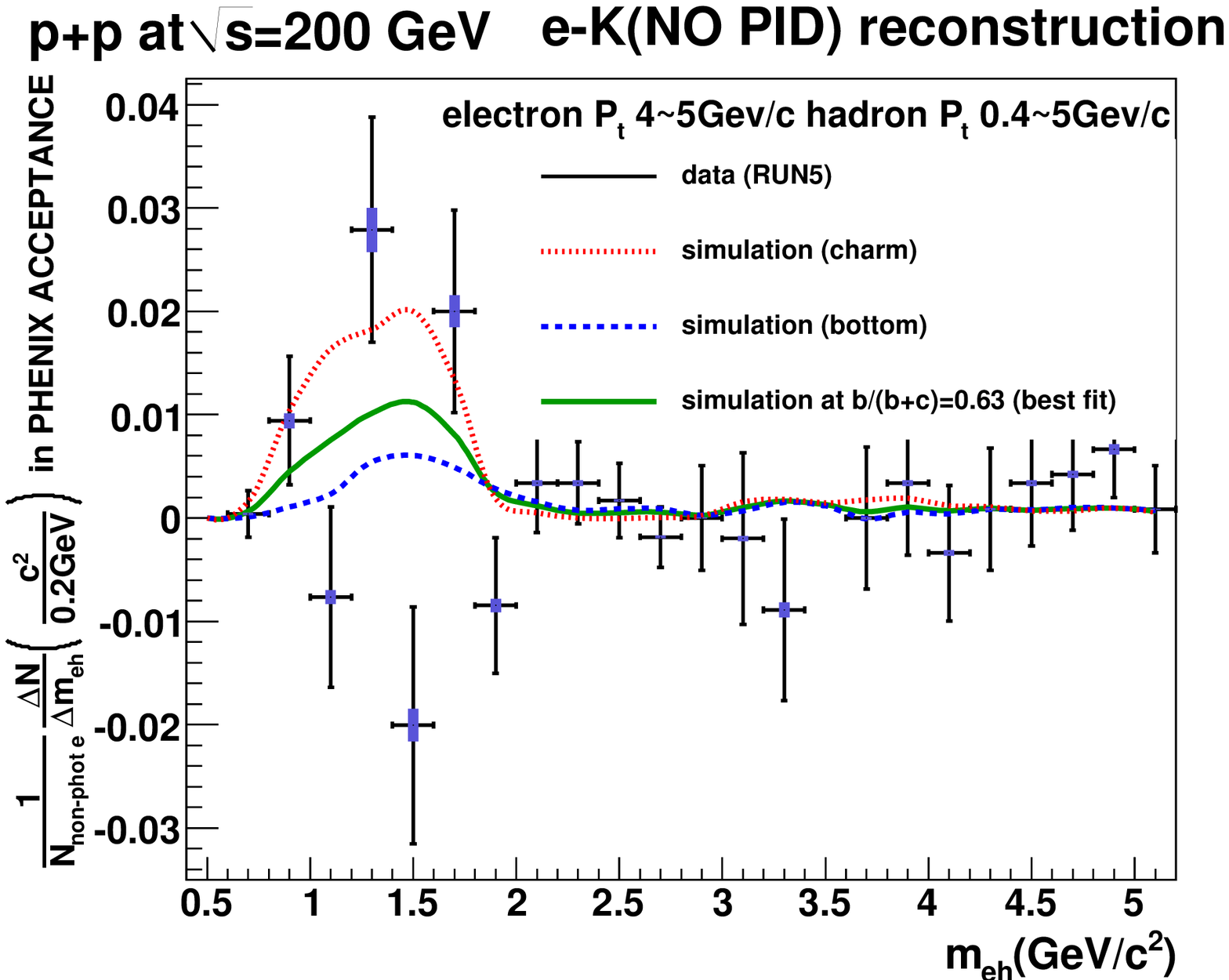}
    \caption{Comparison of the data with PYTHIA and EvtGen simulation 
      about subtracted invariant mass distributions in RUN5.
      Electron $p_{\mathrm{T}}$ range is 4.0-5.0~GeV/$c$.
    }
    \label{chap7_fig10}
  \end{center}
\end{figure}

\begin{figure}[htb]
  \begin{center}
    \includegraphics[width=10.5cm]{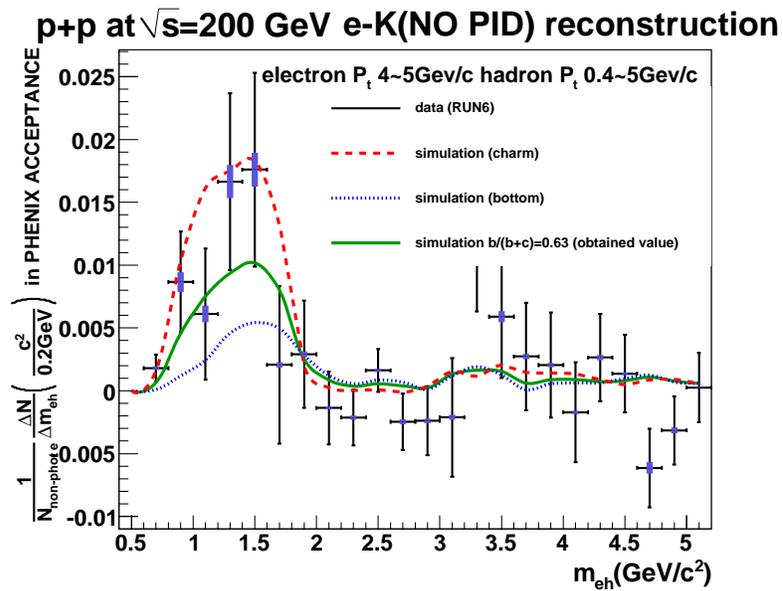}
    \caption{Comparison of the data with PYTHIA and EvtGen simulation 
      about subtracted invariant mass distributions in RUN6.
      Electron $p_{\mathrm{T}}$ range is 4.0-5.0~GeV/$c$.
    }
    \label{chap7_fig11}
  \end{center}
\end{figure}

Figure~\ref{chap7_fig6} to \ref{chap7_fig11} show the invariant mass distributions of extracted signals 
in the data and the simulation at each electron $p_{\mathrm{T}}$ range in RUN5 and RUN6.
In these figures, black points are the data points.
Red points show the result of PYTHIA and EvtGen simulation only for charm production and 
blue points show the result of PYTHIA and EvtGen simulation only for bottom production.
Green points show the result of the simulation which is obtained by combining 
the charm and bottom contributions according to the 
obtained $(b\rightarrow e)/(c\rightarrow e+ b\rightarrow e)$ values.

The agreement of simulation~(green) and real data is good.
$\chi^2/ndf$ values which are calculated in $0.4<M_{eh}<5.0$~GeV/$c^2$ are summarized 
in Table~\ref{chap7_table2}.
Uncertainty of simulation is NOT included in these ($\chi^2/ndf$).
\begin{table}[hbt]
    \begin{center}
     \caption{$\chi^2/ndf$ (theoretical uncertainty is NOT included)}
     \label{chap7_table2}
     \begin{tabular}{|c|cc|}
       \hline
       electron $p_{\mathrm{T}}$ & $\chi^2/ndf$(RUN5) & $\chi^2/ndf$(RUN6)  \\
       \hline 
       2.0-3.0~GeV/$c$ &  20.3/22 &  17.2/22\\
       3.0-4.0~GeV/$c$ &  15.5/22 & 21.2/22\\
       4.0-5.0~GeV/$c$ &  28.0/22 & 23.1/22\\
       \hline
     \end{tabular}
    \end{center}
\end{table}


\clearpage
\section{Cross Section of Bottom}
Cross section of bottom is obtained using the spectrum of the electrons from heavy flavor
and the ratio, $(b\rightarrow e)/(c\rightarrow e+ b\rightarrow e)$.
\subsection{Invariant Cross Section of Electrons from Charm and Bottom}
The differential invariant cross section of the electrons from semi-leptonic decay of charm and 
that from bottom are obtained by (electron spectrum from heavy flavor) $\times$  
$(c(b)\rightarrow e)/(c\rightarrow e+ b\rightarrow e)$.
$(b\rightarrow e)/(c\rightarrow e+ b\rightarrow e)$ is obtained in the four electron $p_{\mathrm{T}}$
range, 2-3GeV/$c$, 3-4GeV/$c$, 4-5GeV/$c$ and 5-7GeV/$c$.
The spectrum of the single non-photonic electrons  is  merged into 
these electron $p_{\mathrm{T}}$ range to make the same bin width as 
$(b\rightarrow e)/(c\rightarrow e+ b\rightarrow e)$.
The yield at the electron $p_{\mathrm{T}}$, where $(b\rightarrow e)/(c\rightarrow e+ b\rightarrow e)$ 
is obtained, is calculated as follows.
\begin{equation}
Y(p_{\mathrm{T}}^{real}) = \frac{f(p_{\mathrm{T}}^{real}) }
{f(p_{\mathrm{T}}^{0}) } \times Y(p_{\mathrm{T}}^{0}).
\end{equation}
Here,   $p_{\mathrm{T}}^{real}$ is the bin values of 
  electron $p_{\mathrm{T}}$ where $(b\rightarrow e)/(c\rightarrow e+ b\rightarrow e)$ is obtained,
  $p_{\mathrm{T}}^{0}$ is the bin values where the electron spectrum is rebined,
  $f(p_{\mathrm{T}})$ is the fit function of the electron spectrum and 
 $Y(p_{\mathrm{T}})$ is the electron yield.



\begin{figure}[htb]
  \begin{center}
    \includegraphics[width=17.cm]{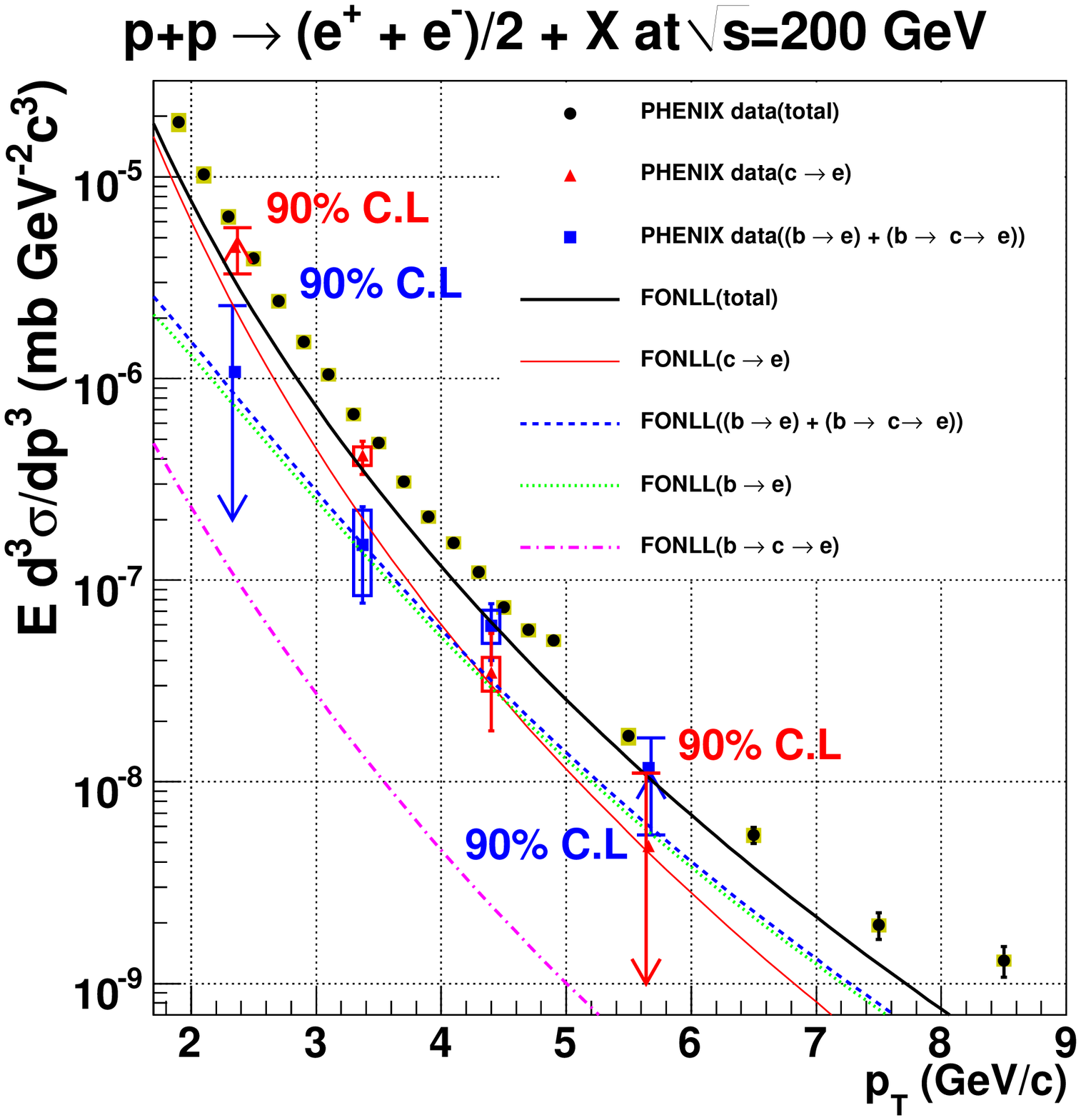}
    \caption{Invariant cross sections 
      of electrons from charm and bottom with FONLL calculation.
    }
    \label{chap7_fig132}
  \end{center}
\end{figure}

Figure~\ref{chap7_fig132} shows the invariant cross section 
of single electrons from charm and those from bottom with FONLL calculation.
The spectrum of single electrons~(circles) is also shown as a reference.
The results are also summarized in Table~\ref{chap7_table51}.

\begin{table}[hbt]
    \begin{center}
     \caption{Invariant cross section of electrons from charm and bottom}
     \label{chap7_table51}
     \begin{tabular}{|c|c|c|}
       \hline
       electron $p_{\mathrm{T}}$ & cross section~(mb~GeV$^{-2}c^3$)& data/FONLL \\
       \hline \hline
       2.35~GeV/$c$ & $>3.30$~(90\% C.L) 4.52~(50\%) $\times 10^{-6}$ & $>1.49$~(90\% C.L) 2.03~(50\%)\\
       3.37~GeV/$c$ & $4.17^{+0.73}_{-0.83} {}^{+0.41}_{-0.46} \times 10^{-7}$ & $2.05^{+0.36}_{-0.41} {}^{+0.20}_{-0.22}$\\
       4.40~GeV/$c$ & $3.49^{+1.95}_{-1.70} \pm 0.66 \times 10^{-8}$ & $1.16^{+0.65}_{-0.56} \pm 0.22 $\\
       5.66~GeV/$c$ & $<1.11$~(90\% C.L) 0.48~(50\%) $\times 10^{-8}$& $<2.48$~(90\% C.L) 1.08~(50\%) \\
       \hline\hline
       bottom & &  \\
       2.35~GeV/$c$ &  $<2.30$~(90\% C.L) 1.08~(50\%) $\times 10^{-6}$&  $<2.74$~(90\% C.L) 1.29~(50\%)\\
       3.37~GeV/$c$ & $1.49^{+0.83}_{-0.73} {}^{+0.73}_{-0.66} \times 10^{-7}$ & $0.99^{+0.55}_{-0.48} {}^{+0.48}_{-0.43}$\\
       4.40~GeV/$c$ & $5.95^{+1.70}_{-1.95} \pm 1.10 \times 10^{-8}$ & $1.87^{+0.54}_{-0.61} \pm 0.34$\\
       5.66~GeV/$c$ & $>0.54$~(90\% C.L) 1.17~(50\%) $\times 10^{-8}$&  $>0.90$~(90\% C.L) 1.93~(50\%) \\
       \hline
     \end{tabular}
    \end{center}
\end{table}

\subsection{Total Cross Section of  Bottom}

Total cross section of bottom is obtained from the spectrum of the electron from
bottom.
The procedure to get total cross section of bottom can be written by following equations.
\begin{eqnarray}
  \left. \frac{d\sigma_{b\bar{b}}}{dy}\right|_{y=0} &=& \frac{1}{BR(b \rightarrow e)}
  \frac{1}{C_{e/B}} \frac{d\sigma_{b \rightarrow e}}{dy},\\
  \sigma_{b\bar{b} } &=& \int_y dy \frac{d\sigma_{b\bar{b}}}{dy} \sim 
  R \times \left.\frac{d\sigma_{b\bar{b} }}{dy}\right|_{y=0}.
\end{eqnarray}
The procedures to calculate above equations are following.
\begin{itemize}
\item $b \rightarrow c \rightarrow  e$ subtraction and 
  $p_{\mathrm{T}}$ extrapolation to obtain $d\sigma_{b \rightarrow e}/dy$.
\item  Kinematical correction~($C_{e/B}$)
\item  Branching ratio correction.
\item  Rapidity extrapolation. $R$ is a correction factor for 
  rapidity extrapolation
\end{itemize} 
Differential cross section of single electrons from bottom shown in Fig~\ref{chap7_fig132} 
is integrated from $p_{\mathrm{T}}=3$~GeV/$c$ to $p_{\mathrm{T}}=5$~GeV/$c$.
The points at $2<p_{\mathrm{T}}<3$~GeV/$c$ and $5<p_{\mathrm{T}}<7$~GeV/$c$ is dropped off for this integral, 
since confidence level is only determined at these region.
\begin{equation}
\left. \frac{d\sigma^{b\rightarrow e+b\rightarrow c \rightarrow e}}{dy} \right|_{y=0}(3<p_{\mathrm{T}}<5~GeV/c)
 = 0.0048^{+0.0018}_{-0.0016}({\rm stat}) {}^{+0.019}_{-0.018}({\rm sys}) \mu {\rm b}.
\end{equation}

\subsubsection{$p_{\mathrm{T}}$ Extrapolation}
\begin{figure}[htb]
  \begin{center}
    \includegraphics[width=11.cm]{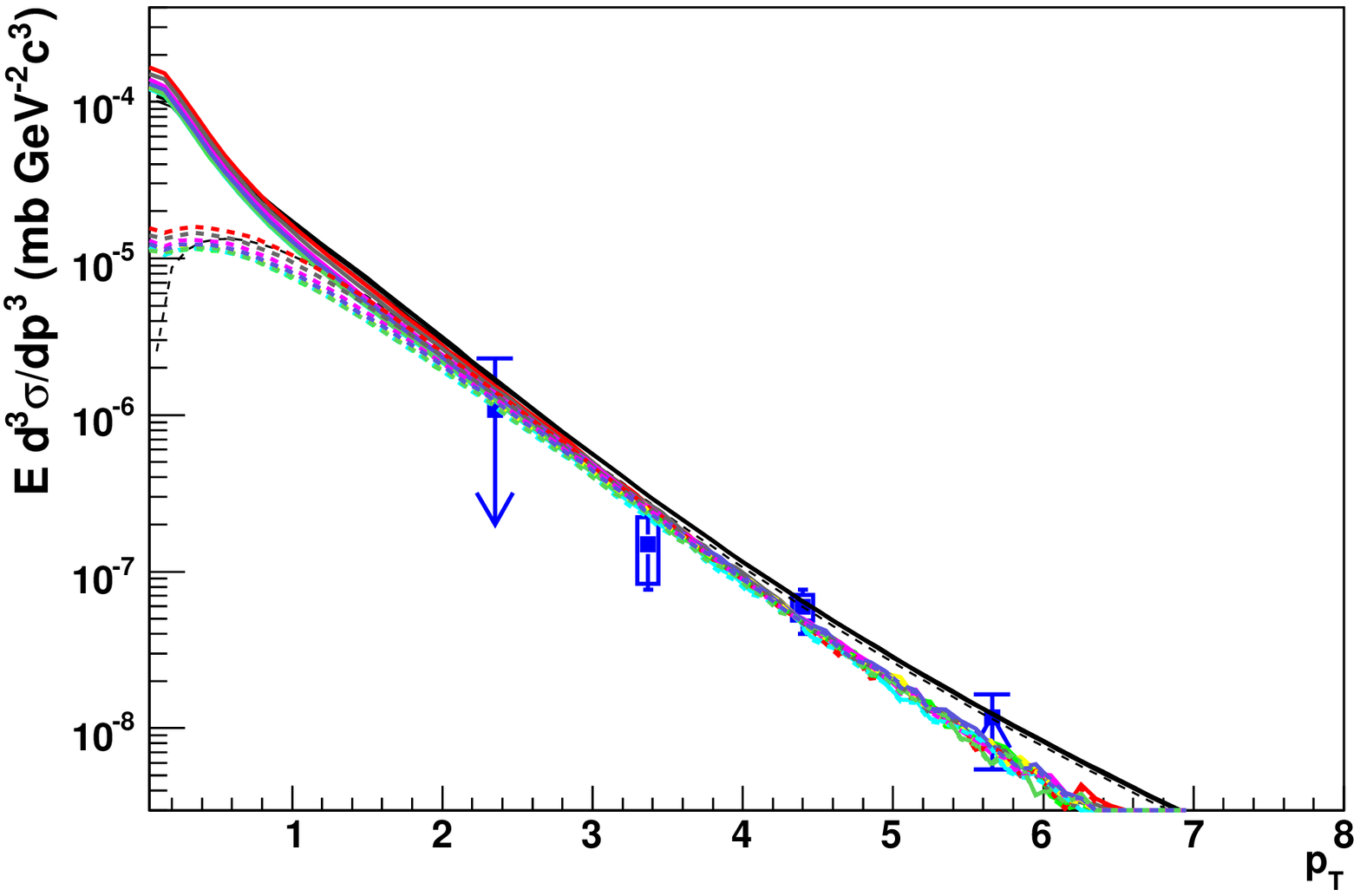}
    \caption{
      Invariant cross sections 
      of electrons from bottom from FONLL calculation and PYTHIA.
      The simulations include electron from $b \rightarrow c \rightarrow  e$
      and $b \rightarrow e$.
      Black line show FONLL calculation.
      Other lines show PYTHIA with $1.5<k_{\mathrm{T}}<10$~GeV/$c$.
      Dotted lines show electron from $b \rightarrow e$.
    }
    \label{chap7_fig32}
  \end{center}
\end{figure}

$b \rightarrow c \rightarrow  e$ subtraction and $p_{\mathrm{T}}$ extrapolation
are done by using PYTHIA and FONLL calculation.
Figure~\ref{chap7_fig32} shows invariant cross sections of the electrons from bottom 
with FONLL calculation and PYTHIA  with $1.5<k_{\mathrm{T}}<10$~GeV/$c$.
Solid lines show the electron from $b \rightarrow c \rightarrow  e$
and $b \rightarrow e$ and dotted lines show the electron from $b \rightarrow e$.
In Fig~\ref{chap7_fig32} black line show FONLL calculation and
other lines show PYTHIA with $1.5<k_{\mathrm{T}}<10$~GeV/$c$. 
The distribution of the  simulations are normalized at 4-5~GeV/$c$ points.
Correction factor is determined from the simulation as follows. \\
\begin{equation}
\frac{ (d\sigma^{b\rightarrow e})/(dy)_{\mid y=0}} 
  {(d\sigma^{b\rightarrow e+b\rightarrow c \rightarrow e})/(dy)_{\mid y=0}
    (3<p_{\mathrm{T}}<5~GeV/c)}.
\end{equation}
Obtained correlation factors from simulation are summarized 
in Table~\ref{chap7_table52}.
\begin{table}[hbt]
    \begin{center}
     \caption{Correction factors for $p_{\mathrm{T}}$ extrapolation and 
       $b \rightarrow c \rightarrow  e$ subtraction}
     \label{chap7_table52}
     \begin{tabular}{cc}
       simulation & Correction factor \\
       \hline \hline
       PYTHIA $k_{\mathrm{T}}$1.5  & 18.6\\
       PYTHIA $k_{\mathrm{T}}$2.5  & 16.9\\
       PYTHIA $k_{\mathrm{T}}$3.5  & 15.9\\
       PYTHIA $k_{\mathrm{T}}$5.0  & 15.3\\
       PYTHIA $k_{\mathrm{T}}$7.5  & 14.9\\
       PYTHIA $k_{\mathrm{T}}$10.0  & 14.7\\
       FONLL $p_{\mathrm{T}}$ scaling &  18.1 \\
       FONLL $p_{\mathrm{T}}$ scaling~(max) &  18.8 \\
       FONLL $p_{\mathrm{T}}$ scaling~(min) &  17.0 \\
     \end{tabular}
    \end{center}
\end{table}
We take 16.8 as the correction factor for the $p_{\mathrm{T}}$ extrapolation 
and $b \rightarrow c \rightarrow  e$ subtraction.
We assigned 2.0 as systematic error to cover PYTHIA simulation and FONLL  results.

\subsubsection{Kinematical Correction}
\begin{figure}[htb]
  \begin{tabular}{c c}
    \begin{minipage}{\minitwocolumn}
      \begin{center}
	\includegraphics[width=6.5cm]{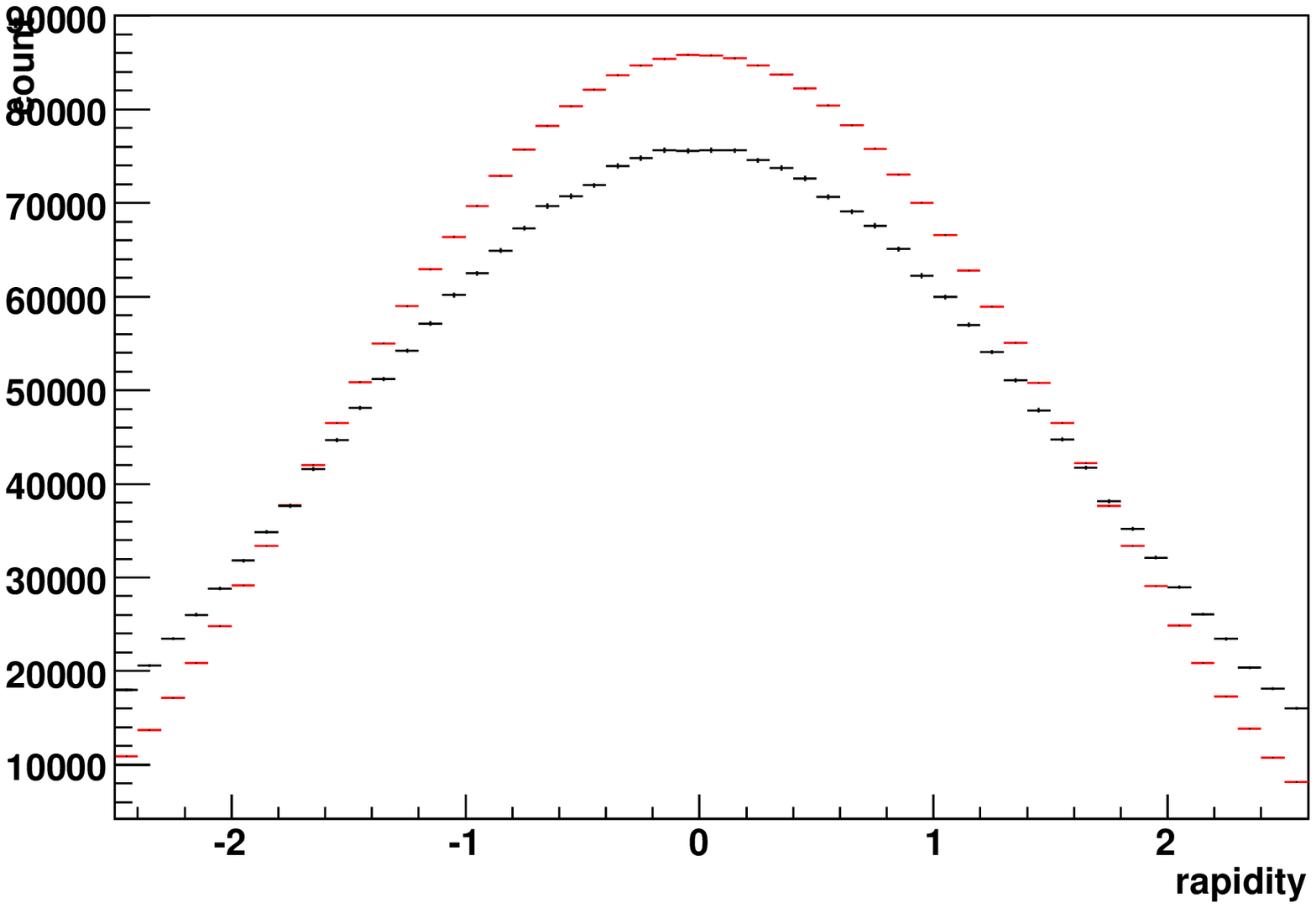}
        \caption{
	  Rapidity distribution of B hadron and electrons from
	  bottom at PYTHIA.
	  Black points show the distribution of  electrons and 
	  blue points show that of  B hadron.
	}
        \label{chap7_fig33}
      \end{center}
    \end{minipage}
    &
    \begin{minipage}{\minitwocolumn}
      \begin{center}
	\includegraphics[width=6.5cm]{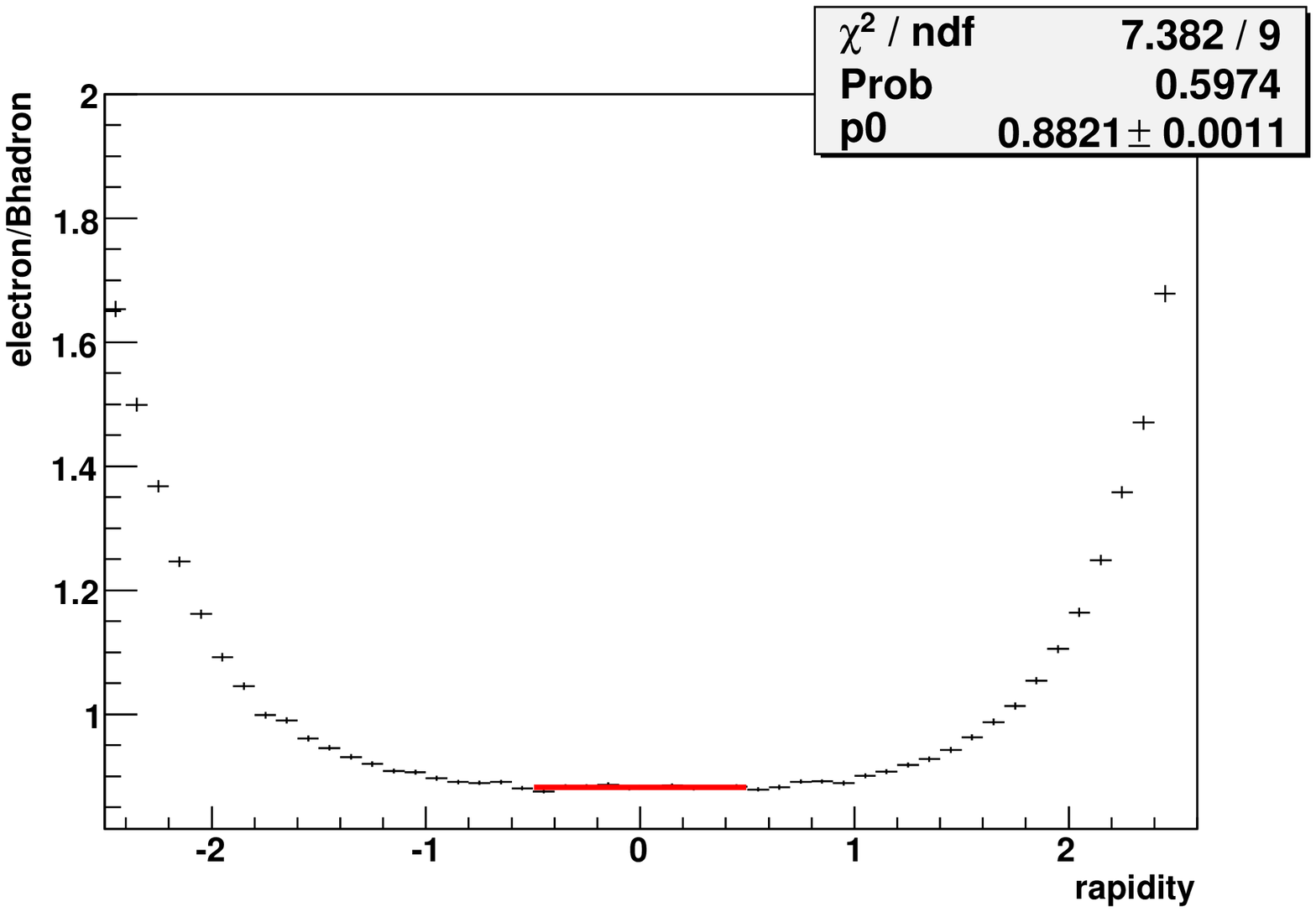}
        \caption{
	  Ratios of the rapidity distributions shown at Fig.\ref{chap7_fig33}
	  }
        \label{chap7_fig34}
      \end{center}
    \end{minipage}
  \end{tabular}
\end{figure}

The kinematical correlation factor, $C_{e/B}$ is applied to account for 
the difference in rapidity distribution of the electron from bottom and B hadron.\\
PYTHIA simulation is used to determine a kinematical correction factor, $C_{e/B}$.\\
Figure~\ref{chap7_fig33} shows the rapidity distribution of B hadron 
and the electrons from bottom at PYTHIA.
Black points show the rapidity distribution of  the electrons and 
red points show that of  B hadron.
Figure~\ref{chap7_fig34} shows the ratios of the rapidity distributions of
the electrons over that of B hadron shown at Fig.\ref{chap7_fig33}.
$C_{e/B}$ is determined by straight line fit of this ratios.
We use 0.88 as $C_{e/B}$.\\
Systematic error is not assigned for this correction, since 
this correction factor is determined by pure kinematics.

\subsubsection{Branching ratio}
\begin{table}[tbt]
    \begin{center}
     \caption{Electron branching ratios of bottom hadrons}
     \label{chap7_table54}
     \begin{tabular}{|c|c|}
       \hline
       hadron & BR(e) \\
       \hline
       $B^{+(-)}$  & 10.8$\pm$ 0.4       \\

       $B^{0}$  & 10.1$\pm$ 0.4       \\

       $B_s$   & 7.9$\pm$ 2.4       \\

       B baryons & 8.6$\pm$ 2.5       \\
       \hline
     \end{tabular}
    \end{center}
\end{table}

Inclusive $BR(b \rightarrow e)$ is calculated from the production ratios of B hadron
and their exclusive electron branching ratios.
Their exclusive electron branching ratios are summarized in Table~\ref{chap7_table54}.
As a result, inclusive $BR(b \rightarrow e)$ is determined to be 10.0\% $\pm$ 1\%.
This assignment is conservative and also cover LEP result, 10.8\%.\\
$d\sigma_{b\bar{b}}/dy_{\mid y=0}$ and 
$d\sigma^{b \rightarrow e}/dy_{\mid y=0}$ 
are obtained from above correction factors.
The results are as follows.

\begin{equation}
  \left.\frac{d\sigma^{b \rightarrow e}}{dy}\right|_{y=0} =
 0.081 ^{+0.030}_{-0.027}({\rm stat})  {}^{+0.034}_{-0.027}({\rm sys}) \mu {\rm b}.
\end{equation}

\begin{equation}
  \left.\frac{d\sigma_{b\bar{b}}}{dy}\right|_{y=0}  =
 0.92 ^{+0.035}_{-0.031}({\rm stat}) {}^{+0.39}_{-0.36}({\rm sys}) \mu {\rm b}.
\end{equation}

\begin{figure}[htb]
  \begin{center}
    \includegraphics[width=12.cm]{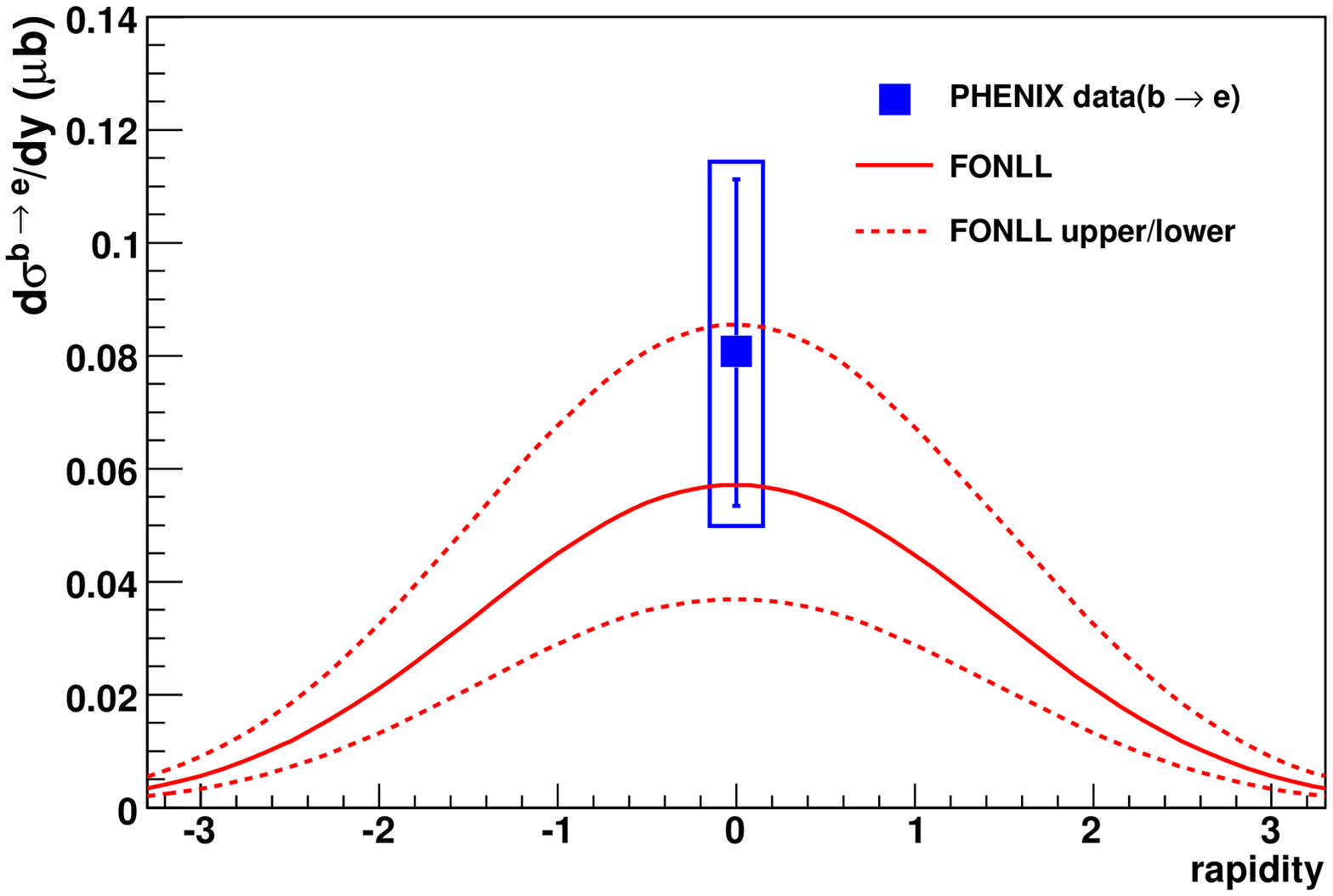}
    \caption{
      $d\sigma^{b \rightarrow e}/dy_{\mid y=0}$
      with FONLL prediction as a function of rapidity.
    }
    \label{chap7_fig201}
  \end{center}
\end{figure}

\begin{figure}[htb]
  \begin{center}
    \includegraphics[width=12.cm]{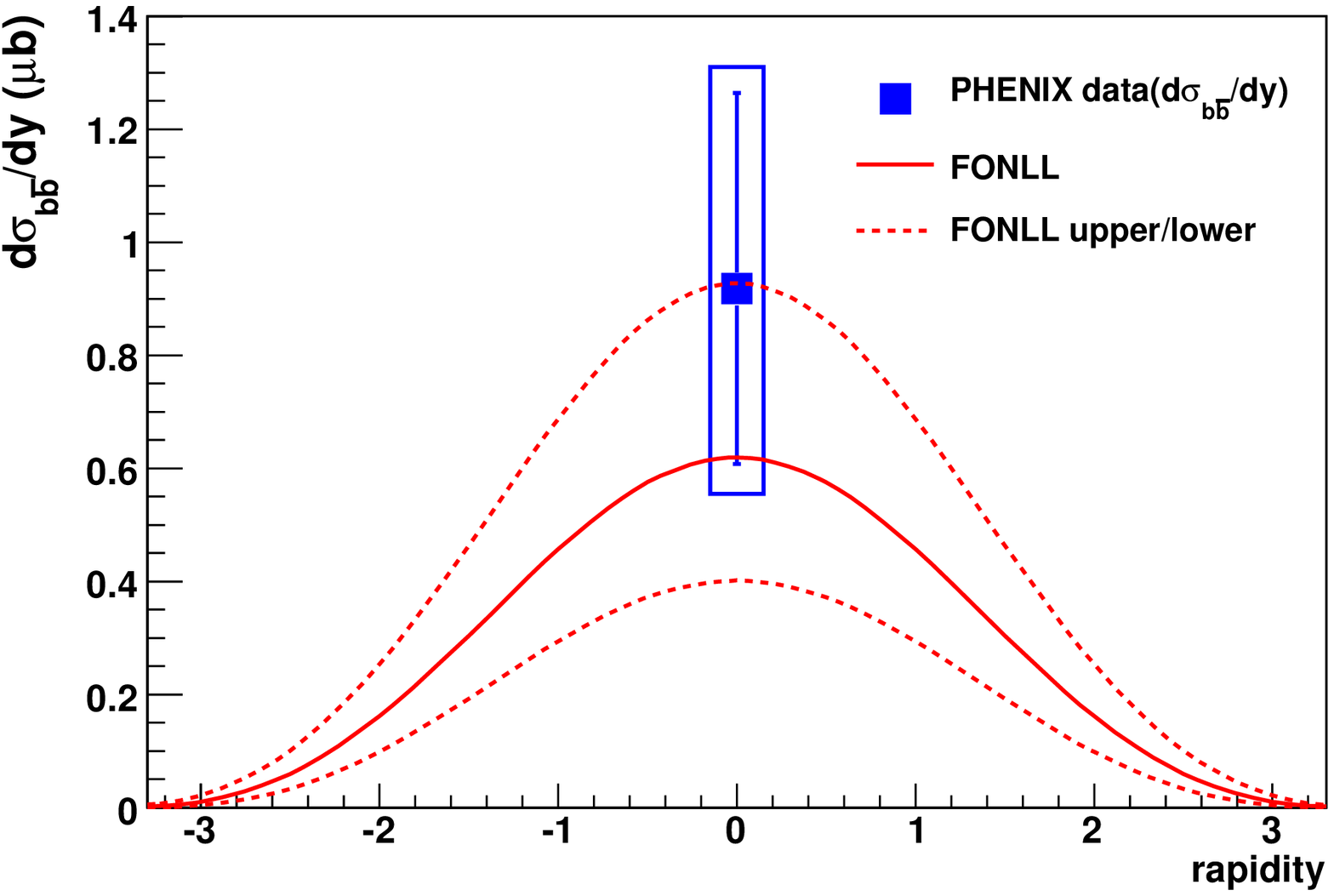}
    \caption{
      $d\sigma_{b\bar{b}}/dy_{\mid y=0}$
      with FONLL prediction as a function of rapidity.
    }
    \label{chap7_fig202}
  \end{center}
\end{figure}

Figure~\ref{chap7_fig201} shows $d\sigma^{b \rightarrow e}/dy_{\mid y=0}$
with FONLL prediction as a function of rapidity.
Figure~\ref{chap7_fig202} shows $d\sigma_{b\bar{b}}/dy_{\mid y=0}$
with FONLL prediction as a function of rapidity.
The FONLL prediction is very consistent with the experimental result.

\subsubsection{Rapidity Extrapolation}
\begin{figure}[htb]
  \begin{center}
    \includegraphics[width=10.cm]{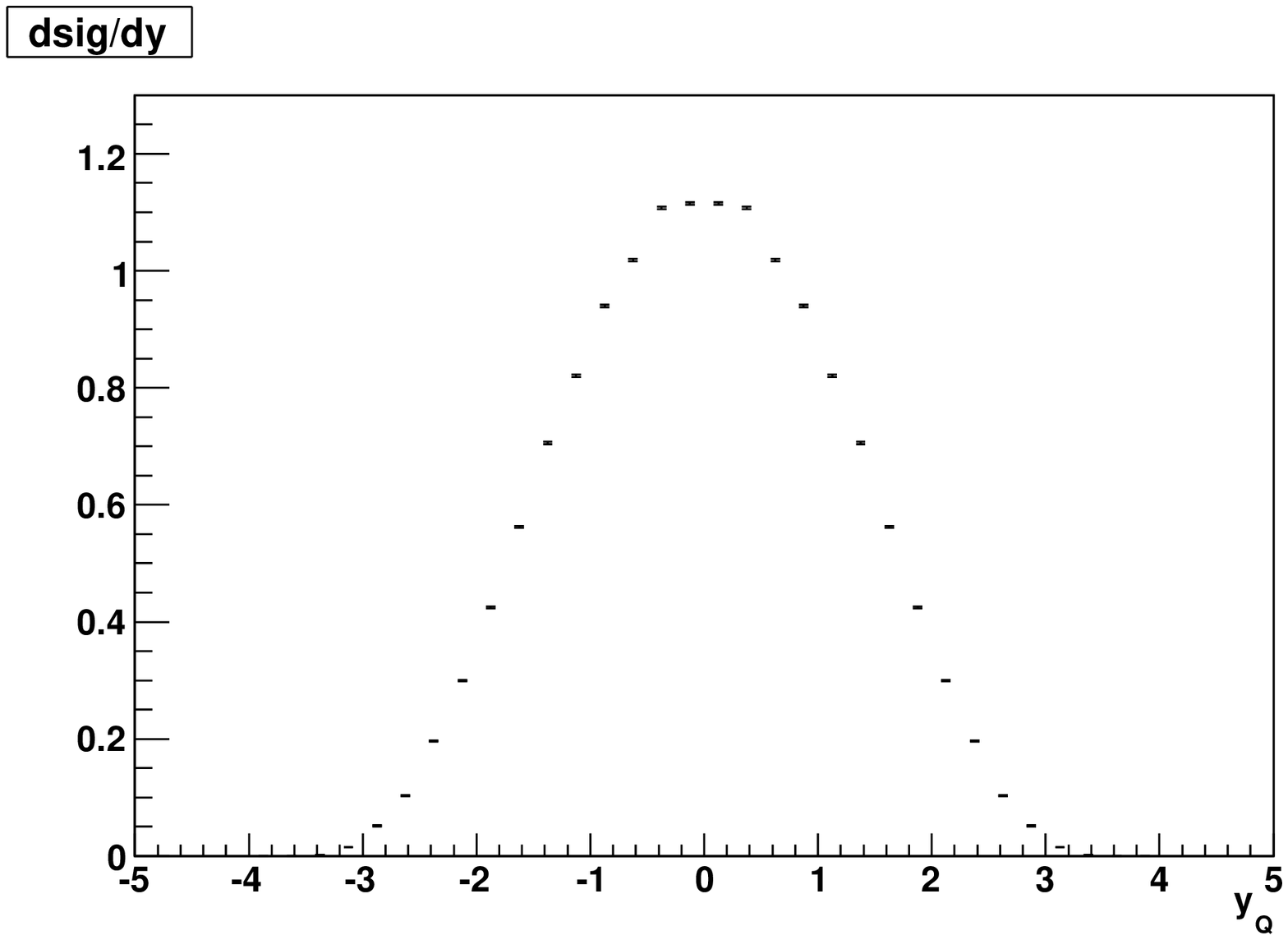}
    \caption{
      Rapidity distribution of 
      bottom quark at HVQMNR using CTEQ5M as parton distribution function.
    }
    \label{chap7_fig35}
  \end{center}
\end{figure}

Total cross section of bottom is obtained by extrapolating
rapidity of B hadron.
The correction factor for rapidity extrapolation is determined by using the simulation
as follows
\begin{equation}
R = \frac{\int dy \frac{d\sigma}{dy}(B\quad hadron)} 
{\int_{-0.5}^{0.5} dy \frac{d\sigma}{dy}(B\quad hadron)}.
\end{equation}
The correction factor by rapidity distribution of B hadron at PYTHIA 
is 3.30.
NLO calculation for heavy quark production~(HVQMNR) is also used for 
rapidity extrapolation~\cite{bib10}, since PYTHIA is just LO calculation.
Figure~\ref{chap7_fig35} show rapidity distribution of 
bottom quark at HVQMNR using CTEQ5M as parton distribution function for example.

Generated rapidity distribution at HVQMNR is that of bare b quark, 
while we should integrate cross section for rapidity of B hadron.
Generated rapidity distribution of bare b quark is expected to differ from 
that of bare b quark slightly by the fragmentation process.\\
The correction factor for the difference in the rapidity distribution between bare b quarks 
and B hadrons is estimated by the similar way to determine $C_{e/B}$.
0.96 is used for the correction for the fragmentation of bare b into  B hadrons.\\
Therefore the rapidity correction is done as follows.\\
\begin{equation}
R=\frac{\int dy \frac{d\sigma}{dy}(B\quad hadron)} 
{\int_{-0.5}^{0.5} dy \frac{d\sigma}{dy}(B\quad hadron)}
=\frac{\int dy \frac{d\sigma}{dy}(bare\quad b)} 
{\int_{-0.5}^{0.5} dy \frac{d\sigma}{dy}(bare\quad b)} \times \frac{1}{0.96}.
\end{equation}

The correction factor is calculated by using HVQMNR at various conditions.
Results are summarized in Table~\ref{chap7_table55}.
CTEQ5M value, 3.44 is used as the correction factor.
Systematic error for the correction factor is assigned 0.25 to 
cover the results at various conditions.
$\sigma_{bb}$ is obtained from the correction factor for 
rapidity extrapolation.
The result is:
\begin{equation}
   \sigma_{b\bar{b}}=
  3.16 ^{+1.19}_{-1.07}({\rm stat}) {}^{+1.37}_{-1.27}({\rm sys}) \mu {\rm b}.
\end{equation}
The FONLL predicts $\sigma_{b\bar{b}}=1.87^{+0.99}_{-0.67} \mu {\rm b}$ and agrees with the experimental result.

%

\begin{table}[hbt]
    \begin{center}
     \caption{Correction factors for rapidity extrapolation }
     \label{chap7_table55}
     \begin{tabular}{c|cc}
       simulation condition & Correction factor(bare b) & Correction factor(B hadron)\\
       \hline \hline
       CTEQ4M(PDF) & 3.37 & 3.51 \\
       CTEQ5M1(PDF) & 3.22& 3.35 \\
       CTEQ5M(PDF) & 3.30 &3.44\\
       CTEQ5HQ(PDF) &3.37 & 3.51\\
       GRVHO(PDF) &3.44 & 3.58\\
       CTEQ5M(PDF) b mass 4.5~GeV &3.20 &  3.33\\
       PYTHIA   & &  3.30\\
     \end{tabular}
    \end{center}
\end{table}
   \chapter{Discussion}
In this chapter, we discuss implications of the experimental results
on the electrons from semi-leptonic decay of heavy flavors~(single non-photonic electrons) 
in $p+p$ and Au+Au collisions.
The measured $p_{\mathrm{T}}$ distributions of single electrons from charm and bottom
are compared with pQCD calculation in Sec.~7.1.
Total cross section of bottom is compared with the measured result via di-electron spectrum in PHENIX
in Sec.~7.2.
Total cross sections of charm and bottom in world data measured in hadron colliders are presented in Sec.~7.3.
An implications of the $R_{AA}$ and $v_2$ in Au+Au collisions are discussed based on 
the measured ratio,  $(b\rightarrow e)/(c\rightarrow e+ b\rightarrow e)$, in Sec.~7.4.
The $R_{AA}$ and $v_2$ in Au+Au collisions  are compared with latest theoretical models in Sec.~7.5
Finally, important measurements about heavy flavor in the near future are discussed in Sec.~7.6.

\section{Comparison with Perturbative QCD}
The measured yield of heavy flavor in $p+p$ collisions provides a good test of perturbative QCD.
For this purpose, the ratios of measured yield over the FONLL prediction are studied.
\begin{figure}[htb]
  \begin{center}
    \includegraphics[width=14cm]{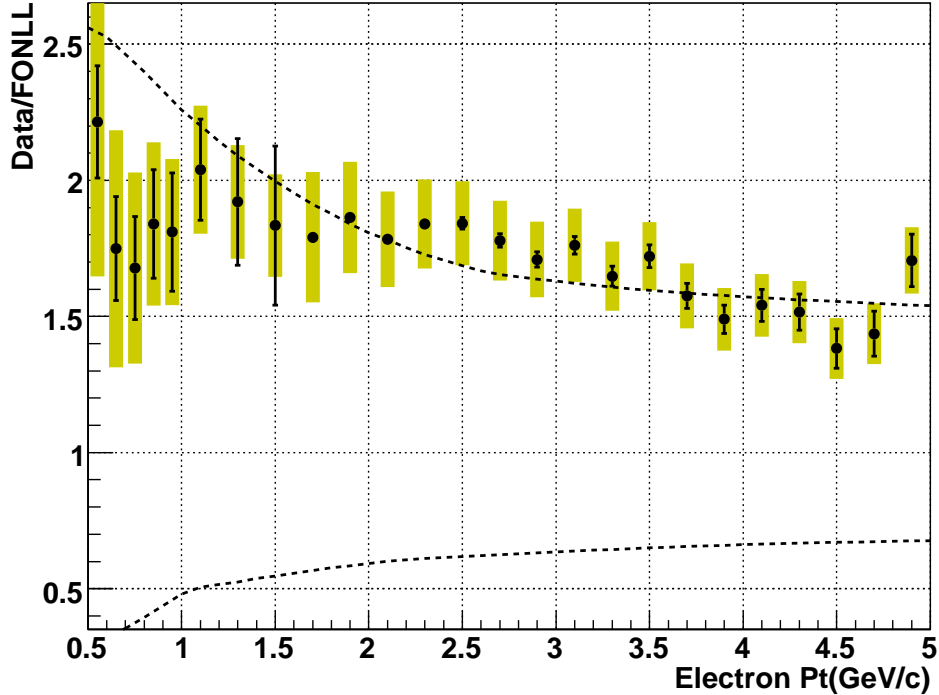}
    \caption{The ratios, data/FONLL of single non-photonic electrons as a function 
      of electron  $p_{\mathrm{T}}$.
      The dotted line represent uncertainties in the FONLL calculations.
    }
    \label{fig:chap7_hf_ratio1}
  \end{center}
\end{figure}
Figure~\ref{fig:chap7_hf_ratio1} the ratios, data/FONLL of single non-photonic electrons as a function 
of electron  $p_{\mathrm{T}}$.
The dotted lines in Fig~\ref{fig:chap7_hf_ratio1} represent uncertainties in the FONLL calculations.
The FONLL calculation reproduce the measured yield of single non-photonic electrons within its uncertainty.
Most of single non-photonic electrons for $p_{\mathrm{T}}<2$~GeV/c originates from charm as shown 
in Sec.~\ref{sec:bc_r}.
Therefore, Fig~\ref{fig:chap7_hf_ratio1} indicates the ratio data/FONLL for charm production
is $\sim 2$.

\begin{figure}[htb]
  \begin{center}
    \includegraphics[width=14cm]{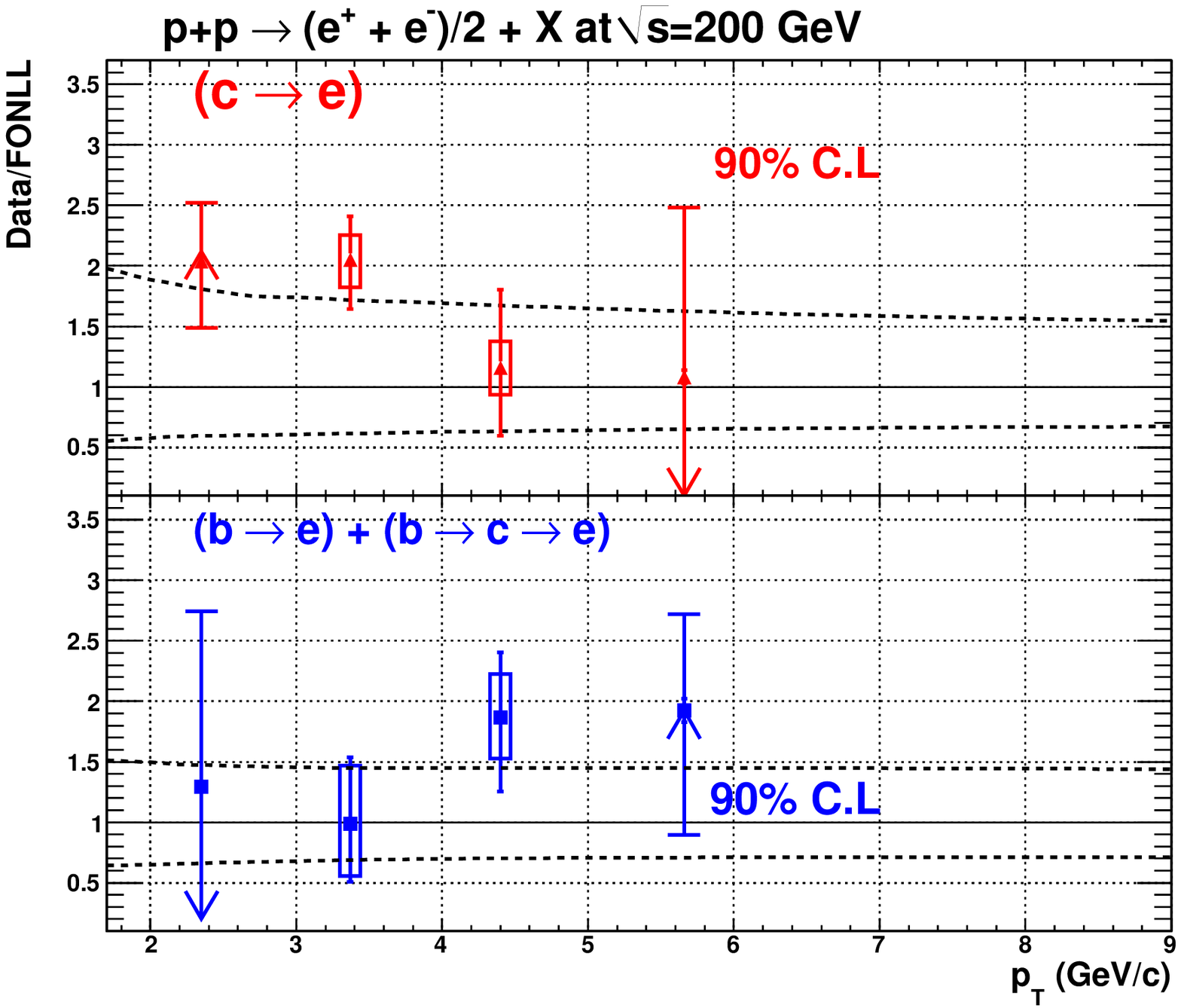}
    \caption{The ratios, data/FONLL of single non-photonic electrons from charm~(upper panel) 
      and bottom~(lower panel) as a function of electron  $p_{\mathrm{T}}$.
      The dotted lines represent uncertainties in the FONLL calculations.
    }
    \label{fig:chap7_hf_ratio2}
  \end{center}
\end{figure}
Figure~\ref{fig:chap7_hf_ratio2} the ratios, data/FONLL of the electrons from charm~(upper panel) and 
bottom~(lower panel) separately as a function of electron  $p_{\mathrm{T}}$.
The FONLL prediction for charm production agrees with the data within the theoretical uncertainty
and the ratio of data/FONLL is $\sim$ 2.
The uncertainty of FONLL for bottom production is less than that for charm production 
due to large mass of bottom.
The FONLL prediction for bottom production also agrees with the data and the ratio of data/FONLL is $\sim$ 1.

The similar tendency can be found in $p+\bar{p}$ collisions at $\sqrt{s}=$1.96~GeV at Tevatron.
\begin{figure}[htb]
  \begin{center}
    \includegraphics[width=14cm]{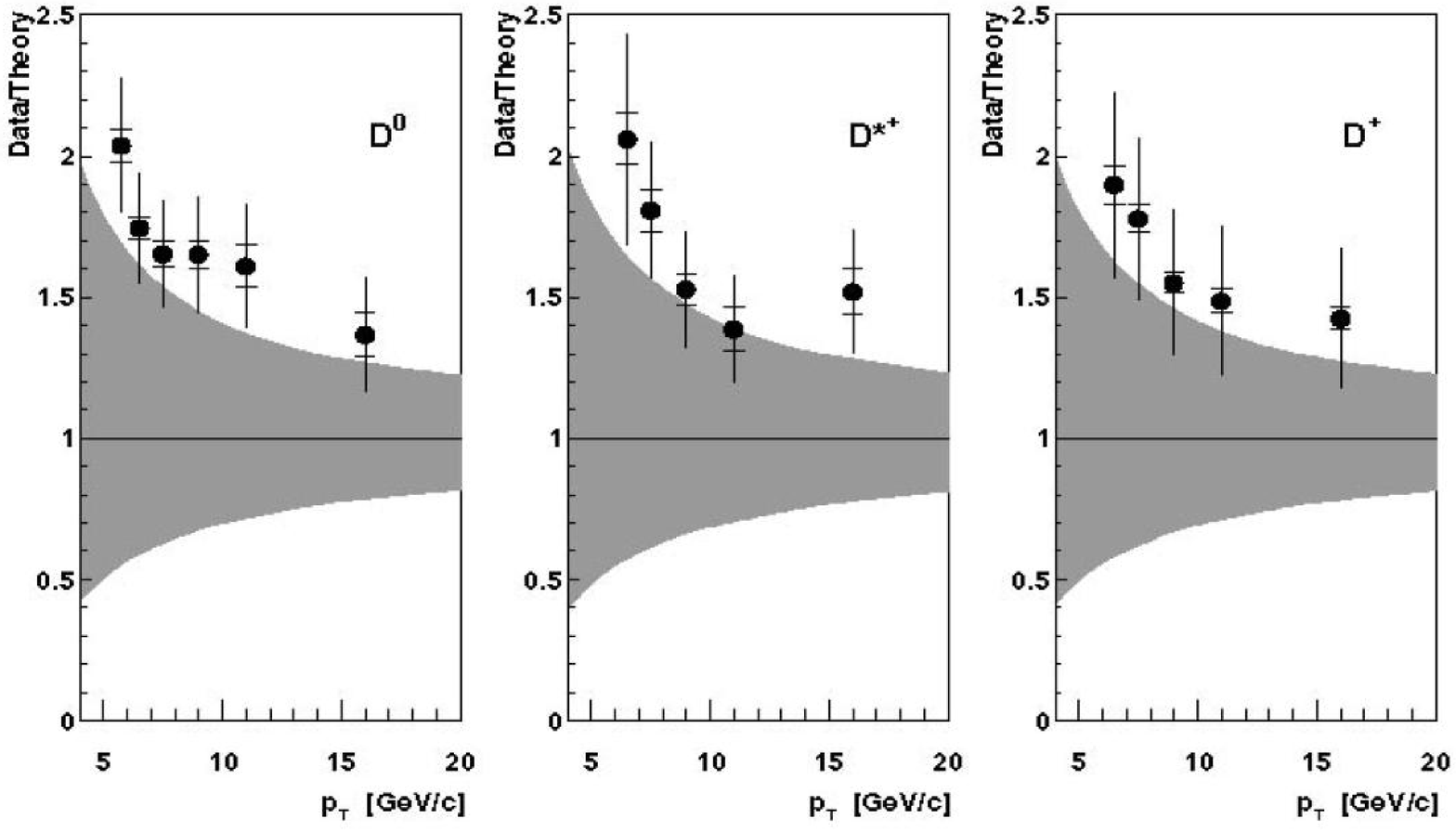}
    \caption{The ratios, (CDF data)/(FONLL) of $D^{0}$~(left panel), $D^{\ast+}$~(middle panel) and $D^{+}$~(right panel).
      as a function of hadron  $p_{\mathrm{T}}$.
      The shadow area represent uncertainties in the FONLL calculations~\cite{bib:mateo1}.
    }
    \label{fig:chap7_hf_ratio3}
  \end{center}
\end{figure}
\begin{figure}[hbt]
  \begin{center}
    \includegraphics[width=14cm]{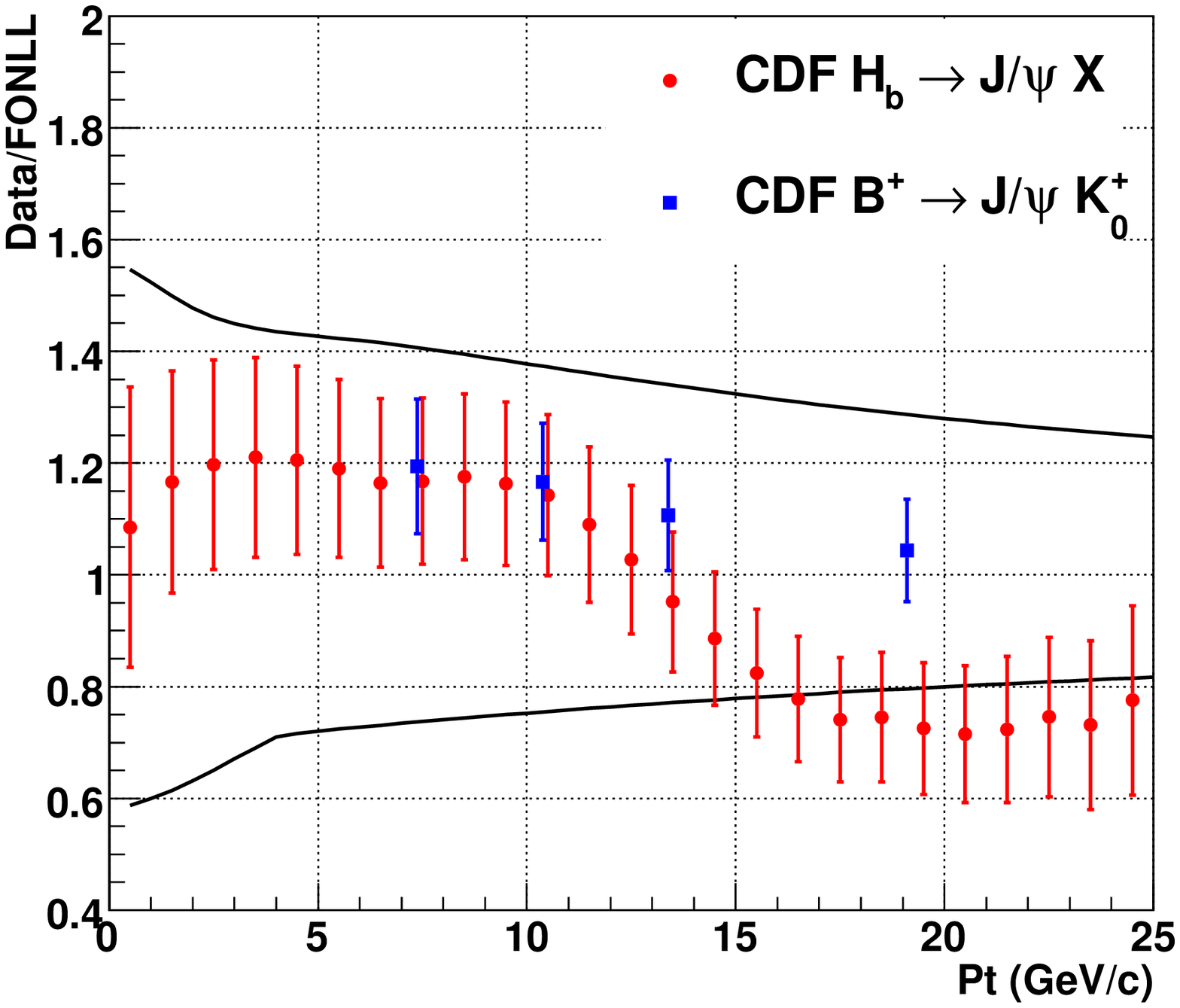}
    \caption{The ratios, data/FONLL of single non-photonic electrons from charm~(upper panel) 
      and bottom~(lower panel) as a function of electron  $p_{\mathrm{T}}$.
      The dotted lines represent uncertainties in the FONLL calculations~\cite{bib:cdf3,bib:cdf4,bib:fonll3}.
    }
    \label{fig:chap7_hf_ratio4}
  \end{center}
\end{figure}
Figure~\ref{fig:chap7_hf_ratio3} and \ref{fig:chap7_hf_ratio4} shows the ratio, (measured results at CDF)/FONLL
, of D and B hadrons as a function of heavy flavored hadron $p_{\mathrm{T}}$, 
respectively~\cite{bib:mateo1,bib:cdf3,bib:cdf4,bib:fonll3}.
Maximum values of FONLL calculations for charm production are consistent with the measured results at CDF
and the ratio is $\sim$ 1.5-2.0.
FONLL calculations for bottom production agrees with the measured results at CDF within its uncertainty and 
The ratio of data/FONLL is $\sim$ 0.8-1.2.

The ratios of (experimental results)/FONLL are $\sim$ 2 for charm production and $\sim$ 1 for bottom production
at RHIC and Tevatron.
Such difference can be understood in terms of a better convergence of the pQCD calculation 
for bottom production.
It is worth to note that compared variables are not bare quarks but hadrons or these decay electrons and 
$p_{\mathrm{T}}$ distribution of hadrons~(these decay electrons) in FONLL agrees with the measured results.
This fact supports not only pQCD but also theoretical treatment of fragmentation process works well.

\section{Comparison with Di-electron Measurement}
PHENIX has measured the electron-positron pair mass spectrum in
$p+p$ collisions at $\sqrt{s}=200$~GeV, from which
the production cross section of heavy flavors is also obtained~\cite{bib:diele_pp}.
This measurement provides a good cross check for measurements to the results from single non-photonic electrons.

\begin{figure}[thb]
 \begin{center}
   \includegraphics[angle=0,width=12cm]{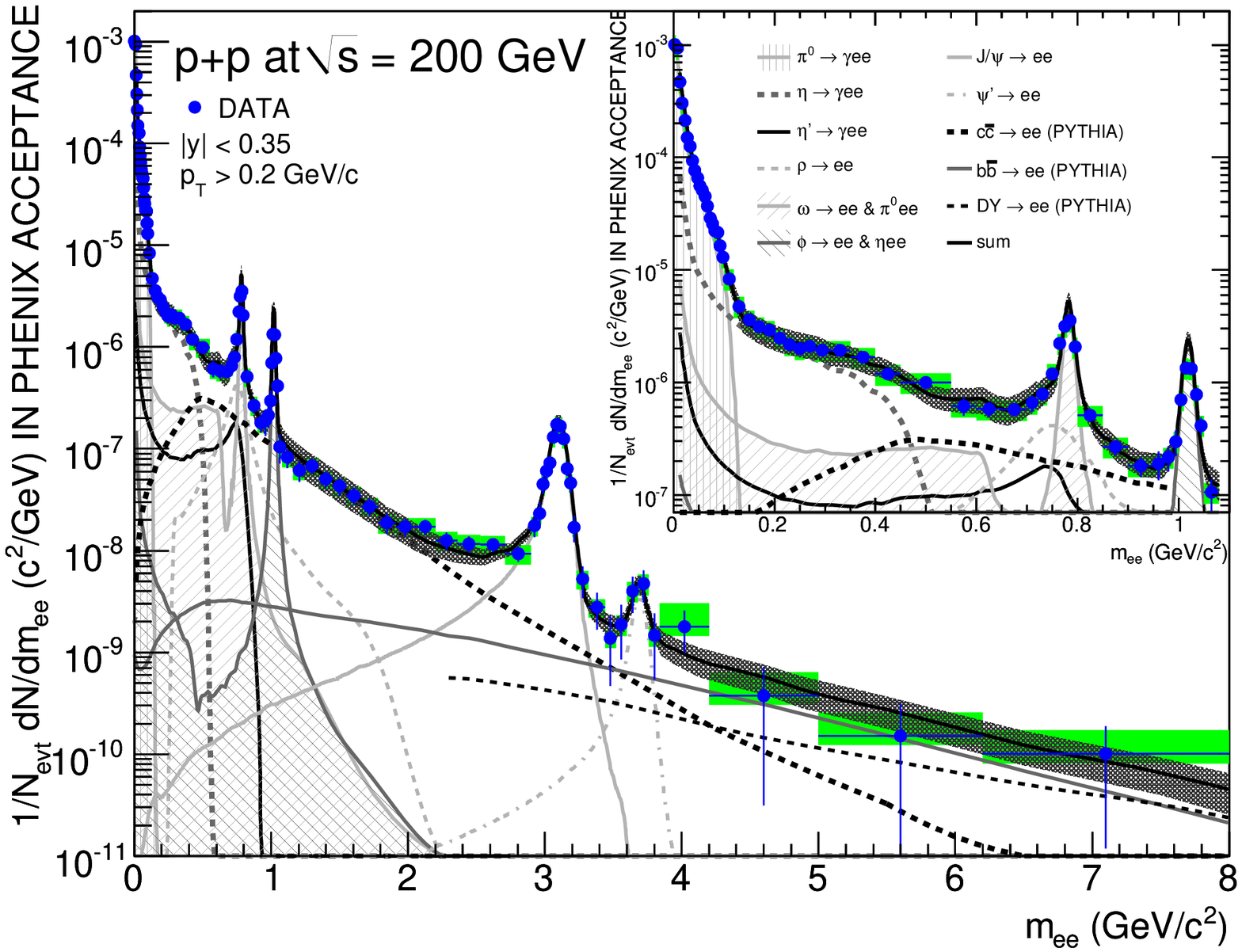}
   \caption{The $e^-e^+$ pair yield per $p+p$ collision 
    in PHENIX acceptance with a cocktail of known sources.}
   \label{fig4a}
 \end{center}
\end{figure}

Figure~\ref{fig4a} shows measured $e^+e^-$ pair yield per $p+p$ collision 
in PHENIX acceptance. 
Combinatorial and correlated background are already subtracted.
In Fig.~\ref{fig4a} a cocktail of known sources discussed in Sec.~\ref{sec:cock} is also shown. 
The cocktail calculations account for the continuum in the mass region below $\sim$
1~GeV/$c^2$ and vector meson peaks. Except for the quarkonium peaks, the $e^+e^-$ pair 
in the mass range above 1.1~GeV/$c^2$ is dominated by single non-photonic electron pairs
correlated through flavor conservation.

\begin{figure}[thb]
 \begin{center}
   \includegraphics[angle=0,width=12cm]{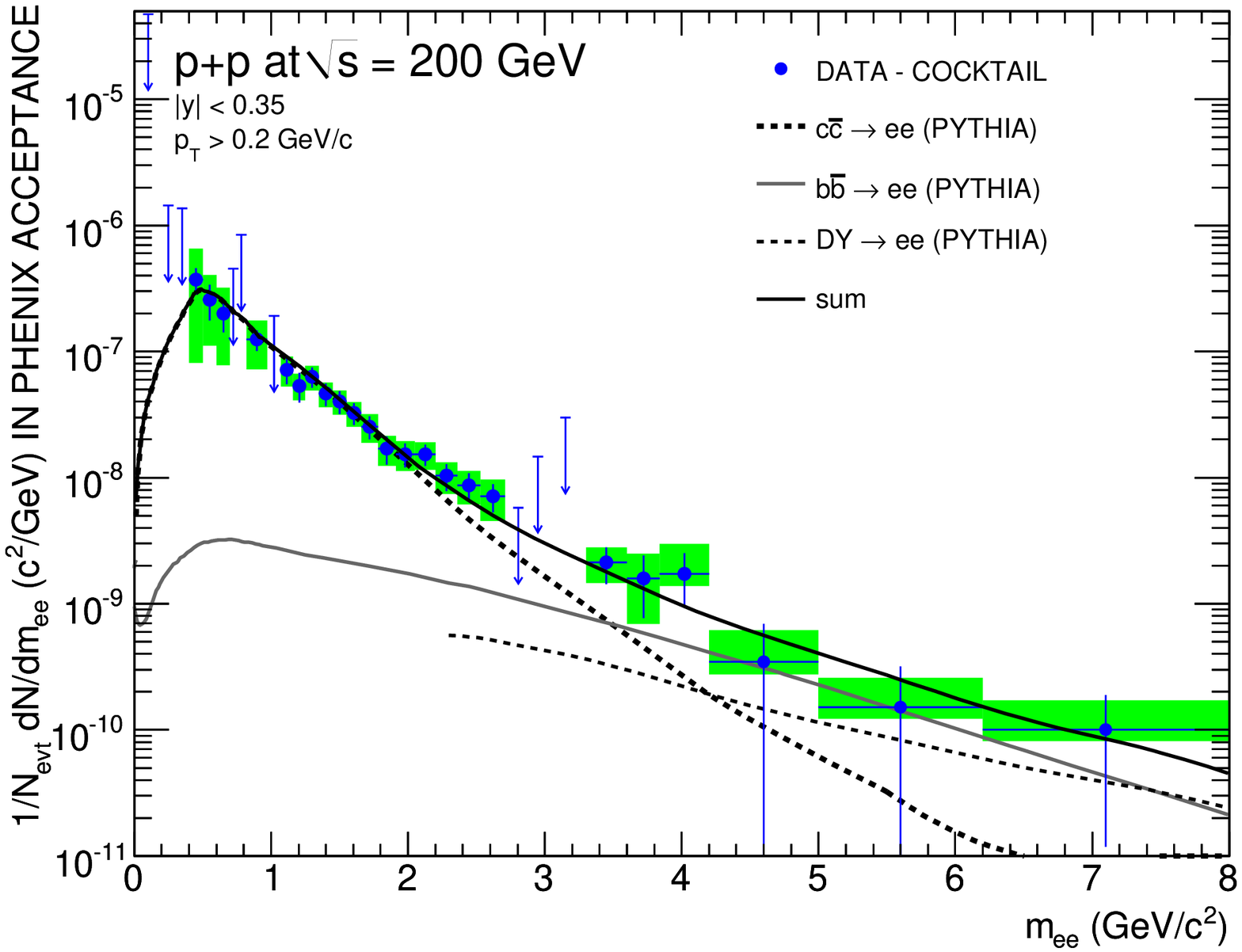}
   \caption{The $e^-e^+$ pair yield remaining after subtraction of 
    the cocktail.}
   \label{fig:figdiele}
 \end{center}
\end{figure}
Figure~\ref{fig:figdiele} shows the $e^-e^+$ pair yield remaining after the subtraction of 
the cocktail originated from the light particles and the quarkonium. The remaining contributions 
are from $c\bar{c} \rightarrow e^+e^-$, $b\bar{b}\rightarrow e^+e^-$ and Drell-Yan process.
To extract contribution from $c\bar{c}$, the $e^+e^-$ pair yield in the range from 1.1 to 2.5
GeV/$c^2$ is integrated.
Integrated yield is extrapolated to zero $e^+e^-$ pair mass by using the di-electron spectrum from 
$c\bar{c}$ simulated by PYTHIA.
Contributions from $b\bar{b}$ and Drell-Yan process are estimated and subtracted. 
The extrapolated $e^+e^-$ yield is converted cross section of charm.
Total cross section of charm is obtained as  $\sigma_{c\bar{c}}=
   544 \pm 39({\rm stat}) \pm 142({\rm sys}) \pm 200(model) \mu {\rm b}$, 
by using rapidity distribution from NLO pQCD calculation~\cite{bib:NLO}.
This result is compatible with the result from single non-photonic electron 
at PHENIX which gives $\sigma_{c\bar{c}}=
   567 \pm 57({\rm stat}) \pm 224({\rm sys}) \mu {\rm b}$~\cite{bib:hq3}.

Cross section of $b\bar{b}$ is also evaluated from  di-electron spectrum as follows.
The $e^+e^-$ pair distribution after subtraction of Drell-Yan is fitted
using the $e^+e^-$ pair distributions from charm and bottom which are produced by PYTHIA.
The obtained total cross section of bottom is $\sigma_{b\bar{b}}=
   3.9 \pm 2.5({\rm stat}) ^{+3}_{-2}({\rm sys}) \mu {\rm b}$.
This result is consistent with the result from the spectrum of the single electrons 
obtained in this thesis.

\section{Heavy Flavor Production in Hadron Collider }
\begin{figure}[htb]
  \begin{center}
    \includegraphics[width=15.cm]{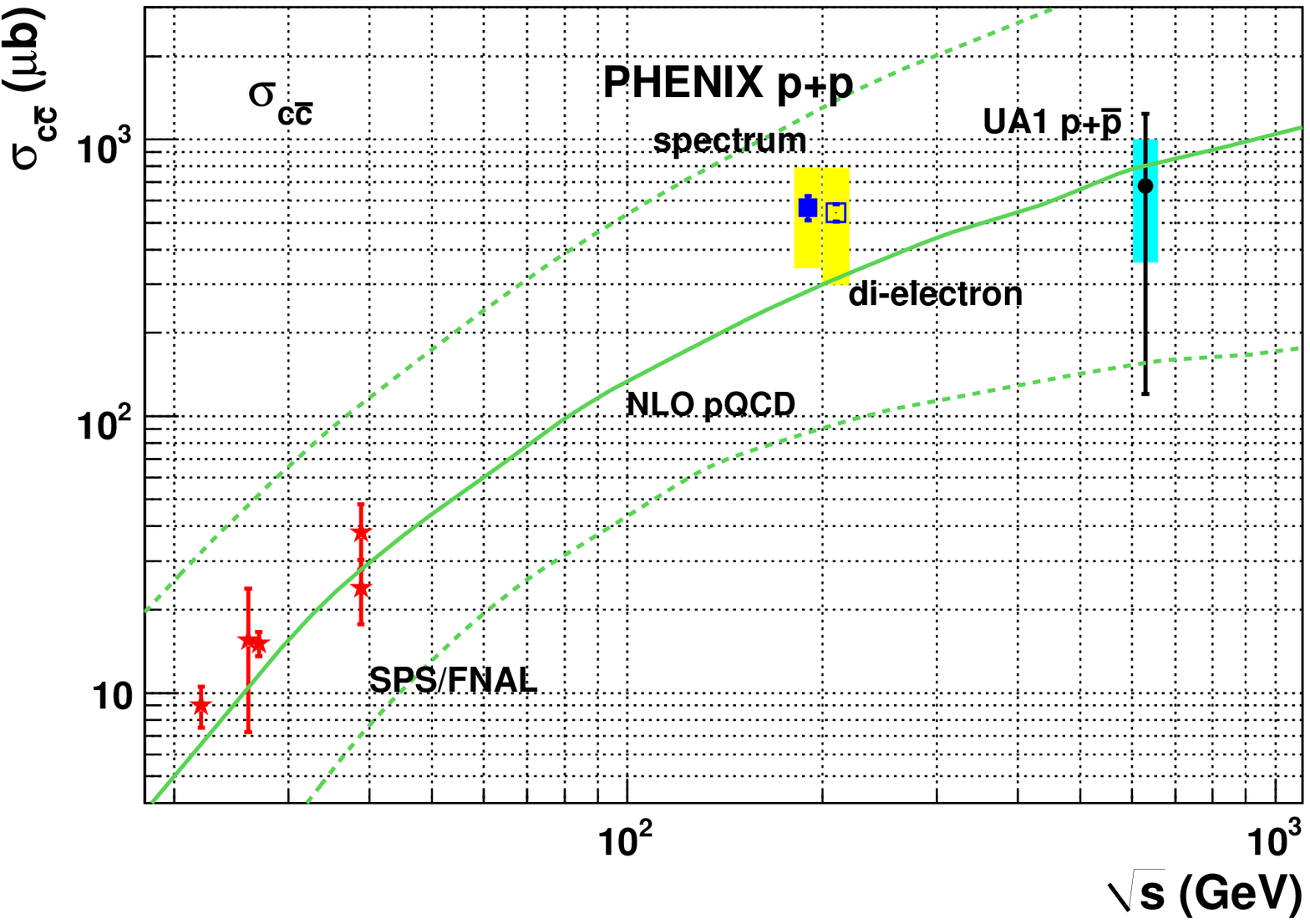}
    \caption{
      Total cross section of charm 
      with other experiments as a function of $\sqrt{s}$~\cite{bib:hq2}.
    }
    \label{chap7_cross}
  \end{center}
\end{figure}
\begin{figure}[htb]
  \begin{center}
    \includegraphics[width=17.cm]{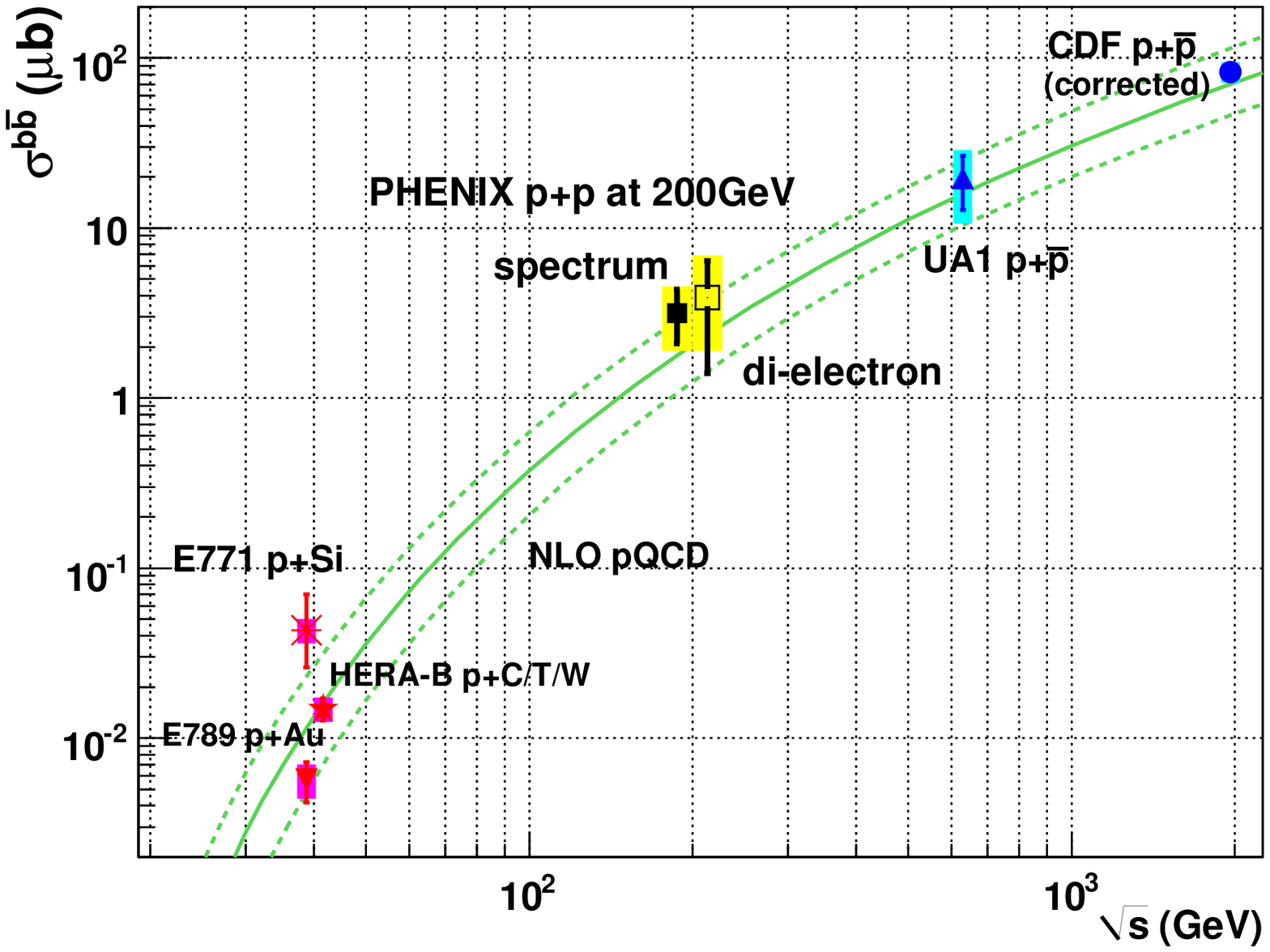}
    \caption{
      Total cross section of bottom 
      with other experiments as a function of $\sqrt{s}$~\cite{bib5,bib6,bib7,bib8,bib9}.
    }
    \label{chap7_fig203}
  \end{center}
\end{figure}

Figure~\ref{chap7_cross} and \ref{chap7_fig203} show the total cross sections of charm and bottom from the spectrum and 
di-electron analysis with other experiments in hadron collider as a function of 
$\sqrt{s}$~\cite{bib5,bib6,bib7,bib8,bib9,bib:hq2}, respectively.
The CDF experiment for bottom production published to the limited rapidity range, $\sigma_{b\bar{b}}(\mid y \mid <0.6)$.
The result of CDF is extrapolated to the cross section, assuming the rapidity distribution given by
HVQMNR.
In Fig.~\ref{chap7_cross} and Fig.~\ref{chap7_fig203}, smooth lines represent cross section of charm and bottom 
calculated by NLO pQCD and dotted lines represent uncertainties in NLO pQCD.

Dependence of the cross section of heavy flavor production on  $\sqrt{s}$ predicted by NLO pQCD agrees with
world data including the result obtained in this thesis.
This agreement indicates charm and bottom production in hadron collider is well understood by pQCD.
The understanding of heavy flavor production supports the expectation that heavy flavors are only produced
in the initial stage in the heavy ion collision.
Therefore, measurements in $p+p$ collisions in this thesis provide important baselines for the study 
of the medium created in RHIC.



\section{Interpretation of Results in Au+Au Collisions}
In this section, the existence of the energy loss and flow of bottom in the hot and dense medium created at RHIC based on 
the $R_{AA}$ and $v_{2}$ of the single electrons in Au+Au collisions and the measured 
$(b\rightarrow e)/(c\rightarrow e+b\rightarrow e)$.

The $R_{AA}$ and $v_{2}$ of non-photonic electrons reported by PHENIX include 
the contribution from J/$\psi$, $\Upsilon$ and Drell-Yan process~\cite{bib:hq1}.
Their contributions are small but not negligible as discussed in Sec.~\ref{sec:cock}.
The medium modification of these contribution should be different from that of bare heavy quarks
due to the different physical process in the medium.
Therefore, the contribution from J/$\psi$, $\Upsilon$ and Drell-Yan process should be subtracted from
the experimental result for the apple to apple comparison.
The procedure to correct $R_{AA}$ and $v_{2}$ of non-photonic electrons is described in Appendix.~\ref{sec:craav2}.
Figure~\ref{fig:chap7_raav2_new} shows the corrected $R_{AA}$ and $v_{2}$ of single non-photonic 
electrons in Au+Au collisions.
Upper panel in Figure~\ref{fig:chap7_raav2_new} shows the $R_{AA}$ of the single non-photonic electrons in
0-10\% central Au+Au collisions.
Lower panel in Fig.~\ref{fig:chap7_raav2_new} shows the $v_{2}$ of the single non-photonic electrons.

\begin{figure}[htb]
  \begin{center}
    \includegraphics[width=13.cm]{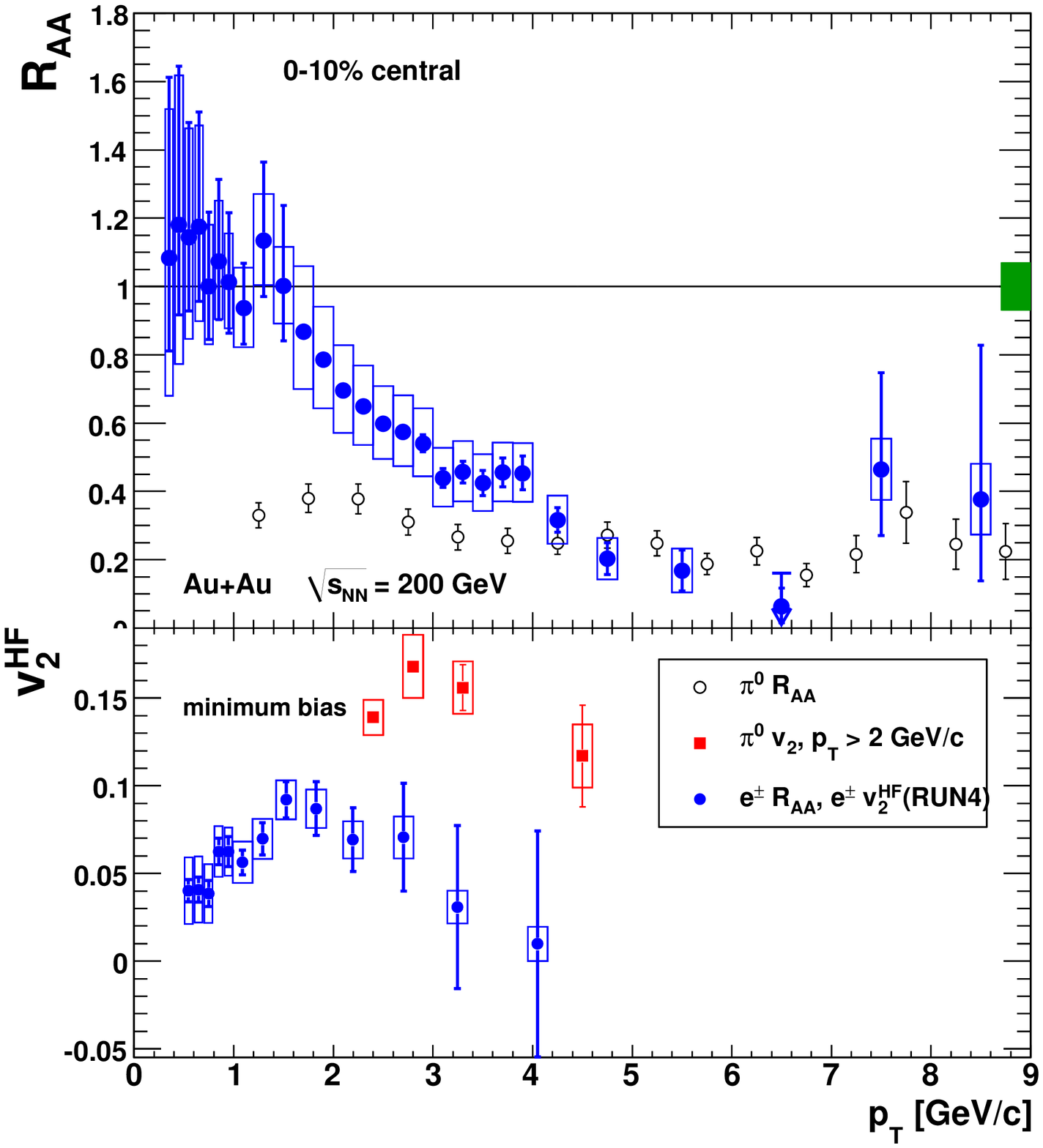}
    \caption{
      Upper panel:$R_{AA}$ of the electrons from semi-leptonic decay of heavy flavor.
      Lower panel:$v_{2}$ of the electrons from semi-leptonic decay of heavy flavor.
    }
    \label{fig:chap7_raav2_new}
  \end{center}
\end{figure}

\subsection{Ratio of Bottom over Charm} \label{sec:bcr}
The fraction of bottom in single non-photonic electrons, $(b\rightarrow e)/(c\rightarrow e+ b\rightarrow e)$, 
is parameterized as a function of electron $p_{{\rm T}}$ assuming the shape of $p_{\mathrm{T}}$ distribution
in FONLL.
The absolute value is determined by the least-square fitting to 
the experimental results of  $(b\rightarrow e)/(c\rightarrow e+ b\rightarrow e)$ using Eq.~\ref{eq:chi}.
The used fit method is described in Appendix.~\ref{sec:fit}.
Figure~\ref{fig:chap7_bcr} shows the resulting ratios of the single electrons from bottom over charm.
Red line represents the best fit result and blue lines show $1\sigma$ uncertainties.
The obtained absolute value of cross section from the fit is defined as the initial state of heavy flavor 
in heavy ion collisions for the comparison of $R_{AA}(p_{\mathrm{T}})$ and $v_2(p_{\mathrm{T}})$.


\begin{figure}[htb]
  \begin{center}
    \includegraphics[width=14cm]{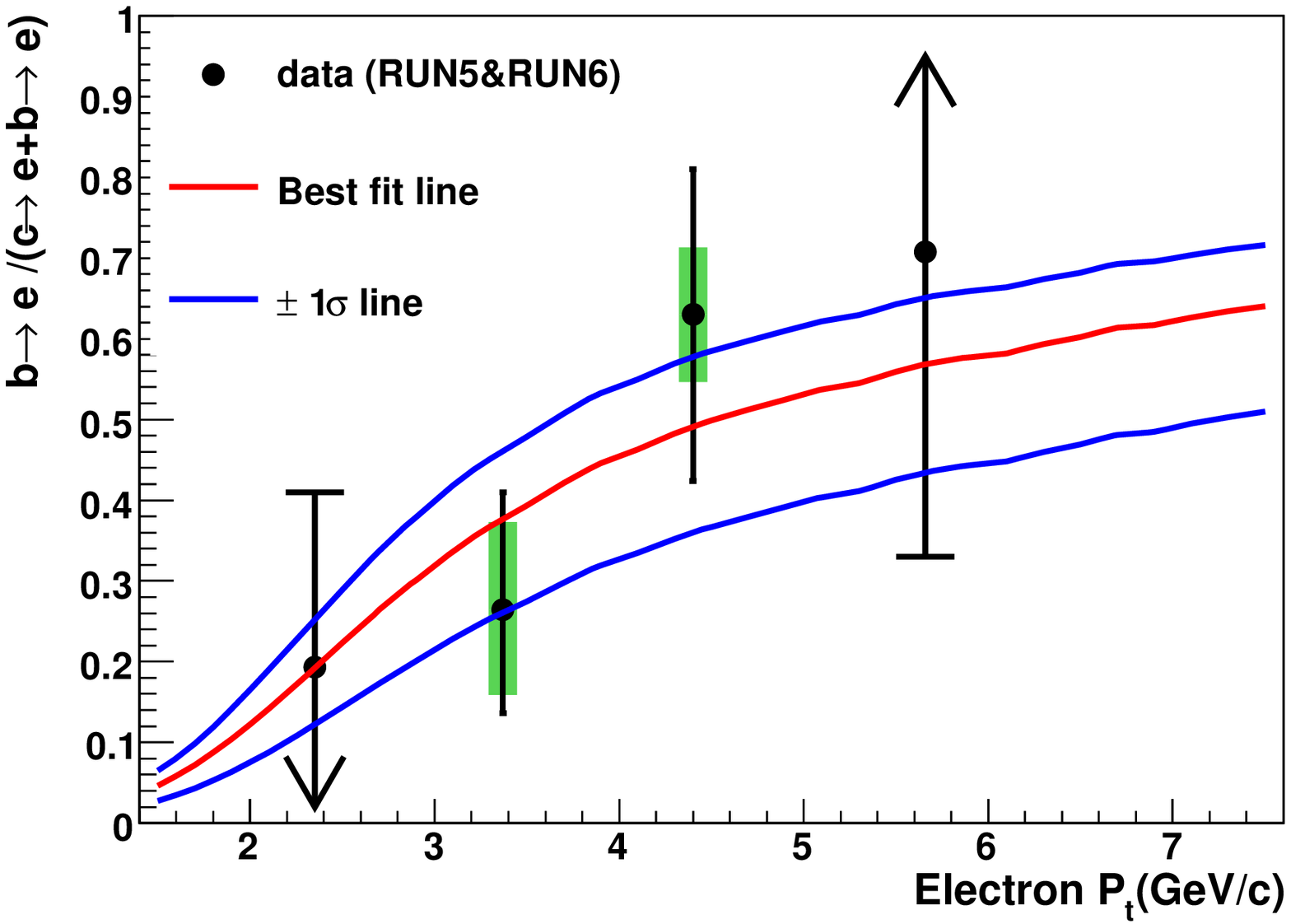}
    \caption{ The ratio of the electrons from bottom over charm to use
      the comparison of $R_{AA}(p_{\mathrm{T}})$ and $v_2(p_{\mathrm{T}})$ which is obtained by the 
      least-square fit using Eq.~\ref{eq:chi}.
    }
    \label{fig:chap7_bcr}
  \end{center}
\end{figure}


\subsection{Modification of Bottom in the Medium}
The magnitude of energy loss of bottom in the medium is expected to be much smaller than that of charm 
due to the large mass of bottom.
The existence of significant bottom modification in the medium  has been an open question when 
$(b\rightarrow e)/(c\rightarrow e+ b\rightarrow e)$ has not been measured.
In this subsection, the existence of significant bottom modification in the medium is shown from the measured
$(b\rightarrow e)/(c\rightarrow e+ b\rightarrow e)$.

\begin{figure}[htb]
  \begin{center}
    \includegraphics[width=14cm]{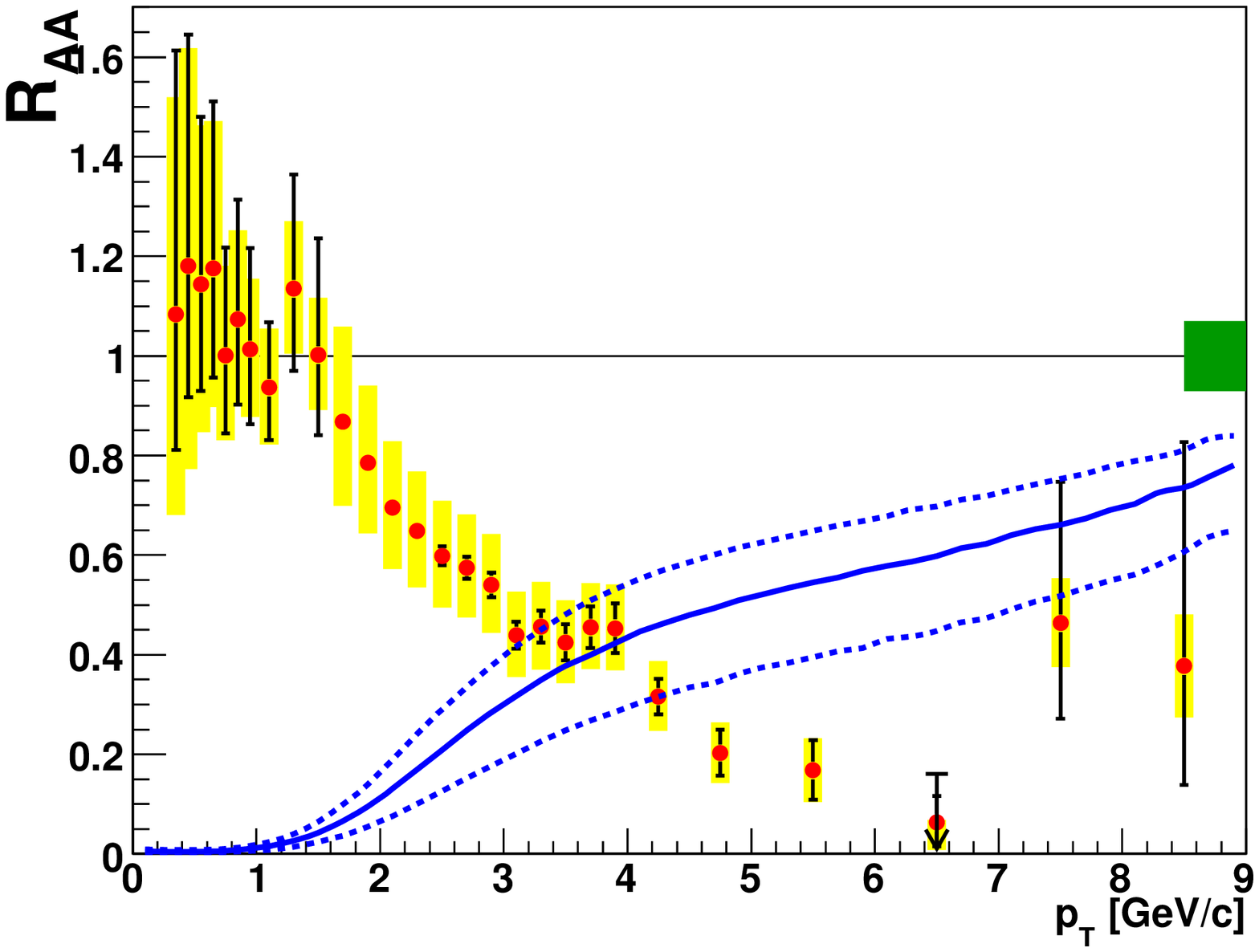}
    \caption{$R_{AA}$ of single non-photonic electrons in Au+Au collisions compared to
      possible minimum values, $(b\rightarrow e)/(c\rightarrow e+ b\rightarrow e)$~(solid line).
      Dotted lines represent the uncertainties of 
      $(b\rightarrow e)/(c\rightarrow e+ b\rightarrow e)$
    }
    \label{fig:chap7_charmonly}
  \end{center}
\end{figure}

For the purpose, we consider only energy loss of charm, that is, we assume that bottom quarks do not 
lose their energy in the medium.
In this case, possible minimum values of $R_{AA}$ of single non-photonic electrons are 
$(b\rightarrow e)/(c\rightarrow e+ b\rightarrow e)$ in $p+p$ collisions, corresponding to the case 
which $R_{AA}$ of the electrons from charm $\rightarrow$ 0.
Figure~\ref{fig:chap7_charmonly} shows $R_{AA}$ of single non-photonic electrons in Au+Au collisions compared to
possible minimum values, $(b\rightarrow e)/(c\rightarrow e+ b\rightarrow e)$~(solid line).
Dotted lines in Fig~\ref{fig:chap7_charmonly} represent the uncertainties of 
$(b\rightarrow e)/(c\rightarrow e+ b\rightarrow e)$ which is obtained in the previous subsection.
Above 4.5~GeV/$c$, significant deviation~($>3\sigma$) exists between the measured  $R_{AA}$ and possible minimum values
where bottom do not lose energy.
This deviation indicates not only charm but also bottom lose a certain fraction of their energy in the medium.

\subsection{Contribution from Bottom in Au+Au Collisions} \label{sec:adsmodel}
The contribution from bottom in Au+Au collisions becomes important 
for the discussion of $v_2(p_{\mathrm{T}})$ of in single non-photonic electrons.
The contribution from bottom in single non-photonic electrons in Au+Au collisions should be larger than
the measured value in $p+p$ collisions, since the suppression magnitude of bottom in the medium is expected 
to be much smaller than that of charm.

To evaluate the contribution of bottom in Au+Au collisions from the measured 
$(b\rightarrow e)/(c\rightarrow e+ b\rightarrow e)$ in this thesis, a model calculation based on Langevin 
equation is done. 
A detailed implementation and procedure for the model calculation are described in Appendix.~\ref{sec:model}.
In this subsection, the procedure is briefly summarized and the result is shown.

The procedure in the model calculation is as follows.
\begin{itemize}
\item{Generation of heavy quarks}
\item{Simulation of space time evolution of heavy quarks in the medium}
\item{Hadronization of bare heavy quarks}
\item{Semi-leptonic decay of heavy flavored hadrons}
\end{itemize}
The shape of $p_{\mathrm{T}}$ distribution of generated charm and bottom quarks are taken from the FONLL calculations.
The absolute value of cross section is normalized according to $(b\rightarrow e)/(c\rightarrow e+ b\rightarrow e)$ 
as described in Sec.~\ref{sec:bcr}.
 
Especially, understanding of space time evolution of heavy quarks in the medium is important 
since the difference of the medium modification between charm and bottom is generated in this stage.
Monte-Carlo simulation using Langevin equation is applied for the description of the space time evolution.
Heavy quarks are produced only in the initial hard collisions and it takes many collisions 
to change the momentum of heavy quark substantially due to their large mass compared with
temperature of the medium~($\sim 200$~MeV).
Therefore, heavy quarks can be described as Brownian particle and the Langevin equation
is a good approximation to model the motion of the heavy quarks 
in the medium~\cite{bib:svet,bib:hees,bib:moore}.

The interaction between heavy quarks and the medium is represented in terms of drag force 
and diffusion coefficients in Langevin equation.
The following relation is assumed for the drag coefficient.
\begin{eqnarray}
  \gamma &=& \alpha \frac{T^2}{M} \label{eq:ads1}, 
\end{eqnarray}
where $M$ is mass of heavy quark and $\alpha$ is the dimensionless free parameter which
is independent of other parameters.
Eq.~\ref{eq:ads1} is motivated by the result from the AdS/CFT correspondence.
This drag coefficient represents the strongly coupling limit of QGP because the AdS/CFT correspondence is valid under
such condition as discussed in Sec.~\ref{sec:coal}.

The magnitude of drag force~(free parameter $\alpha$) is constrained by the fit for the experimental 
$R_{AA}(p_{\mathrm{T}})$ and $v_2(p_{\mathrm{T}})$.
The fit method is described in Appendix.~\ref{sec:fit} and \ref{sec:aucomp}

Figure~\ref{fig:chap7_best_ads1} shows $R_{AA}(p_{\mathrm{T}})$ and $v_2(p_{\mathrm{T}})$
of the decay electrons with the drag force defined in Eq.~\ref{eq:ads1}~(AdS/CFT) 
which are the results of the fit at the 3 ratios of $(b\rightarrow e)/(c\rightarrow e+ b\rightarrow e)$, 
best fit and $\pm 1\sigma$.
The blue solid line shows the result of $R_{AA}(p_{\mathrm{T}})$ and $v_2(p_{\mathrm{T}})$ at the best fit ratio 
of $(b\rightarrow e)/(c\rightarrow e+ b\rightarrow e)$.
The green dashed line and magenta dotted line show the result at the $\pm 1\sigma$ ratios
of $(b\rightarrow e)/(c\rightarrow e+ b\rightarrow e)$.
The fit results  are summarized in Table~\ref{chap7_fit1}.
As a result, the experimental results are successfully reproduced with $\gamma =2.1_{-0.6}^{+0.4} \frac{T^2}{M}$ 
including the uncertainty of the ratio of bottom over charm.
\begin{figure}[htb]
  \begin{center}
    \includegraphics[width=15cm]{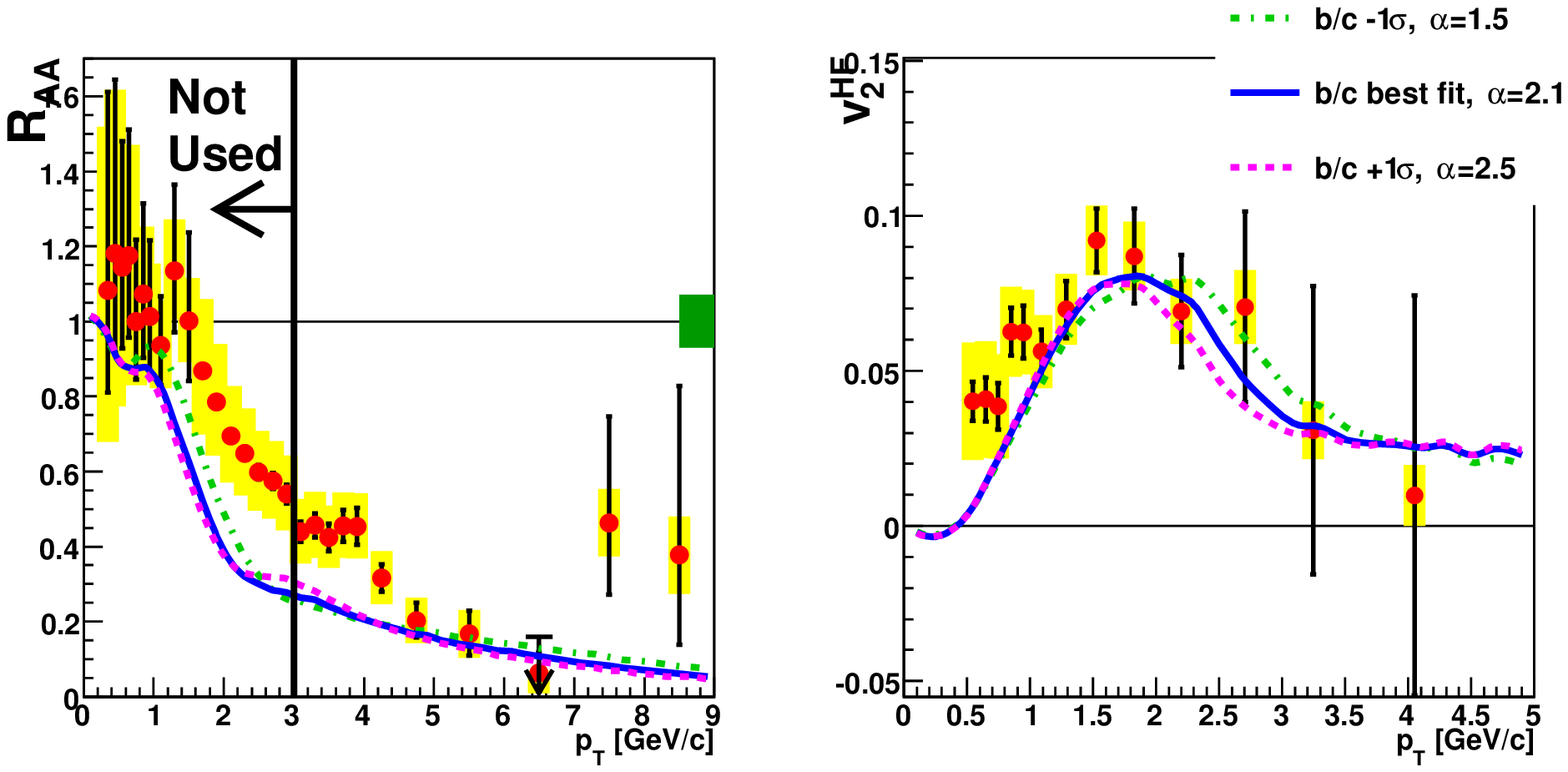}
    \caption{$R_{AA}(p_{\mathrm{T}})$ and $v_2(p_{\mathrm{T}})$
      of single electrons which are the fit results for $R_{AA}(p_{\mathrm{T}})$ and $v_2(p_{\mathrm{T}})$ at the 3 ratios 
      of $(b\rightarrow e)/(c\rightarrow e+ b\rightarrow e)$, best fit and $\pm 1\sigma$.
      The drag force is defined in Eq.~\ref{eq:ads1}~(AdS/CFT).
      Data points of $R_{AA}(p_{\mathrm{T}})$ with $p_{\mathrm{T}}<3$~GeV/$c$ are not used for the fit.
    }
    \label{fig:chap7_best_ads1}
  \end{center}
\end{figure}

\begin{figure}[htb]
  \begin{center}
    \includegraphics[width=13cm]{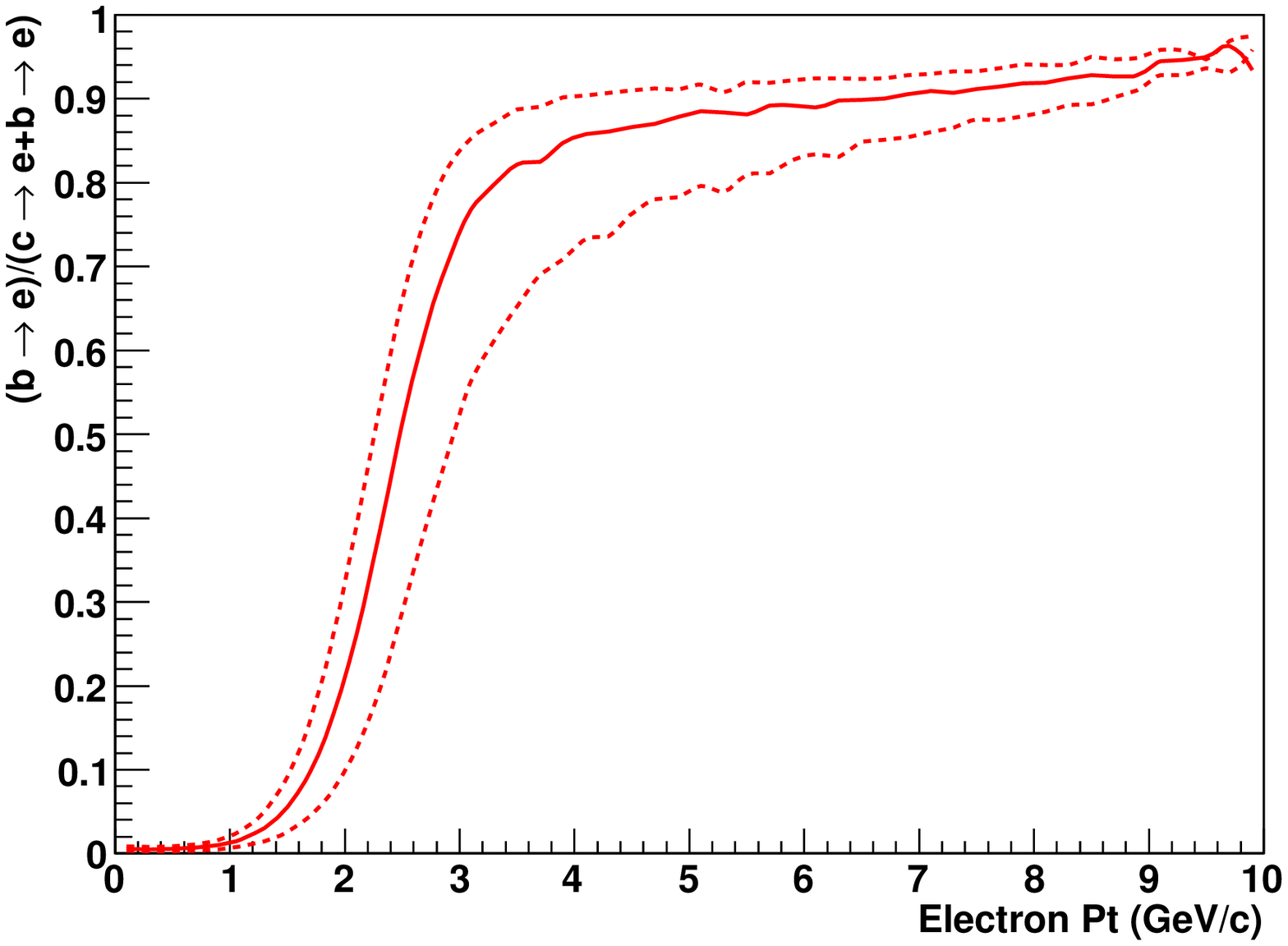}
    \caption{The evaluated $(b\rightarrow e)/(c\rightarrow e+ b\rightarrow e)$ in Au+Au collisions.
      The dotted lines represent $1\sigma$ uncertainties.
    }
    \label{fig:chap7_btoe_ads}
  \end{center}
\end{figure}

$(b\rightarrow e)/(c\rightarrow e+ b\rightarrow e)$ in Au+Au collisions is evaluated as a function of 
electron $p_{\mathrm{T}}$ from the fit result of the magnitude of drag force~($\alpha$).
Figure~\ref{fig:chap7_btoe_ads} shows the evaluated $(b\rightarrow e)/(c\rightarrow e+ b\rightarrow e)$ in Au+Au collisions.
The dotted lines represent $1\sigma$ uncertainties.

Fig.~\ref{fig:chap7_btoe_ads} indicates the contribution from bottom dominates in single non-photonic
electrons above $p_{\mathrm{T}}\sim2$~GeV/$c$.
The decreasing $v_2(p_{\mathrm{T}})$ of single non-photonic electrons at $p_{\mathrm{T}}>2$~GeV/$c$ can be understood 
as the effect of increasing contribution of bottom, since  $v_2(p_{\mathrm{T}})$ of bottom is much smaller than
that of charm.
In addition, Fig.~\ref{fig:chap7_btoe_ads} suggests most of single non-photonic electrons is originated from bottom at 
$p_{\mathrm{T}}>3$~GeV/$c$.
This fact directly leads to that the existence and magnitude of bottom flow can be studied by 
measurement of $v_2(p_{\mathrm{T}})$ at $p_{\mathrm{T}}>3$~GeV/$c$.
It is difficult to obtain a strong physics message about bottom flow from the current measured $v_2(p_{\mathrm{T}})$,
due to limited statistics in Au+Au collisions.
However, more precise result about $v_2(p_{\mathrm{T}})$ of single non-photonic electrons will be obtained at the 
PHENIX experiment in the near future.
The existence and magnitude of bottom flow will be revealed based on the coming result and 
the measured $(b\rightarrow e)/(c\rightarrow e+ b\rightarrow e)$.
The measured $(b\rightarrow e)/(c\rightarrow e+ b\rightarrow e)$ in $p+p$ collisions provides an important 
base line for such discussion.

\clearpage
\section{The Latest Theoretical Calculations} \label{sec:theory}
There are latest theoretical predictions which are based on the microscopic interaction and are compatible with the
experimental results.
Such models  provide not only the information of the medium, but also the possible scenarios of 
the microscopic interaction of heavy quarks.
\subsection{Resonance Model} \label{sec:reso}
\begin{figure}[htb]
  \begin{center}
    \includegraphics[width=14cm]{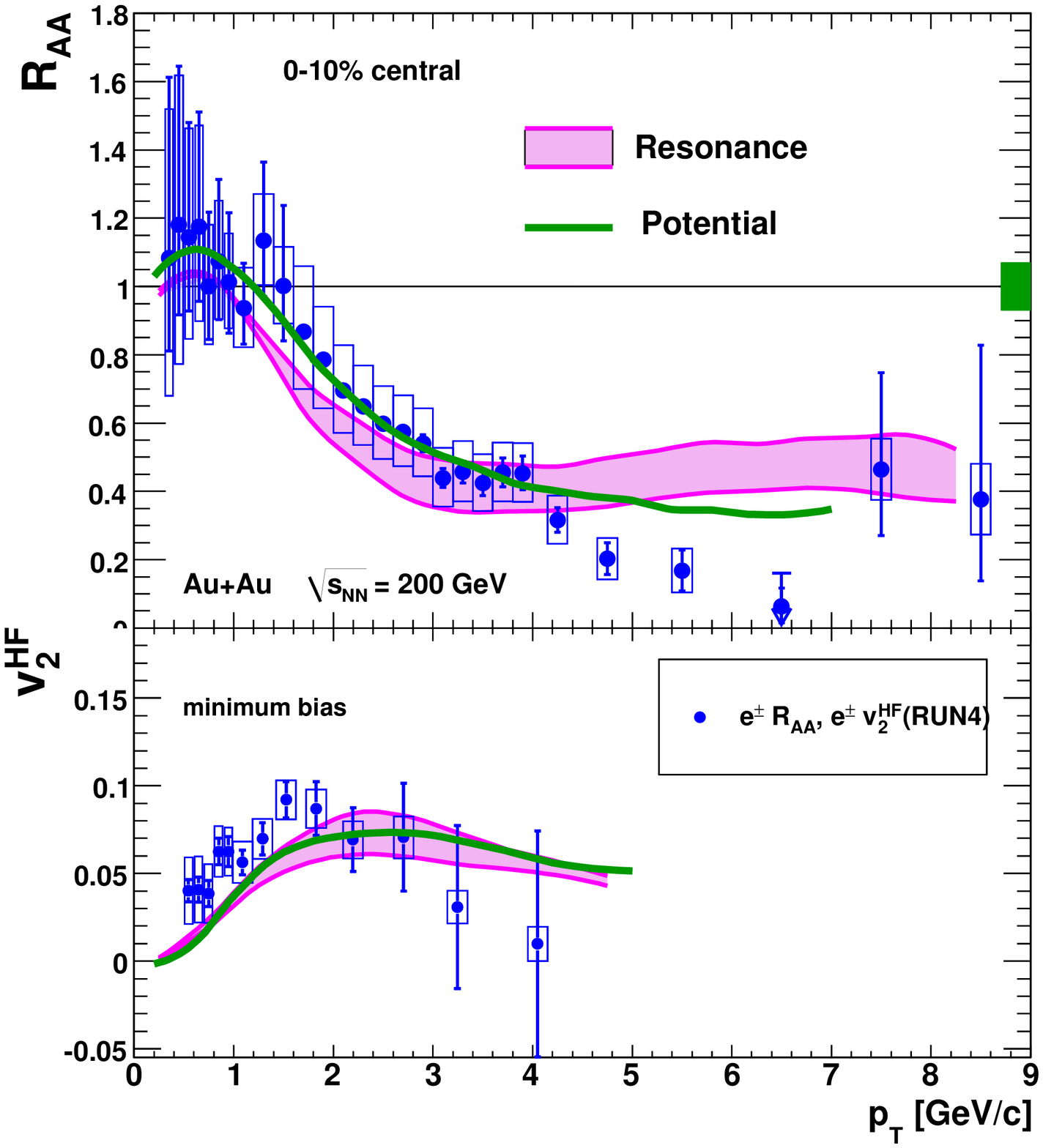}
    \caption{$R_{AA}(p_{\mathrm{T}})$ and $v_2(p_{\mathrm{T}})$ of the electrons from heavy flavor
      in resonance and potential models~\cite{bib:hees,bib:tmat} with the experimental result.
    }
    \label{fig:chap7_rapcomp}
  \end{center}
\end{figure}

The main feature of this model is to assume that resonant D- and B-like states exist in QGP in the temperature 
of 1-2$T_c$~\cite{bib:hees,bib:hees2}.
The existence of such resonance state is motivated by the existence of hadronic resonances in QGP 
suggested by the lattice QCD calculation~\cite{bib:lqcd1,bib:lqcd2}.
The interaction of heavy quarks with light quarks in QGP is strongly enhanced via the resonance state
compared with pQCD.
Treatment of heavy quark transport in the medium is similar with the calculation described in 
Sec.~\ref{sec:lang}.
The drag and diffusion coefficients are given in the microscopic calculation by resonance model.
The charm and bottom quarks are hadronized including coalescence process at the freeze-out time.
Then the heavy flavored hadrons are decayed into the electrons via semi-leptonic decay.
The treatment of these process are also similar with the calculation described in Sec.~\ref{sec:hadro}.
$R_{AA}(p_{\mathrm{T}})$ and $v_2(p_{\mathrm{T}})$ of decay electrons are calculated with this model.
Figure~\ref{fig:chap7_rapcomp} shows the calculated $R_{AA}(p_{\mathrm{T}})$ and $v_2(p_{\mathrm{T}})$ of 
decay electrons with the experimental result.

\subsection{Potential Model} \label{sec:tmat}
The feature of this model is to adopt the results from the lattice QCD calculation to treat the interaction of heavy quarks.
Although lattice QCD is a powerful tool in the non-perturbative QCD calculation, lattice QCD can not 
be used for the dynamical process.
Therefore, a combination of a reduced T-matrix approach and lattice QCD calculation is used to treat the 
dynamical scattering process of heavy quarks~\cite{bib:tmat}.
The reduced interaction kernel in T-matrix approach is identified as a 2-body potential
which is extracted from lattice QCD calculation.
There is still open issues about the extraction method of the 2-body potential from the lattice 
calculation~\cite{bib:tmat}. 
The largest uncertainty in the model is this extraction method.
\begin{figure}[htb]
  \begin{center}
    \includegraphics[width=14cm]{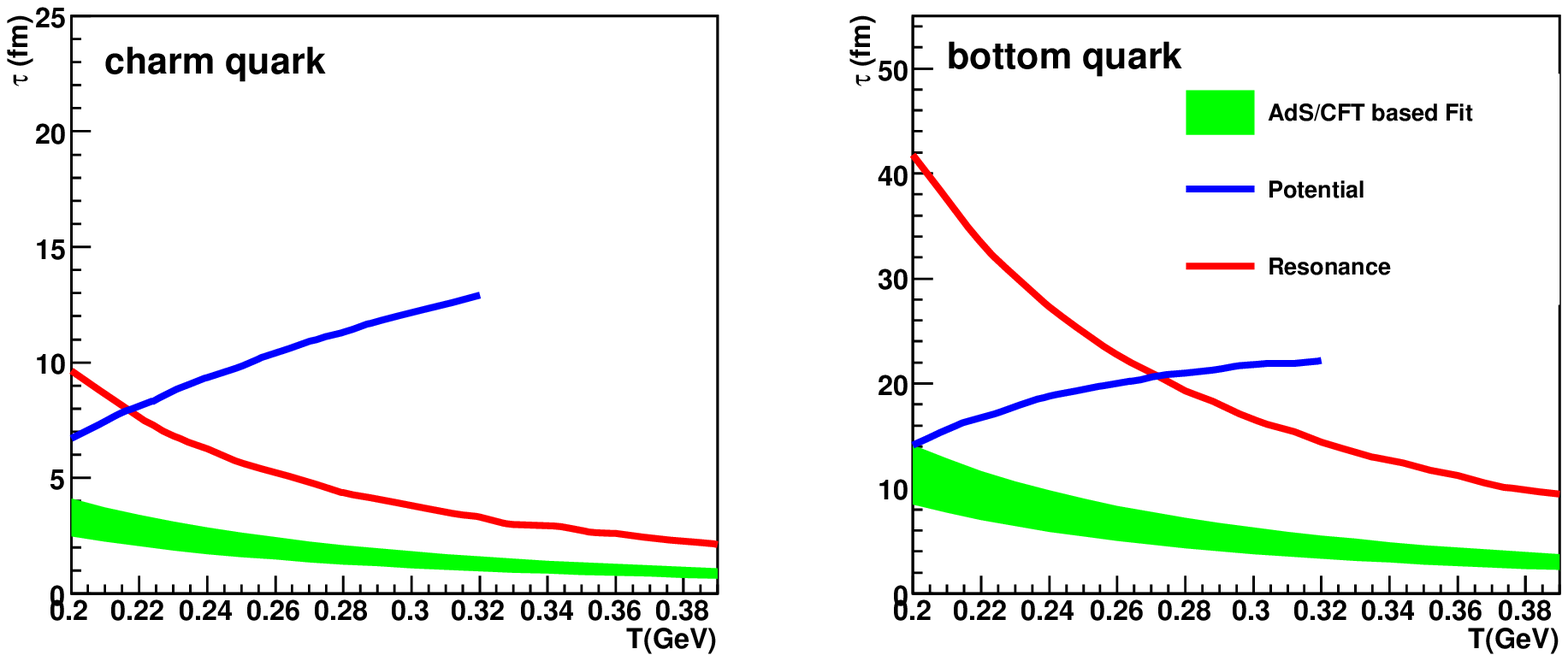}
    \caption{Thermalization time $\tau_{HQ}$ of charm and bottom quark calculated by  
      resonance model and potential model with $\tau_{HQ}$ given by Eq.~\ref{eq:tau_hq} as a function of 
      temperature.
    }
    \label{fig:chap7_taucomp}
  \end{center}
\end{figure}
\begin{figure}[htb]
  \begin{center}
    \includegraphics[width=7cm]{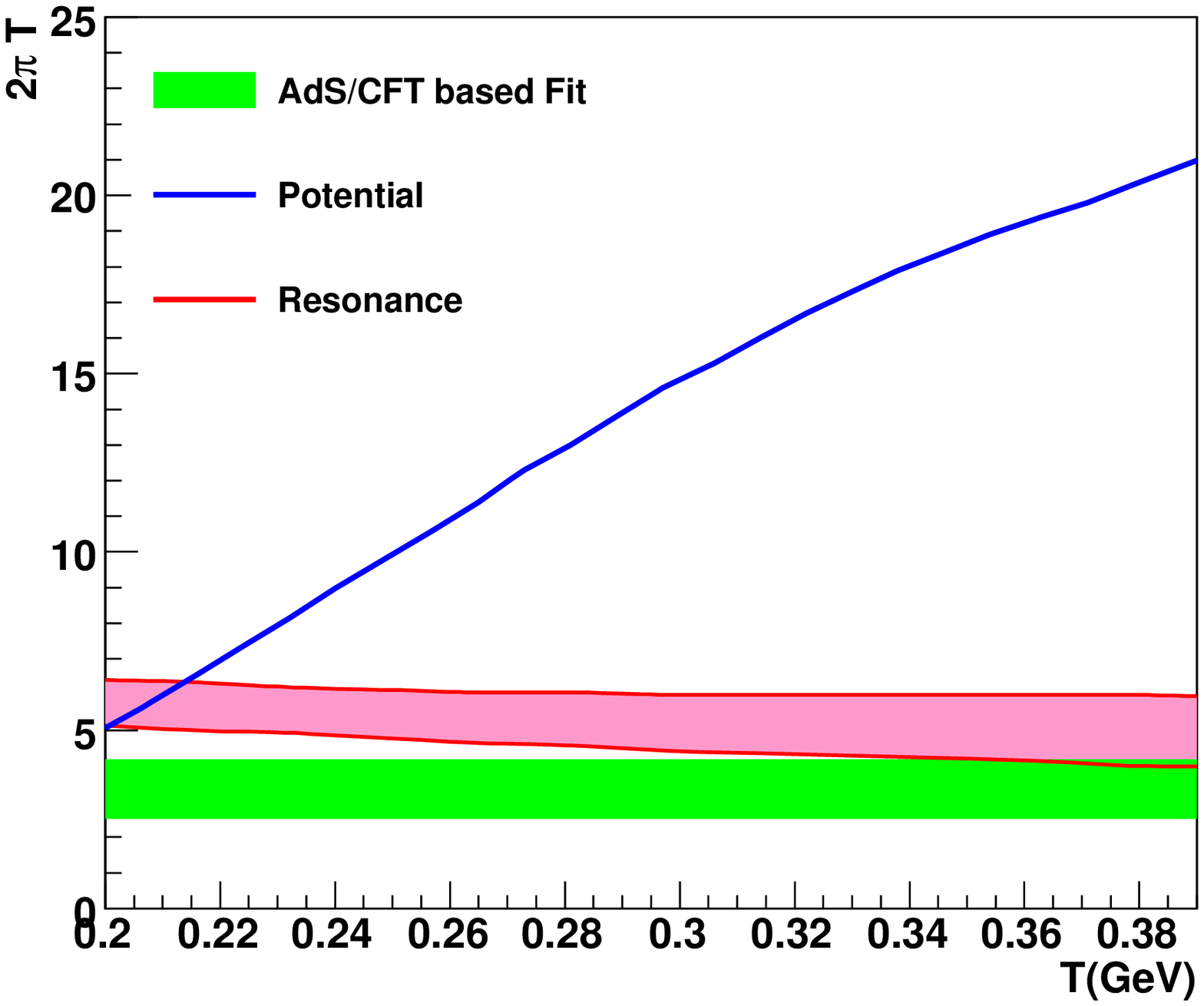}
    \caption{The spatial diffusion constant in units of the thermal wave length, 1/$2\pi T$ 
      calculated by resonance model and potential model with $D_{s}$ given by Eq.~\ref{eq:sdif_res} as a function of 
      temperature. 
    }
    \label{fig:chap7_difcomp}
  \end{center}
\end{figure}
The  drag and diffusion coefficients are given by T-matrix calculation.
Treatment of quark transport, hadronization and decay processes is the same as in Sec.~\ref{sec:reso}.
$R_{AA}(p_{\mathrm{T}})$ and $v_2(p_{\mathrm{T}})$ of decay electrons are calculated from this model.
The calculated $R_{AA}(p_{\mathrm{T}})$ and $v_2(p_{\mathrm{T}})$ are shown in Fig~\ref{fig:chap7_rapcomp}.
The $R_{AA}(p_{\mathrm{T}})$ and $v_2(p_{\mathrm{T}})$ calculated by resonance model and potential model
almost agree with the data.
\subsection{Thermalization Time}
Figure~\ref{fig:chap7_taucomp} shows thermalization time $\tau_{HQ}$ of charm and bottom quark calculated using
resonance model, potential model and the used model in Sec.~\ref{sec:adsmodel}~(AdS/CFT based fit) as a function of 
temperature. 
Figure~\ref{fig:chap7_difcomp} shows the spatial diffusion constant in units of the thermal wave length, 1/$2\pi T$ 
calculated by resonance model and potential model with the used model in Sec.~\ref{sec:adsmodel}~(AdS/CFT based fit) 
as a function of temperature. 
In Fig.~\ref{fig:chap7_taucomp} and \ref{fig:chap7_difcomp}, the results based on AdS/CFT fit are obtained in 
Appendix.~\ref{sec:adsresult}.

Contrary to both AdS/CFT based fit and the resonance model, 
the lattice QCD based potential model provides an increase of the thermalization time and the spatial 
diffusion constant of heavy quarks with increasing temperature.
This indicates the potential model predicts QGP becomes weak coupling state with increasing temperature.
The potential model seems to be the most realistic approach for the interaction of the heavy quark in the medium
among the three models, and the temperature dependence agrees with the picture where QGP becomes a weakly 
coupled gas at sufficiently large temperature due to color screening effect.
The AdS/CFT correspondence is valid at a strongly coupled region as already described and the assumed 
D- and B-like resonance will dissolve at high temperature.
Therefore, the reason of the contracted temperature dependence of the AdS/CFT correspondence and the resonance model
could be considered to be that the AdS/CFT correspondence and the resonance model are plausible only at the near
critical temperature.

There is an interesting discussion about the nature of the medium, $\eta/s$ related to the spatial diffusion constant. 
This consideration is described in Appendix.~\ref{sec:vis}.

\subsection{Dissociation Model}
\begin{figure}[htb]
  \begin{center}
    \includegraphics[width=14cm]{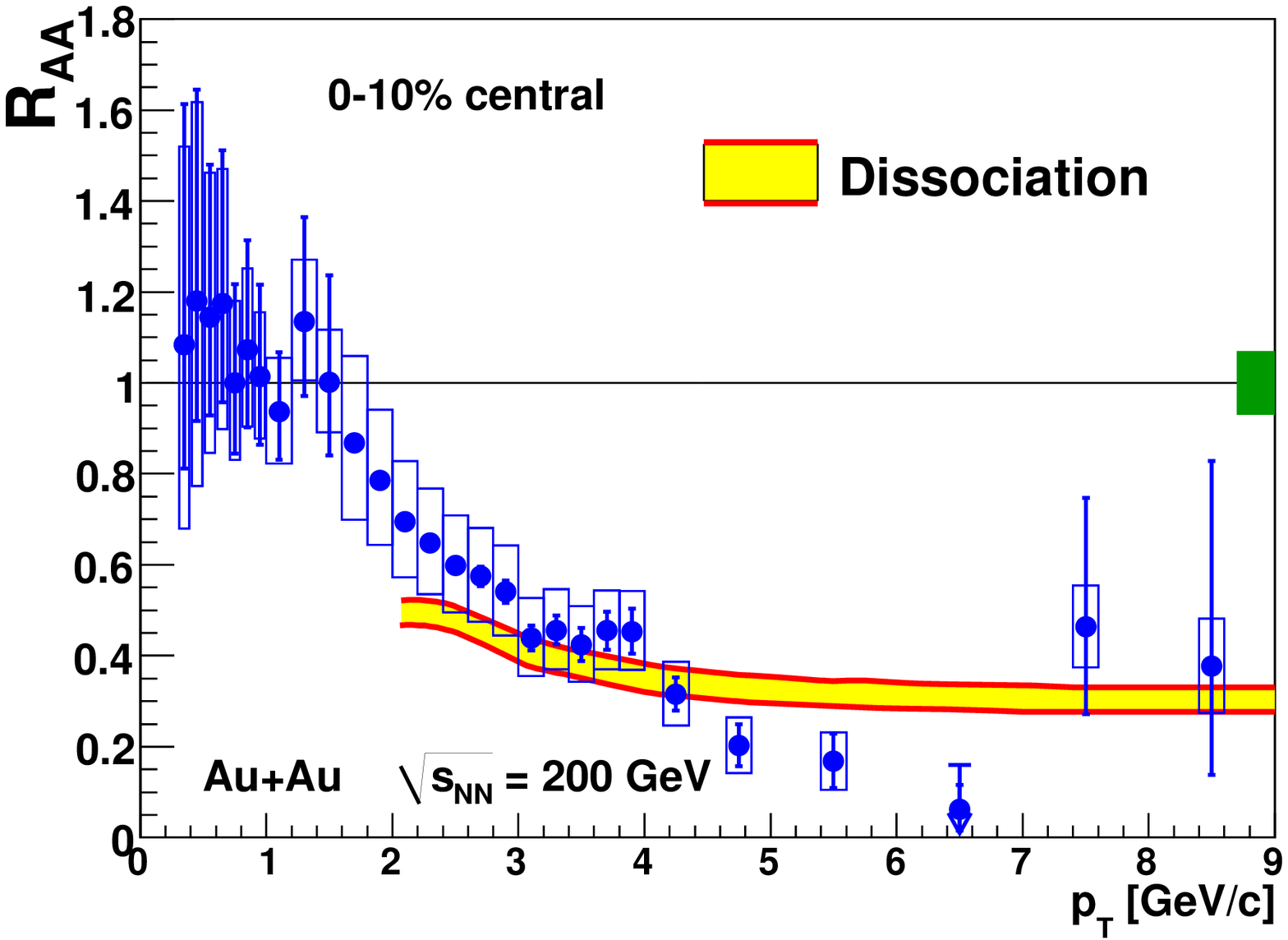}
    \caption{The calculated $R_{AA}(p_{\mathrm{T}})$ and $v_2(p_{\mathrm{T}})$ of 
      decay electrons in dissociation model with the experimental result.
    }
    \label{fig:chap7_disso}
  \end{center}
\end{figure}

In Ref.~\cite{bib:disloss}, a conceptually different approach is introduced to reproduce the magnitude of
$R_{AA}(p_{\mathrm{T}})$ of decay electrons.
Since the formation time~($\tau_{form}$) of D and B meson~(1.6~fm for D meson with 10~GeV/$c$ and 0.4~fm 
for B meson with 10~GeV/$c$) is less than the lifetime of QGP, 
c and b quarks are assumed to hadronize to D and B meson in QGP in this model.
The D and B meson in QGP dissociate into c and b quarks via the collisional interaction in QGP.
Heavy quarks lose a large fraction of their energy via these fragmentation and dissociation process.
Such mechanism becomes more important for heavy quarks with low $p_{\mathrm{T}}$, since 
$\tau_{form} \propto p_{\mathrm{T}}$.
The largest uncertainty in the model is the treatment of the fragmentation process, which may be modified
in the medium~(e.g. coalescence process, modified fragmentation functions etc).
Figure~\ref{fig:chap7_disso} shows the calculated $R_{AA}(p_{\mathrm{T}})$ of 
decay electrons in dissociation model with the experimental result.
This model have the remarkable feature that the magnitude of the suppression of B meson becomes larger
than that of D meson above $\sim$10~GeV/$c$.

\section{Toward Further Understanding}
The measured  $(b\rightarrow e)/(c\rightarrow e+ b\rightarrow e)$ suggests
more precise result about $v_2(p_{\mathrm{T}})$ of single non-photonic electrons at high $p_{\mathrm{T}}$ 
will reveal the existence and magnitude of bottom flow as described in Sec.~\ref{sec:adsmodel}.
Therefore, more statistics in Au+Au collisions is important to the understand the behavior of bottom in the medium.

Various theoretical models are proposed for the interpretation of the interaction of heavy quarks in the medium 
and some of these succeed to reproduce the experimental results as discussed in Sec.~\ref{sec:theory}.
More theoretical and experimental studies are necessary to build up a unified picture of interaction 
of heavy quarks in the medium.
Especially, separate measurement of the B- and D-hadrons in Au+Au collisions, more precise measurement
in d+Au collisions, more statistics at high $v_2(p_{\mathrm{T}})$ and measurement at the medium with higher 
temperature are most important topics. 
These measurement will be done at PHENIX with the new detector, Silicon Vertex Detector and the experiments at LHC
in the near future.

  \chapter{Conclusion}
Measurement of the electrons from semi-leptonic decay of heavy flavor~(single non-photonic electrons) 
in $p+p$ collisions at $\sqrt{s}$=200~GeV 
has been carried out with the PHENIX detector in the RHIC Year-2005 and Year-2006 run.
It provides a good test of pQCD due to the large mass and a test of theoretical treatments of fragmentation process. 
Measurement of heavy flavor in $p+p$ collisions also provides the important base line of the interpretation of the
result of heavy flavor in Au+Au collisions at RHIC, since heavy flavor is only produced in
the initial collisions. 

A strong suppression at high $p_{\mathrm{T}}$ and azimuthal anisotropy of the single electrons have 
been observed in central Au+Au collisions.
Measured single electrons include contribution from both charm and bottom.
The magnitude of energy loss of bottom in the hot and dense medium is expected to be much smaller
than that of charm due to the large difference of their masses.
In addition, since the thermalization time of bottom should be larger than that of charm, 
the magnitude of flow of bottom is expected to be much smaller than that of charm.
This fact indicates charm quarks lose a large fraction of their energy and flow in the matter created in Au+Au collisions.
On the other hands, the existence of bottom modification~(energy loss and flow) in the medium 
has been an open question without the determination of $(b\rightarrow e)/(c\rightarrow e+ b\rightarrow e)$.

The measurement of the fraction of bottom contribution in the single non-photonic electrons provides
a precise test of perturbative QCD, since the fraction allows us to compare $p_{\rm{T}}$ distributions of
charmed and bottomed hadrons with these in pQCD separately. 
Especially, $p_{\rm{T}}$ distribution about bottomed hadrons is important due the a better 
convergence for bottom production.
The fraction also provides the important base line to discuss at the behavior
of bottom quarks in the medium created in Au+Au collisions.

The first measurement aiming to determine the fraction of charm and bottom in single non-photonic electrons
via a new method, partial reconstruction of $D/\bar{D} \rightarrow e^{\pm} K^{\mp} X$ decay, has been also 
carried out in $p+p$ collisions in the RHIC Year-2005 and Year-2006 run.
The fraction of bottom contribution in the single non-photonic electrons is determined experimentally.
It is found that there is the considerable contribution from bottom in the single non-photonic electrons above 3~GeV/$c$.
The first spectra of single non-photonic electrons from charm and bottom are measured based on the fraction of bottom
at RHIC.
$p_{\rm{T}}$ distribution predicted in pQCD agrees with the measured spectra within its uncertainty and 
the ratio, (data/pQCD) is $\sim$ 2 for charm production and  $\sim$ 1 for bottom production.
The same tendency can be found at Tevatron.
The total cross section of bottom is also determined 
to  $\sigma_{b\bar{b}}= 3.16 ^{+1.19}_{-1.07}({\rm stat}) {}^{+1.37}_{-1.27}({\rm sys}) \mu {\rm b}$.
It is found that the perturbative QCD predictions of charm and bottom production are consistent with the measurement
at PHENIX.

The existence of energy loss of bottom quarks in the medium created in Au+Au collisions is found
based on the measured $(b\rightarrow e)/(c\rightarrow e+ b\rightarrow e)$ in $p+p$ collisions.
The contribution from bottom in single non-photonic electrons in Au+Au collisions should be larger than
the measured value in $p+p$ collisions due to the difference of expected suppression patterns between charm and bottom.
The contribution of bottom in Au+Au collisions is evaluated from the measured 
$(b\rightarrow e)/(c\rightarrow e+ b\rightarrow e)$ and a model calculation based on Langevin equation.
As a result, most of single non-photonic electrons in Au+Au collisions 
may be originated from bottom at $p_{\mathrm{T}}>3$~GeV/$c$. 
This result directly leads to that the existence and magnitude of bottom flow can 
be studied by measurement of $v_2(p_{\mathrm{T}})$ at $p_{\mathrm{T}}>3$~GeV/$c$.
The existence and magnitude of bottom flow will be revealed from the coming result in the near future.
The measured $(b\rightarrow e)/(c\rightarrow e+ b\rightarrow e)$ in $p+p$ collisions provides an important 
base line for such discussion.

The mechanism of the such strong interaction of heavy quarks in the medium is still under debate.
The initial nuclear effect, the temperature dependence of the magnitude of the interaction and 
separate measurement of $R_{AA}(p_{\mathrm{T}})$ and $v_2(p_{\mathrm{T}})$ for D and B mesons
are necessary to understand the mechanism.
Therefore, separate measurement of the B and D hadrons in Au+Au collisions, more precise measurement
at d+Au collisions, more statistics at high $v_2(p_{\mathrm{T}})$ and measurement at the medium with higher 
temperature are most important subjects. 
These measurement will be done at PHENIX with the new detector, Silicon Vertex Detector and the experiments at LHC
in the near future.

  \chapter*{Acknowledgment}
First of all, I would like to thank my supervisor, Prof.~H.~Hamagaki, for his essential advice and comments
about the experiment and physics. His abundant knowledge and precise suggestion always guided me.
He has helped me not only with his advice but also with support to concentrate on the research.
Without his help, I could not accomplish this work.
I would like to express my sincere thanks to Prof.~K.~Ozawa for his kind supports for the analysis 
and a great deal of encouragements.
I could learn many techniques of the experiments and analysis from him 
and could enjoy the research and my life in BNL with his support.
I would like to express my deep thanks to Prof.~Y.~Akiba for his many incisive comments and suggestions.
He introduced me to single electron analysis and this work benefited especially from his idea and suggestions.
I wish to thank Dr.~T.~Gunji for his encouragement for the completion of this work. 
His abundant knowledge of the RICH detector and electron analysis helped me.

I would like to express my thanks to Prof.~T.~Sakaguchi for his great help with the operation 
of the RICH detector and thoughtful advice on the data analysis.
I am obliged to Dr.~S.~Kametani for the help with my early work and computing environment.
I with to thank Dr.~F.~Kajihara for his abundant experience of the single electron analysis.
This work is largely motivated with the his work on the single electron.
I with to thank Dr.~T.~Isobe for his abundant knowledge of the EMCal detector.
I with to thank Dr.~S.~Oda for his kind help and abundant experience of electron analysis.

I wish to acknowledge for  all the collaborators of the PHENIX experiment.
I am much obliged to the previous and present spokespersons, Prof.~W.~A.~Zajc and 
Prof.~B.~V.~Jacak for their heartful administration of the collaboration.
I am grateful to the Year-2005 and Year-2006 RUN coordinators, Prof.~J.~Lajoie and Prof.~A.~Deshpande
for their great leadership.
I would like to thank Dr.~J.~Haggerty for his great work with the experiment.
I would like to express my thanks to the conveners of the Heavy/Light Working
Group, Prof.~M.~J.~Leitch, Dr.~R.~G.~de~Cassagnac, Prof. T.~K.~Hemmick, Dr.~A.~Lebedev, Dr.~A.~Toia and 
Dr.~X.~Wei for many constructive comments and suggestions on the analysis.
I would like to appreciate the other members of Paper Preparation Group~094 and single electron analysis group, 
Dr.~S.~Butsyk, Dr.~R.~Averbeck, Dr.~A.~Dion, Dr.~S.~Sakai and Dr.~D.~Hornback for their help
and valuable physics discussion on this analysis.
I am obliged to Dr.~C.~Pinkenburg, Dr.~D.~P.~Morrison and Dr. J.~T.~Mitchel
for their marvelous assistance on the analysis environment at BNL. 

I wish to express my appreciation to the past and present member of our group, Dr.~M.~Inuzuka,
Dr.~K.~Kino, Dr.~T.~Horaguchi, Mr.~N.~Kurihara, Mr.~S.~Saito, Mr.~Y.~Yamaguchi, Mr.~Y.~Aramaki,
Mr.~S.~Sano, Mr.~A.~Takahara, Mr.~R.~Akimoto and Mr.~Y.~Hori for their help, interesting discussions and 
the friendship.
I could really enjoy my daily life at Japan and BNL with them.
I wish to express my thanks to the CNS secretaries, Ms.~M.~Hirano, Ms.~T.~Endo, Ms.~I.~Yamamoto, 
Ms.~Y.~Kishi, Ms.~K.~Takeuchi, Ms.~T.~Itagaki, Ms.~Y.~Fujiwara and Y.~Soma.
I could concentrate on the research with their support.

I hope to express my thanks to all the staff of the RHIC project, Collider-Accelerator and 
Physics Department at BNL.
I am also grateful to all the members of Radiation Laboratory in RIKEN.
This analysis is largely supported by the RIKEN PHENIX Computing Center in Japan~(CCJ).
I wish to my special thank to Dr.~S.~Yokkaichi, Dr.~T.~Ichihara, Dr.~Y.~Watanabe
and Dr.~T.~Nakamura for their steady operation of CCJ.
I wish to acknowledge all the members of the PHENIX-J group for their help with the experiment,
the analysis and my life at BNL.
I  wish to acknowledge JSPS for its financial support to concentrate on the research.

I am obliged to Prof.~H.~Aihara, Prof.~S.~Shimoura, Prof.~T.~Hatsuda and Prof.~M.~Fukushima
for their valuable comments and constructive advice. I would like to express my
appreciation to Prof.~H.~Sakurai for his truly relevant advice.

I wish to express my deepest gratitude to my parents, Yuzo and Noriko, my brother and sister,
Koshiro and Ichiko, and my grand parents, Michio, Tatsuko, Isamu and Sachiko for
their support and encouragements of my continuing this work. 
I could not continue and finish this work without their strong support.

Last but not least, I am most grateful to my fiancee, Kyoko, for all her moral support.
I would like to dedicate this thesis to her.
  \appendix
   \chapter{Data Tables}
\begin{table}[htb]
   \begin{center}
        \caption{Invariant cross section of single non-photonic electrons 
     from heavy flavor decays at y=0.}
    \begin{tabular}{c|ccc}
      \hline
     $p_{\mathrm{T}}$~(GeV/$c$)& cross section~(mb~GeV$^{-2}$$c^3$)&stat error &sys error \\
      \hline
      0.55 & 0.00433 & 0.000404 & 0.00111 \\
0.65 & 0.00218 & 0.000237 & 0.000542 \\
0.75 & 0.00132 & 0.000148 & 0.000275 \\
0.85 & 0.000911 & 9.87e-005 & 0.000149 \\
0.95 & 0.000572 & 6.87e-005 & 8.47e-005 \\
1.1 & 0.00033 & 3e-005 & 3.82e-005 \\
1.3 & 0.000141 & 1.71e-005 & 1.53e-005 \\
1.5 & 6.51e-005 & 1.04e-005 & 6.7e-006 \\
1.7 & 3.29e-005 & 2.92e-007 & 4.41e-006 \\
1.9 & 1.87e-005 & 1.55e-007 & 2.06e-006 \\
2.1 & 1.03e-005 & 9.17e-008 & 1.01e-006 \\
2.3 & 6.36e-006 & 6.32e-008 & 5.66e-007 \\
2.5 & 3.94e-006 & 4.56e-008 & 3.31e-007 \\
2.7 & 2.43e-006 & 3.36e-008 & 2e-007 \\
2.9 & 1.52e-006 & 2.52e-008 & 1.23e-007 \\
3.1 & 1.05e-006 & 1.93e-008 & 7.97e-008 \\
3.3 & 6.65e-007 & 1.48e-008 & 5.12e-008 \\
3.5 & 4.8e-007 & 1.16e-008 & 3.5e-008 \\
3.7 & 3.08e-007 & 9.05e-009 & 2.32e-008 \\
3.9 & 2.07e-007 & 7.22e-009 & 1.6e-008 \\
4.1 & 1.54e-007 & 5.87e-009 & 1.15e-008 \\
4.3 & 1.1e-007 & 4.82e-009 & 8.28e-009 \\
4.5 & 7.35e-008 & 3.84e-009 & 5.91e-009 \\
4.7 & 5.66e-008 & 3.28e-009 & 4.42e-009 \\
4.9 & 5.03e-008 & 2.85e-009 & 3.58e-009 \\
5.5 & 1.69e-008 & 1.02e-009 & 1.69e-009 \\
6.5 & 5.45e-009 & 5.09e-010 & 4.64e-010 \\
7.5 & 1.95e-009 & 2.91e-010 & 1.54e-010 \\
8.5 & 1.3e-009 & 2.24e-010 & 8.82e-011 \\
      \hline
    \end{tabular}
   \end{center}
\end{table}

\begin{table}[htb]
  \begin{center}
    \caption{The Data/FONLL ratios of single non-photonic electron yield at y=0.}
    \begin{tabular}{c|ccc}
      \hline
     $p_{\mathrm{T}}$~(GeV/$c$)& cross section&stat error &sys error \\
      \hline
      0.55 & 2.21 & 0.206 & 0.567 \\
0.65 & 1.75 & 0.19 & 0.435 \\
0.75 & 1.68 & 0.189 & 0.35 \\
0.85 & 1.84 & 0.199 & 0.3 \\
0.95 & 1.81 & 0.217 & 0.268 \\
1.1 & 2.04 & 0.185 & 0.236 \\
1.3 & 1.92 & 0.233 & 0.208 \\
1.5 & 1.83 & 0.293 & 0.189 \\
1.7 & 1.79 & 0.0159 & 0.24 \\
1.9 & 1.86 & 0.0155 & 0.205 \\
2.1 & 1.78 & 0.0159 & 0.175 \\
2.3 & 1.84 & 0.0183 & 0.164 \\
2.5 & 1.84 & 0.0213 & 0.155 \\
2.7 & 1.78 & 0.0246 & 0.146 \\
2.9 & 1.71 & 0.0283 & 0.138 \\
3.1 & 1.76 & 0.0324 & 0.134 \\
3.3 & 1.65 & 0.0367 & 0.127 \\
3.5 & 1.72 & 0.0416 & 0.126 \\
3.7 & 1.58 & 0.0463 & 0.119 \\
3.9 & 1.49 & 0.0521 & 0.115 \\
4.1 & 1.54 & 0.0589 & 0.115 \\
4.3 & 1.52 & 0.0666 & 0.114 \\
4.5 & 1.38 & 0.0722 & 0.111 \\
4.7 & 1.44 & 0.0831 & 0.112 \\
4.9 & 1.71 & 0.0967 & 0.121 \\
5.5 & 1.3 & 0.0788 & 0.13 \\
6.5 & 1.45 & 0.135 & 0.123 \\
7.5 & 1.56 & 0.233 & 0.124 \\
8.5 & 2.83 & 0.488 & 0.192 \\
      \hline
    \end{tabular}
   \end{center}
  
\end{table}

\begin{table}[hbt]
    \begin{center}
     \caption{Result of $(b\rightarrow e)/(c\rightarrow e+ b\rightarrow e)$ in RUN5 and RUN6 }
     \begin{tabular}{|c|c|}
       \hline
       electron $p_{\mathrm{T}}$ & $(b\rightarrow e)/(c\rightarrow e+ b\rightarrow e)$ \\
       \hline \hline
       2.35~GeV/$c$ & $<0.41$~(90\% C.L) 0.19~(50\% point) \\
       \hline
       3.37~GeV/$c$ & $0.26^{+0.14}_{0.13}(stat) ^{+0.11}_{0.11} (sys)  $     \\
       \hline
       4.40~GeV/$c$ & $0.63^{+0.18}_{-0.21} (stat) \pm 0.08 (sys)  $     \\
       \hline
       5.66~GeV/$c$ & $>0.33$~(90\% C.L) 0.71~(50\% point)  \\
       \hline
     \end{tabular}
    \end{center}
\end{table}

\begin{table}[hbt]
    \begin{center}
     \caption{Invariant cross section of electrons from charm and bottom}
     \begin{tabular}{|c|c|c|}
       \hline
       electron $p_{\mathrm{T}}$ & cross section~(mb~GeV$^{-2}c^3$)& data/FONLL \\
       \hline \hline
       charm & &  \\
       2.35~GeV/$c$ & $>3.30$~(90\% C.L) 4.52~(50\%) $\times 10^{-6}$ & $>1.49$~(90\% C.L) 2.03~(50\%)\\
       3.37~GeV/$c$ & $4.17^{+0.73}_{-0.83} {}^{+0.41}_{-0.46} \times 10^{-7}$ & $2.05^{+0.36}_{-0.41} {}^{+0.20}_{-0.22}$\\
       4.40~GeV/$c$ & $3.49^{+1.95}_{-1.70} \pm 0.66 \times 10^{-8}$ & $1.16^{+0.65}_{-0.56} \pm 0.22 $\\
       5.66~GeV/$c$ & $<1.11$~(90\% C.L) 0.48~(50\%) $\times 10^{-8}$& $<2.48$~(90\% C.L) 1.08~(50\%) \\
       \hline\hline
       bottom & &  \\
       2.35~GeV/$c$ &  $<2.30$~(90\% C.L) 1.08~(50\%) $\times 10^{-6}$&  $<2.74$~(90\% C.L) 1.29~(50\%)\\
       3.37~GeV/$c$ & $1.49^{+0.83}_{-0.73} {}^{+0.73}_{-0.66} \times 10^{-7}$ & $0.99^{+0.55}_{-0.48} {}^{+0.48}_{-0.43}$\\
       4.40~GeV/$c$ & $5.95^{+1.70}_{-1.95} \pm 1.10 \times 10^{-8}$ & $1.87^{+0.54}_{-0.61} \pm 0.34$\\
       5.66~GeV/$c$ & $>0.54$~(90\% C.L) 1.17~(50\%) $\times 10^{-8}$&  $>0.90$~(90\% C.L) 1.93~(50\%) \\
       \hline
     \end{tabular}
    \end{center}
\end{table}

   \chapter{Comparison Between Real Data and Simulation in RUN6}
\section{eID varibles}\label{sec:eidrun6}
The distributions of the variables used for the electron identification from the 
PISA simulation are compared to these of the real data in RUN6.
The used cuts for each variable comparison are described in Sec.~\ref{sec:pisa}.

\begin{figure}[htb]
     \begin{center}
       \epsfig{figure=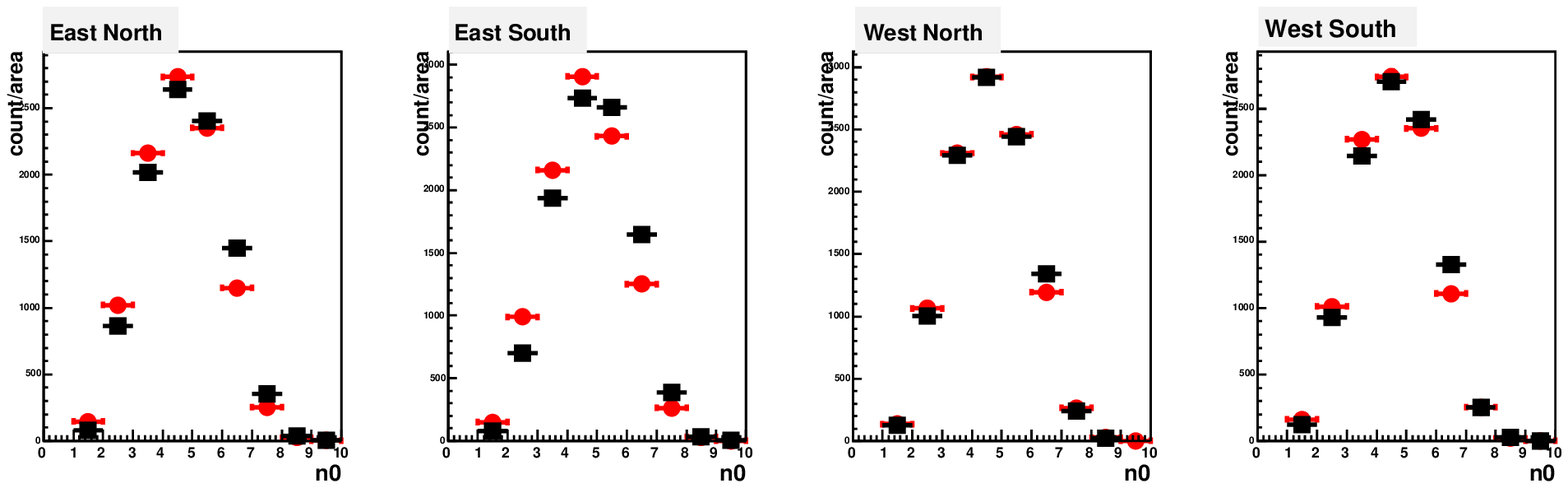,width=13cm}
       \caption{
	 The distribution of {\sf n0} with the standard eID cut without {\sf n0} cut 
	 and the 0.5$<p_{\mathrm{T}}<$5 GeV/$c$ cut in the real data~(square) and 
	 simulation~(circle).
	 \label{fig:n0_run6}}
     \end{center}
   \end{figure}

\begin{figure}[htb]
     \begin{center}
       \epsfig{figure=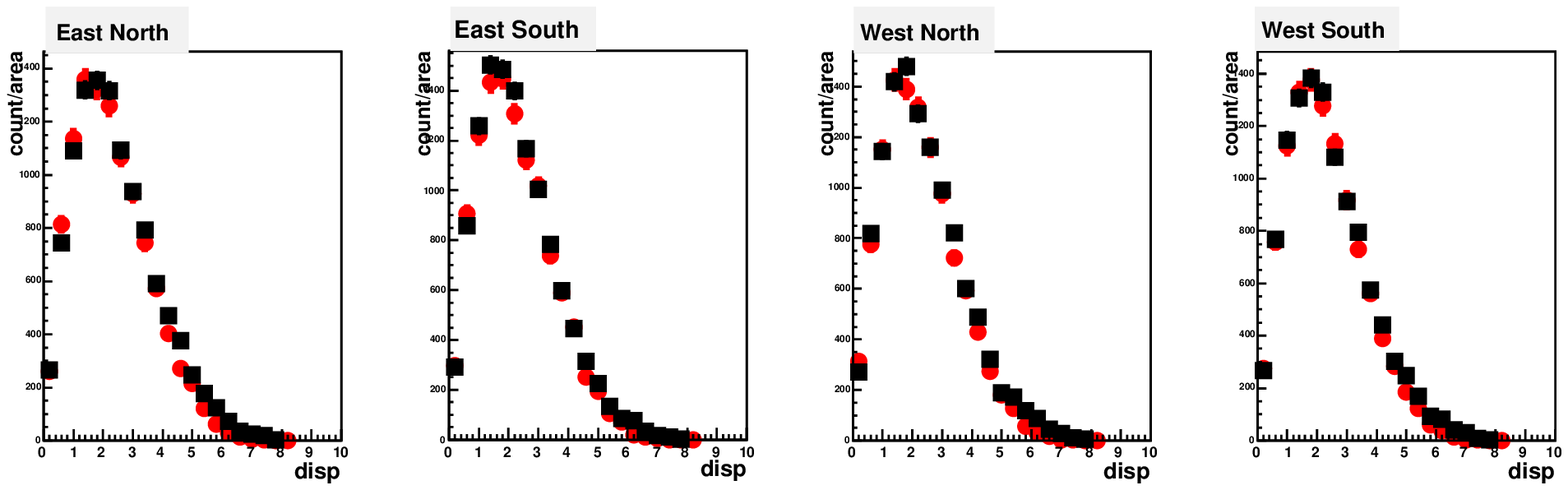,width=13cm}
       \caption{
	 The distribution of {\sf disp} with the standard eID cut without {\sf disp} cut 
	 and the 0.5$<p_{\mathrm{T}}<$5 GeV/$c$ cut in the real data~(square) and 
	 simulation~(circle).
	 \label{fig:disp_run6}}
     \end{center}
   \end{figure}

\begin{figure}[htb]
     \begin{center}
       \epsfig{figure=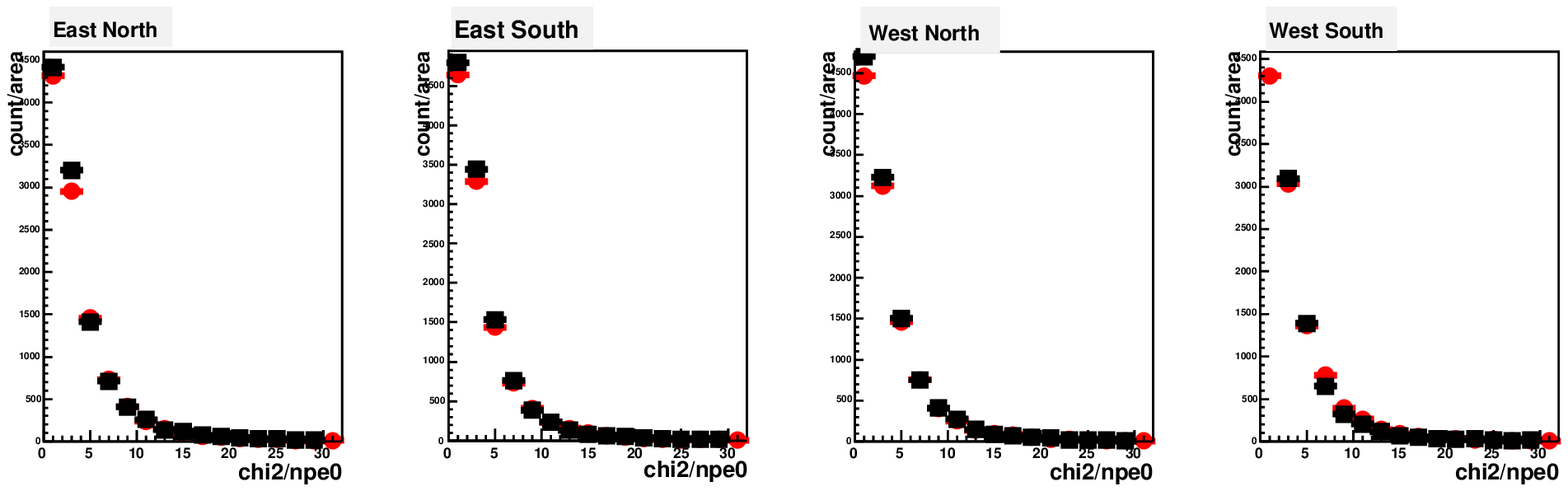,width=13cm}
       \caption{
	 The distribution of {\sf chi2/npe0} with the standard eID cut without {\sf chi2/npe0} 
	 cut and the 0.5$<p_{\mathrm{T}}<$5 GeV/$c$ cut in the real data~(square) and 
	 simulation~(circle).
	 \label{fig:chi_run6}}
     \end{center}
   \end{figure}

\begin{figure}[htb]
     \begin{center}
       \epsfig{figure=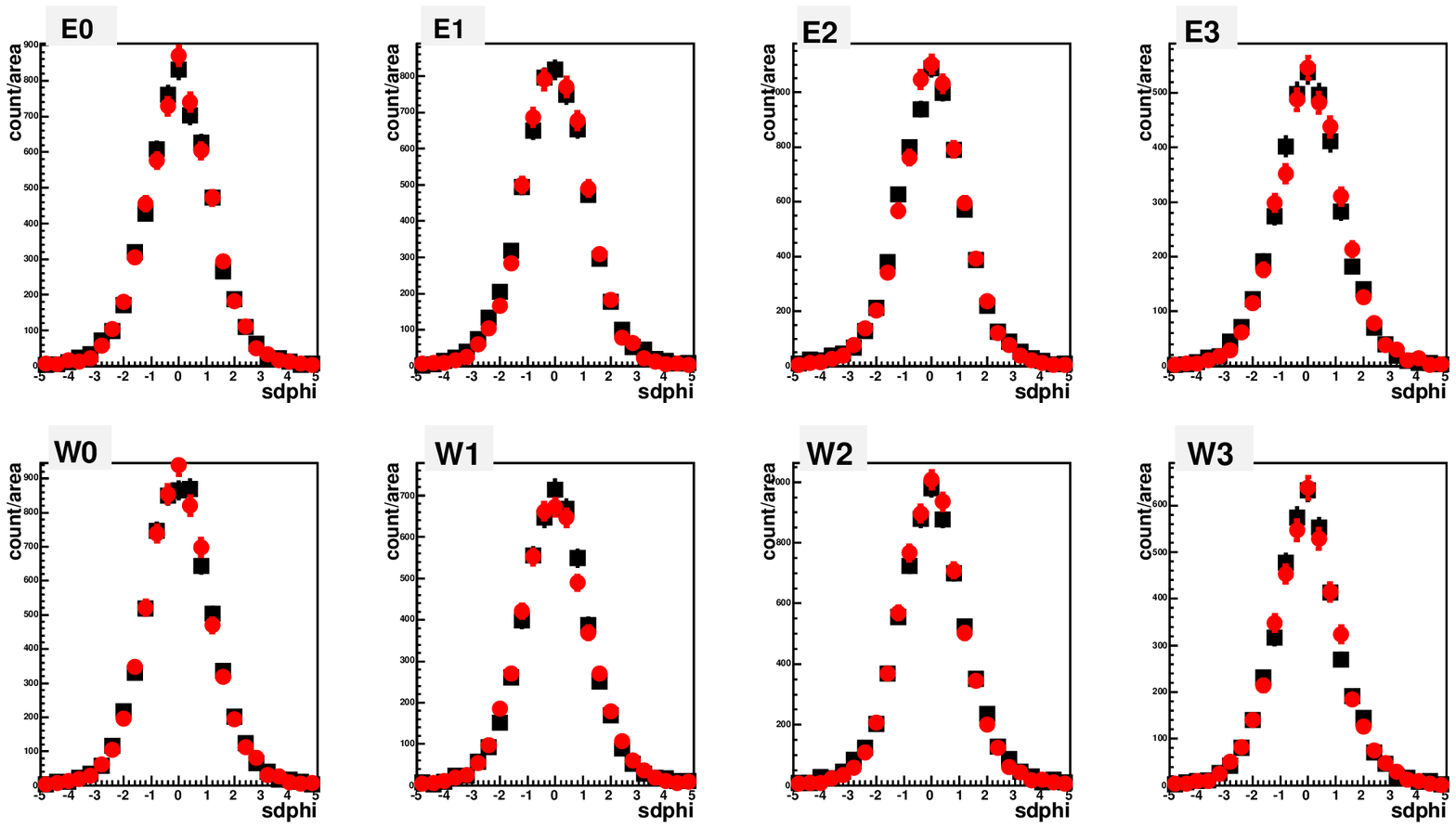,width=13cm}
       \caption{
	 The distribution of {\sf emcsdphi\_e} with the standard eID cut without 
	 {\sf  emcsdphi(z)\_e} cut and the 0.5$<p_{\mathrm{T}}<$5 GeV/$c$ 
	 cut in the real data~(square) and simulation~(circle).
	 \label{fig:sdphi_run6}}
     \end{center}
   \end{figure}

\begin{figure}[htb]
     \begin{center}
       \epsfig{figure=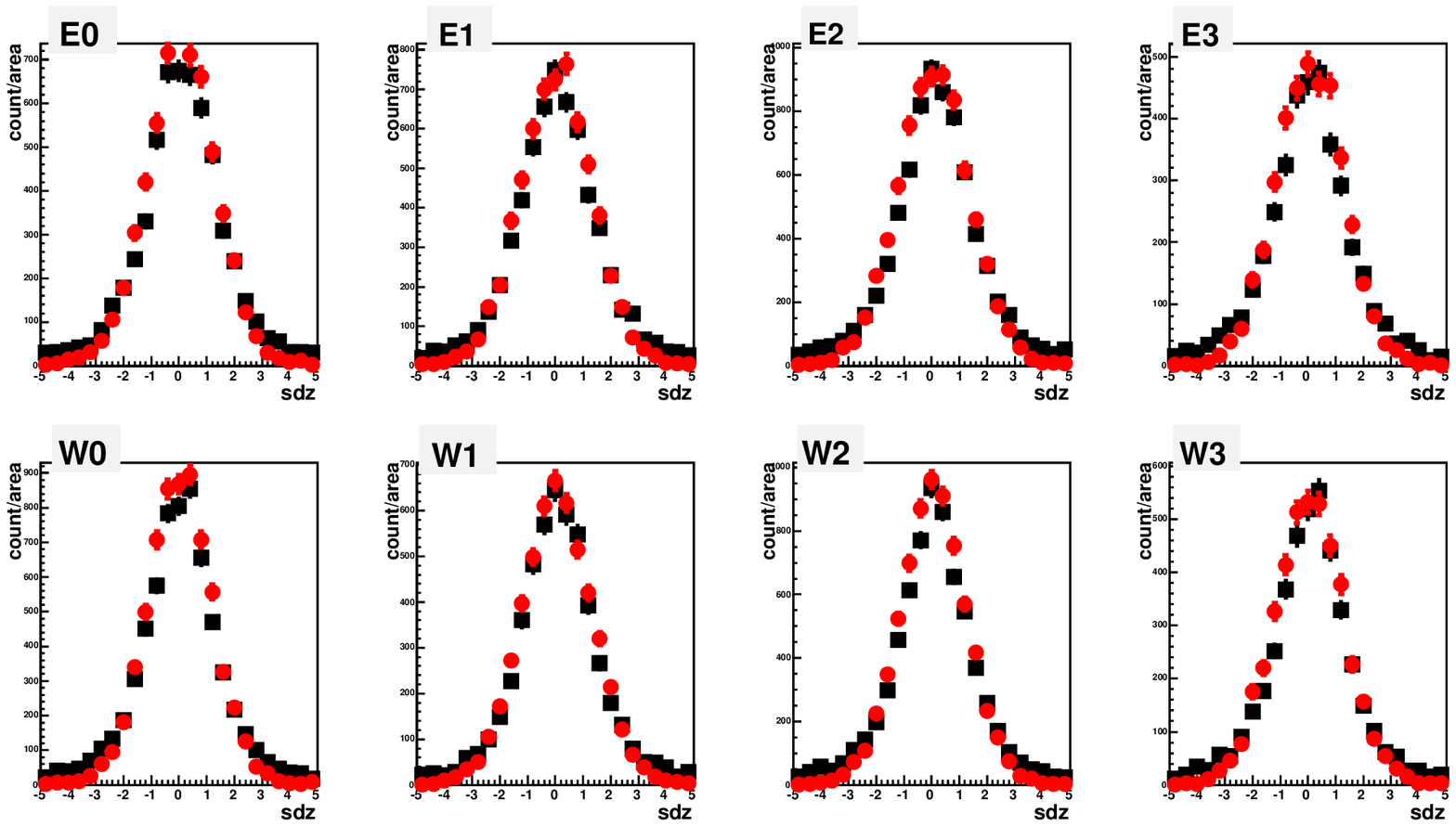,width=13cm}
       \caption{
	 The distribution of {\sf emcsdz\_e} with the standard eID cut without 
	 {\sf  emcsdphi(z)\_e}  cut 
	 and the 0.5$<p_{\mathrm{T}}<$5 GeV/$c$ cut in the real data~(square) and 
	 simulation~(circle).
	 \label{fig:sdz_run6}}
     \end{center}
   \end{figure}

\begin{figure}[htb]
     \begin{center}
       \epsfig{figure=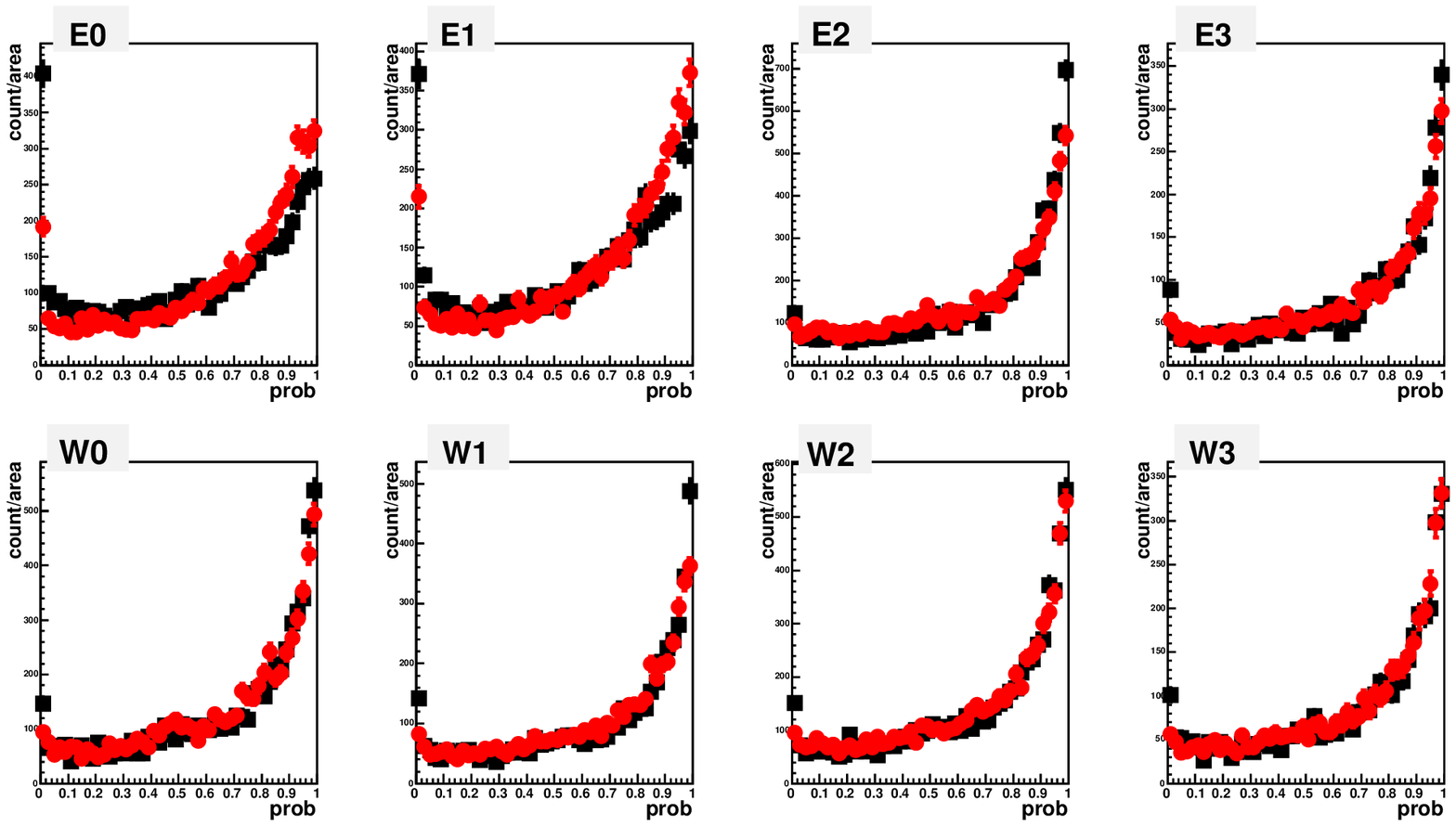,width=13cm}
       \caption{
	 The distribution of {\sf prob} with the standard eID cut without {\sf prob} cut 
	 and the 0.5$<p_{\mathrm{T}}<$5 GeV/$c$ cut in the real data~(square) and 
	 simulation~(circle).
	 \label{fig:prob_run6}}
     \end{center}
   \end{figure}

\begin{figure}[htb]
       \begin{center}
	 \epsfig{figure=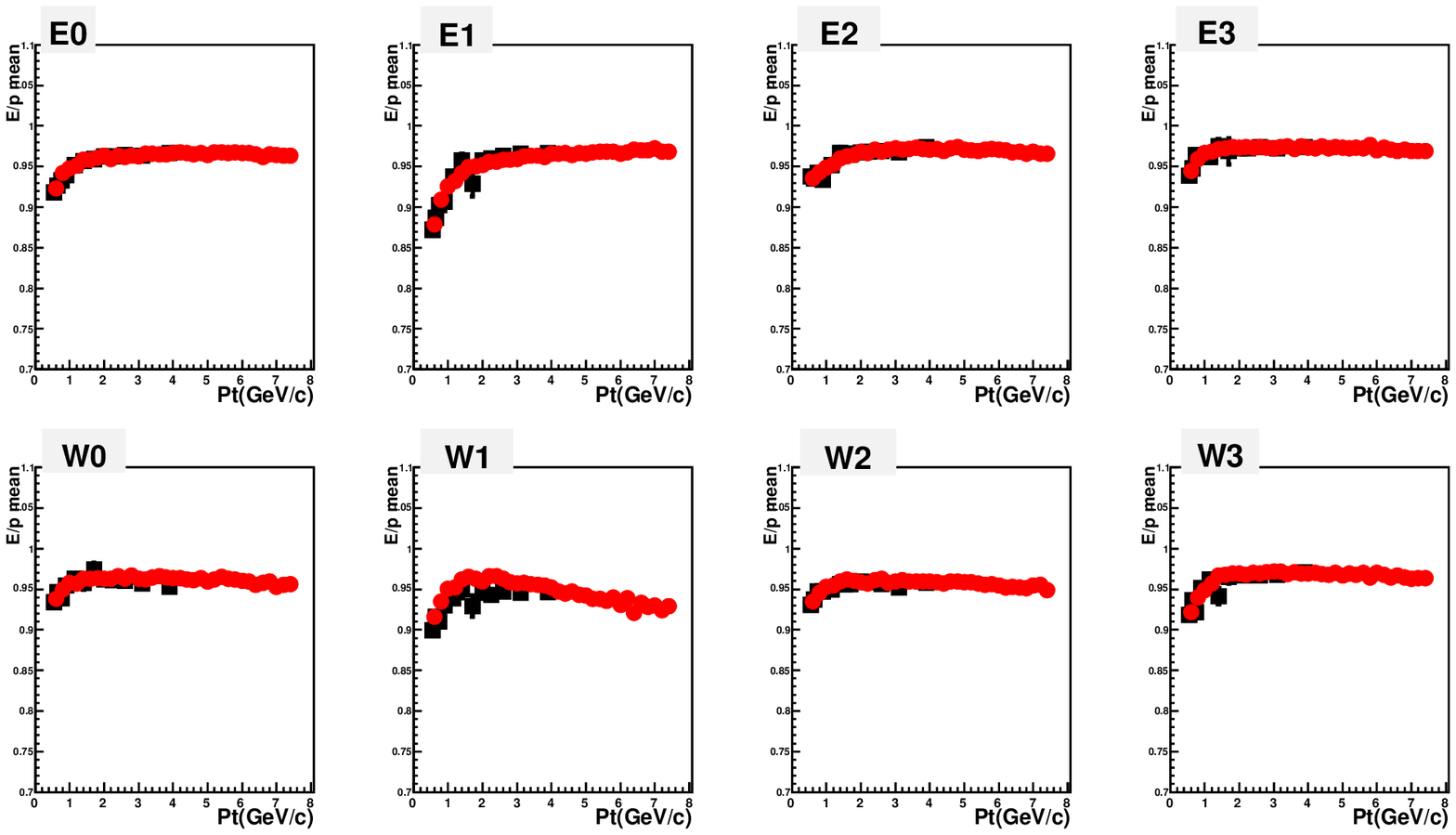,width=13cm}
	 \caption{
	   The mean value of {\sf ecore/mom} distribution with the standard eID cut 
	   as a fiction of electron $p_{\mathrm{T}}$ in the real data~(square) and 
	   simulation~(circle).
	   \label{fig:epmean_run6}}
       \end{center}
     \end{figure}

\begin{figure}[htb]
       \begin{center}
       \epsfig{figure=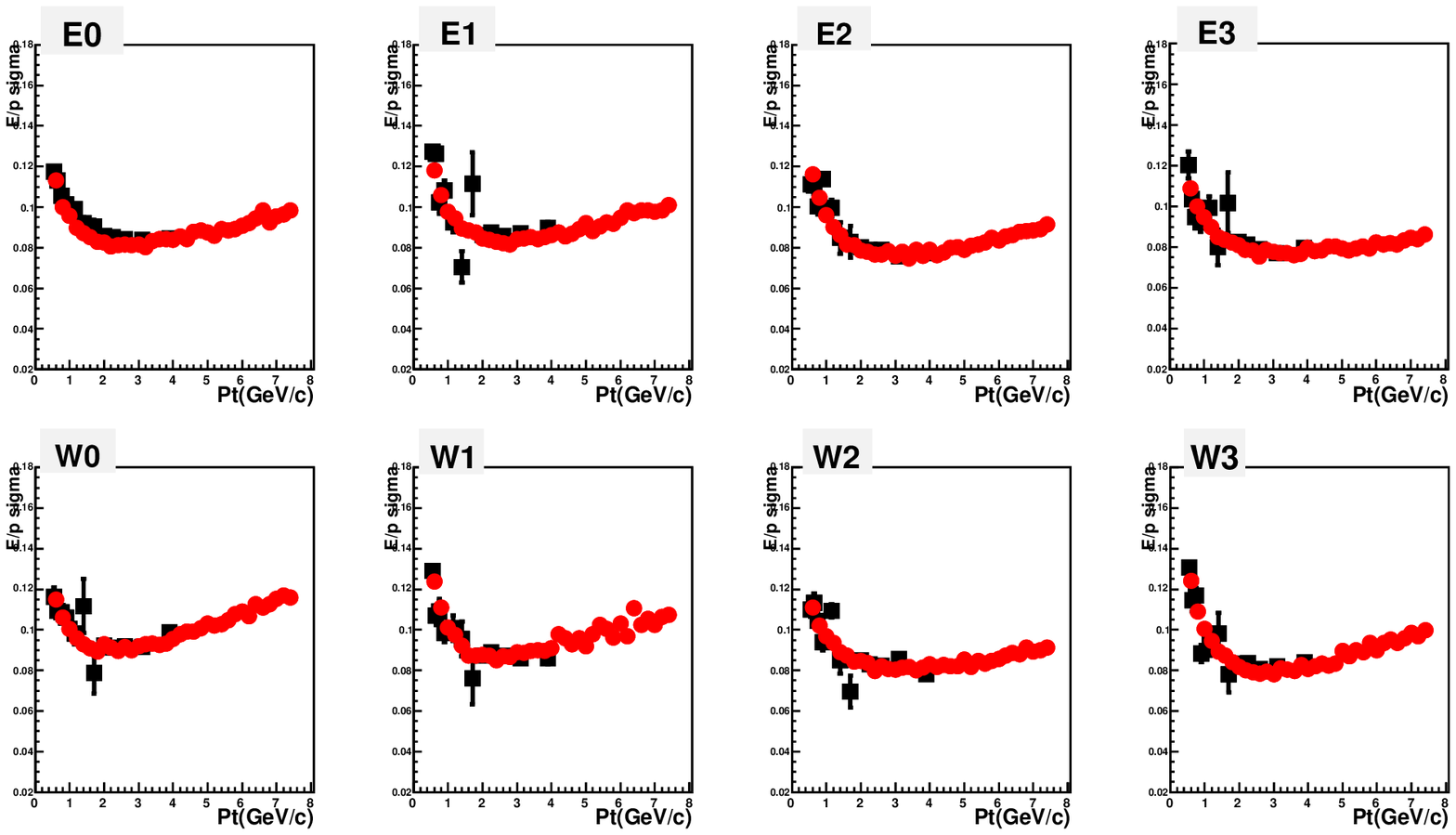,width=13cm}
       \caption{
	 The sigma value of {\sf ecore/mom} distribution with the standard eID cut 
	 as a fiction of electron $p_{\mathrm{T}}$ in the real data~(square) and 
	 simulation~(circle).
	 \label{fig:epsigma_run6}}
     \end{center}
   \end{figure}

 Figure~\ref{fig:n0_run6}, \ref{fig:disp_run6} and \ref{fig:chi_run6} show 
   the distributions of RICH variables, {\sf n0}, {\sf disp} and {\sf chi2/npe0} 
   at each RICH sector, respectively.
   In addition, Figure~\ref{fig:sdphi_run6}, ~\ref{fig:sdz_run6} and 
   ~\ref{fig:prob_run6} show the distributions of EMCal variables at each sector, 
   {\sf emcsdphi\_e}, {\sf emcsdz\_e} and {\sf prob}, respectively.
   Figure~\ref{fig:epmean_run6} and \ref{fig:epsigma_run6} show mean and sigma values
   of {\sf ecore/mom} distributions as a function of electron $p_{\mathrm{T}}$.
   In Fig.~\ref{fig:n0_run6}-\ref{fig:epsigma_run6}, black squares show the results from the real data
   in RUN6 and red circles show these from the PISA simulation with RUN6 tuning parameters and 
   CM++ field.
   The distribution in simulation is normalized by the number of entries at 
   each sector.
   The distributions of the simulation and these of the real data match well.

\section{Geometrical Acceptance}\label{sec:georun6}
\begin{figure}[htb]
       \begin{center}
	 \epsfig{figure=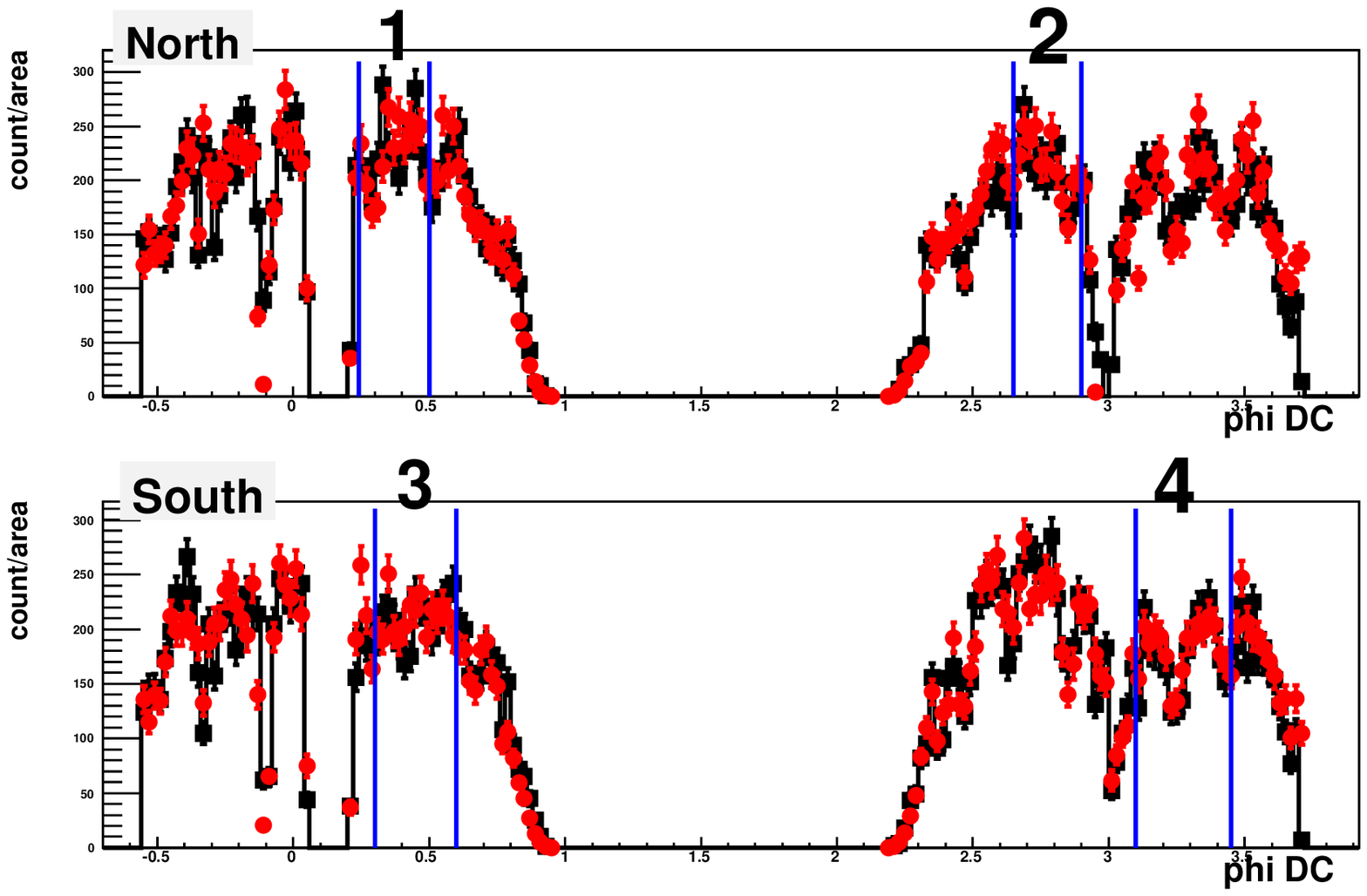,width=13cm}
	 \caption{
	   The distribution of {\sf phi} with the standard eID cut and 
	   the 0.5$<p_{\mathrm{T}}<$5 GeV/$c$ cut in the real data in RUN6~(square) and 
	   simulation~(circle).
	   \label{fig:dcphi_run6}}
       \end{center}
     \end{figure}

\begin{figure}[htb]
       \begin{center}
       \epsfig{figure=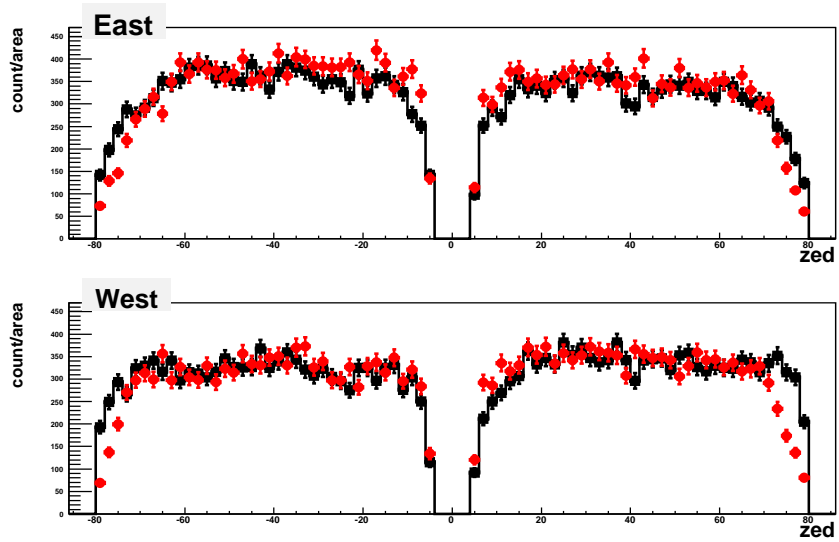,width=13cm}
       \caption{
	 The distribution of {\sf zed} with the standard eID cut and 
	 the 0.5$<p_{\mathrm{T}}<$5 GeV/$c$ cut in the real data in RUN6~(square) and 
	 simulation~(circle).
	 \label{fig:dczed_run6}}
     \end{center}
   \end{figure}

The distributions of {\sf phi}, {\sf zed} of the simulation are compared 
   with these of the real data for the electron samples selected by the standard eID 
   and a transverse momentum  with  0.5$<p_{\mathrm{T}}<$5 GeV/$c$.
   Figure~\ref{fig:dcphi_run6} shows the distributions of  {\sf phi} at North~(top panel)
   and South~(bottom panel)sector, and 
   Figure~\ref{fig:dczed_run6} shows the distributions of  {\sf zed} at East~(top panel)
   and West~(bottom panel) sector.
   In Fig.~\ref{fig:dcphi_run6} and \ref{fig:dczed_run6}, black squares show 
   the real data in RUN6 and red circles show the PISA simulation with 
   RUN6 tuning parameters and CM++ field.
   The distributions of simulation are normalized by number of entries in 
   the reference regions, where are little low efficiency, dead or noisy area.
   In Fig~\ref{fig:dcphi_run6} and \ref{fig:dczed_run6}, the used reference region  
   to normalize is region 1.
     
   \chapter{Correlation Function of Electron and Hadron} \label{sec:c_hq}
\section{Correlation Function in Real Data}
{\bf Acceptance filter is NOT applied when we study the correlation function }\\
\\
Two particle correlations with respect to the azimuthal angular difference, which is called 
'correlation function' is studied in this chapter.
The one of two particle is called 'trigger particle' and the other is called 'associated particle'.
Correlation function  is defined as the measured yield of the assocated particles as a 
function of the azimuthal angular difference.
Correlation function of trigger electrons and associated hadrons is studied to compare with PYTHIA simulation.
Since charge asymmetry of electing hadron pairs have little PYTHIA parameter dependence, 
the comparison with real data between PYTHIA should be done 
for other observables to study the tuning status of PYTHIA.
For this purpose, we choose the correlation function between the trigger non-photonic electrons and 
the associated hadrons.

\begin{figure}[h]
  \begin{tabular}{c c}
    \begin{minipage}{\minitwocolumn}
      \begin{center}
	\includegraphics[width=7.5cm]{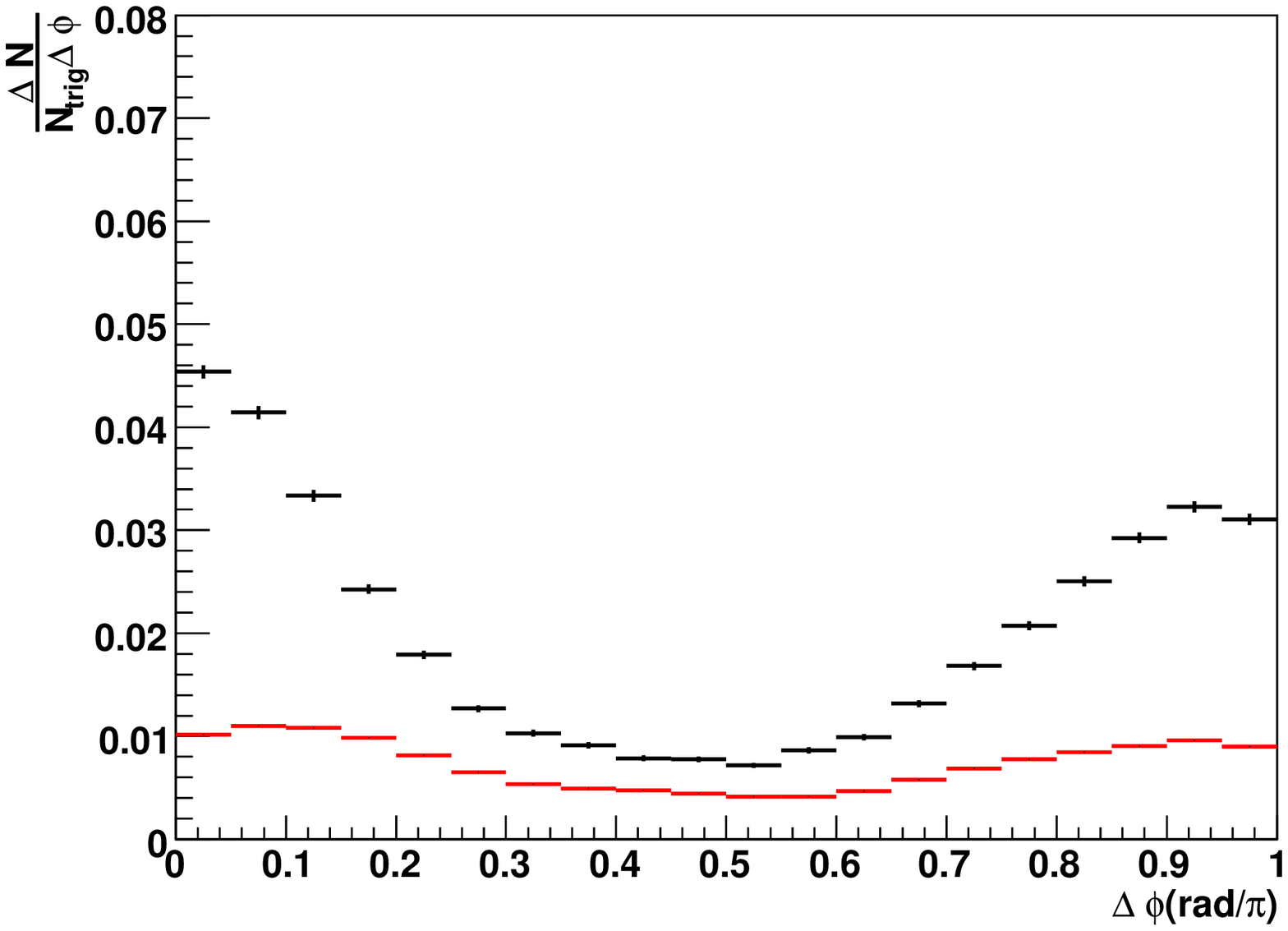}
        \caption{The $\Delta$N/$\Delta \phi$ distribution per
	  the number of the trigger electrons, when the trigger electrons were
	  the inclusive electrons.Black points shows the $\Delta$N/$\Delta \phi$ in
	  real events and red points show the background.
	}
        \label{ap2_fig16}
      \end{center}
    \end{minipage}
    &
    \begin{minipage}{\minitwocolumn}
      \begin{center}
	\includegraphics[width=7.5cm]{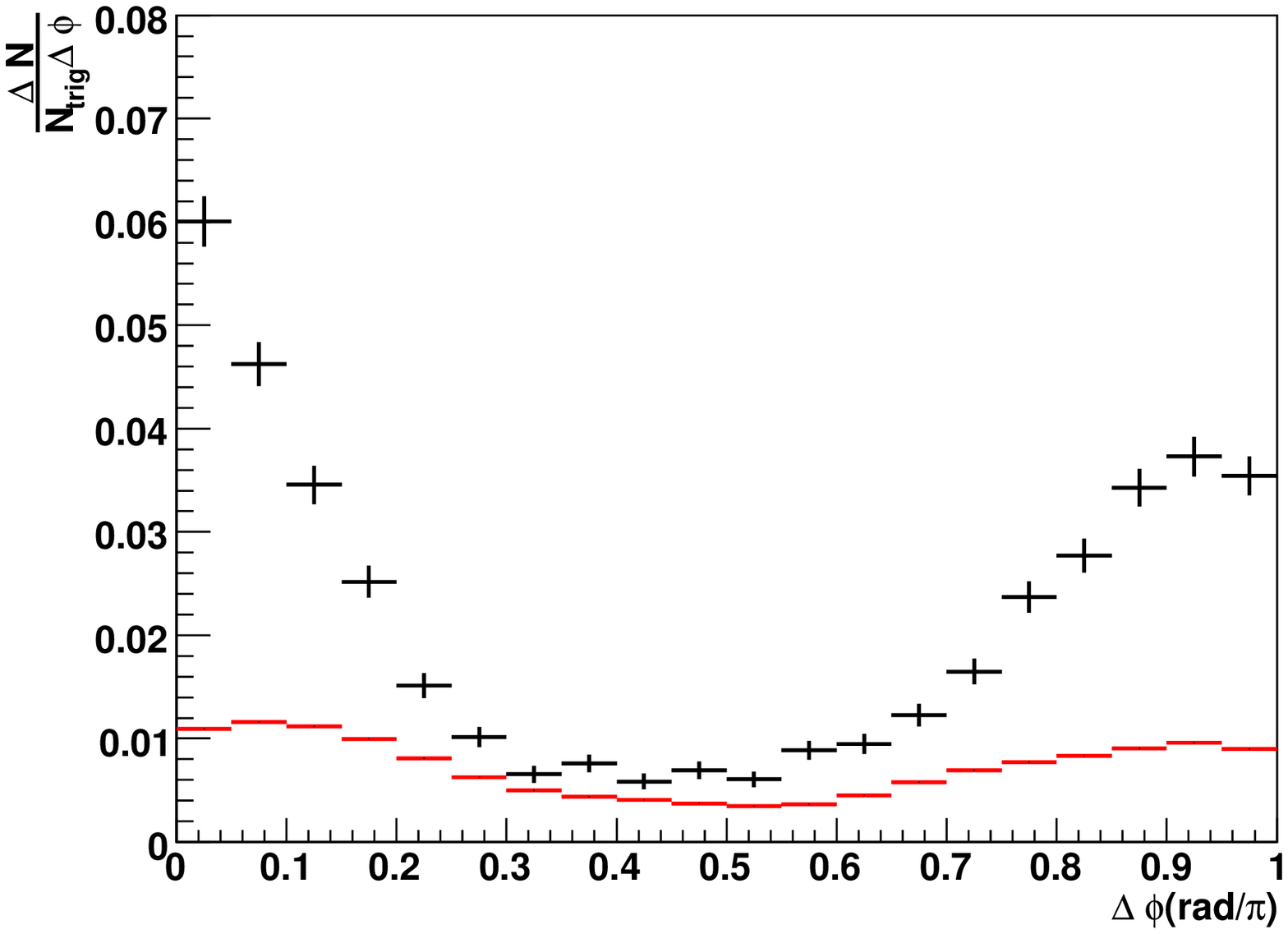}
        \caption{The $\Delta$N/$\Delta \phi$ distribution per
	  the number of the trigger electrons, when the trigger electrons were 
	  the photonic electrons.Black points shows the $\Delta$N/$\Delta \phi$ in
	  real events and red points show the background
	}
        \label{ap2_fig17}
      \end{center}
    \end{minipage}
  \end{tabular}
\end{figure}

The condition of analysis cut for trigger electrons is described at Sec~\ref{sec:ecut}.
$p_{\mathrm{T}}$ range of the selected trigger electrons is $2.0<p_{\mathrm{T}}<5.0$~GeV/$c$ and 
that of the selected associated hadrons is $0.4<p_{\mathrm{T}}<5.0$~GeV/$c$.
Figure~\ref{ap2_fig16} shows the raw number of the associated hadrons as a funtion of the azimuthal angle with 
respect to the trigger electron, $\Delta$N/$\Delta \phi$, per
the number of the trigger electrons, where the trigger electrons are inclusive electrons.
Black points shows the $\Delta$N/$\Delta \phi$ in ERT triggered events in RUN5.
Red points show the $\Delta$N/$\Delta \phi$ in mixing events.
$\Delta$N/$\Delta \phi$ in mixing events is considered as the background to take into account
the effect of geometrical acceptance.
$\Delta$N/$\Delta \phi$ in mixing events is normalized by the number of the trigger electrons.

Figure~\ref{ap2_fig17} shows the $\Delta$N/$\Delta \phi$ distribution per
the number of the trigger electrons, where the trigger electrons are the photonic electrons.
Black points show the $\Delta$N/$\Delta \phi$ in ERT triggered events in RUN5 and red points show 
the background used by the mixing event mothod.
The photonic electrons are identified by the invariant mass distribution of di-electron.
When the associated electron is found in the associated particles, we calculate
the invariant mass of the trigger electrons and the associated electrons.
The trigger electrons is identified as the photonic electron trigger, when the invariant mass
is bellow 0.08~GeV/$c^2$.
When we make the background for the photonic electron trigger by mixing events, 
the photonic trigger electrons are selected by the above method to take account for 
the acceptance bias by the selection of the photonic electrons.

\begin{figure}[htb]
  \begin{center}
    \includegraphics[width=14cm]{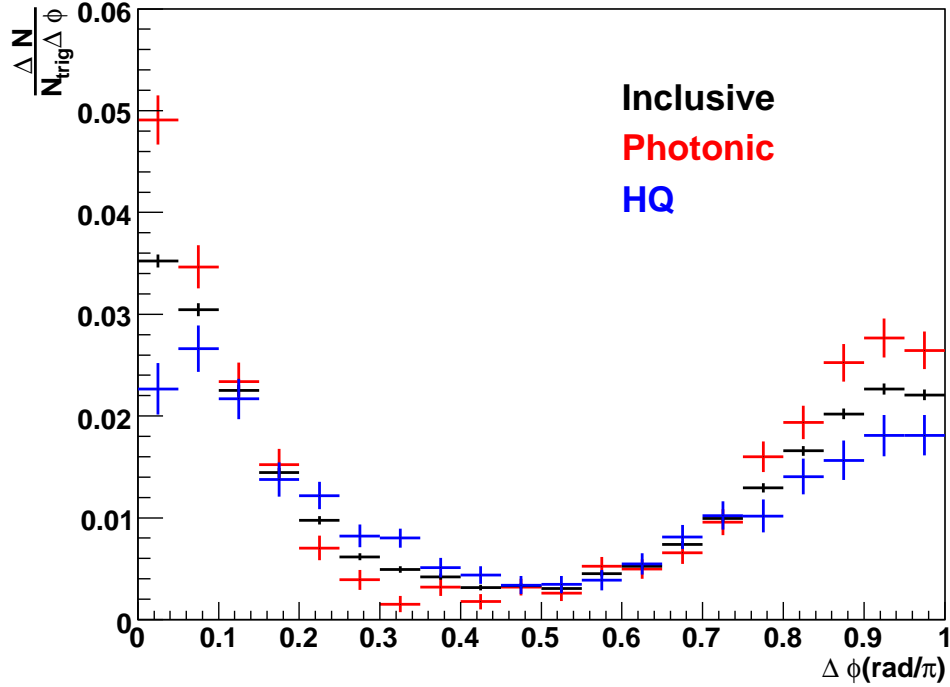}
    \caption{The correlation function of electron-hadrons when the trigger electrons
      are inclusive~(black), photonic~(red) and heavy flavor~(blue).}
    \label{ap2_fig18}
  \end{center}
\end{figure}

The correlation functions of the inclusive and the photonic electron-hadrons are obtained 
by the subtraction of the background $\Delta$N/$\Delta \phi$ distribution.
The correlation function of electrons from heavy flavor - hadrons is obtained 
as follows.
\[
C_{HQ}(\Delta \phi) = (C_{incl}(\Delta \phi) - (1-R_{HQ})\times C_{phot}(\Delta \phi))/R_{HQ}.
\]
Here,\\
\begin{itemize}
\item $C_{HQ}(\Delta \phi)$ is the correlation function, where the trigger electrons are 
  from heavy flavor.
\item $C_{incl}(\Delta \phi)$ is the correlation function, where the trigger electrons are 
  inclusive electrons
\item  $C_{phot}(\Delta \phi)$ is the correlation function, where the trigger electrons are 
  photonic electrons
\item $R_{HQ}$ is the fraction of electrons from heavy flavor in the inclusive electrons.
\end{itemize}
Figure~\ref{ap2_fig18} shows the correlation functions of electron-hadrons, where the trigger electrons
are inclusive~(black), photonic~(red) and heavy flavor~(blue).

\section{PYTHIA tuning status}
The PYTHIA tuning status is studied by the comparison of the correlation function of
heavy flavor electrons and hadrons between in RUN5 data and PYTHIA simulation.
The correlation function is obtained in PYTHIA simulation by the similar way in
the analysis at real data.

\begin{figure}[htb]
  \begin{center}
    \includegraphics[width=11cm]{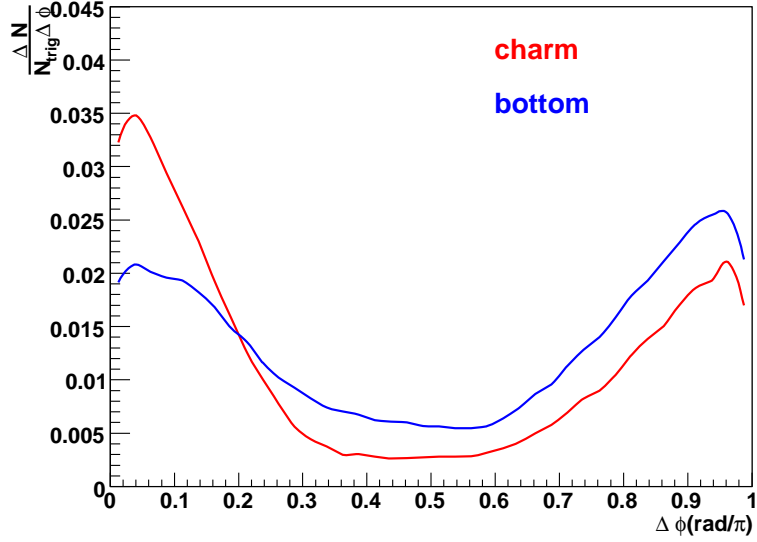}
    \caption{The correlation function of electrons and hadrons at 
      the tuned PYTHIA, when the trigger electrons were from charm and bottom.
      Red line shows the correlation function in the case of charm production and 
      blue line shows that in the case of bottom production.}
    \label{ap2_fig101}
  \end{center}
\end{figure}

\begin{figure}[htb]
  \begin{center}
    \includegraphics[width=11cm]{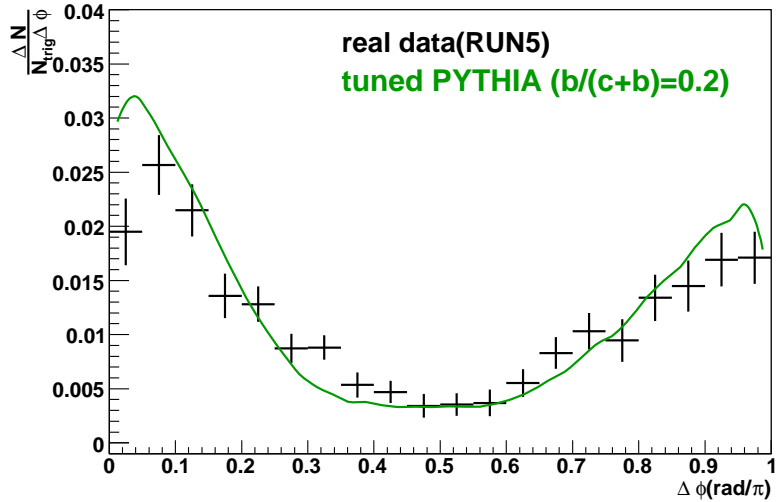}
    \caption{The correlation function of electrons and hadrons, where
      the trigger electrons are from heavy flavor.
      Black points show the result at RUN5 data obtained  and 
      green line shows the result at PYTHIA when we set $R_{b}$ to 0.2.}
    \label{ap2_fig201}
  \end{center}
\end{figure}

Figure~\ref{ap2_fig101} shows the correlation function of electrons and hadrons in
the PYTHIA simulation, where the trigger electrons are from charm and bottom.
Red line shows the correlation function in the case of charm production and 
blue line shows that in the case of bottom production.
For the comparison with real data and PYTHIA, the obtained PYTHIA results are mixed up
as follows.
\[
C_{HQ}(\Delta \phi) = R_{b}(C_{b}(\Delta \phi) + (1-R_{b})C_{c}(\Delta \phi)).
\]
Here,
\begin{itemize}
\item $C_{HQ}(\Delta \phi)$ is the correlation function, where the trigger electrons are 
  from heavy flavor.
\item $C_{c~(b)}(\Delta \phi)$ is the correlation function, where the trigger electrons
 are   from charm~(bottom)
\item $R_{b}$ is $\frac{N_{b\rightarrow e}}{N_{c\rightarrow e}+N_{b\rightarrow e}}$.
\end{itemize}
We set $R_{b}$ to 0.2 from this analysis for the comparison with real data.
Figure~\ref{ap2_fig201} shows the correlation function of electrons and hadrons, where
the trigger electrons are from heavy flavor.
Black points show the result in RUN5 data and 
green line shows the result in PYTHIA.
The PYTHIA simulation agrees with the real data.
This indicated the tuning for PYTHIA simulation is well.

   \chapter{Centrality Information} \label{sec:gtable}
Collision centrality, which is an observable corresponded to impact parameter $b$, is determined by 
information by BBC and ZDC in PHENIX.
Au+Au collisions at $\sqrt{s_{NN}}=200$~GeV are categorized into 9 centrality groups, 0\%-10\%,
10\%-20\%, 20\%-30\%, 30\%-40\%, 50\%-60\%, 60\%-70\%, 70\%-80\% and 80\%-92.3\%.
0\%-10\% collisions are most central collisions and 80\%-92.3\% collisions are most pheripheral collisions.
Impact parameter~($b$), the number of binary collisions~($N_{coll}$) and the number of participants~($N_{part}$), 
which corresponds to each centrality group, are determined according to the Glauber Model~\cite{bib:phgrau}.
The results are summarized in Table~\ref{tab:ap6}.

\begin{table}[htb]
   \begin{center}
      \caption{Results of Glauber Calculations for Au+Au collisions at $\sqrt{s_{NN}}=200$~GeV.
     }
   \label{tab:ap6}
    \begin{tabular}{c|ccc}
      \hline
      Centrality~(\%) & $b$~(fm) & $N_{part}$ & $N_{coll}$ \\
      \hline
      0-10    & 3.1  & 325 & 955 \\
      10-20   & 5.7  & 234 & 603 \\
      20-30   & 7.4  & 167 & 374 \\
      30-40   & 8.7  & 114 & 220 \\
      40-50   & 9.9  &  74 & 120 \\
      50-60   & 11.0 &  46 &  61 \\
      60-70   & 11.9 &  26 &  29 \\
      70-80   & 12.8 &  13 &  12 \\
      80-92.3 & 14.1 &   6 &  5 \\
      \hline
    \end{tabular}
   \end{center}
  
\end{table}
   \chapter{Details of $\epsilon_{c}$ and  $\epsilon_{b}$} \label{sec:detail}

	\begin{table}[hbt]
	  \begin{center}
	    \caption{Detail of charm and bottom decay for  electron $p_{\mathrm{T}}$ 3-4 GeV/c}
	    \label{chap6_table7}
	    \begin{tabular}{c|c c c}
	      \hline
	      channel & $N_{tag}$ (part)/(all) & $N_{ele}$ (part)/(all) & $\epsilon$ \\
	      \hline
	      $D^0 $ & & & \\
	      $D^0\rightarrow e^+ K^- \nu_e$&46.71\%&29.46\%&8.06 $\pm$ 0.18\% \\
	      $D^0\rightarrow e^+ K^{\ast -} \nu_e$&11.84\%&3.11\%&19.31 $\pm$ 0.73\% \\
	      $D^0\rightarrow e^+ \pi^- \nu_e$&4.65\%&5.77\%&4.10 $\pm$ 0.36\% \\
	      $D^0\rightarrow e^+ \rho^- \nu_e$&1.29\%&0.45\%&14.63 $\pm$ 1.70\% \\
	      $D^0\rightarrow e^+ other$&0.91\%&0.44\%&10.39 $\pm$ 1.74\% \\
	      \hline
       $D^+ $ & & & \\
       $D^+\rightarrow e^+ \bar{K^0} \nu_e$&21.57\%&39.47\%&2.78 $\pm$ 0.13\% \\
       $D^+\rightarrow e^+ \bar{K^{\ast 0}} \nu_e$&6.08\%&4.05\%&7.64 $\pm$ 0.64\% \\
       $D^+\rightarrow e^+ \pi^0 \nu_e$&1.84\%&3.79\%&2.47 $\pm$ 0.35\% \\
       $D^+\rightarrow e^+ \rho^0 \nu_e$&0.14\%&0.32\%&2.20 $\pm$ 2.31\% \\
       $D^+\rightarrow e^+ other$&1.60\%&1.92\%&4.23 $\pm$ 0.72\% \\
       \hline
       $D_s $ & & & \\
       $D_s\rightarrow e^+ \phi \nu_e$&0.44\%&0.73\%&3.04 $\pm$ 1.56\% \\
       $D_s\rightarrow e^+ \eta \nu_e$&3.91\%&7.80\%&2.55 $\pm$ 0.26\% \\
       $D_s\rightarrow e^+ \eta' \nu_e$&0.58\%&0.57\%&5.18 $\pm$ 1.49\% \\
       $D_s\rightarrow e^+ other$&0.16\%&0.82\%&0.98 $\pm$ 0.90\% \\
       \hline
       $\Lambda_c $ & & & \\
       $\Lambda_c\rightarrow e^+ \Lambda \nu_e$&-0.57\%&0.26\%&-11.24 $\pm$ 5.15\% \\
       $\Lambda_c\rightarrow e^+ other$&-1.15\%&1.04\%&-5.63 $\pm$ 2.34\% \\
       \hline
       $B^0 $ & & & \\
       $B^0\rightarrow e^+ D^- \nu_e$&4.56\%&5.93\%&1.41 $\pm$ 0.22\% \\
       $B^0\rightarrow e^+ D^{\ast -} \nu_e$&0.76\%&23.28\%&0.06 $\pm$ 0.12\% \\
       $B^0\rightarrow e^+ other$&5.48\%&5.29\%&1.90 $\pm$ 0.26\% \\
       \hline
       $B^+ $ & & & \\
       $B^+\rightarrow e^+ D^0 \nu_e$&8.31\%&6.55\%&2.33 $\pm$ 0.22\% \\
       $B^+\rightarrow e^+ D^{\ast 0} \nu_e$&29.43\%&25.17\%&2.14 $\pm$ 0.12\% \\
       $B^+\rightarrow e^+ other$&3.64\%&5.86\%&1.14 $\pm$ 0.26\% \\
       \hline
       $B_s $ & & & \\
       $B_s \rightarrow e^+ $total  &16.69 \%&11.62\%&2.63 $\pm$ 0.16\% \\
       \hline
      $B had\rightarrow e^+ others $&11.41\%&10.82\%&1.93 $\pm$ 0.17\% \\
       \hline
      $B \rightarrow c \rightarrow e $&19.71\%&5.48\%&6.60 $\pm$ 0.34\% \\
       \hline
     \end{tabular}
    \end{center}
\end{table}

\begin{table}[hbt]
    \begin{center}
     \caption{Detail of charm and bottom decay for  electron $p_{\mathrm{T}}$ 4-5 GeV/c}
      \label{chap6_table9}
     \begin{tabular}{c|c c c}
       \hline
       channel & $N_{tag}$ (part)/(all) & $N_{ele}$ (part)/(all) & $\epsilon$ \\
       \hline
       $D^0 $ & & & \\
       $D^0\rightarrow e^+ K^- \nu_e$&46.18\%&29.62\%&10.79 $\pm$ 0.18\% \\
       $D^0\rightarrow e^+ K^{\ast -} \nu_e$&10.54\%&3.42\%&21.30 $\pm$ 0.69\% \\
       $D^0\rightarrow e^+ \pi^- \nu_e$&3.97\%&5.52\%&4.97 $\pm$ 0.38\% \\
       $D^0\rightarrow e^+ \rho^- \nu_e$&1.02\%&0.49\%&14.53 $\pm$ 1.62\% \\
       $D^0\rightarrow e^+ other$&1.14\%&0.47\%&16.78 $\pm$ 1.76\% \\
       \hline
       $D^+ $ & & & \\
       $D^+\rightarrow e^+ \bar{K^0} \nu_e$&23.71\%&39.07\%&4.20 $\pm$ 0.13\% \\
       $D^+\rightarrow e^+ \bar{K^{\ast 0}} \nu_e$&5.62\%&4.40\%&8.83 $\pm$ 0.60\% \\
       $D^+\rightarrow e^+ \pi^0 \nu_e$&2.03\%&3.58\%&3.93 $\pm$ 0.34\% \\
       $D^+\rightarrow e^+ \rho^0 \nu_e$&0.21\%&0.34\%&4.33 $\pm$ 2.23\% \\
       $D^+\rightarrow e^+ other$&1.33\%&1.95\%&4.73 $\pm$ 0.71\% \\
       \hline
       $D_s $ & & & \\
       $D_s\rightarrow e^+ \phi \nu_e$&0.37\%&0.80\%&3.20 $\pm$ 1.49\% \\
       $D_s\rightarrow e^+ \eta \nu_e$&3.71\%&7.70\%&3.33 $\pm$ 0.27\% \\
       $D_s\rightarrow e^+ \eta' \nu_e$&0.32\%&0.62\%&3.63 $\pm$ 1.49\% \\
       $D_s\rightarrow e^+ other$&0.37\%&0.81\%&3.16 $\pm$ 0.90\% \\
       \hline
       $\Lambda_c $ & & & \\
       $\Lambda_c\rightarrow e^+ \Lambda \nu_e$&-0.36\%&0.32\%&-7.89 $\pm$ 4.81\% \\
       $\Lambda_c\rightarrow e^+ other$&-0.16\%&0.88\%&-1.30 $\pm$ 2.55\% \\
       \hline
       $B^0 $ & & & \\
       $B^0\rightarrow e^+ D^- \nu_e$&5.79\%&5.84\%&2.08 $\pm$ 0.28\% \\
       $B^0\rightarrow e^+ D^{\ast -} \nu_e$&-0.48\%&23.89\%&-0.04 $\pm$ 0.15\% \\
       $B^0\rightarrow e^+ other$&5.34\%&5.09\%&2.20 $\pm$ 0.34\% \\
       \hline
       $B^+ $ & & & \\
       $B^+\rightarrow e^+ D^0 \nu_e$&7.40\%&6.40\%&2.43 $\pm$ 0.28\% \\
       $B^+\rightarrow e^+ D^{\ast 0} \nu_e$&31.15\%&25.89\%&2.53 $\pm$ 0.14\% \\
       $B^+\rightarrow e^+ other$&1.61\%&5.65\%&0.60 $\pm$ 0.33\% \\
       \hline
       $B_s $ & & & \\
       $B_s \rightarrow e^+ $total  &19.83 \%&11.97\%&3.48 $\pm$ 0.20\% \\
       \hline
       $B had\rightarrow e^+ others $&14.67\%&10.79\%&2.86 $\pm$ 0.21\% \\
       \hline
       $B \rightarrow c \rightarrow e $&14.69\%&4.47\%&6.91 $\pm$ 0.46\% \\
       \hline
     \end{tabular}
    \end{center}
\end{table}

\begin{table}[hbt]
    \begin{center}
     \caption{Detail of charm and bottom decay for  electron $p_{\mathrm{T}}$ 5-7 GeV/c}
      \label{chap6_table11}
     \begin{tabular}{c|c c c}
       \hline
       channel & $N_{tag}$ (part)/(all) & $N_{ele}$ (part)/(all) & $\epsilon$ \\
       \hline
       $D^0 $ & & & \\
       $D^0\rightarrow e^+ K^- \nu_e$&50.38\%&29.51\%&13.60 $\pm$ 0.22\% \\
       $D^0\rightarrow e^+ K^{\ast -} \nu_e$&9.46\%&3.20\%&23.53 $\pm$ 0.84\% \\
       $D^0\rightarrow e^+ \pi^- \nu_e$&2.72\%&5.77\%&3.75 $\pm$ 0.44\% \\
       $D^0\rightarrow e^+ \rho^- \nu_e$&1.27\%&0.48\%&21.16 $\pm$ 1.89\% \\
       $D^0\rightarrow e^+ other$&1.03\%&0.43\%&19.32 $\pm$ 2.15\% \\
       \hline
       $D^+ $ & & & \\
       $D^+\rightarrow e^+ \bar{K^0} \nu_e$&22.03\%&39.60\%&4.43 $\pm$ 0.15\% \\
       $D^+\rightarrow e^+ \bar{K^{\ast 0}} \nu_e$&4.38\%&4.10\%&8.51 $\pm$ 0.73\% \\
       $D^+\rightarrow e^+ \pi^0 \nu_e$&2.11\%&3.77\%&4.46 $\pm$ 0.37\% \\
       $D^+\rightarrow e^+ \rho^0 \nu_e$&0.20\%&0.33\%&4.87 $\pm$ 2.67\% \\
       $D^+\rightarrow e^+ other$&1.33\%&1.91\%&5.53 $\pm$ 0.85\% \\
       \hline
       $D_s $ & & & \\
       $D_s\rightarrow e^+ \phi \nu_e$&0.41\%&0.74\%&4.42 $\pm$ 1.81\% \\
       $D_s\rightarrow e^+ \eta \nu_e$&3.46\%&7.66\%&3.59 $\pm$ 0.32\% \\
       $D_s\rightarrow e^+ \eta' \nu_e$&0.46\%&0.57\%&6.32 $\pm$ 1.90\% \\
       $D_s\rightarrow e^+ other$&0.58\%&0.80\%&5.79 $\pm$ 1.07\% \\
       \hline
       $\Lambda_c $ & & & \\
       $\Lambda_c\rightarrow e^+ \Lambda \nu_e$&-0.32\%&0.28\%&-9.09 $\pm$ 6.70\% \\
       $\Lambda_c\rightarrow e^+ other$&0.51\%&0.85\%&4.79 $\pm$ 3.30\% \\
       $\Lambda_c \rightarrow e^+ $total  &0.19 \%&1.13\%&1.37 $\pm$ 2.98\% \\
       \hline
       $B^0 $ & & & \\
       $B^0\rightarrow e^+ D^- \nu_e$&5.55\%&5.81\%&2.59 $\pm$ 0.25\% \\
       $B^0\rightarrow e^+ D^{\ast -} \nu_e$&2.80\%&24.29\%&0.31 $\pm$ 0.13\% \\
       $B^0\rightarrow e^+ other$&7.30\%&5.08\%&3.89 $\pm$ 0.30\% \\
       \hline
       $B^+ $ & & & \\
       $B^+\rightarrow e^+ D^0 \nu_e$&8.80\%&6.34\%&3.75 $\pm$ 0.24\% \\
       $B^+\rightarrow e^+ D^{\ast 0} \nu_e$&27.02\%&26.10\%&2.80 $\pm$ 0.13\% \\
       $B^+\rightarrow e^+ other$&1.47\%&5.55\%&0.71 $\pm$ 0.29\% \\
       \hline
       $B_s $ & & & \\
       $B_s \rightarrow e^+ $total  &21.74 \%&12.21\%&4.82 $\pm$ 0.17\% \\
       \hline
       $B had\rightarrow e^+ others $&14.94\%&10.57\%&3.83 $\pm$ 0.20\% \\
       \hline
       $B \rightarrow c \rightarrow e $&10.38\%&4.04\%&6.95 $\pm$ 0.42\% \\
       \hline
     \end{tabular}
    \end{center}
\end{table}

   \chapter{Fit Method Including Systemateic Uncertainies} \label{sec:fit}
This chapeter describes appiled fit method to take into account the correlation of systematic uncertainties.

The uncertainties of the measured points are categorized  into 
type~A~($p_{{\rm T}}$-uncorrelated, statistical + systematic, $\sigma_i$ ), 
type~B~($p_{{\rm T}}$-correlated,  $\sigma_{bi}$) and type~C~(normalization, $T_{AB}$ in this calculation
$\sigma_{c}$) as discussed in Sec.~\ref{hq_spe}, where the $\sigma$'s represent 
the standard deviations of the assumed Gaussian distributed uncertainties. 
In order to consider the such correlated uncertainties, the fit is done by minimizing a
following variable~\cite{bib:phfit}.
\begin{equation}
\label{eq:chi}
\chi^2 \equiv \left[ \sum_{i=1}^n \frac{(y_i+\sum_j \epsilon_{bj}\sigma_{bj}
+\epsilon_c \sigma_c - \mu_i({\bf q}))^2} {\tilde{\sigma_i}^2} \right] + \sum_j \epsilon_{bj}^2 +  \epsilon_{c}^2,
\end{equation}
where $y_1,y_2,...,y_n$ are the experimental results  and $\mu_1({\bf q}),\mu_2({\bf q}),...,\mu_n({\bf q})$
are the theoretical predictions with parameter sets, ${\bf q}$.
$\sigma_{bj}$ represent jth standard deviation in type~B systematic uncertainties.
$\epsilon_{bj}$ and $\epsilon_c$ are the fractions of the type~B and C systematic
uncertainties that all points are moved together, that is, $\epsilon_{bj}$ and $\epsilon_c$ are
normalized uncertainty to have Gaussian form with $\sigma=1$.
$\tilde{\sigma_i}$ is a quadrature sum of statistical error and type~A systematic error basically.
When the statistical error and $p_{{\rm T}}$-correlated errors have some relation, $\tilde{\sigma_i}$
is modified to take into account the such relation.
The minimum $\chi^2$~($\chi^2_{min}({\bf q})$) is searched for each  ${\bf q}$ in Eq.~\ref{eq:lan} 
by varying $\epsilon_{bj},\epsilon_{c}$.
Then, the parameter set which gives the minimum $\chi^2_{min}({\bf q})$ is regarded as the best fit values.

For any fixed values of $\epsilon_{bj}$ and $\epsilon_{c}$, Eq.~\ref{eq:chi} follows the $\chi^2$
distribution with n+1+j degrees of freedom, for testing the
theoretical predictions $\mu_i({\bf q})$, because it is the sum of n+1+j Gaussian distributed random variables. 
The best fit, $\chi^2_{min}$, the minimum of   $\chi^2$ should follow the $\chi^2$ distribution
with n-1  degrees of freedom. 

\section{$R_{AA}(p_{\mathrm{T}})$ and $v_2(p_{\mathrm{T}})$ Comparison}
We try to find the drag force which reproduces the experimental $R_{AA}(p_{\mathrm{T}})$ 
and $v_2(p_{\mathrm{T}})$ by the least-square fit using Eq.~\ref{eq:chi}.

In this fit, we consider 2 type~B uncertainties and 1 type~C uncertainty:
\begin{itemize}
\item{$\sigma_{b1}$: The type~B systematic uncertainty of $R_{AA}$}
\item{$\sigma_{b2}$: The type~B systematic uncertainty of $v_{2}$}
\item{$\sigma_{c}$:  The type~C systematic uncertainty of $R_{AA}$}
\end{itemize}
The uncertainties in $R_{AA}(p_{\mathrm{T}})$ and $v_2(p_{\mathrm{T}})$ are correlated via the type~B 
uncertainties in the electron spectrum in Au+Au collisions.
Since it is found this correlated uncertainty is small, this correlation is neglected for the simplicity.
$\sigma_{b1(2)}$ is assumed to have 100\% correlation with $p_{{\rm T}}$ for the simplicity.
For the $v_2(p_{\mathrm{T}})$ fitting, we use statistical error as  $\tilde{\sigma_i}$.

\section{Fit Method for $R_{AA}$}
There is the relation between the statistical error and $p_{{\rm T}}$ correlated 
systematic error from the nature of the statistical error.
To take into account the relation, the following $\tilde{\sigma_i}$is applied.

Let us consider a simple example, where the experimental observable can be expressed as follows.
\begin{equation} \label{eq:ob1}
y = C_1 \times C_2 \times ... \times C_n \times y_0 = a\times y_0,
\end{equation}
where, $C_1...C_n$ are systematic correction factors, $a$ is the overall correction factors,
$y_0$ is the raw yield and $y$ is  the experimental observable.
In this case, statistical error of $y$, $\sigma_{y}$ can be written as follows.
\begin{equation}
 \sigma_y = a\times \sigma_{y_0},
\end{equation}
Where, $\sigma_{y_0}$ is the statistical error of $y_0$.
When $y$ move to $y'$~($=y+\sum_j \epsilon_{bj}\sigma_{bj}+\epsilon_c \sigma_c$) due to 
the change of systematic correction~($a \rightarrow a'$) 
as described in Eq.~\ref{eq:chi}, statistical error of $y'$ becomes as follows.
\begin{eqnarray}
 \sigma_{y'} &=& a'\times \sigma_{y_0} = \frac{a'}{a}\sigma_y \\
 &=& \frac{y+\sum_j \epsilon_{bj}\sigma_{bj}+\epsilon_c \sigma_c}{y} \sigma_y.
\end{eqnarray}
Therefore, in the case which the experimental observable can be expressed in Eq.~\ref{eq:ob1},
we should use $\tilde{\sigma_y}=(y+\sum_j \epsilon_{bj}\sigma_{bj}+\epsilon_c \sigma_c)/y \sigma_y$ 
in Eq.~\ref{eq:chi} instead of $\sigma_y$.

As a nest step,   $\tilde{\sigma_y}$ of $R_{AA}(p_{\mathrm{T}})$ is considered.
Statistical error of $R_{AA}(p_{\mathrm{T}})$ above 3~GeV/c is determined by 
the spectrum of single non-photonic electrons in Au+Au collisions.
The spectrum of single non-photonic electrons is obtained as follows.
\begin{equation}
 Y = a\times N -B,
\end{equation}
where $N$ is raw yield of the inclusive electrons, $a$ is the overall correction factor,
$B$ is the yield of background electrons and $Y$ is the value of the spectrum of single non-photonic electrons.
We assume the change of the value of the spectrum, $Y \rightarrow Y'$ by varying $\epsilon_{b1}$ in Eq.~\ref{eq:chi}
is the change of the correlation factor, $a \rightarrow a'$ for the simplicity.
This assumption is almost justified in the $R_{AA}(p_{\mathrm{T}})$ analysis.
Therefore, when the spectrum $Y$ is moved to $Y'$, the statistical error becomes as follows.
\begin{equation} \label{eq:raasigma}
  \frac{Y'+B}{Y+B} a(\Delta N)_{stat} = \frac{Y'+B}{Y+B} (\Delta Y)_{stat} \label{stat_ap}
\end{equation}
The statistical error of $R_{AA}$ can be written from the scaling error, $\epsilon_c$ and Eq.~\ref{eq:raasigma}
with the signal to the background ratio in the measured electrons~($R_{NP} = Y/B$) as follows.
\begin{equation} \label{eq:raasigma2}
  \sigma_{Y_{stat}} \times \left(\frac{\frac{R_{AA}(i)}{R_{AA}(i)+\epsilon_{b1}\sigma_{b1}}
    R_{NP}+1}{R_{NP}+1}+\frac{\epsilon_c \sigma_c}{R_{AA}} \right) 
\end{equation}
Therefore, $\tilde{\sigma_Y}$ defined in Eq.~\ref{eq:raasigma2} is used instead of $\sigma_Y$.

   \chapter{$R_{AA}$ and $v_{2}$ Correction} \label{sec:craav2}
\begin{figure}[htb]
  \begin{center}
    \includegraphics[width=14.cm]{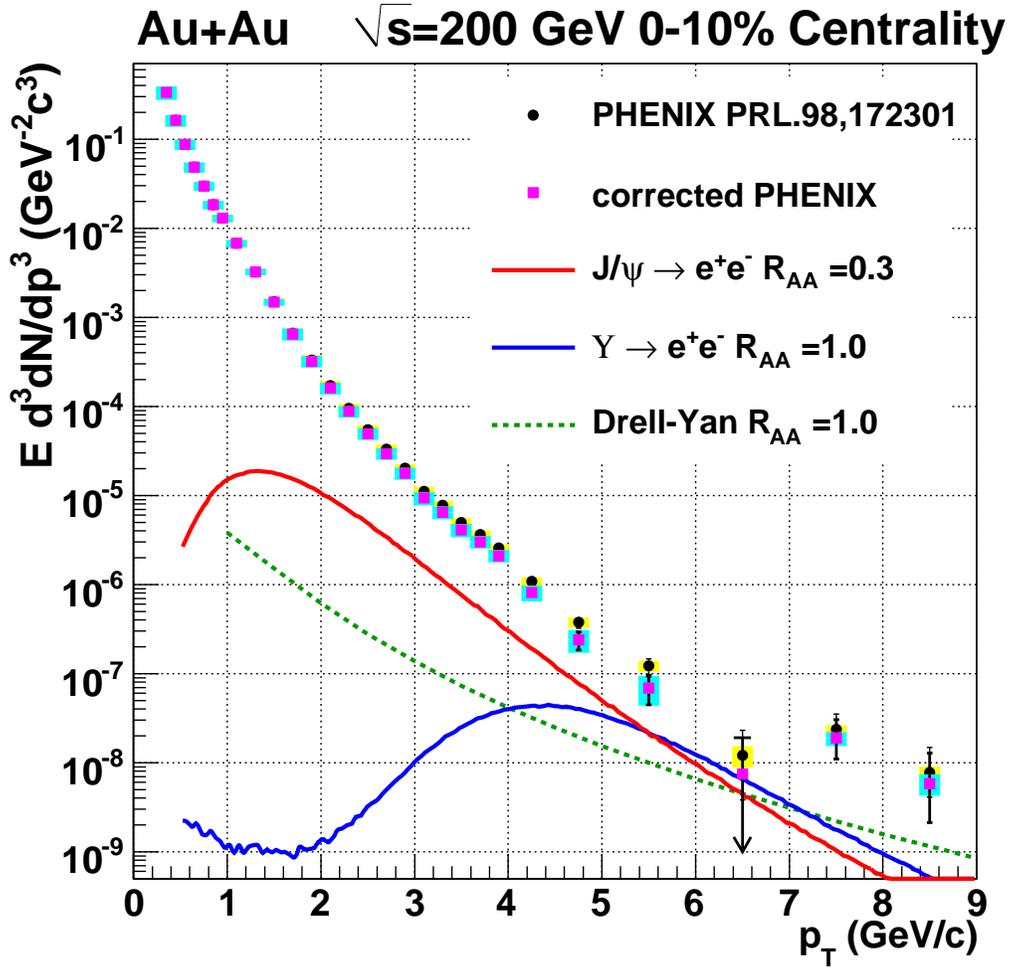}
    \caption{
      The spectra of the electrons from heavy flavor measured in PHENIX~\cite{bib:hq1},
      additional non-photonic background~(J/$\psi$, $\Upsilon$ and Drell-Yan process), and the electrons from semi-leptonic
      decay of heavy flavor in 0-10\% central Au+Au collisions.
    }
    \label{fig:chap7_aucent}
  \end{center}
\end{figure}

\begin{figure}[htb]
  \begin{center}
    \includegraphics[width=14.cm]{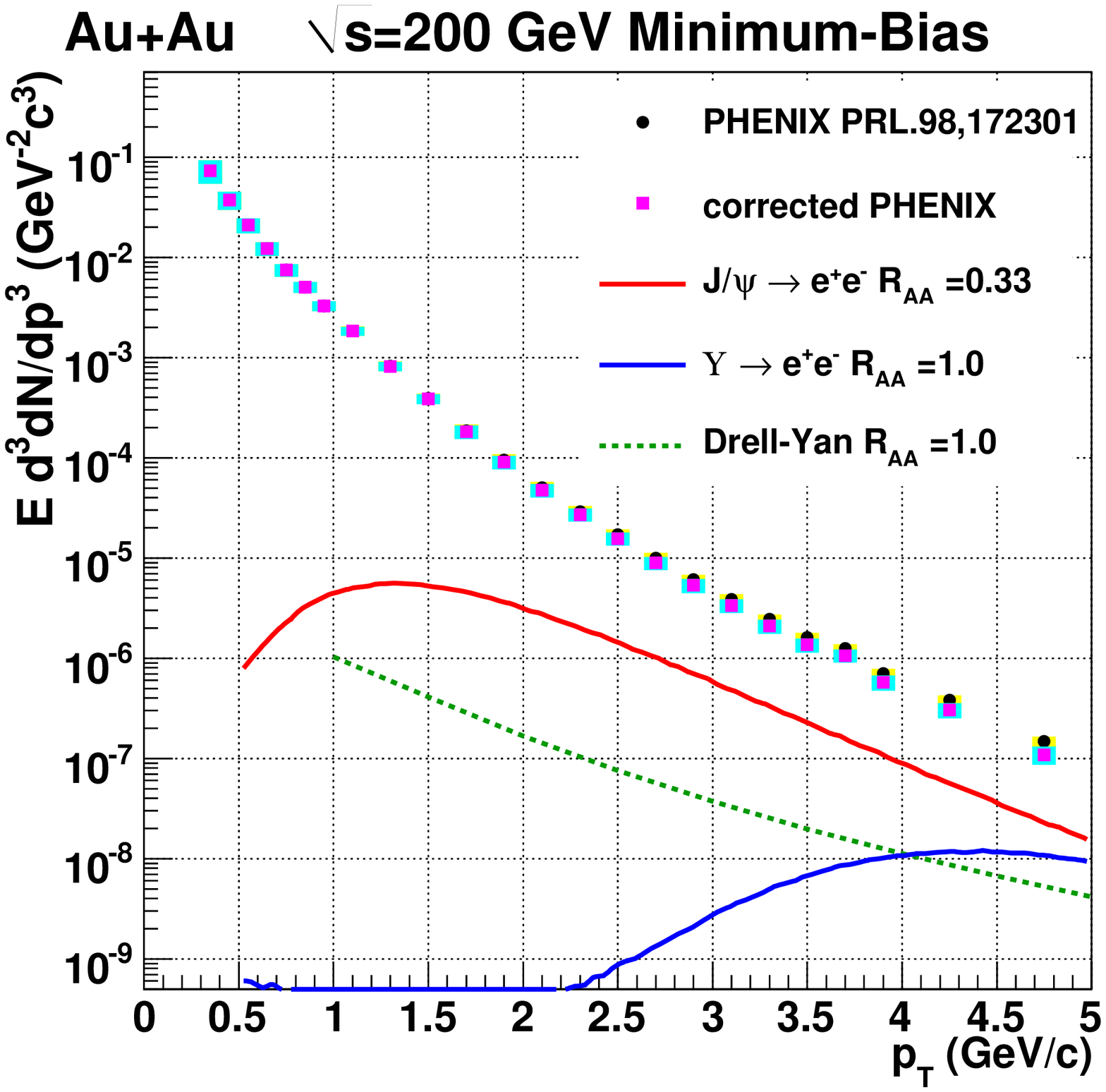}
    \caption{
      The spectra of the electrons from heavy flavor measured in PHENIX~\cite{bib:hq1},
      additional non-photonic background~(J/$\psi$, $\Upsilon$ and Drell-Yan process), and the electrons from semi-leptonic
      decay of heavy flavor in minimum bias Au+Au collisions.
    }
    \label{fig:chap7_auminb}
  \end{center}
\end{figure}

The $R_{AA}$ and $v_{2}$ of non-photonic electrons reported by PHENIX include 
the contribution of the electrons from J/$\psi$, $\Upsilon$ and Drell-Yan process~\cite{bib:hq1}.
The contribution from J/$\psi$, $\Upsilon$ and Drell-Yan process should be subtracted from
the experimental result in this chapter.

Figure~\ref{fig:chap7_aucent} shows the spectrum of the non-photonic electrons in 0-10\% central Au+Au collisions measured 
in PHENIX~\cite{bib:hq1}, which includes single non-photonic electrons from heavy flavor and the contributions from 
non-photonic background~(J/$\psi$, $\Upsilon$ and Drell-Yan process).
It has been observed the yield of J/$\psi$ is suppressed in Au+Au collisions compared with that in p+p collisions 
at PHENIX~\cite{bib:jpsi}.
In Fig.~\ref{fig:chap7_aucent}, $R_{AA}$ of  J/$\psi$ is assumed to be 0.3 according to the experimental result.
50\% systematic error is assinged for the contribution of J/$\psi$ to cover $p_{\rm T}$ dependance of $R_{AA}$ measured in
PHENIX~\cite{bib:jpsi}. 
While the yields of $\Upsilon$ and Drell-Yan process have not been measured at RHIC in Au+Au collisions,
the yields of $\Upsilon$ and Drell-Yan process are expected not to be suppressed~\cite{bib:latquark}.
$R_{AA}$ of $\Upsilon$ and Drell-Yan process is assumed to be one.
70\% and 75\% systematic error are assinged for contribution of $\Upsilon$ and Drell-Yan process including the uncertainty
of the yield of $\Upsilon$ and Drell-Yan in p+p collisions and the uncertainty of $R_{AA}$.
The spectrum of single non-photonic electrons in 0-10\% central Au+Au collisions is obtrained by the subtraction of
non-photonic background~(J/$\psi$, $\Upsilon$ and Drell-Yan process) from the spectrum of the non-photonic electrons
reported by PHENIX.
$R_{AA}$ of the single non-photonic electrons is obtained from the spectrum of the single non-photonic 
electrons in p+p and Au+Au collisions.

To obtain $v_{2}$ of single non-photonic electrons from that of  non-photonic electrons in 
minimum-bias Au+Au collisions reported by PHENIX, the spectrum of single non-photonic electrons 
in minimum-bias Au+Au collisions is also obtained in similar way in 0-10\% central collisions.
Figure~\ref{fig:chap7_auminb} shows the spectrum of the non-photonic electrons in minimum-bias Au+Au collisions measured 
in PHENIX~\cite{bib:hq1}, which includes single non-photonic electrons from heavy flavor and the contributions from 
non-photonic background~(J/$\psi$, $\Upsilon$ and Drell-Yan process).
$R_{AA}$ of  J/$\psi$ is assumed to be 0.33 according to the experimental result~\cite{bib:jpsi}.
$R_{AA}$ of $\Upsilon$ and Drell-Yan process is assumed to be one in the same case of 0-10\% central collisions.
There is small contribution of J/$\psi$ and $\Upsilon$ above 3~GeV/$c$.
$v_{2}$ of non-photonic electrons measured in PHENIX~\cite{bib:hq1} is also corrected to take account to the effect of
electrons from J/$\psi$ and $\Upsilon$.
The $v_{2}$ of the electrons from J/$\psi$ and $\Upsilon$ is assumed to be zero, since the $v_{2}$ of J/$\psi$ is 
currently consistent with zero~\cite{bib:jpsiv3}.
The deviation of the corrected $v_{2}$ from the original $v_{2}$ is included into systematic error of the corrected $v_{2}$.

   \chapter{Model Calculation Based on Langevin Equation} \label{sec:model}
A model calculation is done to study the medium property and underling physics from the 
$R_{AA}(p_{\mathrm{T}})$ and $v_2(p_{\mathrm{T}})$ of the single non-photonic electrons from charm and bottom, where 
the obtained ration of b/c+b is used in 
this model calculation.

The procedure is as follows.
\begin{itemize}
\item{Generation of heavy quarks}
\item{Simulation of space time evolution of heavy quarks in the medium}
\item{Hadronization of bare heavy quarks}
\item{Semi-leptonic decay of heavy flavored hadrons}
\end{itemize}
Especially, understanding of space time evolution of heavy quarks in the medium is important 
since what we want to know is the information of the medium via the interaction of the heavy quark and the medium.
Monte-Carlo simulation using Langevin equation is applied for the description of the space time evolution.
Heavy quarks are produced only in the initial hard collisions and it takes many collisions 
to change the momentum of heavy quark substantially due to their large mass compared with
temperature of the medium~($\sim 200$~MeV).
Therefore, heavy quarks can be described as Brownian particle and the Langevin equation
is a good approximation to model the motion of the heavy quarks 
in the medium~\cite{bib:svet,bib:hees,bib:moore}.
The interaction between heavy quarks and the medium is represented in terms of drag force 
and diffusion coefficients in Langevin equation.
The magnitude of drag force and diffusion coefficients will be constrained to reproduce the 
experimental $R_{AA}(p_{\mathrm{T}})$ and $v_2(p_{\mathrm{T}})$.
In this section, drag force and diffusion coefficients which reproduce the results of the single non-photonic
electrons are studied.
\section{Initial Condition of Heavy Quarks}
The bulk thermalization time is assumed to 
be $\tau_0$~($=\sqrt{t^2-z^2}$) $= 0.6$~fm~\cite{bib:hirano4,bib:hirano5}.
The space and momentum distribution of heavy quarks at $\tau_0$ is defined as 
the initial condition.
The momentum distribution of heavy quarks at $\tau_0$ is assumed to be the distribution  
at which heavy quarks are generated, since there is not significant interaction for heavy quarks before $\tau_0$.
Heavy quarks are assumed to stream freely in the transverse direction before  $\tau_0$.

\begin{figure}[htb]
  \begin{center}
    \includegraphics[width=8cm]{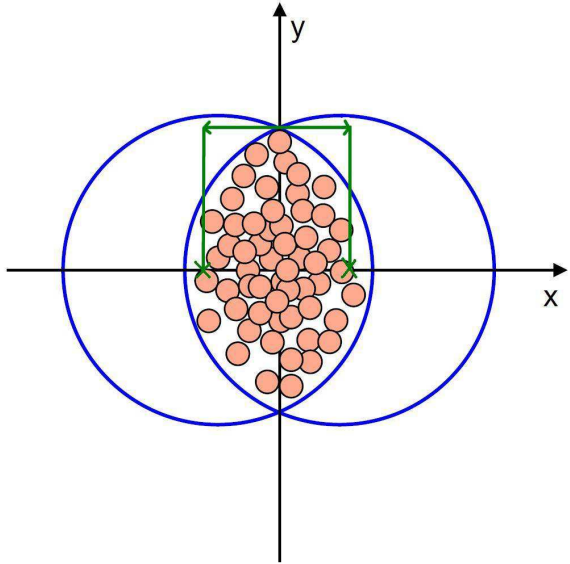}
    \caption{Conceptual view of the heavy ion collision and coordinate system.
    }
    \label{fig:chap7_glau}
  \end{center}
\end{figure}

\begin{figure}[htb]
  \begin{center}
    \includegraphics[width=13cm]{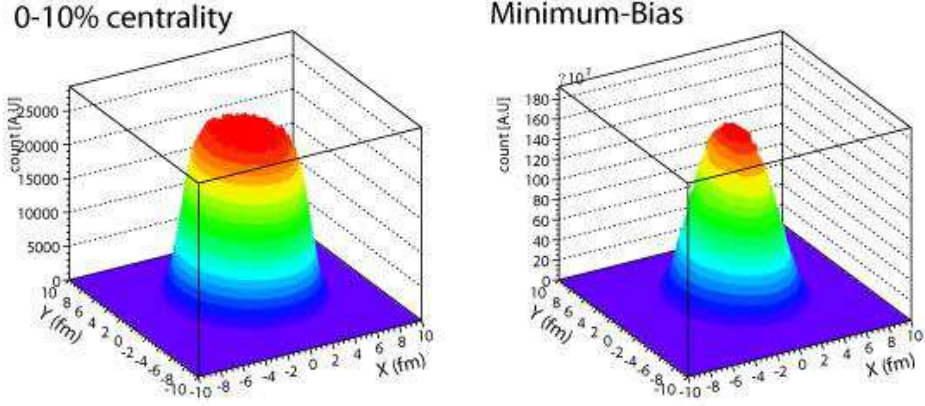}
    \caption{The distribution in transverse plane of the initial position of heavy quark 
      generated according to Eq.\ref{eq:tab} for 0-10\% centrality events and Minimum-Bias events.
    }
    \label{fig:chap7_inip}
  \end{center}
\end{figure}

In the heavy ion collisions, the space distribution of generated heavy quarks
is determined by the geometry of the collisions, because heavy quark is produced only in
the initial hard collision and 
the yield is scaled by the number of nucleon-nucleon collisions as described in Sec.~\ref{sec:heavy_mod}.
The initial positions of heavy quarks are calculated according to the Glauber 
model~\cite{bib:glau} discussed in Sec.~\ref{sec:Glauber}.
The space distribution of heavy quarks in transverse plane~($n^{r}_{in}(x,y)$)
is represented as 
\begin{equation}
\label{eq:tab}
n^{r}_{in}(x,y)  \propto \rho_{A}(x-b/2,y)\times \rho_{B}(x+b/2,y).
\end{equation}
Here, $\rho_{A(B)}$ is nuclear density function and $b$ is the impact parameter of the 
collision. Reaction plane of the collision is defined as (x,z) plane with
the impact parameter to be parallel to the $x$-axis.
Figure~\ref{fig:chap7_glau} shows the conceptual view of the heavy ion collision and coordinate system. 
Impact parameter~($b$) is used $3.1$~fm for 0-10\% centrality events.
For Minimum Bias events, events with the impact parameter corresponding to each centrality are merged accoring
to Table~\ref{tab:ap6} in Appendix~\ref{sec:gtable}, where the production of heavy qurak is assumed to be $N_{coll}$
scaling.
The nuclear density function is parameterized by a Woods-Saxon function defined in Eq.~\ref{eq:wood-saxon}.
Figure~\ref{fig:chap7_inip} shows the space distribution in transverse plane of 
heavy quarks generated according to Eq.\ref{eq:tab} for 0-10\% centrality events and 
Minimum-Bias events.
For the distribution in z direction, it is assumed that spatial rapidity, 
$y_{S}=\frac{1}{2}\ln \frac{t+z}{t-z}$, has the same value with the rapidity in momentum space.

\begin{figure}[htb]
  \begin{center}
    \includegraphics[width=11cm]{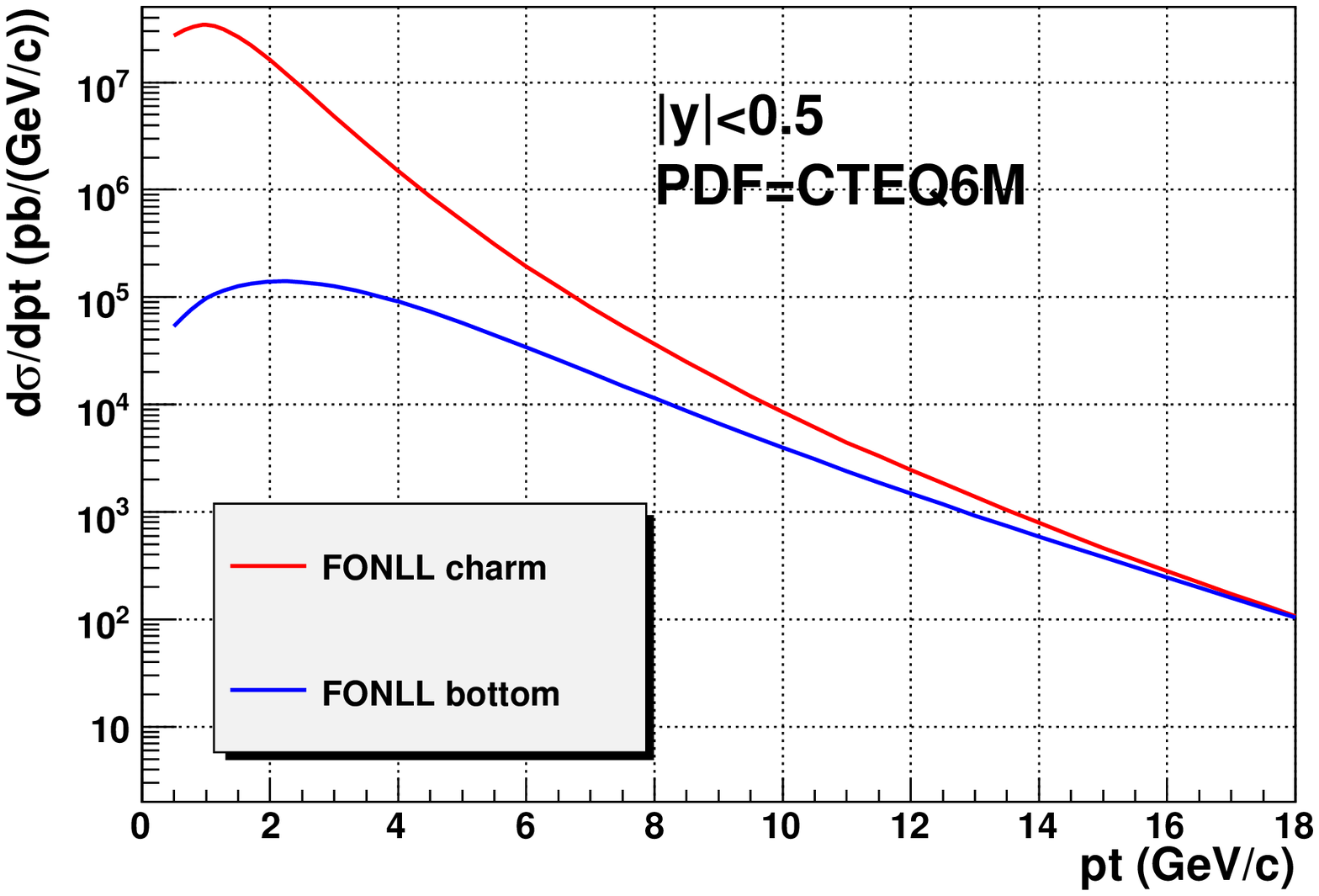}
    \caption{FONLL calculation of the $p_{\mathrm{T}}$ distribution of 
      charm and bottom at mid-rapidity~($\mid y \mid<0.5$) in p+p $\sqrt{s}=200GeV$ 
collisions~\cite{bib:fonll5}.
    }
    \label{fig:chap7_fonll}
  \end{center}
\end{figure}

The initial momentum spectra of charm and bottom are determined by FONLL calculation~\cite{bib:fonll1}.
The absolute values of cross section of charm and bottom are normalized according to the experimental results
obtained in this thesis, which is discussed in Sec.~\ref{sec:bcr}.
Figure~\ref{fig:chap7_fonll} shows FONLL calculation of the $p_{\mathrm{T}}$ distribution of 
charm and bottom at mid-rapidity~($\mid y \mid<0.5$) in p+p $\sqrt{s}=200$~GeV collisions~\cite{bib:fonll5}.
The azimuthal distribution is assumed to be flat.

Initial nuclear effects such as shadowing and Cronin effect are not considered for simplicity in
this calculation. 
Such effects are not large as discussed in Sec.~\ref{sec:ini}, while these should be 
considered for the precise description.

\section{Space Time Evolution of Heavy Quarks}\label{sec:lang}
The hot and dense medium created with heavy ion collisions is assumed to be thermalized
at $\tau_0= 0.6$~fm and then the medium follows the space-time evolution~\cite{bib:hirano4,bib:hirano5}.
Heavy quarks propagate the hot and dense medium.
Relativistic hydrodynamics has been applied for the description of the space-time evolution of the medium
from $\tau_0$ to the freeze-out, which is the background of the motion of heavy quarks.
Hydrodynamical calculations succeed in reproduction of the observed light particle spectra
in low $p_{\mathrm{T}}$, azimuthal anisotropy of the light particles and multiplicity 
well~\cite{bib:hirano2,bib:hirano3,bib:hirano4,bib:hirano5}.
In this model calculation, the results of the relativistic hydrodynamical calculation are used to 
describe the evolution of the medium, which are taken from Ref.~\cite{bib:hirano4,bib:hirano5}.

\begin{figure}[htb]
  \begin{center}
    \includegraphics[width=15cm]{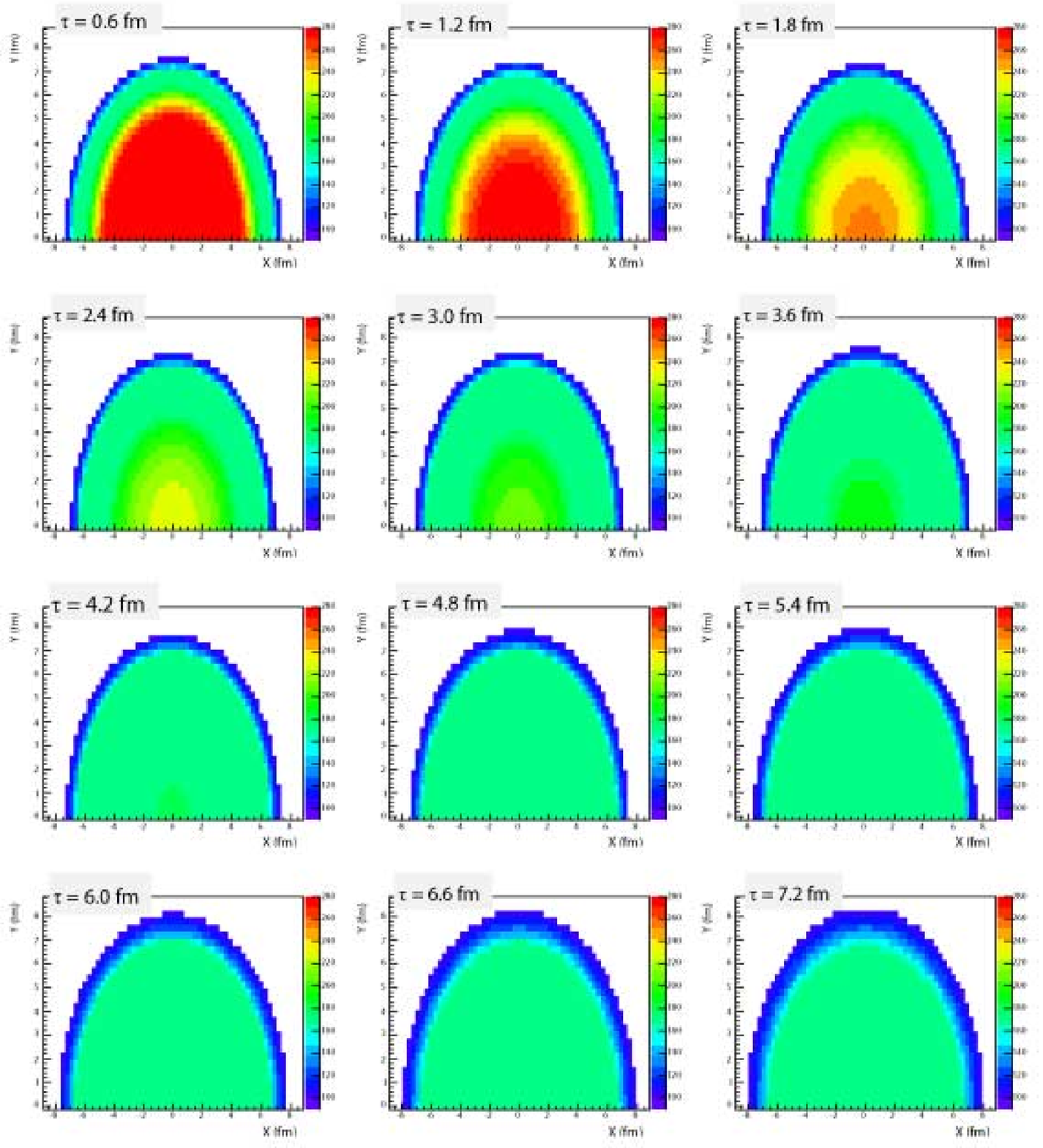}
    \caption{The calculated temperature profile at $z=0$  in 0-10\%
      central Au+Au collisions~\cite{bib:hirano4,bib:hirano5}.
      Each panel indicates time evolution of the matter.
    }
    \label{fig:chap7_hydro}
  \end{center}
\end{figure}

The relativistic hydrodynamical calculation is characterized by local temperature~($T({\bf x},t)$),
energy density~($e({\bf x},t)$), flow velocity~(${\bf v}({\bf x},t)$) and fraction of QGP
~($f_{qgp}({\bf x},t)$).
The critical temperature and the thermal freeze-out temperature are assumed to be 
170~MeV and 100~MeV, respectively.
Figure~\ref{fig:chap7_hydro} shows the calculated temperature profile at $z=0$  in 0-10\%
central Au+Au collisions.
Each panel corresponds to the temperature field at each proper time.
The hydrodynamical calculation in the centrality corresponding to the impact parameter are used as the model calculation
for Minimum Bias events.

The end time of space time evolution of heavy quarks is determined by the fraction of QGP~($f_{qgp}({\bf x},t)$).
$f_{qgp}({\bf x},t)$ is s calculated by the relativistic hydrodynamics at each local point.
The  fraction of QGP is defined as,
\begin{equation}
f_{qgp}({\bf x},t) \equiv \frac{e-e_{had}}{e_{QGP}-e_{had}}.
\end{equation}
Here, $e_{QGP}$~($e_{had}$) is the maximum (minimum) value of the energy density 
at local grid.
The evolution of heavy quark is stopped when $f_{qgp}(\bf{x},t)$ becomes bellow 0.5 and then heavy quarks are
hadronized.

Space time evolution of generated heavy quarks in the medium is simulated by Monte-Carlo calculation according to
relativistic Langevin equation.
The relativistic Langevin equation in the rest frame of the background fluid 
is written as,
\begin{eqnarray}
  \Delta {\bf x}(t) &=& \frac{{\bf p}}{E} \Delta t,  \label{eq:lan} \\
  \Delta {\bf p}(t) &=& -\gamma  {\bf p} \Delta t + {\bf \xi}(t) \Delta t. \label{eq:lanp}
\end{eqnarray}
Here, $\Delta {\bf x}$ and $\Delta {\bf p}$ are deviations of space and momentum of 
heavy quarks in a discrete step of time, $\Delta t$.
$\gamma$ is the drag coefficient and ${\bf \xi}$ represents random momentum kick that is uncorrelated in time.
${\bf \xi}$ is assumed to have Gaussian shape and is characterized by the following equations.
\begin{eqnarray}
  \label{eq:lan2}
  <\xi_i(t)> &=&0, \\
  <\xi_i(t)\xi_j(t')> &=& D_{p} \delta_{ij} \delta(t-t'),
\end{eqnarray}
where $D_{p}$ is the diffusion coefficient in momentum space.
$\gamma$ and $D_{p}$ are related from the relativistic Kramers equation 
and the requirement for the equation to have the Jutter distribution 
as a stationary solution~\cite{bib:akamatu,bib:relan}.
The relation is as follows.
\begin{equation}
\gamma + \frac{dD_{p}}{d(p^2)} = \frac{D_{p}}{2ET} \label{eq:rela}.
\end{equation}
The relation between $\gamma$ and $D_{p}$ is reduced to more simple form called as
Einstein's fluctuation-dissipation relation when $D_{p}$ is independent of momentum.
\begin{eqnarray}
\gamma &=& \frac{D_{p}}{2ET} \nonumber \\
       &\sim&  \frac{D_{p}}{2MT}(E\rightarrow M) .
\end{eqnarray}
In the non-relativistic limit, the spatial diffusion 
constant~($D_{sx}\equiv (<x^2>-<x>^2)/2\Delta t$) is written in terms of $\gamma$  as follows.
\begin{equation} \label{eq:sdif}
D_{s} = \frac{T}{M\gamma}.
\end{equation}
The thermal relaxation time of heavy quarks~($\tau_{HQ}$) is obtained as the inverse of
drag coefficient~($\gamma$).
\begin{equation} \label{eq:tau}
\tau_{HQ} \sim \frac{1}{\gamma}
\end{equation}

The parametric dependence of drag coefficient~(momentum, temperature and so on) is governed by
the microscopic interaction of heavy quark with the medium, and it has not been 
understood.
In this model calculation, we concentrate on the phenomenological understanding of the interaction and
the medium property.
Therefore,  we assume the parametric dependence of drag coefficient with one free parameter and 
the free parameter is constrained with the experimental results.
The following two relations are assumed for the drag coefficient.
\begin{eqnarray}
  \gamma &=& \alpha \frac{T^2}{M} \label{eq:ads}, \\
  \gamma &=& \alpha \frac{T^2}{E} \label{eq:col},
\end{eqnarray}
where $M(E)$ is mass~(energy) of heavy quark and $\alpha$ is the dimensionless free parameter which
is independent of other parameters.

The two relations represent the two extreme cases.
Eq.~\ref{eq:ads} is motivated by the result from the AdS/CFT correspondence.
The drag coefficient can be expressed as bellow according to the AdS/CFT 
correspondence~\cite{bib:akamatu,bib:adsdrag0,bib:adsdrag1,bib:adsdrag2}.
\begin{equation}
 \gamma \sim 2\frac{T^2}{M} \propto \frac{T^2}{M}.
\end{equation}
This drag coefficient represents the strongly coupling limit of QGP because the AdS/CFT correspondence is valid under
such condition as discussed in Sec.~\ref{sec:coal}.

In the other hand, Eq.~\ref{eq:col} is motivated by the perturbative QCD calculation of 
collisional process~\cite{bib:coldrag1,bib:moore}.
This drag coefficient represents the week coupling regime of QGP,
where perturbative collisional process is dominant.

The corresponding diffusion coefficient $D_{p}$ is obtained from the generalized fluctuation-dissipation 
relation in Eq~\ref{eq:rela} as follows.
\begin{eqnarray}
  D_{p} &=& \alpha \frac{2T^3}{M}(E+T) \label{eq:ads_d} \quad \quad ({\rm AdS/CFT}), \\
  D_{p} &=& \alpha 2T^3 \label{eq:col_d} \quad \quad ({\rm pQCD}).
\end{eqnarray}

The evolution of heavy quarks is analyzed numerically according to Eq.~\ref{eq:lan} and \ref{eq:lanp}
in the rest frame of the  medium fluid.
And then heavy quarks are transformed in the laboratory frame, to take into the account the 
medium fluid.
This process is continued until the end time of space time evolution determined above.

\section{Hadronization Process} \label{sec:hadro}
Since the experimental observable is not bare quark but hadron, the hadronization process should 
be considered in the model calculation to compare with the experimental result.
As discussed in Sec.~\ref{sec:coal}, the quark coalescence~(recombination) model has received
renewed interest in the context of RHIC data, by providing a successful explanation of
two phenomena observed in intermediate $p_{\mathrm{T}}$ hadron spectra, the constituent 
quark number scaling of the elliptic flow and the large baryon-to-meson 
ratios~\cite{bib:coal1,bib:coal2}. 
It is natural to apply it to the hadronization of heavy quarks as well.
Since the fraction of baryons containing heavy quarks over meson is rather small, 
the hadronization to heavy flavored meson is only considered here. 

Since the probability of the hadronization of heavy quarks via coalescence process becomes 
small at high $p_{\mathrm{T}}$ where the phase-space density of light-quark to coalesce becomes very small,
the hadronization of heavy quarks via usual fragmentation process discussed in Sec~.\ref{sec:ufrag} becomes
dominant process at high $p_{\mathrm{T}}$~($p_{\mathrm{T}}>3$~GeV/$c$).
Thus, the total D~(B)-meson spectrum takes the following form
\begin{equation}
\frac{dN_{D(B)}^{tot}}{d^2{\bf p}_{{\rm T}}} = \frac{dN_{D(B)}^{coal}}{d^2{\bf p}_{{\rm T}}}
+\frac{dN_{D(B)}^{frag}}{d^2{\bf p}_{{\rm T}}}
\end{equation}

The yield of heavy flavored mesons producted via quark coalescence process is modeled 
according to the following equation~\cite{bib:dcoal1,bib:dcoal2}.
\begin{eqnarray}
\label{eq:coal}
\frac{dN_{M}}{d^2{\bf p}_{{\rm T}}} &=& \frac{g_M}{(\Delta y)^2}\int d^2 {\bf r}_{Q{\rm T}} 
d^2 {\bf r}_{q{\rm T}} d^2 {\bf p}_{Q{\rm T}} d^2 {\bf p}_{q{\rm T}} \times \left. \frac{dN_{Q}}
{d^2 {\bf r}_{Q{\rm T}}d^2{\bf p}_{Q{\rm T}}}\right|_{\mid y \mid< \Delta y/2}
\left.\frac{dN_{q}} {d^2 {\bf r}_{q{\rm T}}d^2{\bf p}_{q{\rm T}}}\right|_{\mid y \mid< \Delta y/2} \nonumber \\
&\times& \int dy_{SQ} dy_{Q} dy_{Sq} dy_{q} \delta(y_{SQ} -y_{Q})   \delta(y_{Sq} -y_{q})  
f_{M}(x_{Q},x_{q};p_{Q},p_{q}) 
\delta({\bf p}_{{\rm T}} - {\bf p}_{Q{\rm T}} - {\bf p}_{q{\rm T}}), \nonumber\\
& &
\end{eqnarray}
where $Q$ and $q$ denote heavy quark~(c,b) and light quark~(u,d,s), respectively.
$\Delta y$ is rapidity range and $g_M$ is statistical factor to take into account 
the internal quantum numbers~(spin and color).
$f_{M}(x_{Q},x_{q};p_{Q},p_{q})$ is the coalescence probability function which
depends on the overlap of heavy and light quark in momentum and coordinate space.
Here, we assume a uniform distribution as follows~\cite{bib:dcoal3},
\begin{equation}
f_{M}(x_{Q},x_{q};p_{Q},p_{q}) = \frac{9\pi}{2\Delta_x^3 \Delta_p^3}
\Theta(\Delta_x^2-(x_{Q}-x_{q})^2)
\Theta(\Delta_p^2-\frac{(p_{Q}-p_{q})^2}{4}\frac{(M_{Q}-m_{q})^2}{4}).
\end{equation}
Here, $\Delta_x, \Delta_p$ are the covariant spatial and momentum
coalescence radii, and they are related with each other by the uncertainty relation $\Delta_x \Delta_p \sim 1$.
In this calculation, we take 0.24~GeV as $\Delta_p$~\cite{bib:dcoal1}.\\
The heavy quark distribution function~($\frac{dN_{Q}}
{d^2 {\bf r}_{Q{\rm T}}d^2{\bf p}_{Q{\rm T}}}$) is directly taken from the output of
the Langevin simulations discussed in the previous section.

\begin{figure}[htb]
  \begin{center}
    \includegraphics[width=12cm]{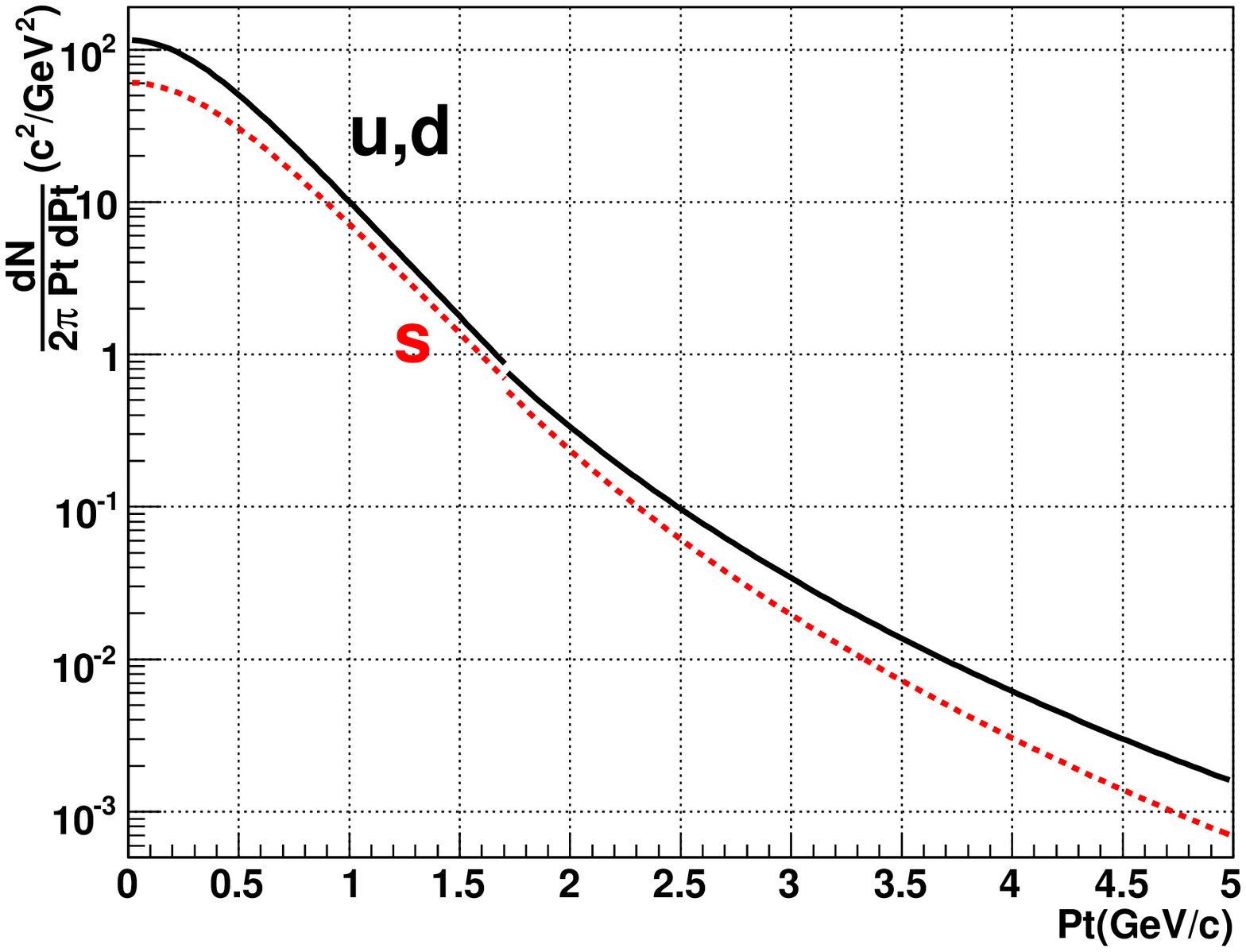}
    \caption{The light-quark distributions to use the calculation
      of Eq.~\ref{eq:coal}. Black line shows the distribution of u and d quark and 
      red dotted line shows that of s quark.
    }
    \label{fig:chap7_coal}
  \end{center}
\end{figure}
The light-quark distributions~($\frac{dN_{q}}{d^2 {\bf r}_{q{\rm T}}d^2{\bf p}_{q{\rm T}}}$) are 
taken from the successful application 
of the coalescence model of Ref.~\cite{bib:dcoal4} to light hadron observables at RHIC.
Figure~\ref{fig:chap7_coal} shows the light-quark distributions in momentum 
space~($\frac{dN_{q}}{d^2{\bf p}_{q{\rm T}}}$) to be use in the calculation
of Eq.~\ref{eq:coal}. Black line shows the distribution of u and d quark and 
red dotted line shows that of s quark.
The light-quark above 1.7~GeV/$c$ is considered as mini-jet parton and the  $p_{\mathrm{T}}$ distribution is 
assumed to have power-law form, where that $p_{\mathrm{T}}$ distribution is tuned to reproduce observed 
jet quenching for high $p_{\mathrm{T}}$ charged particles~\cite{bib:dcoal3,bib:dcoal4}. 
The light-quark bellow 1.7~GeV/$c$ is considered as thermalized partons in QGP
and the  $p_{\mathrm{T}}$ distribution is assumed to have exponential form.
To take into account collective flow of QGP, these partons are boosted by a flow velocity 
${\bf v}_{{\rm T}} = \beta({\bf r}_{{\rm T}}/R) $, depending on their transverse radial 
positions  ${\bf r}_{{\rm T}}$. Here, $R$ is the transverse size of the QGP~(8fm) at 
hadronization, and $\beta$ is the collective flow velocity of QGP and is taken to be 0.5$c$~\cite{bib:dcoal4}.

\begin{figure}[htb]
  \begin{center}
    \includegraphics[width=12cm]{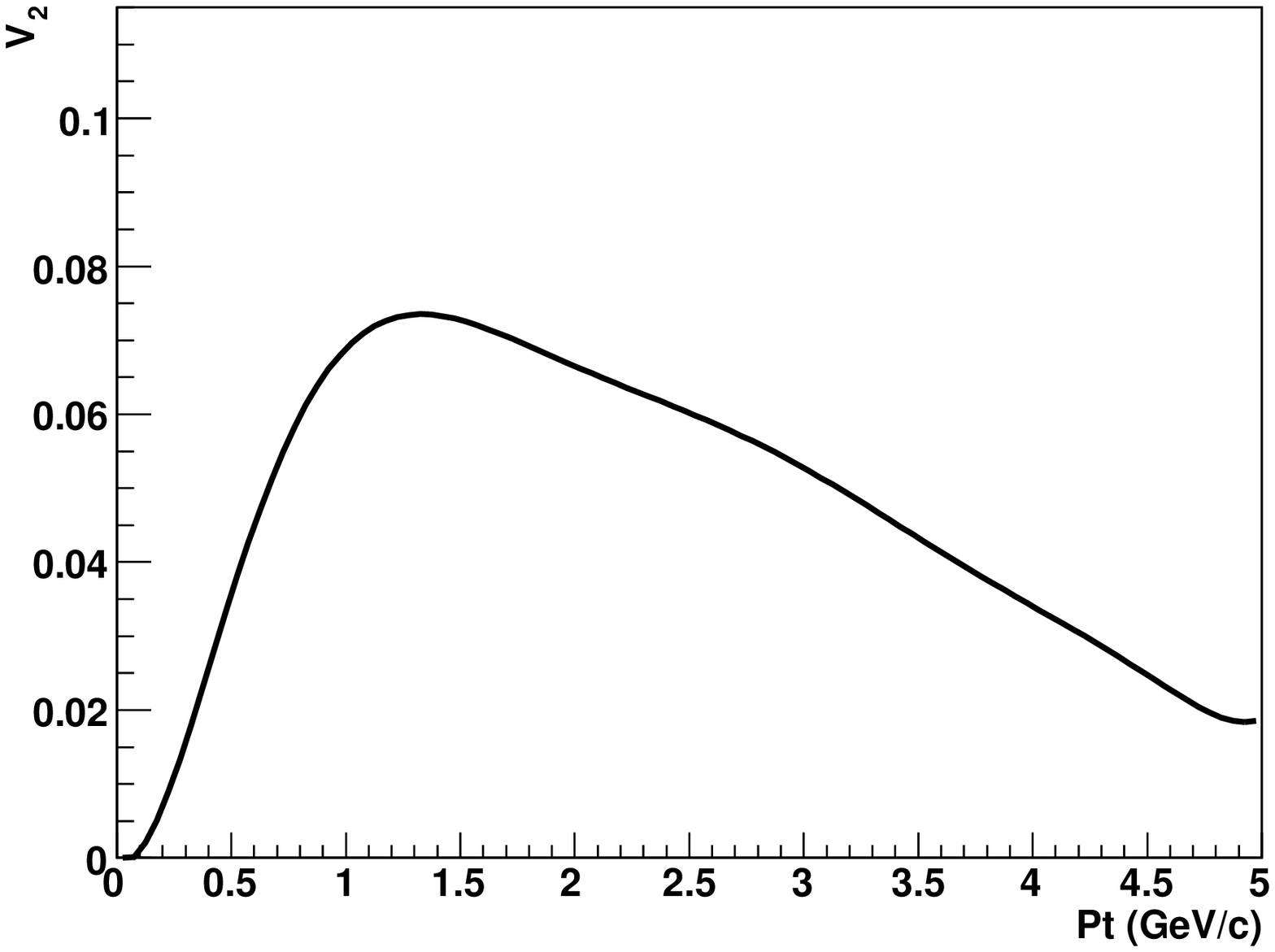}
    \caption{The assumed $v_2(p_{{\rm T}})$ of light-quarks to use
      the calculation~\cite{bib:sakai}.
    }
    \label{fig:chap7_lv2}
  \end{center}
\end{figure}

In addition, the $p_{\mathrm{T}}$ distribution of light-quarks is assumed to have the following 
relation with respect to the azimuthal direction in momentum space and the magnitude of momentum to take into account 
the effect of the elliptic flow of light quarks in the medium.
\begin{equation}
\frac{dN_{q}} {d^2{\bf p}_{q{\rm T}}} \propto 1+2v_2(p_{{\rm T}}) \cos(2\phi)
\end{equation}
$v_2(p_{{\rm T}})$ is taken from Ref.~\cite{bib:sakai} which reproduces the elliptic flow of
light flavored hadron.
Figure~\ref{fig:chap7_lv2} shows the assumed $v_2(p_{{\rm T}})$ of light-quarks to use
the calculation.

The remaining c and b quarks from coalescence process are
hadronized using $\delta$-function fragmentation~\cite{bib:hees}.
While $\delta$-function fragmentation is not accurate treatment for the fragmentation,
the result via $\delta$-function fragmentation almost agrees with the experimental 
result.

Finally, the D and B mesons  are decayed into electrons via semi-leptonic decay.
This process is done by using EvtGen simulation as done in the analysis of experimental data.

\section{Numerical Result}\label{sec:mod_res}
The numerical results of the model calculation are shown in this section. 
 $R_{AA}(p_{\mathrm{T}})$ and $v_2(p_{\mathrm{T}})$ are determined in this calculation 
as follows.
\begin{eqnarray}
  R_{AA}(p_{\mathrm{T}}) &=& \frac{N_{out}(p_{\mathrm{T}})}{N_{in}(p_{\mathrm{T}})},  \\
  v_2(p_{\mathrm{T}})    &=& <\cos2\phi>(p_{\mathrm{T}}).
\end{eqnarray}
Here, $N_{in(out)}(p_{\mathrm{T}})$ is the input~(output) yield of the calculation.
Input electron yield is determined as the yield of decay electrons of D~(B) mesons from 
hadronization of input heavy quark via $\delta$-function fragmentation.

\begin{figure}[htb]
  \begin{center}
    \includegraphics[width=15cm]{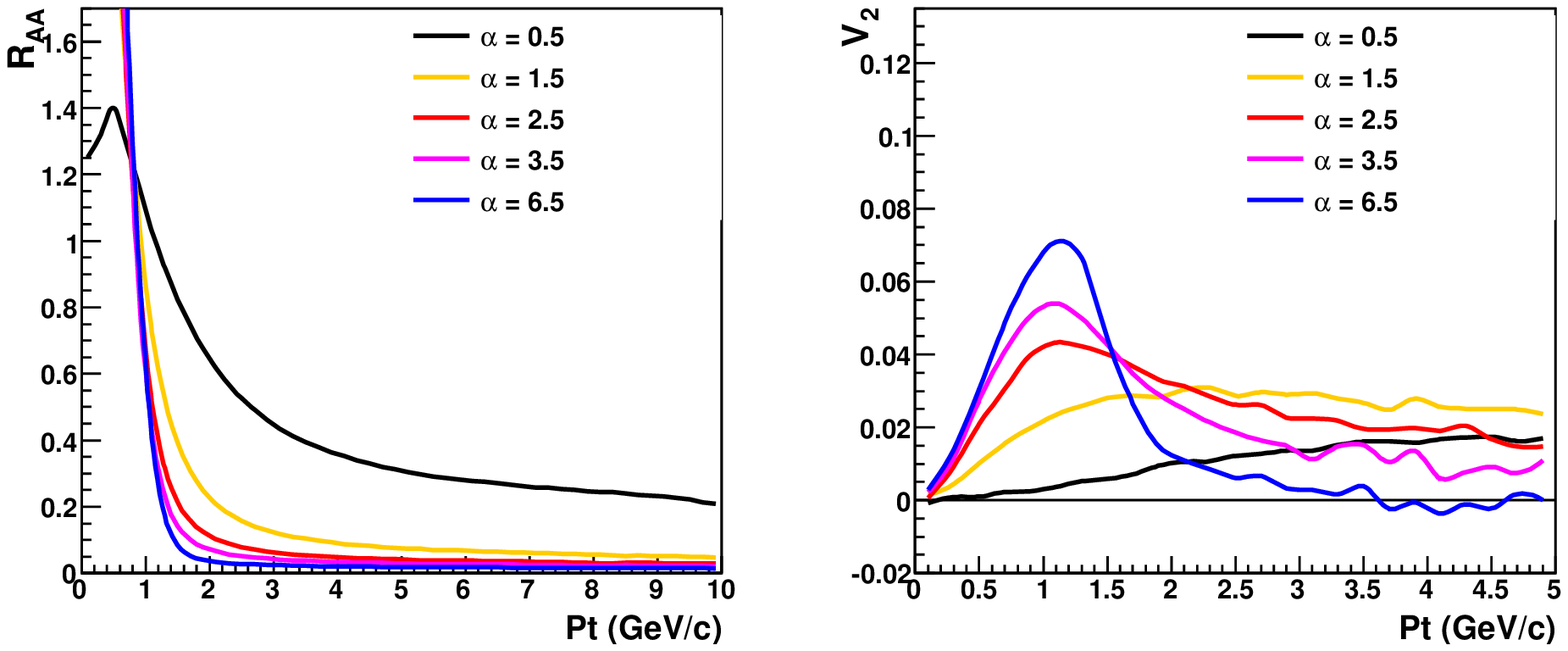}
    \caption{$R_{AA}(p_{\mathrm{T}})$ and $v_2(p_{\mathrm{T}})$ of charm quarks 
      calculated by the drag force defined in Eq.~\ref{eq:ads}~(AdS/CFT).
      Calculated conditions correspond 0-10\% centrality events for $R_{AA}(p_{\mathrm{T}})$
      and Minimum Bias events for $v_2(p_{\mathrm{T}})$.
    }
    \label{fig:chap7_c_v2raa}
  \end{center}
\end{figure}

\begin{figure}[htb]
  \begin{center}
    \includegraphics[width=15cm]{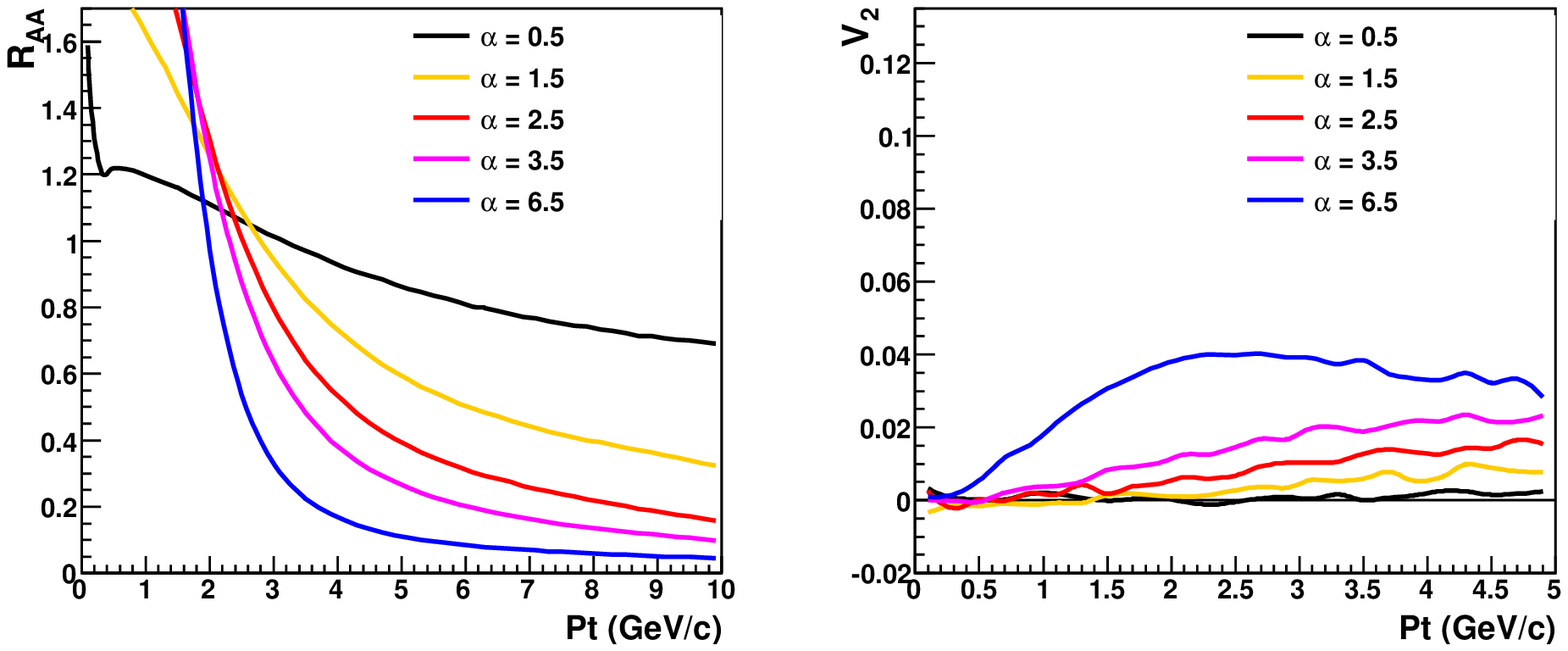}
    \caption{$R_{AA}(p_{\mathrm{T}})$ and $v_2(p_{\mathrm{T}})$ of bottom quarks 
      calculated by the drag force defined in Eq.~\ref{eq:ads}~(AdS/CFT).
      Calculated conditions correspond 0-10\% centrality events for $R_{AA}(p_{\mathrm{T}})$
      and Minimum Bias events for $v_2(p_{\mathrm{T}})$.
    }
    \label{fig:chap7_b_v2raa}
  \end{center}
\end{figure}

Figure~\ref{fig:chap7_c_v2raa} and \ref{fig:chap7_b_v2raa} show $R_{AA}(p_{\mathrm{T}})$ and 
$v_2(p_{\mathrm{T}})$ of charm and bottom quarks calculated by the drag force defined 
in Eq.~\ref{eq:ads} with various values of one free parameter, respectively.
$R_{AA}(p_{\mathrm{T}})$ results are calculated for 0-10\% centrality events~($b=3.1$~fm)
and $v_2(p_{\mathrm{T}})$ results are calculated for Minium-Bias events
to compare the experimental results.
The magnitude of suppression at high $p_{\mathrm{T}}$ and  $v_2(p_{\mathrm{T}})$ become larger as
the drag force become stronger.
In addition the magnitude of suppression at high $p_{\mathrm{T}}$ and  $v_2(p_{\mathrm{T}})$ of charm quark are
larger than these of bottom quark.
These tendencies are expected.

\begin{figure}[htb]
  \begin{center}
    \includegraphics[width=15cm]{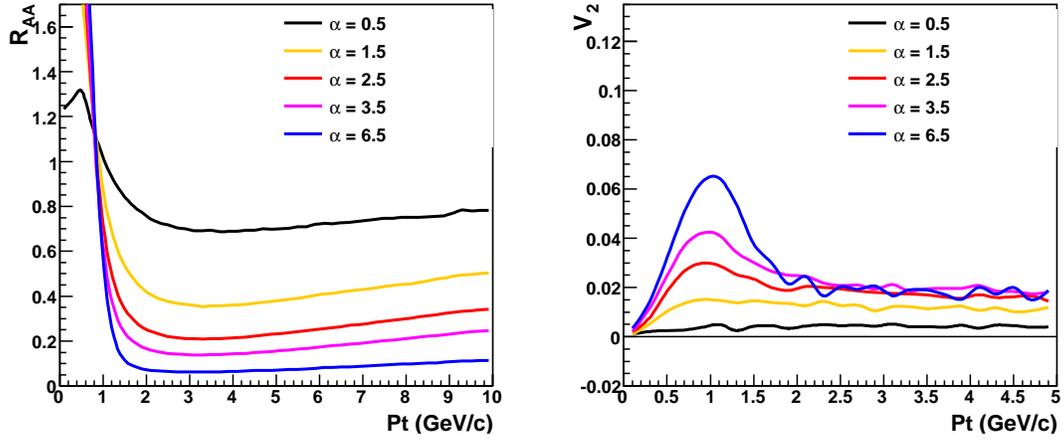}
    \caption{$R_{AA}(p_{\mathrm{T}})$ and $v_2(p_{\mathrm{T}})$ of charm quarks 
      calculated by the drag force defined in Eq.~\ref{eq:col}~(pQCD).
    }
    \label{fig:chap7_cw_v2raa}
  \end{center}
\end{figure}

\begin{figure}[htb]
  \begin{center}
    \includegraphics[width=15cm]{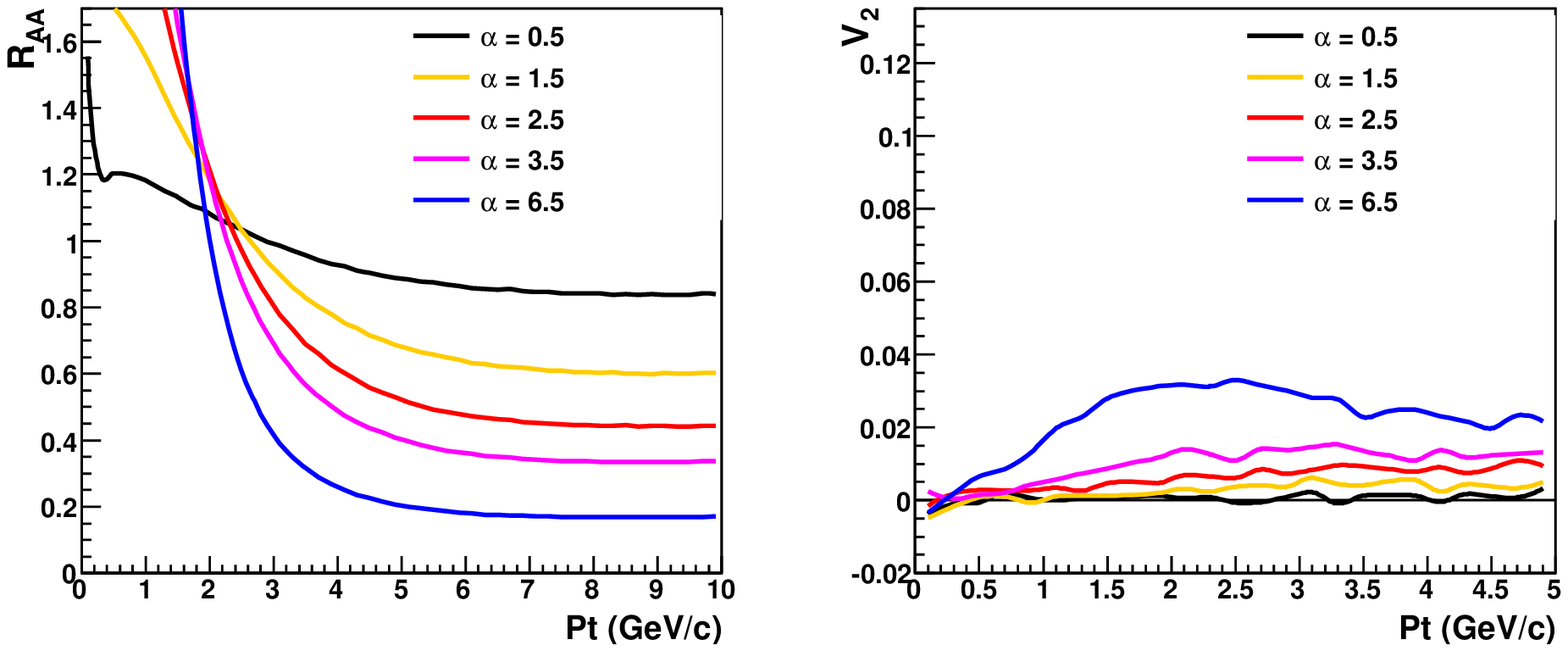}
    \caption{$R_{AA}(p_{\mathrm{T}})$ and $v_2(p_{\mathrm{T}})$ of bottom quarks 
      calculated by the drag force defined in Eq.~\ref{eq:col}~(pQCD).
    }
    \label{fig:chap7_bw_v2raa}
  \end{center}
\end{figure}
Figure~\ref{fig:chap7_cw_v2raa} and \ref{fig:chap7_bw_v2raa} are the same as 
Fig.~\ref{fig:chap7_c_v2raa} and \ref{fig:chap7_b_v2raa}, however the used drag force is 
defined in Eq.~\ref{eq:col}.
The tendency of $R_{AA}(p_{\mathrm{T}})$ and $v_2(p_{\mathrm{T}})$ with the drag force defined in
Eq.~\ref{eq:col} is the almost same as the result with the drag force in Eq.~\ref{eq:ads}.

\begin{figure}[htb]
  \begin{center}
    \includegraphics[width=15cm]{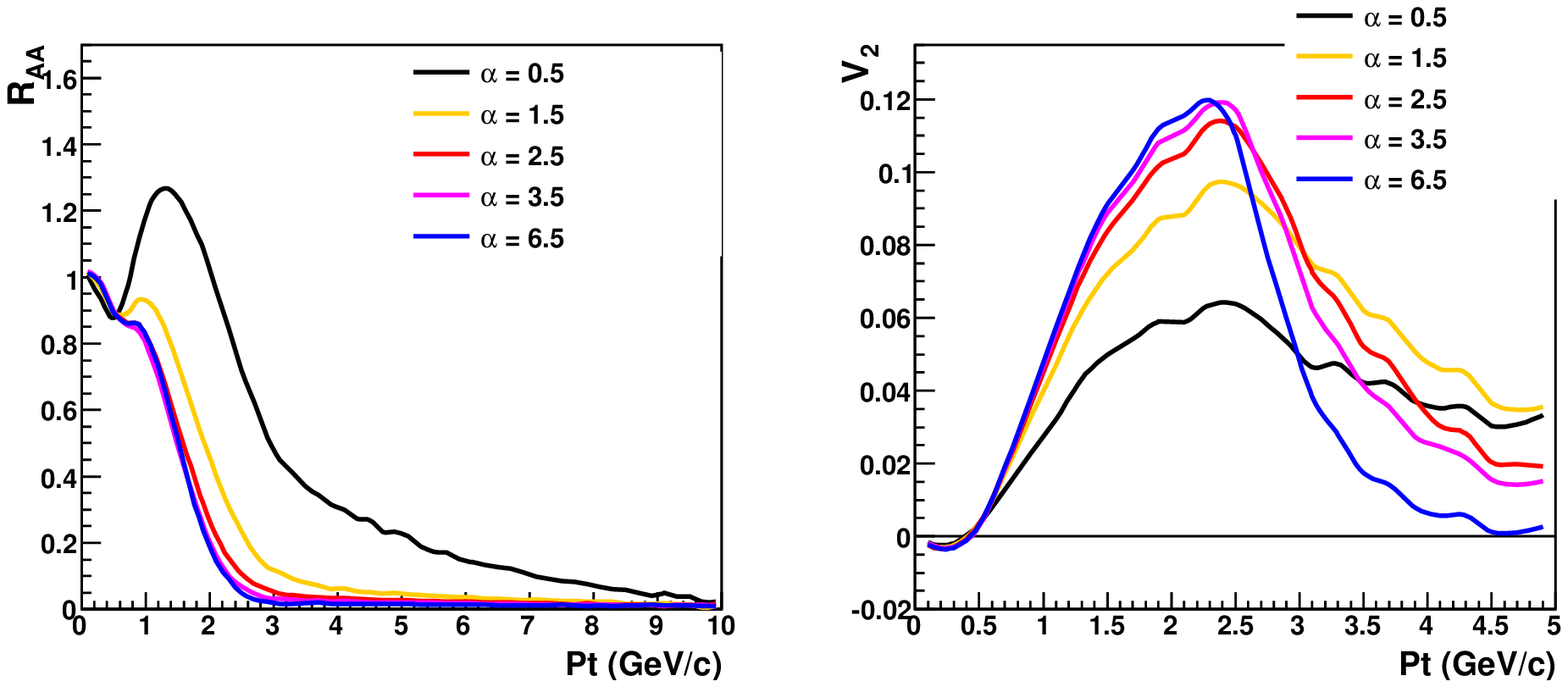}
    \caption{$R_{AA}(p_{\mathrm{T}})$ and $v_2(p_{\mathrm{T}})$ of electrons from charm quarks 
      calculated by the drag force defined in Eq.~\ref{eq:ads}~(AdS/CFT).
      Calculated conditions correspond 0-10\% centrality events for $R_{AA}(p_{\mathrm{T}})$
      and Minimum Bias events for $v_2(p_{\mathrm{T}})$.
    }
    \label{fig:chap7_ce_v2raa}
  \end{center}
\end{figure}

\begin{figure}[htb]
  \begin{center}
    \includegraphics[width=15cm]{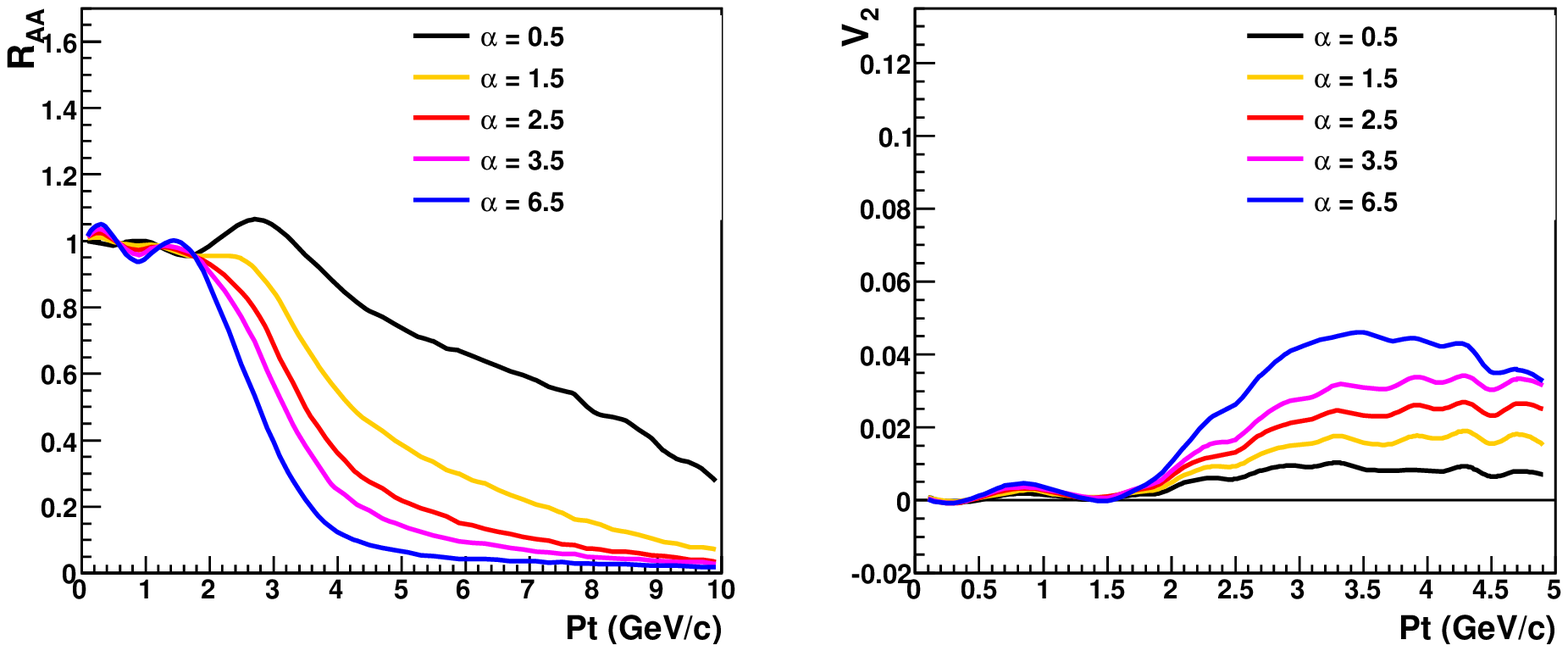}
    \caption{$R_{AA}(p_{\mathrm{T}})$ and $v_2(p_{\mathrm{T}})$ of electrons from bottom quarks 
      calculated by the drag force defined in Eq.~\ref{eq:ads}~(AdS/CFT).
      Calculated conditions correspond 0-10\% centrality events for $R_{AA}(p_{\mathrm{T}})$
      and Minimum Bias events for $v_2(p_{\mathrm{T}})$.
    }
    \label{fig:chap7_be_v2raa}
  \end{center}
\end{figure}

\begin{figure}[htb]
  \begin{center}
    \includegraphics[width=15cm]{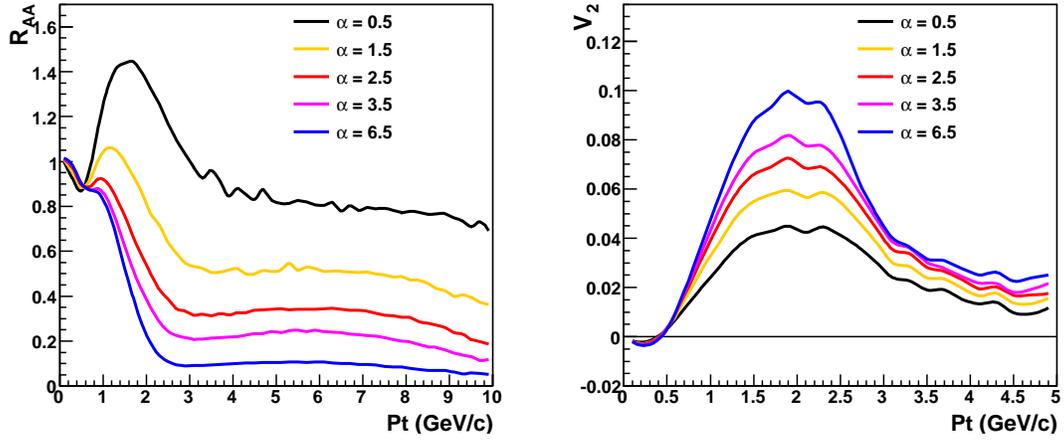}
    \caption{$R_{AA}(p_{\mathrm{T}})$ and $v_2(p_{\mathrm{T}})$ of electrons from charm quarks 
      calculated by the drag force defined in Eq.~\ref{eq:col}~(pQCD).
    }
    \label{fig:chap7_cew_v2raa}
  \end{center}
\end{figure}

\begin{figure}[htb]
  \begin{center}
    \includegraphics[width=15cm]{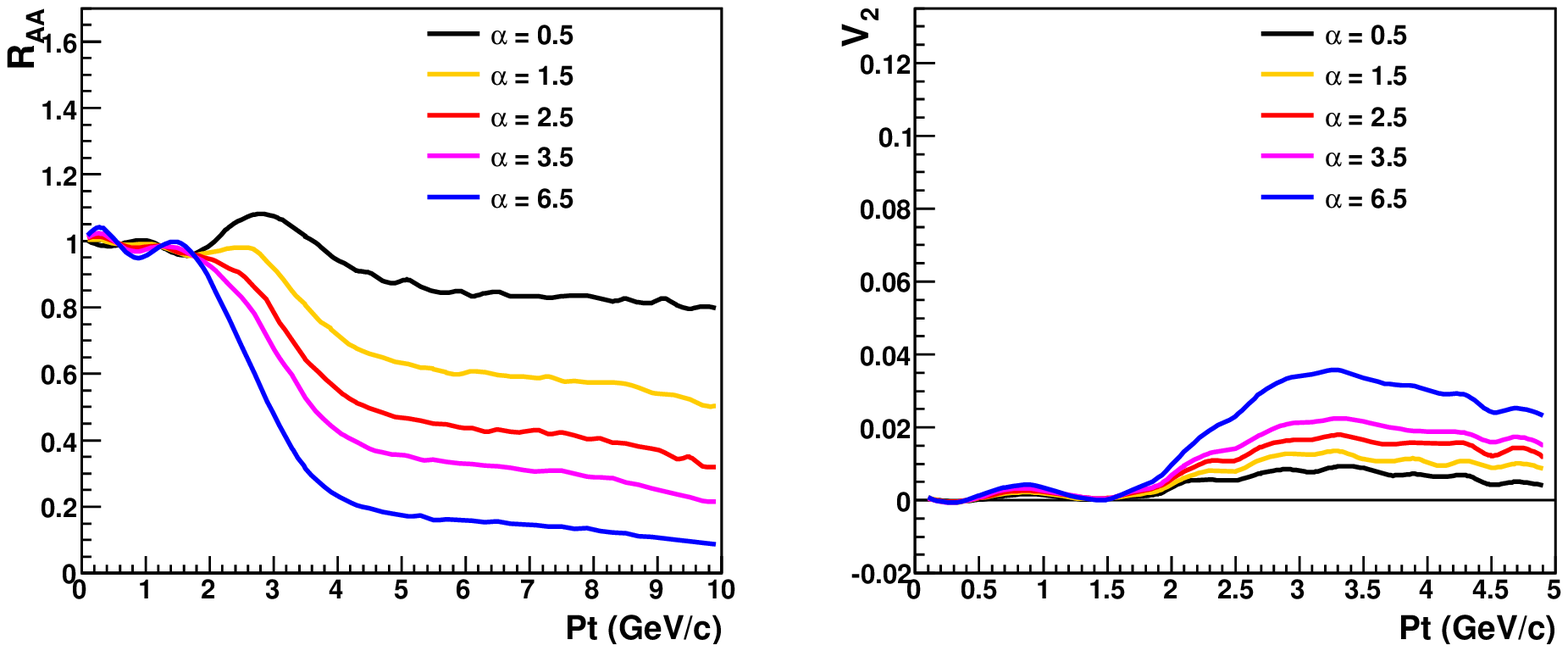}
    \caption{$R_{AA}(p_{\mathrm{T}})$ and $v_2(p_{\mathrm{T}})$ of electrons from bottom quarks 
      calculated by the drag force defined in Eq.~\ref{eq:col}~(pQCD).
    }
    \label{fig:chap7_bew_v2raa}
  \end{center}
\end{figure}

Figure~\ref{fig:chap7_ce_v2raa} and \ref{fig:chap7_be_v2raa} show $R_{AA}(p_{\mathrm{T}})$ and 
$v_2(p_{\mathrm{T}})$ of single electrons from charm and bottom quarks calculated by the drag force defined 
in Eq.~\ref{eq:ads}, respectively.
Figure~\ref{fig:chap7_cew_v2raa} and \ref{fig:chap7_bew_v2raa} are the same as 
Fig.~\ref{fig:chap7_ce_v2raa} and \ref{fig:chap7_be_v2raa}, however the used drag force is 
defined in Eq.~\ref{eq:col}.
$R_{AA}(p_{\mathrm{T}})$ of electrons below 3~GeV/$c$ is strongly enhanced than $R_{AA}(p_{\mathrm{T}})$ 
of quarks. This behavior is the effect of coalescence process which make 
$p_{\mathrm{T}}$ of hadrons larger by $0.5\sim 1$~GeV/$c$ than that of quarks. 
The magnitude of $v_2(p_{\mathrm{T}})$ is also enhanced than that of quarks and 
this is also from coalescence process.

These numerical results nicely demonstrate the magnitude of suppression and $v_2(p_{\mathrm{T}})$ of charm quarks
are much larger than these of bottom quarks.


\section{Comparison between Experimental Results and Model Calculation} \label{sec:aucomp}
In this section, free parameters in the model calculation are constrained 
within the obtained experimental uncertainties.
First, the ratio of absolute value of cross section of input bottom over that of charm is constrained
according to the experimental result, which is obtained in this thesis.
The fit result for $(b\rightarrow e)/(c\rightarrow e+ b\rightarrow e)$ is described in Sec.~\ref{sec:bcr}.
Then, we perform a simultaneous least-squares fit about $R_{AA}(p_{\mathrm{T}})$ and $v_2(p_{\mathrm{T}})$
to the model calculation~($R_{AA}(p_{\mathrm{T}},\gamma),v_2(p_{\mathrm{T}},\gamma)$).
Since the data points of $R_{AA}(p_{\mathrm{T}})$ in  $1.7<p_{\mathrm{T}}<3$~GeV/$c$ have a very small 
statistical uncertainty, the free parameter in the model which gives the minimized $\chi^2$~($\chi^2_{min}$) 
is almost determined by the shape of $R_{AA}(p_{\mathrm{T}})$ in  $1.7<p_{\mathrm{T}}<3$~GeV/$c$.
However, the effect of coalescence process is dominant in $R_{AA}(p_{\mathrm{T}})$ in  $1.7<p_{\mathrm{T}}<3$~GeV/$c$,
which is not from the medium property and what we want to know.
Therefore, the data points above $p_{\mathrm{T}}>3$~GeV/$c$ are used for the $R_{AA}(p_{\mathrm{T}})$ fit to
extract the medium property.

$R_{AA}(p_{\mathrm{T}})$ and $v_2(p_{\mathrm{T}})$ of single non-photonic 
electrons~($R_{AA}^{mix}(p_{\mathrm{T}})$ and $v_2^{mix}(p_{\mathrm{T}})$) to compare with the experimental results 
are determined according to the constrained ratios, $(b\rightarrow e)/(c\rightarrow e+ b\rightarrow e)$.

To take into the account the uncertainty of the absolute value of charm and bottom cross section, we prepare three kinds of 
sets of ($R_{AA}^{mix}(p_{\mathrm{T}})$ and $v_2^{mix}(p_{\mathrm{T}})$), which correspond to the results with three kinds 
of values of $(b\rightarrow e)/(c\rightarrow e+ b\rightarrow e)$ that are best fit value and $\pm 1\sigma$ as shown
in Fig.~\ref{fig:chap7_bcr}.
The least square fit using Eq.~\ref{eq:chi} are performed for these 3 $R_{AA}^{mix}(p_{\mathrm{T}})$ 
and $v_2^{mix}(p_{\mathrm{T}})$ set.

\begin{figure}[htb]
  \begin{center}
    \includegraphics[width=15cm]{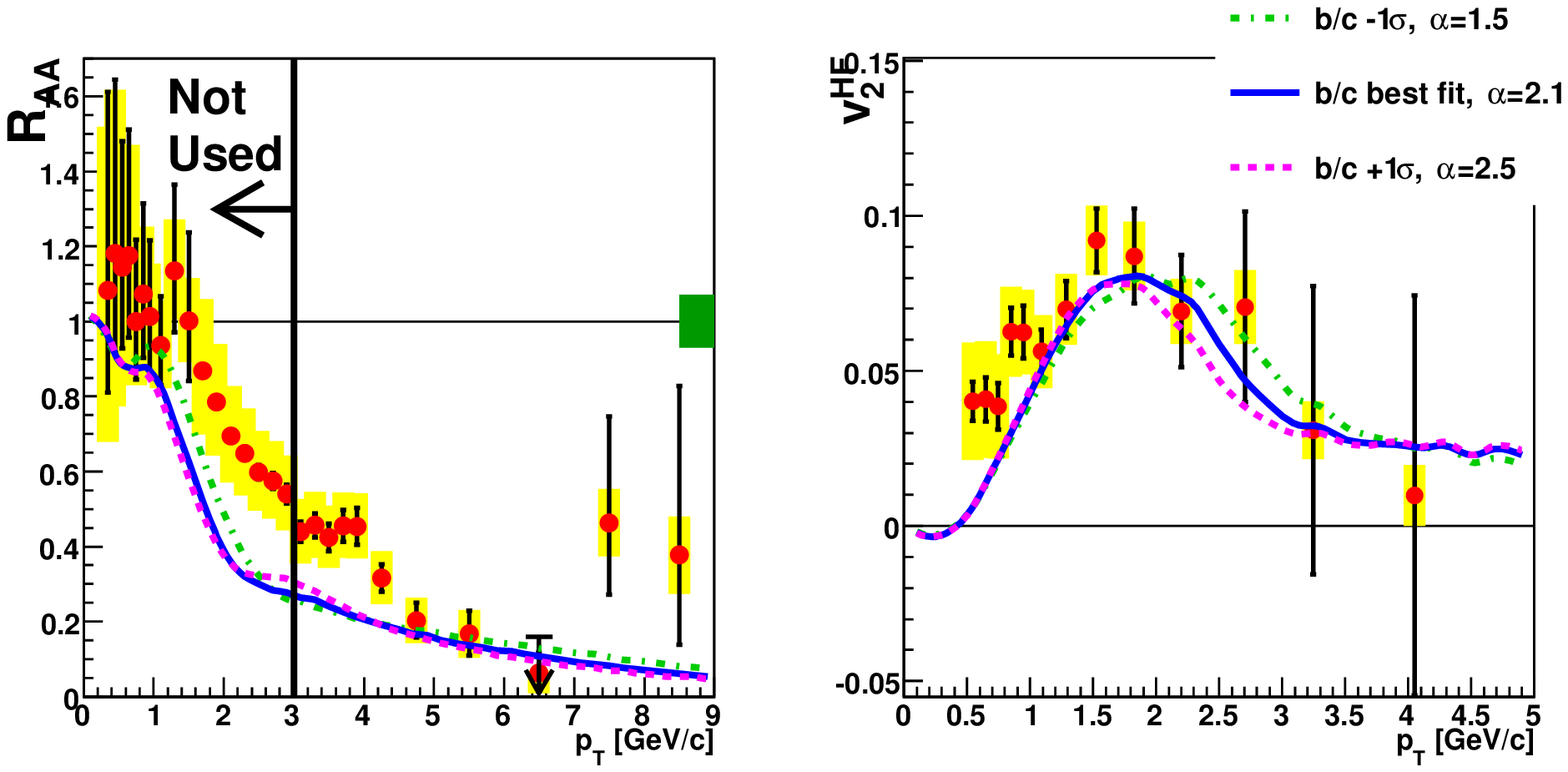}
    \caption{$R_{AA}(p_{\mathrm{T}})$ and $v_2(p_{\mathrm{T}})$
      of single electrons which are the fit results for $R_{AA}(p_{\mathrm{T}})$ and $v_2(p_{\mathrm{T}})$ at the 3 ratios 
      of $(b\rightarrow e)/(c\rightarrow e+ b\rightarrow e)$, best fit and $\pm 1\sigma$.
      The drag force is defined in Eq.~\ref{eq:ads}~(AdS/CFT).
      Data points of $R_{AA}(p_{\mathrm{T}})$ with $p_{\mathrm{T}}<3$~GeV/$c$ are not used for the fit.
    }
    \label{fig:chap7_best_ads}
  \end{center}
\end{figure}

Figure~\ref{fig:chap7_best_ads} shows $R_{AA}(p_{\mathrm{T}})$ and $v_2(p_{\mathrm{T}})$
of the decay electrons with the drag force defined in Eq.~\ref{eq:ads}~(AdS/CFT) 
which are the results of the fit at the 3 ratios of $(b\rightarrow e)/(c\rightarrow e+ b\rightarrow e)$, 
best fit and $\pm 1\sigma$.
The blue solid line shows the result of $R_{AA}(p_{\mathrm{T}})$ and $v_2(p_{\mathrm{T}})$ at the best fit ratio 
of $(b\rightarrow e)/(c\rightarrow e+ b\rightarrow e)$.
The green dashed line and magenta dotted line show the result at the $\pm 1\sigma$ ratios
of $(b\rightarrow e)/(c\rightarrow e+ b\rightarrow e)$.
The fit results  are summarized in Table~\ref{chap7_fit1}.
As a result, the experimental results are successfully reproduced with $\gamma =2.1_{-0.6}^{+0.4} \frac{T^2}{M}$ 
including the uncertainty of the ratio of bottom over charm.
\begin{table}[hbt]
    \begin{center}
     \caption{Fit result using the drag force defined in Eq.~\ref{eq:ads}~(AdS/CFT).}
     \label{chap7_fit1}
     \begin{tabular}{c|cc|ccc}
       $(b\rightarrow e)/(c\rightarrow e+ b\rightarrow e)$ & $\alpha$ & $\chi^2$/NDF & $\epsilon_{b1}$ & $\epsilon_{b2}$ & $\epsilon_{c}$\\ 
       \hline \hline
       best fit   & 2.1 & 32.2/23& -2.0 &-1.4&-0.7\\
       -1$\sigma$ & 1.5 & 37.4/23& -2.2 &-1.5&-0.6\\
       +1$\sigma$ & 2.5 & 28.8/23& -1.8 &-1.3&-0.6\\
     \end{tabular}
    \end{center}
\end{table}

\begin{figure}[htb]
  \begin{center}
    \includegraphics[width=15cm]{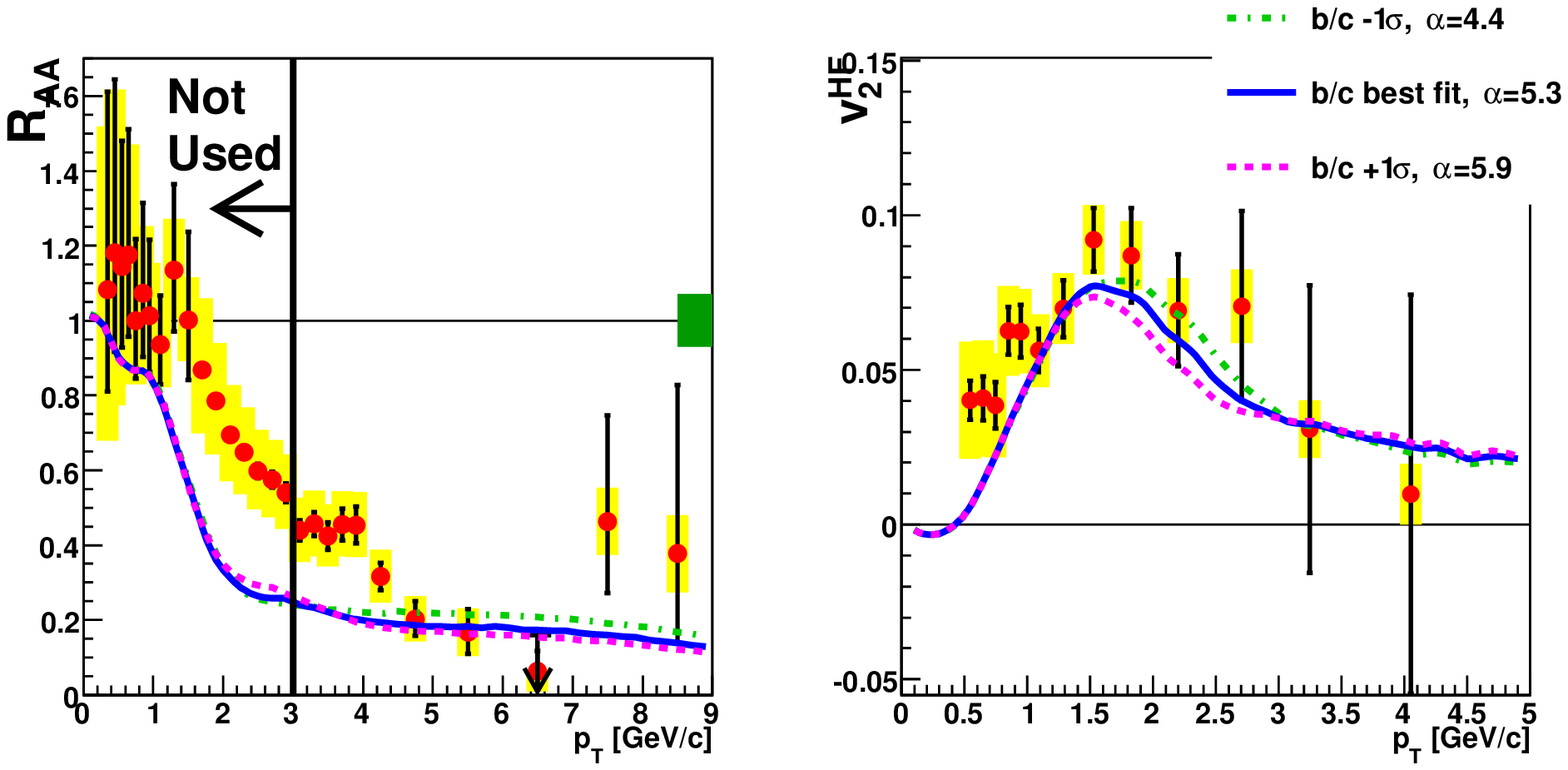}
    \caption{$R_{AA}(p_{\mathrm{T}})$ and $v_2(p_{\mathrm{T}})$
      of single electrons which are the fit results for $R_{AA}(p_{\mathrm{T}})$ and $v_2(p_{\mathrm{T}})$ at the 3 ratios 
      of $(b\rightarrow e)/(c\rightarrow e+ b\rightarrow e)$, best fit and $\pm 1\sigma$.
      The drag force is defined in Eq.~\ref{eq:ads}~(pQCD).
      Data points of $R_{AA}(p_{\mathrm{T}})$ with $p_{\mathrm{T}}<3$~GeV/$c$ are not used for the fit.
    }
    \label{fig:chap7_best_col}
  \end{center}
\end{figure}

Figure~\ref{fig:chap7_best_col} also shows the calculated $R_{AA}(p_{\mathrm{T}})$ and $v_2(p_{\mathrm{T}})$
with drag force is defined in Eq.~\ref{eq:col}~(pQCD).
In Fig.~\ref{fig:chap7_best_col}, the meaning of the each line is the same in Fig.~\ref{fig:chap7_best_ads}.
Table~\ref{chap7_fit2} is the same as Table~\ref{chap7_fit1}, however the used drag force is defined at
Eq.~\ref{eq:col}~(pQCD).
As a result, the experimental results are successfully reproduced with $\gamma =5.3_{-0.9}^{+0.6} \frac{T^2}{E}$.
About factor 2 large $\alpha$ is necessary with the drag force defined in Eq.~\ref{eq:col} to 
reproduce the experimental results, compared with the drag force defined in Eq.~\ref{eq:ads}.

In Fig.~\ref{fig:chap7_best_ads} and Fig.~\ref{fig:chap7_best_col}, there is a deviation of $R_{AA}(p_{\mathrm{T}})$
below 3~GeV/$c$ between the model calculation and the experimental result.
More accurate treatment of coalescence process, fragmentation process and initial nuclear effect are needed
for the description of heavy flavor in the medium below 3~GeV/$c$.
In addition, the more realistic parametric form of the drag force will be necessary for the complete reproduction
of the experimental results.

\begin{table}[hbt]
    \begin{center}
     \caption{Fit results using the drag force defined in Eq.~\ref{eq:col}~(pQCD).}
     \label{chap7_fit2}
     \begin{tabular}{c|cc|ccc}
       $(b\rightarrow e)/(c\rightarrow e+ b\rightarrow e)$ & $\alpha$ & $\chi^2$/NDF & $\epsilon_{b1}$ & $\epsilon_{b2}$ & $\epsilon_{c}$\\ 
       \hline \hline
       best fit   & 5.3 & 45.8/23& -2.1 &-1.3&-0.2 \\
       -1$\sigma$ & 4.4 & 54.7/23& -2.2 &-1.3&-0.6 \\
       +1$\sigma$ & 5.9 & 40.3/23& -2.2 &-1.4&-0.6 \\
     \end{tabular}
    \end{center}
\end{table}

\section{Result of Model Calculation} \label{sec:adsresult}
Based on the Langevin calculations, the drag force is constrained within the experimental uncertainties as follows.
\begin{eqnarray}
  \gamma &=&  2.1_{-0.6}^{+0.4}\frac{T^2}{M} \label{eq:ads_r}, \\
  \gamma &=& 5.3_{-0.9}^{+0.6} \frac{T^2}{E} \label{eq:col_r}.
\end{eqnarray}
The result of Eq.~\ref{eq:ads_r} is compatible with the result, $\gamma \sim 2\frac{T^2}{M}$, from AdS/CFT 
correspondence~\cite{bib:akamatu,bib:adsdrag1,bib:adsdrag2}.
This fact may suggest the medium is deconfined and strongly coupled matter, since the AdS/CFT correspondence
is only valid in such condition~\cite{bib:adsdrag3}.

\begin{figure}[htb]
  \begin{center}
    \includegraphics[width=12cm]{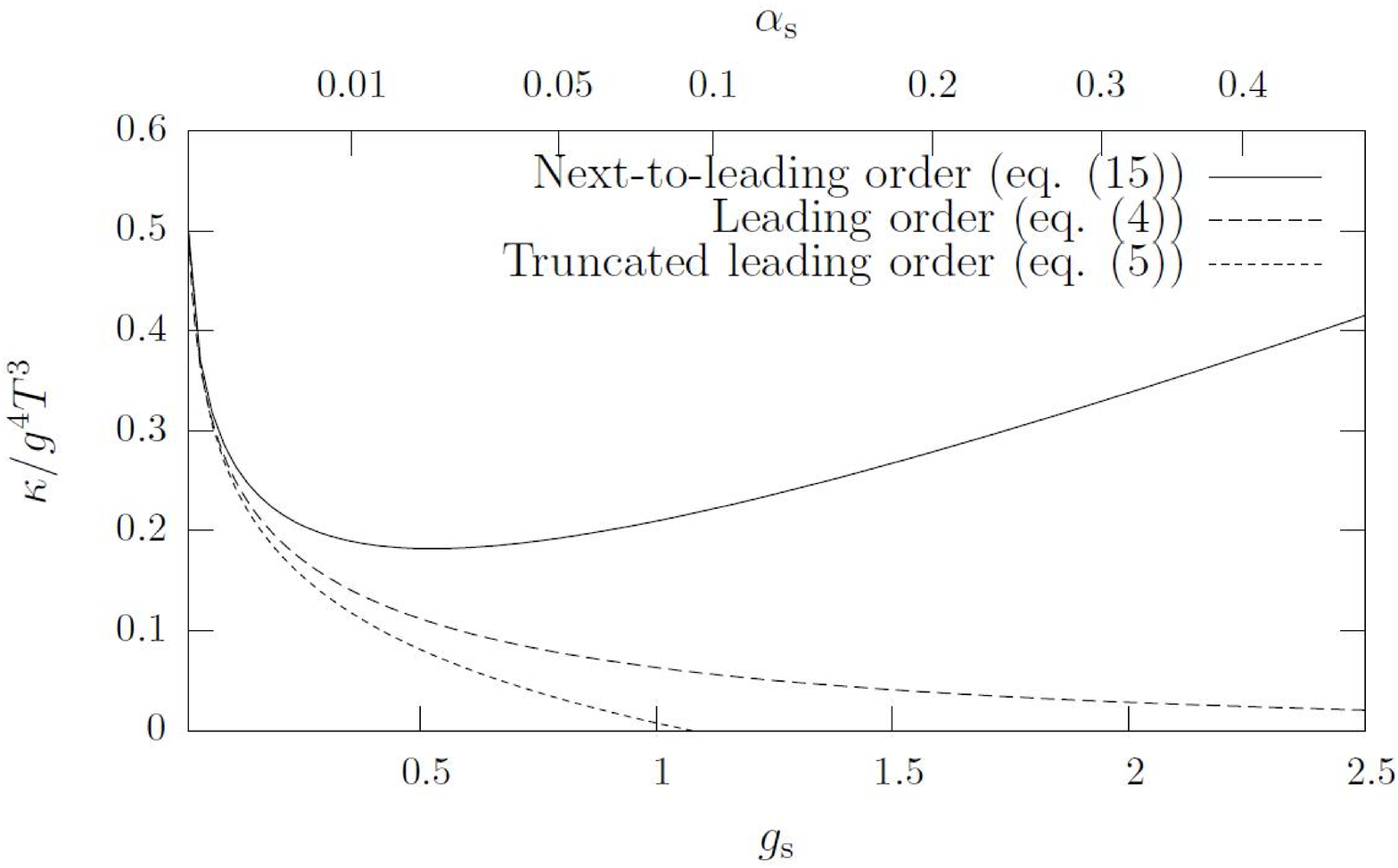}
    \caption{The diffusion constant in momentum space as a function of $g_s$ a calculated by 
      3-color, 3-flavor NLO pQCD with leading order calculation~\cite{bib:colnlo}
    }
    \label{fig:chap7_dragnlo}
  \end{center}
\end{figure}

Comparison between the result of Eq.~\ref{eq:col_r} with pQCD calculation is interesting 
because Eq.~\ref{eq:col_r} is motivated by the perturbative QCD calculation of collisional process~(weak coupling region).
The drag force calculated by 3-color, 3-flavor NLO pQCD is given as follows~\cite{bib:colnlo}
\begin{equation}
\gamma = \frac{g_{s}^{4}}{3\pi}\frac{T^2}{M}\left( \ln\frac{1}{g_s}+0.07428+1.9026g_s+ O(g_s^2)\right) ,
\label{eq:nlo_d}
\end{equation}
where $g_s$ is the coupling constant in QCD.
Figure~\ref{fig:chap7_dragnlo} shows the diffusion constant in momentum space as a function of $g_s$ where the results are
in the leading order calculation and the next-leading order calculation~\cite{bib:colnlo}.
The result indicates the convergence of the pQCD calculation is poor above $g_s \sim 0.1$.
$g_s \sim 2$ is required from the comparison between Eq.~\ref{eq:col_r} and Eq.~\ref{eq:nlo_d}.
Eq.~\ref{eq:nlo_d} becomes clearly unreliable for $g_s \sim 2$.
This fact means that higher order pQCD or the non-perturbative treatment are necessary to describe the experimental 
result.
                                                        
The obtained drag force in Eq.~\ref{eq:ads_r} is converted to thermalization time~($\tau_{HQ}$) and 
the spatial diffusion constant~($D_{s}$) of heavy quarks using Eq.~\ref{eq:tau} and Eq.~\ref{eq:sdif}.

\begin{eqnarray}
D_{s} &=& \frac{1}{2.1_{-0.6}^{+0.4}T} \label{eq:sdif_res}\\
\tau_{HQ} &=& \frac{M}{2.1_{-0.6}^{+0.4} T^2} \label{eq:tau_hq}
\end{eqnarray}
With typical value $T=200$~MeV, $M_{charm}=1.25$~GeV and $M_{bottom}$=4.3~GeV,
$\tau_{charm}$ and $\tau_{bottom}$ are $2.9_{-0.5}^{+1.2}$~fm and $10.1_{-1.6}^{+4.0}$~fm, respectively.
$\tau_{charm}$ is shorter than the typical lifetime of QGP, $\tau_{QGP}\sim 6$~fm. 
Therefore, the experimental results indicate charm quarks are thermalized in the QGP.
In addition, the obtained $\tau_{charm}$ is much smaller than LO pQCD calculation(~$\sim35$fm)~\cite{bib:hees}.
The non-perturbative treatment is necessary to reproduce such $\tau_{charm}$ from the microscopic interaction.\\
   \chapter{Shear Viscosity} \label{sec:vis}
The ratio of shear viscosity to entropy density, which characterize the hydrodynamical property of the medium,
is related to the spatial diffusion constant~$D_{s}$.
A kinetic theory for an ultra-relativistic gas provides the relation of $\eta/s$~\cite{bib:rapp_review,bib:kine_es}.
\begin{equation}\label{eq:es_rao}
\frac{\eta}{s} \sim \frac{1}{5} T D_{s}.
\end{equation}
Leading-order pQCD calculation of elastic scattering also provides the relation as 
follows~\cite{bib:moore,bib:pqcd_es}.
\begin{equation}\label{eq:es_pqcd}
\frac{\eta}{s} \sim \frac{1}{6} T D_{s}
\end{equation}
Eq.~\ref{eq:es_rao} and Eq.~\ref{eq:es_pqcd} are compatible and these equations are believed to be valid 
in the weak-coupling limit.
It is not clear whether Eq.~\ref{eq:es_rao} and Eq.~\ref{eq:es_pqcd} are plausible in the strongly coupled matter,
while in Ref.~\cite{bib:moore}, Eq.~\ref{eq:es_pqcd} are expected to be largely independent of its perturbative 
assumption due to the cancellation  of the higher order correction.
Using Eq.~\ref{eq:es_rao}, Eq.~\ref{eq:es_pqcd} and  Eq.~\ref{eq:sdif_res}, 
the ratio of shear viscosity to entropy density, $\eta/s$ is given as bellow.
\begin{equation} \label{eq:es_weak}
\frac{\eta}{s} \sim \frac{0.8-1.7}{4\pi}.
\end{equation}
This result is very close to  the conjectured quantum lower bound $1/4\pi$.
Such $\eta/s$ indicates the created medium at RHIC is near a prefect fluid and strongly coupled matter.

On the other hand, the relation between $\eta/s$ and $D_{s}$ in the strong coupling limit is estimated 
using the AdS/CFT correspondence and the assumed proportionality $\eta/s \propto D_{s}$ from the kinetic 
theory in Ref~\cite{bib:rapp_review}. 
\begin{equation} \label{eq:es_sqgp}
\frac{\eta}{s} \sim \frac{1}{2} T D_{s}
\end{equation}
The significantly smaller coefficient in  Eq.~\ref{eq:es_rao} and Eq.~\ref{eq:es_pqcd} compared to that in 
Eq.~\ref{eq:es_sqgp} could be understood to reflect the expected underestimation of the shear viscosity 
if a gas estimate is applied in a liquid-like regime~\cite{bib:kine_es}.
Using Eq.~\ref{eq:es_sqgp}, $\eta/s$ becomes as bellow.
\begin{equation} \label{eq:es_strong}
\frac{\eta}{s} \sim \frac{2.5-4.2}{4\pi}
\end{equation}
This result is also close to the conjectured quantum lower bound $1/4\pi$.

Finally, by combining Eq.~\ref{eq:es_weak} and Eq.~\ref{eq:es_strong}, we obtain,
\begin{equation} \label{eq:es_result}
\frac{\eta}{s} \sim \frac{0.8-4.2}{4\pi}
\end{equation}
When we find more realistic relation between $\eta/s$ and $D_{s}$, 
more precise $\eta/s$ will be obtained.

\begin{figure}[htb]
  \begin{center}
    \includegraphics[width=12cm]{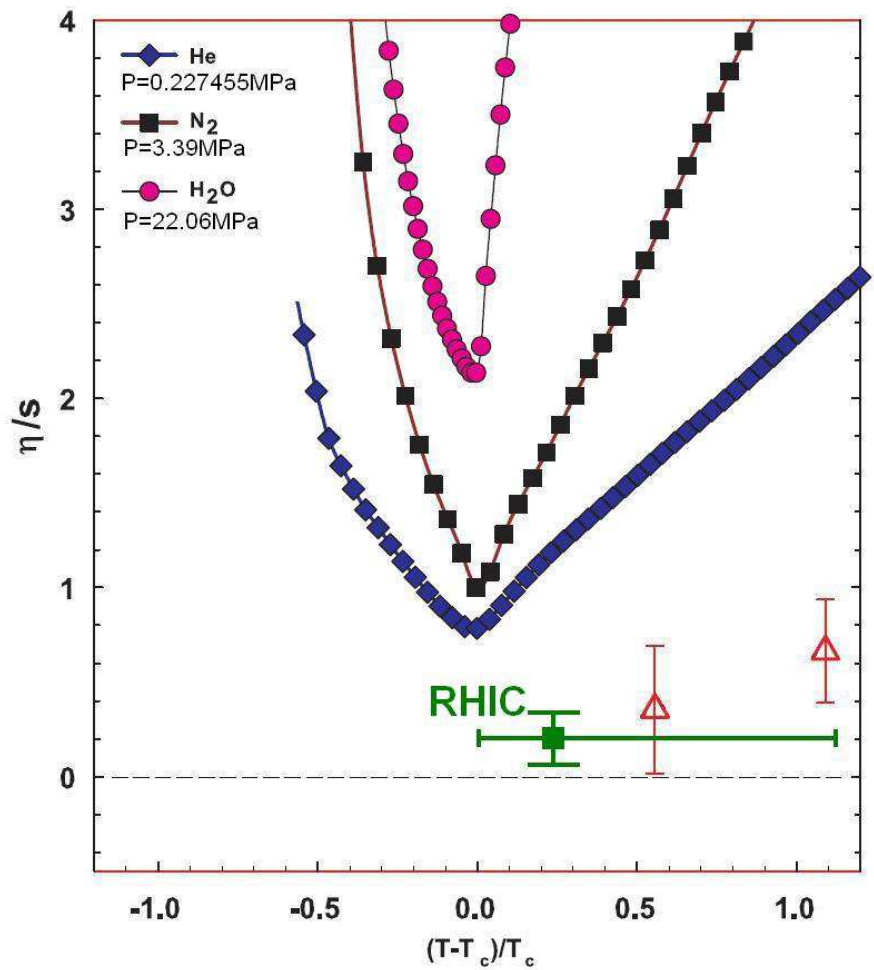}
    \caption{The compilation of the ratio of shear viscosity to entropy density 
      for various substances~\cite{bib:es_lat,bib:es_mol}.
    }
    \label{fig:chap7_es}
  \end{center}
\end{figure}

Figure~\ref{fig:chap7_es} shows the compilation of $\eta/s$ density 
for various substances~\cite{bib:lacey}. Green line shows the result in Eq.~\ref{eq:es_result}.
$T_c=170$~MeV and $T=210$~MeV are assumed according to the used hydrodynamical calculations for the 
(T-T$_c$)/T$_c$. $T=210$~MeV is the mean temperature of the matter in which heavy quarks moved.
Since the temperature of the matter depends on the time, the error of (T-T$_c$)/T$_c$ is determined 
from the highest  temperature  and T$_c$.
Upper 3 symbols show $\eta/s$ of atomic He, molecular N$_2$ and H$_2$O~\cite{bib:es_mol}
as a function of temperature with the fixed pressures.
The fixed pressures are their critical pressures, that is, 0.227455~MPa for He, 3.39~MPa for   N$_2$
and 22.06~MPa for H$_2$O~\cite{bib:es_mol}.
Red triangles shows
the case for pure-glue lattice QCD calculations~\cite{bib:es_lat}.
$\eta/s$ of the matter created at RHIC is remarkably small compared to other matters and the very small  $\eta/s$ is 
one of most striking features of the matter.

The result in Eq.~\ref{eq:es_result} is compatible with the lattice QCD calculation.
The lattice QCD calculation predicts $\eta/s$ becomes large with increasing temperature, which is 
the expected behavior from the discussion in Sec.~\ref{sec:tmat}.
The temperature dependence of $\eta/s$ may be revealed from the experiments at LHC and the interaction of 
heavy quarks in the medium will be studied more precisely.

    \addcontentsline{toc}{chapter}{Bibliography}
\end{document}